\documentclass[11pt]{article}
\usepackage{amssymb}
\usepackage{amsmath}
\usepackage{amstext}
\usepackage{graphicx,epsfig}
\usepackage{epsfig}
\usepackage{verbatim} 
\usepackage{fancybox}
\usepackage{color}
\usepackage{ulem}
\usepackage{enumitem}
\usepackage{subfigure}
\usepackage{bbm}
\usepackage{parskip}
\usepackage{dsfont}
\usepackage[numbers,sort&compress]{natbib}

\newcommand{\Comment}[1]{{}}
\definecolor{darkblue}{rgb}{0.15,0.35,0.55}
\definecolor{reddish}{rgb}{0.65, 0.2, 0.2}
\definecolor{darkgreen}{RGB}{50,150,0}
\usepackage[linktocpage=true]{hyperref}
\hypersetup{
colorlinks=true,
citecolor=darkblue,
linkcolor=reddish,
urlcolor=darkblue,
pdfauthor={},
pdftitle={},
pdfsubject={}
}

\newcommand{\be}{\begin{equation}}
\newcommand{\ee}{\end{equation}}
\newcommand{\bea}{\begin{eqnarray}}
\newcommand{\eea}{\end{eqnarray}}
\newcommand{\beas}{\begin{eqnarray*}}
\newcommand{\eeas}{\end{eqnarray*}}

\def\({\left(}
\def\){\right)}

\newcommand{\rd}{{\rm d}}
\newcommand{\vp}{\varphi}

\def\gsim{ \lower .75ex \hbox{$\sim$} \llap{\raise .27ex \hbox{$>$}} }
\def\lsim{ \lower .75ex \hbox{$\sim$} \llap{\raise .27ex \hbox{$<$}} }

\newcommand{\bmat}{\xi}  

\newcommand{\Mpl}{M_\mathrm{Pl}} 
\newcommand{\rholab}{\rho_\mathrm{lab}}

\title{}
\author{}

\numberwithin{equation}{section}

\begin{document}

\renewcommand{\thefootnote}{\fnsymbol{footnote}}
\begin{center}
{\Huge \bf{Beyond the Cosmological\\ Standard Model}}
\end{center} 

\vspace{1.2truecm}
\thispagestyle{empty}
\centerline{{\Large Austin Joyce,${}^{\rm a,b,}$\footnote{\tt ajoy@uchicago.edu} Bhuvnesh Jain,${}^{\rm b,}$\footnote{\tt bjain@physics.upenn.edu} Justin Khoury${}^{\rm b,}$\footnote{\tt jkhoury@sas.upenn.edu} and Mark Trodden${}^{\rm b,}$}\footnote{\tt trodden@physics.upenn.edu}}
\vspace{.5cm}

\centerline{\it${}^{\rm a}$Enrico Fermi Institute and Kavli Institute for Cosmological Physics}
\centerline{\it University of Chicago, Chicago, IL 60637}

\vspace{.5cm}

\centerline{\it${}^{\rm b}$Center for Particle Cosmology, Department of Physics and Astronomy}
\centerline{\it University of Pennsylvania, Philadelphia, PA 19104}

\vspace{-.2cm}
\begin{abstract}
\vspace{-.6cm}
\noindent
After a decade and a half of research motivated by the accelerating universe, theory and experiment have a reached a certain level of maturity. The development of theoretical models beyond $\Lambda$ or smooth dark energy, often called modified gravity, has led to broader insights into a path forward, and a host of observational and experimental tests have been developed. In this review we present the current state of the field and describe a framework for anticipating developments in the next decade. We identify the guiding principles for rigorous and consistent modifications of the standard model, and discuss the prospects for empirical tests.

We begin by reviewing recent attempts to consistently modify Einstein gravity in the infrared, focusing on the notion that additional degrees of freedom introduced by the modification must ``screen'' themselves from local tests of gravity. We categorize screening mechanisms into three broad classes: mechanisms which become active in regions of high Newtonian potential, those in which first derivatives of the field become important, and those for which second derivatives of the field are important. Examples of the first class, such as $f(R)$ gravity, employ the familiar chameleon or symmetron mechanisms,  whereas examples of the last class are galileon and massive gravity theories, employing the Vainshtein mechanism.
In each case, we describe the theories as effective theories and discuss prospects for completion in a more fundamental theory. 
We describe experimental tests of each class of theories, summarizing laboratory and solar system tests and describing in some detail astrophysical and cosmological tests. Finally, we discuss prospects for future tests which will be sensitive to different signatures of new physics in the gravitational sector.  

The review is structured so that those parts that are more relevant to theorists {\it vs}. observers/experimentalists are clearly indicated, in the hope that this will serve as a useful reference for both audiences, as well as helping those interested in bridging the gap between them.
\end{abstract}

\newpage

\tableofcontents
\newpage
\renewcommand*{\thefootnote}{\arabic{footnote}}
\setcounter{footnote}{0}

\part{Introduction}
\parskip=1.5pt
\normalsize

\section{The cosmological constant and its discontents}
\label{introsection}
There is overwhelming observational evidence that the universe is undergoing accelerated expansion from observations of type Ia supernovae~\cite{Riess:1998cb,Schmidt:1998ys, Perlmutter:1998np, Garnavich:1998th,Freedman:2000cf,Riess:2004nr,Astier:2005qq,Kowalski:2008ez,Kessler:2009ys,Amanullah:2010vv,Suzuki:2011hu,Sako:2014qmj}, from Cosmic Microwave Background measurements~\cite{deBernardis:2000gy,Lange:2000iq,Balbi:2000tg,Pryke:2001yz,Spergel:2003cb,Hinshaw:2012aka, Hou:2012xq,Ade:2013zuv,2013ApJ...779...86S,Sievers:2013ica}, and from detailed studies of large-scale structure~\cite{Tegmark:2003ud,Seljak:2004xh,Eisenstein:2005su,Blake:2011en,Beutler:2011hx,Dawson:2012va,2012MNRAS.427.3435A,Samushia:2012iq}. All of the measurements are in good agreement, with all data consistent with a $\Lambda$-cold dark matter ($\Lambda$CDM) cosmology~\cite{Ostriker:1995rn, Bahcall:1999xn}, with a value of the cosmological constant of about $\Lambda_{\rm obs.} \sim (10^{-3} {\rm eV})^4$.

From a theoretical viewpoint, this concordance cosmology is somewhat troubling. Expressed in Planck units---which seem to be the most natural units to use when talking about gravity---this observed cosmological constant is absurdly small
\be
\Lambda_{\rm obs.}  \sim (10^{-30}M_{\rm Pl})^4~.
\ee
This is the	 {\it cosmological constant problem}. On its own this might not seem like much of a problem; small numbers appear in various places in physics all the time. For example, the electron mass, $m_e \sim 10^{-7}~ {\rm TeV}$, is quite small when measured in the units natural to the Standard Model, but this does not worry us greatly. The reason that we are comfortable with such a small parameter is that it is stable under quantum corrections---in field theory language, the small parameter is {\it technically natural}.\footnote{We mean this in the 't Hooft sense~\cite{'tHooft:1979bh}: the theory enjoys an enhanced symmetry in the limit where the electron mass goes to zero (chiral symmetry), which tells us that quantum corrections to the electron mass must be proportional to the mass itself: $\delta m_e \propto m_e$. So, if we set the electron mass to be small, it {\it stays} small. No such symmetry is known for the cosmological constant.}
What makes the cosmological constant problem particularly worrisome is that we expect a large contribution to the cosmological constant from particles that we know to exist in the Standard Model (SM). Indeed, we expect a contribution\footnote{The form of this contribution may be deduced by noting that on flat space, Lorentz invariance forces $\langle T_{\mu\nu}\rangle \propto \eta_{\mu\nu}$. Then, we invoke the equivalence principle to promote $\eta_{\mu\nu}\mapsto g_{\mu\nu}$.} to the cosmological constant of the form~\cite{Weinberg:1988cp}
\be
\langle T_{\mu\nu}\rangle \sim -\langle\rho\rangle g_{\mu\nu}~,
\ee
from quantum-mechanical processes involving SM fields. We can heuristically estimate its size by modeling SM fields as a collection of independent harmonic oscillators at each point in space and then summing over their zero-point energies\footnote{The alert reader might complain that we have chosen to keep the leading divergence of the integral, which is sensitive to ultraviolet physics. More conservatively we could focus on the logarithmically divergent piece, which is universal and makes the most optimistic assumptions about ultraviolet (UV) physics, which would lead to $\langle\rho\rangle \sim -m^4\log\left(\Lambda_{\rm UV}/m\right)$. However, this does not appreciably change the results, plugging in the top quark mass $m_t \sim 0.1$ TeV and $\Lambda_{\rm UV} \sim 1$ TeV, we obtain $\Lambda_{\rm theory} \sim (0.1~{\rm TeV})^4 $.
}
\be
\langle\rho\rangle \sim \int_0^{\Lambda_{\rm UV}}\frac{\rd^3 k}{(2\pi)^3}\frac{1}{2}\hbar E_k \sim \int_0^{\Lambda_{\rm UV}}\rd k~k^2\sqrt{k^2+m^2} \sim \Lambda^4_{\rm UV}~.
\ee
Here $\Lambda_{\rm UV}$ is the cutoff of our theory---the energy scale up to which we can trust predictions. Most conservatively, the Standard Model has been extremely well tested up to energies around the weak scale, $\Lambda_{\rm UV} \sim 1~{\rm TeV}$. Plugging in this value, we find a theoretical expectation for the cosmological constant to be around
\be
\Lambda_{\rm theory} \sim ({\rm TeV})^4 \sim ~10^{-60}~M_{\rm Pl}^4~,
\ee
while the observed value is
\be
\Lambda_{\rm obs.} \sim M_{\rm Pl}^2H_0^2 \sim 10^{-60}({\rm TeV})^4 \sim 10^{-120}~M_{\rm Pl}^4~.
\ee
This discrepancy of 60 orders of magnitude is somewhat disconcerting. Of course it could be that the bare value of the CC (the number we put in the Lagrangian) is such that it precisely cancels $\Lambda_{\rm theory}$ to $\sim60$ decimal places, but such a scenario appears incredibly {\it fine-tuned}.\footnote{To some degree, this is an aesthetic judgment; ideally we would want the observed value of the CC to be ``generic," but failing that, we would settle for technical naturalness.} 

\begin{table}[float]
\centerline{
\small
\begin{tabular}{| l | c | c | c | c |}
	\hline
	{\bf Constant} & {\bf HEP units} (eV)  & {\bf Planck units} ($M_{\rm Pl}$) & {\bf SI units} (kg) & {\bf Length scale} (m) \\ \hline
	Planck mass ($M_{\rm Pl})$& $10^{27 }$ &1 & 4$\times 10^{-9}$ & $10^{-33}$\\		\hline
	Weak scale $(\sim1~{\rm TeV})$ & $10^{12}$ & $10^{-16}$ & $10^{-24}$& $10^{-18}$ \\\hline
	Electron mass $(m_e)$ & $5\times 10^5$ & $2\times 10^{-22}$ & 9$\times 10^{-31}$&  $10^{-12}$\\ \hline
	Electron volt (eV)&  $1$& 4$\times 10^{-28}$ & $2\times 10^{-36}$ &  $10^{-6}$\\		\hline
	Hubble parameter ($H_0$) & $10^{-33}$ & $10^{-60} $ & $10^{-69}$& $10^{26}$  \\ 
	\hline
\end{tabular}
}
\caption{\small Approximate conversion between different units and their corresponding length scales for some commonly appearing constants.}
\end{table}

It is this tension between theory and observation that has led to numerous attempts to explain the smallness of the observed cosmological constant. Within the standard $\Lambda$CDM paradigm, one possibility is that the value of the cosmological constant is selected via some mechanism from a distribution of values and that we live in a universe with a small value for essentially {\it anthropic} reasons---if the cosmological constant were much larger, cosmological structures could not form~\cite{Weinberg:1987dv,Weinberg:1988cp}. This idea has recently seen a resurgence in the context of the string theory landscape~\cite{Susskind:2003kw, Bousso:2000xa, Kachru:2003aw,Douglas:2003um,Douglas:2006es}.

In many senses, the problem posed by the accelerating universe is to the standard model of cosmology what the weak hierarchy problem is to the Standard Model of particle physics. Both are fine-tuning problems, arising because of the quantum instability of a physical scale whose existence is directly verified through observations or experiments. However, the naturalness problem afflicting cosmology is to some degree more robust than its particle physics counterpart. Indeed, the weak hierarchy problem is a result of radiative corrections to the Higgs mass from hypothetical particles {\it beyond} the weak
scale.\footnote{Dimensional regularization, which makes optimistic assumptions about the UV physics, forbids the presence of quadratically-divergent contributions to the Higgs mass.}
However, in the case of the CC, vacuum energy contributions from {\it known} particles, such as the electron, are already problematic. Furthermore, the required solution for the CC problem is arguably more radical. While the proposed solutions to the weak hierarchy problem---for example supersymmetry or technicolor---are by no means trivial, they fit within the standard framework of local quantum field theory.  On the other hand, no dynamical solution to the CC problem is possible within General Relativity (GR), as was shown by~\cite{Weinberg:1988cp}. The argument of~\cite{Weinberg:1988cp} is reviewed in Appendix~\ref{Weinbergnogo}.

As in particle physics, the search for a compelling resolution to the cosmological constant problem has also led to many interesting extensions to the $\Lambda$CDM model.
However, the progress that has been made can also be seen in numerous proposals for novel physics which are interesting in their own right. In this review, we focus on these proposals, considering models of new physics in the gravitational sector and their observational consequences. Often, this program goes by the name of {\it modified gravity}. Though these models are inspired by the CC problem, we prefer to ask a subtly different and broader question: {\it in what ways can new physics appear in the gravitational sector and how can they be tested}?

\noindent
{\bf Conventions:} Throughout the review we will use the mostly plus metric convention {\it i.e.}, $\eta_{\mu\nu} = {\rm diag}(-1, 1, 1, 1)$ and define the reduced Planck mass by $M_{\rm Pl}^2 \equiv (8\pi G_{\rm N})^{-1}$. We (anti-)symmetrize with weight one, {\it e.g.}, $S_{(\mu\nu)} = \frac{1}{2}(S_{\mu\nu}+S_{\nu\mu})$. Commas denote partial derivatives {\it e.g.}, $V_{,\phi} = \partial_\phi V$, overdots denote derivatives with respect to coordinate time: $\dot a = \partial_t a$ and primes refer to radial derivatives: $\phi' = \partial_r\phi$.

\section{Looking beyond $\Lambda$CDM}
\label{lookbeyondlcdm}

Einstein gravity is remarkably robust. It represents the unique interacting theory of a Lorentz invariant massless helicity-2 particle~\cite{Papapetrou:1948jw,Gupta:1952zz,Kraichnan:1955zz,Weinberg:1965rz, Feynman:1996kb, Deser:1969wk}.\footnote{Remarkably, the assumption of Lorentz invariance is not even necessary. As shown recently in~\cite{Khoury:2011ay,Khoury:2013oqa}, the weaker assumption of spatial Lorentz covariance is sufficient to show that GR is the unique theory of 2 transverse, traceless degrees of freedom. This hints  at Lorentz invariance being an emergent symmetry of the gravitational sector.}
(In Appendix~\ref{deserweinbergproof}, we give a self-contained derivation of this result, following Deser's proof~\cite{Deser:1969wk}.) Therefore, essentially all new physics in the gravitational sector introduces new degrees of freedom, which are typically Lorentz scalars. 

The main focus of this review is to show how the interactions of these putative new fields with matter are both constraining and a source of novel physics. In particular, we will see that a number of {screening mechanisms} exist, allowing such fields to remain unseen by local tests of gravity. Indeed, we will take the broader point of view that, independent of the motivations provided by cosmic acceleration, if such light scalars are present in nature (arising {\it e.g.}, from string theory or some other more fundamental theory), they must appear in some screened form, or they would have already been discovered. Before embarking on our study of this possible new physics, it is worth first exploring how the cosmological constant problem itself might find a satisfactory resolution without the addition of new degrees of freedom.

\subsection{The anthropic perspective}

The smallness of $\Lambda$ introduces two related, but essentially distinct, problems. The first is what is usually called the cosmological constant problem. This can be most succinctly phrased as the question: {\it why is the observed value of $\Lambda$ so small in Planck units?} Related to this is the question, usually referred to as the {\it coincidence problem}, which may be phrased as: {\it why is the energy density of this CC so close to the present matter density?} It is attempting to answer one or both of these questions that has led to various proposals for new physics.

One possible solution is that the cosmological constant is just small because a universe with a larger cosmological constant would not be able to support the formation of large-scale structures and the presence of life capable of asking why the cosmological constant is so small. Weinberg famously argued  that the cosmological constant should be within a few orders of magnitude of the upper bound which still allows cosmological structures to form, which is approximately $\Omega_\Lambda/\Omega_{\rm m}\sim10-100$~\cite{Weinberg:1988cp,Weinberg:1987dv}. This is of course very tantalizing, as the observed value is $\Omega_\Lambda/\Omega_{\rm m}\sim2-3$. For this idea to make sense, it is necessary for the underlying laws of physics to allow for multiple realizations of the universe---in time, space, or both---with different values of the cosmological constant. Recently these anthropic ideas have gained new life, particularly within the context of the string theory landscape~\cite{Susskind:2003kw, Bousso:2000xa, Kachru:2003aw,Douglas:2003um,Douglas:2006es}. Roughly, the idea is that the vast number of compactifications of string theory to 4 dimensions gives rise to a rich landscape of (meta-)stable de Sitter vacua, each with a different value of the cosmological constant. Dynamical transitions between the various vacua are possible via Coleman--De Luccia tunneling~\cite{Coleman:1977py,Callan:1977pt,Coleman:1980aw} and it is therefore not surprising that we find ourselves in a vacuum with a small CC, as ones with larger values of $\Lambda$ would not be hospitable to life. However, to date it has been extremely difficult to find satisfactory constructions of de Sitter space in string theory. 

Even assuming that many such stable constructions exist, we are a long way from having established the anthropic explanation as the solution to the problem of cosmic acceleration. String theory remains perhaps the most promising candidate for the correct theory of quantum gravity, but it is yet to connect with testable phenomena. Eternal inflation seems to be a reasonable consequence of generic inflationary models~\cite{Vilenkin:1983xq,Linde:1986fd,Vilenkin:1994ua,Guth:2000ka,Creminelli:2008es}, but it has yet to be definitively demonstrated that this effect, taking place out of the usual semi-classical regime we consider in inflation, must occur.\footnote{See~\cite{Martinec:2014uva,Boddy:2014eba} for recent cautionary takes on this question.} If all of the above ingredients can be established, it may nevertheless prove extremely difficult to show that the anthropic explanation is the correct one, and, of course, any experimental evidence that cosmic acceleration has a dynamical source would rule out such an answer immediately. Thus, while the string landscape and eternal inflation provides one logical way to address the cosmological constant problem, it would be extremely premature to abandon the search for other theoretical explanations for cosmic acceleration.

\subsection{Dynamical dark energy}
\label{desec}

Another intriguing possibility is that the ultimate value of the cosmological constant is zero, and that cosmic acceleration is due to the potential energy of a field, with some sort of mechanism to dynamically relax it to a small value. This notion naturally leads to models of {\it dark energy} which invoke a slowly-rolling cosmological scalar field to source accelerated expansion, akin to cosmological inflation~\cite{Starobinsky:1979ty,Guth:1980zm,Albrecht:1982wi,Linde:1981mu}.

In such models it is difficult to see how to avoid fine tuning at the same level as just tuning the bare CC~\cite{Weinberg:1988cp}. Nevertheless, these dynamical dark energy models offer the simplest extension of $\Lambda$CDM and have been extensively studied (see for example~\cite{Fujii:1982ms,Ford:1987de,Wetterich:1987fm,Peebles:1987ek,Ratra:1987rm,Caldwell:1997ii,Liddle:1998xm,Kolda:1998wq,Amendola:1999er,Wang:1999fa,Cooray:1999da,Barreiro:1999zs,Huterer:2000mj,Chevallier:2000qy,Boyle:2001du,Melchiorri:2002ux,Pilo:2003gu,Simon:2004tf,Huterer:2006mv,Lim:2010yk,Mortonson:2010mj,Marsh:2014xoa}), and often go under the name {\it quintessence}. For reviews and further references, see~\cite{Sahni:1999gb,Peebles:2002gy,Padmanabhan:2002ji,Copeland:2006wr,Linder:2007wa,Frieman:2008sn,Li:2011sd,Tsujikawa:2013fta}. 

To understand how a field can drive cosmic acceleration, consider the action of a scalar field minimally coupled to Einstein gravity
\be
S = \int\rd^4x\sqrt{-g}\left(\frac{M_{\rm Pl}^2R}{2}-\frac{1}{2}(\partial\phi)^2-V(\phi)\right)~.
\ee
The stress-energy tensor for the scalar field is given by
\be
T_{\mu\nu}^\phi = \partial_\mu\phi\partial_\nu\phi-g_{\mu\nu}\left(\frac{1}{2}(\partial\phi)^2+V(\phi)\right)~.
\label{scalarTmunu}
\ee
For a homogeneous profile, $\phi = \phi(t)$, a cosmological scalar acts like a perfect fluid with equation of state $w = P/\rho$ given by
\be
w_\phi = \frac{\frac{1}{2}\dot\phi^2-V(\phi)}{\frac{1}{2}\dot\phi^2+V(\phi)}~.
\ee
If we want to source the observed expansion, we must have $w_\phi \simeq -1$, which requires a very slowly-rolling field: $\dot\phi^2\ll V(\phi)$. There has also been a lot of interest in constructing quintessence models which can produce an equation of state of the ``phantom" type ($w_\phi <-1$)~\cite{Caldwell:1999ew,Faraoni:2001tq,Caldwell:2003vq,Carroll:2003st, Nojiri:2003vn,Singh:2003vx,Elizalde:2004mq,Feng:2004ad,Guo:2004fq,Vikman:2004dc,Hu:2004kh,Nojiri:2005sx,Nojiri:2005sr,Cai:2009zp}. Similar to natural inflation~\cite{Freese:1990rb}, it has been proposed that the flatness of the quintessnce potential could be protected by the field being a pseudo-Nambu--Goldstone boson~\cite{Frieman:1995pm,Kaloper:2005aj}. Additionally, it has been proposed that dark energy and inflation can both be driven by the same field~\cite{Peebles:1998qn,Peloso:1999dm,Kaganovich:2000fc,Dimopoulos:2001ix,Majumdar:2001mm,Rosenfeld:2005mt,Neupane:2007mu,Neupane:2007jm,Bose:2008ew}

Quintessence models have a rich phenomenology; one of the more interesting phenomena is that---by suitably choosing the potential---the energy density in the quintessence field can be made to ``track'' the energy density in radiation/matter at early times and then grow to dominate the energy budget at late times~\cite{Wetterich:1994bg,Copeland:1997et,Ferreira:1997hj,Ferreira:1997au,Zlatev:1998tr,Steinhardt:1999nw,Zimdahl:2001ar,Chimento:2003iea}. The canonical example of a potential that produces this behavior is the Ratra--Peebles potential~\cite{Ratra:1987rm}
\be
V(\phi) = \frac{M^{4+n}}{\phi^{n}}\,,
\label{RatraPeebles}
\ee
where $n > 0$ is a constant.  In~\cite{Caldwell:2005tm}, a useful division was introduced which separates quintessence theories into two types. In the first, the field is frozen by Hubble friction at early times and only recently has ``thawed" and begun to roll down its potential. In these models $w_\phi \simeq -1$ until very recently in the cosmological evolution. The second type of models consist of a scalar field which is slowing down while it rolls, causing the equation of state to approach $w \simeq -1$ at present day and ``freeze"---the tracker models discussed above are examples of this behavior. These two types of models fill out different areas in the $(w_\phi, \dot w_\phi)$ plane and have qualitatively different features~\cite{Chiba:2005tj,Scherrer:2005je,Barger:2005sb,Linder:2006sv,Scherrer:2007pu,Chiba:2012cb}.

An important generalization of quintessence models is to theories which are derivatively-coupled: so-called $P(X)$ models. Applications to cosmic acceleration go under the name $K$-essence~\cite{Chiba:1999ka,ArmendarizPicon:2000dh,ArmendarizPicon:2000ah,Chiba:2002mw,Padmanabhan:2002cp,Malquarti:2003nn,Malquarti:2003hn,Chimento:2003zf,Silverstein:2003hf,GonzalezDiaz:2003rf,Scherrer:2004au,Aguirregabiria:2004te,Piazza:2004df,Rendall:2005fv,Bonvin:2006vc,Babichev:2007dw,dePutter:2007ny,Kang:2007vs,Bilic:2008zk,Martin:2008xw,Myrzakulov:2010tc}, and were first considered within the context of cosmological inflation~\cite{ArmendarizPicon:1999rj, Garriga:1999vw}. There have also been efforts to embed quintessence in some more fundamental theory; for example supersymmetry~\cite{Brax:1999yv,Brax:1999gp,Masiero:1999sq,Copeland:2000vh,Kallosh:2002gf} or string theory~\cite{Choi:1999xn,Gasperini:2001pc,Townsend:2001ea,Damour:2002nv,Damour:2002mi,Kim:2002tq,Kaloper:2008qs,Panda:2010uq,Cicoli:2012tz}.

Importantly, quintessence models must be very weakly coupled to matter, otherwise the scalar field will mediate a fifth force (see Section~\ref{scalarforcesec}), in conflict with tests of gravity.\footnote{However, if the quintessence field {\it does} couple to electromagnetism, interesting experimental signatures are possible~\cite{Carroll:1998zi,Lue:1998mq,Copeland:2003cv}.} Absent some symmetry principle, setting the coupling to matter to be small is a fine tuning beyond that required to make the CC itself small (which is, itself, also required here), since if the coupling is set to zero classically, it will be generated through loops albeit with its resulting value suppressed by powers of $M_{\rm Pl}$. In this review, we will not focus on pure quintessence models. However, the lesson that a scalar field can drive cosmic acceleration is one that we will take to heart.

\subsection{Modifications to Einstein's equations}
Another possibility---which we will discuss only briefly---is that the CC problem may be addressed through some modification of Einstein's equations which does not introduce new degrees of freedom. One approach along these lines is {\it unimodular gravity}~\cite{Anderson:1971pn}; in this theory, the determinant of the metric is fixed to be $-1$. A starting point to understand unimodular gravity is to consider the traceless Einstein equations following Weinberg~\cite{Weinberg:1988cp}
\be
R_{\mu\nu} - \frac{1}{4}Rg_{\mu\nu} = \frac{1}{M_{\rm Pl}^2}\left(T_{\mu\nu}-\frac{1}{4}Tg_{\mu\nu}\right)~.
\label{einsteineqsunimod}
\ee
Taking a covariant divergence and utilizing the Bianchi identity, one finds the following
\be
\nabla_\mu R = -\nabla_\mu T~,
\ee
which upon integrating yields
\be
T = -R +\Lambda~,
\ee
where $\Lambda$ is an integration constant. Plugging this expression back into~\eqref{einsteineqsunimod} yields precisely the full Einstein equations, but where $\Lambda$ is now an integration constant, unrelated to any bare CC in the action. How then, can we derive the equations~\eqref{einsteineqsunimod} from an action principle? In fact, if we consider the Einstein--Hilbert action coupled to matter
\be
S = \frac{M_{\rm Pl}^2}{2}\int\rd^4x\sqrt{-g}R + S_{\rm matter}[g_{\mu\nu}, \psi]~,
\ee
and demand that they are stationary under variations which keep the determinant fixed, we recover precisely~\eqref{einsteineqsunimod} (see, {\it e.g},~\cite{Weinberg:1988cp}). Alternatively, the same equations can be obtained from an action where the determinant is fixed to be non-dynamical via a Lagrance multiplier (see for example~\cite{Padilla:2014yea}). It is for this reason the theory is called {\it unimodular}. It is clear that at the classical level, unimodular gravity and Einstein gravity are equivalent, making it somewhat unclear if we have made any progress toward resolving the CC problem. Further, it is not clear to what degree Einstein gravity and unimodular gravity are equivalent at the quantum-mechanical level. Nevertheless, it remains an interesting approach. For a nice introduction and survey of references, we direct the reader to~\cite{Padilla:2014yea,Fiol:2008vk}.

Another possibility is to attempt to modify the gravitational equations in a non-local way, various proposals for which have been considered in~\cite{Kaloper:2013zca,Kaloper:2014dqa} and~\cite{Gabadadze:2014rwa}. One other approach is to modify Einstein's equations by introducing auxiliary fields (see {\it e.g.}~\cite{Carroll:2006jn,Pani:2013qfa}).

\section{How can new physics appear?}
\label{section:intro-to-screening}

Although no compelling mechanism has been found to date which solves the CC problem by introducing new degrees of freedom, we can robustly infer two  properties they must possess. For concreteness, we assume they are scalars for the time being, which we denote collectively as $\phi$.
First, to neutralize $\Lambda$ to an accuracy of order $\sim H_0^2M_{\rm Pl}^2 \sim ({\rm meV})^4$, the scalars must have a mass at most comparable to the present-day Hubble parameter,
\begin{equation}
m_\phi ~\lower .75ex \hbox{$\sim$} \llap{\raise .27ex \hbox{$<$}} ~H_0\,.
\label{mphi}
\end{equation}
If they were much more massive, they could be integrated out and would be irrelevant to the low energy dynamics. 

\begin{figure}[t]
\centering
\includegraphics[width=3.9in]{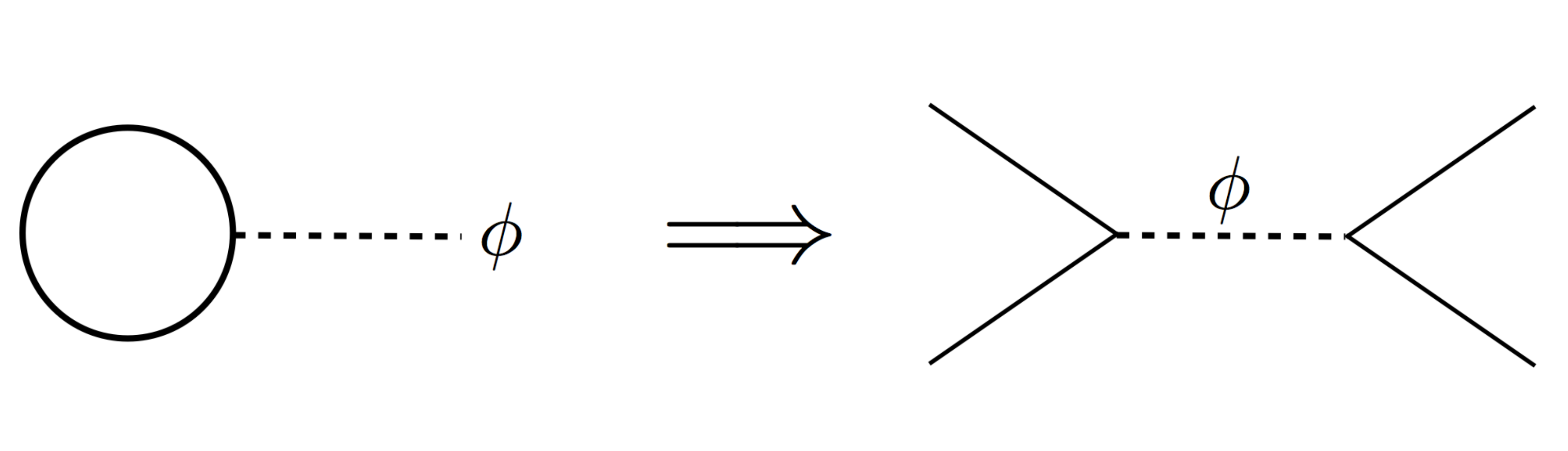}
\caption{\label{CCdiags} \small If additional scalars are to neutralize the Standard Model vacuum energy contribution, the tadpole diagram (left) involving the scalar field (dotted line) attached to Standard Model fields (solid line) running in the loop is necessary. By unitarity, the tree-level diagram (right) with a scalar exchanged by Standard Model fields is also allowed, implying that the scalar field mediates a $5^{\rm th}$ force that must therefore be screened in the local environment. }
\end{figure}

Secondly, these scalars must couple to Standard Model fields since, as we saw in Section~\ref{introsection}, SM fields contribute ${\cal O}({\rm TeV}^4)$ to the vacuum energy. Another way of saying this is that the tadpole diagram shown in Figure~\ref{CCdiags} must be present in the theory. However, by unitarity, so must the exchange diagram in Figure~\ref{CCdiags}. Hence $\phi$ mediates a force between Standard Model fields, whose range is $\sim m_\phi^{-1}$. Given~\eqref{mphi}, this is comparable to the present Hubble radius. Thus, the scalar mediates a fifth force, both at cosmological distances and within the solar system. However, gravity is exquisitely well tested within the solar system (see {\it e.g.},~\cite{Will:2005va,Will:2014kxa}) so there must necessarily be some mechanism which hides these new fields from local observations. This can be achieved through {\it screening mechanisms}, which rely on the high density of the local environment (relative to the mean cosmological density) to suppress deviations from GR. In what follows we will describe three general classes of screening mechanisms and highlight some of their characteristic properties and observational signatures.

Apart from cosmology and the cosmological constant problem, screening mechanisms are also motivated by the vast experimental effort aimed at testing the fundamental nature of gravity on a wide range of scales, from laboratory to solar system to galactic and extra-galactic scales. (For a review, see~\cite{Will:2014kxa}.) As we will see later in this review, viable screening theories often make novel predictions for local gravitational experiments. The subtle nature of these signals has forced experimentalists to re-think the implications of their data and inspired the design of novel experimental tests. The theories of interest thus offer a rich phenomenology and a spectrum of testable predictions for ongoing and near-future tests of gravity.

\subsection{Screening mechanisms: force-law classification}
\label{scalarforcesec}

One way to classify the different kinds of screening mechanisms is to study the connection between fields that appear in a Lagrangian and our classical physics notions of force and potential. The derivation of the force between two objects due to the scalar will naturally present some mechanisms by which we can hide this force in dense environments, like our solar system.

To begin, consider a general theory of a scalar field coupled conformally to matter:
\be
{\cal L} = -\frac{1}{2}Z^{\mu\nu}(\phi, \partial\phi,\ldots)\partial_\mu\phi\partial_\nu\phi-V(\phi)+g(\phi)T^\mu_{\;\mu}~,
\label{Lintro}
\ee
where $Z^{\mu\nu}$ schematically encodes derivative self-interactions of the field, and $T^\mu_{\;\mu}$ is the trace of the matter stress-energy tensor.\footnote{For the purpose of this schematic argument, we assume universal coupling for simplicity. With chameleons~\cite{Khoury:2003aq,Khoury:2003rn}, one can more generally assume different couplings to different matter species, thereby explicitly violating the weak equivalence principle. We also ignore the possibility of derivative interactions with matter, such as the disformal screening mechanism~\cite{Koivisto:2012za,Noller:2012sv,Brax:2012ie,Zumalacarregui:2012us,vandeBruck:2013yxa,Brax:2013nsa}.} For non-relativistic sources, we can as usual make the replacement $T^\mu_{\;\mu}\rightarrow -\rho$. In the presence of a point source, $\rho =  {\cal M}\delta^3(\vec{x})$, we can then expand the field about its background solution $\bar\phi$ as $\phi = \bar\phi+\vp$ to obtain the equation of motion for the perturbation: 
\be
Z(\bar\phi)\Big(\ddot\vp-c_s^2(\bar\phi)\nabla^2\vp\Big)+m^2(\bar\phi)\vp = g(\bar\phi){\cal M} \delta^3(\vec{x}) \,,
\label{generalscalareom}
\ee
where $c_s$ is an effective sound speed. In general, we have in mind that the background value $\bar\phi$ is set by other background quantities, such the local density $\bar\rho$ or the Newtonian potential $\Phi$. Neglecting the spatial variation of $\bar\phi$ over the scales of interest, the resulting static potential is
\be
V(r) = -\frac{g^2(\bar\phi)}{Z(\bar\phi) c_s^2(\bar\phi)}\frac{e^{-\frac{m(\bar\phi)}{\sqrt{Z(\bar\phi)}c_s(\bar\phi)}r}}{4\pi r}{\cal M} ~.
\label{scalarpoteqn}
\ee
The corresponding force is therefore {\it attractive}, as it should be for scalar mediation.

Now the problem is clear: for a light scalar, and with the other parameters ${\cal O}(1)$, we see that $\vp$ mediates a gravitational-strength long range force $F_\vp\sim 1/r^2$. Local tests of GR forbid any such force to high precision. The question we are then led to ask is: {\it How can we make this force sufficiently weak in the local environment to reproduce the successful phenomenology of Einstein gravity in the solar system, while allowing for significant deviations from GR on astrophysical or cosmological scales?}\footnote{A trivial way to approximately recover GR is of course to make the scalar-mediated force sufficiently weak on {\it all} scales, {\it e.g.}, by making the coupling of $\vp$ to matter, $g$, to be very small universally. This is an entirely reasonable thing to do (modulo fine-tuning) and leads to {\it dark energy} models touched upon in Section~\ref{desec} where the dark sector does not interact significantly with visible matter. We will not dwell much on theories of this type, preferring to focus on the phenomenologically richer possibility of the scalar force being {\it environmentally}---as opposed to universally---weak.}

The fact that the various parameters $g$, $Z$, $c_s$ and $m$ appearing in~\eqref{scalarpoteqn} depend on the background value of the field provides a clue as to how this could work.
Screening mechanisms can be understood as making each of these parameters (or a combination thereof) depend on the environment:

\begin{itemize}

\item {\it Weak coupling:} One possibility is to let the coupling to matter, $g$, depend on the environment. In regions of high density---where local tests of gravity are performed---the coupling is very small, and the fifth force sufficiently weak to satisfy all of the constraints. For example, in regions of low density, such as in the cosmos, $g$ can be of order unity, resulting in a fifth force of gravitational strength. Examples include the {\it symmetron}~\cite{Hinterbichler:2010es, Pietroni:2005pv,Olive:2007aj} or {\it varying-dilaton}~\cite{Damour:1994zq, Brax:2011ja} theories.

\item {\it Large mass:} Another option is to let the mass of fluctuations, $m(\bar\phi)$, depend on the ambient matter density. In regions of high density, such as on Earth, the field acquires a large mass, making its effects short range and hence unobservable. Deep in space, where the mass density is low, the scalar is light and mediates a fifth force of gravitational strength. This idea leads quite naturally to screening of the {\it chameleon} type~\cite{Khoury:2003aq, Khoury:2003rn}. 

\item {\it Large inertia:}  We may also imagine making the kinetic function, $Z(\bar\phi)$, large environmentally. This leads us to screening of the {\it kinetic} type, either with first derivatives becoming important~\cite{Babichev:2009ee, Babichev:2011kq,Brax:2012jr,Burrage:2014uwa}, or with second derivatives being relevant~\cite{Vainshtein:1972sx, Deffayet:2001uk,Nicolis:2004qq,Nicolis:2008in}. This latter mechanism, where second derivatives are important, is also known as {\it Vainshtein} screening. 

\end{itemize}

A fourth possibility suggests itself, namely screening by making $c_s(\bar\phi)$ very large environmentally. Putting aside the issue that this manifestly relies on superluminality, it is important to realize that~\eqref{scalarpoteqn} only applies to {\it static} sources. For time-dependent sources, on the other hand, the sound speed will only multiply spatial gradients and therefore screening of this type will not be very efficient.\footnote{Indeed, even Vainshtein screening is less efficient in the presence of sources which evolve in time. One way to understand this is by noting that time-dependent sources allow for cancellations in $Z(\bar\phi)$ between $\dot{\bar\phi}$ and $\vec\nabla\bar\phi$ terms due to the indefinite signature of the metric, making $\lvert Z\rvert$ smaller than na\"ive estimates.} This raises a challenge, in particular, for reproducing standard cosmological evolution at early times.

\subsection{Screening mechanisms: phenomenological classification}

Another way to classify the screening mechanisms, and one we will adopt in this review, is based on the nature of the screening criterion. We distinguish three cases: $i)$ screening set by the local field value $\phi$; $ii)$ screening set by the first derivative $\partial\phi$; and $iii)$ screening by the second derivative $\partial^2\phi$. This classification is more phenomenological and better suited for astrophysical and cosmological observations.

\begin{itemize}

\item {\it Screening based on $\phi$}: The first class of screening mechanisms correspond to scalar self-interactions being governed by a potential $V(\phi)$. Hence, whether or not the scalar develops non-linearities depends on the local value of $\phi$. This includes  the symmetron, chameleon and dilaton screening mechanisms~\cite{Khoury:2003aq, Khoury:2003rn,Hinterbichler:2010es, Pietroni:2005pv,Olive:2007aj}. These are all examples of screening where the additional degrees of freedom develop weak couplings to matter, a high mass, or strong self-interactions in regions of high Newtonian potential. These are mechanisms which shut off the fifth force when the gravitational potential $\Phi$ exceeds some critical value,
\be
\Phi ~\gsim ~\Lambda\,. 
\ee
A useful rule of thumb to ascertain which regions of the universe are screened or unscreened is to map out  the {\it gravitational potential} smoothed on some scale. 

\item  {\it Screening based on $\partial\phi$}: The second class of screening mechanisms relies on first derivatives of the field becoming important; these are mechanisms which operate when $\partial\phi~\gsim~\Lambda^2$. An example of a theory of this type is $k$-mouflage~\cite{Babichev:2009ee, Babichev:2011kq}, but it can also be present in generic models with kinetic interactions (so-called $P(X)$ models)~\cite{Brax:2012jr}, for example DBI can exhibit this type of screening~\cite{Burrage:2014uwa}. These are mechanisms which shut off the fifth force when the local gravitational acceleration, $\vec{a} = -\vec{\nabla}\Phi$, exceeds some critical value,
\be
|\nabla\Phi|~\gsim ~\Lambda^2\,.
\ee
Here, the heuristic rule to ascertain which regions of the universe are screened or unscreened is to map out the {\it gravitational acceleration} smoothed on some scale. 
This mechanism is useful for constructing theories of MOdified Newtonian Dynamics (MOND)~\cite{Milgrom:1983ca}. See~\cite{Sanders:2002pf} and references therein.

\item  {\it Screening based on $\partial^2\phi$}: The third broad class of theories which screen relies on second derivatives of the field becoming important to the dynamics, $\partial^2\phi~\gsim~\Lambda^3$, while higher-order derivatives remain small. Screening in these theories is essentially a variant of the {\it Vainshtein mechanism}\footnote{This mechanism first appeared in massive gravity to resolve the van Dam--Veltman--Zakharov (vDVZ) discontinuinty as the mass of the graviton is taken to zero~\cite{vanDam:1970vg, Zakharov:1970cc}.}~\cite{Vainshtein:1972sx} (see also~\cite{Deffayet:2001uk,Gruzinov:2001hp,Porrati:2002cp,Babichev:2009jt,Babichev:2009us,Babichev:2010jd}). The best-known example of a theory exhibiting the Vainshtein mechanism is the galileon~\cite{Nicolis:2008in}, which we will review and show explicitly how screening occurs around spherical sources. The Vainshtein mechanism has also been of great interest recently in connection to massive gravity, where recent theoretical developments have made it possible to write down a classically consistent nonlinear theory of a massive graviton~\cite{ArkaniHamed:2002sp, Creminelli:2005qk,Nibbelink:2006sz,deRham:2010ik,deRham:2010kj}. This mechanism shuts off the fifth force when the local curvature or density, $R \sim \vec\nabla^2 \Phi$, exceeds some critical value, 
\be
|\nabla^2\Phi| ~\gsim ~\Lambda^3\,.
\ee
To determine which regions of the universe are screened or unscreened, here it is useful to map out the {\it curvature} smoothed on some scale.

\end{itemize}

One might be tempted to go further and posit a fourth class of screening mechanisms---the obvious generalization---which depends on {\it third} derivatives of the field becoming large: $\partial^3\phi~\gsim~\Lambda^4$. But it can be seen that this notion will fail for consistency reasons. A (Lorentz-invariant) theory which has third derivatives in the action will necessarily have third (or higher) order equations of motion.\footnote{One loophole is to drop the assumption of Lorentz invariance~\cite{Hinterbichler:2014cwa}.} This means that to specify initial data it will not suffice to specify only the initial configuration of the field and its first derivative, but also the second derivative of the field. This additional piece of required initial data tells us that we are propagating more than the degree of freedom associated with the scalar. This new degree of freedom generically has the wrong-sign kinetic term---there is a {\it ghost} in the theory. This notion can be formalized by going to the Hamiltonian formulation: the theory will describe more than one canonical pair of a field and its momentum, and the Hamiltonian will be unbounded from below. This classic result is known as Ostrogradsky's theorem~\cite{Ostrogradsky} (see Appendix~\ref{OstrogradskyApp} for a review).

\subsection{Effective field theory as a unifying language}

Over the last decade, the theoretical cosmology community has increasingly shifted towards the viewpoint that the effective field theory (EFT) framework offers the most promising and theoretically robust arena in which to
explore extensions to the standard $\Lambda$CDM model. The basic rules of effective field theory are simple: identity the low energy fields, specify the symmetries of the theory, and write down all possible terms in the action consistent with these symmetries.\footnote{For nice introductions to the tools and some applications of effective theories, see~\cite{Polchinski:1992ed,Burgess:2003jk,Kaplan:2005es,Burgess:2007pt}.} The power of EFT lies in the fact that there is typically a systematic expansion in some small parameter (often a derivative expansion) which tells us what terms we can ignore to a given order of accuracy. As the name might suggest, EFTs are only {\it effective} descriptions of low-energy physics, valid up to some energy/length scale, called the {\it cutoff} of the theory. Beyond this energy scale, the theory ceases to be predictive, and some more fundamental theory must take over an describe the physics. Typically such a high-energy parent theory is called a {\it UV completion}. Requiring such a UV completion is not a deficiency of a theory; often times at low energies or long distances some degrees of freedom are not important to the dynamics, but become important when moving to shorter distances. The EFT breaking down is just a reflection of the fact that degrees of freedom we are not keeping track of are becoming important. An illustrative example is that of hydrodynamics: at macroscopic scales, the physics of fluids is well-described by the Navier--Stokes equations, but if we attempt to describe the interactions of water molecules with these equations, we will surely fail.

Of course, not every EFT we write down is well defined, and powerful questions of theoretical consistency constrain the structure of the theory. Specifically, the theory should be ghost-free, contain no gradient instabilities, and any tachyonic ones should be understood in a sensible way. (For completeness, these pathologies are reviewed in Appendix~\ref{ghostgradtachyon}.) A ghost (field with wrong-sign kinetic term) spells doom for the effective field theory, unless its mass lies above the cutoff. A gradient instability (wrong-sign gradient term) is also a show-stopper---the rate of instability is of order of the cutoff, so there is no regime where the theory makes sense. A tachyon, on the other hand, can be dealt with in a systematic way---perturbations around the (unstable) vacuum offer a good effective description on time scales smaller than the inverse tachyon mass, which sets the instability time scale. 

There are other, more subtle properties of an EFT which, although certainly less pathological than ghost or gradient instabilities, are generally considered to be undesirable. Namely, it is possible for a seemingly Lorentz-invariant EFT to allow superluminal propagation around certain backgrounds. Relatedly, the S-matrix may not satisfy the standard analyticity properties that follow from locality. (See Appendix~\ref{superlumapp} for a detailed discussion of superluminality and S-matrix non-analyticity.) Superluminality by itself does not necessarily imply the existence of closed time-like curves (CTCs) and their associated pathologies. As in GR~\cite{Hawking:1991nk}, there may be Chronology Protection forbidding the formation of CTCs from healthy initial conditions within the EFT description. (This has been argued to be the case for galileons in~\cite{Burrage:2011cr,Evslin:2011rj}. See~\cite{Babichev:2007dw} for earlier arguments in the context of $P(X)$ theories.) But superluminality (and/or a non-analytic S-matrix) does imply that the UV completion of the EFT is non-standard, {\it i.e.}, it cannot be a local quantum field theory or perturbative string theory~\cite{Adams:2006sv}.

All screening mechanisms by definition rely on a scalar field developing non-linearities in certain environments. Generically, this is triggered by higher-dimensional
operators becoming large, which na\"ively would imply a breakdown of the EFT expansion. Contrary to this na\"ive expectation, there are in fact examples (such
as the galileons) where---in certain regimes---a subset of terms in the EFT can become large classically while all other operators remain negligible. A familiar example
of this occurs in GR: gravity becomes classically non-linear at a (macroscopic) black hole horizon, but Planck-suppressed corrections to Einstein's theory remain small. This is
by no means guaranteed for the scalar theories under consideration, and successful examples, as we will see, rely on special symmetries. 

\begin{figure}[t]
\centering
\includegraphics[width=2.8in]{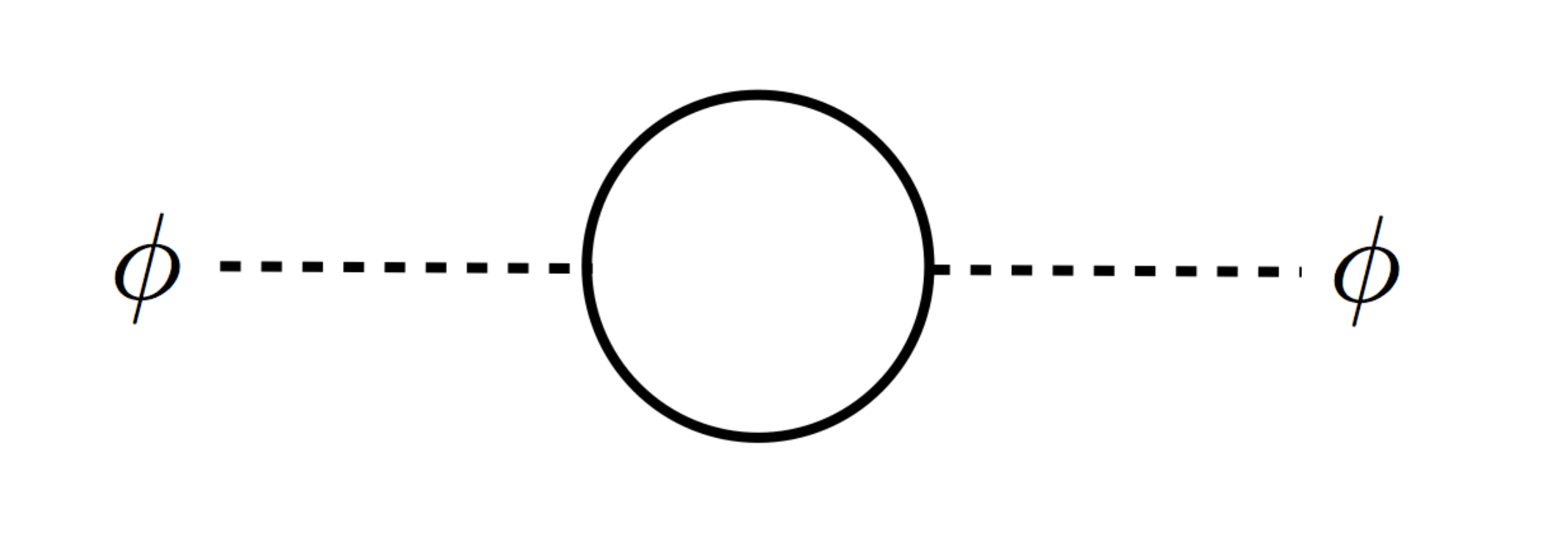}
\caption{\label{1loopmass} \small This 1-loop diagram with Standard Model fields (solid) running in the loop renormalizes the mass of the scalar field (dashed line).} 
\end{figure}

Another common feature of screening mechanisms is the conformal coupling to matter, $g(\phi)T^\mu_{\;\mu}$, depicted in~\eqref{Lintro}. 
An immediate concern is whether this coupling will generate a large mass for $\phi$ from matter fields running in loops. See Figure~\ref{1loopmass}
for an illustration. Indeed, as is well-known from the hierarchy problem for the Higgs, scalar fields are notoriously difficult to keep light. Assuming
for concreteness a linear coupling $\xi \phi T^\mu_{\;\mu}/M_{\rm Pl}$, with $\xi \sim {\cal O}(1)$ for gravitational-strength mediation, the
quantum correction to the scalar mass from the 1-loop diagram in Figure~\ref{1loopmass} is~\cite{Upadhye:2012vh}
\be
\delta m_\phi \sim \xi \frac{\Lambda^2_{\rm UV}}{M_{\rm Pl}}\,,
\ee
where $\Lambda_{\rm UV}$ is the cutoff of the EFT. Fortunately, this mass correction is harmless, for the simple reason that $\Lambda_{\rm UV}$ is
generally very small. Indeed, in order for $\phi$ to become non-linear in regions of typical density $\rho \ll M_{\rm Pl}^4$ requires a
low strong coupling scale. In practice, we will find cutoffs ranging from $\Lambda_{\rm UV} \sim (1000~{\rm km})^{-1}\sim 10^{-41}~M_{\rm Pl}$ for Vainshtein to $\Lambda_{\rm UV} \sim {\rm mm}^{-1}\sim 10^{-32}~M_{\rm Pl}$
for chameleon and kinetic screening. In all cases, $\delta m_\phi \;\lsim \; H_0$.

Throughout the review, the effective field theories we write down are for fields which describe the background dynamics. For a general parameterization of theories of this type (albeit in the weak-field regime), see~\cite{Park:2010cw}. There is a complementary viewpoint where an Friedmann--Lema\^{i}tre--Robertson--Walker (FLRW) background is assumed and the effective field theory for perturbations about this background is constructed~\cite{Creminelli:2008wc,Bloomfield:2012ff,Gubitosi:2012hu,Gleyzes:2013ooa,Piazza:2013coa,Frusciante:2013zop,Bloomfield:2013efa,Gergely:2014rna}. This {\it effective field theory of dark energy} is closely related to the effective field theory of inflation formalism~\cite{Creminelli:2006xe,Cheung:2007st,Cheung:2007sv}.

\section{Organization of the review}
This review is broken into two  parts that follow. In the next part we discuss the various theoretical developments that have emerged from attempts to modify gravity, motivated in part by the goal of explaining the accelerated expansion of the universe. We organize the material by the types of screening mechanisms employed, and discuss topics in discrete pieces. An effort has been made to keep the various sections self-contained in the hope that readers looking to educate themselves only about a specific topic will be able to find what they need more easily.

In the final part of the review we move on to experimental tests, categorized as to whether the relevant phenomena can be captured in the laboratory, in astrophysical systems or in cosmology.  Again, the material is organized so that readers who are interested in the implications of a given test should be able to dip into the article and find what they need, without reading the whole review.

A number of technical results required for the topics in both parts are relegated to appendices, where they can be found if needed, but where their details will not distract from the main discussion. These results represent part of the ``folklore" in the cosmology community and can be found scattered throughout the literature, but we felt it worthwhile to reproduce them all in one place here.

Finally, we note that there are many other reviews which overlap in varying degrees with the content of this review.
For general reviews of the cosmological constant problem see~\cite{Weinberg:1988cp,Carroll:1991mt,Carroll:2000fy,Weinberg:2000yb,Padmanabhan:2002ji,Peebles:2002gy,Copeland:2006wr,Frieman:2008sn,Silvestri:2009hh,Caldwell:2009ix,Li:2011sd}. There are also many reviews of various aspects of modified gravity approaches to the cosmological constant and new physics in the gravitational sector. General reviews of modified gravity are given in~\cite{Jain:2010ka,Tsujikawa:2010zza,Clifton:2011jh, Brax:2012bsa,Khoury:2013tda}. More specialized reviews include reviews of $f(R)$ gravity~\cite{Woodard:2006nt,Sotiriou:2008rp,DeFelice:2010aj,Nojiri:2010wj}; Chameleon screening~\cite{Khoury:2013yya,Khoury:2010xi}; galileons~\cite{Trodden:2011xh,deRham:2012az,Deffayet:2013lga,Babichev:2013usa} and massive gravity~\cite{Hinterbichler:2011tt,deRham:2014zqa}. For reviews of observational tests of GR and dark energy, see~\cite{Will:2005va,Will:2014kxa} and~\cite{Sahni:2006pa,Weinberg:2012es}, respectively. A topic we do not discuss in great depth is brane-world gravity and cosmology, but reviews can be found in
~\cite{Rubakov:2001kp,Langlois:2002bb,Brax:2003fv,Maartens:2003tw,Brax:2004xh,Lue:2005ya,Maartens:2010ar}.

\part{New physics in the gravitational sector: theoretical developments} 

There are, in principle, a number of different ways to classify modified gravity theories. In fact, it is not completely clear how to perform such a classification distinct from dark energy theories in general, since in many cases they both involve new degrees of freedom. Given this, one way to begin is with the realization that any new gravitationally coupled degrees of freedom run the risk of being ruled out by local tests of gravity unless some new physics comes into play. Thus, it seems natural to classify theories according to the mechanism through which these new degrees of freedom are screened in the solar system, where most precision tests hold. See Section \ref{section:intro-to-screening} for an introduction to screening. 

In this Part we discuss a number of different approaches to modifying gravity organized according to this classification.

\section{Screening by deep potentials: $\Phi~\gsim~\Lambda$}

The first broad class of modified gravity models we will consider are those which are screened in regions of high Newtonian potential ({\it i.e.}, for $\Phi$ larger than some critical value). The value of the Newtonian potential above which the field is screened is set by the parameters in the theory. Roughly speaking, theories of this type can be cast as a scalar-tensor theory
\be
S = \int\rd^4x\sqrt{-g}\left(\frac{M^2_{\rm Pl}}{2}R-\frac{1}{2}(\partial\phi)^2-V(\phi)\right)+S_{\rm matter}\left[A^2(\phi)g_{\mu\nu}, \psi\right]~,
\label{scalartensoraction}
\ee
where $R$ is the Ricci scalar of the metric $g_{\mu\nu}$, and $V(\phi)$ is a (for now) arbitrary potential. The matter action is a functional of the Jordan-frame metric $\tilde g_{\mu\nu} = A^2(\phi)g_{\mu\nu}$ and the various matter fields in the theory, $\psi$. Matter fields described by $S_{\rm matter}$ couple to $\phi$ through the conformal factor $A(\phi)$ implicit in $\tilde{g}_{\mu\nu}$. The acceleration of a {\it test particle} is influenced by the scalar field via
\be
\vec{a} = -\vec{\nabla}\Phi - \frac{{\rm d}\ln A(\phi)}{{\rm d}\phi} \vec{\nabla}\phi = -\vec{\nabla}\bigg(\Phi + \ln A(\phi)\bigg)\,,
\label{acc}
\ee
where it is important to remember that $\Phi$ is the {\it Einstein-frame} Newtonian potential. 

By suitably choosing the potential for the scalar $V(\phi)$ and its coupling to matter, $A(\phi)$, it is possible to arrange a situation where the scalar propagates freely and mediates a fifth force in regions of low Newtonian potential, but where the force is shut off in high density regions, such as in the solar system. To see how this works, we note that the equation of motion for $\phi$ following from~\eqref{scalartensoraction} is
\be
\square\phi = V,_{\phi}-A^3(\phi)A,_{\phi}\tilde T~,
\label{newtonianscreeningscalareom}
\ee 
where $\tilde T$ is the trace of the Jordan frame matter stress tensor, $\tilde T = \tilde g^{\mu\nu}\tilde T_{\mu\nu}$. Since the Jordan frame metric couples minimally to matter, it is this stress tensor that is covariantly conserved\footnote{Note that this follows straightforwardly from diffeomorphism invariance.}
\be
\tilde\nabla_\mu \tilde T^{\mu\nu} = 0~.
\ee
The gravitational part of the action is governed by the Einstein equations
\be
R_{\mu\nu}-\frac{1}{2}Rg_{\mu\nu} = \frac{1}{M_{\rm Pl}^2}\left( T_{\mu\nu}^{\rm matter}+T_{\mu\nu}^\phi\right)~.
\ee
Taking a divergence and using the contracted Bianchi identity ($\nabla_\mu G^{\mu\nu} = 0$) we see that the matter stress tensor in the Einstein frame is not conserved
\be
\nabla_\mu T^{\mu\nu}_{\rm matter} =\frac{A,_{\phi}}{A} T_{\rm matter} \partial^\nu\phi~,
\label{nonconservedtmunu}
\ee
where we have used the scalar equation of motion~\eqref{newtonianscreeningscalareom}, the fact that $\tilde T = A^{-4}T_{\rm matter}$, and the expression for the scalar stress-energy tensor~\eqref{scalarTmunu}.
If we consider a non-relativistic matter source ($T_{\rm matter} = -\tilde\rho$), specialize to an FLRW ansatz for the metric, and define the energy density $\rho\equiv A^{-1}T_{\rm matter}$, the $0$-component of equation~\eqref{nonconservedtmunu} reads
\be
\dot{\rho}+3H\rho = 0~.
\ee
so this density {\it is} conserved in the Einstein frame. Defining this quantity is
a useful thing to do, because in terms of $\rho$ the equation of motion~\eqref{newtonianscreeningscalareom} for the scalar takes a particularly simple form in terms of an {\it effective} potential
\be
\square\phi = V_{{\rm eff}\, ,\phi} (\phi)~~~~~~~~~~~~{\rm where}~~~~~~~~V_{\rm eff}(\phi) = V(\phi)+A(\phi)\rho~.
\label{effpoteom}
\ee
The crucial point here is that the matter density $\rho$ appears in the effective potential to which the scalar $\phi$ responds. Therefore, by suitably choosing the potential and coupling to matter, we can create a situation where the force due to the scalar field is hidden in regions of high density. In the following, we will consider two concrete realizations of this idea. The first is the {\it chameleon} mechanism, where the scalar field develops a large mass in regions of high density, and mediates a gravitational-strength force elsewhere. The other incarnation of this idea we will describe is the {\it symmetron} mechanism, where the effective potential is chosen such that the field $\phi$ develops a nonzero vacuum expectation value in regions of low density.

\subsection{Chameleon mechanism}
\label{cham}

\begin{figure}
\centering
\includegraphics[width=3in]{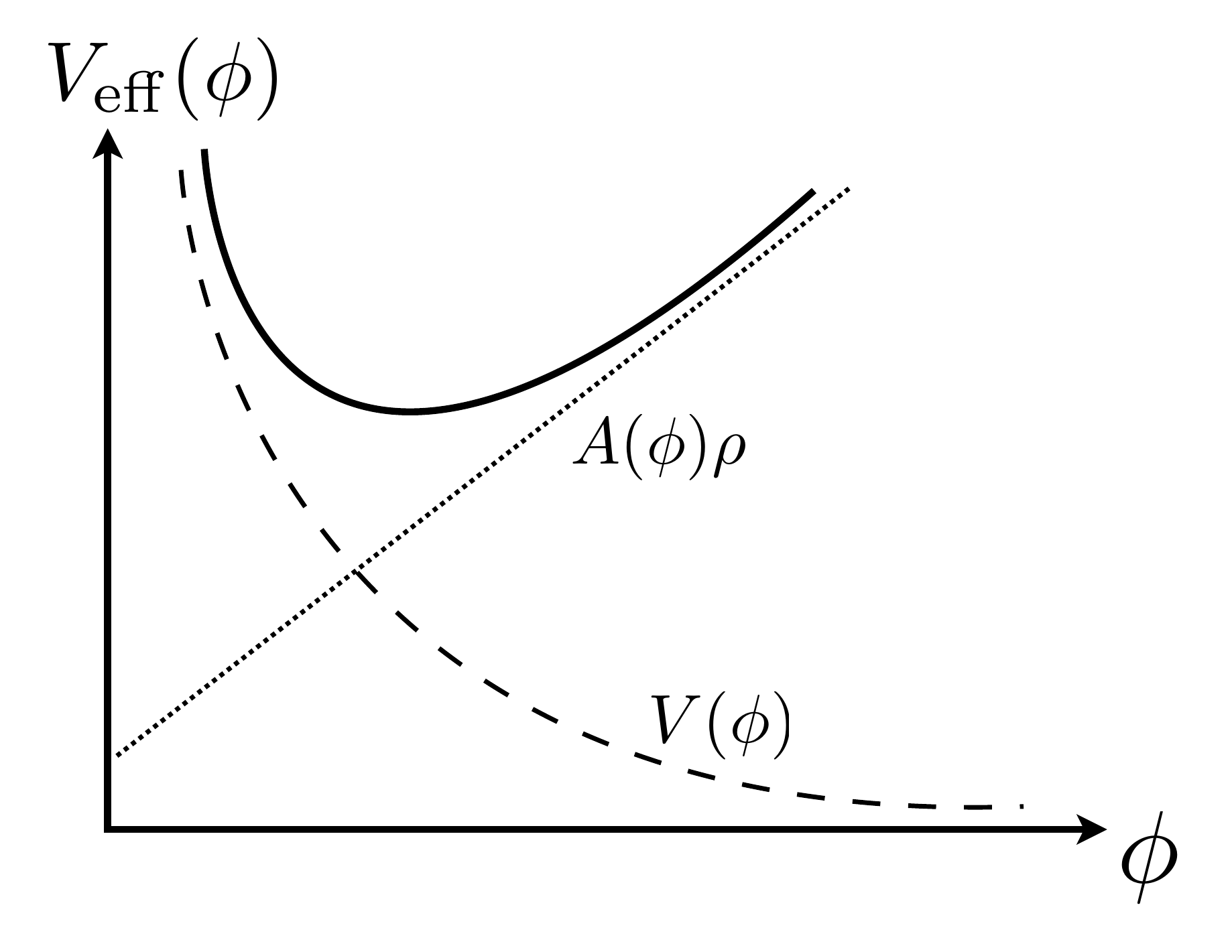}
\caption{\label{chameffpot} \small Sketch of the effective potential felt by a chameleon field (solid line). The effective potential is a sum of the bare potential of runaway form, $V(\phi)$ (dashed line) and a density-dependent piece, from coupling to matter (dotted line). Reproduced from~\cite{Jain:2010ka}.}
\end{figure}

We first consider the situation where a scalar field develops a density-dependent mass. This possibility was first explored in~\cite{Khoury:2003aq, Khoury:2003rn, Brax:2004qh}. Heuristically, the aim is to suitably choose $V(\phi)$ and $A(\phi)$ such that the mass coming from the effective potential
\be
m^2_{\rm eff}(\bar\phi) = V,_{\phi\phi}^{\rm eff}(\bar\phi) = V,_{\phi\phi}(\bar\phi) + A,_{\phi\phi}(\bar\phi)\rho~,
\label{meff}
\ee
is small in regions of low density and large in regions of high density, hiding the force from view. Various considerations constrain the functional form of the potential $V(\phi)$ and the coupling function $A(\phi)$~\cite{Jain:2010ka, Khoury:2003aq, Khoury:2003rn, Brax:2008hh}:\footnote{Of course these conditions need only be satisfied over the field range of interest, and not globally.}

\begin{itemize}

\item In order to have interesting effects, we want to balance the two contributions to the effective potential in~\eqref{effpoteom}. Without loss of generality, we assume that $A(\phi)$ is monotonically increasing ($A_{,\phi} > 0$), and that
$V(\phi)$ is monotonically decreasing ($V,_{\phi} < 0$) over the relevant field values.

\item Typically, the dominant contribution to the chameleon effective mass~\eqref{meff} comes from $V(\phi)$. For stability, we therefore require $V,_{\phi\phi} > 0$ over the relevant field range.

\item Finally, in order for the chameleon effective mass to increase with density, we must have $V,_{\phi\phi\phi} < 0$. 

\end{itemize}

We will see later (Sec.~\ref{cosmoeffects}) that chameleon field excursions, both temporally and spatially, are constrained by phenomenology to be much smaller than the Planck mass, $\Delta \phi \ll M_{\rm Pl}$. 
Hence the coupling function can be well-approximated by the linear form
\be
A(\phi) \simeq 1 + \xi \frac{\phi}{M_{\rm Pl}}\,.
\label{Alinear}
\ee
The constant $\xi$ must be positive, consistent with our aforementioned assumptions, and ${\cal O}(1)$ for a gravitational-strength scalar force. The linearized approximation cannot be applied to the potential $V(\phi)$, on the other hand, since this must necessarily involve a much smaller mass scale to achieve a large range in $m_{\rm eff}$.

We will focus on a few concrete models where Chameleon screening arises, but the phenomenon is fairly general and has been extensively studied beyond the contexts we discuss~\cite{Mota:2006fz,Mota:2006ed,Feldman:2006wg,Brax:2007ak,Das:2008iq,Davis:2009vk,Brax:2010tj,Brax:2010kv,Cannata:2010qd,Boddy:2012xs,Nastase:2013los, Erickcek:2013oma,Erickcek:2013dea}. For example, Chameleons have been extended to $K$-essence theories~\cite{Wei:2004rw}, used for inflation~\cite{Hinterbichler:2013we,Nastase:2013ik} and embedded in supersymmetry and string theory~\cite{Brax:2004ym,Hinterbichler:2010wu,Brax:2011qs,Brax:2012mq,Brax:2013yja}.

\subsubsection{A simple example}

\begin{figure}
\centering
\includegraphics[width=5in]{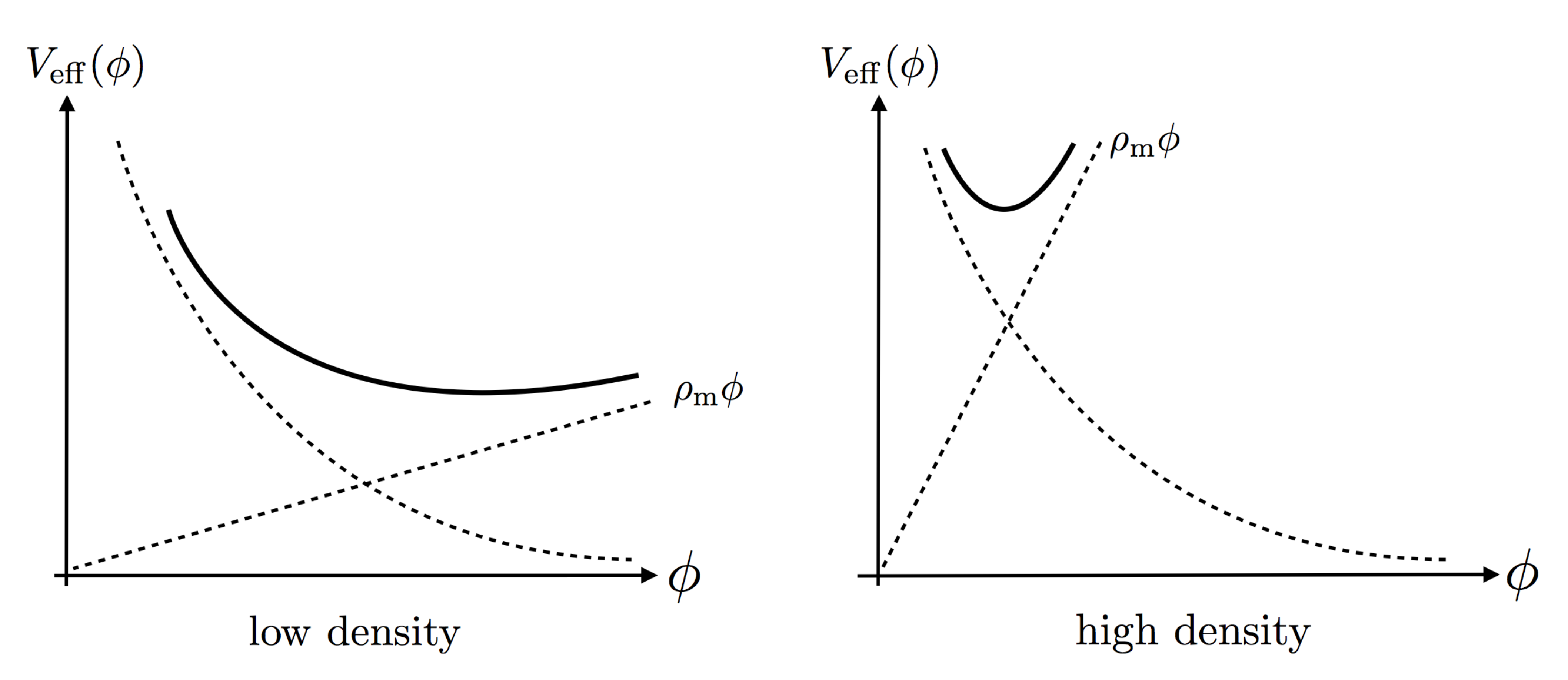}
\caption{\label{champotcomp} \small Comparison of chameleon effective potential in regions of low and high density. In regions of low density, the curvature of the potential is much shallower, corresponding to a light scalar that mediates a long range force. In regions of high density, the scalar acquires a large mass, and the force shuts off.}
\end{figure}

One potential which satisfies all of these constraints---the one originally considered in~\cite{Khoury:2003aq, Khoury:2003rn}---is the Ratra--Peebles inverse power-law potential~\eqref{RatraPeebles}, which is also and example of tracker quintessence models~\cite{Zlatev:1998tr, Steinhardt:1999nw}:
\be
V(\phi) = \frac{M^{4+n}}{\phi^n}~,
\label{trackerpot}
\ee
where $n > 0$ is a constant. With the coupling function \eqref{Alinear}, the minimum of the effective potential lies at
\be
\bar\phi(\rho) \approx \left(\frac{nM^{4+n}M_{\rm Pl}}{\xi \rho}\right)^\frac{1}{n+1}~,
\ee
and the effective mass of the $\phi$ fluctuations is given by
\be
m_{\rm eff}^2(\rho) \approx n(n+1)M^{-\frac{4+n}{1+n}}\left(\frac{\xi\rho}{nM_{\rm Pl}}\right)^\frac{n+2}{n+1}~.
\ee
Here it is manifest that the effective mass is a function of the ambient density, $\rho$, and that it increases with increasing density, exactly as desired.

The tightest constraint on the model comes from the so-called E$\ddot{{\rm o}}$t-Wash~\cite{Adelberger:2006dh} laboratory tests of the inverse square law, which set an upper limit of $\approx 50\;\mu$m
on the fifth-force range, assuming a gravitational-strength coupling. (In Section~\ref{chamlabtests} we comment more on local tests of chameleon theories.) Modeling the chameleon profile in the E$\ddot{{\rm o}}$t-Wash set-up, and taking into account
that torsion-balance measurements are performed in vacuum, this constraint translates into an upper bound on $M$~\cite{Khoury:2003aq,Khoury:2003rn}
\begin{equation}
M\; \lower .75ex \hbox{$\sim$} \llap{\raise .27ex \hbox{$<$}}\; 10^{-3}\;\;{\rm eV}\,,
\label{Mlim}
\end{equation}
which, remarkably, coincides with the dark energy scale. This also ensures consistency with all known constraints on deviations from GR, including post-Newtonian tests in the solar system and binary pulsar observations~\cite{Khoury:2003aq,Khoury:2003rn}.

\subsubsection{Spherically symmetric source and the thin-shell effect}
\label{thinshellsec}
In order to understand the details of how the chameleon force is suppressed in the presence of high ambient density, we solve for the field profile in the presence of a massive compact object, following~\cite{Khoury:2003rn}. We consider a spherically symmetric object with radius $R$, density $\rho_{\rm obj.}$ and total mass $M$, all of which are assumed to be constant for simplicity. Further, we imagine that this object exists in the presence of a homogeneous background of density $\rho_{\rm amb.}$. We denote by $\bar\phi_{\rm obj.}$ and $\bar\phi_{\rm amb.}$ the minima of the effective potential at the object and ambient density, respectively. Assuming that the coupling function takes the form~\eqref{Alinear}, the scalar equation of motion~\eqref{effpoteom} for a static and spherically-symmetric background reduces to
\be
\phi'' +\frac{2}{r}\phi' = V_{,\phi}  + \xi \frac{\rho}{M_{\rm Pl}}~,
\ee
where the density $\rho(r)$ is given by
\be
\rho(r) = \left\{\begin{array}{l}
\rho_{\rm obj.}~,~~~~~~~~r < R
\\
\rho_{\rm amb.}~,~~~~~~~r > R
\end{array}\right.~.
\ee
This is a second-order differential equation, and as such we must impose two boundary conditions. The first is that the solution be regular at the origin, that is $\rd\phi/\rd r = 0$ at $r=0$, and the second is that the field approach its ambient-density minimum $\phi_{\rm amb.}$ as $r\to \infty$.

For a given potential, such as the inverse power-law form~\eqref{trackerpot}, one can of course resort to numerical integration. However, it is instructive to derive the general solution through simple analytical arguments~\cite{Khoury:2003aq,Khoury:2003rn}. For a sufficiently large body---in a sense that will be made precise below---the field approaches the minimum of its effective potential deep in its interior:
\begin{figure}
\centering
\includegraphics[width=4.5in]{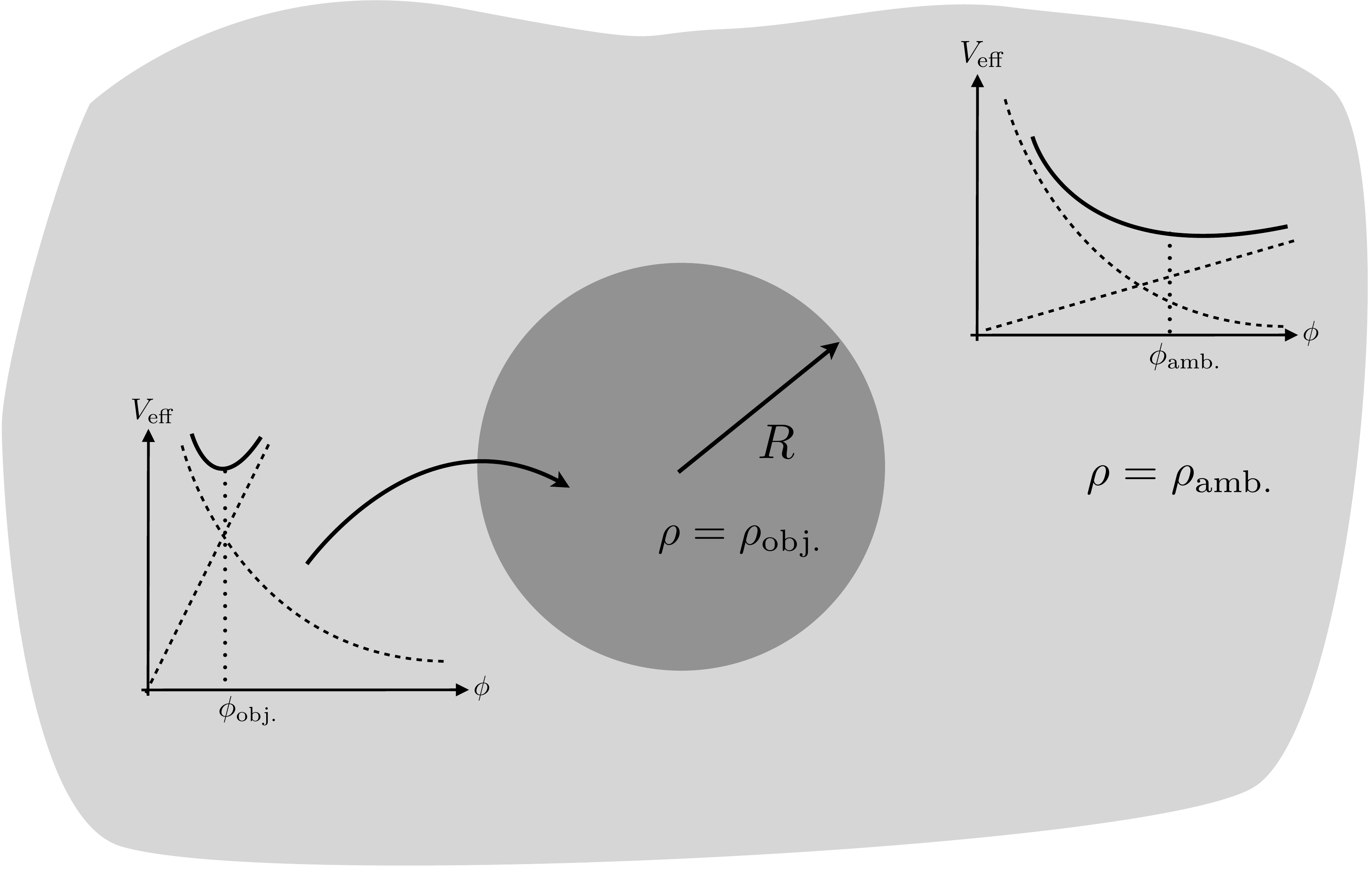}
\caption{\label{chambodyfig} \small Setup for the computation of the thin-shell effect of Section~\ref{thinshellsec}.}
\end{figure}
\be
\phi \simeq \phi_{\rm obj.}~;~~~~~~~~~~~~r < R \,.
\ee
Outside of the object, but still within an ambient Compton wavelength ($ r < m^{-1}_{\rm amb.}$) away, the field profile goes approximately as $1/r$:
\be
\phi \simeq \frac{A}{r} + B~;~~~~~~~~~~~~R < r < m^{-1}_{\rm amb.}\,.
\ee
The constants $A$ and $B$ are fixed by boundary conditions: 
\begin{itemize}
\item
imposing $\phi \to \phi_{\rm amb.}$ as $r\to\infty$ implies $B = \phi_{\rm amb.}$; 
\item
setting $\phi(R) = \phi_{\rm obj.}$ fixes $A = -R(\phi_{\rm amb.} - \phi_{\rm obj.})$.
\end{itemize}
The exterior solution therefore takes the form
\be
\phi \simeq - \frac{R}{r} (\phi_{\rm amb.} - \phi_{\rm obj.}) + \phi_{\rm amb.}\,.
\label{phiext}
\ee

There is a nice analogy between the above solution and electrostatics~\cite{JonesSmith:2011tn,Pourhasan:2011sm}. Indeed, since Laplace's equation, $\nabla^2 \phi\simeq 0$, approximately holds both inside and outside the source, the body acts as a conducting sphere---any chameleon charge is confined to a thin shell of thickness $\Delta R$ near the surface. The surface ``charge density" given by $\xi \rho \Delta R/M_{\rm Pl}$ supports the discontinuity in field gradients, as in electrostatics:
\be
\left.\frac{{\rm d} \phi}{{\rm d}r}\right\vert_{r=R_+} = \frac{\xi \rho}{M_{\rm Pl}} \Delta R\,.
\label{fielddisc}
\ee
Combining~\eqref{phiext} and~\eqref{fielddisc}, we can then solve for the shell thickness:
\be
\frac{\Delta R}{R} = \frac{\phi_{\rm amb.} - \phi_{\rm obj.}}{6\xi M_{\rm Pl}\Phi} \,,
\label{screencond}
\ee
where $\Phi \equiv M/8\pi M_{\rm Pl}^2R$ is the surface gravitational potential. The exterior field profile can thus be written as
\be
\phi(r> R) \simeq  -\frac{3\xi}{4\pi M_{\rm Pl}} \frac{\Delta R}{R} \frac{Me^{-m_{\rm amb.}(r-R)}}{r}+\phi_{\rm amb.}~,~~~~~~~~ ({\rm screened})\,,
\ee
where we have restored the Yukawa exponential factor since the field is massive. This profile is identical to that of a massive scalar of mass $m_{\rm amb.}$, except
that the coupling is reduced by the thin-shell factor $\Delta R/R \ll 1$. In this case the source is said to be {\it screened}.

Clearly, the above approximations break down if $\frac{\phi_{\rm amb.} - \phi_{\rm obj.}}{6\xi M_{\rm Pl}\Phi}\, \gsim\, 1 $. For fixed density contrast, this
corresponds to a weak source ({\it i.e.}, one with small $\Phi$). In this regime, the coupling is not suppressed by a thin-shell factor, and the object is said to be {\it unscreened}:
\be
\phi (r>R) \simeq  -\frac{\xi}{4\pi M_{\rm Pl}} \frac{Me^{-m_{\rm amb.}(r-R)}}{r}~,~~~~~~~~~~~~~~~~~~({\rm unscreened})\,.
\ee

This chameleon screening effect can be understood qualitatively as follows. For sufficiently massive objects, deep inside the object the chameleon will minimize the effective potential for the interior density. Correspondingly, the mass of fluctuations of the chameleon field is large inside the object. Therefore, there is a Yukawa suppression of the contribution from the core of the object to the exterior field profile. Only a thin shell beneath the surface contributes significantly to the exterior field profile. Another way of saying this is that the chameleon field effectively only couples to a thin shell beneath the surface of an object. In contrast, gravity couples to the entire mass of the object, therefore the chameleon force on an exterior test mass will be suppressed compared to the force due to gravity. The analogy of the thin shell effect to electrostatics~\cite{JonesSmith:2011tn,Pourhasan:2011sm} can also be exploited to compute the chameleon profile in the presence of more general distributions of matter using familiar methods.

\subsubsection{Example: $f(R)$ gravity}
A modification to gravity, which can exhibit chameleon screening, that has received a lot of attention is so-called $f(R)$ gravity; in these models, the Ricci scalar in the Einstein--Hilbert action is replaced by an arbitrary function of the Ricci scalar. Models involving higher curvature invariants have been interesting to theorists since early investigations by Stelle~\cite{Stelle:1976gc,Stelle:1977ry}, which showed in particular that a theory with quadratic curvature terms is renormalizable. In the context of inflation, Starobinsky wrote down a Lagrangian of the form ${\cal L} \sim R+R^2$ which was shown to be able to drive cosmic acceleration~\cite{Starobinsky:1980te} in the early universe. In~\cite{Capozziello:2002rd,Capozziello:2003tk,Carroll:2003wy}, this idea was adapted to explain the late-time acceleration of the universe, without invoking a cosmological scalar field. Since then, there have been numerous investigations involving models of this type. 

As was argued in Section~\ref{lookbeyondlcdm}, this modification introduces an additional scalar degree of freedom---indeed, we will see that the theory is equivalent to a scalar-tensor theory~\cite{Chiba:2003ir,Nunez:2004ji}. As such, there are various constraints on the functional form of $f(R)$, coming from theoretical considerations~\cite{Soussa:2003re,Dolgov:2003px,DeFelice:2006pg,Sawicki:2007tf}, cosmology~\cite{Amarzguioui:2005zq,Amendola:2006kh,Amendola:2006we} and solar system constraints~\cite{Flanagan:2003rb,Olmo:2005zr,Chiba:2006jp,Erickcek:2006vf,Jin:2006if,Multamaki:2006zb}. Indeed, many of the earliest studied models were ruled on out some of these grounds, with solar system constraints being the most stringent. However, classes of models were found which are compatible with solar system constraints, essentially by invoking the chameleon mechanism to screen the additional degree of freedom~\cite{Faulkner:2006ub,Navarro:2006mw,Hu:2007nk,Starobinsky:2007hu,Capozziello:2007eu,Tsujikawa:2007xu,Brax:2008hh}. The cosmologies of various models are studied in~\cite{Vollick:2003aw,Carloni:2004kp,Allemandi:2004wn,Capozziello:2006dj,Nojiri:2006gh,delaCruzDombriz:2006fj,Capozziello:2007ec,Motohashi:2010zz,Motohashi:2012wc}. Additionally, the behavior of perturbations and large-scale structure has been carefully studied in $f(R)$ gravity~\cite{Zhang:2005vt,Bean:2006up,Song:2006ej,Koivisto:2006ie,Hu:2007pj,Song:2007da,Pogosian:2007sw,Carloni:2007yv,Koyama:2009me,Motohashi:2009qn,Li:2011uw,Li:2011pj}. One intriguing possibility is that modifications of the $f(R)$ type could explain cosmic inflation and late-time acceleration in a unified way, which is investigated in~\cite{Nojiri:2003ni,Nojiri:2003ft,Nojiri:2007as,Nojiri:2007cq,Nojiri:2007uq,Cognola:2007zu,Nojiri:2008fk,Nojiri:2008nt,Bamba:2008xa}. It has been argued that higher-curvature theories can arise naturally from string theory under appropriate assumptons~\cite{Nojiri:2003rz,Biswas:2005qr}. Like Einstein gravity, these theories admit a Palatini formulation, which is quite subtle and has been extensive studied~\cite{Flanagan:2003iw,Meng:2003uv,Vollick:2003ic,Olmo:2005hc,Allemandi:2005qs,Koivisto:2005yc,Allemandi:2005tg,Sotiriou:2006qn, Sotiriou:2006hs,Sotiriou:2006sf,Carroll:2006jn,Iglesias:2007nv,Bertolami:2007gv,Fay:2007gg,Tsujikawa:2007tg,Olmo:2011uz}. Models which can cross the phantom divide and have effective equations of state $w < -1$ have been studied~\cite{Amendola:2007nt,Motohashi:2010qj,Motohashi:2010tb,Bamba:2010iy}, along with investigations into energy conditions in $f(R)$ gravity~\cite{Jacobson:1995uq,PerezBergliaffa:2006ni,Santos:2007bs}. In~\cite{Frolov:2008uf} it was pointed out that some models would be incompatible with the existence of relativistic stars, this issue is inverstigated in~\cite{Kainulainen:2007bt,Kobayashi:2008tq,Kobayashi:2008wc,Miranda:2009rs,Babichev:2009td,Upadhye:2009kt,Babichev:2009fi,Cooney:2009rr}. For some investigations into the quantum properties of these theories, see~\cite{Cognola:2005de,Cognola:2005sg,Machado:2007ea}. In~\cite{Dyer:2008hb}, the variational principle of higher-derivative theories is examined and formulated systematically. Additionally, N-body simulations of gravitational collapse and clustering have been carried out in~\cite{Schmidt:2008tn,Oyaizu:2008sr,Oyaizu:2008tb,Khoury:2009tk,Schmidt:2009sv,Chan:2009ew,Wyman:2013jaa,Zhao:2010qy,Ferraro:2010gh,Li:2012by,Li:2013tda,Baldi:2013iza,He:2014eva}. Functions of other curvature invariants have also been considered~\cite{Carroll:2004de,Nunez:2004ts,Nojiri:2005jg,Cognola:2006eg,Capozziello:2006uv}, but $f(R)$ theories are distinguished in that they (and $f$(Gauss--Bonnet) theories) do not have ghosts. For reviews of all of these topics and more, see~\cite{Woodard:2006nt,Sotiriou:2008rp,DeFelice:2010aj,Nojiri:2010wj}.

The action for this modification to Einstein gravity is of the form
\be
S = \frac{M^2_{\rm Pl}}{2}\int\rd^4x\sqrt{-g}\Big(R+f(R)\Big) + S_{\rm matter}[g_{\mu\nu}, \psi]~,
\label{fRaction}
\ee
where we have assumed that the matter fields, $\psi$, couple minimally to the metric $g_{\mu\nu}$, which has Ricci scalar $R$. In fact this theory is classically equivalent to a scalar-tensor theory~\cite{Chiba:2003ir,Nunez:2004ji}. To see this, consider the alternate action
\be
S = \frac{M^2_{\rm Pl}}{2}\int \rd^4x\sqrt{-g}\left(R+f(\Phi)+\frac{\rd f}{\rd\Phi}(R-\Phi)\right)+ S_{\rm matter}[g_{\mu\nu}, \psi] \ ,
\label{frscalaraction}
\ee
with equations of motion~\cite{Song:2006ej,Hu:2007nk,Silvestri:2009hh}\footnote{Note that we have assumed that $f_{,\Phi\Phi} \neq 0$ in the $\Phi$ equation of motion.}
\begin{align}
\label{frgeom}
&(1+f_R)R_{\mu\nu}-\frac{1}{2}\left(R+f-2\square f_R\right)g_{\mu\nu}-\nabla_\mu\nabla_\nu f_R = \frac{1}{M_{\rm Pl}^2}T_{\mu\nu}^{\rm matter}\\\label{frphieom}
&\Phi = R~,
\end{align}
where $f_R\equiv \rd f/\rd R = \rd f/\rd\Phi$.
From equation~\eqref{frphieom}, we see that $\Phi$ is an auxiliary field---its equation of motion is non-dynamical (it does not involve time derivatives of $\Phi$). At the classical level, we may therefore use this equation to eliminate $\Phi$ from the action and reproduce the $f(R)$ action~\eqref{fRaction}.\footnote{In field theory language, we are {\it integrating out} $\Phi$ at tree level.} We also note that the trace of~\eqref{frgeom} can be cast as an equation for the scalar degree of freedom $f_R$:
\be
\square f_R = \frac{1}{3}\left(R+2f-f_RR+\frac{1}{M_{\rm Pl}^2}T\right) \equiv \frac{\rd V_{\rm eff}(f_R)}{\rd f_R}~,
\ee
where $V_{\rm eff}(f_R)$ is an effective potential. Taking another derivative, we obtain the effective mass of the scalar $f_R$
\be
m_{\rm eff}^2(f_R) = \frac{1}{3}\left(\frac{1+f_R}{f_{RR}}-R\right)~,
\label{frscalarmass}
\ee
where $f_{RR}\equiv  \rd^2f/\rd R^2$.

In fact the action~\eqref{frscalaraction} is nothing more than Einstein gravity plus a canonical scalar non-minimally coupled, albeit in disguise. To make this explicit, we simultaneously make a conformal transformation and a field redefinition
\be
\tilde g_{\mu\nu} = \left(1+\frac{\rd f}{\rd\Phi}\right) g_{\mu\nu}~,~~~~~~~~~~~~~~\phi = -\sqrt{\frac{3}{2}}M_{\rm Pl}\log\left(1+\frac{\rd f}{\rd\Phi}\right)~.
\ee
This leads to the action
\be
S =\int \rd^4x\sqrt{-\tilde g}\left( \frac{M^2_{\rm Pl}}{2}\tilde{R}-\frac{1}{2}\tilde g^{\mu\nu}\partial_\mu\phi\partial_\nu\phi - V(\phi)\right)+S_{\rm matter}[e^{\sqrt{2/3}\phi/M_{\rm Pl}}\tilde g_{\mu\nu}, \psi]~,
\ee
where we have defined
\be
V(\phi) = \frac{M_{\rm Pl}^2}{2}\frac{\left(\phi\frac{\rd f}{\rd\phi}-f(\phi)\right)}{\left(1+\frac{\rd f}{\rd\phi}\right)^2}~.
\ee
The action now takes the form~\eqref{scalartensoraction} with $A^2(\phi)=e^{\sqrt{2/3}\phi/M_{\rm Pl}}$, and for a suitable choice of $V(\phi)$ will exhibit chameleon screening. The potential for the scalar field is set by our choice of the function $f$, and an important thing to note is that theories that look extremely complicated in one description may be simple from the other perspective. For example, simple functions of $R$ often correspond to non-analytic potentials for the scalar $\phi$, and vice versa.

Such a modification to gravity is interesting because the additional scalar degree of freedom acts as another source in Einstein's equations, which can drive cosmic acceleration. Indeed, specializing to an FLRW background, the Friedmann and acceleration equations take the form~\cite{Song:2006ej,Hu:2007nk,Silvestri:2009hh}
\begin{align}
H^2 +\frac{f}{6}-\frac{\ddot a}{a}f_{R}+H\dot f_R &= \frac{1}{3 M_{\rm Pl}^2}\rho\\
\frac{\ddot a}{a}-f_RH^2+\frac{f}{6}+\frac{\ddot f_R}{2} &= -\frac{1}{6M_{\rm Pl}^2}(\rho+3P)~.
\end{align}
Here, when $f(R) \neq 0$, the additional contributions can be interpreted as a perfect fluid with equation of state
\be
w_{\rm eff} = -\frac{1}{3}-\frac{2}{3}\frac{\big(H^2 f_R-f/6-H\dot f_R-\ddot f_R/2\big)}{\big(-H^2 f_R-f/6-H\dot f_R+f_R R/6\big)}~.
\ee
By suitably choosing the functional form of $f(R)$, one can then reproduce any expansion history desired.

The function $f(R)$ in~\eqref{fRaction} is not a completely free function. There are various constraints on its form coming both from theoretical consistency and phenomenological viability~\cite{Silvestri:2009hh}:

\begin{itemize}
\item In regions of high curvature, where $\lvert Rf_{RR}\rvert \ll 1$ and $f_R \sim 0$, the equation for the mass of the scalar~\eqref{frscalarmass} reduces to
\be
m_{\rm eff}^2(f_R) \approx \frac{1}{3f_{RR}}~.
\ee
In order to avoid tachyons, we must have $f_{RR} > 0$ in this regime~\cite{Dolgov:2003px, Sawicki:2007tf,Song:2006ej}.

\item To keep the graviton from becoming a ghost, we must have $1+f_R > 0$ everywhere~\cite{Nunez:2004ji}.

\item Gravitation is very well-tested in the early universe through, for example, primordial nucleosynthesis, and so we would like to recover Einstein gravity at early times (high curvatures). Formally, this means we must have
\be
\frac{f(R)}{R} \longrightarrow 0~~~~~{\rm and}~~~~~f_R\longrightarrow 0~~~~~{\rm as}~~~~~R\longrightarrow\infty~.
\ee
Since we have $f_{RR} > 0$ everywhere, $f_R$ must approach $0$ from below as $R\to\infty$ so $f_R$ is a monotonically increasing {\it negative} function, $f_R < 0$~\cite{Silvestri:2009hh,Hu:2007nk}.

\item Finally, we want to satisfy solar system constraints on fifth forces. In order for this to be true, we must have $\lvert f_R\rvert \ll 1$ in the present universe. This is because this quantity is what sources the fifth force~\cite{Khoury:2003rn}. In~\cite{Hu:2007nk} the bound of $\lvert f_R\rvert~\lsim~ 10^{-6}$ was obtained. We will discuss more stringent bounds later on.
\end{itemize}
With these constraints in mind, two well-studied choices for the form of the function $f(R)$ are the {\it Hu--Sawicki} model~\cite{Hu:2007nk}\footnote{We have adopted the conventions of~\cite{Jain:2010ka} for the HS model, which takes $M^2 = m^2/c_2^{1/n}$ and $a = c_1c_2^{1/n-1}$ relative to~\cite{Hu:2007nk}.}
\be
f(R) = - \frac{aM^2}{1+\left(\frac{R}{M^2}\right)^{-\alpha}}~,
\ee
and the {\it Starobinsky} model~\cite{Starobinsky:2007hu}
\be
f(R) = aM^2\left[\left(1+\frac{R^2}{M^4}\right)^{-\frac{\alpha}{2}}-1\right]~,
\ee
where both $a, \alpha >0$. The cosmologically relevant regime is $R \gg M^2$, where both models take the same form~\cite{Jain:2010ka}
\be
f(R) \approx aM^2\left[ \left(\frac{R}{M^2}\right)^{-\alpha}-1\right]~.
\ee
For completeness, we note that in the Einstein frame, this corresponds to a potential for the scalar of the form~\cite{Jain:2010ka}
\be
V(\phi)\approx \frac{a}{2}M^2M_{\rm Pl}^2\left[1-(\alpha+1)\left(\sqrt{\frac{2}{3}}\frac{1}{a\alpha}\frac{\phi}{M_{\rm Pl}}\right)^\frac{\alpha}{1+\alpha}\right]~.
\ee

\subsection{Symmetron mechanism}
\label{symm}

Another method of screening long range forces is to weaken the coupling to matter in regions of high density or Newtonian potential. One way of doing this is the {\it symmetron} mechanism~\cite{Hinterbichler:2010es,Hinterbichler:2011ca}, in which the coupling of the scalar to matter is proportional to the vacuum expectation value (VEV) of the field. (For earlier related work, see~\cite{Pietroni:2005pv,Olive:2007aj}.) The effective potential is chosen so that the field acquires a nonzero VEV in low-density regions, but the symmetry is restored in high density regions. Therefore, in such high density regions, the field has zero VEV and so does not couple to matter. In regions of low density, the field spontaneously breaks some symmetry and acquires a VEV, allowing it to couple to matter and mediate a force. The cosmology of this theory is studied in~\cite{Hinterbichler:2011ca}, while perturbations and large-scale structure are examined in~\cite{Clampitt:2011mx,Brax:2011pk,Llinares:2012ds,Taddei:2013bsk}. For N-body simulations of the matter power spectrum and halo mass function, see~\cite{Davis:2011pj,Winther:2011qb}. If the symmetron couples to electromagnetism, there can be interesting signatures of cosmological domain walls~\cite{Olive:2010vh,Olive:2012ck}. Symmetron fields have also been used for inflation~\cite{Dong:2013swa}.

\subsubsection{A ${\mathbb Z}_2$ example}

Here we consider the simplest incarnation of this mechanism, studied in~\cite{Hinterbichler:2010es}: a ${\mathbb Z}_2$-symmetric action of the symmetry-breaking form
\be
S = \int\rd^4x\sqrt{-g}\left(\frac{M_{\rm Pl}^2}{2}R-\frac{1}{2}(\partial\phi)^2+\frac{\mu^2}{2}\phi^2-\frac{\lambda}{4}\phi^4\right)+S_{\rm matter}\left[\left(1+\frac{1}{2M^2}\phi^2\right)^2 g_{\mu\nu}, \psi\right]~.
\ee
Clearly, this is of the form~\eqref{scalartensoraction} with the potential
\be
V(\phi) = -\frac{\mu^2}{2}\phi^2+\frac{\lambda}{4}\phi^4~,
\ee
and the coupling to matter given by
\be
A(\phi) = 1+\frac{1}{2M^2}\phi^2+{\cal O}\left(\frac{\phi^4}{M^4}\right)~.
\label{symmAfunct}
\ee
Here, $M$ is some high mass scale, so that the terms we have ignored are negligible ($\phi\ll M$). It is then clear from~\eqref{effpoteom} that the effective potential felt by $\phi$ in the presence of a non-relativistic source is given by
\be
V_{\rm eff}(\phi) = \frac{1}{2}\left(\frac{\rho}{M^2}-\mu^2\right)\phi^2+\frac{\lambda}{4}\phi^4~.
\ee
So we see that $\rho$ acts precisely like an effective mass! Whether or not the ${\mathbb Z}_2$ symmetry is spontaneously broken depends on the ambient matter density. In regions of low density, the symmetry is spontaneously broken and $\phi$ acquires a vacuum expectation value
\be
\bar\phi = \sqrt\frac{\mu^2}{\lambda}~.
\ee
However, in regions of high density, $\rho \gg M^2\mu^2$, and the symmetry is restored. The key insight here is that the force due to $\phi$ is mediated by fluctuations about these background values $(\phi = \bar\phi+\delta\phi)$. Since it is quadratic in $\phi$, the coupling to matter is {\it proportional} to the vacuum expectation value:
\be
\frac{\phi^2}{M^2} T\sim \frac{\bar\phi}{M^2}\delta\phi ~\rho~.
\ee
Therefore in high density regions, where $\phi$ has {\it no} VEV, the fluctuations $\delta\phi$ do not couple to matter! Thus, $\phi$ does not mediate a fifth force in these regions.

The symmetron has a thin-shell effect similar to that of the chameleon~\cite{Hinterbichler:2010es}. Consider once again the ideal case of a static, spherically-symmetric source of homogeneous density $\rho >  \mu^2M^2$. For simplicity, we assume that the object lies in vacuum, so that the symmetron tends to its symmetry-breaking VEV far away: $\phi\rightarrow\bar\phi$ as $r\rightarrow \infty$. 

For a sufficiently massive source, in a sense that will be made precise shortly, the solution has the following qualitative behavior. Deep in the core of the object, 
the symmetron is weakly coupled to matter, since the matter density forces $\phi\approx 0$ there. Near the surface, meanwhile, the field must grow
away from $\phi = 0$ in order to asymptote to the symmetry-breaking VEV far away. The symmetron is thus weakly coupled to the core of the object,
and its exterior profile is dominated by the surface contribution. In other words, analogously to chameleon models, there is a thin-shell screening effect
suppressing the symmetron force on an external probe.

Explicit calculations show that whether screening occurs or not depends on the parameter~\cite{Hinterbichler:2010es}
\be
\alpha \equiv \frac{\rho R^2}{M^2} = 6\frac{M_{\rm Pl}^2}{M^2}\Phi \,.
\label{alp}
\ee
Objects with $\alpha\gg 1$ display thin-shell screening, and the resulting symmetron-mediated force on a test particle
is suppressed by $1/\alpha$ compared to the gravitational force. Objects with $\alpha \ll 1$, on the other hand, do not have a thin shell---the symmetron gives an ${\cal O}(1)$ correction to the gravitational attraction in this case.

\subsubsection{Dilaton screening}

A third type of density-dependent screening mechanism, which we will mention only briefly here, is the environmentally-dependent dilation mechanism of~\cite{Damour:1994zq, Brax:2011ja}. Conceptually, this mechanism is similar to the symmetron, but here the potential and coupling to matter are given by
\be
V(\phi) = V_0e^{-\phi/M_{\rm Pl}} ~,~~~~~~{\rm and}~~~~~A(\phi) = 1+\frac{1}{2M}(\phi-\phi_\star)^2~.
\ee
In dense regions, where $\phi \approx \phi_\star$, the coupling to matter is negligible, while in low density regions the field $\phi$ mediates a gravitational-strength force.

\subsection{Cosmological effects}
\label{cosmoeffects}

A general limitation of the  chameleon, symmetron and varying-dilaton mechanisms discussed in this Section---and more generally of any mechanism whose screening condition
is set by the local Newtonian potential---is that the range of the scalar-mediated force on cosmological scales is bounded:
\be
m_{\rm cosmo}^{-1} \lesssim {\rm Mpc}\,.
\label{mbound}
\ee
Hence, these mechanisms have negligible effect on density perturbations on linear scales today. Correspondingly, the deviation from the $\Lambda$CDM background cosmology due to the scalar energy density is negligible. That said, these mechanisms remain interesting as a way to hide light scalars suggested by fundamental theories. The way to test these theories is to study small scale phenomena,
as we will review below.

The bound~\eqref{mbound} on the Compton wavelength was shown in detail and under very general conditions in~\cite{Wang:2012kj}.
We sketch the proof for the simplest case of the chameleon theory discussed in Section~\ref{thinshellsec} with monotonically-decreasing $V(\phi)$ and
$\xi \sim {\cal O}(1)$. The starting point is to require that the Milky Way galaxy be screened:
\be
\frac{\phi_{0}}{6M_{\rm Pl}\Phi_{\rm MW}} \lesssim 1\,,
\ee
where $\phi_0$ is the present cosmological value of the chameleon, and where we have assumed the coupling is $\xi \sim {\cal O}(1)$. Clearly, this should be a necessary condition to satisfy local tests of gravity. Since the gravitational potential of the Milky Way is $\Phi_{\rm MW} \sim 10^{-6}$,
this implies a bound on the field excursion on cosmological scales:
\be
\phi_0 \lesssim 10^{-6}\; M_{\rm Pl}\,.
\label{phi0bound}
\ee
On the other hand, consider the scalar evolution equation on cosmological scales:
\be
\ddot{\phi} + 3H\dot{\phi} = -V_{,\phi} -   \xi\frac{\rho}{M_{\rm Pl}}\,.
\ee
The source term $\sim \rho/M_{\rm Pl}$ exerts a significant pull on the chameleon. If left unabated, it would drive the field over a distance $\sim M_{\rm Pl}$ in a Hubble time,
in conflict with~\eqref{phi0bound}. Of course, this is prevented by the potential term, provided that
\be
V_{,\phi} \simeq -   \xi\frac{\rho}{M_{\rm Pl}}\,.
\ee
This condition must hold at least for a Hubble time, over which time the density changes by $\Delta\rho \sim H^2_0 M_{\rm Pl}^2$. Using the relation $m_{\rm cosmo}^2 = V_{,\phi\phi}$, we have
\be
\Delta V_{,\phi} \simeq m_{\rm cosmo}^2 \Delta\phi \sim H^2_0 M_{\rm Pl}\,.
\ee
From~\eqref{phi0bound}, it follows that $\Delta\phi \lesssim 10^{-6}\; M_{\rm Pl}$, and hence
\be
 m_{\rm cosmo}^{-1}  \,\lsim\, 10^{-3} \;H_0^{-1} \sim {\rm Mpc}\,,
 \ee
as advertised. This bound was also derived in~\cite{Brax:2011aw} using slightly different arguments.\footnote{For a similar argument in the context of the symmetron, see~\cite{Bamba:2012yf}.}

The bound of $10^{-6}\;M_{\rm Pl}$ on the field excursion also implies that the scalar energy density must be negligible, modulo a cosmological constant contribution. Indeed, using $\dot{\phi} \sim \phi_0 H_0$, the kinetic energy is suppressed:
\be
\Omega_{\rm kin} \sim  \frac{\dot{\phi}^2}{H^2_0 M_{\rm Pl}^2} \sim \frac{\phi_0^2}{M_{\rm Pl}^2} \lesssim 10^{-12} \,.
\ee
The potential energy (ignoring a constant contribution) is also suppressed:
\be
\Omega_{\rm pot} \sim \frac{|\Delta V|}{H^2_0M_{\rm Pl}^2} \sim \frac{|V_{,\phi}| \Delta\phi}{H^2_0M_{\rm Pl}^2} \sim \xi \frac{\rho\phi_0}{H^2_0 M_{\rm Pl}^3} \lesssim 10^{-6}\,,
\ee
where we have used $\rho \sim H^2_0M_{\rm Pl}^2$. Thus the expansion history is indistinguishable from the $\Lambda$CDM model. A related fact is that the conformal factor $A(\phi)\simeq 1+ \xi\frac{\phi}{M_{\rm Pl}}$ relating the Einstein-frame and Jordan-frame metrics is approximately constant:
\be
\frac{\Delta A}{A} \sim \frac{\Delta \phi}{M_{\rm Pl}} \lesssim 10^{-6}\,.
\ee
(This confirms the claim made earlier, and justifies the linear approximation of~\eqref{Alinear} for chameleons.) This precludes the possibility of {\it self-acceleration}, {\it i.e.}, acceleration not due to a form of dark energy,  but instead from a varying conformal factor~\cite{Wang:2012kj}. 

\subsection{Radiative stability: 1-loop considerations}

While much of the work on chameleon theories has focused on their classical description, it is crucial to understand the robustness of the screening
mechanism to quantum corrections. At first sight, this question appears to be trivial: chameleons couple to matter fields and gravitons, hence matter or graviton
loops should generate quadratically-divergent radiative corrections to the chameleon mass: $\Delta m_{\rm eff} \sim \Lambda^2_{\rm UV} /M_{\rm Pl}^2$. However, as
we will see below, the cutoff $\Lambda_{\rm UV}$ is generally so small ($\Lambda_{\rm UV} \sim {\rm meV}$) that these corrections are completely negligible. 

Following~\cite{Upadhye:2012vh} we focus on quantum corrections due to the one-loop Coleman--Weinberg correction (see, for example,~\cite{Weinberg:1996kr}):
\begin{equation}
\Delta V(\phi)
=
\frac{m^4_{\rm eff} (\phi)}{64\pi^2}
\ln \left(\frac{m^2_{\rm eff} (\phi)}{\mu^2}\right)\,,
\label{e:DVloop}
\end{equation}
where $\mu$ is an arbitrary mass scale. Since this correction grows with increasing chameleon mass as $m^4_{\rm eff}$, it is immediately clear that quantum corrections can present problems for chameleon theories.  Chameleon screening of fifth forces operates by increasing $m_{\rm eff}$, so quantum corrections must become important above some effective mass. On the other hand, laboratory measurements place a lower bound on the effective mass. This causes tension between a model's classical predictivity and the predictions that it makes. Viable chameleons must tiptoe between being heavy enough to avoid fifth force constraints and remaining light enough to keep quantum corrections under control. 

\begin{figure}[tb]
\centering
\includegraphics[width=4.2in]{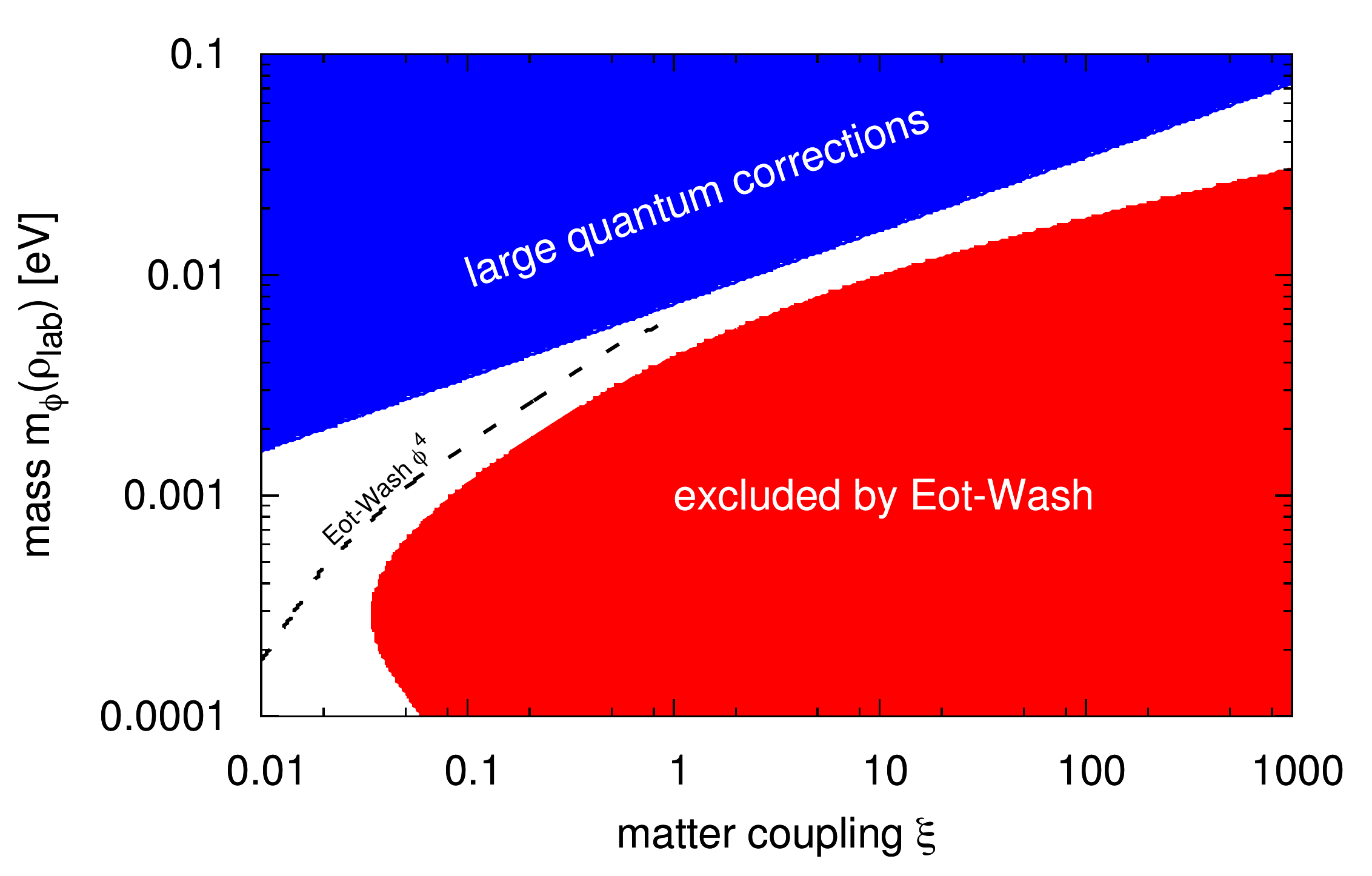}
\caption{\small Model-independent constraints on chameleon fields in the $\xi$,~$m_\phi$ plane with $\rholab = 10$~g/cm$^3$, reproduced from~\cite{Upadhye:2012vh}.  Shaded regions show loop bounds from \eqref{e:mbound} and experimental constraints from E\"ot-Wash~\cite{Kapner:2006si}.  The dashed curve shows the direct bound on the $\phi^4$ model  for $\xi<1$~\cite{Adelberger:2006dh}, converted to $m_{\rm eff}$.
  \label{f:mloop_and_eotwash}}
\end{figure}

Quantitatively, it is possible to derive the following model-independent bound on the Compton wavelength $m_{\rm eff}^{-1}$ at laboratory density in order for quantum corrections to be under control:
\begin{equation}
m_{\rm eff}
\lesssim
\left(\frac{48\pi^2 \bmat^2 \rholab^2}{\Mpl^2}\right)^\frac{1}{6}
=
0.0073 \left( \frac{\bmat \rholab}{10~{\rm g\,  cm}^{-3}} \right)^\frac{1}{3}  \,  {\rm eV}\,.
\label{e:mbound}
\end{equation}
Remarkably, this bound holds quite generally, independent of the detailed form of $V(\phi)$. For gravitational coupling ($\bmat \sim 1$) and typical densities, this mass scale is close to the dark energy scale of $\rho_\Lambda^{1/4} = 0.0024$\,eV. This results from the numerical coincidence that $(\rholab/\Mpl)^{4/3} \sim \rho_\Lambda$. Most importantly, the Compton wavelength corresponding to this maximum mass, $0.027 (\bmat \rholab/10~{\rm g\, cm}^{-3})^{-1/3}$~mm, is comparable to the length scales probed by the smallest-scale torsion pendulum experiments. 

This tension is shown in Figure~\ref{f:mloop_and_eotwash}, reproduced from~\cite{Upadhye:2012vh}. The E\"ot-Wash constraint~\cite{Kapner:2006si} assumes Yukawa potentials (with constant mass $m$). To
translate these Yukawa constraints to the chameleon case,~\cite{Upadhye:2012vh} made the conservative approximation of substituting for $m$ the maximum chameleon mass achieved in the experimental set-up.
This is conservative because the range of the chameleon-mediated force is in actuality {\it longer} in parts of the experiment, which should result in a tighter constraint. This is shown explicitly for a $\phi^4$ potential as the dashed curved in Figure~\ref{f:mloop_and_eotwash}---the excluded region widens, as expected, but still leaves a narrow window of allowed models. An improvement of a factor of $\simeq 2$ in the Yukawa range could close the gap and eliminate {\it all} chameleon theories around  $\bmat=1$ whose quantum corrections are under control.

\section{Screening with higher-derivative interactions: $\partial\phi; \partial^2\phi~\gsim~ \Lambda$}

As we discussed earlier, around a particular  background the behavior of new degrees of freedom---be they new fields, or part of the metric---can be captured in the effective field theory formalism. In the previous Section, we studied terms that either self-couple such fields, or couple them in nontrivial ways to matter, and showed how the na\"ive dynamics can be substantially altered in regions of high density. In this subsection we explore the other logical possibility.

In the effective Lagrangian, the only other types of terms that are generated beyond self couplings and couplings to matter are those terms involving derivatives of the new fields. When we write down our first field theories, we immediately encounter the simplest example of such a term as the kinetic term that is a quadratic term in derivatives of the field, and thus generates the evolution part of the resulting second order equations of motion. Indeed, even before thinking of field theories, it is the presence of the kinetic energy term in the Lagrangian for a point particle that gives rise to the acceleration part of $F=ma$. 

At first glance, there is a multitude of Lorentz-invariant terms involving derivatives that could form an infinite sequence of contributions to the effective theory just as, for example, higher and higher powers of fields are possible in the potential and coupling parts of the theory. However, there are two important reasons that---in many cases---such terms are not expected to be important. The first reason is simply the point of the effective field theory approach. Kinetic terms are already marginal operators in the Lagrangian, and any higher powers of them must therefore be irrelevant operators, and as such will come suppressed by an appropriate power of the UV cutoff of the theory, $\Lambda$. Since effective field theories are useful within their regimes of validity (organized as an expansion in energy over the cutoff), it is easy to see that, in general, whenever such higher order terms become important---that is $E/\Lambda\sim 1$---the entire infinite tower of such terms becomes equally important, and the field theory will break down.

The second reason that theorists are wary of Lagrangians with higher order derivatives is because, as we have mentioned before, in general if we go beyond first derivatives in the action, then, aside from a few special cases (that we will discuss), systems with more than two time
derivatives in their equations of motion admit an equivalent classical description in terms of a Lagrangian that generically contains ghosts. In the effective field theory approach such terms are, of course, expected in the Lagrangian, but, again, terms that contain higher
derivatives appear from an expansion of an unknown UV--complete theory. This expansion provides an accurate description of the full theory only at
low energies, and the physical degrees of freedom are assumed to be only those that appear in the ground state of the theory.
Thus, classically, the presence of extra solutions to the field equations that correspond to the existence of ghosts is an artifact of the effective theory, due to the truncation of an infinite series, and if it is possible to push the masses of these degrees of freedom beyond the cutoff, then the ghosts can be ignored. As we have also discussed, ghosts, and the validity of the effective field theory are not the only things to be worried about in theories with nontrivial derivative interactions---such theories may pose new challenges, such as the existence of superluminally propagating modes around certain backgrounds. Furthermore, while we will work within the effective field theory approach, there is an implicit assumption that it is possible, at least in principle, to find a UV completion of the theory, for which our approach describes the correct infrared (IR) physics. It turns out that even this cannot always be guaranteed, providing yet another powerful technical constraint on the existence of viable theories with higher-derivative interactions. (See Appendix~\ref{superlumapp} for a detailed discussion.)

The considerations mentioned above mean that a great deal of care is required when constructing theories in which derivative interaction terms play an important part. Nevertheless, a number of important effects are possible in models that evade the above worries in interesting ways, and which provide a new way to implement screening.

There are essentially two closely-related ways in which screening through derivatives can take place. The first is what we term {\it kinetic screening}, in which only first derivatives of the relevant fields enter the Lagrangian, thus ensuring second order equations of motion and the absence of ghosts in the most obvious manner. The second possibility is through the {\it Vainshtein effect}, in which higher derivative terms enter the Lagrangian, ghosts are avoided in a more subtle way, and, roughly speaking, the relevant physics is sensitive to the local field-space curvature $\partial\partial\Phi$. In what follows we shall consider these separately for pedagogical reasons, and ultimately will pay particular attention to models in which the Vainshtein effect is active.

\subsection{Kinetic screening}
For simplicity, let us focus on scalar fields, generically written as $\phi$. Lagrangians with exotic kinetic terms, but involving only first derivatives of fields are often referred to as $P(X, \phi)$ models, with $X\equiv -\frac{1}{2}\partial_{\mu} \phi\partial^{\mu} \phi$, since the requirement that the Lagrangian be a Lorentz scalar means that the first derivatives must appear in this combination. 

Models of this type have been widely applied to cosmology. They first appeared in the so-called {\it K-inflation} models~\cite{ArmendarizPicon:1999rj,Garriga:1999vw}, for other studies of kinetic-driven inflation, see~\cite{ArkaniHamed:2003uz,Senatore:2004rj,Mukhanov:2005bu,Langlois:2008wt} and for discussion of non-Gaussianity in these models, see~\cite{Seery:2005wm,Chen:2006nt,Chen:2006xjb,Huang:2006eha,Langlois:2008qf,Arroja:2008yy,Langlois:2008mn,Khoury:2008wj,RenauxPetel:2008gi,Chen:2009bc,Mizuno:2010ag,RenauxPetel:2011dv,Ribeiro:2012ar}. Models with non-canonical kinetic structure have also been much studied for cosmic acceleration, where they go by the name {\it K-essence}~\cite{Chiba:1999ka,ArmendarizPicon:2000dh,ArmendarizPicon:2000ah,Chiba:2002mw,Padmanabhan:2002cp,Malquarti:2003nn,Malquarti:2003hn,Chimento:2003zf,Silverstein:2003hf,GonzalezDiaz:2003rf,Scherrer:2004au,Aguirregabiria:2004te,Piazza:2004df,Rendall:2005fv,Bonvin:2006vc,Babichev:2007dw,dePutter:2007ny,Kang:2007vs,Bilic:2008zk,Martin:2008xw,Myrzakulov:2010tc}. A particular functional form of $P(X, \phi)$ which has attracted considerable attention are Dirac--Born--Infeld (DBI) models, which arise in the action for $D$-branes in string theory~\cite{Padmanabhan:2002cp,Silverstein:2003hf,Alishahiha:2004eh,Chen:2005ad}. Another model which has attracted considerable attention is the ghost condensate~\cite{ArkaniHamed:2003uy}, which can be used to violate the null energy condition and constuction bouncing cosmologies~\cite{Creminelli:2006xe,Buchbinder:2007ad,Creminelli:2007aq}. It has been argued that the ghost condensate is in conflict with the $2^{\rm nd}$ law of black hole thermodynamics~\cite{Dubovsky:2006vk,Eling:2007qd}. (See~\cite{Mukohyama:2009rk} for a dissenting viewpoint.) A postivie energy theorem for $P(X, \phi)$ theories coupled to Einstein gravity has been proved in~\cite{Nozawa:2013maa,Elder:2014fea}. Futher, $P(X,\phi)$ models have been supersymmetrized~\cite{Khoury:2010gb,Baumann:2011nm}, and topological defects and solitons are studied in~\cite{Babichev:2006cy,Bazeia:2007df,Babichev:2007tn,Jin:2007fz,Adam:2007ij,Adam:2007ag,Babichev:2008qv,Andrews:2010eh,Bazeia:2010vb,Amin:2013ika}. Analyses of screening in these models can be found in~\cite{Babichev:2009ee,Babichev:2011kq,Goon:2010xh,Brax:2012jr,Brax:2014wla,Burrage:2014uwa}, along with studies of large-scale sctructure~\cite{ArmendarizPicon:2005nz,Brax:2014yla}. Away from cosmology, theories of this type have been used to model fluids and solids~\cite{Comer:1993zfa,Dubovsky:2005xd,DiezTejedor:2005fz,Arroja:2010wy,Endlich:2010hf,Dubovsky:2011sj,Dubovsky:2011sk,Nicolis:2011cs,Nicolis:2011ey,Endlich:2012pz,Endlich:2012vt,Nicolis:2013lma}.

The simplest model of this type is a theory with a shift symmetry
\be
\phi(x)\longmapsto\phi(x)+c \ ,
\ee
where $c$ is a constant. For simplicity, we also imagine that the theory is invariant under the discrete symmetry $\phi \mapsto -\phi$, so that the lowest-order Lagrangian is 
\be
{\cal L} = -\frac{1}{2}(\partial\phi)^2+\frac{\alpha}{4\Lambda^4}(\partial\phi)^4+\frac{g}{M_{\rm Pl}}\phi T \ ,
\label{goldstonelag}
\ee
where $\Lambda$ has units of mass, $\alpha$ and $g$ are dimensionless numbers, and we may absorb the magnitude of $\alpha$ into $\Lambda$, so we need only consider $\alpha = \pm 1$. The $\phi T$ coupling to matter breaks the shift symmetry, but so long as $M_{\rm Pl} \gg \Lambda$, this breaking will be soft. The equation of motion descending from this Lagrangian is
\be
\square\phi-\frac{\alpha}{\Lambda^4}\partial_\mu\left((\partial\phi)^2\partial^\mu\phi\right)+\frac{g}{M_{\rm Pl}}T = 0~.
\label{PXeom}
\ee
\subsubsection{Spherically-symmetric source}
To see screening at work in this model, we consider a point source
\be
T = - M\delta^{(3)}(\vec x)~,
\ee
and search for static, spherically-symmetric solutions. The equation of motion~\eqref{PXeom} reduces to~\cite{Gabadadze:2012sm}
\be
\vec\nabla\cdot\left(\vec\nabla\phi-\frac{\alpha}{\Lambda^4}(\vec\nabla\phi)^2\vec\nabla\phi\right) = \frac{g M}{M_{\rm Pl}}\delta^{(3)}(\vec x)~.
\ee
Due to the shift symmetry of $\phi$, the left hand side appears as a total divergence
 and thus admits a first integral; integrating both sides, we obtain~\cite{Brax:2012jr,Gabadadze:2012sm}
\be
\phi' -\frac{\alpha}{\Lambda^4}\phi'^{\,3} = \frac{1}{4\pi r^2} \frac{g M}{M_{\rm Pl}} \ .
\label{phiprimecubic}
\ee
This is a cubic equation in $\phi'$, and may be solved exactly by radicals. In the case $\alpha=-1$, for example, it reads
\be
\frac{\phi'(r)}{\Lambda^2} = \frac{\left(\frac{8\pi}{3}\right)^{1/3} \left(\frac{r}{r_\star}\right)^{2/3}}{\left(-9+\sqrt{81+192\pi^2\left(\frac{r}{r_\star}\right)^4}\right)^{1/3}} - (72\pi)^{-\frac{1}{3}}\left(-9+\sqrt{81+192\pi^2\left(\frac{r}{r_\star}\right)^4}\right)^{1/3}\left(\frac{r_\star}{r}\right)^{2/3} \ .
\ee
This full solution is not particularly enlightening, it is much more useful to consider two asymptotic regimes: far from and close to the source. Far from the source, the term linear in $\phi'$ is dominant, while close to the source, the cubic term in $\phi'$ is more important. The crossover scale between these two regimes occurs when $\phi'\sim\Lambda^2$, giving
\be
r_\star = \frac{1}{\Lambda}\left(\frac{g M}{M_{\rm Pl}}\right)^{1/2}~.
\label{xx2crossover}
\ee
In the asymptotic regimes, we have\footnote{Since $\alpha = \pm1$, it follows that $\alpha = \alpha^{-1}$.}
\be
\phi'(r) \sim \left\{\begin{array}{lr} \frac{\Lambda^2}{4\pi}\left(\frac{r_\star}{r}\right)^2&~~~~~~~~~~~~~~~~~~{\rm for}~~~~~~r\gg r_\star \\
 (-\alpha)^{1/3}\Lambda^2 \left(\frac{r_\star}{r}\right)^{2/3} &~~~~~~~~~~~{\rm for}~~~~~~r\ll r_\star \end{array} \right. \ .
 \label{phiprimeeqn}
\ee
Note that it is clear from these forms that a consistent continuous solution is only possible if we choose $\alpha  =-1$, and we shall use this value from now on.\footnote{Indeed, looking at the exact solutions to~\eqref{phiprimecubic}, one finds that a solution for all $r>0$, where $\phi$ dies off at infinity, only exists for $\alpha=-1$~\cite{Dvali:2012zc}.}

\begin{figure}
\centering
\includegraphics[width=3.2in]{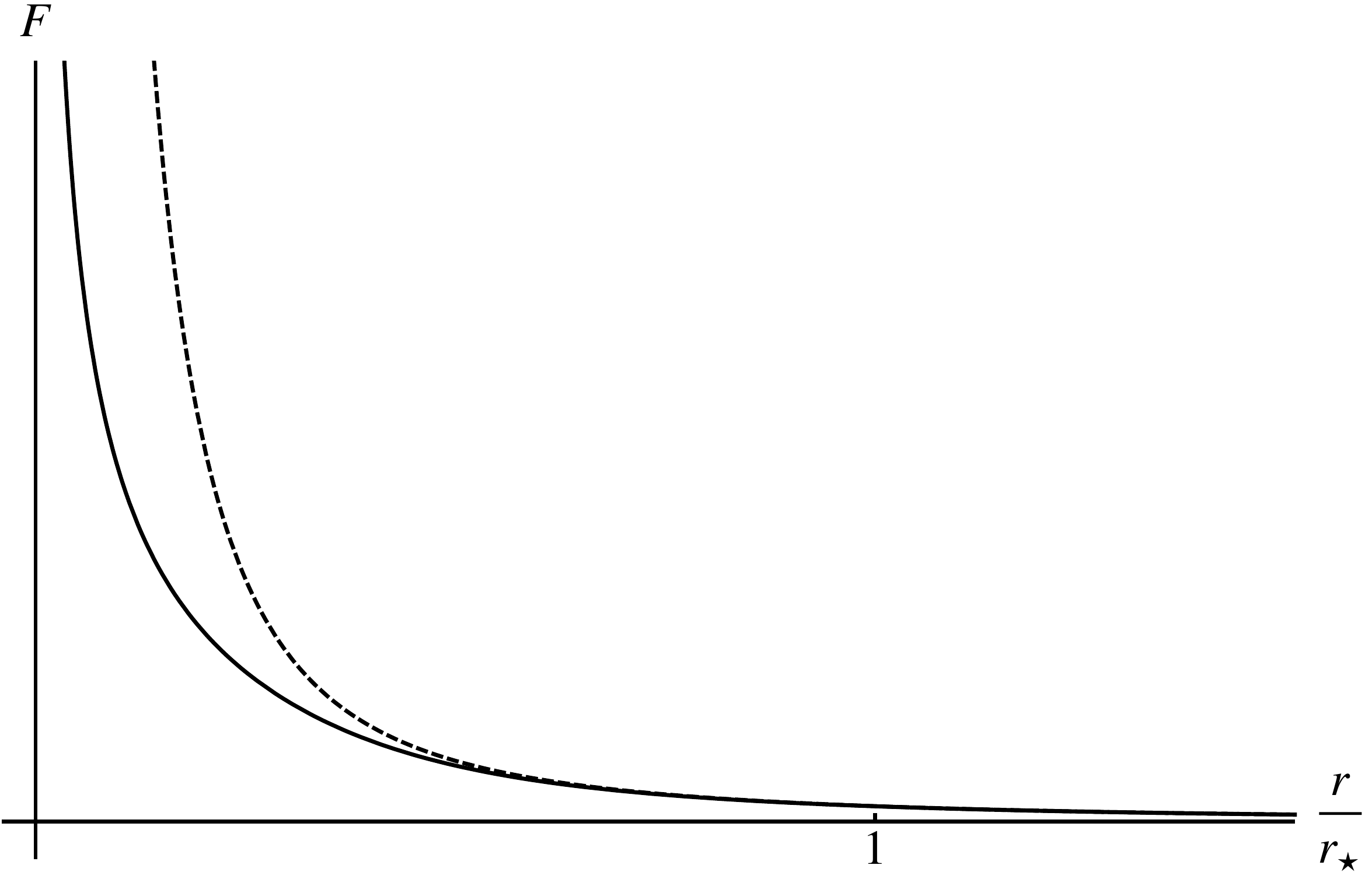}
\includegraphics[width=3.2in]{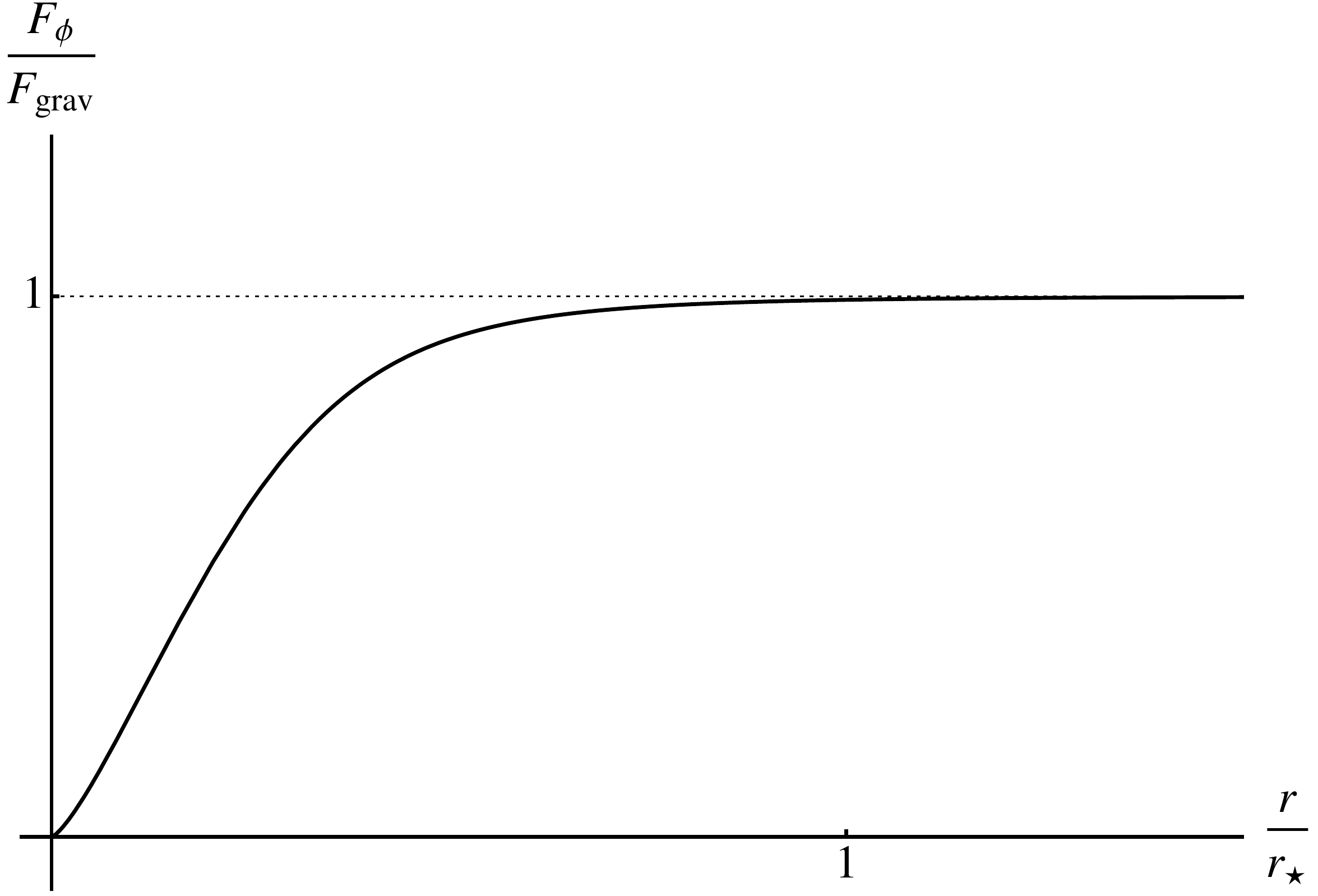}
\caption{\label{pxforces}\small Kinetic screening in the $P(X)=X+X^2$ model. \small {\it Left:} Plot of the scalar force $F_\phi = \frac{g}{M_{\rm Pl}}\phi'(r)$ (solid) and gravitational force $F_{\rm grav} = M/(8\pi M_{\rm Pl}^2r^2)$ (dashed) versus $r/r_\star$. {\it Right}: Ratio of scalar to gravitational force. Note that $\phi'$ is normalized so that $F_\phi/F_{\rm grav} \to 1$ as $r\to\infty$ (corresponding to $g=1/2$). Inside the screening radius, $r_\star$, the ratio of the force mediated by the scalar relative to that of gravity goes to zero sharply.}
\end{figure}

We are now in a position to understand screening of the force mediated by $\phi$. First, we recall that the force due to the scalar field is $\vec F_\phi(x) = \frac{g}{M_{\rm Pl}} \vec\nabla\phi =  \hat r\frac{g}{M_{\rm Pl}}\phi'(r)$, with $\phi'(r)$ given by~\eqref{phiprimeeqn},
while the force due to gravity around a heavy source is of course
\be
F_{\rm grav}(r) = \frac{M}{8\pi M_{\rm Pl}^2}\frac{1}{r^2} = \frac{\Lambda^2}{8\pi M_{\rm Pl}}\left(\frac{r_\star}{r}\right)^2 \ .
\label{pxgrav}
\ee
Thus, we see that far from a heavy source $\phi$ mediates a gravitational-strength force, but close to the source the ratio of the $\phi$ force to that of gravity goes to zero as $\sim r^{4/3}$. This is the essence of kinetic screening. In Figure~\ref{pxforces} we plot the full analytic solution for $\phi'$ along with the force due to gravity. We have arbitrarily chosen $\Lambda = 10^{-1}M_{\rm Pl}$ and rescaled the radial coordinate by the crossover scale $r_\star$.

\subsubsection{Generalizations}
So far in this Section, we have focused on the simplest $P(X)$ theory which admits kinetic-type screening, namely
\be
P(X) = X-\frac{1}{\Lambda^4}X^2~,
\ee
but the phenomenon is of course much more general~\cite{Babichev:2009ee,deRham:2014wfa}. Consider an arbitrary (analytic) function $P(X)$, coupled to a point source:
\be
{\cal L} = P(X) - \frac{g M}{M_{\rm Pl}}\delta^{(3)}(\vec x) \phi= \sum_{n=1}^\infty c_n \frac{X^n}{\Lambda^{4n-4}} - \frac{g M}{M_{\rm Pl}}\delta^{(3)}(\vec x)\phi~.
\label{generallag}
\ee
Notice that here the scalar Lagrangian still enjoys a symmetry under shifts by a constant, meaning that the equation of motion can be written as a total divergence and is therefore a polynomial in $\phi'$.\footnote{We assume that the $c_n$ are ${\cal O}(1)$ and are chosen in such a way that there exists a root to this equation which is real everywhere and falls off at infinity. A concrete example of a function which satisfied the desired properties is the (D)BI Lagrangian~\eqref{bioniclag}.}
As in the $X-X^2$ case, there is a crossover between two qualitatively different behaviors; far from the source the field profile will have the usual Coulomb $\sim 1/r$ form, while near the source, the force will be suppressed relative to gravity. This crossover between these two behaviors happens when $X/\Lambda^4  \sim 1$, the same scale as~\eqref{xx2crossover}
\be
r_\star  = \frac{1}{\Lambda}\left(\frac{g M}{M_{\rm Pl}}\right)^{1/2}~.
\label{rstaragain}
\ee
Far inside $r_\star$, we have $X/\Lambda^4 \gg 1$, so it is clear that the term in~\eqref{generallag} with the highest exponent will dominate. Indeed, for a fixed $n$, it is straightforward to show that the field profile takes the form~\cite{deRham:2014wfa}
\be
\phi'(r) \sim \left\{\begin{array}{lr} \Lambda^2\left(\frac{r_\star}{r}\right)^2&~~~~~~~~~~~~~~~~~~{\rm for}~~~~~~r\gg r_\star \\
 \Lambda^2 \left(\frac{r_\star}{r}\right)^{2/(2n-1)} &~~~~~~~~~~~{\rm for}~~~~~~r\ll r_\star \end{array} \right. \ ,
 \label{phiprimearbitraryn}
\ee
and the ratio of this fifth force to the force due to gravity~\eqref{pxgrav} scales as
\be
\frac{F_\phi}{F_{\rm grav}} \sim \left(\frac{r}{r_\star}\right)^\frac{4n-4}{2n-1}~.
\label{forceratios}
\ee

Note that as we take the limit $n\to\infty$, the force asymptotes to scaling like $\sim r^2$ relative to that of gravity~\cite{deRham:2014wfa}. This may seem like a somewhat artificial case, but there is a well-motivated example of a $P(X)$ theory which includes such an infinite number of terms, the {\it Dirac--Born--Infeld} Lagrangian. This theory has been extensively studied in many contexts including inflation~\cite{Silverstein:2003hf, Alishahiha:2004eh, Mizuno:2010ag} and late time acceleration. The DBI theory arises naturally as the world volume theory of a brane probing a higher-dimensional space and therefore inherits the isometries of the target space as global symmetries~\cite{deRham:2010eu,Goon:2011uw,Goon:2011qf,Burrage:2011bt}. 

In~\cite{Burrage:2014uwa}, the (D)BI theory of a (negative tension) brane embedded in a $5$-dimensional space which has two time-like directions was considered, the lowest order Lagrangian is of the $P(X)$ form
\be
{\cal L} = \Lambda^4\sqrt{1-X/\Lambda^4}~.
\label{bioniclag}
\ee
Notice that expanding out this function reproduces the $X -X^2$ example studied earlier at lowest order. In particular, this means that the earlier conclusions about superluminality~\cite{Goon:2010xh} and (non)-analyticity persist.
While this Lagrangian has an infinite number of terms when thought of as a series as in~\eqref{generallag}, the relative coefficients of these terms are fixed by symmetry. Indeed, in addition to conventional $4d$ Lorentz invariance, the action~\eqref{bioniclag} is invariant under the following non-linearly realized symmetries
\be
\delta_{P_5} \phi = 1~,~~~~~~~~~~~~~~~~
\delta _{J_{\mu 5}}\phi = x^\mu - \phi\partial^\mu\phi~,
\ee
which correspond to 5-dimensional translations and boosts. Coupling the theory~\eqref{bioniclag} to a massive source, one again finds that the equation of motion admits a first integral, which can be solved for $\phi'(r)$ as~\cite{Dvali:2010jz, Burrage:2014uwa}
\be
\phi'(r) = \frac{\Lambda^2}{\sqrt{1+16\pi^2 \left(r/r_\star\right)^4}}~.
\ee
As expected, far from the source $r \gg r_\star$, we see that the force goes as $\sim (r_\star/r)^2$ as expected, while near the source, the ratio between the force due to the scalar and that due to gravity scales as
\be
\frac{F_{\rm BI}}{F_{\rm grav}} \sim \left(\frac{r}{r_\star}\right)^2~,
\ee
which is precisely the $n\to\infty$ limit of~\eqref{forceratios}. In~\cite{Burrage:2014uwa}, screening of this (D)BI-type was dubbed {\it BIonic} screening, and it can be thought of as an edge case of screening intermediate between finite-order kinetic screening and Vainshtein screening, which we will discuss in the next section. This particular choice of action has many interesting properties: for example, it is possible to find an exact solution for the field sourced by $N$ different point masses.

\subsubsection{Radiative stability}
Up to this point, we have been considering purely classical dynamics; we found that in regions where $X/\Lambda^4~\gsim~1$, interesting screening effects are possible. However, as effective field theorists, this should worry us. Recall that the organizing principle of EFT is that there is some expansion which tells us that we only need to consider a finite number of operators to model a given phenomenon. However, we are interested precisely in the regime where this expansion appears to be breaking down! We should be concerned that quantum corrections will spoil the nice behaviors we have just uncovered.

There are, roughly speaking, two types of correction we should worry about, the first is that if we start with a Lagrangian of the form
\be
{\cal L} = X - \frac{1}{\Lambda^4}X^2~,
\ee
we expect that quantum corrections will generate terms of the form ${\cal L} \sim X^n/\Lambda^{2n-4}$, which are higher powers in $X$. In the regimes where $X/\Lambda^4 \ll 1$, these do not concern us, as they will be negligible corrections to the background dynamics. However, screening takes place in the regime where $X/\Lambda^4~\gsim~1$, so it is not so clear that these corrections will be small---if they are large, we should not trust the background we are considering, and the theory is not predictive.

The second worry is that operators with more derivatives will be generated by quantum corrections; that is, operators of the form ${\cal L} \sim \partial^\ell X^n/\Lambda^{2n+\ell-4}$. Similar to before---since we have ignored all operators of this type---if they become important we should not trust the conclusions we have drawn.

Although the situation looks somewhat bleak, there are indications that things may turn out better than expected and it is possible that $P(X)$ theories can be made to be radiatively stable, but accounting for quantum effects properly is subtle. The corrections are handled carefully and clearly in~\cite{deRham:2014wfa}; here we just summarize the bottom line. One key point is that the scale $\Lambda$ which we put into the Lagrangian is {\it not} the cutoff of the theory (the scale beyond which new physics enters), but rather is the scale of  strong-coupling (where loop effects become important). By suitably re-summing the loop effects, we can still trust the theory.

Consider a $P(X)$ theory, where we define $X \equiv -\frac{1}{2}(\partial\phi)^2/\Lambda^4$ so that the Lagrangian takes the form
\be
{\cal L} = \Lambda^4 P(X)~.
\label{PXgen}
\ee
Here $\Lambda$ is the strong-coupling scale of the theory, we will refer to the cutoff as $\Lambda_{\rm c}$. A careful analysis\footnote{Here we consider only corrections due to logarithmic divergences. These divergences are the most robust, in the sense that they are insensitive to UV physics, but this makes an optimistic assumption that we can find a UV completion where the would-be effects of power-law divergences can be canceled.} reveals that quantum corrections to the operators we have written down in $P(X)$ scale at worst as~\cite{deRham:2014wfa}
\be
\Delta P(X) \sim \frac{\Lambda_{\rm c}^4}{{\rm max}(Z^{\mu\nu})}~,
\ee
Where the notation ${\rm max}(Z^{\mu\nu})$ means the largest eigenvalue of the matrix
\be
Z^{\mu\nu} = 2P'(X)\delta^{\mu\nu}-\frac{4}{\Lambda^4}P''(X)\partial^\mu\phi\partial^\nu\phi~,
\ee
which is built from the function $P$. In order for these loop corrections to be small, we must demand that $\Delta P(X) \ll P(X)$. For a given term, $X^n$, and assuming that ${\rm max}(Z^{\mu\nu}) \sim P' \sim P''/\Lambda^4$  this boils down to demanding that $\lvert X\rvert$ be sufficiently large~\cite{deRham:2014wfa}
\be
\lvert X\rvert \gg \left(\frac{\Lambda_{\rm c}}{\Lambda}\right)^\frac{4}{2n-1}~,
\ee
which it is indeed possible to satisfy, and in fact gets easier as $n$ gets larger.

Next we want also to make sure that corrections of the form $\sim \partial^\ell X^n/\Lambda^{2n+\ell-4}$ can also be made negligible. It turns out that this is also possible, provided that the logarithmic derivatives of $Z$ are small~\cite{deRham:2014wfa}, schematically this constraint takes the form
\be
\left\lvert\frac{\partial Z}{Z}\right\rvert^4 \ll \Lambda^4 P(X)~,
\ee
which is satisfied if $\partial/\Lambda \ll 1$. This is in line with out EFT reasoning, in the $P(X)$ theories we are considering, only $X$ is becoming large, the additional terms that we are not writing down are suppressed by additional powers of $\partial/\Lambda$. in the language of DBI, this is a situation where the velocity of the brane is allowed to be very large so long as the acceleration remains small. We will see in the next section that very similar reasoning holds for the galileons, they can develop large non-linearities but remain quantum-mechanically stable.

A surprising byproduct of this analysis---which is emphasized in~\cite{deRham:2014wfa}---is that the previous discussion does not rely on symmetry in any essential way. An arbitrary functional form for $P(X)$ is radiatively stable in the regime where $X$ becomes very large. This is sort of a quantum-mechanical analog of screening.

\subsubsection{Signs and superluminality}
While the screening effect discussed above is novel, and would be crucial for models such as this one to be consistent with observations in, for example, the solar system, the above model also provides a clear example of problems that can arise in models with nontrivial kinetic interactions. In particular, having made the choice $\alpha = -1$ to obtain a solution close to the source that smoothly matches onto the solution at long distances, this necessarily introduces an oddity into the theory: apparently {\it superluminal} propagation of perturbations. 

\begin{figure}
\centering
\includegraphics[width=3.3in]{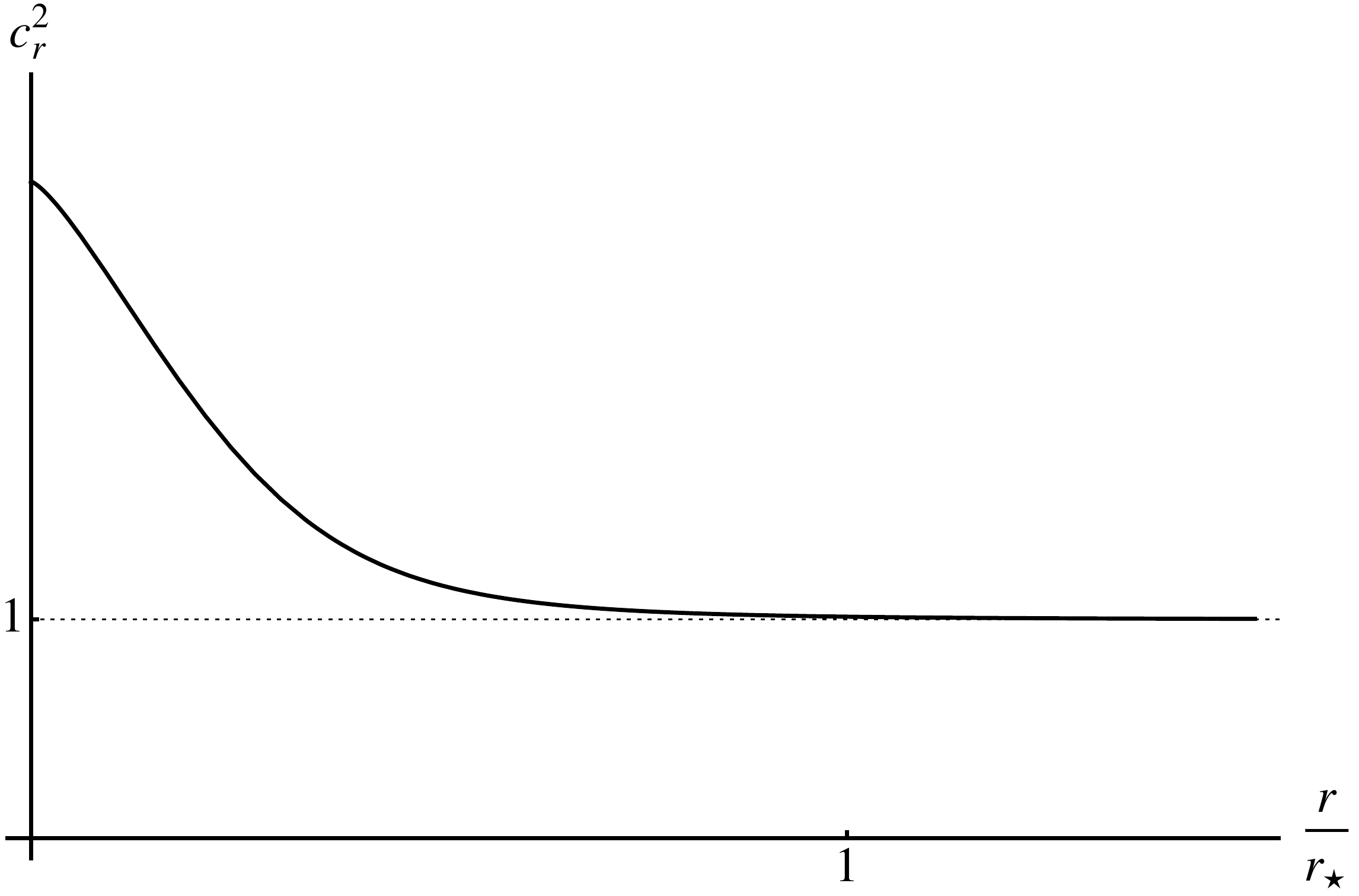}
\caption{\label{pXsuperluminal}\small Plot of speed of radial fluctuations versus distance from a spherically symmetric source, in units of $r_\star$ for the $P(X) = X-X^2$ example. Note that as we approach the source, the fluctuations propagate superluminally in the radial direction.}
\end{figure}

To see this, consider perturbing the $P(X)$ theory~\eqref{PXgen} around a spherically-symmetric background $\bar{\phi}(r )$. Expanding to quadratic order in perturbations
$\varphi = \phi - \bar{\phi}$, we obtain
\be
{\cal L}_\vp = \frac{1}{2}\bar{P}_{,X} \Big(\dot\vp^2- (\partial_\Omega\vp)^2\Big) -\frac{1}{2}\Big(\bar{P}_{,X} + 2\bar{X}\bar{P}_{,XX}\Big) \vp'^{\,2}~,
\label{pxspherics}
\ee
where $(\partial_\Omega\vp)^2 = r^{-2}(\partial_\theta\vp)^2+(r\sin\theta)^{-2}(\partial_\phi \vp)^2$ is the standard angular derivative term. The radial and angular speeds of perturbations 
can be immediately read off:
\begin{align}
\nonumber
c_r^2 &= 1 + 2\frac{\bar{X}\bar{P}_{,XX}}{\bar{P}_{,X}}~; \label{crsquared}\\
c_\Omega^2 &= 1~.
\end{align}
To avoid ghosts, it is clear from~\eqref{pxspherics} we must demand $\bar{P}_{,X}  > 0$. To have screening at arbitrarily large values of $X < 0$ (radial profile), we must have $\bar{P}_{,XX} < 0$.
It follows that the radial sound speed is superluminal at all distances. For example, the simple $P(X) = X +\alpha X^2$ considered earlier gives
\be
c_r^2  = 1 - \frac{2\alpha \bar{\phi}'^{\,2}}{\Lambda^4 - \alpha \bar{\phi}'^{\,2}}\,.
\ee
Since $\alpha=-1$ is required for screening, the second term in~\eqref{crsquared} gives a positive contribution to the radial speed of sound, making it superluminal at all distances. This is plotted in 
Figure~\ref{pXsuperluminal}. 

We can also make a connection with the analyticity arguments discussed in Appendix~\ref{superlumapp}. In this case, the superluminality and (lack of) analyticity of the $2\to2$ scattering amplitude appear to be closely related. To see this, consider again the $P(X) = X +\alpha X^2$ Lagrangian. In this theory, it is straightforward to compute the $2\to2$ scattering amplitude using standard techniques. The result is the expression
\be
{\cal A}_{2\to2}(s,t) = \frac{\alpha}{2\Lambda^4}(s^2+t^2+u^2) = \frac{\alpha}{\Lambda^4}(s^2+t^2-st)~,
\ee
where the Mandelstam $s,t,u$ variables we are using are defined in Appendix~\ref{superlumapp}, and where in the last equality we have used $s+t+u=0$, because the theory is massless. If we look at this amplitude in the forward limit ($t\to0$), we then obtain
\be
{\cal A}_{2\to2}(s, 0) = \frac{\alpha}{\Lambda^4}s^2~.
\ee
The dispersion relation in equation~\eqref{dispersionsumrule} tells us that the coefficient $\alpha/\Lambda^4$ should be positive in a Lorentz-invariant theory with an analytic S-matrix. However, we see that this is violated precisely for the choice of $\alpha$ which admits screening (and also superluminality). This is circumstantial evidence that theories which admit kinetic screening cannot arise as a low-energy effective description from integrating out degrees of freedom in a {\it local} theory, but this issue is far from settled.

\subsection{The Vainshtein mechanism: galileons}
\label{galileonsection}

A related kinetic screening mechanism that has been of particular interest in recent years is the Vainshtein effect. This phenomenon is seen in some models arising from brane constructions in higher dimensions, and in limits of  theories of massive gravity~\cite{Vainshtein:1972sx,Deffayet:2001uk,Gruzinov:2001hp,Babichev:2009jt,Babichev:2010jd}. We will describe briefly a bit later how this effect is relevant to massive gravity. However, first, it is pedagogically simplest to describe the galileon theories, and how the Vainshtein effect manifests in them.

There are special classes of scalar field theories possessing higher derivative Lagrangians, but which nonetheless have second order equations of motion. It is now known that there are a large number of different examples of such theories, as we shall briefly discuss later, but the first to be discovered---and the canonical example---is the simple galileon theory. Galileons are higher-derivative field theories which are both invariant under the galilean shift symmetry
\be
\phi(x) \longmapsto \phi(x)+c+b_\mu x^\mu~,
\label{galileansymmetry}
\ee
and which have second order equations of motion. 

This symmetry initially appeared in the decoupling limit of the DGP (Dvali--Gabadadze--Porrati) brane-world\footnote{We do not discuss brane-worlds~\cite{Rubakov:1983bb,Rubakov:1983bz,Randall:1999ee,Randall:1999vf,Binetruy:1999ut,Shiromizu:1999wj,Dvali:2000hr,Dvali:2000xg} in any great depth, but studies of cosmology can be found in~\cite{Garriga:1999yh,Csaki:1999jh,Csaki:1999mp,ArkaniHamed:2000eg,Gregory:2000jc,Bowcock:2000cq,Deffayet:2000uy,Deffayet:2001pu,Deffayet:2002sp,Sahni:2002dx,Maeda:2003ar,Lue:2004rj,Song:2006jk,Chan:2009ew}. Of particular interest was the fact that the DGP model admits a self-accelerated branch of solutions. Unfortunately, perturbations about this branch are ghost-like~\cite{Luty:2003vm,Nicolis:2004qq,Koyama:2005tx,Gorbunov:2005zk,Charmousis:2006pn,Koyama:2007zz}. For reviews see~\cite{Rubakov:2001kp,Langlois:2002bb,Brax:2003fv,Maartens:2003tw,Brax:2004xh,Lue:2005ya,Maartens:2010ar}.}  model~\cite{Dvali:2000hr}, where it is inherited from higher-dimensional Poincar\'e invariance~\cite{Luty:2003vm, Nicolis:2004qq}. This symmetry was abstracted in~\cite{Nicolis:2008in} to a more general scalar field model. The restriction to terms which have second-order equations of motion is non-trivial---for example terms of the form $(\square\phi)^n$ are invariant under the symmetries, but have higher-order equations of motion.\footnote{As reviewed in Appendix~\ref{OstrogradskyApp}, higher-order equations of motion can lead to Ostrogradsky-type instabilities and the propagation of additional ghost-like degrees of freedom.}

Remarkably, there are a finite number of terms ($d+1$ in $d$-dimensions) that satisfy these combined requirements of invariance under \eqref{galileansymmetry} and second order equations of motion. 
Also interesting for our purposes is that each galileon term in the Lagrangian is not strictly invariant under the symmetries \eqref{galileansymmetry}, but rather  shifts by a total derivative, leaving the action invariant. It was shown in \cite{Goon:2012dy} that this is a natural consequence of the fact that the galileons are Wess--Zumino terms for spontaneously broken space-time symmetries, as we will discuss briefly.

Although galileons originally arose in brane-world modifications of gravity, they have appeared in other well-behaved modifications of gravity, for instance massive gravity \cite{deRham:2010ik,deRham:2010kj}. Further, models termed ``galileons" in the literature has grown far beyond theories invariant under the symmetry~\eqref{galileansymmetry}; indeed, galileon has become nearly synonymous with well-behaved derivatively-coupled theory.

The galileons have been used to address cosmic acceleration \cite{Chow:2009fm,Silva:2009km,Kobayashi:2009wr,DeFelice:2010as,DeFelice:2010gb,DeFelice:2010pv,Gannouji:2010au,Ali:2010gr,Mota:2010bs,Barreira:2012kk} and the origin of density perturbations in the early universe through inflation \cite{Anisimov:2005ne,Kobayashi:2010cm,Burrage:2010cu,Creminelli:2010qf,Kamada:2010qe,DeFelice:2010nf,Kobayashi:2010wa,DeFelice:2011zh,Kobayashi:2011pc,RenauxPetel:2011uk,Fasiello:2013dla} and inflationary alternatives \cite{Creminelli:2010ba,Hinterbichler:2011qk,LevasseurPerreault:2011mw,Wang:2012bq,Liu:2011ns,Hinterbichler:2012mv,Hinterbichler:2012fr,Hinterbichler:2012yn,Creminelli:2012my}. They have also been used to violate the null energy condition~\cite{Nicolis:2009qm,Creminelli:2010ba,Qiu:2011cy,Easson:2011zy,Hinterbichler:2012yn,Creminelli:2012my,Easson:2013bda,Rubakov:2013kaa,Elder:2013gya,Rubakov:2014jja,Battefeld:2014uga}. Covariantizing the galileons for such applications is subtle, it requires introducing non-minimal couplings to curvature, which generically destroys the shift symmetry \cite{Deffayet:2009wt,Deffayet:2009mn,Deffayet:2010qz,Deffayet:2011gz,Pujolas:2011he} (for a construction which couples galileons covariantly to massive gravity while retaining galilean symmetry, see \cite{Gabadadze:2012tr,Andrews:2013ora,Goon:2014ywa}). Galileons have been generalized in various directions: they have been embedded in supersymmetry and supergravity~\cite{Khoury:2011da,Koehn:2013hk,Koehn:2013upa,Farakos:2013fne}, extended to $p$-forms \cite{Deffayet:2010zh}, extended to multi-galileons \cite{Padilla:2010de,Padilla:2010tj,Padilla:2010ir,Hinterbichler:2010xn,Andrews:2010km,Garcia-Saenz:2013gya} and coupled consistently to gauge fields \cite{Zhou:2011ix,Goon:2012mu}. The shift symmetry itself has even been generalized to a shift by an arbitrary polynomial~\cite{Hinterbichler:2014cwa}. Solitons in galileon systems are studied in~\cite{Endlich:2010zj,Padilla:2010ir,Masoumi:2012np,Babichev:2012qs,Zhou:2012fk}. Galileons can also be made to appear in the decoupling limit of nonlinear theories of massive vector fields~\cite{Tasinato:2014eka,Heisenberg:2014rta,Tasinato:2014mia}. See~\cite{Hiramatsu:2012xj,Li:2013nua} for numerical simulation of the Vainshtein effect in galileon theories.

A powerful technique for deriving galileon theories was introduced in \cite{deRham:2010eu}. By considering a 3-brane probing a non-dynamical bulk, the actions for galileons, conformal galileons and covariant galileons appear as the non-relativistic limit of world-volume and bulk Lovelock invariants \cite{Lovelock:1971yv}. This probe brane construction has been greatly generalized---it has been applied to all cases of maximally symmetric branes probing maximally symmetric bulks \cite{Goon:2011qf,Burrage:2011bt,Goon:2011uw}, to higher co-dimensions~\cite{Hinterbichler:2010xn,Goon:2012mu} and to cosmological backgrounds \cite{Goon:2011xf}, each leading to novel scalar theories. See 
\cite{Trodden:2011xh} for a review of these extensions. Recently this construction has been applied to cases where the bulk metric itself is dynamical, this results in a theory of coupled galileons and massive gravitons \cite{Gabadadze:2012tr,Andrews:2013ora,Goon:2014ywa}.

\subsubsection{Simplest example: the cubic galileon}

The simplest non-trivial theory exhibiting the Vainshtein mechanism is the cubic galileon theory:
\be
{\cal L} = -3(\partial\phi)^2 - \frac{1}{\Lambda^3}\Box\phi(\partial\phi)^2 +\frac{g}{M_{\rm Pl}}\phi T^\mu_{\;\mu}\ ,
\label{L3}
\ee
where $g \sim {\cal O}(1)$ for gravitational strength coupling and $\Lambda$ is the strong-coupling scale of the theory. (The funny normalization of the kinetic term has been chosen so that the decoupling limit of the DGP model corresponds to exactly $g = 1$ for convenience.) The first two terms are manifestly strictly invariant under the ordinary shift symmetry $\phi\mapsto \phi + c$. They also shift by a total derivative under the galilean shift $\phi \mapsto \phi + b^\mu x_\mu$. The coupling to matter explicitly breaks these symmetries, but only very softly (since $M_{\rm Pl} \gg \Lambda$). Ignoring this mild breaking, the theory is invariant under the galileon symmetries. It remains to check the second defining feature of a galileon---that the equation of motion is second-order. Varying~\eqref{L3} gives
\be
6\Box\phi + \frac{2}{\Lambda^3} \bigg( (\Box\phi)^2 - (\partial_\mu\partial_\nu\phi)^2\bigg) = - \frac{g}{M_{\rm Pl}} T^\mu_{\;\mu} \,.
\label{L3eom}
\ee
This equation is of course non-linear, but nevertheless second-order---we need only supply the same amount of initial data as for an ordinary scalar field to obtain a unique solution. 

The cubic interaction term is non-renormalizable, so we should treat~\eqref{L3} as an effective field theory and write down all possible operators consistent with the symmetries, which take the schematic form\footnote{We ignore higher-order galileon terms here for simplicity. This choice is technically natural; these terms do not get generated by quantum corrections, as we will discuss in Section~\ref{generalgals}.}
\be
{\cal L} = -3(\partial\phi)^2 - \frac{1}{\Lambda^3}\Box\phi(\partial\phi)^2  + \sum_{n = 2}^\infty\sum_{\ell=0}^\infty \frac{c_{n, \ell}}{\Lambda^{3n+2\ell-4}} \partial^{2\ell}(\partial^2\phi)^n + \frac{g}{M_{\rm Pl}}\phi T^\mu_{\;\mu} \,,
\label{L3EFT}
\ee
where we have suppressed the Lorentz structure, and assume the $c_{n, \ell}$s are all of ${\cal O}(1)$ since, even if we set them to zero classically, they will be generated quantum-mechanically. The Vainshtein mechanism~\cite{Vainshtein:1972sx,Deffayet:2001uk} relies on the $(\partial\phi)^2\Box\phi/\Lambda^3$ term becoming large compared to the kinetic term $(\partial\phi)^2$ near massive objects so that $\partial^2\phi \gg \Lambda^3$. In this regime, the expectation is that higher-order operators should become important as well, signaling that the effective field theory is breaking down.

However---contrary to this expectation---there is in fact a regime in which the galileon term can be large with the other operators remaining negligible. This is because~\eqref{L3EFT} is actually an expansion with two parameters~\cite{Luty:2003vm,Nicolis:2004qq}: a {\it classical} expansion parameter,
\be
\alpha_{\rm cl} \equiv \frac{\partial^2\phi}{\Lambda^3}\,,
\ee
which measures the strength of classical non-linearities; there is also a {\it quantum} expansion parameter,
\be
\alpha_{\rm q} \equiv \frac{\partial^2}{\Lambda^2}\,,
\ee
which measures the size of quantum effects (this also measures the relevance of the additional non-galileon operators in~\eqref{L3EFT}). We will see that it is possible for classical non-linearities to be important ($\alpha_{\rm cl} \gg 1$) while
quantum corrections are small ($\alpha_{\rm q} \ll 1$). This is because the dangerous operators of the form $\partial^{2\ell}(\partial^2\phi)^n$  have at 2 least derivatives per field, and hence are suppressed
by powers of $\alpha_{\rm q}$ relative to galileon terms. Actually, this argument is somewhat subtle for the following reason: as we take $n$ large, terms with exactly 2 derivatives per field $\sim(\partial^2\phi)^n$ will be enhanced by many powers of $\alpha_{\rm cl}$ and only suppressed by 2 powers of $\alpha_{\rm q}$ relative to the cubic galileon---and can thus become important. In the strict $n\to\infty$ limit, this occurs infinitesimally close to the Vainshtein radius. Therefore, in order to trust the theory inside the Vainshtein radius, the coefficients of these dangerous terms must be tuned to be very small. However, if we ignore power-law divergences this tuning is {\it technically natural}, as loops involving galileon operators only generate terms with 3 or more derivatives per field (see {\it e.g.},~\cite{Nicolis:2004qq,deRham:2012az}). However, ignoring the power laws (which are dependent on UV physics) is an optimistic choice, which assumes that the UV physics will be well-behaved~\cite{Nicolis:2004qq}.

\subsubsection{Solution around spherically-symmetric source}

This is perhaps best illustrated by considering the behavior of the galileon theory, and the emergence of Vainshtein screening, near a  static point source of mass $M$, so that $T^\mu_{\;\mu} =-M\delta^{(3)}(\vec{x})$. (For an investigation of screening around other matter distribution shapes, see~\cite{Bloomfield:2014zfa}.) Assuming the field profile is static and  spherically-symmetric, ($\phi = \phi(r)$),~\eqref{L3eom} reduces to~\cite{Nicolis:2004qq}
\be
\vec{\nabla} \cdot \left(6\vec\nabla\phi + \hat{r} \frac{4}{\Lambda^3}\frac{(\vec\nabla\phi)^2}{r}\right) = \frac{gM}{M_{\rm Pl}} \delta^{(3)}(\vec{x})\,.
\ee
This equation can be integrated, choosing to integrate over a sphere centered at the origin, we obtain
\be
6\phi' + \frac{4}{\Lambda^3}\frac{\phi'^2}{r} = \frac{gM}{4\pi r^2 M_{\rm Pl}}\, .
\ee
This equation is now algebraic in $\phi'$, and so admits a solution by radicals. Focusing on the branch for which $\phi'\rightarrow 0$ at spatial infinity,\footnote{The other branch matches asymptotically to a self-accelerated solution
and has unstable ({\it i.e.}, ghost-like) perturbations.}
\be
\phi'(r) = \frac{3\Lambda^3  r}{4} \left( -1+ \sqrt{1 + \frac{1}{9\pi}\left(\frac{r_{\rm V}}{r}\right)^3 }\right)\, ,
\label{galfieldprofile}
\ee
where we have introduced the {\it Vainshtein radius}
\be
r_{\rm V} \equiv \frac{1}{\Lambda } \left(\frac{gM}{M_{\rm Pl}}\right)^{1/3}\ .
\ee
\begin{figure}[tb]
\begin{center}
\includegraphics[width=6in]{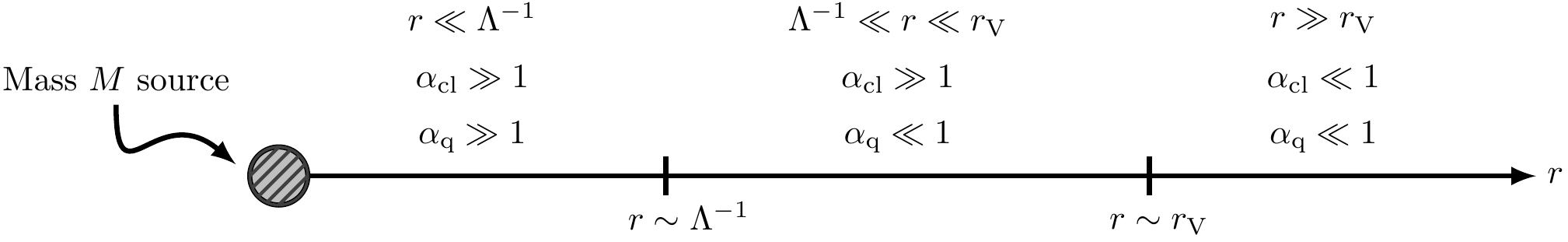}
\caption{\small Various regimes of the galileon theory around a spherically symmetric source. Beyond the Vainshtein radius, $r_{\rm V} \sim (M/\Lambda^3 M_{\rm Pl})^{1/3}$, the field mediates a long range force and both the classical, $\alpha_{\rm cl} \sim (r_{\rm V}/r)^3$, and quantum, $\alpha_{\rm q}\sim (r\Lambda)^{-2}$, non-linearity parameters are small. Very close to the source---$r\ll \Lambda^{-1}$, the inverse cutoff---both $\alpha_{\rm cl}$ and $\alpha_{\rm q}$ are large and the theory is not predictive. However, there is an intermediate regime, $\Lambda\ll r\ll r_{\rm V}$, where classical non-linearities are important ($\alpha_{\rm cl}\gg1$), but quantum effects can consistently be neglected ($\alpha_{\rm q}\ll 1$).
}
  \label{vainshteinregime}
\end{center}
\end{figure}
This nontrivial radial profile is crucial to the operation of Vainshtein screening. We consider two regimes:

\begin{itemize}

\item $r \gg r_{\rm V}$: Far away from the source, the solution is approximately a $1/r^2$ profile,
\be
\phi'(r\gg r_{\rm V}) \simeq \frac{g}{3} \cdot \frac{M}{8\pi M_{\rm Pl} r^2}\,.
\ee
In this regime, the galileon force relative to gravity is given by
\be
\left. \frac{F_{\phi}}{F_{\rm gravity}}\right\vert_{r\gg r_{\rm V}}\simeq \frac{g^2}{3}\,.
\ee
(For DGP, where $g=1$, this reproduces the famous $1/3$ enhancement.)
In this regime, both the classical and quantum expansion parameters are small (we assume $M \gg M_{\rm Pl}$, so that $\Lambda^{-1} \ll r_{\rm V}$):
\be
\alpha_{\rm cl} \sim \left(\frac{r_{\rm V}}{r}\right)^3 \ll 1\,~~~~~~~~~~~~~\alpha_{\rm q} \sim \frac{1}{(r\Lambda)^2} \ll 1\,.
\ee
This tells us that both classical non-linearities and quantum corrections are unimportant.

\item $r \ll r_{\rm V}$: Near the source,~\eqref{galfieldprofile} reduces to
\be
\phi'(r\ll r_{\rm V})  \simeq \frac{\Lambda^3r_{\rm V}}{2} \sqrt{\frac{r_{\rm V}}{r}}\sim \frac{1}{\sqrt{r}} \,.
\label{Er<<rV}
\ee
The force due to the galileon relative to that of gravity is now given by
\be
\left. \frac{F_{\phi}}{F_{\rm gravity}}\right\vert_{r\ll r_{\rm V}} \sim \left(\frac{r}{r_{\rm V}}\right)^{3/2}\ll 1\,,
\label{galforcesup}
\ee
so the scalar force is strongly suppressed at distances much less than the Vainshtein radius.
In this regime, the classical non-linearity parameter is very large (as it must be, this is the source of the screening),
\be
\alpha_{\rm cl} \sim  \left(\frac{r_{\rm V}}{r}\right)^{3/2} \gg 1\,.
\ee
But notice that the quantum parameter is {\it not}; it takes the same form as before:
\be
\alpha_{\rm q} \sim \frac{1}{(r\Lambda)^2} \,.
\ee
At distances $r \gg \Lambda^{-1}$, this is small and quantum corrections are under control meaning that the classical solution can be trusted. Of course, sufficiently close to the source,
$r \ll \Lambda^{-1}$, the quantum parameter becomes ${\cal O}(1)$, radiative corrections become important, and the effective field theory breaks down. (In fact, this statement is too conservative---we will see shortly that perturbations
acquire a large kinetic term scaling as $\sim (r_{\rm V}/r)^{3/2}$. Upon canonical normalization, this translates to a higher strong coupling scale. Even ignoring this fact, the scale $\Lambda$ is only the {\it strong-coupling} scale, it may be possible to re-sum the quantum corrections into a predictive theory.)

\end{itemize}
Therefore, as advertised, we see that there exists a regime,  $ \Lambda^{-1} \ll r \ll r_{\rm V}$, where classical non-linearities are important while quantum effects remain small. 

This situation is not so alien, an analogous situation occurs in GR~\cite{Hinterbichler:2010xn,Hinterbichler:2011tt}. The Einstein--Hilbert action, expanded in terms of the canonically-normalized metric perturbation, $g_{\mu\nu}\sim \eta_{\mu\nu}+ h_{\mu\nu}/M_{\rm Pl}$, takes the schematic form
\be
{\cal L}_{\rm GR} = M_{\rm Pl}^2 \sqrt{-g} R = h\partial^2 h + \sum_{n\geq 2} \frac{h^{n} \partial^2 h}{M_{\rm Pl}^{n-1}}\,.
\ee
In other words, the action consists of a kinetic term, $h\partial^2 h$,  plus an infinite number of interaction terms each of which has exactly two derivatives but arbitrary powers of $h/M_{\rm Pl}$.\footnote{Similar to the galileon interactions, the relative coefficients of these terms are not renormalized, but here it is due to diffeomorphism invariance.} Therefore we see that the measure of classical non-linearity is
\be
\alpha_{\rm cl}^{\rm grav.} = \frac{h}{M_{\rm Pl}}\,.
\ee
Quantum effects generate higher-curvature terms in this theory, which can also be expanded in $h$, to take the form
\be
{\cal L}_{\rm higher-curv.} = \sqrt{-g} R^2,~\sqrt{-g}R_{\mu\nu}R^{\mu\nu} \ldots = \sum_{n\geq 2,~m\geq 4} \frac{\partial^m h^n}{M_{\rm Pl}^{m+n-4}}\,.
\ee
The suppression of these terms relative to the classical operators is by powers of the factor
\be
\alpha_{\rm q}^{\rm grav.} \sim{\partial^2\over M_{\rm Pl}^2}\,.
\ee
Around a point source, the field takes the spherically-symmetric profile $h\sim \frac{M}{M_{\rm Pl}r}$, for which the non-linearity parameters scale as
\be
\alpha_{\rm cl}^{\rm grav.} \sim \frac{M}{M_{\rm Pl}^2 r} \sim \frac{r_{\rm Sch}}{r}\,,~~~~~~~~~~~~~~~~~~ \alpha_{\rm q}^{\rm grav.} \sim \frac{1}{M_{\rm Pl}^2 r^2}\,,
\ee
where $r_{\rm Sch} \equiv M/4\pi M_{\rm Pl}^2$ is the Schwarzschild radius of the massive source. Therefore, for $r\gg r_{\rm Sch}$ (such as in the solar system), classical non-linearities are negligible, whereas for $r\ll r_{\rm Sch}$ (such as inside or near the horizon of a black hole) $\alpha_{\rm cl}$ is large and they dominate. Quantum effects are negligible at distances larger than the Planck length, $r\gg 1/ M_{\rm Pl}$, but of course become important as we approach Planck scale. The black hole horizon is the interesting middle regime---analogous to the Vainshtein radius---where classical non-linearities are large and can produce important effects which can be trusted in light of quantum corrections.

\subsubsection{Perturbations around the spherically-symmetric background}

Above we have considered the background field profile around a massive source, but the Vainshtein mechanism can be further understood by considering perturbations about this solution. We can consider linearized perturbations\footnote{See~\cite{Brito:2014ifa} for a numerical investigation of nonlinear perturbations to spherical solutions in the galileon model.} by expanding~\eqref{L3} as $\varphi = \phi - \bar{\phi}$, $T_{\mu\nu} = T_{\mu\nu}+\delta T_{\mu\nu}$ gives
\be
{\cal L}_{\varphi} = \left[ 3+  \frac{2}{\Lambda^3}\left(\bar\phi'' + \frac{2\bar\phi'}{r} \right)\right] \left(\dot{\varphi}^2 - (\partial_\Omega \varphi)^2\right) - \left[3 + \frac{4}{\Lambda^3}\frac{\bar\phi'}{r} \right] (\partial_r\varphi)^2 - \frac{1}{\Lambda^3}\Box\varphi(\partial\varphi)^2 + \frac{g}{M_{\rm Pl}} \varphi \delta T^\mu_{\;\mu} \ ,
\ee
where $\partial_\Omega$ denotes the usual angular derivatives. If we then look deep inside the Vainshtein radius ($r \ll r_{\rm V}$), by substituting the expression~\eqref{Er<<rV} for $\phi'$, we obtain
\be
{\cal L}_{\varphi}\sim \left(\frac{r_{\rm V}}{r}\right)^{3/2} \left(\dot{\varphi}^2 - (\partial_\Omega \varphi)^2 - \frac{4}{3} (\partial_r\varphi)^2 \right) - \frac{1}{\Lambda^3}\Box\varphi(\partial\varphi)^2 + \frac{g}{M_{\rm Pl}} \varphi \delta T^\mu_{\;\mu}.
\label{L3varphi}
\ee
The key thing to notice in this expression is that an enhancement factor of  $(r_{\rm V}/r)^{3/2} \gg 1$ multiplies the kinetic term, telling us that perturbations acquire a large inertia near a massive source.
Said differently, performing the canonical normalization $\varphi_{\rm c} \equiv \left(\frac{r_{\rm V}}{r}\right)^{3/4} \varphi$, the effective coupling to matter is reduced to
\be
g_{\rm eff} \sim \left(\frac{r}{r_{\rm V}}\right)^{3/4} g \ll g\,;
\ee
this indicates that galileon perturbations decouple from matter. Further, the strong coupling scale $\Lambda$ is dressed to a higher scale
\be
\Lambda^{\rm eff} \sim \left(\frac{r_{\rm V}}{r}\right)^{3/4}  \Lambda~\gg \Lambda\,,
\ee
which leads the perturbations to have weaker self-interactions.  

\begin{figure}
\centering
\includegraphics[width=3.3in]{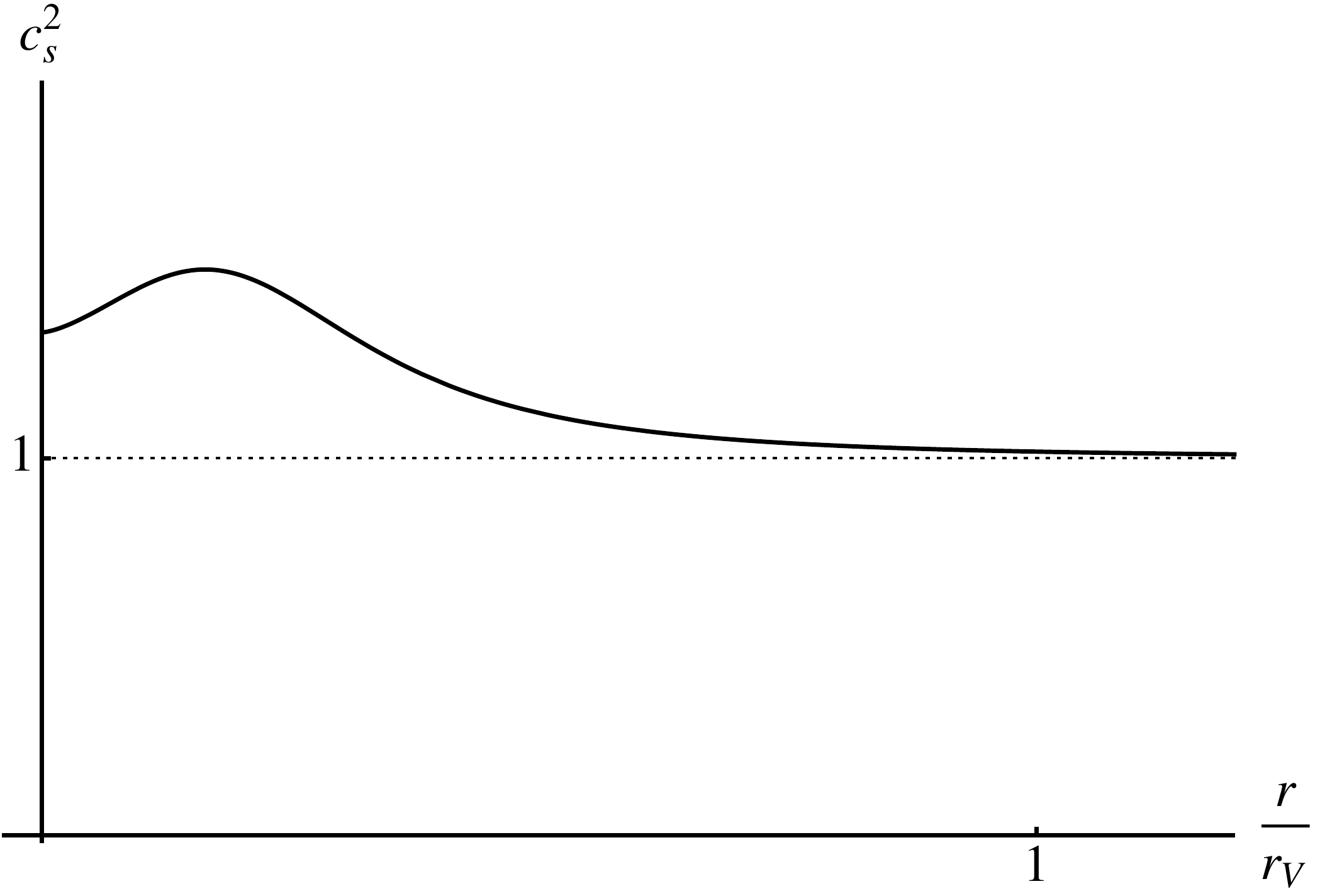}
\caption{\label{DGPsuperluminal}\small Plot of speed of radial fluctuations versus distance from a spherically symmetric source, in units of $r_{\rm V}$ for the cubic galileon.}
\end{figure}

Another thing to notice from~\eqref{L3varphi} is that the radial speed of propagation is {\it superluminal}:
\be
c_{\rm s}^{\rm radial} = \sqrt{\frac{4}{3}}\,.
\ee 
This superluminality is a generic feature of galileons---galileon interactions are derivative interactions, so a galileon background (even one which is arbitrarily weak) deforms the
light-cone for perturbations in such a way that there is always a direction in which the speed of propagation is superluminal~\cite{Nicolis:2009qm}.\footnote{This conclusion only holds for asymptotically flat solutions, by going to asymptotically cosmological solutions, superluminality may be avoided while retaining Vainshtein screening~\cite{Babichev:2012re,Berezhiani:2013dw, Berezhiani:2013dca}. Superluminality is also present in DGP~\cite{Hinterbichler:2009kq} and multi-galileons~\cite{deFromont:2013iwa}.} With superluminality
comes the risk of the standard ghastly paradoxes, such as traveling back in time to kill one's grandfather, which arise if closed time-like curves (CTC) are allowed to form.
The situation is actually not as bad as one might fear. To start with, galileons by themselves are completely fine---the effective light-cone for the metric governing perturbations, albeit wider than the Minkowskian light-cone, admits a well-defined causal structure. In other words, galileons by themselves cannot generate CTCs. On the other hand, CTCs become possible when considering galileons coupled to (Lorentz-invariant) matter. However, it was conjectured in~\cite{Burrage:2011cr} that galileons are protected from the formation of CTCs by a Chronology Protection Criterion, analogous to that of GR~\cite{Hawking:1991nk}. If one starts with healthy initial conditions and tries to construct a CTC, the galileon effective field theory will break down before it can form~\cite{Burrage:2011cr}. As discussed in Appendix~\ref{superlumapp}, however, the existence of superluminal propagation around certain backgrounds signals the UV completion of galileons, if one exists, is not a local (Lorentz-invariant) quantum field theory but something more exotic (or more interesting, depending on one's point of view). However, recent arguments suggest that this apparent superluminality might be an artifact of trusting a tree-level computation in a regime where it is unreliable~\cite{deRham:2013hsa, deRham:2014lqa}, so the severity of this peculiarity is far from settled.
A related tension is that the galileon terms lead to scattering amplitudes which do not obey dispersion relations obtained from arguments about S-matrix analyticity, discussed in Appendix~\ref{superlumapp}. On the other hand, this argument relies on the existence of an S-matrix for galileons, which has been questioned recently in~\cite{Berezhiani:2013dca,Berezhiani:2013dw}.

Here we have focused on Vainshtein screening around static sources. The mechanism is expected also to operate in time-dependent situations. In these cases, things are substantially more difficult to treat analytically, but initial investigations of binary systems indicate that the mechanism does indeed operate, however it is somewhat less efficient than might be na\"ively estimated~\cite{deRham:2012fw,Chu:2012kz,deRham:2012fg}. One way to understand this is that the indefinite signature of the spacetime metric allows for cancellations between $\dot{\bar\phi}$ and $\vec\nabla\bar\phi$ in the kinetic matrix. This should be viewed as an opportunity---it may prove possible to find some time-dependent astrophysical situation where screening is weak, leading to large deviations from GR predictions.

\subsubsection{General galileons}
\label{generalgals}

Above we investigated the simplest galileon theory, the cubic galileon, but
in $d$ dimensions, there are $(d+1)$ galileon terms. In four dimensions, they take the form
\bea  
\label{galileonterms}
{\cal L}_1&=&\phi\ , \nonumber \\
{\cal L}_2&=&\frac{1}{2}(\partial\phi)^2 \ ,\nonumber \\
{\cal L}_3&=&\frac{1}{2}\square {\phi}(\partial {\phi})^2 \ ,\\
{\cal L}_4&=&\frac{1}{4}(\partial\phi)^2\left((\square\phi)^2-(\partial_\mu\partial_\nu\phi)^2\right) , \nonumber \\
{\cal L}_5&=& \frac{1}{3}(\partial\phi)^2\left((\square\phi)^3+2(\partial_\mu\partial_\nu\phi)^3-3\square\phi(\partial_\mu\partial_\nu\phi)^2\right)\nonumber\ .
\eea
A more compact way of expressing this is that (in $d$ dimensions) the $n$-th Lagrangian is given by the expression (up to overall normalization)
\be
{\cal L}_{n} = (n-1)\eta^{\mu_1\nu_1\mu_2\nu_2\cdots\mu_{n-1}\nu_{n-1}}\, \phi\partial_{\mu_1}\partial_{\nu_1}\phi\partial_{\mu_2}\partial_{\nu_2}\phi\cdots\partial_{\mu_{n-1}}\partial_{\nu_{n-1}}\phi\,.
\label{galotherparameterization}
\ee 
Here we have defined $\eta^{\mu_1\nu_1\mu_2\nu_2\cdots\mu_n\nu_n}\equiv{1\over n!}\sum_p\left(-1\right)^{p}\eta^{\mu_1p(\nu_1)}\eta^{\mu_2p(\nu_2)}\cdots\eta^{\mu_np(\nu_n)}$, with the sum running over all permutations of the $\nu$ indices, with $(-1)^p$ the sign of the permutation.\footnote{An equivalent representation in terms of Levi--Civita symbols is
\begin{equation*}
\mathcal{L}_n \sim \epsilon_{\mu_1\cdots\mu_{n-1}\alpha_{n}\cdots\alpha_{d}}\epsilon^{\nu_1\cdots\nu_{n-1}\alpha_{n}\cdots\alpha_{d}}\phi \partial_{\nu_1}\partial^{\mu_1}\phi\cdots\partial_{\nu_{n-1}}\partial^{\mu_{n-1}}\phi\sim\delta_{\mu_1}^{[\nu_1}\cdots\delta_{\mu_{n-1}}^{\nu_{n-1}]} \phi \partial_{\nu_1}\partial^{\mu_1}\phi\cdots\partial_{\nu_{n-1}}\partial^{\mu_{n-1}}\phi~.
\end{equation*}
}
This $\eta$ tensor is anti-symmetric in the $\mu$ indices, anti-symmetric in the $\nu$ indices, and symmetric under swapping any pair of $\mu$, $\nu$ indices with any other pair.

In this representation, the symmetries of the $\eta$ tensor make it straightforward to derive the equations of motion by varying the Lagrangian with respect to $\phi$. The symmetries forbid three derivatives acting on any field, so all of the terms give the same contribution to the equations of motion after integration by parts, leading to the Euler--Lagrange equation
\be
{\cal E}_n = n(n-1)\eta^{\mu_1\nu_1\mu_2\nu_2\cdots\mu_{n-1}\nu_{n-1}}\partial_{\mu_1}\partial_{\nu_1}\phi\partial_{\mu_2}\partial_{\nu_2}\phi\cdots\partial_{\mu_{n-1}}\partial_{\nu_{n-1}}\phi = 0~,
\ee
which is manifestly second-order (only second derivatives of the field appear). We will now discuss some interesting properties of the general galileon theory.

\noindent{\bf Euler hierarchy:}\\
\indent The galileons have the interesting property that the $(n+1)$-th galileon Lagrangian is just $(\partial\phi)^2$ times the $n$-th galileon equation of motion. To see this, we employ an identity satisfied by the $\eta$ tensor~\cite{Hinterbichler:2010xn}
\be
\eta^{\mu_1\nu_1\mu_2\nu_2\cdots\mu_{n}\nu_{n}} = \frac{1}{n}\Big(\eta^{\mu_1\nu_1}\eta^{\mu_2\nu_2\cdots\mu_{n}\nu_{n}}-\eta^{\mu_1\nu_2}\eta^{\mu_2\nu_1\mu_2\nu_3\cdots\mu_{n}\nu_{n}}+\cdots+(-1)^n\eta^{\mu_1\nu_n}\eta^{\mu_2\nu_1\cdots\mu_n\nu_{n-1}}\Big)~.
\ee
This allows us to express ${\cal L}_{n+1}$ in terms of the equation of motion for ${\cal L}_n$:
\be
{\cal L}_{n+1} = -\frac{n+1}{2n(n-1)}(\partial\phi)^2{\cal E}_n+\frac{n-1}{2}\partial_{\mu_1}\Big((\partial\phi)^2\eta^{\mu_1\nu_1\cdots\mu_{n-1}\nu_{n-1}}\partial_{\nu_1}\phi\partial_{\mu_2}\partial_{\nu_2}\phi\cdots\partial_{\mu_{n-1}}\partial_{\nu_{n-1}}\phi\Big)~.
\ee
Theories of this type---where the $(n+1)$-th order Lagrangian can be built from the equation of motion of the $n$-th order Lagrangian plus a total derivative---are well studied, and are known as {\it Euler hierarchies}~\cite{Fairlie:1991qe,Fairlie:1992nb, Fairlie:1992he, Curtright:2012gx}.

\noindent{\bf Non-renormalization:}\\
\indent Another remarkable property enjoyed by the galileons is that {\it they do not receive quantum corrections at any order in perturbation theory}~\cite{Luty:2003vm,Hinterbichler:2010xn}. This holds rather generally, for any number of galileons in any number of dimensions~\cite{Hinterbichler:2010xn}, but here we will focus on the quintic galileon in four dimensions, following the arguments of~\cite{Hinterbichler:2010xn}. To begin, we consider the most general Lagrangian compatible with the symmetries: a linear combination of the various galileon Lagrangians~\eqref{galileonterms} and higher-derivative terms 
\be
{\cal L} = \sum_{n=2}^5 \frac{c_n}{\Lambda^{3(n-2)}}{\cal L}_n+ \sum_{n = 2}^\infty\sum_{\ell=0}^\infty \frac{c_{n, \ell}}{\Lambda^{3n+2\ell-4}_{\rm s}} \partial^{2\ell}(\partial^2\phi)^n ~,
\ee
where we have set $c_1=0$ so that there is no tadpole and to guarantee that $\bar\phi=0$ is a solution to the equations of motion. In order to study the quantum structure of the theory, we compute the effective action
\be
\Gamma(\phi_c) = \Gamma^{(2)} \phi_c\phi_c+\Gamma^{(3)}\phi_c\phi_c\phi_c+\ldots~,
\label{galeffectiveaction}
\ee
where $\Gamma^{(n)}$ is the $n$-point 1-particle irreducible (1PI) vertex. As we will see, when we compute~\eqref{galeffectiveaction} the only terms that get generated have at least two derivatives per field---the galileon terms receive no quantum corrections. This can be understood intuitively as follows: the symmetries of the $\eta$-tensor forbid three derivatives acting on any one field, so we can integrate by parts freely. Therefore, when we compute the quantum effective action, we are free to integrate by parts to put at least two derivatives on each external line.

\begin{figure}[tb]
\centering
\includegraphics[width=5.5in]{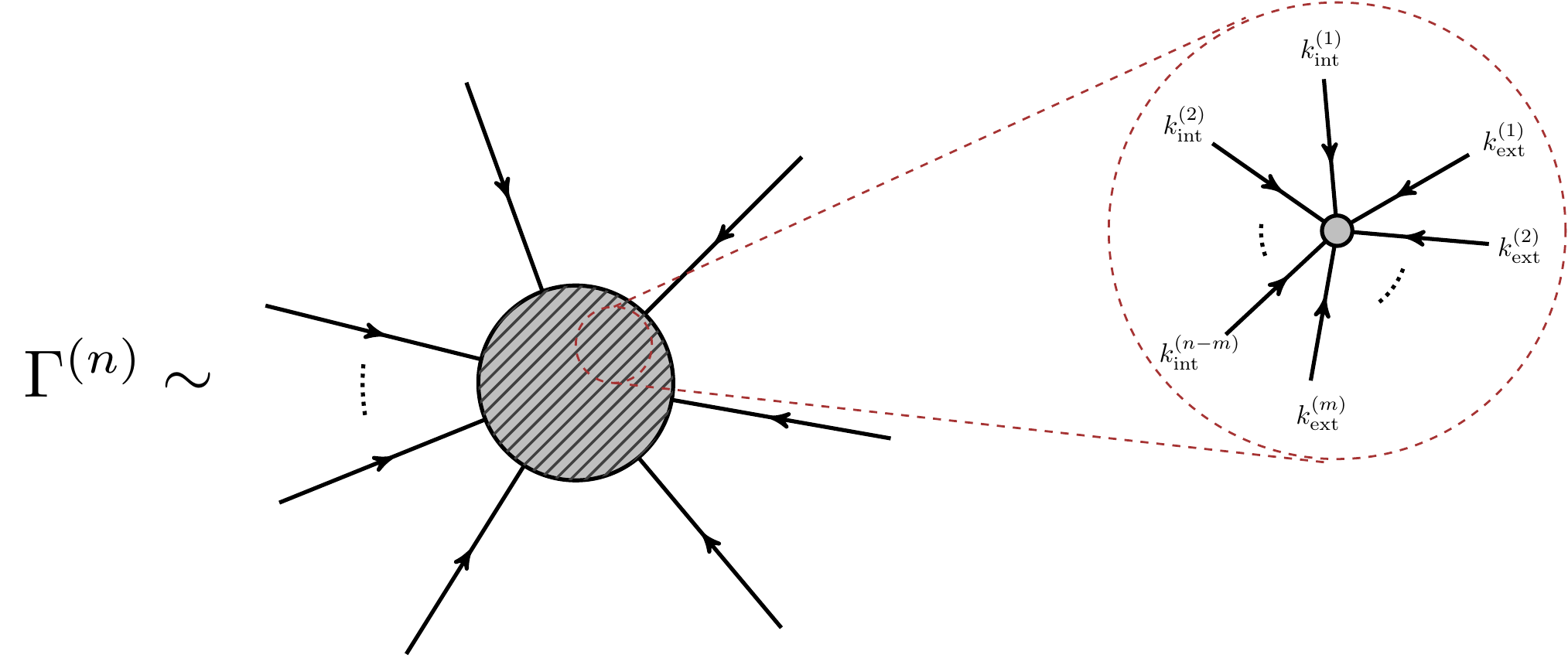}
\caption{\small A contribution to the quantum effective action in the theory of a galileon. For a given 1PI graph, focusing on a sub-diagram with $m$ external lines allows us to deduce that only terms with two derivatives per field are generated quantum-mechanically. Therefore, the galileons are {\it not} renormalized.
}
  \label{nonrenormfig}
\end{figure}

We now formalize this line of reasoning. Within a given 1PI diagram, focus on one of its constituent vertices which connects to $m$ external lines. This situation is depicted in Figure~\ref{nonrenormfig}. If the vertex comes from a non-galileon term, then clearly each external line will have at least two derivatives acting on it, and the graph will not generate a galileon term in $\Gamma$. We can then focus on the case where the vertex comes from one of the galileons. There are two cases: if an external line comes from a $\partial^2\phi$ factor in the galileon, then it clearly has two derivatives acting on it. Therefore, the only worrisome case is if the external line comes from the undifferentiated $\phi$ in ${\cal L}_n$, and the contraction takes the form
\be
{\cal L}_{n} \sim \eta^{\mu_1\nu_1\mu_2\nu_2\cdots\mu_{n-1}\nu_{n-1}}\left( \phi_{\rm ext}\partial_{\mu_1}\partial_{\nu_1}\phi_{\rm ext}\cdots\partial_{\mu_{m-1}}\partial_{\nu_{m-1}}\phi_{\rm ext}\cdots\partial_{\mu_{m}}\partial_{\nu_{m}}\phi_{\rm int}\cdots\partial_{\mu_{n-1}}\partial_{\nu_{n-1}}\phi_{\rm int}\right).
\label{internalvertex}
\ee
We now use the symmetries of the $\eta$-tensor to write the piece containing internal fields as a double total derivative:
\be
{\cal L}_{n} \sim \eta^{\mu_1\nu_1\mu_2\nu_2\cdots\mu_{n-1}\nu_{n-1}}\left( \phi_{\rm ext}\partial_{\mu_1}\partial_{\nu_1}\phi_{\rm ext}\cdots\partial_{\mu_{m-1}}\partial_{\nu_{m-1}}\phi_{\rm ext}\cdots\partial_{\mu_{m}}\partial_{\nu_{m}}\left[\phi_{\rm int}\cdots\partial_{\mu_{n-1}}\partial_{\nu_{n-1}}\phi_{\rm int}\right]\right).
\ee
This means that the Feynman rule for this vertex has two factors of the sum of internal momenta, $\sum k_{\rm int}$, which we may trade for external momenta $\sum k_{\rm int} = -\sum k_{\rm ext}$. This means that the Feynman rules for the vertex~\eqref{internalvertex} have two powers of external momentum for each external field.

Since every external line comes with two powers of momentum in every 1PI vertex, only terms with two derivatives per field get generated in the quantum effective action---the coefficients of the galileon terms do not receive quantum corrections. This holds to {\it all loop orders}. As an example, the 1-loop quantum effective action is of the form~\cite{Nicolis:2004qq}
\be
\Gamma \sim \sum_m\left[\Lambda^4+\Lambda^2\partial^2+\partial^4\log\left(\frac{\partial^2}{\Lambda^2}\right)\right]\left(\frac{\partial^2\phi}{\Lambda^3}\right)^m~,
\ee
which clearly only has contributions to terms with at least two derivatives per field.

\noindent{\bf Topological nature:}\\
\indent A surprising fact about galileons is that they are {\it topological} in a suitably understood sense. If we interpret the galileon as a goldstone boson, non-linearly realizing the symmetries~\eqref{galileansymmetry}, the five galileon terms appear as Wess--Zumino terms~\cite{Goon:2012dy}. Here we give a brief sketch of the construction. To begin, we note that the galileons parameterize the coset space
\be
{\rm Gal}\big((d-1)+1, 1\big)/{\rm SO}((d-1),1)~,
\label{galcoset}
\ee
where ${\rm Gal}\big((d-1)+1, 1\big)$ is the {\it galileon group} (distinct from the galilean group).\footnote{
The commutation relations for the {\it galileon algebra} are
\begin{equation*}
\left[P_\mu, B_\nu\right] = \eta_{\mu\nu}C~;~~~~~~~~~~~~~~~\left[J_{\rho\sigma}, B_\nu\right] = \eta_{\rho\nu}B_\sigma-\eta_{\sigma\nu}B_\rho~,
\end{equation*}
plus the commutation relations of the Poincar\'e algebra.
Here, $C$ and $B_\nu$ generate the constant and linear-gradient shifts in the scalar field, while $P_\mu$, $J_{\mu\nu}$ are the standard generators of the Poincar\'e algebra.
} 
 It is convenient to coordinatize this space by $(x^\mu, \phi, \xi^\mu)$. It turns out that the additional parameter $\xi^\mu$ is redundant and can be traded for derivatives of $\phi$ via
\be
\xi_\mu = -\partial_\mu\phi~,
\ee
but it is most straightforward to not make this substitution until the very end of our calculation.\footnote{The fact that the $\xi^\mu$ can be eliminated is a consequence of the fact that $\phi$ is a goldstone field for a broken {\it space-time} symmetry. In this case, $\phi$ is able to non-linearly realize all five symmetries~\eqref{galileansymmetry}---and thus parameterize the coset~\eqref{galcoset} by itself. In the literature, this often goes by the name {\it inverse Higgs effect}~\cite{Ivanov:1975zq}. There is a vast literature on non-linear realizations of space-time symmetries in general and the inverse Higgs effect in particular. For various perspectives, see~\cite{Volkov:1973vd,Nielsen:1975hm,Low:2001bw,McArthur:2010zm, Hidaka:2012ym, Watanabe:2012hr, Nicolis:2012vf,Nicolis:2013sga}.
}

On this coset space, the vector fields
\be
C= \partial_\phi~;~~~~~~~~~~~~~B_\mu = \partial_{\xi_\mu}+x_\mu\partial_\phi~,
\label{LIVFSgalcoset}
\ee
generate the transformations~\eqref{galileansymmetry}. Further, we can also construct a basis of left-invariant 1-forms:
\be
\omega_P^\mu = \rd x^\mu~;~~~~~~~~~~~~~
\omega_B^\mu = \rd\xi^\mu~;~~~~~~~~~~~~~
\omega_C = \rd \phi+\xi_\mu\rd x^\mu~.
\label{MC1forms}
\ee

Since these 1-forms have nice transformation properties under the galileon symmetries, it is straightforward to construct invariant actions using them. Indeed this is the entire basis of the {\it coset construction} of Callan, Coleman, Wess and Zumino~\cite{Coleman:1969sm,Callan:1969sn} and Volkov~\cite{Volkov:1973vd}, which is familiar to particle physicists. The procedure is to combine the 1-forms~\eqref{MC1forms} with the exterior product into a $d$-form which is then integrated over space-time to produce an action. However, when we impose the constraint $\xi_\mu = -\partial_\mu\phi$, we find that $\omega_C=0$, so the only building block at our disposal is $\omega_B^\mu = -\rd x^\nu\partial_\nu\partial^\mu\phi$. If we construct actions from this building block, clearly every field $\phi$ will come with at least two derivatives, so it will be impossible to construct the galileon terms~\eqref{galileonterms}. This is not surprising, as this procedure generates {\it strictly invariant} Lagrangians, while we know the galileons shift by a total derivative under the symmetries.

The resolution is that the galileon terms do not correspond to $d$-forms built from~\eqref{MC1forms}, but rather come from $(d+1)$-forms pulled back to the physical space-time. Terms that must be constructed in this way are known as {\it Wess--Zumino} terms~\cite{Wess:1971yu,Witten:1983tw, D'Hoker:1994ti}. To see how this works, we consider the $(d+1)$-form
\be
\omega_n^{\rm wz}=\epsilon_{\mu_1\cdots\mu_d}\omega_C\wedge\omega_B^{\mu_1}\wedge\cdots\wedge\omega_B^{\mu_{n-1}}\wedge\omega_P^{\mu_n}\wedge\cdots\wedge\omega_P^{\mu_d}~,
\ee
which is {\it exact}:
\be
\omega_n^{\rm wz} = \rd \beta^{\rm wz}_n~.
\ee
(We will not concern ourselves with the explicit form of $\beta_n^{\rm wz}$; it can be found in~\cite{Goon:2012dy}.) The correspondence between these objects and the galileon Lagrangians is~\cite{Goon:2012dy}
\be
S = \int_M \omega^{\rm wz}_n= \int_{\partial M}\beta^{\rm wz}_n = (-1)^{n-1}\frac{1}{n}\int\rd^dx~{\cal L}_n~,
\ee
where we have defined the $n$-th galileon as
\be
{\cal L}_n = (d-n+1)!(n-1)!\delta_{\mu_1}^{[\nu_1}\cdots\delta_{\mu_{n-1}}^{\nu_{n-1}]} \phi \partial_{\nu_1}\partial^{\mu_1}\phi\cdots\partial_{\nu_{n-1}}\partial^{\mu_{n-1}}\phi~.
\label{wzlags1}
\ee

In order to understand in what sense the galileons are topological, we note that while $\omega_n^{\rm wz}$ is left-invariant under the action of the vector fields~\eqref{LIVFSgalcoset}, the $d$-form  $\beta^{\rm wz}_n$ is {\it not}. Rather, it shifts by a total derivative under this transformation, so we see that the form $\omega_n^{\rm wz}$ is left-invariant under the symmetries, but {\it cannot} be written as the exterior derivative of a form which is itself left-invariant. The form $\omega_n^{\rm wz}$ is therefore a nontrivial element of what is known as {\it Chevalley--Eilenberg cohomology}~\cite{Chevalley:1948zz}.\footnote{This cohomology theory also sometimes goes by the name {\it relative Lie algebra cohomology}. For a nice introduction to physical applications of these ideas, see~\cite{deAzcarraga:1995jw, deAzcarraga:1998uy, deAzcarraga:1997gn}.
}

So, we see that the galileons are topological in that they are nontrivial elements in a particular cohomology theory. This can be seen as a generalization of Witten's construction of the WZ term in the chiral Lagrangian~\cite{Witten:1983tw}. Note that here, in contrast to the situation in the chiral Lagrangian, there does not exist a global topological criterion which quantizes the coefficients of the galileon terms. In the chiral Lagrangian, this is the underlying reason for the non-renormalization of the Wess--Zumino--Witten term. However, the topological nature of the galileons does give us some insight into their non-renormalization. Being Wess--Zumino terms, the galileons only shift by a total derivative under the relevant symmetries. However, all the building blocks of the quantum effective action are {\it strictly} invariant, so it is not surprising that we cannot build an operator that renormalizes the galileons using them.

\noindent{\bf Duality:}\\
\indent A final, and unexpected property exhibited by the galileons is that they are {\it self-dual}. As was shown in~\cite{Fasiello:2013woa, deRham:2013hsa}, by performing simultaneously a field-dependent coordinate transformation and a field redefinition
\begin{align}
y^\mu &\longmapsto x^\mu +\frac{1}{\Lambda}\partial^\mu\rho\\
\pi &\longmapsto -\rho-\frac{1}{2\Lambda^2}(\partial\rho)^2\,,
\end{align}
a galileon theory 
\be
{\cal L}[\pi] = \sum_{n=2}^5 c_n{\cal L}_n[\pi]~,
\ee
where ${\cal L}_n$ is defined as in~\eqref{wzlags1}, and where we have set the tadpole term to zero for simplicity, is mapped to to a dual galileon theory
\be
{\cal L}[\rho] = \sum_{n=2}^5 p_n{\cal L}_n[\rho]~,
\ee
where the relation between the coefficients is
\be
\left(
\begin{array}{c}
p_2\\
p_3\\
p_4\\
p_5
\end{array}
\right)
=
\left(
\begin{array}{ccccc}
1 & 0 & 0 & 0\\
2 & -1 & 0 & 0\\
\frac{3}{2} & -\frac{3}{2} & 1 & 0\\
\frac{2}{5} & -\frac{3}{5} & \frac{4}{5} & -1\\
\end{array}
\right)
\left(
\begin{array}{c}
c_2\\
c_3\\
c_4\\
c_5
\end{array}
\right)~.
\ee
Here we have focused on four dimensions, but an equivalent duality holds in any number of dimensions (the relations between the coefficients just become more intricate~\cite{deRham:2013hsa}). The key fact that makes such a duality possible is that the derivative of the scalar field transforms as a {\it scalar} under the simultaneous redefinition and coordinate transformation:
\be
\frac{\partial \pi}{\partial y^\mu} \longmapsto \frac{\partial \rho}{\partial x^\mu} ~.
\ee
This duality is actually much more general than just mapping galileon theories into each other, recently it has been generalized to a wide class of scalar field theories and even to vector fields~\cite{deRham:2014lqa, Kampf:2014rka}. There also exists a similar mapping between theories of conformal galileons and conformal DBI galileons~\cite{Bellucci:2002ji,Creminelli:2013fxa,Creminelli:2014zxa}

This duality is interesting for a number of reasons, but perhaps the most interesting is investigating the dual of a free theory:
\be
{\cal L} = -\frac{1}{2}(\partial\pi)^2~,
\ee
which corresponds to taking $c_2 = -1/12$ and $c_3=c_4=c_5=0$. The dual theory has $p_2= -1/12$, $p_3 = -1/6$, $p_4=-1/8$ and $p_5=-1/30$. One can check explicitly that the S-matrix elements on both side of the duality agree~\cite{deRham:2013hsa,Kampf:2014rka}. Furthermore, the specific quintic galileon theory to which the free particle maps admits superluminal solutions {\it despite} having a precisely analytic S-matrix (it is trivial in fact). In addition to further obfuscating the relationship between superluminality and analyticity, it forces us to question to what extent tree-level superluminality is a problem. On one side of the duality, we have a free theory, which is perfectly causal and UV-complete, while on the other side we have a theory which has apparent superluminality. However, the equivalence between the theories indicates that if we were to properly calculate everything on the ``hard'' side, this theory would not have superluminality either.

\subsection{The Vainshtein mechanism: massive gravity}
\label{massivegravsec}

Recall that Einstein's theory of gravity is the theory of a massless spin-2 particle, the graviton. The notion of giving this particle a mass and considering a theory of {\it massive gravity} has been intriguing to theorists since Fierz and Pauli's discovery of a unique ghost-free linear Lagrangian for a massive graviton, but until recently progress had come to a halt due to the powerful no-go theorem of Boulware and Deser~\cite{Boulware:1973my}, who had described the obstacles to finding a nonlinear completion of this theory. However, in the last few years, a loophole in these objections has been found, and an interacting theory of a massive graviton, free of the Boulware--Deser ghost, has emerged~\cite{deRham:2010ik,deRham:2010kj}.\footnote{See~\cite{Gabadadze:2009ja,deRham:2009rm,deRham:2010gu,Hassan:2011zr,Berezhiani:2011nc} for work on a related model of massive gravity constructed from an auxiliary extra dimension.} (See \cite{Hinterbichler:2011tt,deRham:2014zqa} for excellent reviews of all aspects of massive gravity.) Interestingly,  massive gravity admits self-accelerating solutions in which the de Sitter Hubble factor is of order the mass of the graviton. Since having a light graviton is technically natural \cite{ArkaniHamed:2002sp,deRham:2012ew,deRham:2013qqa}, such a solution is of great interest in the late-time universe as a candidate explanation for cosmic acceleration.

\subsubsection{A brief history of massive gravity}

The history of massive gravity begins when Fierz and Pauli first wrote down the quadratic action for a massive spin-2 particle in 1939~\cite{Fierz:1939ix}. The theory remained somewhat of a theoretical oddity until van Dam, Veltman and Zakharov \cite{vanDam:1970vg, Zakharov:1970cc} noticed around 1970 that the Fierz--Pauli theory does not reduce to linearized GR in the limit in which the mass of the graviton goes to zero; instead, in this limit, there is an additional scalar polarization mode which does not decouple and leads to an ${\cal O}(1)$ departure from GR predictions for the bending of light. This fact is known as the {\it vDVZ discontinuity}, because it indicates that the Fierz--Pauli theory does not have a smooth $m\to 0$ limit. This discrepancy was resolved shortly thereafter by Vainshtein~\cite{Vainshtein:1972sx}, who noted that completing the non-interacting Fierz--Pauli theory to an interacting non-linear theory makes the limit smooth. In an interacting theory, this additional scalar field becomes strongly self-interacting as the graviton mass is taken to zero, causing it to be screened near heavy sources; this is the {\it Vainshtein mechanism} we have discussed above. However, shortly after this discovery of Vainshtein, Boulware and Deser argued that a generic theory of an interacting massive graviton would not propagate the expected 5 polarization states, but would have an additional ghostly polarization~\cite{Boulware:1973my}. These arguments were further solidified in~\cite{ArkaniHamed:2002sp,Creminelli:2005qk}. In~\cite{deRham:2010ik, deRham:2010kj}, de Rham, Gabadadze and Tolley (dRGT) found a loophole and constructed a theory that is free of this pathological sixth mode (this was shown conclusively in~\cite{Hassan:2011hr}). Much work has gone into studying this massive gravity theory: for example studies of cosmological solutions~\cite{deRham:2010tw,D'Amico:2011jj,Koyama:2011yg,Nieuwenhuizen:2011sq,Chamseddine:2011bu,Gumrukcuoglu:2011ew,Gratia:2012wt,Kobayashi:2012fz,Gumrukcuoglu:2012aa,Langlois:2012hk,Motohashi:2012jd,Maeda:2013bha} and black holes~\cite{Nieuwenhuizen:2011sq,Berezhiani:2011mt,Deffayet:2011rh,Volkov:2012wp,Mirbabayi:2013sva,Brito:2013wya,Babichev:2013una} have been undertaken. The theory can also be cast in a vielbein language, as opposed to as a theory of a metric~\cite{Zumino:1970tu,Nibbelink:2006sz,Chamseddine:2011mu,Hinterbichler:2012cn,Deffayet:2012nr,Deffayet:2012zc}. Further, it has been realized that the dRGT construction opens the door to building theories with multiple interacting spin-2 particles~\cite{Hassan:2011zd, Hinterbichler:2012cn}. In what follows, we will consider Lorentz invariant theories of massive gravity, but much work has been done on Lorentz violating theories~\cite{Rubakov:2004eb,Dubovsky:2004sg,Dubovsky:2004ud,Gabadadze:2004iv,Kirsch:2005st,Libanov:2005vu,ArkaniHamed:2005gu,Cheng:2006us,Berezhiani:2007zf,Blas:2007ep,Dubovsky:2007zi,Rubakov:2008nh,Grisa:2008um,Blas:2009my,Lin:2013aha,Langlois:2014jba}.

\subsubsection{Linearized massive gravity}
\label{linearizedmassiveg}

We begin our study of massive gravity by considering the linearized theory of Fierz and Pauli~\cite{Fierz:1939ix}. Throughout we will restrict to the case of $d=4$, but generalizations to other dimensions are possible and relatively straightforward. The Fierz--Pauli theory corresponds to the action
\be
S = M_{\rm Pl}^2\int\rd^4x\left[-\frac{1}{2}\partial_\alpha h_{\mu\nu}\partial^\alpha h^{\mu\nu}+\partial_\mu h_{\nu\alpha}\partial^\nu h^{\mu\alpha}-\partial_\mu h^{\mu\nu}\partial_\nu h+\frac{1}{2}\partial_\alpha h\partial^\alpha h-
\frac{m^2}{2}\left(h_{\mu\nu}h^{\mu\nu}-h^2\right)\right]~.
\label{fierzpaulilag}
\ee
The derivative terms here are the same as those of linearized Einstein gravity, and the mass term has a particular relative coefficient between the two allowed terms (Fierz and Pauli showed that this is the only consistent choice).
The equation of motion descending from this action is then
\be
\square h_{\mu\nu}+\partial_\mu\partial_\nu h-\partial_\mu\partial_\alpha h_\nu^{~\alpha}-\partial_\nu\partial_\alpha h_{\mu}^{~\alpha}+\eta_{\mu\nu}\left(\partial_\alpha\partial_\beta h^{\alpha\beta}-\square h\right)- m^2\left(h_{\mu\nu}-\eta_{\mu\nu}h\right)=0~.
\ee
By taking the divergence of this equation and various traces, one can verify that this equation of motion is equivalent to the following three equations~\cite{Hinterbichler:2011tt}:
\be
\left(\square-m^2\right)h_{\mu\nu}=0~;~~~~~~~~~~~~~~\partial^\mu h_{\mu\nu} = 0~;~~~~~~~~~~~~~~h=0~.
\ee
Analyzing these equations, we can determine that the field $h_{\mu\nu}$ propagates the expected five polarizations of a massive spin-2 particle. A symmetric tensor $h_{\mu\nu}$ na\"ively has 10 independent components. Demanding that it be traceless removes one degree of freedom, and demanding that it be divergence-less provides 4 more constraints, leaving 5 independent components. The remaining equation is a wave equation for these 5 propagating polarizations. 

The kinetic terms in~\eqref{fierzpaulilag} are invariant under linearized diffeomorphisms, which act as
\be
\delta_\xi h_{\mu\nu} = \partial_\mu\xi_\nu+\partial_\nu\xi_\mu~,
\ee
but the mass term is not. This is easily fixed though: gauge invariance is not fundamental, it is merely a redundancy of description in a system, so we are free to reintroduce it. It turns out to be extremely useful to make the action~\eqref{fierzpaulilag} gauge invariant by introducing additional fields using the {\it St\"uckelberg trick}. This introduction of gauge invariance will make the additional degrees of freedom carried by a massive graviton manifest, and isolate the origin of the vDVZ discontinuity in the $m\to0$ limit. This analysis was first applied to massive gravity in~\cite{ArkaniHamed:2002sp}.

The St\"uckelberg trick is to make what looks like a gauge transformation
\be
h_{\mu\nu} \longmapsto h_{\mu\nu}+\partial_\mu A_\nu+\partial_\nu A_\mu~,
\label{stuckel}
\ee
under which the derivative terms are unchanged, while the mass terms shift to give
\begin{align}
\nonumber
S = M^2_{\rm Pl}&\int\rd^dx \left[-\frac{1}{2}h^{\mu\nu}{\cal E}^{\alpha\beta}_{\mu\nu}h_{\alpha\beta}-
\frac{1}{2}m^2\left(h_{\mu\nu}h^{\mu\nu}-h^2\right)+\frac{1}{M_{\rm Pl}^2}h^{\mu\nu}T_{\mu\nu} \right.\\
&~~~~~~~~~~~~~~~~~~~~~~~~~~~~~~~~~~\left.+m^2\left(-\frac{1}{2}F_{\mu\nu}F^{\mu\nu}-2h^{\mu\nu}\partial_\mu A_\nu+2 h\partial_\mu A^\mu\right)\right]~.
\label{astuckelberg}
\end{align}
Here we have written the kinetic term using the Lichnerowicz operator\footnote{The Lichnerowicz operator is given by
\begin{equation*}
{\cal E}^{\alpha\beta}_{\mu\nu}=-\frac{1}{2} \left[\delta_\mu^\alpha\delta_\nu^\beta\square-\delta_\mu^\beta\partial_\nu\partial^\alpha-\delta_\nu^\beta\partial_\mu\partial^\alpha+\eta^{\alpha\beta}\partial_\mu\partial_\nu+\eta_{\mu\nu}\left(\partial^\alpha\partial^\beta-\eta^{\alpha\beta}\square\right)\right]~.
\end{equation*}
} and defined the tensor $F_{\mu\nu} = \partial_\mu A_\nu -\partial_\nu A_\mu$. We have also included a coupling of the field $h_{\mu\nu}$ to an external source, $h^{\mu\nu} T_{\mu\nu}$, which we assume is conserved: $\partial^\mu T_{\mu\nu} = 0$. This replacement makes the action invariant under a gauge transformation where both fields shift
\be
\delta_\xi h_{\mu\nu} = \partial_\mu\xi_\nu+\partial_\nu\xi_\mu~;~~~~~~~~~~~~~~~\delta_\xi A_\mu = -\xi_\mu~.
\ee
Notice that fixing the gauge $A_\mu = 0$ recovers the Fierz--Pauli action. This is progress, but we still have not made all the degrees of freedom manifest; in order to do this, we introduce another gauge symmetry by performing a further St\"uckelberg decomposition
\be
A_\mu \longmapsto A_\mu +\partial_\mu\phi~.
\ee
Along with this replacement, we make one additional transformation to de-mix the graviton and scalar degrees of freedom. We first rescale the fields as $h_{\mu\nu}\mapsto \frac{1}{M_{\rm Pl}}h_{\mu\nu}$, $A_\mu \mapsto \frac{1}{mM_{\rm Pl}}A_\mu$ and $\phi\mapsto \frac{1}{m^2M_{\rm Pl}}\phi$, and then perform the field redefinition
\be
h_{\mu\nu}\longmapsto h_{\mu\nu}+\phi\eta_{\mu\nu}~.
\label{linearconformal}
\ee
After these manipulations, the action~\eqref{astuckelberg} takes the form (see~\cite{Hinterbichler:2011tt} for details)
\begin{align}
\nonumber
S = \int\rd^4x&\left[-\frac{1}{2}h^{\mu\nu}{\cal E}^{\alpha\beta}_{\mu\nu}h_{\alpha\beta}-
\frac{1}{2}m^2\left(h_{\mu\nu}h^{\mu\nu}-h^2\right)-\frac{1}{2}F_{\mu\nu}F^{\mu\nu}+3\phi\bigg(\square+2m^2\bigg)\phi\right.\\\label{linearizedmassivegravity}
&~~-2m\Big(h_{\mu\nu}\partial^\mu A^\nu-h\partial_\mu A^\mu\Big)+3m\Big(2\phi\partial_\mu A^\mu+mh\phi\Big)+\left.\frac{1}{M_{\rm Pl}}h^{\mu\nu}T_{\mu\nu}+\frac{\phi}{M_{\rm Pl}}T\right]~.
\end{align}
This action is now invariant under a pair of gauge transformations:
\begin{align}
&\delta_\xi h_{\mu\nu} = \partial_\mu\xi_\nu+\partial_\nu\xi_\mu~,~~~~~~~~\delta_\xi A_\mu = -m\xi_\mu~,~~~~~~~~\delta_\xi\phi=0~;\\
&\delta_\Lambda h_{\mu\nu} =m\Lambda\eta_{\mu\nu}~,~~~~~~~~~~~~~\delta_\Lambda A_\mu = \partial_\mu\Lambda~,~~~~~~~~~~~\delta_\Lambda \phi = -m\Lambda~.
\end{align}
We are now poised to take the $m\to0$ limit in the action~\eqref{linearizedmassivegravity}. In this limit, all of the degrees of freedom decouple from each other, and we are left with the action
\be
S = \int\rd^4x\left(-\frac{1}{2}h^{\mu\nu}{\cal E}^{\alpha\beta}_{\mu\nu}h_{\alpha\beta}-\frac{1}{2}F_{\mu\nu}F^{\mu\nu}+3\phi\square\phi+\frac{1}{M_{\rm Pl}}h^{\mu\nu}T_{\mu\nu}+\frac{\phi}{M_{\rm Pl}}T\right)~,
\ee
which describes a massless spin-2 particle, a massless spin-1 particle and a massless scalar.\footnote{The decoupling limit can also be used to see why the Fierz--Pauli tuning of the mass term is necessary.} The gauge invariances of this action are linearized diffeomorphisms of $h_{\mu\nu}$ and a U(1) symmetry acting on $A_\mu$:
\be
\delta_\xi h_{\mu\nu} = \partial_\mu\xi_\nu+\partial_\nu\xi_\mu~,~~~~~~~~~~~~~~\delta_\Lambda A_\mu = \partial_\mu\Lambda~.
\ee
Notice that the vector degrees of freedom do not couple to external sources, but the scalar $\phi$ does couple to $T$, and so mediates a gravitational strength fifth force. This tells us that the $m\to0$ limit of linearized massive gravity is {\it not} linearized Einstein gravity. This is the famous {\it van Dam--Veltman--Zakharov (vDVZ) discontinuity}~\cite{vanDam:1970vg, Zakharov:1970cc}.\footnote{Perhaps only of theoretical interest is the fact that the vDVZ discontinuity is absent in (A)dS space~\cite{Higuchi:1986py,Kogan:2000uy,Porrati:2000cp}.}

\subsubsection{Nonlinear massive gravity}
The presence of this fifth force would seem to rule out massive gravity on observational grounds. However, this is only the linear theory; by going beyond Fierz--Pauli theory~\eqref{fierzpaulilag} and considering a theory of a self-interacting massive graviton, we will find that this additional degree of freedom screens itself. 

In order to move to an interacting theory, we are free to introduce non-linearities in both the kinetic structure and in the potential.\footnote{For easy reference, we adopt the notational conventions of~\cite{Hinterbichler:2011tt}.} We choose to promote the kinetic term to be the Ricci scalar\footnote{This is apparently a choice, na\"ively we could also introduce new derivative interactions. Indeed, there do exist derivative interactions whose leading terms in a small-field expansion (`pseudo-linear' terms) are ghost free~\cite{Folkerts:2011ev,Hinterbichler:2013eza,Zinoviev:2013hac,Gao:2014jja}, but they cannot be consistently extended to the full non-linear theory~\cite{Kimura:2013ika, deRham:2013tfa}. Therefore, the Einstein--Hilbert term is the unique nonlinear interaction term involving derivatives.} and for the potential we take all contractions of the tensor $h_{\mu\nu}$~\cite{ArkaniHamed:2002sp,Creminelli:2005qk,Nibbelink:2006sz,deRham:2010ik,Hinterbichler:2011tt}
\be
S = \frac{M_{\rm Pl}^2}{2}\int\rd^4x\sqrt{-g}\bigg(R-\frac{m^2}{2}{\cal V}(g, h)\bigg)~.
\label{generalmassivespin2}
\ee
Here $h_{\mu\nu}$ is defined as the perturbation about some fiducial background metric $g_{\mu\nu} = \bar g_{\mu\nu}+h_{\mu\nu}$. The potential ${\cal V}(g, h)$ consists of all possible contractions of $h_{\mu\nu}$ with the full metric $g_{\mu\nu}$. That is, the potential is of the form ${\cal V}(g, h) = \sum_{n=2}^\infty {\cal V}_n(g, h)$, where ${\cal V}_n$ consists of all contractions with $n$ factors of $h_{\mu\nu}$.\footnote{For $n > d$, the number of space-time dimensions, not all of the possible terms are linearly independent, at each order a single linear combination vanishes, so one of the parameters is redundant.} The first few terms are~\cite{deRham:2010ik,Hinterbichler:2011tt}
\begin{align}
\nonumber
{\cal V}_2(g, h) &= \langle h^2\rangle-\langle h\rangle^2\\\nonumber
{\cal V}_3(g, h) &= c_1\langle h^3\rangle+c_2\langle h^2\rangle\langle h\rangle+c_3\langle h\rangle^3\\\nonumber
{\cal V}_4(g, h) &= d_1\langle h^4\rangle+d_2\langle h^3\rangle\langle h\rangle+d_3\langle h^2\rangle^2+d_4\langle h^2\rangle\langle h\rangle^2+d_5\langle h\rangle^4\\\nonumber
&~~\vdots 
\label{hcontractions}
\end{align}
where we have already enforced the Fierz--Pauli tuning in ${\cal V}_2$, and the $\langle\cdots\rangle$ brackets indicate traces using the full metric $g_{\mu\nu}$. It will turn out that we cannot arbitrarily choose the $c_i, d_i, \ldots$ coefficients. Theoretical consistency forces upon us a rather rigid structure for the theory. This is similar to the linear theory, where $ \langle h^2\rangle$ and $ \langle h\rangle^2$ must appear in a particular combination.

To elucidate this point, we want to introduce diffeomorphism invariance into this theory---as we did in the linear theory---via the St\"uckelberg trick. However, this is slightly more involved in the nonlinear case, because we must introduce full diffeomorphism invariance.

\noindent
{\bf St\"uckelberg trick in the nonlinear theory:}\\\indent
In order make the action~\eqref{generalmassivespin2} diffeomorphism invariant, we introduce 4 St\"uckelberg fields by replacing the {\it background} metric as in~\cite{Hinterbichler:2011tt}
\be
\bar g_{\mu\nu} \longmapsto \bar g_{\alpha\beta}\big(Y(x)\big)\partial_\mu Y^\alpha \partial_\nu Y^\beta~.
\ee
Here it is important that under diffeomorphisms, the St\"uckelberg fields transform as scalar functions $Y^\alpha(x) \mapsto Y^\alpha\left(f(x)\right)$, so that $\bar g_{\mu\nu}$ transforms as a tensor. We also allow the physical field $g_{\mu\nu}$ to transform as a tensor, which allows us to construct diffeomorphism invariants by contracting indices in the usual way. This makes it easy to introduce diffeomorphism invariance into the action~\eqref{generalmassivespin2}: the only place where the background metric appears is in $h_{\mu\nu} = g_{\mu\nu}-\bar g_{\mu\nu}$, and so we therefore replace $h_{\mu\nu}$ everywhere by
\be
h_{\mu\nu} \longmapsto H_{\mu\nu} = g_{\mu\nu} - \bar g_{\alpha\beta}\big(Y(x)\big)\partial_\mu Y^\alpha \partial_\nu Y^\beta~,
\ee
which transforms covariantly.
Often, it is useful to parameterize the St\"uckelberg fields as a deviation from unitary gauge (in which $Y^\alpha =x^\alpha$) and write
\be
Y^\alpha = x^\alpha -A^\alpha~,
\ee
so that the field $H_{\mu\nu}$ takes the form (here we assume that the background metric is flat)
\be
H_{\mu\nu} = h_{\mu\nu} +\partial_\mu A_\nu+\partial_\nu A_\mu - \partial_\mu A^\alpha \partial_\nu A_\alpha~,
\label{nonlinearvecstuckelberg}
\ee
where we have raised and lowered the indices on $A^\alpha$ with $\eta_{\mu\nu}$.
Using the fact that the $Y^\alpha$ transform as diffeomorphism scalars (infinitesimally $\delta_\xi Y^\alpha = \xi^\mu\partial_\mu Y^\alpha$), we can deduce the transformation rules
\bea
\nonumber
\delta_\xi A^\alpha &=& -\xi^\alpha+\xi^\mu\partial_\mu A^\alpha \\
\delta_\xi h_{\mu\nu} &=& \partial_\mu\xi_\nu+\partial_\nu\xi_\mu +\pounds_\xi h_{\mu\nu}~.
\eea
As we did in the linear theory, we isolate the longitudinal mode by performing the further decomposition $A_\mu \mapsto A_\mu +\partial_\mu \phi$, under which~\eqref{nonlinearvecstuckelberg} becomes
\be
H_{\mu\nu} = h_{\mu\nu} +\partial_\mu A_\nu+\partial_\nu A_\mu - \partial_\mu A^\alpha \partial_\nu A_\alpha -\Phi_\mu^\alpha \partial_\nu A_\alpha -\partial_\mu A^\alpha\Phi_{\nu\alpha}  +2\Phi_{\mu\nu}-\Phi_\mu^\alpha\Phi_{\alpha_\nu}~,
\label{flatstuckelberg}
\ee
where, as before $\Phi_{\mu\nu} = \partial_\mu\partial_\nu\phi$. The action is now invariant under the following gauge symmetries
\bea
\label{massivegravgauge1}
\nonumber
\delta_\xi h_{\mu\nu} &=& \partial_\mu\xi_\nu+\partial_\nu\xi_\mu+\pounds_\xi h_{\mu\nu}~,~~~~~~~~~~\delta_\xi A_\mu = -\xi_\mu+\xi^\nu\partial_\nu A_\mu~,~~~~~~~~~~\delta_\xi\phi =0~;\\\label{massivegravgauge2}
\delta_\Lambda h_{\mu\nu} &=& 0~,~~~~~~~~~~~~~~~~~~~~~~~~~~~~~~~~~~~\delta_\Lambda A_\mu = \partial_\mu\Lambda~,~~~~~~~~~~~~~~~~~~~~~~~\delta_\Lambda \phi = -\Lambda~.
\eea
With this St\"uckelberg decomposition in hand, we are now prepared to analyze the nonlinear massive gravity theory~\eqref{generalmassivespin2}.

\noindent
{\bf The effective field theory of a massive graviton:}\\\indent
We now want to consider the theory~\eqref{generalmassivespin2}, but with the St\"uckelberg replacement $h_{\mu\nu}\mapsto H_{\mu\nu}$, where $H_{\mu\nu}$ is defined in~\eqref{flatstuckelberg}. Schematically, the theory is given by
\be
S = \frac{M_{\rm Pl}^2}{2}\int\rd^4x\sqrt{-g}\bigg(R-\frac{m^2}{2}{\cal V}(g, H)\bigg)
\label{massivegraveft}
\ee
where now the potential ${\cal V}$ consists of all possible contractions of $H_{\mu\nu}$. This action has been obtained from~\eqref{generalmassivespin2} by lowering all the indices on $h_{\mu\nu}$ and then everywhere replacing $h_{\mu\nu}\mapsto H_{\mu\nu}$. Note that we do {\it not} do anything to the factors of $g^{\mu\nu}$ or to the measure. We first want to power-count this theory, to get an idea of the form that interactions take. A general interaction term generated by the St\"uckelberg replacement~\eqref{flatstuckelberg} will be of the form~\cite{ArkaniHamed:2002sp,Hinterbichler:2011tt}
\be
{\cal L} \sim M_{\rm Pl}^2m^2 h^{n_h} (\partial A)^{n_A} (\partial^2\phi)^{n_\phi}~,
\label{genop}
\ee
where $n_{h, A, \phi}$ count the number of fields of each type that appear. Notice that the number of derivatives acting on each field is fixed by the St\"uckelberg decomposition. In order to identify the scales suppressing the various operators, we must first canonically normalize the fields through
\be
\hat h_{\mu\nu} = M_{\rm Pl} h_{\mu\nu}~,~~~~~~~~~~~\hat A_\mu = m M_{\rm Pl} A_\mu~,~~~~~~~~~~~\hat\phi = m^2M_{\rm Pl}\phi~,
\ee
where the mass scales have been chosen to agree with the linear analysis. In terms of the redefined fields, the general operator~\eqref{genop} becomes
\be
{\cal L} \sim \frac{\hat h^{n_h}(\partial \hat A)^{n_A} (\partial^2\hat\phi)^{n_\phi}}{M_{\rm Pl}^{n_h+n_A+n_\phi-2}m^{n_A+2n_\phi-2}}\sim \frac{1}{\Lambda_{\rm s}^{n_h+2n_A+3n_\phi-4}}\hat h^{n_h}(\partial \hat A)^{n_A} (\partial^2\hat\phi)^{n_\phi}~,
\ee
where we have defined the strong coupling scale
\be
\Lambda_{\rm s} = \left(M_{\rm Pl}m^\frac{n_A+2n_\phi-2}{n_h+n_A+n_\phi-2}\right)^\frac{n_h+n_A+n_\phi-2}{n_h+2n_A+3n_\phi-4}~.
\ee
From this, assuming $m < M_{\rm Pl}$, we see that the operator with the lowest scale in the theory is
\be
{\cal L} \sim \frac{1}{\Lambda_5^5} (\partial^2\hat\phi)^3~,~~~~~~{\rm where}~~~~~~\Lambda_5 \equiv (M_{\rm Pl}m^4)^{1/5}~.
\ee
Just above this is the scale $\Lambda_4 \equiv (M_{\rm Pl}m^3)^{1/4}$, which suppresses two types of operator:
\be
{\cal L} \sim \frac{1}{\Lambda_4^8}(\partial^2\hat\phi)^4~;~~~~~~~~~~~~~~{\cal L} \sim\frac{1}{\Lambda_4^4}\partial\hat A(\partial^2\hat\phi)^2~.
\ee
The problem is that at both of these scales, a ghost enters the theory. Indeed, terms of the form $(\partial^2\hat\phi)^n$ all lead to ghosts when they become important. Thus, we are motivated to try to cancel these interactions and raise the cutoff so that the strong coupling scale of the theory is $\Lambda_3 \equiv (M_{\rm Pl}m^2)^{1/3}$, which is the scale suppressing the operators
\be
{\cal L} \sim \frac{1}{\Lambda_3^{3n-3}}\hat h(\partial^2\hat\phi)^n~;~~~~~~~~~~~~{\cal L} \sim\frac{1}{\Lambda_3^{3n}}(\partial \hat A)^2(\partial^2\hat\phi)^n~.
\ee
We will see that raising the strong-coupling scale to $\Lambda_3$ leads to a theory with many desirable properties.

\noindent
{\bf Decoupling limit:}\\\indent
We can isolate the interactions which are suppressed by $\Lambda_3 = (M_{\rm Pl}m^2)^{1/3}$ by taking a suitable {\it decoupling limit}. This corresponds to taking the Planck mass to infinity and the graviton mass to zero while keeping the scale $\Lambda_3$ fixed:
\be
M_{\rm Pl} \longrightarrow \infty~,~~~~~~~~~~~~m\longrightarrow 0~,~~~~~~~~~~~~\Lambda_3 = \left(M_{\rm Pl}m^2\right)^{1/3}~ ~~~{\rm fixed}~.
\ee
In this limit, the theory~\eqref{massivegraveft} reduces to (up to quartic order in the fields)\footnote{For simplicity we ignore vector interactions. This is a consistent choice: since vector modes do not couple at linear order to $T_{\mu\nu}$, they appear quadratically in the action and may be consistently set to zero.}
\cite{deRham:2010ik,Hinterbichler:2011tt}
\begin{align}
\label{lambda5theory}
\nonumber
{\cal L}=&-\frac{1}{4}h^{\mu\nu}{\cal E}^{\alpha\beta}_{\mu\nu}h_{\alpha\beta} +h^{\mu\nu}X_{\mu\nu}^{(1)}(\phi)\\\nonumber
&-\frac{1}{\Lambda_5^5}\left[\left(2c_1-1\right)\left[\Phi^3\right]+\left(2c_2+1\right)\left[\Phi\right]\left[\Phi^2\right]+2c_3\left[\Phi\right]^3\right]+\frac{1}{2\Lambda_3^3}h^{\mu\nu}X^{(2)}_{\mu\nu}(\phi)\\
&+\frac{1}{\Lambda_4^8}\left[\left(3c_1-4d_1-\frac{1}{4}\right)\left[\Phi^4\right]+\left(c_2-4d_3+\frac{1}{4}\right)\left[\Phi^2\right]^2\right.\\\nonumber
&~~~~~~~~~~~~~+(2c_2-4d_2)\left[\Phi\right]\left[\Phi^3\right]+(3c_3-4d_4)\left[\Phi^2\right]\left[\Phi\right]^2-4d_5\left[\Phi\right]^4\bigg]+\frac{1}{\Lambda_3^6}h^{\mu\nu}X_{\mu\nu}^{(3)}(\phi)~,
\end{align}
where we the brackets denote a trace of the enclosed tensor.\footnote{For example, we have $\left[\Phi\right] = \square\phi$ and $\left[\Phi^3\right] = \partial_\mu\partial_\nu\phi\partial^\nu\partial^\rho\phi\partial_{\rho}\partial^\mu\phi$.} The $X_{\mu\nu}^{(n)}$ tensors are $n$-th order in the scalar $\phi$. As an effective theory below the cutoff $\Lambda_5$, this theory is perfectly fine, but as we approach the scale $\Lambda_5$, higher-derivative scalar terms of the form $(\partial^2\phi)^3$ become important and the theory propagates a ghost. Similar to the Fierz--Pauli tuning, we want to choose the $c_i$ coefficients so that the terms suppressed by the scale $\Lambda_5$ are absent from the theory. We cannot choose the coefficients to make these terms identically vanish, but we {\it can} arrange for  the pure scalar self-interactions to appear in total derivative combinations. At each order in the fields, there is a unique total derivative combination, which is given by the terms in the characteristic polynomial of the matrix $\Phi_{\mu\nu}$~\cite{deRham:2010ik,Hinterbichler:2011tt}:
\be
{\rm det}({\mathds 1}+\Phi) = 1+ {\cal L}_1^{\rm TD}(\Phi)+\frac{1}{2}{\cal L}_2^{\rm TD}(\Phi)+\cdots+\frac{1}{n!}{\cal L}_n^{\rm TD}(\Phi)+\cdots
\label{characteristicpolynomial}
\ee
This expansion truncates for $n > d$, the space-time dimension. In the case of interest, the total derivative combinations are given by
\begin{align}
\label{ltd1}
{\cal L}_1^{\rm TD}(\Phi) &= \left[\Phi\right]\\
{\cal L}_2^{\rm TD}(\Phi) &= \left[\Phi\right]^2-\left[\Phi^2\right]\\\
\label{ltd3}
{\cal L}_3^{\rm TD}(\Phi) &= \left[\Phi\right]^3-3\left[\Phi\right]\left[\Phi^2\right]+2\left[\Phi^3\right]\\
\label{ltd4}
{\cal L}_4^{\rm TD}(\Phi) &= \left[\Phi\right]^4-6\left[\Phi^2\right]\left[\Phi\right]^2+8\left[\Phi^3\right]\left[\Phi\right]+3\left[\Phi^2\right]^2-6\left[\Phi^4\right]~.
\end{align}
Armed with this information, we can make the terms cubic in $\phi$ in~\eqref{lambda5theory} appear in the total derivative combination~\eqref{ltd3} by choosing~\cite{deRham:2010ik}
\be
c_1 = 2c_3+\frac{1}{2}~;~~~~~~~~~~~~~~c_2 = -3c_3-\frac{1}{2}~.
\ee
With these choices, the leading interactions are now the scalar self-interactions suppressed by the scale $\Lambda_4$.\footnote{There are also terms of the form $\partial A(\partial^2\phi)^n$, but once we arrange to cancel the scalar self interactions, they will always be of the form $\partial A X^{(n)}$ which is a total derivative~\cite{Hinterbichler:2011tt}.} In the same way we may make these interactions appear in the total derivative combination~\eqref{ltd4} by choosing~\cite{deRham:2010ik}
\begin{align}
\nonumber
d_1 &= -6d_5+\frac{1}{16}(24c_3+5)~,~~~~~~~~~d_2 = 8d_5-\frac{1}{4}(6c_3+1)\\
d_3 &= 3d_5-\frac{1}{16}(12c_3+1)~,~~~~~~~~~~~d_4 = -6d_5+\frac{3}{4}c_3~.
\end{align}
This does not remove all of the scalar self-interactions, since there remain terms quintic in the scalar $\phi$ of the schematic form
\be
{\cal L}_5(\phi) \sim \frac{1}{M_{\rm Pl}^3m^8}(\partial^2\phi)^5~.
\ee
These interactions can be chosen to appear in the combination ${\cal L}_5^{\rm TD}(\Phi)$ (which is identically zero in 4 dimensions) by a suitable choice of coefficients of the $5^{\rm th}$ order potential~\cite{deRham:2010ik}.
With these choices of the parameters, the scalar self-interactions drop out, and the decoupling limit Lagrangian is given  (up to quintic order) by~\cite{deRham:2010ik}
\be
\label{declimlag}
{\cal L} = -\frac{1}{4}h^{\mu\nu}{\cal E}^{\alpha\beta}_{\mu\nu}h_{\alpha\beta} +h^{\mu\nu}X_{\mu\nu}^{(1)}-\frac{(6c_3-1)}{2\Lambda_3^3}h^{\mu\nu}X_{\mu\nu}^{(2)}(\phi)-\frac{(8d_5+c_3)}{\Lambda_3^6}h^{\mu\nu}X_{\mu\nu}^{(3)}(\phi)~,
\ee
where the $X^{(n)}$ tensors are given by~\cite{deRham:2010ik,Hinterbichler:2011tt}
\begin{align}
\nonumber
X_{\mu\nu}^{(1)}(\phi) &= \left[\Phi\right]\eta_{\mu\nu}-\Phi_{\mu\nu} \\
X_{\mu\nu}^{(2)}(\phi) &= \Big(\left[\Phi\right]^2-\left[\Phi^2\right]\Big)\eta_{\mu\nu}-2\left[\Phi\right]\Phi_{\mu\nu}+2\Phi_{\mu\alpha}\Phi^\alpha_\nu\\\nonumber
X_{\mu\nu}^{(3)}(\phi) &= \Big(\left[\Phi\right]^3-3\left[\Phi\right]\left[\Phi^2\right]+2\left[\Phi^3\right]\Big)\eta_{\mu\nu}-3\Big(\left[\Phi\right]^2-\left[\Phi^2\right]\Big)\Phi_{\mu\nu}+6\left[\Phi\right]\Phi_{\mu\alpha}\Phi^\alpha_\nu-6\Phi_{\mu\alpha}\Phi^{\alpha\beta}\Phi_{\beta\nu}~.
\end{align}
These are related to the total derivative combinations in~\eqref{characteristicpolynomial} by~\cite{deRham:2010ik,deRham:2010kj,Hinterbichler:2011tt}
\be
X^{(n)}_{\mu\nu}(\phi) = \frac{1}{n+1}\frac{\delta}{\delta \Phi_{\mu\nu}}{\cal L}_{n+1}^{\rm TD}(\Phi)~,
\ee
and have two important properties~\cite{deRham:2010ik}:
\begin{itemize}
\item They are identically conserved, $\partial^\mu X_{\mu\nu}^{(n)}(\phi)=0$. This is true whether or not the equations of motion are satisfied.
\item Their structure guarantees that~\eqref{declimlag} has second-order equations of motion:
\begin{itemize}
\item $X_{ij}^{(n)}(\phi)$ has at most two time derivatives,
\item $X_{0i}^{(n)}(\phi)$ has at most one time derivative,
\item $X_{00}^{(n)}(\phi)$ has no time derivatives.
\end{itemize}
\end{itemize}
Rescaling the gauge parameters as $\xi \to \frac{1}{M_{\rm Pl}}\xi$ and $\Lambda \to \frac{1}{M_{\rm Pl}m}\Lambda$, in the decoupling limit, the gauge symmetries~\eqref{massivegravgauge1} become
\bea
\nonumber
\delta_\xi h_{\mu\nu} &=& \partial_\mu\xi_\nu+\partial_\nu\xi_\mu~,~~~~~~~~~~~~~~~~~~~~\delta_\xi A_\mu = 0~,~~~~~~~~~~~~~~~~~~~~~~~~\delta_\xi\phi =0~;\\
\delta_\Lambda h_{\mu\nu} &=&0~,~~~~~~~~~~~~~~~~~~~~~~~~~~~~~~~~~~\delta_\Lambda A_\mu = \partial_\mu\Lambda~,~~~~~~~~~~~~~~~~~~~~\delta_\Lambda \phi = 0~.
\eea
It is straightforward to see from the conservation of $X_{\mu\nu}^{(n)}$ that the Lagrangian~\eqref{declimlag} is invariant under these symmetries (notice that it is invariant {\it up to a total derivative} under diffeomorphisms). The theory also possesses invariance under galileon shifts of the scalar $\phi$, due to the fact that it always appears with two derivatives on it---we will see later that the galileon terms can be made to appear explicitly.

This theory is quite remarkable: not only has the strong coupling scale increased to $\Lambda_3$, but at this scale---when the higher-derivative operators become important---the theory continues to have second order equations of motion. This is similar to what we saw with the galileons, this theory is capable of having classical nonlinearities become important without propagating a ghost.

Ideally, one would like to continue this procedure to arbitrarily high order---performing the St\"uckelberg replacement on each term in the potential ${\cal V}(g, h)$ in~\eqref{generalmassivespin2} and then canceling the scalar self-interactions order-by-order. Notice that the decoupling limit Lagrangian~\eqref{declimlag} has a very simple form, schematically,
\be
{\cal L}_{\rm dec.} = -\frac{1}{4} h{\cal E} h + \sum_n  \frac{1}{\Lambda_3^{3n-3}}h X^{(n)}~,
\ee
which we might expect to persist at higher orders. However, there are no $X^{(n)}$ tensors for $n > d$ (they all vanish identically!). We are therefore tempted to posit that the decoupling limit Lagrangian~\eqref{declimlag} for the scalar sector is {\it exact}. This turns out to be the case. In~\cite{deRham:2010kj} a useful reorganization of the theory~\eqref{generalmassivespin2} was introduced, which makes this manifest and which is completely ghost free at all orders beyond the decoupling limit, which we now describe.

\subsubsection{de Rham--Gabadadze--Tolley massive gravity}
In~\cite{deRham:2010kj}, de Rham, Gabadadze and Tolley (dRGT) found a way to recast the non-linear theory of a massive graviton in such a way that it is clear that the decoupling limit is free of ghosts. The construction of this theory exploits a structure that, at first glance, appears somewhat unusual. We define the tensor
\be
{\cal K}_{~\nu}^\mu \equiv \delta_{~\nu}^\mu - \sqrt{ \delta_{~\nu}^\mu- H^\mu_{~\nu}}~,
\ee
where, as before,\footnote{We are defining the theory with respect to a flat reference metric $\bar g_{\alpha\beta} = \eta_{\alpha\beta}$, but the extension to a curved reference metric is straightforward.} $H_{\mu\nu} = g_{\mu\nu} - \eta_{\alpha\beta}\partial_\mu Y^\alpha \partial_\nu Y^\beta$. In the tensor ${\cal K}_{~\nu}^\mu$, indices are raised and lowered with $g_{\mu\nu}$. The square root structure should be thought of as a power series:
\be
{\cal K}_{~\nu}^\mu = \delta_{~\nu}^\mu - \sqrt{ \delta_{~\nu}^\mu- H^\mu_{~\nu}} = -\sum_{n=1}^\infty \frac{(2n)!}{(1-2n)(n!)^24^n}(H^n)_{~\nu}^\mu ~,
\ee
where $(H^n)_{~\nu}^\mu = H_{~\nu}^{\alpha_1} H_{~\alpha_1}^{\alpha_2}\cdots H_{~\alpha_{n-1}}^\mu$, and the indices are raised with $g^{\mu\nu}$. This tensor has been defined in this way so that if we perform the scalar St\"uckelberg replacement ($Y^\alpha = x^\alpha-\eta^{\alpha\mu}\partial_\mu\phi$)
\be
H_{\mu\nu} = h_{\mu\nu} + 2\Phi_{\mu\nu}-\eta^{\alpha\beta}\Phi_{\mu\alpha}\Phi_{\nu\beta}~,
\ee
we have\footnote{This can be understood by noting that (restricting to the scalar mode) we can write ${\cal K}_{\mu\nu}$ as
\begin{equation*}
{\cal K}_{\mu\nu} = g_{\mu\nu}  - \sqrt{\eta_{\alpha\beta}\partial_\mu Y^\alpha\partial_\nu Y^\beta} = g_{\mu\nu}  - \sqrt{\eta_{\alpha\beta}(\delta_\mu^\alpha-\Phi^\alpha_\mu)(\delta_\nu^\beta - \Phi_\nu^\beta)}~.
\end{equation*}
}
\be
{\cal K}_{\mu\nu}\Big\rvert_{h_{\mu\nu=0}} = g_{\mu\alpha}{\cal K}^\alpha_{~\nu}\Big\rvert_{h_{\mu\nu}=0} = \Phi_{\mu\nu}~.
\ee
We can now understand the benefit of this reorganization; it will make it simple to determine how the self-interactions of the scalar $\phi$ will appear in the decoupling limit. Equivalent to~\eqref{generalmassivespin2}, we can now write
\be
S = \frac{M_{\rm Pl}^2}{2}\int\rd^4x\sqrt{-g}\bigg(R-\frac{m^2}{2}{\cal W}(g, {\cal K})\bigg)~,
\ee
where ${\cal W}$ is now a general potential built from contractions of the ${\cal K}$ tensor, ${\cal W}(g, {\cal K}) = \sum_{n=2}^\infty {\cal W}_n(g,{\cal K})$, with:
\begin{align}
\nonumber
{\cal W}_2(g, {\cal K}) &= \langle {\cal K}^2\rangle-\langle {\cal K}\rangle^2\\\nonumber
{\cal W}_3(g, {\cal K}) &=\tilde c_1\langle {\cal K}^3\rangle+\tilde c_2\langle {\cal K}^2\rangle\langle {\cal K}\rangle+\tilde c_3\langle {\cal K}\rangle^3\\\nonumber
{\cal W}_4(g, {\cal K}) &= \tilde d_1\langle {\cal K}^4\rangle+\tilde d_2\langle {\cal K}^3\rangle\langle {\cal K}\rangle+\tilde d_3\langle {\cal K}^2\rangle^2+\tilde d_4\langle {\cal K}^2\rangle\langle {\cal K}\rangle^2+\tilde d_5\langle {\cal K}\rangle^4\\
&~~\vdots
\end{align}
It is now straightforward to deduce how the scalar self-interactions will appear in the decoupling limit; they are just given by ${\cal W}(\eta, \Phi)$. In order to arrange their cancellation, we just have to demand that ${\cal K}$ enters the potential in the combinations~\eqref{ltd1}--\eqref{ltd4}. Therefore, we consider the action
\be
S = \frac{M_{\rm P}^2}{2}\int {\rm d}^4 x \, \sqrt{-g}\bigg(R - \frac{m^2}{2}{\cal U}(g, {\cal K})\bigg)~,
\label{dRGTaction}
\ee
where ${\cal U}(g, {\cal K}) = \sum_{n=2}^d \alpha_n {\cal L}^{\rm TD}_n({\cal K})$. Here $\alpha_n$ are free parameters, and the total derivative combinations are~\cite{deRham:2010kj}
\begin{align}
 {\cal L}_2^{\rm TD}({\cal K}) & = \langle{\cal K}\rangle^2-\langle{\cal K}^2\rangle~,\\
 {\cal L}_3^{\rm TD}({\cal K})& = \left\langle{\cal K}\right\rangle^3-3\left\langle{\cal K}\right\rangle\langle{\cal K}^2\rangle+2\langle{\cal K}^3\rangle~,\\
  {\cal L}_4^{\rm TD}({\cal K}) & =  \left\langle{\cal K}\right\rangle^4-6\left\langle{\cal K}\right\rangle^2\langle{\cal K}^2\rangle+3\langle{\cal K}^2\rangle^2+8\left\langle{\cal K}\right\rangle\langle{\cal K}^3\rangle-6\langle{\cal K}^4\rangle~,\\\nonumber
  &~~\vdots 
\end{align}
This form for the potential guarantees that the scalar self-interactions appear in the total derivative combinations~\eqref{ltd1}--\eqref{ltd4}:
\be
\sqrt{-g}~{\cal U}(g, {\cal K})\Big\rvert_{h_{\mu\nu=0}} = \sum_{n=1}^d \alpha_n {\cal L}_n^{\rm TD}(\Phi)~.
\ee
In the decoupling limit, it is possible to go a little further, and determine the mixing terms between $h_{\mu\nu}$ and $\phi$. The action takes the form
\be
S = M_{\rm Pl}^2\int\rd^4x\left(-\frac{1}{4}h^{\mu\nu}{\cal E}^{\alpha\beta}_{\mu\nu}h_{\alpha\beta} - \frac{m^2}{4}h^{\mu\nu}\bar X_{\mu\nu}\right)~,
\ee
where we have defined 
\be
\bar X_{\mu\nu} \equiv \frac{\delta}{\delta h^{\mu\nu}}\left(\sqrt{-g}~{\cal U}(g, {\cal K})\right)\Big\rvert_{h_{\mu\nu}=0}~.
\ee
If we then use the identity~\cite{deRham:2010kj}
\be
 \frac{\delta}{\delta h^{\mu\nu}} \langle{\cal K}^n\rangle\Big\rvert_{h_{\mu\nu}=0} = \frac{n}{2}\left(\Phi_{\mu\nu}^{n-1}-\Phi_{\mu\nu}^n\right)~,
\ee
we obtain the result~\cite{deRham:2010kj}
\be
 \frac{\delta}{\delta h^{\mu\nu}}\left(\sqrt{-g}{\cal L}_n^{\rm TD}({\cal K}) \right)\Big\rvert_{h_{\mu\nu}=0}  = \frac{1}{2}\left(X_{\mu\nu}^{(n)}-nX_{\mu\nu}^{(n-1)}\right) \ .
\ee
Using these facts we can derive the decoupling limit action, which is given by (we define $X^{(0)}_{\mu\nu} = \eta_{\mu\nu}$ and $X^{(-1)}_{\mu\nu} = 0$)
\be
S = M_{\rm Pl}^2\int\rd^4x \left(-\frac{1}{4}h^{\mu\nu}{\cal E}^{\alpha\beta}_{\mu\nu}h_{\alpha\beta} +\frac{\alpha_2}{4} h^{\mu\nu} X^{(1)}_{\mu\nu}-\left(\frac{3\alpha_3-\alpha_2}{8}\right) h^{\mu\nu} X_{\mu\nu}^{(2)} - \left(\frac{4\alpha_4+\alpha_3}{8}\right)h^{\mu\nu}X_{\mu\nu}^{(3)}~.
\right)
\label{dRGTdeclim}
\ee
This agrees with~\eqref{declimlag} upon identifying coefficients as
\be
\alpha_2 = 4~;~~~~~~~~~~~~~\alpha_3 = 2^3 c_3~;~~~~~~~~~~~~~\alpha_4 = 2^4 d_5~.
\ee
The key difference is that the action~\eqref{dRGTdeclim} is the {\it exact} decoupling limit action involving the helicity-2 and helicity-0 degrees of freedom. No truncation at a finite order in fluctuations has been taken. Here we have focused only on the scalar and tensor mixings in the decoupling limit, which is a consistent truncation, but it is possible to derive the full decoupling limit, including vector interactions~\cite{Koyama:2011wx,Gabadadze:2013ria,Ondo:2013wka}.

Although here we have followed the historical route to~\eqref{dRGTaction}, it is actually possible to derive this action from a dimensional reduction of a discrete extra dimension, along the lines of the dimensional deconstruction idea~\cite{Hill:2000mu,ArkaniHamed:2001ca}. This is carried out in~\cite{deRham:2013awa, deRham:2013tfa}.

\noindent
{\bf Galileons in the decoupling limit:}\\\indent
The longitudinal polarization, $\phi$, of a massive graviton turns out to be described by a galileon theory in the decoupling limit (see Section~\ref{galileonsection} for a discussion of galileons). In order to see this, we consider the decoupling limit Lagrangian~\eqref{declimlag},
in which there are mixings $h X^{(1)}$ and $h X^{(2)}$, between $h_{\mu\nu}$ and $\phi$, which can be partially removed\footnote{Note that this works because~\cite{deRham:2010ik,Hinterbichler:2011tt}
\begin{equation*}
{\cal E}_{\mu\nu}^{\alpha\beta}\left(\phi\eta_{\alpha\beta}\right)=-2X^{(1)}_{\mu\nu}~;~~~~~~~~~~{\cal E}_{\mu\nu}^{\alpha\beta}\left(\partial_\alpha\phi\partial_\beta\phi\right) = X^{(2)}_{\mu\nu}~.
\end{equation*}
}
 by the following local field redefinition~\cite{deRham:2010ik}
 \be
 h_{\mu\nu} \longmapsto \hat h_{\mu\nu}+\phi\eta_{\mu\nu}+\frac{(6c_3-1)}{2\Lambda_3^3}\partial_\mu\phi\partial_\nu\phi~.
 \ee
Note also that we have been implicitly assuming that the metric couples to matter through $ h_{\mu\nu}T^{\mu\nu}$, so that this field redefinition induces scalar couplings to $T_{\mu\nu}$ of the form $\sim \phi T$ and $\sim \partial_\mu\phi\partial_\nu T^{\mu\nu}$.
After performing this field redefinition, the decoupling limit Lagrangian is given by
\begin{align}
\nonumber
{\cal L} = -\frac{1}{4}&h^{\mu\nu}{\cal E}^{\alpha\beta}_{\mu\nu}h_{\alpha\beta} +3\phi\square\phi +\frac{3(6c_3-1)}{4\Lambda_3^3}\square\phi(\partial\phi)^2 \\
&+\frac{1}{\Lambda_3^6}\left[\frac{1}{4}(6c_3-1)^2-2(c_3+8d_5)\right](\partial\phi)^2\left(\big[\Phi^2\big]-\big[\Phi\big]^2\right)\\\nonumber
&- \frac{5}{4\Lambda_3^9}(6c_3-1)(c_3+8d_5)(\partial\phi)^2\left(\big[\Phi\big]^3-3\big[\Phi\big]\big[\Phi\big]^2+2\big[\Phi\big]^3\right)-\frac{(8d_5+c_3)}{\Lambda_3^6}h^{\mu\nu}X_{\mu\nu}^{(3)}(\phi)~.
\end{align}
Notice that the self-interactions of the scalar are precisely the galileon terms~\eqref{galileonterms}, and so we can write this action as
\begin{align}
\nonumber
{\cal L} = -\frac{1}{4}h^{\mu\nu}{\cal E}^{\alpha\beta}_{\mu\nu}h_{\alpha\beta} &-6{\cal L}_2^{\rm gal}(\phi)+\frac{3(6c_3-1)}{2\Lambda_3^3}{\cal L}_3^{\rm gal}(\phi)-\frac{4}{\Lambda_3^6}\left[\frac{1}{4}(6c_3-1)^2-2(c_3+8d_5)\right]{\cal L}_4^{\rm gal}(\phi)\\\label{decgalileons}
&- \frac{15}{4\Lambda_3^9}(6c_3-1)(c_3+8d_5){\cal L}_5^{\rm gal}(\phi)-\frac{(8d_5+c_3)}{\Lambda_3^6}h^{\mu\nu}X_{\mu\nu}^{(3)}(\phi)~,
\end{align}
where the galileon terms are normalized as in~\eqref{galileonterms}. This is as simple as things get: there does not exist a local field redefinition which can remove the $hX^{(3)}$ mixing~\cite{deRham:2010ik}. Notice that if we set $c_3=1/6$, $d_5 = -1/48$, the scalar self-interactions disappear and we have a completely free theory of a helicity-2 and a helicity-0 particle.

In light of the fact that the galileons appear in the theory, it should not be surprising that the dRGT theory also exhibits superluminality. Indeed, this has been argued to be the case~\cite{Burrage:2011cr,Gruzinov:2011sq,Deser:2012qx,Deser:2013eua,Deser:2013qza,Yu:2013owa}. However, these calculations are of the classical phase velocity, so it is unclear to what extent this implies acausality in the theory~\cite{Burrage:2011cr,deRham:2011pt}. See Section 10.6.2 in~\cite{deRham:2014zqa} for an extensive discussion of superluminality in massive gravity theories.

\noindent
{\bf Freedom from ghosts:}\\\indent
Although the special form of the $X_{\mu\nu}^{(n)}$ tensors ensures that the dRGT theory~\eqref{dRGTaction} is ghost-free in the decoupling limit, it is still in principle possible that the Boulware--Deser ghost re-emerges in the theory once we look beyond the decoupling limit. However, it has been clearly demonstrated that the full nonlinear theory is free of the Boulware--Deser ghost to all orders beyond the decoupling limit. This has been shown by various authors both from a direct canonical analysis in ADM variables~\cite{Hassan:2011vm,Hassan:2011hr,Hassan:2011ea, Golovnev:2011aa,Hinterbichler:2012cn,Comelli:2012vz,Kluson:2012wf} and in the St\"uckelberg and helicity language~\cite{deRham:2011rn, Mirbabayi:2011aa, deRham:2011qq,Hassan:2012qv}.  While this guarantees that the theory propagates only 5 degrees of freedom, even non-linearly, it nevertheless remains possible for one or more of these 5 physical modes to become ghostly around particular backgrounds, providing further constraints on the theory.

\noindent
{\bf Self--accelerating solutions in the decoupling limit:}\\\indent
We now consider de Sitter solutions in massive gravity without external matter---self-accelerating solutions. As a first step, we search for these solutions in the decoupling limit, following~\cite{deRham:2010tw}.\footnote{See~\cite{deRham:2011by,Heisenberg:2014kea} for a discussion of the cosmology of a related covariantization of this theory.} The decoupling-limit Lagrangian is given by
\be
{\cal L} = -\frac{1}{4}h^{\mu\nu}{\cal E}^{\alpha\beta}_{\mu\nu}h_{\alpha\beta} +\sum_{n=1}^3 \frac{a_n}{\Lambda_3^{3n-3}}h^{\mu\nu}X_{\mu\nu}^{(n)}(\phi)~,
\label{acceldeclim}
\ee
where $a_1= -1/2$ and $a_2, a_3$ are just a re-shuffling of the constants $c_3$ and $d_5$ which appear in, for example,~\eqref{declimlag}. The $X^{(n)}$ tensors are given as before by~\eqref{ltd1}--\eqref{ltd4}, but a particularly convenient way of rewriting them for this purpose (up to overall normalization which can be absorbed into the $a_n$)~\cite{deRham:2010tw,Hinterbichler:2011tt} is
\bea
\nonumber
X_{\mu\nu}^{(1)}(\phi) &=& \epsilon_\mu^{~\alpha\rho\sigma}\epsilon_{\nu~\rho\sigma}^{~\beta}\Phi_{\alpha\beta}~,\\
\nonumber
X_{\mu\nu}^{(2)}(\phi) &=&  \epsilon_\mu^{~\alpha\rho\gamma}\epsilon_{\nu~~~\gamma}^{~~\beta\sigma}\Phi_{\alpha\beta}\Phi_{\rho\sigma}~,\\
X_{\mu\nu}^{(3)}(\phi) & =& \epsilon_{\mu}^{~\alpha\rho\gamma}\epsilon_{\nu}^{~\beta\sigma\delta}\Phi_{\alpha\beta}\Phi_{\rho\sigma}\Phi_{\gamma\delta}~.
\eea
The equations of motion following from~\eqref{acceldeclim} are~\cite{deRham:2010tw}
\bea
\nonumber
& &-\frac{1}{2}{\cal E}^{\alpha\beta}_{\mu\nu}h_{\alpha\beta} +\sum_{n=1}^3 \frac{a_n}{\Lambda_3^{3n-3}}X_{\mu\nu}^{(n)}(\phi) = 0~,\\
& &\left(a_1\epsilon_\mu^{~\alpha\rho\sigma}\epsilon_{\nu~\rho\sigma}^{~\beta}+\frac{2a_2}{\Lambda^3_3}\epsilon_\mu^{~\alpha\rho\gamma}\epsilon_{\nu~~~\gamma}^{~~\beta\sigma}\Phi_{\rho\sigma}+\frac{3a_3}{\Lambda_3^6}\epsilon_{\mu}^{~\alpha\rho\gamma}\epsilon_{\nu}^{~\beta\sigma\delta}\Phi_{\rho\sigma}\Phi_{\gamma\delta}\right)\partial_\alpha\partial_\beta h^{\mu\nu} = 0~.
\eea
To search for de Sitter solutions to these equations, note that, far inside the horizon ($\lvert\vec x\rvert \ll H^{-1}$), the de Sitter metric can be written as~\cite{Nicolis:2008in, deRham:2010tw}
\be
\rd s^2 \simeq \left(1-\frac{1}{2}H^2 x^2\right)\eta_{\mu\nu}\rd x^\mu\rd x^\nu~.
\ee
Thus we look for solutions with $h_{\mu\nu} = -(H^2x^2/2) \eta_{\mu\nu}$, and we make the following ansatz for the scalar field
\be
\phi = q\frac{\Lambda_3^3}{2}x^2 ~,
\ee
where $q$ is a constant. With these assumptions, the equations of motion become~\cite{deRham:2010tw, deRham:2012az}
\bea
\nonumber
& &H^2\left(3a_3 q^2+2a_2 q-\frac{1}{2}\right) = 0~,\\
\label{hubblemassgravdec2}
& &M_{\rm Pl}^2 H^2 = \Lambda_3^3\left(2a_3 q^3+2a_2q^2 -q\right)~.
\eea
The quadratic equation $3a_3 q^2+2a_2 q-\frac{1}{2} =0$ is easily solved for $q$, and inserting the result into the second equation yields $H$~\cite{deRham:2010tw}. Notice that if $q\sim {\cal O}(1)$, then the Hubble constant is parametrically the graviton mass:
\be
H^2 \sim \frac{\Lambda_3^3}{M_{\rm Pl}^2} \sim m^2 \,.
\ee
In order for this to describe a de Sitter solution (with $H >0$) and for scalar and tensor perturbations about this solution to be stable, the parameters $a_2$ and $a_3$ must satisfy~\cite{deRham:2010tw, deRham:2012az}
\be
a_2 < 0~;~~~~~~~~~-\frac{2a_2^2}{3} < a_3 < -\frac{a^2_2}{2}~.
\ee
However, vector perturbations around the self-accelerating background appear to inevitably be either strongly-coupled or ghost-like~\cite{Koyama:2011wx, Tasinato:2012ze}.

\noindent
{\bf Cosmology away from the decoupling limit:}\\\indent
While we have focused on self-accelerating solutions in the decoupling limit, there have also been many studies of cosmological solutions in the full theory~\eqref{dRGTaction}. The situation is rather complicated. In~\cite{D'Amico:2011jj}, it was shown that the non-linear theory does not admit flat FLRW solutions. Nevertheless, there exist other cosmological solutions: open FLRW~\cite{Gumrukcuoglu:2011ew}, and cosmologies with anisotropies in the St\"uckelberg sector~\cite{Gumrukcuoglu:2012aa,Gratia:2012wt,Motohashi:2012jd,Kobayashi:2012fz,Maeda:2013bha}. However, perturbations about these solutions appear to be badly-behaved, both linearly~\cite{Gumrukcuoglu:2011zh,Koyama:2011wx,D'Amico:2012pi,Tasinato:2012ze,Fasiello:2012rw,Chiang:2012vh,Wyman:2012iw,Khosravi:2013axa} and nonlinearly~\cite{Kuhnel:2012gh,DeFelice:2012mx,DeFelice:2013awa}.

In order to remedy this situation and find acceptable cosmological solutions, many extensions of the dRGT theory have been proposed. One possibility is to let the background metric be dynamical, and consider a bi-metric theory~\cite{Hassan:2011zd,Baccetti:2012bk,Hassan:2012wr,Hassan:2012wt} (or even a multi-metric theory~\cite{Khosravi:2011zi,Hinterbichler:2012cn, Hassan:2012wt,Noller:2013yja}), leading to different cosmological solutions~\cite{Comelli:2011zm,Comelli:2011wq,vonStrauss:2011mq,Volkov:2011an,Volkov:2012cf,Volkov:2012zb,Comelli:2012db,Berg:2012kn,Akrami:2012vf,Volkov:2013roa,Tamanini:2013xia,Fasiello:2013woa,Akrami:2013ffa,DeFelice:2014nja,Solomon:2014dua}. Another possibility is to let the graviton mass itself be a field~\cite{Huang:2012pe, Hinterbichler:2013dv, Gumrukcuoglu:2013nza} or to couple in additional fields~\cite{Huang:2013mha,Kluson:2013yaa,Bamba:2013aca,Cai:2014upa,Wu:2014hva} (for instance galileon scalar fields~\cite{Gabadadze:2012tr,Andrews:2013uca,Goon:2014ywa}). One concrete proposal for adding new fields which has attracted some attention is so-called {\it quasi-dilaton} massive gravity, where the theory enjoys a scaling symmetry~\cite{D'Amico:2012zv}. Cosmologies of this theory and its extensions are considered in~\cite{D'Amico:2013kya, Gannouji:2013rwa,DeFelice:2013tsa,DeFelice:2013dua,Gabadadze:2014kaa}. A further enticing possibility is that a particular choice of the potential terms could lead to an extra gauge symmetry about de Sitter space, leading to a theory of {\it partially massless} gravity~\cite{Deser:2001pe,Deser:2001us}. This possibility has been explored~\cite{deRham:2012kf,Hassan:2012gz,Deser:2012qg,Hassan:2012rq,deRham:2013wv}, but it is not clear whether such a theory exists non-linearly~\cite{deRham:2013wv,Deser:2013uy,Joung:2014aba}.

Despite this flurry of activity, to date there does not seem to be an extension of dRGT which possesses a completely satisfactory self-accelerated solution. However, many of these extensions are interesting in their own right.

\noindent
{\bf Vainshtein screening:}\\\indent
One interesting feature of massive gravity is that it too exhibits the Vainshtein effect, with the longitudinal polarization mode of the graviton becoming screened. This is not particularly surprising, since we saw in~\eqref{decgalileons} that in the decoupling limit the helicity-0 mode is described by a galileon theory with an additional mixing between the scalar and the helicity-2 mode. We have already seen that the galileons exhibit Vainshtein screening about a spherical source, and therefore it is immediately clear that in the case where $8d_5+c_3=0$ the theory will exhibit the Vainshtein mechanism. The general case is slightly more delicate, owing to the coupling $h^{\mu\nu}X_{\mu\nu}^{(3)}$, but in this case Vainshtein screening persists. In all cases, the Vainshtein radius is given by
\be
r_{\rm V} \sim \left(\frac{M}{M_{\rm Pl}}\right)^{1/3}\frac{1}{\Lambda_3}~,
\ee
where $M$ is the mass of the source.
See~\cite{Chkareuli:2011te,Koyama:2011xz,Sjors:2011iv,Sbisa:2012zk,Burrage:2012ja,Tasinato:2013rza,Berezhiani:2013dca}  for discussions of Vainsthein screening in the decoupling limit of dRGT. For a discussion of the Vainshtein mechanism in bimetric gravity, see~\cite{Babichev:2013pfa}.

\subsubsection{Degravitation}

Massive gravity also allows us to view the cosmological constant problem in a different light. Rather than asking: {\it why is the observed cosmological constant so small?} we are able to ask the slightly different: {\it why does the cosmological constant not gravitate very strongly?}~\cite{Dvali:2002pe,ArkaniHamed:2002fu,Dvali:2007kt} This is an interesting viewpoint; up to now we have been fighting $\Lambda$, trying to make its value small, but this allows us to take seriously the notion that the large cosmological constant generated via matter loops is physical, and that it does not strongly curve space-time. This phenomenon of {\it degravitation} is intimately tied to massive gravity: as we will see below, any theory that exhibits degravitation must reduce, to a theory of massive/resonance gravity at the linearized level~\cite{Dvali:2007kt}. Moreover, degravitation admits a nice analogy with electromagnetism---it is the gravitational analogue of the screening of charges in a superconductor (Meissner effect)~\cite{Dvali:2007kt}.

A  simple phenomenological modification to Einstein's equations that captures the idea of degravitation is
\be
G_{\rm N}^{-1}(\Box L^2) G_{\mu\nu} = 8\pi T_{\mu\nu}\,.
\label{filter}
\ee
Here Newton's constant, $G_{\rm N}(\Box L^2)$, has been promoted to a high-pass filter which has characteristic scale $L$:
sources with characteristic wavelength $\lambda\ll L$ experiences gravity normally, whereas sources with wavelength $\lambda\gg L$ are degravitated.

Notice that equation~\eqref{filter} violates the Bianchi identity---which follows directly from general covariance---so there must be more to the story.
However, already at the linearized level, where the Bianchi identity is trivially satisfied, we can isolate a problem.
By parameterizing the filter function as $G_{\rm N}^{-1}(\Box L^2) \equiv \left(G_{\rm N}^{(0)}\right)^{-1} \left(1 - \frac{m^2(\Box L^2)}{\Box}\right)$, expanding $g_{\mu\nu} = \eta_{\mu\nu} + \tilde{h}_{\mu\nu}$, and choosing de Donder gauge
$\partial^\mu \tilde{h}_{\mu\nu} = \partial_\nu \tilde{h}/2$,~\eqref{filter} takes the form
\be
\Big(\Box - m^2(\Box L^2)\Big) \left(\tilde{h}_{\mu\nu} - \frac{1}{2}\eta_{\mu\nu}\tilde{h}\right) =  8\pi G_{\rm N}^{(0)}T_{\mu\nu}\,.
\label{filterlin}
\ee
As a special case, we can imagine that $m^2(\Box L^2)\equiv m^2$ is constant. This makes the issue obvious: 
the only allowed (Lorentz-invariant) spin-2 mass term is the
Fierz--Pauli choice~\cite{Fierz:1939ix}: $m^2(h_{\mu\nu} - h\eta_{\mu\nu})$. Other choices propagate a ghost. The mass term in~\eqref{filterlin}
is clearly not of this form, hence it cannot describe a massive spin-2 particle consistently.
Allowing for a more general $m^2(\Box L^2)$ only amplifies the problem; if we write the K\"all\'en--Lehmann spectral representation of the graviton propagator
\be
\frac{1}{\Box - m^2(\Box L^2)} = \int_0^\infty {\rm d}M^2 \frac{\rho(M^2L^2)}{\Box - M^2}\,,
\label{spec}
\ee
we can apply the above argument to each of the massive gravitons in the continuum. 

The resolution of this paradox turns out to be simple: a massive graviton has 5 polarization states (2 with helicity-2, 2 with helicity-1 and 1with helicity-0),
but~\eqref{filterlin} only describes the helicity-2 part of the graviton; it is an effective equation obtained by integrating out the other 3 degrees of freedom. It is worthwhile to briefly review the proof of this fact found in~\cite{Dvali:2007kt,Dvali:2006su}.

We start by generalizing the Fierz--Pauli equation to allow for a momentum-dependent mass for the graviton:
\be
\left({\cal E}h\right)_{\mu\nu} + \frac{m^2(\Box L^2)}{2}(h_{\mu\nu} - \eta_{\mu\nu}h) = 8\pi G_{\rm N}^{(0)} T_{\mu\nu}\,,
\label{filterlin2}
\ee
here the object $\left({\cal E}h\right)_{\mu\nu}  = -\Box h_{\mu\nu}/2 + \ldots$ is the linearized Einstein tensor. As we saw in Section~\ref{linearizedmassiveg}, we can restore the diffeormorphism gauge symmetry by introducing a St\"uckelberg field $A_\mu$~\cite{ArkaniHamed:2002sp}, in the same way as~\eqref{stuckel}:
\be
h_{\mu\nu} = \hat{h}_{\mu\nu} + \partial_\mu A_\nu + \partial_\nu A_\mu\,.
\label{hat}
\ee
This replacement makes $h_{\mu\nu}$ gauge-invariant under the transformations
\be
\delta_\xi\hat{h}_{\mu\nu}  = \partial_\mu \xi_\nu + \partial_\nu \xi_\mu ~~~~~~~~~~~~~~~\delta_\xi A_\mu = -\xi_\mu~.
\ee
This parallels electromagnetism in a Higgs (superconducting) phase, where the photon
is a gauge-invariant observable: $A_\mu = \tilde{A}_\mu + \partial_\mu\phi$. 

The idea is to integrate out $A_\mu$ by solving for it using its equation of motion and then substituting the result back into~\eqref{filterlin2}, giving us an equation for the helicity-2 modes $\hat{h}_{\mu\nu}$. We begin by substituting the decomposition~\eqref{hat} into~\eqref{filterlin2} to obtain
\be
\label{pfs}
({\cal E}\hat{h})_{\mu\nu} +  m^2(\Box  L^2) \left(\hat{h}_{\mu\nu} - \eta_{\mu\nu} \hat{h} + \partial_{\mu}A_{\nu} + \partial_{\nu} A_{\mu}
- 2 \eta_{\mu\nu} \partial^{\alpha}A_{\alpha}\right)  =  8\pi G_{\rm N}^{(0)}T_{\mu\nu} \,.
\ee
In order to isolate an equation for $A_{\mu}$, we take the divergence of this expression, leading to
\be
\label{Aequ}
\partial^{\mu} F_{\mu\nu} =  - \partial^{\mu}  \left(\hat{h}_{\mu\nu} - \eta_{\mu\nu} \hat{h} \right)\,,
\ee
where we have defined $F_{\mu\nu} \equiv \partial_{\mu}A_{\nu}  - \partial_{\nu}A_{\nu}$. Taking another divergence yields the equation
$\partial^{\mu} \partial^{\nu}  \hat{h}_{\mu\nu} -  \Box \hat{h} =  0$, hence $\hat{h}$ can be written as
$\hat{h}_{\mu\nu}  = \tilde{h}_{\mu\nu} - \eta_{\mu\nu} \Pi_{\alpha\beta}\tilde{h}^{\alpha\beta}/3$, where
$\Pi_{\alpha\beta} = \eta_{\alpha\beta} - \partial_{\alpha}\partial_{\beta}/ \Box$ is the transverse projector.
This implies that $\tilde{h}_{\mu\nu}$ carries two degrees of freedom. We now solve~\eqref{Aequ}, for $A_{\mu}$:
\be
\label{asolution}
A_{\nu}  =  -  {1 \over \Box} \partial^{\mu}  \left(\hat{h}_{\mu\nu}  - \eta_{\mu\nu} \hat{h} \right) - \partial_{\nu} \Theta\,, 
\ee
where $\Theta$ is an arbitrary gauge function. By substituting~\eqref{asolution} back into~\eqref{pfs}, and making an appropriate choice for $\Theta$, we find
\be
\left(1-\frac{m^2(\Box L^2)}{\Box}\right) ({\cal E}\tilde{h})_{\mu\nu} =  8\pi G_{\rm N}^{(0)}T_{\mu\nu}\,,
\ee
which is indeed the linearized version of~\eqref{filter}. So we see that, as advocated,~\eqref{filterlin} can be seen as the effective equation for the helicity-2 part $\tilde{h}_{\mu\nu}$, which arises from  integrating out the extra helicities 
of a massive spin-2 representation. 

Therefore, we see that any theory that any theory which degravitates or filters out the cosmological constant must reduce to a theory of massive or resonance gravity in the weak field limit. Note that the implication only goes one way here: not all theory of massive or resonance gravity exhibit degravitation at the non-linear level. 

We now explore how degravitation works at the linearized level. Take~\eqref{filterlin2} with a vacuum energy contribution, $T_{\mu\nu} = -\Lambda\eta_{\mu\nu}$, and for simplicity we will consider the case where $m^2\equiv \frac{1}{L^2}$ is constant:
\be
\left({\cal E}h\right)_{\mu\nu} + \frac{1}{2L^2}(h_{\mu\nu} - \eta_{\mu\nu}h)  = - 8\pi G_{\rm N}^{(0)} \Lambda \eta_{\mu\nu}\,.
\ee
In the absence of a mass term, the solution grows unbounded, $h_{ij} \sim \frac{\Lambda}{6}(t^2\delta_{ij} + x_ix_j)$, which is just the weak-field version of de Sitter space.
However, when we turn on the mass term the solution is just flat space
\be
h_{\mu\nu} = \frac{\Lambda L^2}{3}\eta_{\mu\nu}\,.
\ee
So we see that the gravitational backreaction of $\Lambda$ vanishes in the linearized theory of a massive spin-2 particle! The cosmological constant has been degravitated.

It is reasonable to ask: what forms are allowed for the mass $m^2(\Box L^2)$? One way to parameterize the function which turns out to be useful is in the power law form~\cite{Dvali:2007kt,Dvali:2006su}
\be
m^2(\Box L^2) = L^{-2(1-\alpha)}(- \Box)^\alpha\,.
\label{mparam}
\ee
Various phenomenological and theoretical considerations constrain the constant $\alpha$. In order for the modification to be relevant at large scales (in the infrared), we must have $\alpha < 1$, otherwise the mass term will be unimportant as $\Box\rightarrow 0$. Additionally, we want the graviton propagator to have a positive definite spectral density, $\rho(M^2) \geq 0$ in~\eqref{spec}. If $\alpha < 0$, the left-hand side of~\eqref{spec} will vanish as $\Box\rightarrow 0$, which is impossible if $\rho$ is positive-definite. Therefore, we must have $\alpha\geq 0$. Finally,~\cite{Dvali:2007kt} argue that $\alpha < 1/2$ is necessary for degravitation to be effective in a certain decoupling limit of the theory. Putting these constraints together, the allowed range is
\be
0 \leq \alpha < \frac{1}{2} \,.
\label{alpharange}
\ee
The lower bound corresponds to massive gravity, discussed earlier in this review. The range $\alpha > 0$ corresponds
to having a continuum of massive graviton states. This can be realized explicitly in brane-induced gravity models, where the extra dimensions
have infinite extent. The most-studied example is the DGP model~\cite{Dvali:2000hr}, with one extra dimension. (For a review, see~\cite{Lue:2005ya}.) In the DGP model, the mass
term takes the form $m^2(\Box L^2) = L^{-1}\sqrt{-\Box}$, which corresponds $\alpha = 1/2$ in our parametrization. Interestingly, this just falls
outside the desired range~\eqref{alpharange}. Indeed, the DGP model fails to exhibit degravitation---a large brane tension
generates rapid Hubble expansion on the brane.

Degravitation has been shown to occur in the decoupling limit of dRGT massive gravity~\cite{deRham:2010tw}, which corresponds to $\alpha = 0$.
Indeed, Minkowski space is a solution with an arbitrarily large cosmological constant. There is, however, a tension with phenomenology: through the
Vainshtein mechanism, a large vacuum energy raises the strong coupling scale, which in turn leads to larger deviations from GR. Consistency with solar
system tests puts an upper bound on the largest phenomenologically-acceptable value of $\Lambda$ that can be degravitated in this way, and
unfortunately the answer is a measly meV$^4$. A similar restriction was found in the self-tuning theories of~\cite{Charmousis:2011bf,Charmousis:2011ea,Copeland:2012qf}. 

Another avenue for degravitation is to consider extensions of the DGP model to higher co-dimension. In $D=6$ bulk space-time
dimensions, for instance, a co-dimension 2 brane, like a cosmic string, should create a deficit angle in the extra dimensions
while remaining flat. For $D > 6$, it has been argued that a brane with tension should inflate, but with a Hubble rate that
is {\it inversely} proportional to the brane tension~\cite{Dvali:2002pe}. 

In terms of the $\alpha$ parametrization, $D\geq 6$ brane-induced gravity theories all correspond to $\alpha \approx 0$ in the far infrared~\cite{deRham:2007rw}.
To see this, first note that the gravitational potential on the brane must scale as $1/r^{D-3}$ at large distances, consistent with $D$-dimensional gravity. Expanding in terms of massive states,

\be
\Phi(r)= \int_{0}^{\infty} {\rm d}M^2 \rho(M^2) \frac{e^{- Mr}}{r} \sim \frac{1}{r^{D-3}}\,,
\ee
the spectral density must satisfy $\rho(M^2)\sim M^{D-6}$ as $M\rightarrow 0$. Therefore, we see that in the small momentum limit ($\Box\rightarrow 0$),
\be
\lim_{\Box\rightarrow 0} \left[\frac{1}{\Box - m^2(\Box L^2)}\right]  \sim  \int_0 {\rm d}M^2 \frac{M^{D-6}}{M^2}\,.
\label{spec2}
\ee
This integral is convergent for $D > 6$. Therefore, in the infrared all such theories correspond to $\alpha = 0$. In $D=6$, the integral is
diverges logarithmically, corresponding to $m^2(\Box L^2)\sim \log\Box$. 

The simplest extension of the DGP model---a 3-brane embedded in a $D$-dimensional bulk space-time---had long been thought to be plagued
by ghost instabilities~\cite{Dubovsky:2002jm,Gabadadze:2003ck}. This conclusion was based on the tensorial structure of the graviton
exchange amplitude. Recently, a careful Hamiltonian analysis has revealed that the would-be ghost is in fact constrained,
and therefore non-propagating~\cite{Berkhahn:2012wg}. Much remains to be understood about the theoretical underpinnings
of these higher-dimensional scenarios, but these recent developments open the door to new phenomenological investigations of these theories. 
As a first step in this direction, the cosmology of the $D=6$ model is currently under investigation~\cite{FlorianRobertinprogress}.

Another generalization of the DGP scenario to higher dimensions is {\it cascading gravity}~\cite{deRham:2008zz,deRham:2007rw,deRham:2009wb,deRham:2010rw}. 
In this construction, our 3-brane is embedded within a succession of higher-dimensional branes, each with their own induced gravity term. In this simplest codimension-2 case,
for instance, the 3-brane is embedded in a 4-brane within a flat 6$D$ bulk. As a result, the gravitational force law ``cascades" from 4$D$ ($1/r^2$) to 5$D$ ($1/r^3$) to 6$D$ ($1/r^4$),
as we probe larger distances on the 3-brane. A similar cascading behavior of the force law was also obtained recently in a different codimension-two framework~\cite{Kaloper:2007ap}. Closely related work on intersecting branes was discussed in~\cite{Corradini:2007cz,Corradini:2008tu,Sbisa':2014uza} with somewhat different motivations. See~\cite{Agarwal:2009gy,Agarwal:2011mg,deRham:2009wb} for studies of degravitation in the cascading framework, and~\cite{Minamitsuji:2008fz} for self-accelerated solutions in this context.

\subsection{Horndeski theory}

In Section~\ref{massivegravsec} we considered one way of coupling galileon theories to gravity---having them appear as the helicity-0 component of a massive graviton. This is in a sense, a type of ``covariantization" of galileon theories. However, there is of course another sense in which galileons can be covariantized: the usual way of promoting the background metric to be dynamical. This leads to a line of development orthogonal to that of massive gravity, but no less interesting. 

Covariantizing the galileon Lagrangians~\eqref{galileonterms}, turns out to be rather subtle; the natural thing to do is to promote the background Minkowski metric to be a dynamical field and to promote partial derivatives to covariant derivatives
\be
\eta_{\mu\nu} \longmapsto g_{\mu\nu}~;~~~~~~~~~~~~~~~~~~\partial_\mu\longmapsto \nabla_\mu~.
\ee
For the terms ${\cal L}_1$--${\cal L}_3$, this works fine. However, it turns out that the equations of motion following from the covariantized versions of ${\cal L}_4$ and ${\cal L}_5$ in~\eqref{galileonterms} involve third derivatives of both the metric and the field, indicating that the coupled theory of galileons minimally coupled to gravity propagates a ghost~\cite{Deffayet:2009wt}. This problem can be removed, by introducing non-minimal couplings as~\cite{Deffayet:2009wt}
\bea  
\label{covariantgalileonterms}
{\cal L}_1&=&\phi\ , \nonumber \\
{\cal L}_2&=&\frac{1}{2}(\nabla\phi)^2 \ ,\nonumber \\
{\cal L}_3&=&\frac{1}{2}\square {\phi}(\nabla {\phi})^2 \ ,\\
{\cal L}_4&=&\frac{1}{4}(\nabla\phi)^2\left[(\square\phi)^2-(\nabla_\mu\nabla_\nu\phi)^2 -\frac{1}{4}(\nabla\phi)^2 R\right], \nonumber \\
{\cal L}_5&=& \frac{1}{3}(\nabla\phi)^2\Big[(\square\phi)^3+2(\nabla_\mu\nabla_\nu\phi)^3-3\square\phi(\nabla_\mu\nabla_\nu\phi)^2-6 G_{\nu\rho}\nabla_\mu\phi\nabla^\mu\nabla^\nu\phi\nabla^\rho\phi\Big]\nonumber\,,
\eea
where $R$ and $G_{\mu\nu}$ are the Ricci scalar and Einstein tensor corresponding to the metric $g_{\mu\nu}$, respectively. There are two things to note about these terms: the first is that they break the global galilean symmetry $\delta\phi = b_\mu x^\mu$. Essentially this is because the Lagrangians used to shift by a total derivative under this symmetry, but in the presence of a dynamical background metric, the variations are no longer total derivatives. Second, these are not the unique choices of non-minimal terms which lead to second-order equations of motion, because we can always commute covariant derivatives to induce more non-minimal couplings.\footnote{One unambiguous way to choose a set of non-minimal couplings which preserve second-order equations of motion (but still break the galileon symmetry) is through the brane construction of~\cite{deRham:2010eu}.}

In light of these two facts, we are motivated to ask the following question: {\it what is the most general Lagrangian which couples a scalar field to gravity non-minimally, and which has second-order equations of motion}? This question precisely was asked and answered in~\cite{Deffayet:2009mn}, inspired exactly by this interest in covariantizing the galileon theory, and the resulting theory was referred to as that of {\it generalized galileons}, having the Lagrangian
\begin{align}
{\cal L}_{\rm gen. gal.} &= K(\phi, X)-G_3(\phi, X)\square\phi+G_4(\phi, X)R+G_{4, X}(\phi, X)\big[(\square\phi)^2-(\nabla_\mu\nabla_\nu\phi)^2\Big]\\\nonumber
&+G_5(\phi, X)G_{\mu\nu}\nabla^\mu\nabla^\nu\phi-\frac{1}{6}G_{5, X}(\phi, X)\Big[(\square\phi)^3-3 (\square\phi)(\nabla_\mu\nabla_\nu)^2+2\nabla^\mu\nabla_\alpha\phi\nabla^\alpha\nabla_\beta\phi\nabla^\beta\nabla_\mu\phi\Big],
\end{align}
where $K, G_3, G_4, G_5$ are arbitrary functions of $\phi$ and $X$, where $X = -\frac{1}{2}(\nabla\phi)^2$.

It turns out that the answer to this question was known much earlier, but had gone mostly overlooked in the literature (but was resurrected in~\cite{Charmousis:2011bf}). In 1974, Horndeski~\cite{Horndeski:1974wa} wrote down the most general scalar-tensor Lagrangian which has second order equations of motion. It was pointed out in~\cite{Kobayashi:2011nu} that the generalized galileons and Horndeski's theory are equivalent (see~\cite{Kobayashi:2011nu} for a dictionary translating between the two languages). The Horndeski theory involves 4 free functions of the field and its kinetic term, and can be written as (in the notation of~\cite{Kobayashi:2011nu})
\begin{align}
\nonumber
{\cal L}_{\rm Horndeski} = 3!\delta_\mu^{[\alpha}\delta_\nu^\beta\delta_\sigma^{\gamma]}&\bigg[ \kappa_1(\phi, X)\nabla^\mu\nabla_\alpha\phi R_{\beta\gamma}^{~~\nu\sigma}+\frac{2}{3}\kappa_{1, X}(\phi, X)\nabla^\mu\nabla_\alpha\phi\nabla^\nu\nabla_\beta\phi\nabla^\sigma\nabla_\gamma\phi\\\nonumber
&~~+ \kappa_3(\phi, X)\nabla_\alpha\phi\nabla^\mu\phi R_{\beta\gamma}^{~~\nu\sigma}+2\kappa_{3, X}(\phi, X)\nabla_\alpha\phi\nabla^\mu\phi\nabla^\nu\nabla_\beta\phi\nabla^\sigma\nabla_\gamma\phi\bigg]\\\nonumber
+2!\delta_\mu^{[\alpha}\delta_\nu^{\beta]}&\bigg[(F(\phi, X)+2W(\phi))R_{\alpha\beta}^{~~\mu\nu}+2 F_{,X}(\phi, X)\nabla^\mu\nabla_\alpha\phi\nabla^\nu\nabla_\beta\phi\\\nonumber
&~~+2\kappa_8(\phi, X)\nabla_\alpha\phi\nabla^\mu\phi\nabla^\nu\nabla_\beta\phi\bigg] \\
&-6\Big(F_{,\phi}(\phi, X)+2W_{,\phi}(\phi)-X\kappa_8(\phi, X)\Big)\square\phi+\kappa_9(\phi, X)~,
\end{align}
where $X \equiv -\frac{1}{2}(\nabla\phi)^2$. This Lagrangian contains four completely free functions of $\phi$ and $X$, $\kappa_1, \kappa_3, \kappa_8, \kappa_9$, one constrained function, $F(\phi, X)$, which must satisfy
\be
F_{,X} = 2(\kappa_3+2X\kappa_{3, X}-\kappa_{1, \phi})~,
\ee
and one function of only $\phi$, $W(\phi)$, which may be absorbed by redefining $F$~\cite{Kobayashi:2011nu}. Therefore, we see that in both forms, the theory is described by $4$ arbitrary functions.

This theory has been constructed to have second-order equations of motion, but beyond that there are no symmetry properties which restrict the arbitrary functions. However, it is possible to place restrictions on these terms from their phenomenology. A basic phenomenological requirement is that the fifth force mediated by $\phi$ be screened from solar system tests~\cite{DeFelice:2011th,Koyama:2013paa}. In~\cite{Charmousis:2011bf,Charmousis:2011ea,Copeland:2012qf}, the functional form of the Horndeski theory was restricted by demanding that the resulting theory be {\it self-tuning};\footnote{For related work on the cosmology of this theory, see~\cite{Appleby:2012rx,Bruneton:2012zk,Linder:2013zoa}.} that is, that the resulting theory admit Minkowski space as a solution in the presence of an arbitrary cosmological constant, as a means to address the old CC problem. The result is a Lagrangian theory of the ``Fab Four":
\begin{align}
\nonumber
{\cal L}_{\rm John} &= V_{\rm John}(\phi)G^{\mu\nu}\nabla_\mu\phi\nabla_\nu\phi~,\\\nonumber
{\cal L}_{\rm Paul} &= -\frac{1}{4}V_{\rm Paul}(\phi)\epsilon^{\mu\nu\lambda\sigma}\epsilon^{\alpha\beta\gamma\delta}R_{\lambda\sigma\gamma\delta}\nabla_\mu\phi\nabla_\alpha\phi\nabla_\nu\phi\nabla_\beta\phi~,\\\nonumber
{\cal L}_{\rm George} &= V_{\rm George}(\phi)R~,\\
{\cal L}_{\rm Ringo} &= V_{\rm Ringo}(\phi)\Big(R_{\mu\nu\alpha\beta}R^{\mu\nu\alpha\beta}-4R_{\mu\nu}R^{\mu\nu}+R^2\Big).
\end{align}

Note that this theory evades Weinberg's no-go theorem about self tuning (reviewed in Appendix~\ref{Weinbergnogo}) by allowing for the profile of the field $\phi$ to be time-dependent. However, in order for such a theory to be observationally viable, it must employ the Vainshtein mechanism within the solar system to evade fifth force constrains, and recently it has been argued that the need for solar system screening and weak-coupling constrain the cosmological impact of the Fab Four~\cite{Kaloper:2014xya}.

Horndeski's theory has been extensively studied, for overviews see~\cite{Deffayet:2013lga,Charmousis:2014mia}. It has been used to construct general theories of inflation~\cite{Kobayashi:2011nu, Gao:2011vs,Gao:2011qe,Burrage:2011hd,Ribeiro:2011ax,DeFelice:2011uc,RenauxPetel:2011sb,DeFelice:2013ar} and cosmic acceleration~\cite{DeFelice:2011hq,DeFelice:2011bh,Leon:2012mt,Bloomfield:2011np,Bloomfield:2012ff,Bloomfield:2013efa,Mueller:2012kb,Tsujikawa:2014mba}. Additionally black holes have been studied in this theory in~\cite{Charmousis:2012dw, Babichev:2013cya,Anabalon:2013oea,Sotiriou:2013qea,Kobayashi:2012kh,Kobayashi:2014wsa,Charmousis:2014zaa}, and it has been generalized to describe multiple fields non-minimmally coupled to gravity~\cite{Padilla:2012dx, Sivanesan:2013tba,Padilla:2013jza}.

Finally, note that expanding Horndeski's theory about a cosmological FLRW solution neatly leads to a type of EFT of dark energy, similar to the theories discussed in~\cite{Park:2010cw,Creminelli:2008wc,Bloomfield:2012ff,Gubitosi:2012hu,Gleyzes:2013ooa,Piazza:2013coa,Frusciante:2013zop,Bloomfield:2013efa,Gergely:2014rna}. Additionally, the fact that expanding the general Horndeski theory about a cosmological background in the quasi-static regime requires two functions of time and scale to describe perturbations connects to the parameterized post-Friedmann approach to modified gravity, which we discuss briefly in Sections~\ref{quasistaticsec} and~\ref{section:PCA}.

\newpage
\part{Experimental tests}

We overview experimental tests of gravity on a wide range of scales. We begin with a short summary of the discovery of the accelerating universe and current constraints on its expansion history. 

\section{The accelerating universe}
The discovery of the accelerating universe was made by observing distant type Ia supernovae~\cite{Riess:1998cb,Perlmutter:1998np}, and this remains an important probe of cosmic acceleration~\cite{Schmidt:1998ys, Garnavich:1998th,Freedman:2000cf,Riess:2004nr,Astier:2005qq,Kowalski:2008ez,Kessler:2009ys,Amanullah:2010vv,Suzuki:2011hu,Sako:2014qmj}. Type Ia supernovae are useful because they can be used as {\it standard candles}: this type of supernova occurs when a white dwarf star reaches the Chandrasekhar mass limit and explodes~\cite{Chandrasekhar:1931ih,Hillebrandt:2000ga}. By studying the observed brightness of this explosion {\it vs}. time (the so-called {\it lightcurve}), it is possible to calibrate the intrinsic luminosity of the supernova to better than 10\%~\cite{Phillips:1993ng,Riess:1996pa,Hamuy:1996sq,Perlmutter:1996ds}. We then have an object of known absolute brightness, by measuring its apparent brightness we can infer its distance.

For a given object with an intrinsic luminosity, $L$, the further away it is, the dimmer it appears because the observed flux scales with the distance to the object as $\sim r^{-2}$. In fact, this observation can be used to define the {\it luminosity distance} to an object
\be
d_L = \sqrt{\frac{L}{4\pi F}}~,
\label{eqn:d_L}
\ee
where $L$ is the object's intrinsic luminosity and $F$ is the measured flux. Oftentimes it is useful to express things in terms of {\it redshift}
\be
z\equiv \frac{a_0}{a}-1 = \frac{\lambda_{\rm meas.}}{\lambda_{\rm emit.}}
\ee
which measures by how much the wavelength of light is stretched due to the expansion of the universe between when it was emitted and when it is observed here on Earth. Using this, we can rewrite the luminosity distance as (see, for example~\cite{Weinberg:2008zzc})
\be
d_L(z) = (1+z)\int_{0}^z\frac{\rd z'}{H(z')}~.
\ee
We assume spatial flatness here and in the rest of the following sections. The expression above can be expanded about $z=0$ to obtain
\be
H_0d_L(z) \simeq z+\frac{1}{2}(1-q_0)z^2+\ldots~;
\ee
\begin{figure}
\centering
\includegraphics[width=2.2in]{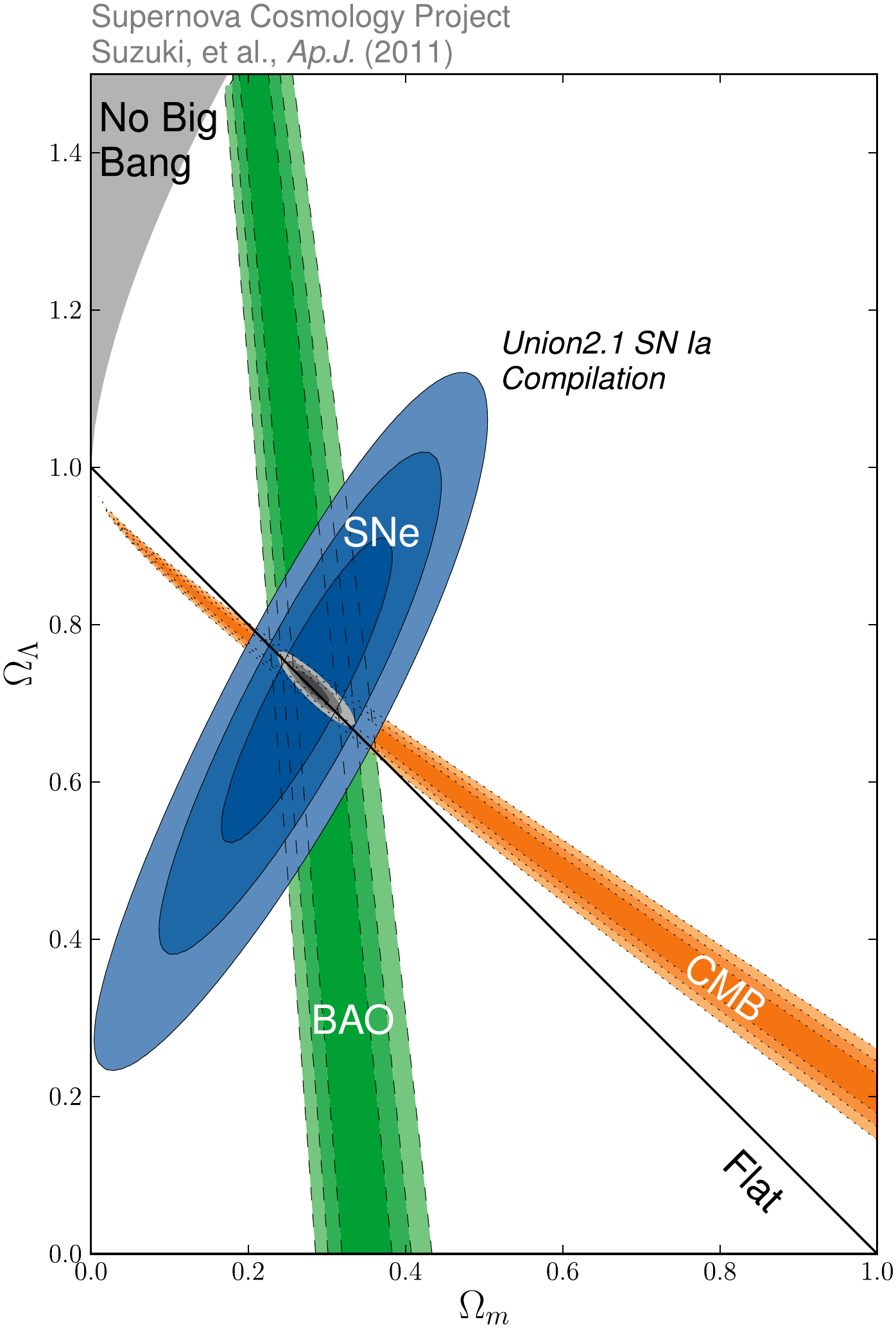}
\includegraphics[width=2.9in]{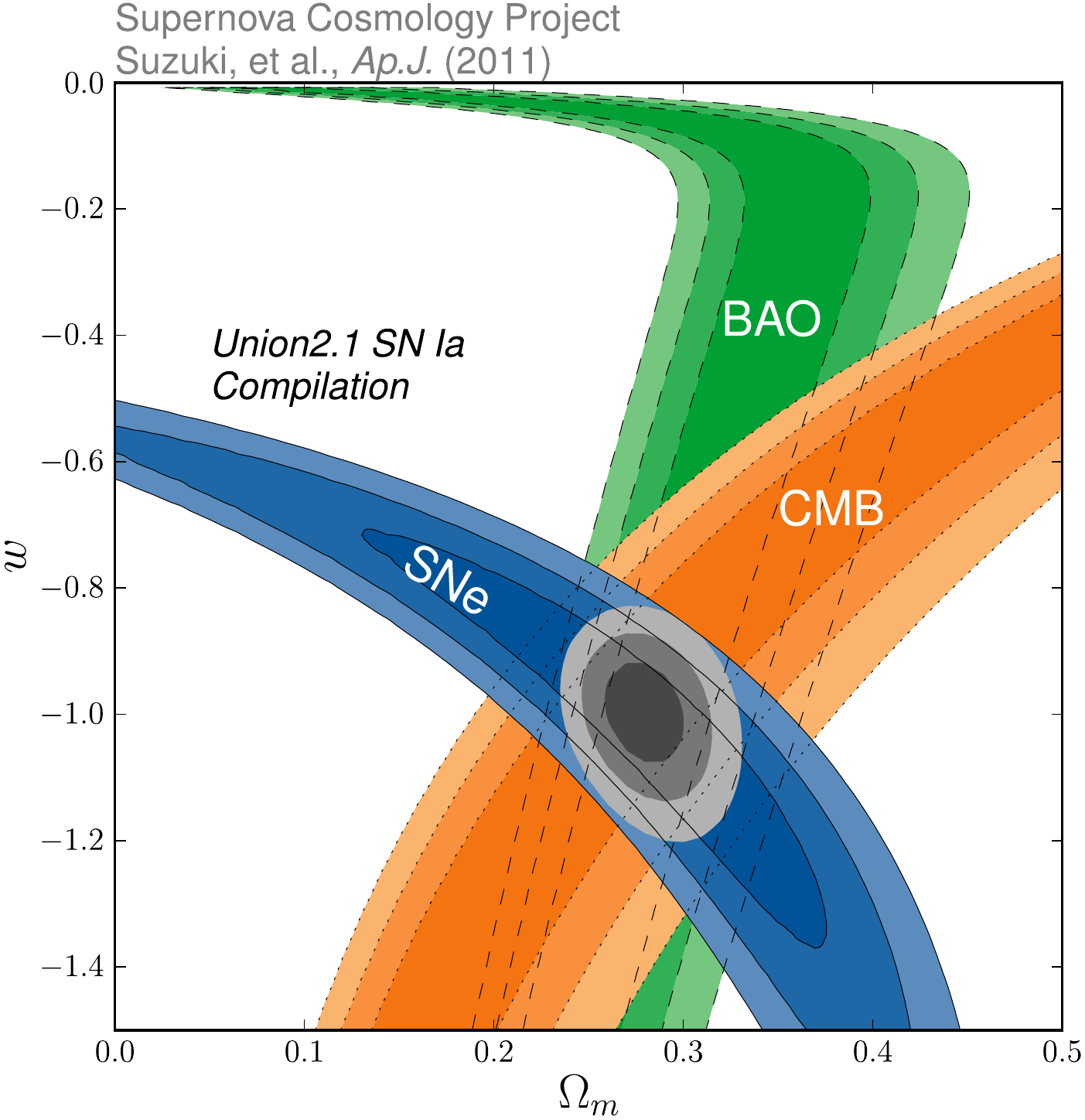}
\caption{\label{SNIaconstraints}\small {\it Left:} Constraints in the $(\Omega_{\rm m}, \Omega_\Lambda)$ plane from type Ia supernovae, assuming a $\Lambda$CDM cosmology. {\it Right:} Constraints in the $(\Omega_{\rm m}, w)$ plane, assuming a homogeneous component with equation of state $w$. Blue ellipses are $1\sigma$, $2\sigma$ and $3\sigma$ confidence regions. Also pictured are complimentary constraints from the cosmic microwave background and baryon acoustic oscillations. Both figures adapted from~\cite{Suzuki:2011hu}.}
\end{figure}
where we have defined the {\it deceleration parameter} $q \equiv - \ddot a/(aH^2)$.
Using measurements of the luminosity distance of standard candles at various redshifts, we can constrain the parameter $q_0$. Often the constraints on $q_0$ are expressed in the $(\Omega_{\rm m}, \Omega_\Lambda)$ plane; by combining the Friedmann equations and assuming that the universe is composed only of matter and $\Lambda$, we can obtain the relation
\be
q_0 = \frac{1}{2}\Omega_{\rm m}-\Omega_\Lambda~.
\ee
In Figure~\ref{SNIaconstraints}, we reproduce the constraints on $\Omega_{\rm m}$ and $\Omega_\Lambda$ from the Supernova Cosmology Project~\cite{Suzuki:2011hu}. Joint constraints between CMB, BAO and supernova data indicate
\be
\Omega_\Lambda \sim 0.7~,
\ee
which implies a value for the cosmological constant of $\Lambda \sim M_{\rm Pl}^2 H_0^2 \sim (10^{-3} {\rm eV})^4$. 

Rather than assuming a $\Lambda$CDM cosmology, we can attempt to fit the data with a perfect fluid component with arbitrary equation of state, $w$. In this way, we obtain a constraint in the $(\Omega_{\rm m}, w)$ plane, and find that $w$ must be rather close to $-1$ at the present day. These constraints are also reproduced in Figure~\ref{SNIaconstraints}. Indeed, in the last decade the basic parameters of the accelerating universe are well measured even without using supernova data---providing a powerful consistency check. 
The two other powerful methods that have been used to measure the geometry of the universe are: the cosmic microwave background (CMB)~\cite{deBernardis:2000gy,Lange:2000iq,Balbi:2000tg,Pryke:2001yz,Spergel:2003cb,Hinshaw:2012aka, Hou:2012xq,Ade:2013zuv,2013ApJ...779...86S,Sievers:2013ica} and the Baryonic Acoustic Oscillation (BAO) feature in the galaxy power spectrum~\cite{Tegmark:2003ud,Seljak:2004xh,Eisenstein:2005su,Blake:2011en,Beutler:2011hx,Dawson:2012va,2012MNRAS.427.3435A,Samushia:2012iq}. In fact, since measurement of $q_0$ only constrains the linear combination $\frac{1}{2}\Omega_{\rm m}-\Omega_\Lambda$, these measurements are a very powerful tool in breaking the degeneracy between these two parameters.

The CMB measurements constrain the total energy density in the universe, $\Omega_{\rm tot.} = \Omega_{\rm m}+\Omega_\Lambda+\Omega_\kappa$. We expect that CMB anisotropies will peak at the Hubble scale at last scattering $R \sim H^{-1}_{\rm cmb}$
~\cite{Bahcall:1999xn,Hu:1996qs,Trodden:2004st}. This gives us a feature of known size, or a {\it standard ruler} on the sky, by measuring its angular size we can infer the angular diameter distance $d_A\equiv d_L/(1+z)^2$, where $d_L$ is given by~\eqref{eqn:d_L}, at the redshift of last scattering $z\approx 1100$. 
For a flat universe, the characteristic domain subtends approximately $\sim1^\circ$ on the sky (corresponding to $\ell\sim 200$ in terms of spherical harmonics). In a positively curved (closed) universe, the feature will appear smaller on the sky and in a negatively curved (open) universe, it will subtend a larger angle~\cite{Kamionkowski:1993aw}. Measurements of the CMB have been used to constrain the combination $\Omega_{\rm m}+\Omega_\Lambda$ to be unity with percent level precision. This gives a constraint on a linear combination of these two parameters orthogonal to that coming from SN Ia measurements.

The BAO measurement typically constrains the angular diameter distance at redshifts of about unity (and more recently up to $z\approx 3$.) Thus, a combination of the distance-redshift relation measured at different redshifts using BAO, CMB and SN distances can probe the evolution of dark energy. See \cite{Weinberg:2012es} for a review. 

In the modified gravity context,  the constraints from geometric methods can be translated into constraints on the distance-redshift relation, without assumptions about dark energy. They can then be compared directly with the predictions of modified gravity theories. Several other cosmological probes test dark energy and gravity using a combination of geometry and growth: galaxy clusters,  gravitational  lensing and the clustering of galaxies in redshift space. These are discussed below in Section \ref{section:cosmology}.

\section{Laboratory and solar system tests}

Solar system measurements have a venerable history as tests of gravity. They strongly constrain the presence of additional light degrees of freedom, necessitating screening mechanisms to render them immune to such tests. In this section we summarize the current status of both solar system and laboratory tests. Tests of gravity have evolved as a search for deviations from the predictions of GR. Therefore, local tests may be separated broadly into general tests of the equivalence principle and inverse square law of gravity, or as measurements of  the Parameterized Post-Newtonian (PPN) parameters.

An important ingredient of Einstein gravity is the weak equivalence principle (WEP), which states that the trajectories of 
freely falling test bodies are independent of internal structure and composition. The WEP is not unique to GR---any theory whose matter fields couple to a unique metric tensor ({\it e.g.}, Brans--Dicke theory~\cite{Brans:1961sx}) satisfies the WEP, independent of the field equations governing this metric.\footnote{Scalar-tensor theories, including Brans--Dicke,
violate the {\it Strong Equivalence Principle} (SEP), which is satisfied in GR and states that gravitational self-energy contributions also do not cause test bodies to fall at different rates.} For a review of the various formulations of the equivalence principle and tests, see~\cite{Will:2005va,Will:2014kxa,Haugan:2001ix}.

To test for small deviations from GR in the solar system, it is useful to employ the PPN framework. This allows us to cast a general metric theory of gravity as a deviation from a Minkowski (or Schwarzschild) background metric and to use local measurements to put constraints on the coefficients which appear.

The early experimental predictions of GR were the anomalous perihelion precession of Mercury and the gravitational deflection of light by the sun, both of which were spectacularly confirmed. In the middle of the 20th century a number of additional weak-field tests of gravity were conducted, including the measurement of gravitational redshift (predicted originally by Einstein), measurement of Shapiro time delay~\cite{Shapiro:1964uw}, and the radiation of gravitational waves by orbiting binaries, which was found to be in excellent agreement with the decrease in the orbital period of the Hulse--Taylor pulsar~\cite{Hulse:1974eb}. Together, these probes test gravity exceptionally well in the weak-field regime, and no deviation from Einstein gravity has been discerned.

\subsection{Tests of the Equivalence Principle and force law}

The weak Equivalence Principle can be tested by measuring the fractional difference in the acceleration of freely falling bodies of different composition. This difference is parameterized by the so-called  $\eta$ parameter, and experiments which measure it are often referred to as E$\ddot{{\rm o}}$tvos-type experiments. The best limit on $\eta$ comes from torsion balance experiments at the University of Washington
(E$\ddot{{\rm o}}$t-Wash), which give $\eta < 2 \times 10^{-13}$~\cite{Wagner:2012ui,Will:2014xja}. 

Another test comes from the fact that alternative theories of gravity do not have to satisfy the strong equivalence principle, which is satisfied by GR. The SEP says that extended objects follow the same trajectories as test masses in a uniform gravitational field. Another way of saying this is that the SEP states that gravitational and inertial masses are the same, even accounting for gravitational self-energy contributions. The SEP is violated in all of the modified gravity theories we consider here. Violations of the SEP result in the Nordtvedt effect~\cite{Nordtvedt:1968qr}---a difference in the free-fall acceleration of the Earth and the Moon towards the Sun---which is detectable by Lunar Laser Ranging (LLR).\footnote{The Earth and Moon have different compositions, so one must be concerned about fluke cancellations between WEP and SEP violations. To disentangle these effects, laboratory tests of the WEP have been carried out using tests masses with Earth-like and Moon-like compositions~\cite{Baessler:1999iv}.} Searches for the Nordtvedt effect in LLR data constrain deviations of PPN parameters from their GR values at the $10^{-4}$ level~\cite{Williams:2004qba, Williams:2003wu}.

Even in cases where the WEP is satisfied, modifications of gravity are constrained by tests of the gravitational inverse-square law~\cite{Fischbach:1999bc}. These can also be thought of as tests of the existence of a fifth force. As a simple example, consider
a scalar field of mass $m$, which mediates a force over a characteristic distance $\lambda$ and which has coupling strength $\alpha$ to normal matter. The Yukawa potential corresponding to this situation is
\begin{equation}
\psi = -\alpha\frac{G_{\rm N}M}{r} e^{-r/\lambda} \ .
\label{eqn:Yukawa}
\end{equation}
Experimental tests can then be viewed as providing limits on $\lambda$ and $\alpha$~\cite{Schlamminger:2007ht,Kapner:2006si,Adelberger:2006dh}. For a gravitational-strength coupling ($\alpha\sim {\cal O}(1)$),
there is no evidence of a fifth force down to a distance of $\lambda = 56\;\mu$m~\cite{Kapner:2006si}. For other searches for deviations from the inverse square law, see~\cite{Long:1998dk,Hoyle:2000cv,Chiaverini:2002cb,Long:2002wn,Hoyle:2004cw,Geraci:2008hb,Bezerra:2011xc,Klimchitskaya:2013rwd} and for a review of these types of experiments, see~\cite{Adelberger:2003zx}

\subsection{Post-Newtonian tests}
\label{postnewtonian}

Solar system tests of gravity are conveniently cast in the language of the 
Parameterized Post-Newtonian (PPN) formalism~\cite{Nordtvedt:1968qs,Will:1971zzb,Will:1972zz}. The PPN expansion can be applied to metric
theories of gravity in regimes where potentials and 
velocities are small: $\Psi, v^2/c^2 \sim \epsilon^2 \ll 1$. In this regime, the metric can be written 
as a perturbation about the Minkowski metric (or, the FLRW metric for an
expanding universe). The metric is then expanded up to second order in the potentials, with coefficients
that are allowed to deviate from their GR values, in order to accommodate modified gravity theories. 

Two commonly used PPN parameters are the so-called $\gamma$ and $\beta$ parameters, which are defined via the metric
\begin{equation}
\rd s^2 = -(1 + 2\Psi - 2\beta\Psi^2)\ {\rm d}t^2 + (1 - 2\gamma\Psi)\ {\rm d}\vec{x}^2~,
\label{eqn:PPN}
\end{equation}
where the potential $\Psi=-G_{\rm N}M/r$ for the Schwarzschild metric. The parameter $\gamma$ describes the space-time curvature induced by a unit mass and the $\beta$ parameter describes the nonlinearity of the superposition law of gravity. A similar approach can be employed for the FLRW metric and an arbitrary (small) potential.
 
In general, for matter described by a fluid, and allowing for generic Poisson-like potentials, the PPN formalism requires ten parameters~\cite{Will:2005va,Will:2014kxa}. Note that the PPN parameters are constant and therefore do not accommodate Yukawa-like modifications with finite $\lambda$. For astrophysical tests we will employ a similar expansion of the metric but allow effective parameters that may have scale and time dependence.
 
The Brans--Dicke theory 
has identical PPN parameter values to GR, except for 
\begin{equation}
\gamma_{\rm BD} = \frac{1+\omega_{\rm BD}}{2+\omega_{\rm BD}}~.
\end{equation}
This parameter $\gamma$ ends up being the most relevant PPN parameter for the theories of interest, and therefore we focus on it for the rest of our discussion.

The tightest constraint on $\gamma$ comes from time-delay measurements in the solar system, specifically from radio waves emitted by the Cassini spacecraft,
which gives $\gamma - 1 =(2.1\pm 2.3)\times 10^{-5}$~\cite{Bertotti:2003rm}. Light deflection measurements, meanwhile, constrain
$\gamma$ at the $10^{-4}$ level~\cite{Shapiro:2004zz}. The $\gamma$  parameter can also be tested by measuring the
perihelion shift of Mercury's orbit, which sets a weaker limit of $10^{-3}$~\cite{mercury}. The PPN nonlinearity parameter $\beta$ is also constrained to be unity at the $10^{-4}$ level. 

All the PPN parameters are constrained by local tests to not deviate from their GR values at the sub-percent level; see Table 4 in \cite{Will:2014kxa}. If modified gravity were characterized only by such scale-independent parameters, we would have no motivation to seek either theoretical descriptions or experimental tests of order unity deviations. However, the inherent nonlinearity of screening mechanisms naturally implies scale-dependent deviations from GR ({\it e.g.}, via the density, potential, or local curvature near macroscopic bodies). This motivates us to search for deviations from Einstein gravity on all scales.

We now discuss how solar system and laboratory tests translate into
specific constraints on the Chameleon and Vainshtein-screened
theories. 

\subsection{Laboratory and solar system tests of chameleon theories}
\label{chamlabtests}

Solar system tests and laboratory tests are useful for some classes of chameleon theories.  
As a concrete example, consider the chameleon potential
\begin{equation}
V(\phi) = {\rm const.} +\lvert\phi\rvert^n~.
\end{equation}
In the range $n~ \lsim -1/2$ or $n>2$, the strongest constraints come from laboratory experiments. This is because torsion pendulum experiments~\cite{Adelberger:2006dh} require the chameleon field to have a Compton wavelength less than a millimeter at laboratory densities $\sim 1$~g/cm$^3$. In this situation each planet in the solar system would have a thin shell even in isolation. This reduces the effective chameleon coupling by many orders of magnitude, so that no effects could be discernible by local probes. As an example, the Earth's coupling is reduced by a factor of $\sim (1\textrm{mm} / 6400 \textrm{ km})^2 \sim 10^{-20}$, and similarly for the other planets. This causes solar system measurements to be weaker than other constraints~\cite{Gubser:2004uf}.

In the regime $-1/2~\lsim ~n < 1$, however, screening requires a large gravitational potential well. In order for the sun to be screened (as it must be from tests of Kepler's Laws), the chameleon self-screening parameter must be smaller than the Sun's gravitational potential, $\Psi\approx 2 \times 10^{-6}$. However, the solar system is not isolated, it sits inside the Galaxy, which itself has a potential well of magnitude $\Psi\approx 10^{-6}$, which in turn sits inside the Local Group and the Virgo Supercluster. Taken together, the added screening effects from the environment likely weakens solar system constraints by a factor of a few~\cite{Hu:2007nk}. These constraints are an order of magnitude weaker for these models than constraints coming from Cehpheid variables and dwarf galaxies~\cite{Jain:2012tn, Vikram:2013uba}.

The idea that the manifestation of a fifth force is sensitive to the environment has spurred a lot of activity in laboratory tests. As mentioned above, gravitational-strength fifth forces are most strongly constrained by torsion pendulum experiments in the laboratory.  In more strongly coupled models, where screening is more powerful,  it is possible to use cold neutron systems as a probe: the presence of the chameleon field both alters the energy levels of neutrons bouncing in the Earth's gravitational field~\cite{Brax:2011hb,Pokotilovski:2013gta,Jenke:2014yel} and can alter the phase of neutrons in interferometry~\cite{Brax:2013cfa}. Intermediate between these two regimes, Casimir force experiments cant be used to probe chameleons of moderate coupling~\cite{Brax:2007vm}. Laboratory experiments also provide strong
constraints on certain types of interactions between dark energy and the Standard Model. Dark energy models which couple directly to electromagnetism can be produced and trapped in afterglow experiments, or produced in the Sun and detected in magnetic helioscopes.

Some of the laboratory experimental efforts aimed at searching for chameleon signatures can be summarized as:
\begin{itemize}

\item
The E\"ot-Wash experiment, which searches for deviations from the inverse-square-law at distances $\mathrel{\mathstrut\smash{\ooalign{\raise2.5pt\hbox{$>$}\cr\lower2.5pt\hbox{$\sim$}}}} 50\;\mu$m. Based on theoretical predictions~\cite{Upadhye:2006vi}, the E\"ot-Wash group was able to constrain part of the chameleon parameter space~\cite{Adelberger:2006dh}.

\item
If a scalar field couples via $e^{\beta_\gamma\phi}F_{\mu\nu}F^{\mu\nu}$ to electromagnetism, then photons traveling in a magnetic field will undergo oscillations between photons and the field $\phi$. The CHameleon Afterglow SEarch (CHASE) experiment~\cite{Chou:2008gr,Steffen:2009sc,Upadhye:2009iv,Steffen:2010ze,Upadhye:2012ar,Steffen:2012rw} has looked for an afterglow from trapped chameleons converting into photons. For a discussion of quantum corrections to chameleon dynamics in these setups, see~\cite{Brax:2013tsa}. Similarly, the  Axion Dark Matter eXperiment (ADMX) resonant microwave cavity was used recently to search for chameleons~\cite{Rybka:2010ah}. Photon-chameleon mixing can also occur deep inside the Sun~\cite{Brax:2010xq} and affect the spectrum of distant astrophysical objects~\cite{Burrage:2008ii}. 

\item 
Through a nice analogy between chameleon screening and electrostatics~\cite{JonesSmith:2011tn,Pourhasan:2011sm}, it was realized that the scalar field would experience an enhancement near the tip of  pointy objects (a``lightning rod" effect), and an experiment has been proposed to exploit this enhancement~\cite{JonesSmith:2011tn}.

\end{itemize}

In addition to these specific tests, there are also possible collider signatures~\cite{Brax:2009ey,Brax:2009aw}. And although we have focused on chameleon searches here, there are also related laboratory signatures of symmetrons~\cite{Upadhye:2012rc}. Finally, the most striking signature of chameleons can be found by testing gravity in space. The screening condition $\frac{\phi_{\rm amb.} - \phi_{\rm obj.}}{6\xi M_{\rm Pl}\Phi} \ll 1$ manifestly depends on the ambient density, so objects that are screened in the laboratory may be unscreened in space. 
This leads to striking predictions for future satellite tests of gravity, such as the planned MicroSCOPE mission\footnote{{\tt http://microscope.onera.fr/}} and STE-QUEST.\footnote{{\tt http://sci.esa.int/ste-quest/}} In particular,
chameleons can result in violations of the (weak) Equivalence Principle in orbit with $\eta \equiv \Delta a/a \gg 10^{-13}$, which would be in blatant conflict with laboratory constraints. Additionally, in space chameleons can mediate a gravitational-strength force, which would appear as ${\cal O}(1)$ deviations
from the value of $G_{\rm N}$ measured on Earth.

\subsection{Laboratory and solar system tests of galileon/Vainshtein theories}

Similar to the case of chameleons, theories which exhibit the Vainshtein mechanism are constrained by solar system and laboratory tests. For concreteness, we focus on the cubic galileon theory
\begin{equation}
{\cal L} = -3(\partial\phi)^2 -\frac{1}{\Lambda^3}\square\phi(\partial\phi)^2 + \frac{g}{M_{\rm Pl}}\phi T \ .
\end{equation}

The strongest solar system constraints on the cubic galileon come from Lunar Laser Ranging (LLR) observations, since ~30 years' worth of data allows LLR monitoring to constrain the Moon's orbit to $\; \lower .75ex \hbox{$\sim$} \llap{\raise .27ex \hbox{$<$}}\;  {\rm cm}$ accuracy. (For a review, see~\cite{Nordtvedt:2003pj}.)

Deep inside the Vainshtein radius, the galileon-mediated force given by~\eqref{galforcesup}, is strongly suppressed, but nevertheless gives a small correction to the Newtonian potential:
\begin{equation}
\frac{\delta\Phi}{\Phi} \simeq \frac{g^2}{2} \left(\frac{r}{r_{\rm V}}\right)^{3/2}~,
\end{equation}
where---as a reminder---we have defined the Vainshtein radius: $r_{\rm V} = \frac{1}{\Lambda}\left(gM/M_{\rm Pl}\right)^{1/3}= (4\pi g r_{\rm Sch} L^2)^{1/3}$, where $r_{\rm Sch} = M/(4\pi M_{\rm Pl}^2)$ is the Schwarzchild radius of the object and $L = (\Lambda^3/M_{\rm Pl})^{1/2}$.

The current constraint from LLR observations is~\cite{Murphy:2012rea}
\begin{equation}
\frac{\delta\Phi}{\Phi} \; \lower .75ex \hbox{$\sim$} \llap{\raise .27ex \hbox{$<$}}\; 2.4\times 10^{-11}\,.
\label{LLRcurrent}
\end{equation}
Substituting $r_{\rm Sch} = 0.886~{\rm cm}$ for the Earth, and $r = 3.84\times 10^{10}~{\rm cm}$ for the Earth-Moon distance, the LLR constraint translates to a bound on $L$~\cite{Dvali:2002vf,Lue:2002sw,Dvali:2007kt,Afshordi:2008rd}: 
\begin{equation}
L \;  \lower .75ex \hbox{$\sim$} \llap{\raise .27ex \hbox{$>$}} \; \frac{H_0^{-1}}{20\,g^{3/2}} \simeq 150 g^{-3/2}~{\rm Mpc}\,,
\end{equation}
where $H_0^{-1} \simeq 10^{28}~{\rm cm} \simeq 3000~{\rm Mpc}$ is the Hubble radius. This bound is expected to be improved by a factor of 10 by the Apache Point Observatory Lunar Laser-ranging Operation (APOLLO)~\cite{Murphy:2012rea}. For $g \sim 1$ this would push $L$ to values larger than $H_0^{-1}$. 

It is also possible to probe Vainshtein-screened systems using planetary orbits. 
In the context of DGP it can be shown~\cite{Lue:2002sw,Dvali:2002vf,Battat:2008bu} that the cubic galileon leads to an angular precession in a planetary orbit of $3 c / (8 r_c) = 25 (1 \textrm{ Gpc}/r_c)~\mu$as/yr, independent of the mass and orbital radius of the planet.  
This yields slightly weaker constraints on $L$ (in the context of the DGP model) than those coming from LLR~\cite{Battat:2008bu}. For more on the effect of Vainshtein-suppressed fifth forces in the solar system, see~\cite{Hiramatsu:2012xj,Andrews:2013qva}

Similar to the chameleon, there exist laboratory tests of models which hide themselves via Vainshtein screening, albeit fewer of them. For a nice review of laboratory tests of galileons (and also solar system tests), see~\cite{Brax:2011sv}. Of particular interest are Casimir force experiments, where the galileon would mediate a fifth force between two parallel plates. In the limit of two infinite perfectly flat and parallel plates, the galileon profile depends only on the direction transverse to the plates $\phi = \phi(z)$. In this situation, all of the higher galileon interactions vanish (in this limit, they are total derivatives) and the theory is described by a free scalar field, which would mediate an un-screened force~\cite{Brax:2011sv}. However, this exact geometric cancellation is extremely difficult to produce in a laboratory setting, and there is no sense in which one can be ``close" to this planar limit. In practice, Casimir force experiments are done by measuring the force between a plate and a sphere. Taking these effects into account weakens the constraints on the galileons coming from these experiments. Another constraint on galileon theories comes from measurements of variations of Newton's constant. In~\cite{Babichev:2011iz}, it was shown that shift-symmetric scalar-tensor theories with a non-minimal coupling between the scalar and matter are strongly constrained by measurements of the (lack of) variation of $G_{\rm N}$.

\section{Astrophysical tests in the nearby universe}
\label{section:astrotests}

Screening mechanisms typically use some measure of the mass distribution
 of halos---for example the density or Newtonian potential---to recover General Relativity deep inside the Milky Way.
However, it is still possible that smaller halos, the outer regions of halos, or even some components of the mass distribution, could experience enhanced forces. For a given mass distribution,
unscreened halos would consequently have both higher internal velocities and higher center of mass
velocity compared to the expectations coming from GR. This can produce deviations on the order of of $\sim$10-100\%  from GR, with
distinct different mechanisms giving disparate predictions for both the size of the effect and the
details of the manner in which transition to GR occurs. Observable effects can be larger on
halo scales than in the linear regime or at high redshift.
Since modified gravity models must recover Einstein gravity at high redshift to be consistent with CMB and Big Bang Nucleosynthesis observations, the effects of enhanced forces manifest primarily at late times. This favors tests
in the nearby universe that rely on objects with short dynamical times. Thus tests in nearby stars and galaxies offer a complementary probe of gravity theories. (See below for a detailed description of cosmological tests.) The account of astrophysical tests in this section follows in part the summary presented in the Snowmass 2013 report~\cite{Jain:2013wgs}. 

In this section we will treat chameleon, symmetron and environmentally dependent dilaton screening mechanisms as a single category, since their qualitative observational signatures are similar. The second category contains kinetic screening and Vainshtein theories, whose signatures are distinct from
chameleon-type theories.  The tests we describe below will contain two fundamental parameters of the theories: the coupling of the  fifth force to matter, and the range of interaction of the fifth force.  

The way that screening works in scalar-tensor
gravity theories implies that on small scales the fifth force impacts some tracers and not others.  Galaxies themselves can have enhanced motions, as discussed below, in tests involving lensing and dynamical masses. Additionally, the components of galaxies---things like stars, gas, neutron stars and black holes---can respond differently to the fifth force because
they can have different levels of screening. Consequently, different components can
acquire different velocities or be displaced in their spatial
distribution, resulting in a variety of observable phenomena. 

In chameleon theories,
galaxies in low-density environments may be unscreened because
Newtonian potential, $\Psi_{\rm N}$, which is the quantity that determines the level of screening,
can be smaller than in the Milky Way. Thus,
dwarf galaxies can exhibit effects of modified or additional forces in both
their infall motions and in their internal dynamics. For Vainshtein-screened theories, the velocities of
galaxies and other tracers of gravity can be enhanced (and can be compared to lensing results) and it is possible for compact objects to separate from stars and gas. Order of magnitude estimates for these observable effects have been constructed for $f(R)$ and galileon/DGP theories. We will first 
describe the physical effects on stars and galaxies and then describe observational constraints and prospects.

\subsection{Stellar evolution in chameleon theories}

Stars which reside within unscreened galaxies may exhibit effects of modified gravity.
In~\cite{Chang:2010xh} and~\cite{Davis:2011qf} the effects on  giant
and main sequence stars, respectively, are described in chameleon-screened theories: in short, the enhanced gravitational force
makes stars of a given mass brighter and hotter than they would be in GR.  Also, since they consume their fuel at a faster rate, they are also more ephemeral.

For the Sun, the surface potential  is $\Psi_{\rm N} \approx 2\times 10^{-6}$;  coincidentally, the potential of the Milky Way is quite close to this value. This is
believed to be sufficient to screen the galaxy, so that all solar system tests of
gravity are satisfied.
Therefore, main sequence stars of roughly a solar mass 
are likely to be either partially or completely screened. The surface potential varies slowly along the main sequence, so large deviations are not expected even at other masses. 
However, red giants may be an exception: they are at least ten times larger in size than the main sequence star from which
they originated, so their surface potential may be estimated to be $\Psi_{\rm N} \sim 10^{-7}$---so it is possible that their envelopes are unscreened.  

To understand the impact of enhanced forces on stellar evolution, 
let us denote by $\alpha_c$ the coupling parameter setting the
strength of the fifth force in unscreened regions. An object which is completely
unscreened  will feel a fifth force which can be described simply by a rescaling
of $G_{\rm N}$ 
\begin{equation}
G_{\rm N}\longrightarrow G_{\rm N}(1+\alpha_c) .
\end{equation}
For
objects that are partially screened, the total force in the region exterior to the screening
radius can be described by a position dependent rescaling of G:
\begin{equation}\label{eq:g(r)}
G_{\rm N}(r)=G_{\rm N}\left[1+\alpha_c\left(1-\frac{M(r_{\rm s})}{M(r)}\right)\right]
\end{equation}
where $M(r)$ is the mass interior to a shell of radius $r$ and $r_{\rm s}$ is the effective
screening radius. 

The structure of a spherically symmetric star is obtained by solving the equations
of stellar structure that at a given radius $r$ relate $M(r)$ to
 $P(r)$, $\rho(r)$ and $T(r)$---respectively the pressure,
density and temperature. As noted by~\cite{Chang:2010xh} and~\cite{Davis:2011qf},
modifications to the gravitational physics are entirely contained in the equation of hydrostatic equilibrium:
\begin{equation}
\frac{\rd P}{\rd r} = -\frac{G_{\rm N}(r) \rho(r) M(r)}{r^2},
\end{equation}
which represents the condition for the outward pressure to balance the (now enhanced) inward gravitational pull and yield a static solution.
Note that the modification is expressed purely as a change in Newton's constant, $G_{\rm N}$,
which becomes dependent on $r$ if the star is partially screened
according to equation~\eqref{eq:g(r)}. The other three equations---the
continuity, radiative transfer and energy generation equations---are all
unaffected by this change in $G_{\rm N}$. The result of this is that unscreened
stars of a given mass are more compact, burn brighter, and have a higher
effective temperature than screened stars of identical mass and
chemical composition. They also have a shorter main
sequence lifetime due to an increased burning rate and their finite fuel
supply.

The complete system of stellar structure equations for main sequence
stars can be solved under certain simplifying assumptions. 
However, if one wants to examine the dynamical
and nuclear properties in addition to the structure of post-main-sequence
stars, then a numerical approach is needed. Fortunately 
the publicly available stellar evolution code
MESA developed by Paxton et al.~\cite{Paxton:2010ji}, used for chameleon gravity in \cite{Davis:2011qf} and 
\cite{Jain:2012tn}, has been 
extremely useful for quantifying the physical effects of modified gravity on stellar evolution. 
Two observationally relevant applications are the impact on the gross properties of galaxies (given a mix of stellar populations)
and  stellar pulsations, as described below. 

\subsection{Pulsating stars and the distance ladder}

Specific stages of the evolution of giants and supergiants are often used to obtain
accurate distance estimates; they also happen to provide useful tests of gravity.
Two commonly used features are  the nearly universal
luminosity of  $\lsim~2 M_\odot$ stars at the tip of the red giant branch (TRGB) and the
period-luminosity
relation of cepheids---giant stars with masses $\sim 3-10 M_\odot$ that
pulsate when their
evolutionary tracks cross a narrow range in temperature known as the
{\it instability strip}.
The tight relation between luminosity and other observables is what makes these stars valuable distance indicators---it also
makes them useful for tests of gravity.

In order to estimate the pulsation period, it is necessary to go beyond hydrostatic equilibrium and consider
the full dynamical radial acceleration of a fluid element at radius $r$, which is described by the momentum equation:
\be
\ddot{r} = -\frac{G_{\rm N}M(r)}{r} - \frac{1}{\rho}\frac{\partial{P}}{\partial{r}}~.
\label{eqn: acceleration}
\ee

\begin{figure}[htb]
\begin{center}
\includegraphics[width=9cm]{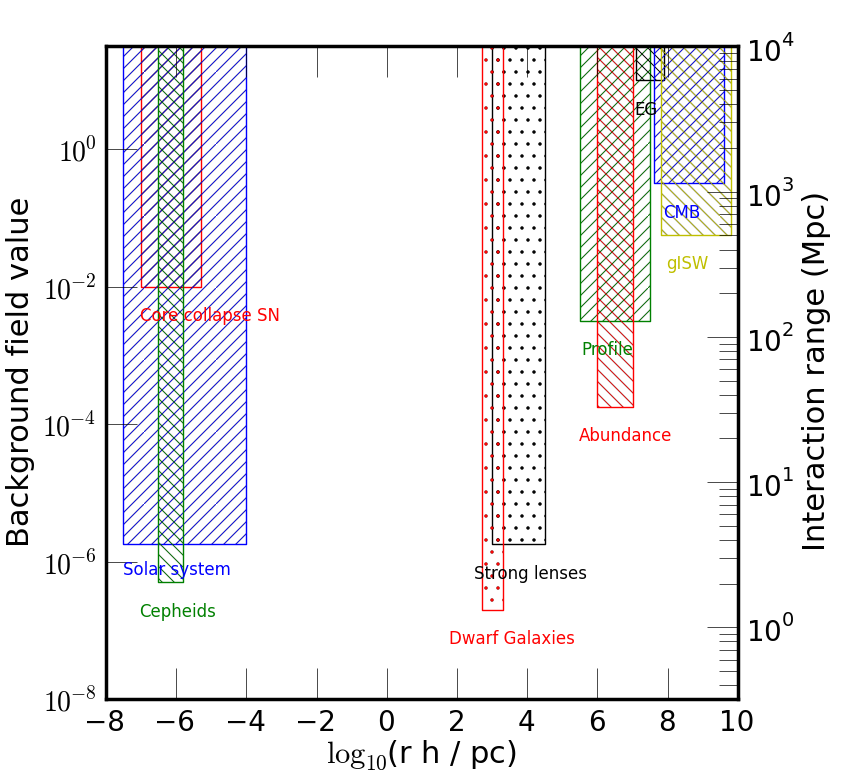}
\caption{\small Limits on chameleon theories coming from astrophysical~\cite{Hu:2007nk,Jain:2012tn,Vikram:2013uba} and cosmological~\cite{Song:2007da,Giannantonio:2009gi,Schmidt:2009am} probes. The $x$-axis gives the range of length scales probed by particular experiments. The parameter on the $y$-axis is the background field value (left hand side $y$-axis label), or the range of the interaction (right hand side $y$-axis label) for an $f(R)$ model of the accelerating universe. The rectangular regions give the regions of parameter space excluded by a particular experiment. All of the constraints except for the solar system measurements have  been obtained within the last 5 years, illustrating the impressive interplay between theory and experiment in the field. The dot-filled rectangles indicate preliminary results from ongoing work. This figure is adapted from Lombriser et al.~\cite{Lombriser:2011zw}. 
\label{fig:chameleonlimits}
}
\end{center}
\end{figure}

The time period of pulsations, $\Pi$, may
be estimated through various approximations; it is sensitive to the
enhancement in $G_{\rm N}$ in chameleon theories, scaling roughly as $1/\sqrt{G_{\rm N}}$. 
For background field values (for $f(R)$ theories, the parameter $f_{R0}$) in the range $10^{-6}$--$10^{-7}$, the
 TRGB luminosity is largely robust to modified gravity while the cepheid period-luminosity
relation is altered. Measurements of these properties
within screened and unscreened galaxies then provide tests of gravity:
the two distance indicators should agree for screened galaxies but not for
unscreened galaxies~\cite{Jain:2012tn}. Using the MESA stellar evolution code, these authors solved for the profile of $G_{\rm N}$ inside the stellar envelope for chameleon theories, and used it to estimate the change in pulsation period and therefore the impact on inferred distances. The resulting constraints using cepheids and TRGB stars are currently the most stringent on chameleon theories, with an upper limit on background field values of $f_{R0}~\lsim~5\times 10^{-7}$ at the 95\% confidence level~\cite{Jain:2012tn}, as shown in Figure \ref{fig:chameleonlimits}. Recently, hydrodynamic effects under MG were incorporated into the pulsation period estimate leading to increased deviations under MG~\cite{Sakstein:2013pda}. 

Cepheids and other variable stars remain promising candidates for tests of gravity. Indeed the full complement of distance indicators can be turned into a test of gravity, including Type Ia supernovae (which are the most compact distance indicators) and masers around supermassive black holes (which provide a purely geometric distance measure, independent of gravity theory). Relevant to the dynamics of cepheids and supernovae is the possibility of scalar radiation as a source of energy loss. Scalar radiation must be accounted for in future high precision tests that use stellar pulsations or explosions, and may also provide additional tests~\cite{Upadhye:2013nfa}.

\subsection{Galaxies in the nearby universe}

In unscreened galaxies, gas, stars and compact objects (neutron stars and black holes)  can respond differently to the scalar force in MG.
For unscreened dwarf galaxies,
the rotation of the stellar disk can be slower than that of the neutral Hydrogen gas disk because the stars are screened for some range of parameters in chameleon theories. 
The rotation rates of the stellar and gas disks can be measured using different optical and radio observations, respectively.
Additionally, the external fifth force on a dwarf galaxy can result in a segregation of the
the stars, giant stars and gas along the direction of the external force. For unscreened
disk galaxies, this also leads to a warping of the shape of the stellar
disk~\cite{Vikram:2013uba}. The difference in forces felt by compact objects can lead to the displacement of supermassive black
holes in Vainshtein theories, which is discussed in the following
subsection. All these effects are potentially
observable via high resolution imaging, and preliminary tests of gravity using data in the literature have been carried out
or are ongoing. Figure \ref{fig:chameleonlimits} shows some of the upper limits obtained from
these tests.

There are many other probes which offer the possibility of testing chameleon theories astrophysically. For example, observations of circularly-polarized starlight in the wavelength range $1-10^3$\AA~could be a strong indication of mixing between chameleons and photons~\cite{Burrage:2008ii}. A difference in the ratio of the electron to proton mass between laboratories on Earth and in space can be measured and would indicate the presence of a chameleon-like scalar particle~\cite{Levshakov:2010cw}. In~\cite{Burrage:2008ii,Levshakov:2010cw}, it was shown that
the scatter in the luminosity of astrophysical objects can be used to search for chameleons, particularly through observing active galactic nuclei.

\subsection{Astrophysical tests of galileons}
An important property of galileons is that black holes carry no galileon hair~\cite{Kaloper:2011qc,Hui:2012qt},\footnote{Whether this is true in general for kinetic/Vainshtein screened systems is a question of active research.} while stars of course couple to the galileon. This leads to an interesting observational signature: in the presence of an external linear gradient, an astrophysical black hole should be offset (possibly by an observable amount) from the center of its host galaxy~\cite{Hui:2012jb}. Cosmologically, the galileon-mediated force becomes important at late times and on large scales. This affects various linear-scale observables~\cite{Song:2007wd,Afshordi:2008rd,Lombriser:2009xg,Barreira:2014ija}, such as enhanced large scale bulk flows~\cite{Wyman:2010jp,Khoury:2009tk}, infall velocities~\cite{Zu:2013joa} and weak-lensing signals~\cite{Wyman:2011mp} as described below. See~\cite{Schmidt:2008tn,Oyaizu:2008sr,Oyaizu:2008tb,Khoury:2009tk,Schmidt:2009sv,Chan:2009ew,Wyman:2013jaa,Zhao:2010qy,Ferraro:2010gh,Li:2012by,Li:2013tda,Baldi:2013iza} for N-body simulations.

The situation is somewhat similar for pulsar tests of GR. The very presence of a field that acts as a Lorentz scalar opens up new channels of gravitational radiation in binary systems. Monopole and dipole radiation are in principle possible in these models, but are highly suppressed by both relativistic corrections and the Vainshtein mechanism which makes such tests uncompetitive compared to solar system tests, let alone cosmological ones~\cite{deRham:2012fw,deRham:2012fg,Chu:2012kz}. However the situation might be different for other models that exhibit a Vainshtein mechanism such as more general galileon models. In these models any breaking of spherical symmetry could lead to potentially large corrections to the solutions and thus to the emitted radiation. This remains to be explored in more depth---both at the theoretical and observational level \cite{deRham:2012fg}.

\subsection{Summary and observational prospects for astrophysical tests}
\subsubsection{Physical effects}
In summary,
screening mechanisms yield distinct physical effects in
many settings in nearby stars and galaxies; for example:  
\begin{itemize}

\item In both Vainshtein and Chameleon theories, 
enhancements velocities of unscreened galaxies of order ten percent---or tens of km/s---are expected (typical peculiar velocities of galaxies are a few hundred km/s).
 
\item Segregation of the different components of galaxies: The gas and stellar components respond differently to the external forces in Chameleon theories. As an example, the stellar disk may be warped relative to the HI disk. In Vainshtein theories, it is possible for a galaxy's central supermassive black hole to be displaced by up to 0.1 kpc from the stellar light, along the direction of the external force. 

\item Differential internal dynamics for different tracers. In Chameleon theories, the stellar disk may rotate more slowly than the gas disk. For dwarf galaxies with circular velocities below 100 km/s, systematic differences of 5-10 km/s  are expected.

\item Altered stellar evolution. Giant stars in particular move more rapidly through their evolutionary tracks in Chameleon theories due to the enhanced forces. The most distinct observable consequences of this are for distance indicators, in particular the comparison of cepheid variables and TRGB stars. Relative 
offsets of approximately 5\%  in the distance ladder are expected from these effects. 

\end{itemize}

Approaches to detecting these effects and distinguishing them from astrophysical sources have been widely discussed in the literature cited above. Figure \ref{fig:chameleonlimits} shows the various regimes and constraints from some of these tests. 
 One key element required for a convincing detection would be the creation of both screened (control) and unscreened samples of galaxies or other tracers, in order to extract environmental dependence. Correlating the observed effect with the direction of the scalar fifth force would rule out the primary sources of astrophysical uncertainty in the measurements. 

\begin{figure}[htb]
\begin{center}
\includegraphics[width=3.5in]{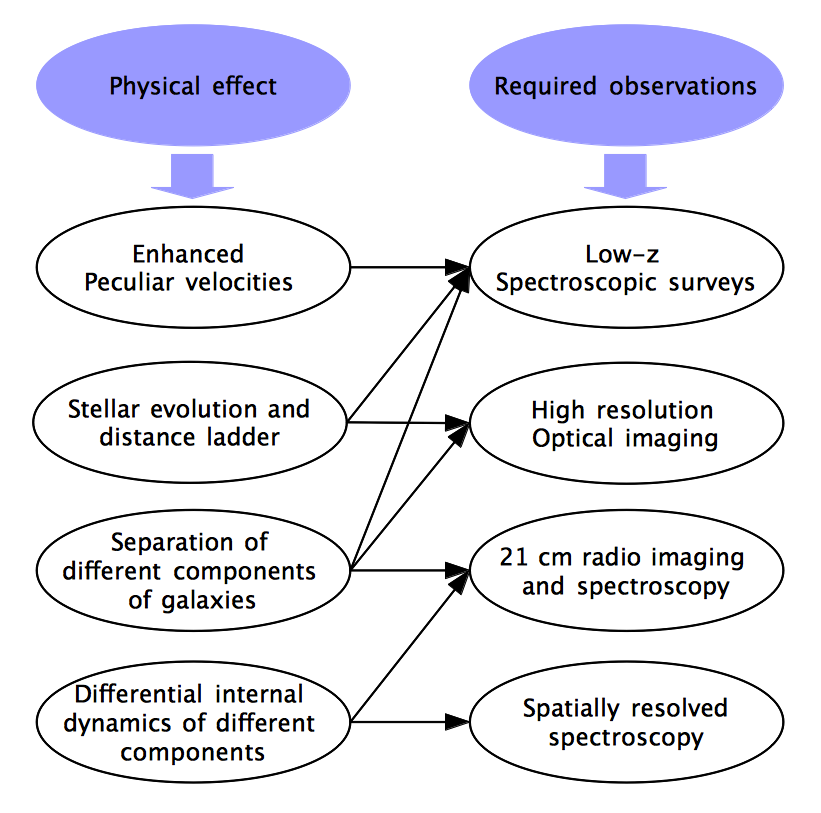}
\caption{\small Astrophysical effects of modified gravity mapped 
to the observations required to carry out the tests. It is clear that multiple 
types of telescope are needed for typical tests. However---as discussed in the text---the sample 
size needed is relatively modest and the galaxies are nearby, so a series of
mini-surveys are feasible with existing facilities. 
}
\label{figure:tests-observations}
\end{center}
\end{figure}

\subsubsection{Observational prospects}

Small scale tests both span a wide range of astrophysical environments and combine multi-wavelength data, often for a single test. However the sample size required for these tests is modest, typically on the order of hundreds of galaxies. Therefore, 
most of the promising tests can piggyback on already planned surveys or existing telescopes. 

The two key parameters of MG theories which are probed by astrophysical tests are the strength of the coupling of the 
scalar to matter and the range of the fifth force. 
For chameleon or symmetron theories, the program described below would improve the bound on the self-screening parameter (related to the range of the force) to $\sim 10^{-8}$. This is the smallest potentials accessible with dwarf galaxies, and would thereby eliminate the entire remaining parameter space of astrophysically-testable models. For theories which screen using the Vainshtein mechanism, the astrophysical tests are less mature, but are possibly more promising, since current limits on theories come only from cosmological scales and are specific to the DGP model. A summary of the some of the astronomical resources needed for the tests described above follows, adapted from~\cite{Jain:2013wgs}. Figure \ref{figure:tests-observations} shows physical effects and their best suited observational probes. 

\begin{itemize}
\item {\it Low redshift spectroscopy}:
Spectroscopic observation of samples of  
galaxies at low redshift is essential for a variety of tests of gravity.  
These observations provide a detailed map of the nearby universe and can be used to extract 
the velocity field traced by galaxies. Surveys of this type can be carried out as part of
cosmological BAO surveys, and in addition, by spectroscopic cameras on other telescopes. Multi-slit or fiber spectrographs on a number of telescopes are suitable for this purpose.

\item {\it Spatially resolved spectroscopy}: Different tracers of galaxies with internal dynamics (stars, ionized and neutral gas clouds) require specialized observations. Some data on dwarf galaxies exist---for the principal purpose of understanding the nature of the inner dark matter profile---which may also be examined for differential motions. Current galaxy surveys using optical telescopes focus on early type, or massive, galaxies. A larger survey of low mass galaxies, for example using the MANGA spectrograph of SDSSIV, could increase the sample by more than an order of magnitude and help address systematics by sampling environments that have both screened and unscreened galaxies. These observations may need
to be supplemented with 21cm radio observations to compare the motions of stars
and gas (see below). 

\item {\it High resolution imaging}:
Imaging surveys require less modification than other gravity tests since targeting is not required: they
 cover large contiguous areas and a wide range
of redshifts. Nevertheless, samples of low-$z$ galaxies will need to be observed with
higher resolution, which is feasible from space or using adaptive optics telescopes from the ground.
Tests that use the distance ladder rely on Hubble Space Telescope observations of dwarf galaxies, in particular
for determining cepheid and TRGB distances. Moving forward, adaptive optics capabilities can extend the range of application some of these techniques ({\it e.g.}, cepheids). Finally, wide field and highly sensitive narrow band imagers can effectively measure properties such as the planetary nebulae luminosity function in local dwarf galaxies. 

\item {\it Radio observations}:
Observations with improved resolution at 21cm are required
to test modified gravity predictions for neutral Hydrogen gas. The spatial resolution needed is up to an order of magnitude 
better than the recent ALFALFA survey, which focused primarily on velocity information. 
These radio observations would be compared to optical data for the
stellar disks. Samples of the order of a hundred galaxies would be sufficient to conduct useful tests and can be obtained with instruments such as the eVLA. 
\end{itemize}
Carrying out specific tests requires two additional elements: 
 
 \begin{itemize}
 \item
In order to quantify the screening levels in  chameleon and Vainshtein theories, a $3d$ map of the gravitational field in the  
 nearby universe is essential. Given a tracer of mass distribution---typically the optical light distribution along with information on the velocity field---it is possible to determine the gravitational field in both Einstein gravity and a modified theory. 
 This map-making exercise for the nearby universe (out to hundreds of Mpc) 
 is an essential part of the observational program.  It is feasible using currently-planned wide 
 area surveys along with additional spectroscopy (described above). 

 \item 
 To determine the level of screening for a MG model, it is necessary to solve the nonlinear equations of the theory. This is a challenging numerical exercise, but some progress has been made for the $f(R)$ and DGP models; we expect future work will be needed to explore new models. This involves 
 collaboration between gravitational theorists and numerical cosmologists and will become more important 
 as detailed connections between specific observations and theories are made via numerical 
 realizations of the survey geometry. This work will also prepare us to determine what a detection of new physics on small scales would mean for cosmological tests and {\it vice versa}. 
  \end{itemize}
 
Thus, while a variety of astrophysical gravity tests have been explored using archival data to date, the next big advances will involve a set of mini-surveys specially designed for such tests. Such surveys will rely on the instrument capabilities described above (most of which already exist) and can be carried out over a 5 year timeframe. In addition to the core data analysis, numerical studies that link theories to detailed predictions for the local universe are necessary and will need to be coupled to the data analysis. Thus the program of astrophysical gravity tests requires a close coupling between theorists, numerical astrophysicists and observers. 

\section{Cosmological tests}
\label{section:cosmology}

Cosmological probes of gravity may be broadly classified as follows:
\begin{enumerate}
\item Tests of the consistency between expansion history and the growth of structure. A discrepancy in the equation of state parameter, $w$, inferred from the
two approaches can signal a breakdown of the GR-based smooth dark energy cosmological
paradigm.
\item Detailed measurements of the linear growth factor across different scales and redshifts.
\item Comparison of the mass distribution inferred from different probes, in particular redshift
space distortions and lensing. The latter is a compelling test of modified gravity, 
since the same test can be carried out over many different scales. Table~\ref{gravitytable} summarizes
the current status of the different tests and future prospects for improvement. 
\end{enumerate}

Similar to tests of dark energy, 
the linear regime offers the twin advantages of ease of prediction and interpretation,
together with a degree of robustness against astrophysical systematics, which typically cannot
alter structure formation on scales larger than 100 Mpc. Further, specific MG models can
produce both scale and redshift-dependent growth that in principle distinguishes them
from dark energy models, given the same expansion history. Cosmological parameter analyses to test MG theories have been performed by several authors
\cite{Saini:1999ba,Song:2007da,Afshordi:2008rd,Schmidt:2009am,Rapetti:2009ri,Giannantonio:2009gi,Lombriser:2009xg,Bean:2010zq,Daniel:2010ky,Zhao:2010dz,Dossett:2011tn}. Specific studies to determine the cosmological impact of galileon theories and to constrain model parameters appear in~\cite{Nesseris:2010pc,Appleby:2011aa,Appleby:2012ba,Weinberg:2012es,Neveu:2013mfa,Barreira:2013jma,Barreira:2014jha,Neveu:2014vua}.

The main limitation
for cosmological tests of gravity, common to those of dark energy, is that the signal is typically small, characterized in many cases by percent level deviations, and may be degenerate with other parameters or physical effects such as scale-dependent galaxy bias. An important area for future work 
is to make explicit the connection between large-scale tests discussed in this section 
and the small scales ones discussed above. Verifying MG effects  over many decades in length scale and environment will enable a more robust approach to 
gravity tests. 

\subsection{Formalism for growth of perturbations}

Structure formation in modified gravity in general differs from that in pure general relativity~\cite{White:2001kt,Sealfon:2004gz,Shirata:2005yr,Stabenau:2006td,Sereno:2006qu,Shirata:2007qk,Wang:2007fsa,Koyama:2009me,Skordis:2005xk,Skordis:2005eu,Dodelson:2006zt,Lue:2004rj,Knox:2005rg,Ishak:2005zs,Koyama:2005kd,Koivisto:2005yc,Amarzguioui:2005zq,Zhang:2005vt,Bean:2006up,Koivisto:2006ie,Li:2006vi,Song:2006ej,Li:2007xn,Linder:2005in,Huterer:2006mva,Uzan:2006mf,Caldwell:2007cw,Amendola:2007rr,Afshordi:2008rd,Khoury:2009tk,Wyman:2010jp}. Perturbative calculations at large scales have shown that promise in connecting predictions in these theories with
observations of large-scale structure (LSS). Nevertheless, in practice carrying out robust tests of MG is
challenging. Broadly, two approaches have been taken, one isto constrain the parameters of a particular model by working out in detail its  predictions for structure growth,  the other is to define 
effective parameters in the spirit of the PPN formalism used
to test GR in the solar system. This Parameterized Post-Friedmann (PPF) framework attempts to parallel the PPN approach, see {\it e.g.}, \cite{Hu:2007pj,Baker:2012zs}.  
Both approaches have their limitations, but we shall see below that there has been much recent progress. In particular in Sections~\ref{quasistaticsec} and~\ref{section:PCA} we describe how two functions of scale and time capture the impact of scalar-tensor MG theories for observational purposes.  

\begin{figure}[htb]
\label{cosmoregimes}
\centering
\includegraphics[width=14cm]{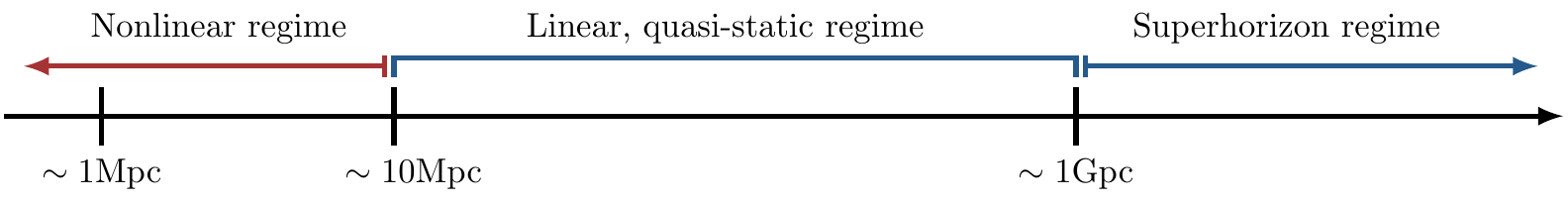}
\caption{\small Scales on which linear perturbation theory is applicable for cosmological perturbations. Outside the horizon ($k\to0$) perturbations are well-described by~\eqref{eqn:superhorizon}. Between $\sim$10--500 Mpc, linear perturbation theory can be used confidently and temporal gradients can be ignored. Below  $\sim$10 Mpc, non-linear gravitational effects cannot be ignored. For scale, the virial radius of a $M\sim 10^{15} M_\odot$ galaxy cluster is $\sim$1 Mpc.
}
\end{figure}

There are three qualitative regimes for the growth of perturbations: the long-wavelength
superhorizon regime, the quasi-static Newtonian regime of where
growth is linear, and  the small scale regime, where things are nonlinear. These three regimes are illustrated in Figure~\ref{cosmoregimes} .
The quasi-static Newtonian regime 
is valid when motions  are non-relativistic and at length scales sufficiently
smaller than the horizon. In this regime (discussed in the next
sub-section) it is sufficient to describe perturbations using the linearized fluid equations in expanding coordinates. In the nonlinear regime---though gravity is still in the weak field limit---density fluctuations
are no longer small. In addition the density/potential fields may
couple to additional scalar fields introduced in modified gravity
theories. Therefore, the nonlinear regime is the hardest to describe in any general way, because the nature of the coupling to scalar fields model dependent. However, this regime may very well be the most discriminatory for some
theories owing to the fact that there can be a rich phenomenology which ranges from galaxy cluster to solar system and laboratory scales. 

\subsubsection{Metric and fluid perturbations}
In order to study the growth of cosmological fluctuations, we first have to briefly introduce the formalism for cosmological perturbation theory. The classic reference is~\cite{Bardeen:1980kt}, and excellent expositions of this formalism can be found in~\cite{Mukhanov:1990me,Ma:1995ey,Dodelson:2003ft,Mukhanov:2005sc,Weinberg:2008zzc,Baumann:2009ds} Our starting point is the flat FLRW metric:
\be
\rd s^2 = \bar g_{\mu\nu}\rd x^\mu\rd x^\nu= -\rd t^2+a^2(t)\rd\vec x^2~,
\ee
which we assume is sourced by a perfect fluid component with stress tensor
\be
\bar T^\mu_{~\nu} = (\bar\rho+\bar P)\bar u^\mu \bar u_\nu+\bar P \delta_\nu^\mu~,
\label{backgroundstresstensor}
\ee
where $\bar u_\mu \bar u^\mu = -1$ and the fluid has equation of state $w = \bar P/\bar\rho$.
We now allow the metric to fluctuate away from this background FLRW geometry
\be
g_{\mu\nu} = \bar g_{\mu\nu}+ h_{\mu\nu}~.
\ee
If we define
\be
h_{\mu\nu} =
\left(
\begin{array}{cc}
h_{00} & a(t) \tilde h_{0i}\\
a(t) \tilde h_{i0} & a^2(t) \tilde h_{ij}
\end{array} 
\right)~,
\ee
the line element can be written as
\be
\rd s^2 = -(1-h_{00})\rd t^2+2 a(t) \tilde h_{i0} \rd x^i \rd t + a^2(t)\left(\delta_{ij}+\tilde h_{ij}\right)\rd x^i\rd x^j~.
\ee
The way we have defined things, indices are raised and lowered with the background metric.
The background FLRW geometry is rotationally invariant, and so we may decompose the 10 degrees of freedom in $h_{\mu\nu}$: $h_{00},  \tilde h_{0i}$ and $\tilde h_{ij}$ into scalar, divergence-less vector and transverse-traceless tensor components~\cite{Mukhanov:1990me,Ma:1995ey,Dodelson:2003ft,Mukhanov:2005sc,Weinberg:2008zzc,Baumann:2009ds}:
\begin{align}
h_{00} &= -2\psi~,\\
\tilde h_{0i} &= \partial_i F+G_i~,\\
\tilde h_{ij} &= -2\phi\delta_{ij}+\partial_i\partial_j B+\partial_iC_j+\partial_jC_i+D_{ij}~,
\end{align}
where $\partial_i G^i = \partial_i C^i = \partial_i D^{ij} = D^{i}_{~i} = 0$. Under a coordinate transformation, $\delta_{\xi}x^\mu = \xi^\mu$,  the fluctuation around the background metric transforms as~\cite{Mukhanov:1990me,Ma:1995ey,Dodelson:2003ft,Mukhanov:2005sc,Weinberg:2008zzc,Baumann:2009ds}
\be
\delta_\xi h_{\mu\nu} = \pounds_\xi \bar g_{\mu\nu} = -\bar g_{\lambda\mu}\partial_\nu\xi^\lambda-\bar g_{\lambda\nu}\partial_\mu\xi^\lambda-\xi^\lambda\partial_\lambda\bar g_{\mu\nu}~.
\ee
Splitting the gauge parameter as $\xi^\mu = (\alpha, \xi^i)$, we therefore deduce
\begin{align}
\delta_\xi h_{00} &= 2 \dot\alpha~,\\
\delta_\xi h_{i0} &= -a^2\dot \xi_i+\partial_i\alpha~,\\
\delta_\xi h_{ij} &= -a^2(\partial_j\xi_i+\partial_i\xi_j)-2a\dot a \alpha\delta_{ij}~,
\end{align}
We can further decompose $\xi_i = \partial_i\xi+\bar \xi_i$ with $\partial_i \bar\xi^i = 0$. From this, we deduce that the perturbations in the scalar-vector-tensor (SVT) decomposition transform as~\cite{Mukhanov:1990me,Ma:1995ey,Dodelson:2003ft,Mukhanov:2005sc,Weinberg:2008zzc,Baumann:2009ds}
\begin{align}
\delta_\xi \psi &= -\dot\alpha~,~~~~~~~~~~~~~~~~~~~~~~\delta_\xi F =\frac{1}{a}\alpha -a\dot\xi~;\\
\delta_\xi \phi &=  H \alpha~,~~~~~~~~~~~~~~~~~~~~~~\delta_\xi B = -2 \xi~;\\
\delta_\xi G_i &= -a\dot{\bar \xi}_i~,~~~~~~~~~~~~~~~~~~~~\delta_\xi C_i  = -\bar\xi_i~;\\
\delta_\xi D_{ij} &= 0~.
\end{align}
Often times, the best way to deal with this gauge ambiguity is to work with {\it gauge-invariant} quantities. For scalar perturbations, the {\it Bardeen variables} are gauge invariant~\cite{Bardeen:1980kt}
\be
\Psi_{\rm B} = \psi - \frac{\rd}{\rd t}\left[a^2\left(\frac{\dot B}{2}-\frac{F}{a}\right)\right]~;~~~~~~~~~\Phi_{\rm B} =\phi - a^2H\left(\frac{\dot B}{2}-\frac{F}{a}\right)~.
\ee
In what follows, we will be concerned only with scalar perturbations, where the line element takes the form
\be
\rd s^2 = -(1+2\psi)\rd t^2+2 a(t) \partial_i F \rd x^i \rd t + a^2(t)\Big((1-2\phi)\delta_{ij}+\partial_i\partial_jB\Big)\rd x^i\rd x^j~,
\ee
and further we will work in {\it Newtonian gauge}, where we use our gauge freedom to set 
\be
B = F = 0~.
\ee 
Note that in this gauge the remaining metric perturbations $\psi$ and $\phi$ coincide with the Bardeen variables, so we write the metric as
\begin{equation}
{\rm d}s^2 = -(1+2\Psi){\rm d}t^2 + (1-2\Phi)a^2(t) {\rm d}{\vec x}^2\,.
\label{eqn:metric}
\end{equation}
This form for the
perturbed metric is fully general for any metric theory of
gravity, other than our having excluded vector and tensor perturbations
(see~\cite{Bertschinger:2006aw} and references therein for justifications). 
Note that $\Psi$ corresponds to the Newtonian potential for
the acceleration of particles, and that in GR
$\Phi=\Psi$ in the absence of anisotropic stresses.  

A metric theory of gravity relates the two potentials above to the
perturbed energy-momentum tensor. We consider perturbing the stress tensor~\eqref{backgroundstresstensor} from its background perfect fluid form:
\be
\delta T^\mu_{~\nu} = (\delta\rho+\delta P)\bar u^\mu \bar u_\nu +(\bar\rho+\bar P) \delta u^\mu \bar u_\nu+(\bar\rho+\bar P)\bar u^\mu \delta u_\nu+ \delta P\delta_\nu^\mu +\sigma^\mu_{~\nu}~,
\ee
where $\sigma^\mu_{~\nu}$ is the {\it anisotropic stress}, which is absent at the background level, and which can be chosen to be transverse to $\bar u^\mu$ and traceless: $\sigma^0_{~\nu} = \sigma^\mu_{~\mu} = 0$. Using the fact that $ g_{\mu\nu}u^\mu u^\nu = -1$ we can deduce that $\delta u^0 = -\psi$. We then write $\delta u^i = v^i/a$. So we have
\be
u^\mu = \left( 1-\psi, v^i/a\right)~;~~~~~~~~~~~~u_\mu = \left(-1-\psi, a v_i+a\tilde h_{0i}\right)~.
\ee
From this, we may compute the components of the perturbed stress tensor:
\begin{align}
T^{0}_{~0} &= -(\bar\rho+\delta\rho)~,\\
T^i_{~0} &= -(\bar\rho+\bar P)v^i/a~,\\
T^0_{~i} &= (\bar\rho+\bar P)(av_i+a\tilde h_{0i})~,\\
T^i_{~j} &= (\bar P+\delta P)\delta^i_j+\sigma^i_j~.
\end{align}
Under a coordinate transform, the fluid perturbations shift as
\be
\delta_\xi \delta T^\mu_{~\nu} = \pounds_\xi \bar T^\mu_{~\nu} = \bar T_{~\nu}^\lambda\partial_\lambda\xi^\mu-\bar T^\mu_{~\lambda}\partial_\nu\xi^\lambda-\xi^\lambda\partial_\lambda\bar T^\mu_{~\nu}~,
\ee
which leads to the transformation rules (again writing $\xi^\mu = (\alpha, \xi^i)$)
\begin{align}
\delta_\xi \delta\rho &= -\dot{\bar\rho}\alpha~;~~~~~~~~~~~~~~~~~\delta_\xi \delta P = -\dot{\bar P}\alpha~,\\
\delta_\xi v^i &= a\dot\xi^i~;~~~~~~~~~~~~~~~~~~~~\delta_\xi \sigma^i_j = 0~.
\end{align}
As before, we can perform an SVT decomposition on the fluid perturbations:
\be
v^i  = \delta^{ij}\partial_j v+\tilde v^i~;~~~~~~~\sigma_{ij} = \left(\partial_i\partial_j-\frac{1}{3}\nabla^2\delta_{ij}\right)\sigma +\frac{1}{2}(\partial_i \sigma_j+\partial_j\sigma_i)+\tilde \sigma_{ij}~,
\ee
where $\partial_i\tilde v^i = \partial_i\sigma^j = \partial_i\tilde\sigma^{ij}=\tilde\sigma^i_{~i}=0$. Two important gauge-invariant combinations of metric and fluid perturbations are the co-moving curvature perturbation, ${\cal R}$, and the curvature perturbation on uniform density hypersurfaces, $\zeta$, given by
\be
{\cal R}= -\phi + \dot a(F+v)~;~~~~~~~~~~~~~~~\zeta = \psi - \frac{\delta\rho}{\dot{\bar\rho}}~.
\ee

We now introduce variables to
characterize the density and
velocity perturbations for a fluid, which we will use to
describe the evolution of matter perturbations. These variables are sufficient to analyze and interpret most cosmological observations, which lie in the quasi-static Newtonian regime discussed below. 
The density fluctuation $\delta$ is given by
\begin{equation}
\delta({\vec x},t) \equiv \frac{\rho({\vec x},t) -
  {\bar\rho(t)} } {\bar\rho(t)} \,,
\label{eqn:delta}
\end{equation}
where $\rho({\vec x},t)$ is the density and ${\bar\rho(t)}$ is the cosmic
mean density. The second fluid variable we introduce is the divergence of the
peculiar velocity, which is given by
\begin{equation}
\theta_v \equiv -\nabla_j T_0^{\;j}/(\bar{\rho}+\bar P)={\vec \nabla} \cdot {\vec v}~.
\end{equation}
Choosing
$\theta_v$ instead of the vector $\vec v$ implies that we have assumed
$\vec v$ to be irrotational ($\tilde v^i=0$). This approximation is sufficiently
accurate in the linear regime for minimally coupled MG models.

In principle, observations of large-scale structure can directly 
measure the four variables we introduced above: 
the scalar potentials $\Psi$ and $\Phi$, and the density and velocity
perturbations $\delta$ and $\theta_v$. 
These variables are the key to distinguishing 
MG models from Einstein gravity plus dark energy. Each has both scale and
redshift dependence, so it is worth noting which variables are probed by different observations and at what
scale and redshift. It is
convenient to work in Fourier space, defining, for example,
\begin{equation}
\hat\delta(\vec k,t) = \int \rd^3 x \ \delta(\vec x,t) \
e^{-i {\vec k} \cdot{\vec x}} \,.
\label{eqn:FT}
\end{equation}
Length scales $\lambda$, correspond to
a statistic such as the power spectrum at wavenumber $k=2\pi/\lambda$. 
From here on, we will work exclusively with the Fourier space quantities 
and drop the hat ($\hat{~}$) symbol for convenience. 

It is possible to calculate the evolution of perturbations in the linear regime. We follow the formalism and notation of~\cite{Ma:1995ey}, except we use physical time $t$ rather than conformal time. We are
interested primarily in the evolution of perturbations after decoupling, so we will neglect radiation and neutrinos as sources of perturbations. 

\subsubsection{Superhorizon perturbations}

The superhorizon regime is the most constrained regime and consequently is the simplest to
describe. In~\cite{Bertschinger:2006aw}, it was pointed out that any metric theory
of gravity that also obeys the equivalence principle must
satisfy a universal evolution equation for metric perturbations. In
conformal Newtonian gauge, assuming adiabatic initial conditions, this
evolution is given by:
\be
\ddot{\Phi} - \frac{\ddot{H}}{\dot{H}}\dot{\Phi} + H\dot{\Psi} +
\left(2\dot{H} - \frac{H\ddot{H}}{\dot{H}}\right)\Psi = 0\,. 
\label{eqn:superhorizon}
\ee
This above equation is equivalent to eq.~(7) of \cite{Hu:2007nk}, which uses a different time variable and the opposite sign convention
for $\Psi$. Treating the ratio of metric potential $\Phi/\Psi$ as a constant
parameter (while this has the virtue of simplicity, it is not justified for generic MG theories), one can solve this equation for a given
background solution $H(t)$. The integrated Sachs--Wolfe effect discussed below extends to  
very large scales and is currently one of the few probes of the superhorizon regime. 

\subsubsection{Quasi-static Newtonian regime}
\label{quasistaticsec}

In the following, we will for the most part make the approximation that
 motion is non-relativistic and restrict ourselves to sub-horizon length
scales. Further, one can self-consistently neglect time
derivatives of the metric potentials relative to spatial
gradients. These approximations will be referred to as the quasi-static,
Newtonian regime. 
The evolution of 
density (or velocity) perturbations can be described by a single second
order differential equation using the linearized fluid equations:  
\be
\ddot{\delta}+2 H \dot{\delta} + \frac{k^2 \Psi}{a^2}
 = 0 \,. 
\label{eqn:lingrowth}
\ee
With $\delta(\vec k,t)\simeq \delta_{\rm initial}(\vec k) D(k,t)$, 
we can use the Poisson equation to substitute for $\Psi$ in terms of $\delta$. Here we write the Poisson equation in two forms, the first is the 
``standard'' Poisson equation
\be
k^2 \Phi=-4\pi G_{\rm N}\bar{\rho} \delta \, ,
\label{eqn:Poisson1}
\ee
 and the second is with the sum of
potentials on the left-hand side. This is convenient for describing 
lensing and the ISW effect. In terms of the generalized gravitational
``constant'' $\tilde{G}$ we then have 
\be
k^2(\Psi+\Phi)=-8\pi \tilde{G}(k,t)\bar{\rho} \delta \, .
\label{eqn:Poisson}
\ee
These equations lead to the expression for the linear growth 
factor $D(k,t)$: 
\be
\ddot{D}+2 H \dot{D} - \frac{4 \pi \tilde G}{(\Phi/\Psi)}
\bar{\rho} \ D = 0 \,. 
\label{eqn:growth}
\ee
From this equation one sees how the combination of $\tilde G$ and $\Phi/\Psi$ alters the linear growth factor. If these parameters have a scale dependence, then even the linear growth factor $D$ becomes scale dependent--- which is a feature not seen in smooth dark energy models. We can also use the relations given above to obtain the linear growth factors for the velocity and the potentials from $D$. 
The growth factor for the velocity divergence is given  in the following sub-Section, while the Poisson equation determines the evolution of the potentials. 

Let us introduce the dimensionless functions 
\be
\mu  \equiv \tilde G/G_{\rm N}, \ \  \gamma \equiv \Phi/\Psi \ ,
\label{eqn:gravparameters}
\ee
where $\mu$ and $\gamma$ are in general functions of space and time. In
 \cite{Zhao:2008bn} and elsewhere, it was argued that a specific scale dependence in these MG parameters is
 expected for scalar-tensor theories. In Fourier space, provided the interaction can be expressed in the
Yukawa form, one expects a correction to the growth factor that is
quadratic in wavenumber $k$. 
With considerations of locality and general covariance, and under the quasi-static approximation, physically acceptable forms of $\mu(a,k)$ and $\gamma(a,k)$ correspond~\cite{Silvestri:2013ne} to ratios of polynomials in $k$, which are even in models with purely scalar extra degrees of freedom, and of second order in most viable models. 
These express $\gamma(a,k)$ and $\mu(a,k)$ as:
\begin{eqnarray}
\label{eqn:gamma}
\gamma &=&  {p_1(a)+p_2(a) k^2 \over 1+p_3(a) k^2}\ , \\
\mu &=& {1 + p_3(a) k^2 \over p_4(a) + p_5(a) k^2} \ .
\label{eqn:mu}
\end{eqnarray}
This leaves five free functions of time to be constrained by data, and we discuss how this is done in practice below in Section \ref{section:PCA}. Note that other choices for the initial functions can be useful in interpreting quasi-static regime observables, in particular, treating $G$ and $\tilde{G}$ defined above. 

\begin{figure}
\centering
\includegraphics[width=6in]{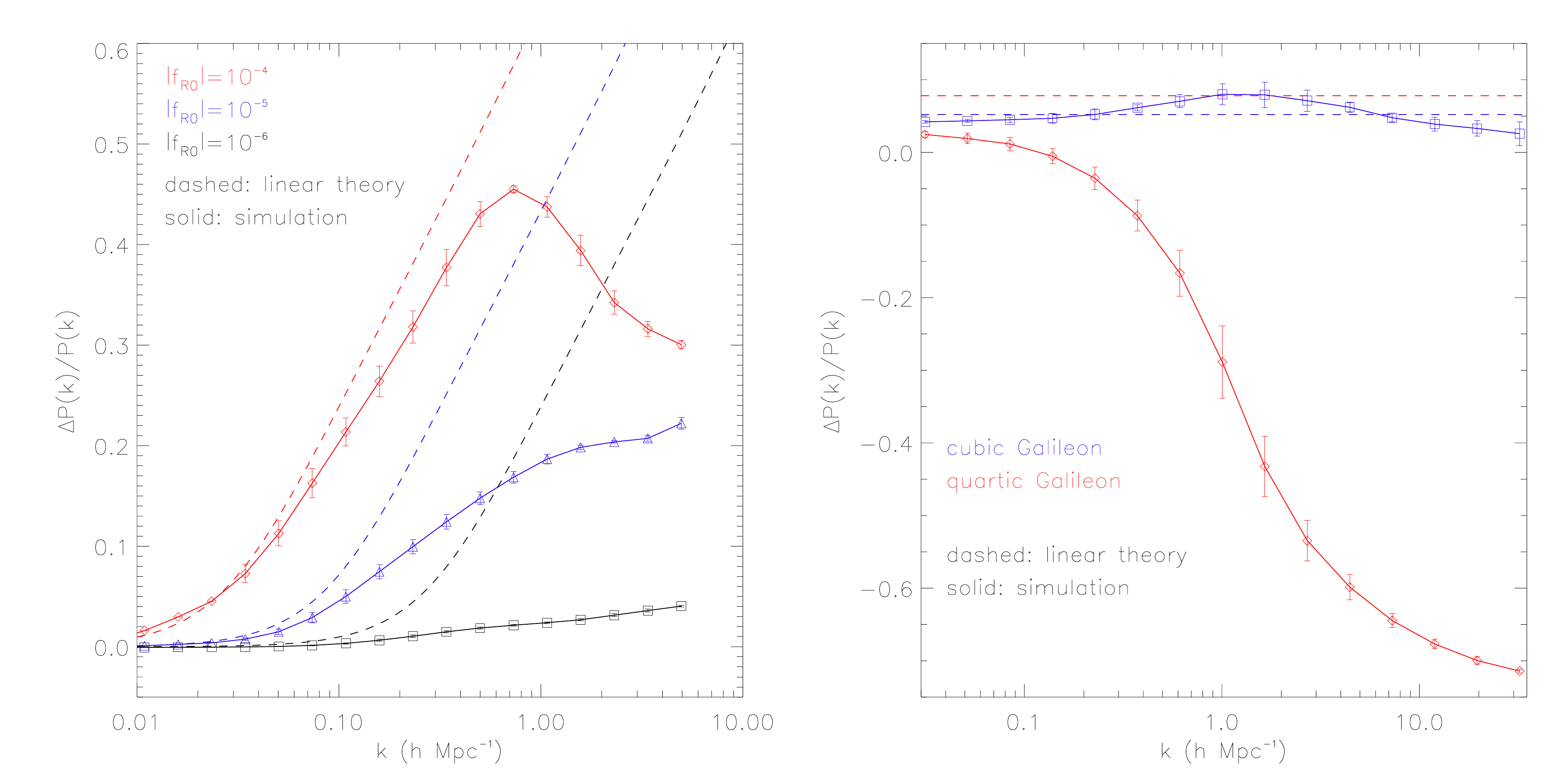}
\caption{\small Power spectra for $f(R)$ (left panel) and galileon (right panel) theories. 
The fractional deviations from $\Lambda$-CDM are shown for the 
present day linear and nonlinear power spectra~\cite{Barreira:2014zza}.  
At high-$k$ (small scales), nonlinear gravitational clustering and the
screening of massive halos alters the power spectrum. 
}
\label{fig:power}
\end{figure}

\subsection{Cosmological observables}

We will use the power spectra of various cosmological observables to describe their scale dependent two-point correlation functions. For example, 
the 3-dimensional power spectrum of the density contrast $\delta(k,z)$  is defined by
\be
\langle \delta({\vec k}, z) \delta({\vec k'}, z) \rangle = 
(2 \pi)^3 \delta^{(3)}({\vec k + k'}) P_{\delta\delta}(k,z)~,
\label{eqn:powerdef}
\ee
where we have traded time variable for observable redshift
$z$. The power spectra of perturbations in other quantities are
defined in an analogous fashion. We will denote the  cross-spectra of two
different variables with appropriate subscripts. For example,
$P_{\delta\Psi}$ denotes the cross-spectrum of the density perturbation $\delta$ 
and the potential $\Psi$.  

Figure \ref{fig:power} shows the linear and nonlinear power spectra 
$P_{\delta\delta}(k,z)$ for $f(R)$ and  two galileon models~\cite{Barreira:2014zza}. The dashed curves show the fractional departures of the linear power spectrum
to $\Lambda$CDM. The symbols show measurements from N-body simulations.  
The strong scale dependence is evident, with significant deviations 
 at wavenumbers $k\gsim 0.1 h$/Mpc.  At higher wavenumbers screening effects suppress MG deviations. 

\subsubsection{Weak gravitational lensing} 

Lensing observables are the result of coherent deflections of light by
mass concentrations. For the metric of eq.~\eqref{eqn:metric}, 
the first order perturbation to a photon trajectory is given by
(generalizing for example eq.~(7.72) of \cite{Carroll:2004st}):  
\begin{equation}
\frac{d^2 x^{(1)\mu}}{d\lambda^2} = -q^2 \vec{\nabla}_\perp(\Psi+\Phi) \ ,
\end{equation}
where $q$ is the norm of the tangent vector along the unperturbed path
and $\vec{\nabla}_\perp$ is the gradient transverse to the unperturbed
path.  This leads to the deflection angle formula
\be
\alpha_i = -\int\rd s~\partial_i(\Psi+\Phi) \,, 
\label{eqn:deflection}
\ee
where $s=q\lambda$ is the
path length and $\alpha_i$ is the $i^{\rm th}$ component of the deflection
angle (which is a two-component vector on the sky). 
Since all lensing observables are obtained by taking derivatives
of the deflection angle, they necessarily depend only on the
linear combination $\Psi+\Phi$ (to first order in the potentials). 
For example, the convergence is given by
the line-of-sight projection: 
\begin{equation}
   \kappa({\bf \theta}) = \frac{1}{2} \int_0^{z_s} \frac{\rd z}{H(z)}
    \frac{r(z) r(z_s,z)}{r(z_s)} \nabla_{{\bf 
    \theta}}^2 (\Psi + \Phi)\, ,
   \label{eqn:kappa}
\end{equation}
where we have taken the sources to lie at redshift $z_s$. 

For the purposes of testing gravity, observables that rely directly on the change in
energy or direction of photons are distinct from those that measure
the clustering or dynamics of tracers such as galaxies or galaxy
clusters, which move non-relativisitically. We summarize the
primary observables that provide tests of gravity on cosmological
scales in this sub-section, and consider galaxy and cluster scale
tests in the following subsection. We  follow the treatment 
of Jain \& Zhang~\cite{Jain:2007yk}.

\begin{figure}[thb]
\centering
\includegraphics[width=13cm]{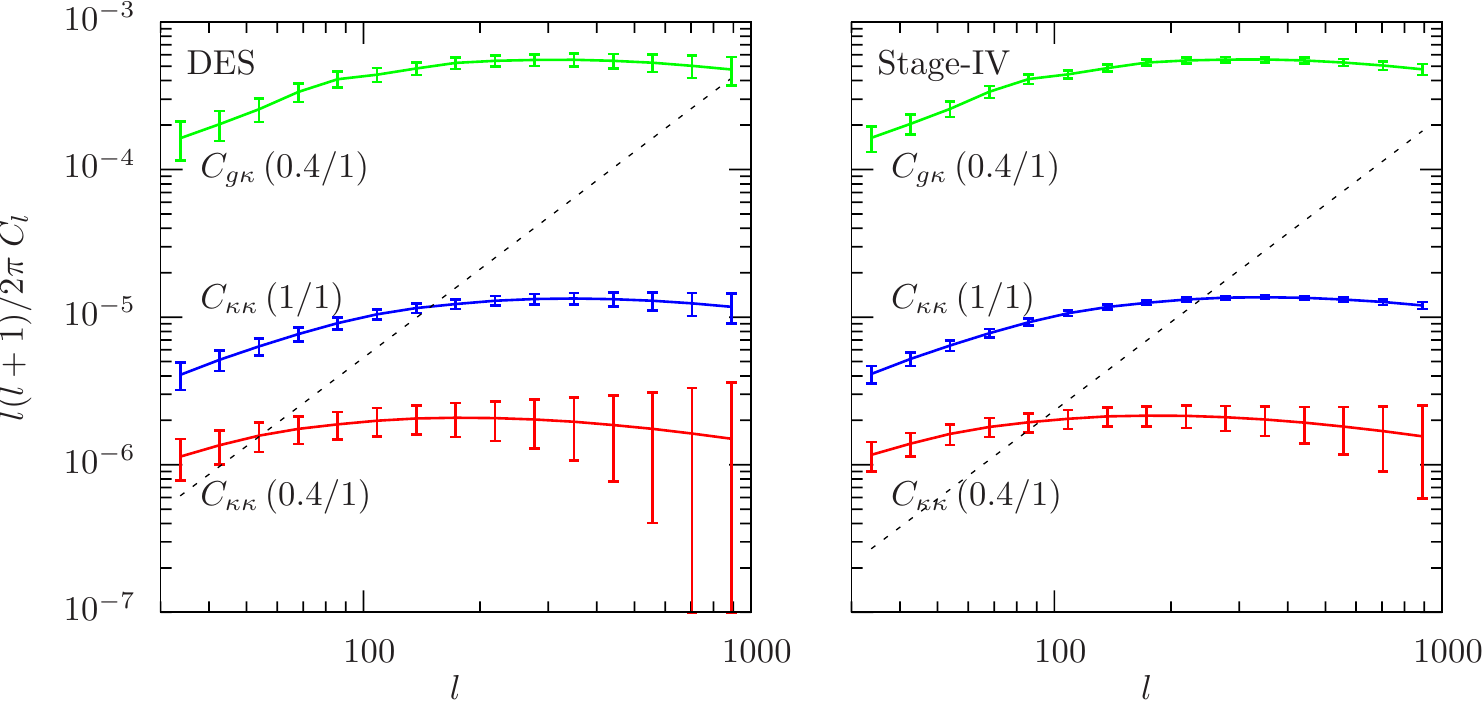}
\caption{\small 
\label{fig:Ckk} 
Examples of shear-shear and galaxy-shear power spectra for 
the DES (left panel) and a Stage-IV survey similar to 
LSST (right panel) \cite{Guzik:2009cm}. 
The upper (green) curves show the galaxy-shear cross power spectrum 
$C_{g\kappa}$, with foreground galaxies at
$z=0.4$ and background galaxies at $z=1$. The lower two curves show  
the shear-shear power 
spectrum  $C_{\kappa \kappa}$ with two choices of redshift bins as
indicated. Error bars include both the sample variance and shape noise 
for the two surveys. 
The contribution to $C_{\kappa \kappa}$ from shape noise for $z=1$ is 
shown separately as well (dashed lines). 
}
\end{figure}

The primary cosmological observables used in weak lensing are the two-point 
correlations of the observed shapes of galaxies and the
cross-correlation of foreground galaxies with the shapes of background
galaxies.  The metric potentials are related to the mass distribution  by
\eqref{eqn:Poisson}
so the lensing power spectra can  be expressed in terms of the 
three-dimensional mass power spectrum $P_{\delta \delta}(k,z)$. 
In the small-sky-patch limit  
the Limber approximation \citep{Limber1953} gives 
\begin{eqnarray}
\label{cls1}
C_{\kappa_i \kappa_j}(l) & = & \frac{9}{4} \Omega_{\rm m}^2 H_0^4   
  \int_0^{\infty} \frac{\rd z}{H(z)a^2} \,   
\zeta^2(k,z) 
  P_{\delta \delta}(k,z) W_L(z,z_i) W_L(z,z_j),
\end{eqnarray}
where the function $\zeta$ contains the modified gravity parameters, and is the Fourier space counterpart of
$\tilde G/G_{\rm N}[(\Phi+\Psi)/\Phi]$. 
The lensing weight function $  W_L(z,z_k) $
depends on the geometry and the redshift 
distribution of lensed galaxies. 
The three-dimensional wavenumber $k$ is given by $k=l/r(z)$. 
By binning the galaxy distribution in redshift \citep{Hu:1999ek}, a number of auto and cross-spectra can 
be measured. The redshift dependence of these lensing spectra  carries
information about the growth of structure that can test gravity
theories. The galaxy-shear cross-spectrum $C_{g\kappa}$ can be defined
in a similar way to $C_{\kappa\kappa}$: it is proportional to $b\ \Omega_{\rm m}
\zeta$, where $b$ is the galaxy bias parameter.  $C_{g\kappa}$ is
easier to measure and can be used to test gravity as discussed below. 
Examples of the two lensing spectra are shown in Figure \ref{fig:Ckk} for 
two different survey parameters \cite{Guzik:2009cm}. 
Statistical errors are shown for the 
different power spectra---it is
evident  that if systematic errors can 
be controlled, upcoming surveys will provide percent level
measurements \cite{Hoekstra:2008db}.  
Simpson et al.~\cite{Simpson:2012ra} have presented a comprehensive analysis of gravity tests using shear-shear correlations measured from the CFHTLenS survey. This represents the state of the art in using weak lensing, but ongoing surveys will reduce the statistical errors by a significant factor~\cite{Simpson:2012ra}.

Lensing observables probe the sum of the metric potentials---this follows from the
geodesic equation applied to photons and is therefore true for any metric theory of gravity. Moreover, the relation of the sum of metric potentials to the mass  distribution is very close to that of GR in scalar-tensor theories that we have considered
\cite{Bruneton:2007si}. Since the Einstein frame is obtained through a
conformal transformation which cannot alter null geodesics, 
the scalar field does not directly alter the geodesics of light rays. 
Thus the deflection angle formula \eqref{eqn:deflection} and 
Poisson equation \eqref{eqn:Poisson}  in the form given above are 
essentially unaltered in these scalar-tensor gravity theories. (However, couplings of the form $\sim\partial_\mu\phi\partial_\nu\phi T^{\mu\nu}$, which arise in theories of massive gravity, can affect lensing~\cite{Wyman:2011mp}.)
Masses of halos
inferred from lensing are the true masses.  It is therefore more useful  to combine lensing with other observations of large-scale structure to carry out robust tests of gravity, as shown in Sections~\ref{crosscorrsec} and~\ref{lensingdynammasssec} below.  
Tests of gravity that
rely solely on lensing measurements  can constrain specific models; alternatively 
lensing measurements with multiple  redshift bins can probe the  the growth of 
structure ({\it e.g.}, \cite{Heavens:2007ka}) which can be compared to the predictions of GR. 

\subsubsection{CMB lensing and the ISW effect}

The CMB power spectrum at angular wavenumber $l$ 
is given by a projection along the line of sight: 
\begin{equation}
C_{TT}(l) = \int \rd k \int \rd z \ F_{\rm CMB}(k, l, z)\ j_l[k r(z)] \,,
\end{equation}
where $r$ denotes the comoving angular diameter distance and 
the spherical Bessel function $j_l$ is the geometric term
through which the CMB power spectrum depends on the distance to the last 
scattering surface. The function $F_{\rm CMB}$ 
combines several terms describing the primordial power
spectrum and the growth of the potential {\it up to} last scattering. 
We will regard $F_{\rm CMB}$ as 
identical to the GR prediction since we do not invoke MG in the early
universe.
In combination with Big Bang Nucleosynthesis, the CMB power spectrum provides a measurement of $G$, and a test of the Friedman equation, at the 10\% level for times 
up to last scattering. 

The CMB anisotropy 
does receive  contributions at redshifts below last scattering, in
particular due to the 
integrated Sachs--Wolfe (ISW) effect~\cite{Sachs:1967er} and from lensing due to 
mass fluctuations along the line of sight~\cite{Lewis:2006fu}.
In the presence of dark
energy or due to modifications in gravity, gravitational potentials
evolve in time and produce a net change in the energy of CMB photons: 
\begin{equation}
\left.\frac{\Delta T}{T}\right|_{\rm ISW}=-\int
\frac{\rd(\Psi+\Phi)}{\rd t} \frac{a(z) \rd z}{H(z)} \,.
\end{equation}
The ISW effect---like gravitational lensing---depends on and probes  the
combination $\Psi+\Phi$. The ISW signal is overwhelmed by the primary CMB at
all scales except for a bump it produces at the largest scales in 
the CMB power
spectrum. Since cosmic variance limits the information available at such large scales, the ISW effect  is more effectively  measured by 
cross-correlation with tracers of large scale structure at low redshift (to which smaller scale modes also contribute). The
resulting cross-correlation signal is a projection of 
$P_{g(\dot\Psi+\dot\Phi)}\left( k,\chi\right)$, the
cross-power spectrum of $(\dot\Psi+\dot\Phi)$ and galaxies (or other
tracers of the LSS such as quasars or clusters).  By cross-correlating
the CMB temperature with the galaxy over-density $\delta_g$, the 
ISW effect has been detected at about the $ 5\sigma$ confidence level
\cite{Cabre:2006qm,Pietrobon:2006gh,Giannantonio:2006du,Ho:2008bz}. This detection provides independent evidence
for the evolution of gravitational potentials, as expected in dark energy models given the prior of a spatially flat universe and GR. It has also provided useful constraints on MG theories as discussed below. 

CMB lensing has been detected via its smearing of the CMB power spectrum and, independently, through the non-Gaussian features it produces in the CMB temperature maps. The latter enable reconstruction of the lensing deflection field, which is dominated by mass fluctuations at redshifts of about 2~\cite{vanEngelen:2012va,Das:2011ak, Ade:2013tyw}.
CMB lensing thus probes  mass fluctuations at higher redshifts ($z\sim 1-5$) than galaxy surveys, providing a valuable addition to tests that use the growth of structure. The measurement accuracy is at the 10\% level currently and is expected to improve significantly in the coming years. 
Cross-correlations of the CMB with foreground tracers due to lensing 
have also been measured~\cite{vanEngelen:2012va, Das:2011ak, Ade:2013tyw,Munshi:2014tua}.
Several applications for MG tests are possible for the future, e.g. Munshi et al.~~\cite{Munshi:2014tua} have shown how 3-point correlations in the CMB temperature maps induced by the ISW effect and CMB lensing can be used to constrain MG theories.
(For a recent review, see~\cite{Namikawa:2014xga}.

\subsubsection{Redshift space galaxy power spectra}

\begin{figure}[htb]
\centering
\includegraphics[width=4in]{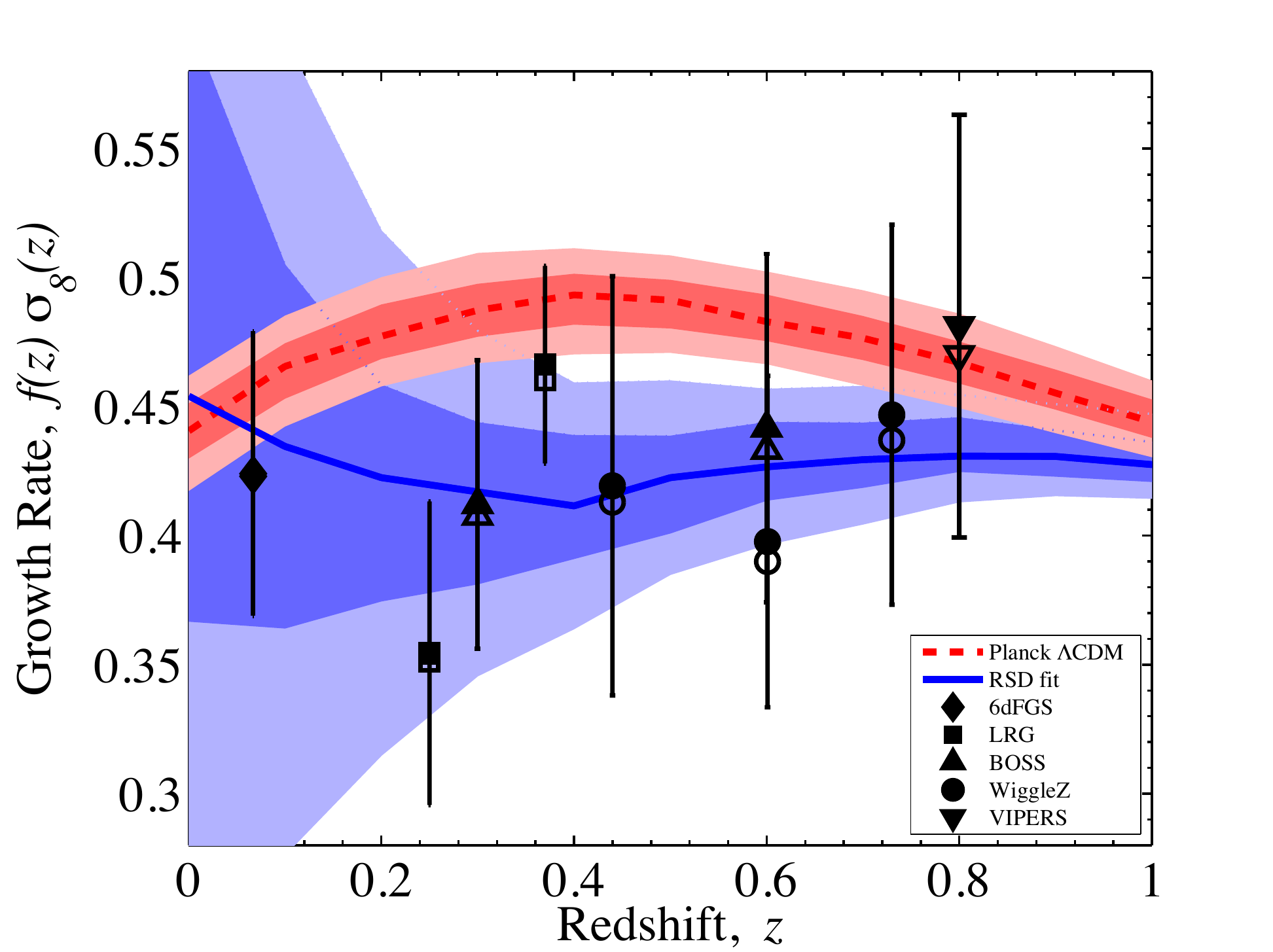}
\caption{\small  
The measured growth rate inferred from redshift space galaxy power spectra is shown for several different surveys. The quantity $f\sigma_8$  is plotted at the redshift of the different survey measurements~\cite{Macaulay:2013swa}. Comparison of the fit to the data, shown in blue, with the Planck best fit in pink shows that low redshift measurements prefer weaker growth, though at low significance so far.
 }
\label{fig:growth}
\end{figure}

On large scales, galaxy clustering depends on the linear growth factor  $D(t)$ given by 
\eqref{eqn:lingrowth}, which determines the 
clustering of matter and is dependent only upon the Newtonian potential $\Psi$. 
The resulting change in the mass power spectrum depends on how much $\Psi$ deviates 
from its GR value and the duration of time this deviation lasts. The results for 
$f(R)$ and DGP models at $z=0$ are shown in
Figure \ref{fig:power}. Redshift surveys of galaxies provide statistical measurements of clustering over 
Mpc-Gpc scales. Cosmological inference from measured galaxy power spectra depends on corrections 
for galaxy bias, the relationship of galaxies to the underlying mass distribution. In the linear regime, it is a 
good approximation to write the galaxy over density as $\delta_g = b \delta$, where $b$ is the bias parameter. 
Several approaches have been developed to mitigate the uncertainty introduced by our limited understanding 
of galaxy bias. We provide next a brief account of the main methodology and recent results inferred from galaxy clustering. 

A relatively direct probe of $\Psi$ at a given
redshift is provided by the distortions of galaxy clustering
in redshift space.  Redshift space distortions in the galaxy power
spectrum arise from motions  along the line-of-sight---on large
scales these are sensitive to  the linear growth factor for
$\theta_v$, denoted $D_{v}$ here, which is related to $D$, the linear density growth factor, via the continuity equation as: 
\begin{equation}
D_{v}\propto a\dot{D}=a f  
H D \,; \ \  f \equiv \frac{\rd\log D}{\rd\log a}\,. 
\label{eqn:loggrowth}
\end{equation}
The line-of-sight component of peculiar velocities causes 
the observable redshift-space power spectrum  $P^{(s)}_{gg}(k, \mu_k)$
to be `squashed' along the line of sight on 
large scales (in the linear regime) and to produce pronounced
`finger-of-God' features on small scales 
(in the nonlinear regime) \citep{Kaiser:1987qv, Hamilton1998}. 
The directional dependence of $P^{(s)}_{gg}$ is given by
$\mu_k \equiv k_{\parallel}/k$, 
which depends on the angle between a wave vector ${\bf k}$ and
the line-of-sight direction. 
Although the picture is more complicated in reality,  
it is a good approximation to decompose the
redshift space power spectrum in terms of three isotropic 
power spectra relating the 
galaxy overdensity  $\delta_g$ and peculiar velocities ${\bf v}$:
the galaxy power spectrum $P_{gg}(k)$,
the velocity power spectrum $P_{vv}(k)$ 
and the cross power spectrum $P_{gv}(k)$ as follows 
\cite{Kaiser:1987qv,Scoccimarro:2004tg}
\be
P^{(s)}_{gg}(k, \mu_k) = \Big[ P_{gg}(k) + 2 \mu_k^2 P_{gv}(k) + \mu_k^4 P_{vv}(k) \Big] F(k^2 \mu^2_k \sigma^2_v)\,, 
        \label{psz:decomp}
\ee
where the term $F(k^2 \mu^2_k \sigma^2_v)$ describes non-linear 
velocity dispersion effects. 

The angular dependence  in the above equation allows us to
obtain the component power spectra from 
$P^{(s)}_{gg}$.  The real space power spectrum of galaxies, $P_{gg}(k)$, 
is the easiest to extract from the measurements, but its interpretation requires knowledge of 
galaxy bias. The pure velocity 
power spectrum $P_{vv}(k)$ has the largest error bars, while 
the cross-spectrum $P_{gv}(k)$ can be estimated more easily. Using the full angular 
dependence both bias and the growth rate of clustering can be estimated. 

Figure \ref{fig:growth} shows the inferred growth rate
(the variable $f$ in the figure is defined in~\eqref{eqn:loggrowth} and $\sigma_8$ is the amplitude of mass fluctuations at $8$ Mpc$/h$).
More recently, in a series of papers, the BOSS survey has presented measurements of growth and tests of GR-based dark energy models~\cite{Beutler:2013yhm, Sanchez:2013tga, Samushia:2013yga, Chuang:2013wga}. The error bars are somewhat smaller but the basic picture is as presented in Figure \ref{fig:growth}: the measured growth rate is 
smaller than extrapolating from the CMB fluctuations 
using the GR-based dark energy models. Needless to say this would be fascinating if it holds up, especially since generic scalar-tensor gravity theories enhance growth relative to GR. We discuss below the prospects for improving on these measurements.

\subsubsection{Model independent approaches to cosmological analysis}
\label{section:PCA}

\begin{figure}[htb]
\centering
\includegraphics[width=9cm]{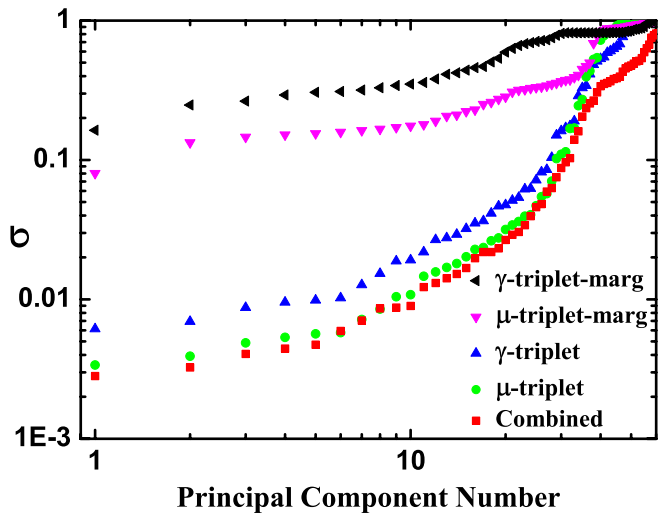}
\caption{\small 
\label{fig:PCA} 
The uncertainties associated with the eigenmodes of the MG functions $\mu$ and $\gamma$. These are the three functions of time that describe $\mu$ and $\gamma$ when the functional form of their scale-dependence is fixed as described in the text \cite{Hojjati:2013xqa}. Three sets of plots are shown:  when all other functions are kept fixed (unmarginalized), when all other functions are marginalized over, and when all the functions are considered in a combined analysis that does not trace back the origin of a potential deviation to one or another function. In all cases the standard cosmological parameters have been marginalized over. An LSST-like survey is assumed. Note that in the case of marginalization, even such a powerful survey is unable to provide constraints stronger than the 10\% level. Figure reproduced from~\cite{Hojjati:2013xqa}.
}
\end{figure}

There is no clear favorite methodology for testing modified gravity theories in the linear regime. The simplest extension of the dark energy program is to add parameters to the growth rate that are sensitive to MG.  However, MG theories generically predict scale and time dependent modifications to the growth rate.  Two functions of time and scale are sufficient to  describe the general dynamics of linear scalar perturbations. But even surveys to be carried out over the next decade will not have the power to provide useful constraints on parameters that are allowed to freely vary with redshift and scale. 
The problem is made more tractable by applying constraints on the scale dependence of MG parameters as discussed above (in particular the parameter $\gamma$ and $\mu$  in   
equations~\eqref{eqn:gamma} and \eqref{eqn:mu}). This takes us from two functions of time and scale to five functions of time, which are significantly easier to constrain from observations (see also \cite{Bellini:2014fua,Baker:2012zs,Hojjati:2013xqa}). 

An approach to observational analysis is to apply a principal component analysis (PCA) to  MG functions that are treated as unknown
\cite{Huterer:2002hy,Zhao:2009fn,Hojjati:2011xd,Asaba:2013xql}. 
Principal component analysis then tells us which observables are more likely to be sensitive to the MG functions; or, given a set of observables, which features of MG will be better constrained, and at which scales or times. It accounts for degeneracies among the functions used to describe MG and cosmological parameters, thus giving a realistic picture of constraints on specific parameters in the context of a full cosmological analysis of the data. 
A PCA approach to the five MG functions of time that determine $\gamma$ and $\mu$, using the theoretical prior of \eqref{eqn:gamma} and \eqref{eqn:mu}, was recently presented in \cite{Hojjati:2013xqa}. The authors consider the  observable modes of MG. (See Figure \ref{fig:PCA}. Eigenvalues of the combined eigenmodes of all $p$'s are also shown.) There is a trade-off between two effects: on one side, imposing the theoretical prior on $k$-dependence reduces the number of degrees of freedom  and the covariance between parameters; on the other side, imaging surveys like LSST are not as sensitive to the $z$-dependence as they are to the $k$-dependence of the modified growth functions.  

Thus a model-independent analysis of observational data to test for generic MG theories remains a work in progress. An interesting recent development is the parameterized post-Friedmann approach implemented  in~\cite{Baker:2013hia},
 where, as above,  the spatial dependence is shown to be restricted to quadratic terms in $k$ for a broad class of theories (see also~\cite{Battye:2013aaa,Battye:2013ida}, and for other approaches to the PPF formalism, see~\cite{Hu:2007pj,Hu:2008zd,Fang:2008sn}).
The authors then show how constraints on theory parameters can be obtained from measurements of the expansion history and the growth of structure using just the solution for the background evolution of the theory. On a practical note, the public Boltzman--Einstein solver CAMB has been upgraded by~\cite{Hu:2013twa} to include MG predictions using an effective field theory approach (see also \cite{Dossett:2011tn}).

\subsubsection{Combining lensing and dynamical cross-correlations}
\label{crosscorrsec}

In theories of modified gravity, the Newtonian potential, $\Psi$, generally differs from its value in GR. It is enhanced stronger by a factor of $4/3$ in a certain regime 
in $f(R)$ gravity (on scales between those of chameleon effects and the Compton
wavelength of the $f_R$ field). This corresponds to the ratio $\Phi/\Psi=1/2$ with the sum 
$\Psi+\Phi$ remaining unaltered as discussed above. Therefore, for a given 
mass distribution, significant force enhancements can occur. For DGP gravity, similar 
force enhancements occur for the normal branch.  

The inequality of metric potentials has been exploited to construct different combinations of the information from weak lensing and
redshift space galaxy clustering as  tests of gravity 
\citep{Zhang:2007nk,Acquaviva:2008qp,Song:2008xd,Zhao:2008bn,Zhao:2009fn,Guzik:2009cm}.  
A comparison of lensing and dynamical cross-power spectra ($C_{g\kappa}$ and $P_{gv}(k)$, which is proportional to $b D_v$) was proposed in~\cite{Zhang:2007nk} as a model independent test of gravity. This test is
in principle immune to galaxy bias, at least to first order, and can also
overcome the limitation of sample variance on large scales. Thus it can
be applied in the linear regime provided both multi-color imaging (for lensing) and
spectroscopy (for dynamics) are available for the same sample of galaxies. 
With appropriate redshift binning, these spectra can
constrain the ratio of metric potentials. 

A recent measurement was performed in \cite{Reyes:2010tr}, 
comparing galaxy-velocity and galaxy-shear
cross-correlations from the
SDSS.  They estimated $E_G\approx 0.4$, 
consistent with its value in GR, given by $E_G = \Omega_{\rm m}(z=0)/\beta(z)$, where
$\beta$ is the logarithmic rate of growth parameter introduced above
in eq.~\eqref{eqn:loggrowth}. 
The $20$\% level measurement of $E_G$ by \cite{Reyes:2010tr} spans 
scales of 10-50 Mpc at redshift $z\simeq 0.3$. Smaller scale versions of this comparison 
around galaxy and cluster halos are discussed below.

\subsection{The halos of galaxies and galaxy clusters}

\subsubsection{Lensing and dynamical masses}
\label{lensingdynammasssec}

\begin{figure}[thb]
\centering
\includegraphics[width=5in]{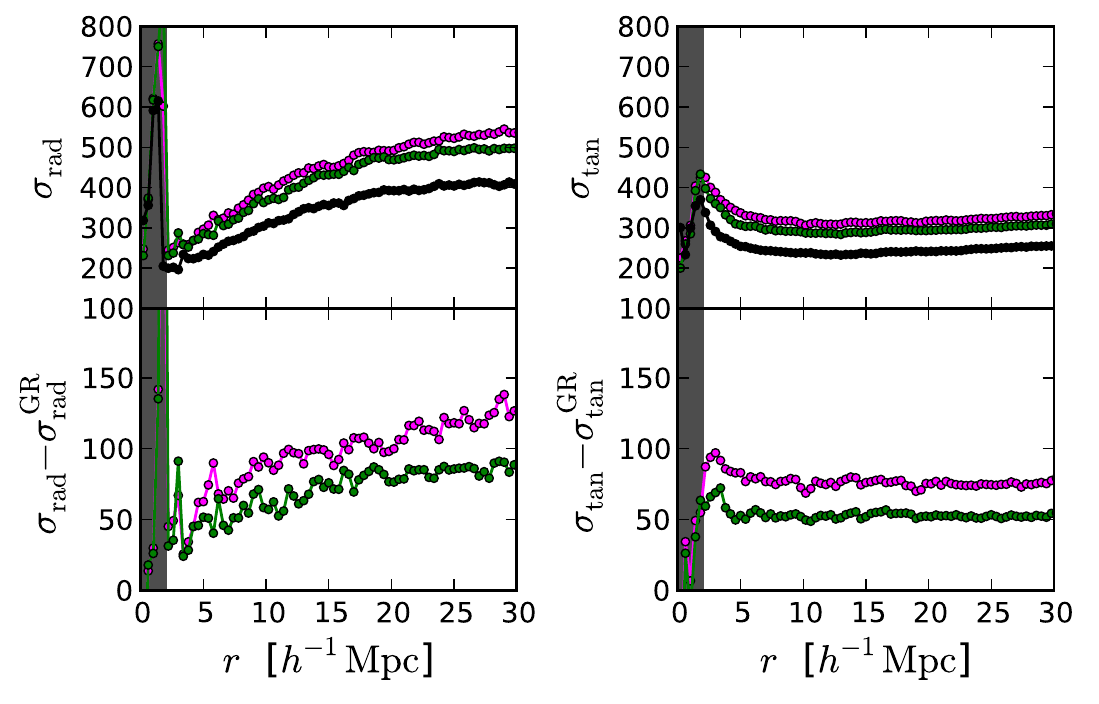}
\caption{\small  
Enhanced velocities around clusters in modified gravity theories. For two galileon models the radial and tangential velocity dispersion profile is compared to the predictions of GR using N-body simulations~\cite{Zu:2013joa}.
 }
\label{fig:infall}
\end{figure}

The comparison of the  lensing and dynamical masses of galaxies
and clusters is a smaller scale version of gravity tests that probe the inequality of the two metric potentials. This test is fairly unique to testing gravity, as it has
little information to add in the dark energy framework.
At least three kinds of tests are available:
the comparison of strong lensing with measured stellar
velocity dispersions in the inner parts of elliptical galaxies~\cite{Bolton:2005nf},
the virial masses of halos from weak lensing and dynamics~\cite{Schmidt:2010jr},
and the infall region that extends to ten or more times the virial radius~\cite{Reyes:2010tr}.
The latter two tests are feasible only for massive clusters or using stacked
measurements of large samples of galaxies binned in luminosity or another observable
that serves as a proxy for halo mass.  Figure \ref{fig:chameleonlimits} 
includes the tests described above as part of a wider set of tests of gravity.

In general the scale force causes the dynamical masses of halos inferred from the 
virial theorem or hydrostatic equilibrium  can be significantly larger than the lensing
(or true) masses (see, for example, \cite{Jain:2007yk,Schmidt:2010jr}).  However, the 
force modifications 
can depend on halo mass and environment, so it is not straightforward
to use dynamical and lensing masses to infer the maximal deviation in $\Phi/\Psi$. 
Figure \ref{fig:infall} shows measurements from N-body simulations of the 
velocity dispersion outside halos~\cite{Zu:2013joa}. 
The comparison with GR for two MG models shows
significant deviations that are worth pursuing with observations. 

Constraints on the Newtonian potential $\Psi$ on small scales are obtained using 
dynamical probes, typically involving galaxy or
cluster velocity measurements. 
On sub-Mpc scales, the Virial theorem for self-gravitating systems in equilibrium 
can be used to constrain $\Psi$ in galaxy 
and cluster halos. Velocity tracers for galaxies include stars and neutral Hydrogen
gas within the halos and satellite galaxies that orbit the outer parts of halos. For galaxy
clusters the tracers are member galaxies and the X-ray emitting hot gas, which is also
mapped using the Sunyaev Zel'dovich (SZ) effect. For relaxed clusters the hot gas is 
assumed to be in hydrostatic equilibrium within the gravitational potential of the halo. 

An interesting  test of gravity has been carried out on galactic scales. In \cite{Bolton:2006yz} and \cite{Schwab:2009nz} a combination of strong lensing observations in SDSS galaxies and the dynamics of stars were used to constrain the ratio of metric potentials. The resulting ratio was found to be consistent with unity, to better than 10\%. While current models of MG do not predict a deviation well inside the virial radii of large galaxies, where the observations were made, it is worth noting the consistency of this relatively accurate measurement with GR. 

\subsubsection{Halo abundances and profiles}

N-body simulations are used to study the nonlinear regime, i.e. once 
the linearized equations for the growth of
perturbations break down. For MG theories, the simulations incorporate 
the coupling of the density
field to a scalar field such as $f_{R}$ for $f(R)$ models. A number
of papers have  reported simulations that include such a
coupling for both $f(R)$ and DGP gravity \cite{Khoury:2009tk,Oyaizu:2008tb,Oyaizu:2008tb,Schmidt:2009sg,Schmidt:2009sv,Schmidt:2009yj,Scoccimarro:2009eu,Chan:2009ew}. 

\begin{figure}[htb]
\centering
\includegraphics[width=6in]{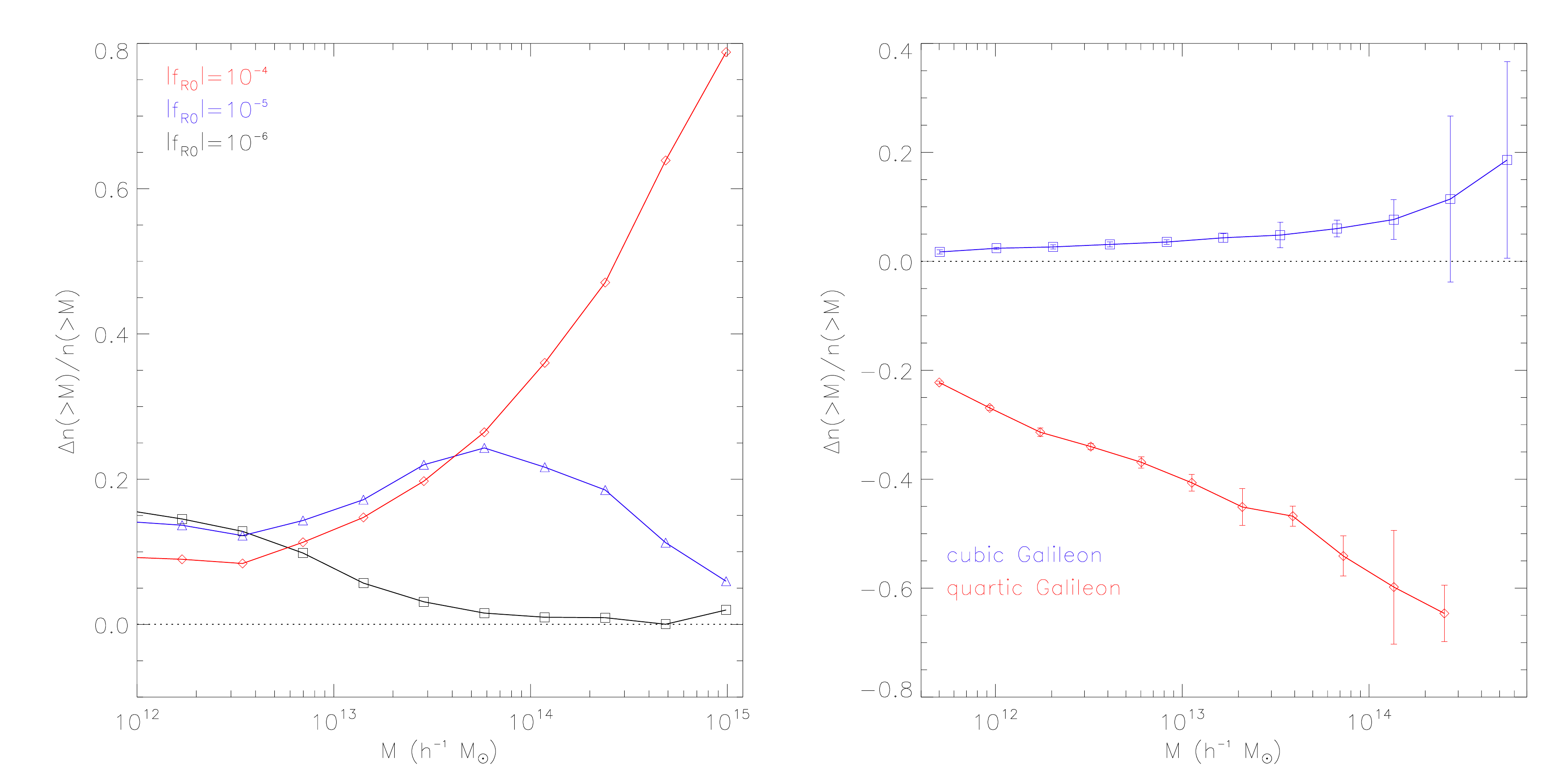}
\caption{\small Halo mass functions for $f(R)$ (left panel) and galileon (right panel) models \cite{Barreira:2014zza}. 
The fractional deviations from $\Lambda$-CDM are shown. The models are the same as in Figures~\ref{fig:power}. See text for details on the use of the mass function as a gravity test. 
 }
\label{fig:massfunction}
\end{figure}

Two essential products from simulations are the nonlinear mass power spectrum 
and the mass function of galaxy and cluster sized halos \cite{Oyaizu:2008tb,Schmidt:2009sg,Schmidt:2009sv,Schmidt:2009yj,Scoccimarro:2009eu,Chan:2009ew,Martino:2008ae}. 
For specific models these predictions allow for comparisons to data. Figures~\ref{fig:power} and 
\ref{fig:massfunction} show the measured power spectra and mass functions from 
N-body simulations for $f(R)$ and  two galileon models~\cite{Barreira:2014zza}.
The simulations incorporate the transition from a
modified gravity regime to GR inside massive halos,  
due to the Vainshtein or chameleon mechanisms.  We have the possibility
of comparing these MG theories to observations in the large-scale regime via the power spectrum and  within and around
galaxy and cluster halos via halo profiles and the mass function. The latter are sensitive to the transition into the screened regime that depends on the screening mechanism. Thus the two regimes provide complementary probes. 

The halo model provides a simple analytical way to quantify the abundance of galaxy clusters as a probe of the growth factor. At some critical threshold $\delta_c$, an evolving perturbation 
gravitationally collapses and virializes to form a halo. Hence, the probability of forming a halo at a given redshift is equivalent to the probability that $\delta \geq
\delta_c$.  Assuming Gaussian random initial conditions, one then finds that the number of collapsed objects $N$ per unit mass $\rd M$ and
comoving volume element $\rd V$ is
\begin{equation}
{\rd N\over \rd M\rd V} = 
F(\sigma)\frac{\bar{\rho}}{M}
\frac{\rd\log\sigma^{-1}}{\rd M} \, ,
\end{equation}
where $\rho_M$ is the matter density in the Universe, and $\sigma^2$ is the
variance of the density perturbations evaluated at some mass scale $M$ (related to the spatial scale $R$ by $M=(4\pi/3) R^3 \bar{\rho}$). Here
$F(\sigma)$, the fraction of mass in collapsed objects, is calibrated from simulations, with its functional form originally motivated by the statistics of the initial Gaussian field of density fluctuations. Since $F(\sigma)$ depends only on the {\it linear} density field via  
$\sigma^2 = [D^2(a)/D^2(a_0)]\sigma_0^2$,  
 the abundance of galaxy clusters is explicitly dependent on the growth
history of the Universe. (See~\cite{Huterer:2013xky} for more details.)

The mass function of cluster sized halos in MG theories can show strong deviations from GR due to enhanced 
gravitational forces from the coupling with the scalar field. Figure \ref{fig:massfunction} shows the fractional deviations in the mass function for $f(R)$ and 
DGP models \cite{Oyaizu:2008tb,Schmidt:2009sg,Schmidt:2009sv,Schmidt:2009yj}. 
For $f(R)$ models significant departures from GR may occur if the scales involved in forming clusters are smaller than the Compton wavelength of the scalar $f_R$ but larger than the scale of chameleon effects that screen halos from the modified forces  and drive the theory to GR. 
Theoretical predictions require careful treatment of spherical collapse 
in MG theories as screening effects are dependent on the halo mass and the
environment \cite{Schmidt:2010jr}. 

For cluster masses approaching $10^{15} M_\odot$, the
deviations are significant, from enhancements of tens of percent for $f(R)$ to about a factor of
2 for DGP  models (note that the DGP model shown in Figure \ref{fig:massfunction} is 
the normal branch of DGP, which has enhanced forces, similar to $f(R)$ gravity). 
These departures are driven by the deviation in the linear
growth rate on a scale of $\sim 10$ Mpc, coupled with the exponential dependence of the mass function at the high mass end. In addition, the mass
inferred from  dynamical measurements differs from the true (lensing) mass 
for unscreened halos; this  amplifies the deviations in the mass function 
as shown by the upper set of curves in Figure \ref{fig:massfunction}. 
The resulting observational
constraints are summarized next.  

The abundance of galaxy clusters from X-ray observations has been used
to constrain the growth factor \cite{Rapetti:2009ri} and hence specific $f(R)$ models \cite{Schmidt:2009am}. The mass function can be significantly enhanced at large masses for $f(R)$ and DGP models 
(see discussion in the previous sub-section). Using
information about the mass function requires nonlinear regime model
predictions that include chameleon dynamics, hence the constraints are
specific to particular models. In \cite{Schmidt:2009am} the
Compton wavelength was constrained to be smaller than $\sim 50\ $Mpc, or equivalently, 
the present day field amplitude to be $f_{R0}<2\times 10^{-4}$ for
a particular $f(R)$ model.  A more recent analysis in \cite{Lombriser:2010mp} 
combines large-scale structure information with galaxy cluster
abundances to find comparable constraints.  

\section{Summary and outlook}
\begin{table*}[th]
\centerline{
\resizebox{17cm}{!}{%
\begin{tabular}{| c | c | c | c | }
\hline
Test &   Length Scale & Theories Probed  & Current Status and Prospects           \\
\hline
Growth~vs.~Expansion &
 100Mpc-1Gpc &
GR + smooth dark energy &
$10$\% accuracy (2-4\%$^1$)  \\
\hline
Lensing~vs.~Dynamical mass$^2$& 
0.01-100Mpc &
Test of GR &
20\% accuracy$^3$  (5\% ) \\
\hline
Astrophysical Tests& 
0.01AU-1Mpc & MG  Screening Mechanisms & $\sim$10\%  (Up to $10\times$  improvement)  \\
\hline
Lab and Solar System Tests & 1mm-1AU & PPN $\rightarrow$ MG parameters$^4$ & 
Constraints are model dependent. \\  
& &  & (Up to 10$\times$ improvement)  \\
\hline
\end{tabular}
}
}
\caption{\small Experimental tests of gravity and dark sector couplings, from mm to Gpc scales. A rough guide to the experimental accuracy is given, comparing current accuracy with improvements expected in the next decade (see text for details.) Footnotes: 1. Combined constraints from BAO, SN, WL, Clusters, RSD and CMB lensing. 2. The test can be done over a range of scales: using strong lensing and stellar velocities inside galaxies, to cosmological scales using cross-correlations. 3. On scales where MG signatures are expected~\cite{Reyes:2010tr}. 4. Also tests dark sector couplings.
\label{gravitytable}}
\end{table*}

We have described experimental probes of gravity that span laboratory to cosmological scales, and a diverse set of environments. Most of the tests are based on observations below redshifts  $z\sim 1$, although a test of the Friedman equation is provided by the early universe as well. 

Cosmological tests of gravity rely on combining probes of the expansion history with the growth of structure. In the literature so far the observations that have been most effectively used for the distance-redshift relation are: CMB, SNIa and BAOs; and for the growth of structure: the CMB power spectrum, ISW cross-correlations, galaxy and CMB  lensing, the redshift space power spectra of galaxies, and the abundance of galaxy clusters (in the nonlinear regime). These tests show consistency with GR although the growth rate is somewhat smaller than predicted by GR-based dark energy models. Specific models that have been constrained by observations included DGP and $f(R)$ models, which are tightly constrained, and galileon models, which are at early stages of being experimentally tested. 

On smaller scales constraints have been obtained via laboratory, solar system and astrophysical tests. 
Current tests of gravity find no indications of departures from
GR. The ratio of dynamical 
to lensing masses  have constrained the ratio of metric potentials at the 
ten  percent level. These tests are restricted
to narrow ranges in mass/length scale and redshift, but upcoming surveys will 
improve on the regimes tested and the precision of the tests. In the nearby universe 
stellar evolution and the dynamics and  morphology of different tracers of galaxies 
have been used to test for the scalar force in MG theories. The tightest experimental limits on chameleon and symmetron theories come from this class of tests. 
Astrophysical  tests are at an early stage of development, but they represent 
 significant progress in both theory and observational
analysis---a decade ago virtually no tests of GR were available on
astrophysical  scales.   

How do we integrate the information from this wide range of tests? And how do we plan a coordinated suite of tests for the future? These questions are the subject of ongoing research. We have given a flavor of the possibilities, but we can be sure that surprises await us---a single new experiment or idea may provide a constraint or discovery that has evaded all the other tests. This is in part due to the subtle ways in which screening mechanisms operate: the tracer and its environment can make a crucial difference in highlighting or suppressing the signature of modified gravity. That said, we summarize the current state of play briefly in Table \ref{gravitytable}. 

\bigskip\bigskip

\noindent{\bf\Large Acknowledgements:} We acknowledge helpful discussions with Lasha Berezhiani, Claudia de Rham, Benjamin Elder, Garrett Goon, Kurt Hinterbichler, Wayne Hu, Elise Jennings, Kazuya Koyama, Eric Linder, Ed Macaulay, Raquel Ribeiro, Jeremy Sakstein, Alessandra Silvestri, Andrew Tolley and Vinu Vikram. We thank Alex Barreira, Alireza Hojjati, Baojiu Li, Alessandra Silvestri, Gongbo Zhao and Amol Upadhye for their help with specific sections of the review. AJ was supported in part by the Kavli Institute for Cosmological Physics at the University of Chicago through grant NSF PHY-1125897, an endowment from the Kavli Foundation and its founder Fred Kavli, and by the Robert R. McCormick Postdoctoral Fellowship. BJ is partially supported by DOE grant grant DE-SC0007901. The work of MT is supported in part by the US Department of Energy and NASA ATP grant NNX11AI95G.

\newpage
\begin{appendix}
\addcontentsline{toc}{section}{Appendix}
\noindent
{\huge\bf Appendix}
\section{Dynamical relaxation---Weinberg's no-go theorem}
\label{Weinbergnogo}
\renewcommand{\theequation}{A-\Roman{equation}}
\setcounter{equation}{0} 

Here we consider an alluring idea---that the cosmological constant may be able to dynamically relax to a small value---and its obstructions. Most prominent among these is a celebrated {\it no-go} result of Weinberg~\cite{Weinberg:1988cp}. In essence, this result says that we cannot achieve anything by tying the value of the CC to the potential of some scalar field---we must fine-tune the potential just as much as we would have had to fine-tune the bare CC.

Here we follow Weinberg's original argument~\cite{Weinberg:1988cp} (see~\cite{Padilla:2010tj} for a different take on the theorem), and consider a set-up in which we have $N$ scalar fields coupled to gravity in any way we like\footnote{Actually, the theorem proved in~\cite{Weinberg:1988cp} is even more general, allowing the $\phi^I$ to be tensors as well, but here we will restrict to scalars.}
\be
S =  \frac{M_{\rm Pl}^2}{2}\int\rd^4x\sqrt{-g}R+S[\phi^I, g_{\mu\nu}]~,
\label{weinbergargact}
\ee
where $S[\phi^I, g_{\mu\nu}]$ can depend arbitrarily on $\phi^I, g_{\mu\nu}$ and their derivatives. We look for a solution where 
\begin{align}
\phi^I &= \bar\phi^I = {\rm constant}\\
g_{\mu\nu} &= \eta_{\mu\nu}~.
\end{align}
With such an ansatz, the Euler--Lagrange equations become very simple
\begin{align}
\label{nogoscalareq}
\frac{\delta{\cal L}}{\delta\phi^I}\bigg\rvert_{g_{\mu\nu};\phi^I ={\rm const.}} &= \frac{\partial{\cal L}}{\partial \phi^I}=0\\
\frac{\delta{\cal L}}{\delta g^{\mu\nu}}\bigg\rvert_{g_{\mu\nu};\phi^I ={\rm const.}} &= \frac{\partial{\cal L}}{\partial g^{\mu\nu}}=0~.
\label{nogogeq}
\end{align}
In order for a solution to~\eqref{nogoscalareq} and~\eqref{nogogeq} to be {\it natural} (as opposed to fine-tuned) we want the trace of the gravitational equation of motion to be satisfied automatically as a consequence of the scalar equations. Another way of saying this is that the trace of the metric equation of motion must be of the form:
\be
g^{\mu\nu}\frac{\partial{\cal L}}{\partial g^{\mu\nu}}+\sum_I f^I(\phi)\frac{\partial{\cal L}}{\partial \phi^I} = 0~.
\label{tracescalareom}
\ee
Demanding this equation be satisfied is equivalent to demanding a particular symmetry of the Lagrangian~\cite{Weinberg:1988cp}. To see this, note that the variation of the action~\eqref{weinbergargact} is
\be
\delta S =\int\rd^4 x\left( \frac{\delta{\cal L}}{\delta g^{\mu\nu}}\delta g^{\mu\nu}+\sum_I\frac{\delta {\cal L}}{\delta \phi^I}\delta\phi^I\right)~.
\ee
If we consider the variations
\be
\delta g^{\mu\nu} = \epsilon g^{\mu\nu}~;~~~~~~~~~\delta\phi^I = \epsilon f^I(\phi)~,
\label{weinbergconformal}
\ee
then~\eqref{tracescalareom} implies that the action is invariant under this symmetry when the fields are taken to be constant. If we start with a Lagrangian invariant under~\eqref{weinbergconformal}, if it admits a solution $\bar\phi^I = {\rm const.}$ with $\frac{\partial{\cal L}}{\partial \phi^I}\big\rvert_{\phi=\bar\phi} = 0$, then the gravitational equation will be satisfied. However, this turns out to be impossible to arrange without some degree of fine tuning.

We rewrite the $N$ fields in terms new scalars $\sigma^a$ ($a=1,\ldots, N-1$) and $\psi$ so that the symmetry transformation~\eqref{weinbergconformal} is now~\cite{Weinberg:1988cp}
\be
\delta g^{\mu\nu} = 2\epsilon g^{\mu\nu}~;~~~~~~~~~\delta\sigma^a = 0~;~~~~~~~~~\delta\psi = -\epsilon~.
\ee
Now, this transformation is nothing but a conformal transformation, with $\psi$ playing the role of a dilaton. This means that when the fields are constant, the Lagrangian can be written as a function of the conformal metric
\be
\hat g_{\mu\nu} = e^{2\psi}g_{\mu\nu}~.
\ee
When all the fields are set to constants, all the curvature invariants of this metric vanish, so the on-shell Lagrangian must be of the form
\be
{\cal L} = \sqrt{-\hat g}{\cal L}( \sigma^a) =  \sqrt{-g}e^{4\psi}{\cal L}(\sigma^a)~.
\ee
However, the equation $\frac{\partial{\cal L}}{\partial g^{\mu\nu}}\big\rvert_{g_{\mu\nu};\sigma^a ={\rm const.}}= 0$ implies that we must have
\be
\sqrt{-g}e^{4\psi}{\cal L}(\sigma^a)\Big\rvert_{g_{\mu\nu};\sigma^a ={\rm const.}} = e^{4\psi}V(\bar\sigma^a)= 0~,
\ee
which is clearly a fine-tuning (we are tuning the potential for the $\sigma^a$ to have a minimum at $V(\bar\sigma^a) = 0$).

It is worthwhile to examine the assumptions which went into this no-go theorem. First, we assumed that there were a finite number of scalar fields---it is possible that the conclusions could be avoided with an infinite number of fields, but to date this loophole has not been exploited. More promising would be to give up the assumption of constant fields; indeed, this is the loophole exploited by `self-tuning' models such as~\cite{Padilla:2010tj,Charmousis:2011bf,Charmousis:2011ea,Copeland:2012qf}, where the scalar sector has non-trivial coordinate dependence.

\section{Einstein gravity is massless spin-2}
\label{deserweinbergproof}
\renewcommand{\theequation}{B-\Roman{equation}}
\setcounter{equation}{0} 

As mentioned in the text, Einstein gravity is the unique low-energy theory of an interacting massless helicity-2 field. This statement has been proven by various authors~\cite{Papapetrou:1948jw,Gupta:1952zz,Kraichnan:1955zz,Weinberg:1965rz, Feynman:1996kb, Deser:1969wk}.
In this Appendix, we review this uniqueness proof. Our starting point is the action for a free massless spin-2 field at lowest order in derivatives\footnote{This structure is imposed upon us by demanding that our Lagrangian be manifestly Lorentz invariant, local and that it describes the two polarizations of a massless spin-2 particle. The field operator, $h_{\mu\nu}$, is {\it not} a tensor under Lorentz transformations, rather it transforms inhomogeneously~\cite{Weinberg:1965rz}
\be
U(\Lambda)h_{\mu\nu}U^{-1}(\Lambda)= \Lambda^{\alpha}_{~\mu}\Lambda^{\beta}_{~\nu}h_{\alpha\beta}(\Lambda^{-1}x)+\partial_\mu\xi_\nu(x, \Lambda)+\partial_\nu\xi_\mu(x, \Lambda)~,
\ee
where the explicit form of $\xi_\mu$ is not important. Therefore, in order to have a Lagrangian which propagates the desired degrees of freedom, we must construct it so that it both {\it looks} Lorentz invariant and is invariant under the additional transformation $\delta h_{\mu\nu} = \partial_\mu\xi_\nu+\partial_\nu\xi_\mu$, which leads us uniquely to~\eqref{linearizedricci}.
}
\be
S = -\frac{1}{2}\int\rd^4x~h^{\mu\nu}\bigg(\square h_{\mu\nu} - \partial_\mu\partial^\alpha h_{\alpha\nu}- \partial_\nu\partial^\alpha h_{\alpha\mu}+\partial_\mu\partial_\nu h+\eta_{\mu\nu}\left(\partial^\alpha\partial^\beta h_{\alpha\beta}-\square h\right)
\bigg)~,
\label{linearizedricci}
\ee
where the field $h_{\mu\nu}$ has dimensions of mass. The action is invariant under the gauge transformation
\be
\delta_\xi h_{\mu\nu} = \partial_\mu\xi_\nu+\partial_\nu\xi_\mu~,
\ee
with gauge parameter $\xi_\mu = \xi_\mu(x)$. These are nothing more than linearized diffeomorphisms. The equation of motion following from this action is
\be
\square h_{\mu\nu} - \partial_\mu\partial^\alpha h_{\alpha\nu}- \partial_\nu\partial^\alpha h_{\alpha\mu}+\partial_\mu\partial_\nu h+\eta_{\mu\nu}\left(\partial^\alpha\partial^\beta h_{\alpha\beta}-\square h\right) = 0~.
\label{lineareom}
\ee

As a theory on its own,~\eqref{linearizedricci} is perfectly fine, but the field $h^{\mu\nu}$ is free; let's see what happens when we try to introduce interactions by coupling it to its own energy momentum tensor. The action is schematically
\be
S \sim \int\rd^4 x\left(h^{\mu\nu}{\cal E}^{\alpha\beta}_{\mu\nu}h_{\alpha\beta}+h^{\mu\nu}S_{\mu\nu}^{(2)}\right)~,
\ee
where ${\cal E}^{\alpha\beta}_{\mu\nu}$ is the Lichnerowicz operator and $S_{\mu\nu}^{(2)}$ is some tensor quadratic in the field $h$, chosen such that
\be
\frac{\delta}{\delta h^{\mu\nu}} \left( h^{\alpha\beta}S_{\alpha\beta}^{(2)}\right) = \Theta^{(2)}_{\mu\nu}~,
\ee
where $\Theta^{(2)}_{\mu\nu}$ is the energy momentum tensor of the quadratic action. The stress tensor can be constructed from the standard Noether procedure. The equations of motion following from this action are
\be
{\cal E}^{\alpha\beta}_{\mu\nu}h_{\alpha\beta} \sim  \Theta_{\mu\nu}^{(2)}~.
\ee
Now, the left hand side is identically divergence-less, but the right hand side is not conserved, because $\Theta_{\mu\nu}^{(2)}$ is {\it not} the full energy-momentum tensor for the field $h$---the cubic piece we added to the action also contributes! However, we may correct for this by adding a quartic piece to the action
\be
S \sim \int\rd^4 x\left(h^{\mu\nu}{\cal E}^{\alpha\beta}_{\mu\nu}h_{\alpha\beta}+h^{\mu\nu}S_{\mu\nu}^{(2)}+h^{\mu\nu}S_{\mu\nu}^{(3)}\right)~,
\ee
so that the equation of motion is 
\be
{\cal E}^{\alpha\beta}_{\mu\nu}h_{\alpha\beta} \sim  \Theta_{\mu\nu}^{(2)}+\Theta^{(3)}_{\mu\nu}~,
\ee
where $\Theta_{\mu\nu}^{(3)}$ is stress tensor of the cubic part of the action. However, this still doesn't fully fix the problem, because now the quartic piece we added contributes to the stress tensor. If we continue to iterate the procedure, we will end up with an infinite number of terms in the action, and the claim is that these re-sum to give Einstein gravity
\be
S_{\rm EH} \sim \int\rd^4 x\left(h{\cal E}h+h \sum_{n=2}^\infty S^{(n)} \right)\sim \int\rd^4x \sqrt{-g}R~.
\ee

This iteration procedure was performed by a shortcut in~\cite{Feynman:1996kb}. Here we will follow an equivalent, but algebraically simpler, derivation of Deser~\cite{Deser:1969wk}. The benefit of this tack is that we will only have to perform a single iteration of the procedure in order to recover Einstein gravity. To begin, we introduce two different fields, $f^{\mu\nu}$ and $\Gamma_{\mu\nu}^\rho$, and write the action as\footnote{Of course, this is nothing more than the linearized version of the Palatini formulation of Einstein gravity:
\be
S = \frac{M_{\rm Pl}^2}{2}\int\rd^4x\sqrt{-g}g^{\mu\nu}\left(\partial_\alpha\Gamma^\alpha_{\mu\nu}-\partial_\nu\Gamma^\alpha_{\mu\alpha}+\Gamma_{\mu\nu}^\alpha\Gamma^\rho_{\alpha\rho}-\Gamma^\alpha_{\rho\mu}\Gamma^{\rho}_{\alpha\nu}\right)~,
\ee
but for now we will pretend that we don't know this.
}
\be
S = \int\rd^4x\left(f^{\mu\nu}\left(\partial_\alpha\Gamma^\alpha_{\mu\nu}-\partial_\nu\Gamma^\alpha_{\mu\alpha}\right)+\eta^{\mu\nu}\left(\Gamma_{\mu\nu}^\alpha\Gamma^\rho_{\alpha\rho}-\Gamma^\alpha_{\rho\mu}\Gamma^{\rho}_{\alpha\nu}\right)
\right)~,
\label{firstorderaction}
\ee
where $f^{\mu\nu}$ is symmetric and $\Gamma^\alpha_{\mu\nu}$ is symmetric in its lower indices. The mass dimensions of the fields are $\left[f\right]=1$, $\left[\Gamma\right] = 2$. This action is completely equivalent to~\eqref{linearizedricci}; this can be seen from the equations of motion obtained by varying with respect to $f^{\mu\nu}$ and $\Gamma^\alpha_{\mu\nu}$:\footnote{Note that in~\eqref{feom}, we have used the facts that $\eta^{\alpha\beta}\Gamma^\mu_{\alpha\beta} = -\partial_\alpha f^{\nu\alpha}$ and $\Gamma_{\mu\rho}^\rho = -\frac{1}{2}\partial_\mu f$,
which can be obtained by taking various traces.}
\begin{align}
\label{gammaeom}
\partial_\alpha\Gamma^\alpha_{\mu\nu}-\frac{1}{2}\partial_\nu\Gamma^\alpha_{\mu\alpha}-\frac{1}{2}\partial_\mu\Gamma^\alpha_{\nu\alpha} &= 0\\
\label{feom}
\partial^\alpha f_{\mu\nu}-\partial_\mu f_{\nu}^{\alpha}-\partial_\nu f_{\mu}^{\alpha}-\frac{1}{2}\eta_{\mu\nu}\partial^\alpha f &= 2\Gamma^\alpha_{\mu\nu}-\delta_\mu^\alpha\Gamma^\rho_{\nu\rho}-\delta_\nu^\alpha\Gamma^\rho_{\mu\rho}~.
\end{align}
We take the derivative $\partial_\alpha$ of~\eqref{feom}, and then insert the result into~\eqref{gammaeom} to obtain:
\be
\square f_{\mu\nu}-\partial_\mu\partial^\alpha f_{\alpha\nu}-\partial_\nu\partial^\alpha f_{\alpha\mu}-\frac{1}{2}\eta_{\mu\nu}\square f = 0~.
\label{feqn}
\ee
Tracing over both sides, we obtain $\square f = -2 \partial_\alpha\partial_\beta f^{\alpha\beta}$. Using this fact, we see that~\eqref{feqn} and~\eqref{lineareom} are equivalent after making the field redefinition
\be
\label{frelatedtoh}
f_{\mu\nu} = \frac{1}{2}\eta_{\mu\nu}h-h_{\mu\nu}~.
\ee
Further, we may solve for $\Gamma_{\mu\nu}^\alpha$ to obtain
\be
\Gamma^\alpha_{\mu\nu} = -\frac{1}{2}\eta^{\alpha\lambda}\left[\partial_\mu\left(f_{\nu\lambda}-\frac{1}{2}\eta_{\nu\lambda}h\right)+\partial_\nu\left(f_{\mu\lambda}-\frac{1}{2}\eta_{\mu\lambda}f\right)-\partial_{\lambda}\left(f_{\mu\nu}-\frac{1}{2}\eta_{\mu\nu}f\right)\right]~.
\ee
If we make the same field redefinition~\eqref{frelatedtoh}, this becomes
\be
\Gamma^\alpha_{\mu\nu} = \frac{1}{2}\eta^{\alpha\lambda}\left(\partial_\mu h_{\nu\lambda}+\partial_\nu h_{\mu\lambda}-\partial_{\lambda} h_{\mu\nu}\right)~,
\ee
which is, of course, the linearized Christoffel symbol.

We know that $f^{\mu\nu}$ should couple to its own stress tensor, so we now want to add a term to the action that, when varied with respect to $f^{\mu\nu}$, gives the stress tensor for the action~\eqref{firstorderaction}. To do this, we must first compute the stress tensor for this action. We will do this in the normal way, by ``covariantizing" the action, varying with respect to the metric and then setting it to be flat.\footnote{You might complain that we are cheating. After all, the point is to obtain Einstein gravity from field theory without appealing to geometric notions. While we are taking a shortcut, the energy-momentum tensor we will obtain is exactly equivalent to the one which would be obtained by purely field-theoretic machinery---that is by constructing it as the conserved current associated with translation invariance using the standard Noether procedure and then symmetrizing.} We covariantize the action by promoting the flat metric $\eta_{\mu\nu}\mapsto G_{\mu\nu}$. We now have to decide how $f^{\mu\nu}$ and $\Gamma_{\mu\nu}^\alpha$ transform with respect to this metric. We choose to let $\Gamma_{\mu\nu}^\alpha$ transform as a tensor while $f^{\mu\nu}$ transforms as a tensor density, $(\sqrt{-G})^{-1}f^{\mu\nu}$. After covariantization, the action~\eqref{firstorderaction} becomes
\be
S = \int\rd^4x\left(f^{\mu\nu}\left(D_\alpha\Gamma^\alpha_{\mu\nu}-D_\nu\Gamma^\alpha_{\mu\alpha}\right)+\sqrt{-G}G^{\mu\nu}\left(\Gamma_{\mu\nu}^\alpha\Gamma^\rho_{\alpha\rho}-\Gamma^\alpha_{\rho\mu}\Gamma^{\rho}_{\alpha\nu}\right)
\right)~,
\label{covariantfirstorderaction}
\ee
where $D_\alpha$ is the $G$-covariant derivative, so that
\be
D_\alpha\Gamma_{\mu\nu}^\rho = \partial_\alpha\Gamma^\rho_{\mu\nu}+C_{\alpha\lambda}^\rho\Gamma_{\mu\nu}^\lambda-C_{\alpha\mu}^\lambda\Gamma_{\lambda\nu}^\rho-C_{\alpha\nu}^\lambda\Gamma_{\mu\lambda}^\rho~,
\ee
where $C_{\mu\nu}^\lambda$ is the Christoffel symbol associated with $G_{\mu\nu}$. The variable $f_{\mu\nu}$ that we are working with is trace-shifted, so we also want to compute the trace-shifted stress tensor
\be
\tau_{\mu\nu} = T_{\mu\nu}-\frac{1}{2}\eta_{\mu\nu}T = \frac{\delta S}{\delta( \sqrt{-G}G^{\mu\nu})}\equiv \frac{\delta S}{\delta\bar G^{\mu\nu}}
\ee
 to couple to it. It is straightforward to vary~\eqref{covariantfirstorderaction}, to obtain
\be
\delta S = \int\rd^4x \left(\frac{\delta}{\delta \bar G^{\mu\nu}}\left[f^{\mu\nu}\left(D_\alpha\Gamma^\alpha_{\mu\nu}-D_\nu\Gamma^\alpha_{\mu\alpha}\right)\right]+\left(\Gamma_{\mu\nu}^\alpha\Gamma^\rho_{\alpha\rho}-\Gamma^\alpha_{\rho\mu}\Gamma^{\rho}_{\alpha\nu}\right)\right)\delta\bar G^{\mu\nu}~,
\label{variationofcovariantaction}
\ee
which we may rewrite as~\cite{Deser:1969wk}
\be
\delta S = \int\rd^4x \left(-\sigma_{\mu\nu}+\Gamma_{\mu\nu}^\alpha\Gamma^\rho_{\alpha\rho}-\Gamma^\alpha_{\rho\mu}\Gamma^{\rho}_{\alpha\nu}\right)\delta\bar G^{\mu\nu}~.
\ee
From this we deduce that the stress tensor for the action~\eqref{firstorderaction} is
\be
\tau_{\mu\nu} = \Gamma_{\mu\nu}^\alpha\Gamma^\rho_{\alpha\rho}-\Gamma^\alpha_{\rho\mu}\Gamma^{\rho}_{\alpha\nu}-\sigma_{\mu\nu}\equiv S_{\mu\nu}-\sigma_{\mu\nu} \ ,
\ee
where $\sigma_{\mu\nu}$ is given by~\cite{Deser:1969wk}
\begin{align}
\nonumber
2\sigma_{\mu\nu} = \partial^\alpha\bigg[\frac{1}{2}\eta_{\mu\nu}\left(f^\lambda_\rho\Gamma_{\lambda\alpha}^\rho-\frac{1}{2}f\Gamma_{\alpha\rho}^\beta\right)+\frac{1}{2}f_{\mu\nu}\Gamma_{\alpha\rho}^\rho-f_{\alpha\mu}\Gamma_{\nu\rho}^\rho+\eta_{\nu\sigma}f_{\alpha}^\beta\Gamma_{\beta\mu}^\sigma+&f_\mu^\rho\left(\eta_{\alpha\beta}\Gamma_{\rho\nu}^\beta-\eta_{\sigma\nu}\Gamma_{\alpha\rho}^\sigma\right)\bigg]\\
&+(\mu\leftrightarrow\nu) \ .
\end{align}
The claim is then that the appropriate coupling to consider in the Lagrangian is $f^{\mu\nu}S_{\mu\nu}$. That is, we consider the action
\be
S = \int\rd^4x\bigg(M_{\rm Pl}\left(\eta^{\mu\nu}+\frac{1}{M_{\rm Pl}}f^{\mu\nu}\right)\left[\partial_\alpha\Gamma^\alpha_{\mu\nu}-\partial_\nu\Gamma^\alpha_{\mu\alpha}\right]+\left(\eta^{\mu\nu}+\frac{1}{M_{\rm Pl}}f^{\mu\nu}\right)\left[\Gamma_{\mu\nu}^\alpha\Gamma^\rho_{\alpha\rho}-\Gamma^\alpha_{\rho\mu}\Gamma^{\rho}_{\alpha\nu}\right]
\bigg)~.
\ee
We have added two terms to the action, one is the coupling $f^{\mu\nu}S_{\mu\nu}$, the other is the total derivative $\partial_\beta[\eta^{\mu\nu}\Gamma^\beta_{\mu\nu}-\eta^{\beta\mu}\Gamma^\alpha_{\mu\alpha}]$. In order to add these terms, it was necessary to introduce a dimensionful parameter, which we have (with foresight) called $M_{\rm Pl}$, with units of mass.
Notice that the procedure we have just carried out does {\it not} have to be iterated---the term we added to the action does not depend on $\eta_{\mu\nu}$, so if we were to covariantize and vary, we would find that it does not contribute to $\tau_{\mu\nu}$.

Therefore, all that remains is for us to derive the equations of motion for this action, and verify that they produce the equations of motion
\be
\square f_{\mu\nu}-\partial_\mu\partial^\alpha f_{\alpha\nu}-\partial_\nu\partial^\alpha f_{\alpha\mu}-\frac{1}{2}\eta_{\mu\nu}\square f = -\tau_{\mu\nu}~.
\ee
This is of course guaranteed to work: note that by performing a field redefinition
\be
\sqrt{-g}g^{\mu\nu} = \eta^{\mu\nu}+\frac{1}{M_{\rm Pl}}f^{\mu\nu}~,
\ee
and re-scaling $\Gamma \mapsto M_{\rm Pl}\Gamma$
the action becomes
\be
S = M_{\rm Pl}^2\int\rd^4x\sqrt{-g}g^{\mu\nu}\left(\partial_\alpha\Gamma^\alpha_{\mu\nu}-\partial_\nu\Gamma^\alpha_{\mu\alpha}+\Gamma_{\mu\nu}^\alpha\Gamma^\rho_{\alpha\rho}-\Gamma^\alpha_{\rho\mu}\Gamma^{\rho}_{\alpha\nu}\right)~,
\ee
which is just the first order (Palatini) formulation of Einstein gravity. The equation of motion for $\Gamma_{\mu\nu}^\alpha$ is
\be
\Gamma_{\mu\nu} = \frac{1}{2}g^{\alpha\lambda}\left(\partial_\mu g_{\nu\lambda}+\partial_\nu g_{\mu\lambda}-\partial_\lambda g_{\mu\nu}\right)~,
\ee
and substituting this back into the action yields the familiar Einstein--Hilbert action
\be
S = M_{\rm Pl}^2\int\rd^4x\sqrt{-g}R~.
\label{deserEHaction}
\ee
It is worth emphasizing that throughout, we have only considered terms which have second derivatives. In this sense, Einstein gravity is the unique {\it low energy} theory of a Lorentz-invariant massless helicity-2 particle.

For completeness, we must also determine how $g_{\mu\nu}$ should couple to external matter. We (now) know that equation of motion for gravity plus matter fields, $\psi$, must take the form (in $M_{\rm Pl}=1$ units)
\be
G_{\mu\nu} = \frac{\delta S_{\rm matter}[\psi, g]}{\delta g^{\mu\nu}}~.
\ee
However, $g_{\mu\nu}$ couples to the {\it total} stress energy, so that we also have
\be
G_{\mu\nu} = T^{\rm matter}_{\mu\nu}~.
\ee
This just implies that we can obtain $T^{\rm matter}_{\mu\nu}$ by varying with respect to $g_{\mu\nu}$, indicating that it couples in the usual way.

\section{Ostrogradsky's theorem}
\label{OstrogradskyApp}
\renewcommand{\theequation}{C-\Roman{equation}}
\setcounter{equation}{0} 

Here we review Ostrogradsky's theorem, following~\cite{Woodard:2006nt}. The theorem states that for a {\it non-degenerate} Lagrangian which depends on higher derivatives, the Hamiltonian is necessarily unbounded. More precisely, all but one of the canonical momenta appear linearly in the Hamiltonian, as we will see. We will derive the theorem in the context of the classical mechanics of a single particle, but the result generalizes readily to field theory~\cite{deUrries:1998bi, deUrries:1995ty}. We consider a Lagrangian which depends on the position of a point particle as a function of time, $q(t)$, and arbitrarily many time derivatives
\be
{\cal L}(q, \dot q, \ldots, q^{(N)})~,
\ee
where $q^{(N)} \equiv \frac{\rd^Nq}{\rd t^N}$. The equation of motion for this Lagrangian is then
\be
\sum_{i=0}^N\left(-\frac{\rd}{\rd t}\right)^i\frac{\partial{\cal L}}{\partial q^{(i)}} = 0 \ .
\ee
If the Lagrangian is non-degenerate, this equation depends on $q^{(2N)}$ and therefore can be rewritten as
\be
q^{(2N)} = f(q, \ldots, q^{(2N-1)})~.
\ee
This differential equation requires $2N$ pieces of initial data, and therefore there are $2N$ canonical coordinates. Ostrogradsky instructs us to define these as~\cite{Woodard:2006nt}
\be
Q_a = q^{(a-1)}~;~~~~~~~~~~~~~~~~P_a = \frac{\delta{\cal L}}{\delta q^{(a)}} = \sum_{i=a}^N\left(-\frac{\rd}{\rd t}\right)^{i-a}\frac{\partial {\cal L}}{\partial q^{(i)}}~,
\label{Ostrogradskycoords}
\ee
where here by $\frac{\delta}{\delta q^{(a)}}$ we mean the Euler--Lagrange derivative with respect to $q^{(a)}$. Now, if the Lagranigan is non-degenerate, we can solve for $q^{(N)}$ in terms of the canonical coordinates, $Q_a$, and the $N$th momentum, $P_N$, as
\be
q^{(N)} = F(Q_1, \ldots, Q_N, P_N) \ .
\ee
We then construct the Hamiltonian in the standard way via a Legendre transform
\be
{\cal H} = \sum_{a=1}^NP_a q^{(a)}-{\cal L} = P_1Q_2+P_2Q_3+\ldots+P_{N-1}Q_N+P_N F-{\cal L}(Q_1, \ldots, Q_{N}, F)~.
\ee
Here we see that every momentum except possibly $P_N$ appears linearly in the Hamiltonian, and hence it is unbounded with respect to each of these momenta. The only thing left to check is that this Hamiltonian accurately captures the dynamics of the system; that is, it should really be the Hamiltonian as we normally think of it. This is relatively straightforward; recall Hamilton's equations
\be
\dot Q_a = \frac{\partial{\cal H}}{\partial P_a}~;~~~~~~~~~~~~~~~~\dot P_a = - \frac{\partial{\cal H}}{\partial Q_a}~.
\ee
Now we notice that the first of these equations tells us\footnote{This is obvious for $a \neq N$, but for $a=N$ it requires a little thought:
\begin{equation*}
\dot Q_N = \frac{\partial{\cal H}}{\partial P_N} = F+P_N\frac{\partial F}{\partial P_N}-\frac{\partial{\cal L}}{\partial F}\frac{\partial F}{\partial P_N} = F \implies q^{(N)} = F
\end{equation*}
}
\be
\dot Q_a = Q_{a+1} \implies Q_i = q^{(i-1)}~,
\ee
while the second equation tells us that
\be
P_a = \frac{\delta{\cal L}}{\delta q^{(a)}}~.
\ee
The $\dot P_1$ equation then yields the Euler--Lagrange equation
\be
\sum_{i=0}^N\left(-\frac{\rd}{\rd t}\right)^{i}\frac{\partial {\cal L}}{\partial q^{(i)}} = 0~.
\ee
This confirms that the Hamiltonian generates time translations in the normal way~\cite{Woodard:2006nt}.

The above discussion is rather general and abstract, so it is useful to apply the formalism to a particular example and construct the Hamiltonian explicitly, showing its unboundedness. The theory we consider is the {\it Pais--Uhlenbeck oscillator}~\cite{Pais:1950za}, a simple classical-mechanical example of a higher-derivative quadratic Lagrangian
\be
{\cal L}_{\rm PU} = \frac{1}{2}\ddot q^2-\frac{1}{2}(m_1^2+m^2_2)\dot q^2+\frac{1}{2}m_1^2m_2^2q^2~.
\ee
This Lagrangian depends on $q, \dot q,$ and $\ddot q$, and its equation of motion is fourth-order in derivatives
\be
\ddddot q+(m_1^2+m^2_2)\dot q+m_1^2m_2^2q = 0~.
\label{PUeom}
\ee
This is a fourth-order equation---requiring four pieces of initial data---indicating that there are four canonical coordinates, we define them as in~\eqref{Ostrogradskycoords}~\cite{Mannheim:2004qz,Bender:2007wu,Chen:2012au} 
\be
\begin{array}{ll}
Q_1 = q & ~~~~~~~~~~~~~~~~~P_1 = -\dddot q - (m_1^2+m_2^2)\dot q\\
Q_2 = \dot q & ~~~~~~~~~~~~~~~~~P_2 = \ddot q ~.
\end{array}
\label{PUcanonicalcoords}
\ee
Recall that the equation of motion~\eqref{PUeom} depended on $\ddddot q$, indicating the Lagrangian is non-degenerate, and we see here that we can trivially solve for $\ddot q$ in terms of the canonical coordinates and $P_2$. We can then construct the Hamiltonian as~\cite{Mannheim:2004qz,Bender:2007wu,Chen:2012au} 
\be
{\cal H}_{\rm PU} = P_1Q_2+P_2^2 - {\cal L}_{\rm PU}(Q_1, Q_2, P_2) = P_1Q_2 +\frac{1}{2}P_2^2+\frac{1}{2}(m_1^2+m_2^2)Q_2^2-\frac{1}{2}m_1^2m_2^2Q_1^2~.
\ee
As advertised, the momentum $P_1$ appears linearly in the Hamiltonian, indicating that it is unbounded.\footnote{For proposals on how to deal with the ghost in this model, see~\cite{Bender:2007wu, Chen:2012au}.} We can also verify that Hamilton's equations reproduce the canonical transformation~\eqref{PUcanonicalcoords} and that the $\dot P_1$ equation gives the Euler--Lagrange equation
\be
\dot P_1 = -\frac{{\cal H}_{\rm PU}}{\partial Q_a} \implies \ddddot q+(m_1^2+m^2_2)\dot q+m_1^2m_2^2q = 0~.
\ee

\section{Effective field theory diagnostics}
\label{ghostgradtachyon}
\renewcommand{\theequation}{D-\Roman{equation}}
\setcounter{equation}{0} 

In this Appendix, we detail some of the more common pathologies at the level of the effective field theory, namely ghosts, gradient instabilities, tachyons and superluminality.

\subsection{Ghosts}

A common pathology of the field theories that arise from modifications of gravity is the existence of {\it ghosts}. Ghosts are fields whose quanta either have negative energy or negative norm, indicating an instability in the theory. Most commonly, ghost instabilities manifest as fields with {\it wrong sign} kinetic term
\be
{\cal L}_{\rm ghost} = \frac{1}{2}(\partial\chi)^2-\frac{m^2}{2}\chi^2~.
\ee
Of course, the sign of the kinetic term is merely a matter of convention---our choice of metric signature. However, what is dangerous is if this field is coupled to other fields with correct sign kinetic terms, for example
\be
{\cal L} = -\frac{1}{2}(\partial\phi)^2-\frac{m_\phi^2}{2}\phi^2 + \frac{1}{2}(\partial\chi)^2-\frac{m_\chi^2}{2}\chi^2+\lambda\phi^2\chi^2~.
\ee
Since the $\chi$ particles have negative energy, the vacuum is unstable to the process $0 \to \phi\phi+\chi\chi$, which costs zero energy. This process will happen copiously (with an infinite rate), and in fact is a sign that the theory is ill-defined~\cite{Carroll:2003st, Cline:2003gs}.

\begin{figure}
\centering
\includegraphics[width=2in]{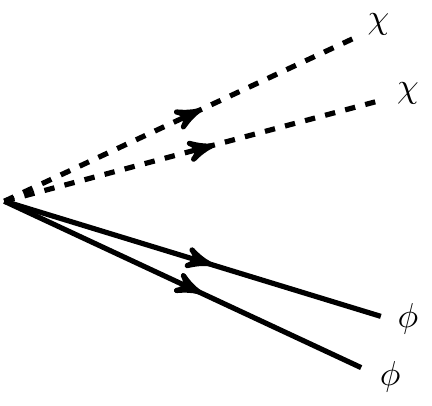}
\caption{\small In theories with a ghostly field, the vacuum is unstable to rapid pair production of ghost particles and healthy particles, causing the theory to be ill-defined (See~\cite{Carroll:2003st,Cline:2003gs}).}
\end{figure}

In constructing theories, we have to ensure that the theory is free from these ghost instabilities. Most often, ghost instabilities arise from {\it higher derivative} terms in the Lagrangian. A powerful theorem, due to Ostrogradsky~\cite{Ostrogradsky} tells us that, in most cases, if the equations of motion are higher than second order in time derivatives, the theory will have a ghost instability (we review this theorem in Appendix~\ref{OstrogradskyApp}).

Let us illustrate this through a simple example of a higher-derivative theory and show how it may be written as a theory of a healthy field and a ghostly field. Consider the Lagrangian
\be
{\cal L} = -\frac{1}{2}(\partial\psi)^2+\frac{1}{2\Lambda^2}(\square\psi)^2-V(\psi)~,
\label{simpleghostlag}
\ee
where $\Lambda$ is the cutoff of the effective theory. As long as we work at energies far below $\Lambda$, this theory is completely well defined. However, as we approach $\Lambda$, all of the terms we have neglected---which are suppressed by additional powers of $\partial ({\rm fields})/\Lambda$---become equally important and we no longer have a well-defined expansion. Note that the equations of motion following from the Lagrangian~\eqref{simpleghostlag} are fourth order. To see explicitly that this implies the system secretly has a ghost, we introduce an auxiliary field, $\chi$, as follows~\cite{Creminelli:2005qk}
\be
{\cal L} = -\frac{1}{2}(\partial\psi)^2+\chi\square\psi-\frac{\Lambda^2}{2}\chi^2-V(\psi)~.
\label{auxchifield}
\ee
Now, the equation of motion for $\chi$ is $\chi = \square\psi/\Lambda^2$, and substituting it back into the Lagrangian, we recover~\eqref{simpleghostlag}, confirming the classical equivalence between the two theories. In order to remove the kinetic mixing and diagonalize the Lagrangian, we make the field redefinition $\psi = \phi-\chi$, in terms of which the Lagrangian becomes (after integration by parts)~\cite{Creminelli:2005qk}
\be
{\cal L} = -\frac{1}{2}(\partial\phi)^2+\frac{1}{2}(\partial\chi)^2-\frac{\Lambda^2}{2}\chi^2-V(\phi, \chi)~.
\ee
Now we see that the Lagrangian~\eqref{simpleghostlag} is equivalent to a theory of two scalar fields, one healthy and one ghostly. It is important to note that the presence of a ghost does not necessarily spell doom for a theory; as long as the mass of the ghostly mode lies above the cutoff of the theory, we can interpret the presence of the ghost as an artifact of truncating the EFT expansion at finite order. We see that this is the case here, the mass of the ghost lies at the cutoff of the theory, so it may be possible to consistently ignore the ghost at energies far below the cutoff and appeal to unknown UV physics to cure the ghost as we approach the cutoff.\footnote{To see how this can work, instead of introducing the auxiliary field $\chi$ as in~\eqref{auxchifield}, consider the Lagrangian~\cite{Creminelli:2005qk}
\be
{\cal L} = -\frac{1}{2}(\partial\psi)^2+\chi\square\psi-(\partial\chi)^2-\frac{\Lambda^2}{2}\chi^2-V(\psi)~.
\label{betterbehavedchilag}
\ee
The equation of motion for $\chi$ is
\be
\chi = \frac{\square\psi}{\Lambda^2\left(1-\frac{\square}{\Lambda^2}\right)}~;
\ee
if we substitute this back in, we recover~\eqref{simpleghostlag}, up to extra terms suppressed by powers of $\partial^2 ({\rm fields})/\Lambda^2$, which we have dropped in the EFT expansion anyway. If we then diagonalize~\eqref{betterbehavedchilag}, we find two healthy scalars
\be
{\cal L} = -\frac{1}{2}(\partial\phi)^2-\frac{1}{2}(\partial\chi)^2-\frac{\Lambda^2}{2}\chi^2-V(\phi,\chi)~.
\ee
} 
What {\it is} dangerous is a theory possessing a ghost within the regime of validity of the effective theory. If this is true, the theory loses its predictive power and we cannot calculate. 

\subsection{Gradient instabilities}

Another pathology which often plagues effective field theories is the presence of {\it gradient} instabilities. Much in the way that a ghost instability is related to wrong sign temporal derivatives, gradient instabilities are due to wrong sign spatial gradients. To see why this is worrisome, consider the simplest (obviously non-Lorentz-invariant) example: a free scalar field with wrong sign spatial gradients
\be
{\cal L} = \frac{1}{2}\dot\phi^2+\frac{1}{2}(\vec\nabla\phi)^2~.
\label{gradinstablag}
\ee
The solutions to the equation of motion following from this Lagrangian are (in Fourier space)
\be
\phi_k(t) \sim e^{\pm kt}~,
\ee
where $k \equiv \sqrt{{\vec k}^2}$. Note that the growing mode solution, $\phi\sim e^{kt}$, grows without bound, signaling an instability in the theory on a timescale
\be
\tau_{\rm inst.} \sim k^{-1}~.
\ee
Therefore, the highest-energy modes contribute most to the instability. This means that the theory does not make sense, even thought of as an effective theory. In general, in a theory with a gradient instability, and a cutoff $\Lambda$, the effective theory cannot consistently describe any energy regime. For modes with $k \ll \Lambda$, the characteristic timescale is $t_k \gg \tau_{\rm inst.}$, and they will be sensitive to the instability in the theory, whereas modes with $k\gg \Lambda$ are beyond the regime of validity of our EFT. We are forced to conclude that an effective theory with a gradient instability is non-predictive.

\subsection{Tachyonic instabilities}

Another instability that sometimes appears in effective theories is the presence of a {\it tachyon}. Most simply, a tachyonic instability appears as a field with a negative mass squared. Unlike the other instabilities, the presence of a tachyon does not indicate any particular pathology in the definition of the theory, but rather is a signal that we are not perturbing about the true vacuum of the theory. For example, this is precisely what happens in the Higgs mechanism---in the Lagrangian, the Higgs field appears as a tachyon, but of course everything is well defined.

Again, we consider a toy example: a scalar field with a negative mass term
\be
{\cal L} = -\frac{1}{2}(\partial\phi)^2+\frac{m^2}{2}\phi^2
\ee
In the long-wavelength ($k\to0$) limit, the solution for the field $\phi$ is
\be
\phi(t) \sim t^{\pm mt}~.
\ee
Again, the growing-mode solution indicates an instability. However, unlike last time, the timescale for this instability is independent of $k$ and is given by the inverse mass of the field
\be
\tau_{\rm inst.} \sim m^{-1}~.
\ee
Therefore, if we focus on modes for which $k\gg m^{-1}$, they will be insensitive to the fact that the system is unstable. This type of thinking is familiar from cosmology---if we go to high momenta, the modes evolve as though they are on Minkowski space and are insensitive to the cosmological evolution. 

This analysis generalizes to theories with a cutoff, $\Lambda$, in which there exists a regime $m \ll k \ll\Lambda$, where the effective field theory is perfectly well defined, provided that there is a hierarchy between the mass, $m$, and the cutoff of the theory.

\subsection{Analyticity, locality and superluminality}
\label{superlumapp}

So far, the pathologies we have discussed manifest themselves in the effective description we consider (for example, ghosts are visible in the low energy effective theory). However, there are also apparent illnesses of an effective theory which are of a more subtle nature, and indicate that a theory---completely well-defined in the IR---may secretly not admit a standard UV completion. Whether or not such pathologies are fatal then depends on whether one can make sense of the relevant theories in the UV while abandoning one or more of the usual requirements, such as Lorentz-invariance. 

By far the most common sickness of this type is the presence of {\it superluminality} in a low-energy effective field theory. To understand why this is a problem, we note that a crucial ingredient of a Lorentz-invariant quantum field theory is {\it microcausality}. This property states that the commutator of two local operators vanishes for spacelike separated points as an operator statement~\cite{Peskin:1995ev}
\be
\left[{\cal O}_1(x), {\cal O}_2(y)\right] = 0~;~~~~~{\rm when}~~~~~(x-y)^2 > 0~.
\label{microcausalitystatement}
\ee
The relation to causality is fairly clear; if two operators are evaluated at points outside each others' lightcones, they should not have an effect on each other. Indeed, in~\cite{Dubovsky:2007ac}, it was shown that~\eqref{microcausalitystatement} can be seen as a consequence of the causal structure of the theory, and hence holds for an arbitrary curved space, as long as the relevant fields have a well-defined Cauchy problem. Therefore, we immediately see that there is an apparent tension between superluminality and causality: in a theory with superluminal propagation, operators outside the light cone do not necessarily commute, indicating that the theory is secretly acausal or non-local.\footnote{Another consequence of particles which propagate with a higher velocity than gravitons is the possible presence of gravi-Cherenkov radiation~\cite{Moore:2001bv}.}

In reality, theories which admit superluminality can be perfectly causal, but just on a widened light-cone. Consider a non-renormalizible higher derivative theory of a scalar
\be
{\cal L} = -\frac{1}{2}(\partial\phi)^2+\frac{1}{\Lambda^3}\partial^2\phi(\partial\phi)^2+\frac{1}{\Lambda^4}(\partial\phi)^4+\cdots
\label{nonlocallightcone}
\ee
expanded about some background $\phi = \bar\phi+\vp$. A theorem due to Leray\footnote{See page 251 of~\cite{Wald:1984rg}.} says that the causal structure is set by an {\it effective} metric~\cite{Wald:1984rg, Bruneton:2006gf,Dubovsky:2007ac,Babichev:2007dw}
\be
{\cal L} = -\frac{1}{2}G^{\mu\nu}(x,\bar\phi, \partial\bar\phi, \partial^2\bar\phi, \ldots)\partial_\mu\vp\partial_\nu\vp + \cdots \ .
\ee
Now, provided that $G^{\mu\nu}$ is globally hyperbolic, this theory will be perfectly causal, but in general it may have directions in which the $\vp$ perturbations propagate outside the lightcone used to define the theory~\eqref{nonlocallightcone}. On the face of it, this might not seem to be much of a worry, but it is vaguely unsettling that the Lorentz-invariant Lagrangian we wrote down secretly is not. If we think of the theory~\eqref{nonlocallightcone} as a low-energy effective field theory, it must have a UV completion at some high energy scale. However, since this theory admits superluminality, it cannot be UV completed by a Lorentz-invariant quantum field theory. Here, by Lorentz-invariant, we mean Lorentz-invariant with respect to the metric $\eta_{\mu\nu}$ used to define~\eqref{nonlocallightcone}.

In many cases this heuristic reasoning can be made precise by exploiting the close relationship between Lorentz invariance and {\it S-matrix analyticity}. In a Lorentz-invariant quantum field theory, the S-matrix is an analytic function of the external momenta, except for branch cuts which correspond to the production of intermediate states and the presence of poles, which correspond to physical particles or bound states. From S-matrix analyticity, we can derive {\it dispersion relations} to establish the positivity of various scattering amplitudes. The connection between this approach, superluminality and UV completions of low energy theories was first explored in~\cite{Adams:2006sv} and here we summarize some of the logic.

We begin by considering the $2\to2$ scattering amplitude in a Lorentz-invariant theory. We focus on this amplitude, because intuitively it should have something to do with superluminality, since we can think of propagation in the effective metric~\eqref{nonlocallightcone} as a sequence of scattering processes with a background field~\cite{Adams:2006sv}.
\begin{figure}
\centering
\includegraphics[width=2.2in]{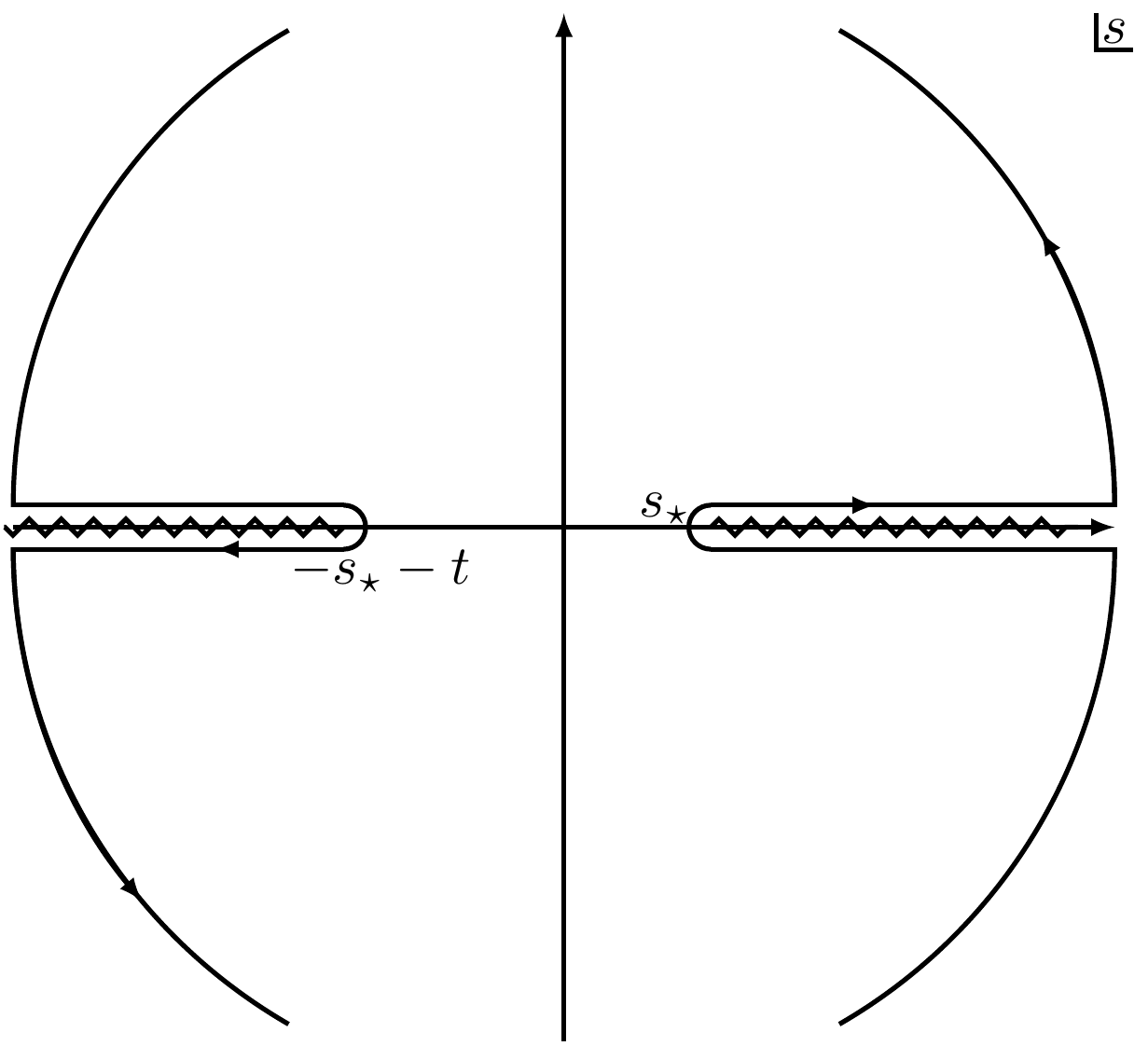}
\caption{\small Integration contour in the complex $s$-plane used to derive the dispersion relation~\eqref{dispersionrelation}. Integration along this contour picks up the discontinuity across the cuts, which correspond to above-threshold particle production.}
\label{dispersionint}
\end{figure}
Consider the $4$-point amplitude ${\cal A}(s, t)$, which is analytic in the $s$-plane, except for a pole at $s=0$ and along cuts on the real axis above some threshold value $\lvert s_\star\rvert < \infty$.\footnote{Here $s$ and $t$ are the usual Mandelstam variables, see for example~\cite{Peskin:1995ev}.}
The parameter $s_\star$ is the energy at which we would expect to start pair producing particles in a scattering process (in a massive theory, this is the energy scale corresponding to the mass of the constituent particles). Considering a closed curve, $\gamma$, around $s=0$ of radius $r < s_\star$ and using Cauchy's integral formula, we find
\be
\left.\frac{\partial^2}{\partial s^2}{\cal A}(s, t)\right\rvert_{s=0} = \frac{1}{i\pi}\oint_\gamma\rd s\frac{{\cal A}(s, t)}{s^3}~.
\ee
Now, we deform the contour into a double-keyhole contour and integrate along the cuts as in Figure~\ref{dispersionint}. On the positive real axis, we obtain the discontinuity along the cut, which is the imaginary part of the analytic function ${\cal A}$.\footnote{This follows from the fact that ${\cal A}(s) = {\cal A}^*(s^*)$, which is an {\it assumption} of S-matrix theory about how amplitudes should behave.} The integral along the cut on the negative real axis can be related to the integral on the positive side via crossing symmetry, $s\to -t-s$, to obtain
\be
\frac{\partial^2}{\partial s^2}{\cal A}(s, t)\Big\rvert_{s=0}  = \frac{2}{\pi}\int_{s_\star}^\infty\rd s\left(\frac{1}{s^3}+\frac{1}{(s+t)^3}\right){\rm Im}{\cal A}(s, t)~.
\label{dispersionrelation}
\ee
In order for this dispersion relation to make sense, we must make sure that the point at infinity gives no contribution to the integral. This contribution will vanish as long as the amplitude is bounded by $s^2$ as $s\to\infty$. That this is true in theories with a mass gap follows from the Froissart bound~\cite{Froissart:1961ux, Martin:1962rt, Adams:2006sv}. Many theories of interest do not have a mass gap\footnote{In massless theories, another subtlety arises: formally $s_\star \to 0$. {\it However}, we expect non-analyticities near $s = 0$ to show up as multi-particle states in our amplitudes; in derivatively-coupled theories, the relation~\eqref{dispersionrelation} can still make sense at tree level: the optical theorem tells us that multi-particle states should in fact only appear at higher order in momenta. For example, in the theory of a dilaton we have
$
{\cal A}^{2\to2}_{1-{\rm loop}} \sim \left\lvert{\cal A}^{2\to2}_{\rm tree}\right\rvert^2 \sim \left(s^2+t^2+u^2\right)^2 \sim {\cal O}(s^4)~,
$
so the dispersion relation holds at tree level~\cite{Komargodski:2011xv, Nicolis:2009qm}.
} (for example the galileon); there are some arguments that amplitudes should be bounded similarly at infinity in these situations~\cite{Adams:2006sv}, but it is not certain.\footnote{One situation where this can be made precise is when the UV completion is a CFT. In this case, we know that at high energies, the amplitude scales as
$
\lim_{s\to\infty}{\cal A}(s, t) \sim s^{2-\epsilon}~,
$
where $\epsilon = 4-\Delta > 0$, and $\Delta$ is the conformal weight of the fields; so the contour at infinity gives no contribution ~\cite{Komargodski:2011xv, Luty:2012ww, Elvang:2012st}.
}
If we now look at the forward limit $(t\to0)$ of the expression~\eqref{dispersionrelation}, the optical theorem tells us ${\rm Im}{\cal A}(s, 0) \geq 0$, which establishes the inequality
\be
\frac{\partial^2}{\partial s^2}{\cal A}(s, 0)\Big\rvert_{s=0}  = \frac{4}{\pi}\int_{s_\star}^\infty\rd s\frac{{\rm Im}{\cal A}(s, 0)}{s^3} \geq 0~.
\label{dispersionsumrule}
\ee
Therefore, in the forward limit, the $2\to2$ amplitude must display a {\it positive} $s^2$ contribution.
This inequality holds in {\it any} Lorentz-invariant theory described by an S-matrix. This includes both local quantum field theory and perturbative string theories~\cite{Adams:2006sv}, making it a very powerful probe. Violation of this dispersion relation indicates a violation of Lorentz invariance in the theory.

\begin{figure}
\centering
\includegraphics[width=3.5in]{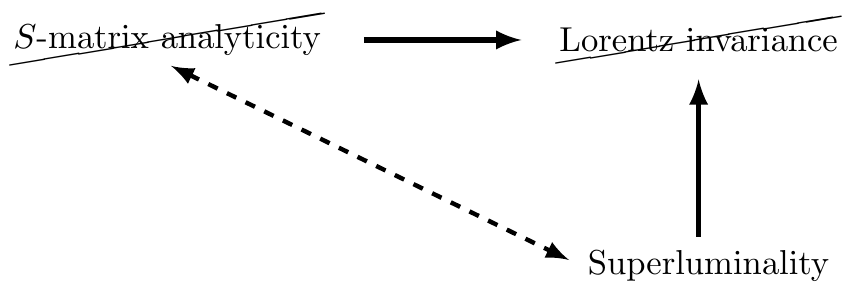}
\caption{\small Relationship between analyticity, locality and Lorentz invariance. Either non-analyticity of the $S$-matrix {\it or} superluminal propagation indicates a violation of Lorentz invariance. However, the implication between the two is less well understood.}
\end{figure}

We have seen that both superluminality and a violation of S-matrix analyticity indicate a violation of Lorentz invariance in a low energy effective theory. We might be tempted to posit that the relationship between these two things is tight, {\it i.e.}, that superluminality always implies that the theory will violate the sum rule~\eqref{dispersionsumrule}, but unfortunately the relationship is somewhat more subtle. For the simplest theories, the relationship is tight, but it is possible to construct theories that admit apparently superluminal signals, but which obey~\eqref{dispersionsumrule}~\cite{Nicolis:2009qm, Hinterbichler:2012yn, deRham:2013hsa}. Here we have only focused on the simplest dispersion relation, it is possible to extract more intricate ones using similar arguments~\cite{Nicolis:2009qm,Elvang:2012st,Bellazzini:2014waa}. It is generally expected that theories which admit superluminality should violate some dispersion relations, as they follow directly from Lorentz invariance and analyticity and we know that superluminal theories are secretly not Lorentz-invariant, but there is no known single dispersion relation that is always violated by a superluminal theory.
In any case, in order for a low-energy effective theory to be UV-completable by a local, Lorentz invariant QFT or perturbative string theory, it must satisfy the dispersion relation~\eqref{dispersionsumrule} (along with an infinite number of other ones) and it must not have superluminal signals in the effective theory.

Even the issue of whether a theory actually exhibits superluminality is somewhat subtle, as is emphasized in~\cite{deRham:2014lqa}. The superluminality that appears in {\it e.g.},~\eqref{pxspherics} and~\eqref{L3varphi} is in the {\it phase} velocity of perturbations:
\be
v_{\rm phase} = \frac{\omega}{k} 
\ee
However, this is not the relevant quantity for determining causality of the theory; it is perfectly acceptable to have superluminal phase velocities (and indeed, even group velocities), so long as the {\it front} velocity is luminal. This constraint derives from demanding that the retarded propagator has support only on the future light cone, which is satisfied provided that the index of refraction $n(\omega)$ defined through
\be
k = n(\omega) \omega~,
\ee
is analytic in the upper complex plane and that its limit is $n(\omega) \to 1$ as $\omega \to \infty$~\cite{Shore:2007um,fastlightslowlight,deRham:2014lqa}. The phase velocity can also be written in terms of the index of refraction as
\be
v_{\rm phase} = \frac{1}{n(\omega)}~,
\ee
which implies that in order for the theory to be causal, the infinite-momentum limit of the phase velocity---or the front velocity---must be luminal~\cite{Shore:2007um,deRham:2014lqa}
\be
\lim_{k\to \infty} v_{\rm phase}(k) = v_{\rm front} = 1~.
\ee

Here we see that the causality of a theory depends on it's high-energy behavior; formally the dispersion relation at infinite momentum. However, we know that as we go to high enough energies, theories which exhibit the Vainshtein mechanism become strongly-coupled, and quantum corrections become important. Therefore, in order to reliably compute the front velocity in a theory of this type, loop effects must be taken into account. The fact that a theory exhibits a superluminal dispersion relation at tree level in the Lagrangian is not sufficient to conclude that the theory is acausal. One concrete example of this type of phenomenon occurs in the context of galileon duality~\cite{deRham:2013hsa}. On one side of the duality, we have a free theory, where the front velocity can reliably computed to be luminal, while on the other side of the duality we have apparent superluminality. Presumably if quantum effects could be suitably re-summed, the front velocity on this side would also be 1.

\end{appendix}

\renewcommand{\em}{}
\bibliographystyle{utphys}
\addcontentsline{toc}{section}{References}
{\scriptsize \bibliography{gravreview48}}

\providecommand{\href}[2]{#2}\begingroup\raggedright\begin{thebibliography}{1000}

\bibitem{Riess:1998cb}
{\bf Supernova Search Team} Collaboration, A.~G. Riess {\em et al.},
  ``{Observational evidence from supernovae for an accelerating universe and a
  cosmological constant},'' \href{http://dx.doi.org/10.1086/300499}{{\em
  Astron.J.} {\bf 116} (1998)  1009--1038},
\href{http://arxiv.org/abs/astro-ph/9805201}{{\tt arXiv:astro-ph/9805201
  [astro-ph]}}.

\bibitem{Schmidt:1998ys}
{\bf Supernova Search Team} Collaboration, B.~P. Schmidt {\em et al.}, ``{The
  High Z supernova search: Measuring cosmic deceleration and global curvature
  of the universe using type Ia supernovae},''
  \href{http://dx.doi.org/10.1086/306308}{{\em Astrophys.J.} {\bf 507} (1998)
  46--63},
\href{http://arxiv.org/abs/astro-ph/9805200}{{\tt arXiv:astro-ph/9805200
  [astro-ph]}}.

\bibitem{Perlmutter:1998np}
{\bf Supernova Cosmology Project} Collaboration, S.~Perlmutter {\em et al.},
  ``{Measurements of Omega and Lambda from 42 high redshift supernovae},''
  \href{http://dx.doi.org/10.1086/307221}{{\em Astrophys.J.} {\bf 517} (1999)
  565--586},
\href{http://arxiv.org/abs/astro-ph/9812133}{{\tt arXiv:astro-ph/9812133
  [astro-ph]}}.

\bibitem{Garnavich:1998th}
{\bf Supernova Search Team} Collaboration, P.~M. Garnavich {\em et al.},
  ``{Supernova limits on the cosmic equation of state},''
  \href{http://dx.doi.org/10.1086/306495}{{\em Astrophys.J.} {\bf 509} (1998)
  74--79},
\href{http://arxiv.org/abs/astro-ph/9806396}{{\tt arXiv:astro-ph/9806396
  [astro-ph]}}.

\bibitem{Freedman:2000cf}
{\bf HST} Collaboration, W.~Freedman {\em et al.}, ``{Final results from the
  Hubble Space Telescope key project to measure the Hubble constant},''
  \href{http://dx.doi.org/10.1086/320638}{{\em Astrophys.J.} {\bf 553} (2001)
  47--72},
\href{http://arxiv.org/abs/astro-ph/0012376}{{\tt arXiv:astro-ph/0012376
  [astro-ph]}}.

\bibitem{Riess:2004nr}
{\bf Supernova Search Team} Collaboration, A.~G. Riess {\em et al.}, ``{Type Ia
  supernova discoveries at z $ >$ 1 from the Hubble Space Telescope: Evidence
  for past deceleration and constraints on dark energy evolution},''
  \href{http://dx.doi.org/10.1086/383612}{{\em Astrophys.J.} {\bf 607} (2004)
  665--687},
\href{http://arxiv.org/abs/astro-ph/0402512}{{\tt arXiv:astro-ph/0402512
  [astro-ph]}}.

\bibitem{Astier:2005qq}
{\bf SNLS} Collaboration, P.~Astier {\em et al.}, ``{The Supernova legacy
  survey: Measurement of omega(m), omega(lambda) and W from the first year data
  set},'' \href{http://dx.doi.org/10.1051/0004-6361:20054185}{{\em
  Astron.Astrophys.} {\bf 447} (2006)  31--48},
\href{http://arxiv.org/abs/astro-ph/0510447}{{\tt arXiv:astro-ph/0510447
  [astro-ph]}}.

\bibitem{Kowalski:2008ez}
{\bf Supernova Cosmology Project} Collaboration, M.~Kowalski {\em et al.},
  ``{Improved Cosmological Constraints from New, Old and Combined Supernova
  Datasets},'' \href{http://dx.doi.org/10.1086/589937}{{\em Astrophys.J.} {\bf
  686} (2008)  749--778},
\href{http://arxiv.org/abs/0804.4142}{{\tt arXiv:0804.4142 [astro-ph]}}.

\bibitem{Kessler:2009ys}
R.~Kessler, A.~Becker, D.~Cinabro, J.~Vanderplas, J.~A. Frieman, {\em et al.},
  ``{First-year Sloan Digital Sky Survey-II (SDSS-II) Supernova Results: Hubble
  Diagram and Cosmological Parameters},''
  \href{http://dx.doi.org/10.1088/0067-0049/185/1/32}{{\em Astrophys.J.Suppl.}
  {\bf 185} (2009)  32--84},
\href{http://arxiv.org/abs/0908.4274}{{\tt arXiv:0908.4274 [astro-ph.CO]}}.

\bibitem{Amanullah:2010vv}
R.~Amanullah, C.~Lidman, D.~Rubin, G.~Aldering, P.~Astier, {\em et al.},
  ``{Spectra and Light Curves of Six Type Ia Supernovae at 0.511 $<$ z $<$ 1.12
  and the Union2 Compilation},''
  \href{http://dx.doi.org/10.1088/0004-637X/716/1/712}{{\em Astrophys.J.} {\bf
  716} (2010)  712--738},
\href{http://arxiv.org/abs/1004.1711}{{\tt arXiv:1004.1711 [astro-ph.CO]}}.

\bibitem{Suzuki:2011hu}
N.~Suzuki, D.~Rubin, C.~Lidman, G.~Aldering, R.~Amanullah, {\em et al.}, ``{The
  Hubble Space Telescope Cluster Supernova Survey: V. Improving the Dark Energy
  Constraints Above z $>$ 1 and Building an Early-Type-Hosted Supernova
  Sample},'' \href{http://dx.doi.org/10.1088/0004-637X/746/1/85}{{\em
  Astrophys.J.} {\bf 746} (2012)  85},
\href{http://arxiv.org/abs/1105.3470}{{\tt arXiv:1105.3470 [astro-ph.CO]}}.

\bibitem{Sako:2014qmj}
{\bf SDSS} Collaboration, M.~Sako {\em et al.}, ``{The Data Release of the
  Sloan Digital Sky Survey-II Supernova Survey},''
\href{http://arxiv.org/abs/1401.3317}{{\tt arXiv:1401.3317 [astro-ph.CO]}}.

\bibitem{deBernardis:2000gy}
{\bf Boomerang} Collaboration, P.~de~Bernardis {\em et al.}, ``{A Flat universe
  from high resolution maps of the cosmic microwave background radiation},''
  \href{http://dx.doi.org/10.1038/35010035}{{\em Nature} {\bf 404} (2000)
  955--959},
\href{http://arxiv.org/abs/astro-ph/0004404}{{\tt arXiv:astro-ph/0004404
  [astro-ph]}}.

\bibitem{Lange:2000iq}
{\bf Boomerang} Collaboration, A.~E. Lange {\em et al.}, ``{Cosmological
  parameters from the first results of BOOMERANG},''
  \href{http://dx.doi.org/10.1103/PhysRevD.63.042001}{{\em Phys.Rev.} {\bf D63}
  (2001)  042001},
\href{http://arxiv.org/abs/astro-ph/0005004}{{\tt arXiv:astro-ph/0005004
  [astro-ph]}}.

\bibitem{Balbi:2000tg}
A.~Balbi, P.~Ade, J.~Bock, J.~Borrill, A.~Boscaleri, {\em et al.},
  ``{Constraints on cosmological parameters from MAXIMA-1},''
  \href{http://dx.doi.org/10.1086/323608}{{\em Astrophys.J.} {\bf 545} (2000)
  L1--L4},
\href{http://arxiv.org/abs/astro-ph/0005124}{{\tt arXiv:astro-ph/0005124
  [astro-ph]}}.

\bibitem{Pryke:2001yz}
C.~Pryke, N.~Halverson, E.~Leitch, J.~Kovac, J.~Carlstrom, {\em et al.},
  ``{Cosmological parameter extraction from the first season of observations
  with DASI},'' \href{http://dx.doi.org/10.1086/338880}{{\em Astrophys.J.} {\bf
  568} (2002)  46--51},
\href{http://arxiv.org/abs/astro-ph/0104490}{{\tt arXiv:astro-ph/0104490
  [astro-ph]}}.

\bibitem{Spergel:2003cb}
{\bf WMAP} Collaboration, D.~Spergel {\em et al.}, ``{First year Wilkinson
  Microwave Anisotropy Probe (WMAP) observations: Determination of cosmological
  parameters},'' \href{http://dx.doi.org/10.1086/377226}{{\em
  Astrophys.J.Suppl.} {\bf 148} (2003)  175--194},
\href{http://arxiv.org/abs/astro-ph/0302209}{{\tt arXiv:astro-ph/0302209
  [astro-ph]}}.

\bibitem{Hinshaw:2012aka}
{\bf WMAP} Collaboration, G.~Hinshaw {\em et al.}, ``{Nine-Year Wilkinson
  Microwave Anisotropy Probe (WMAP) Observations: Cosmological Parameter
  Results},''
\href{http://arxiv.org/abs/1212.5226}{{\tt arXiv:1212.5226 [astro-ph.CO]}}.

\bibitem{Hou:2012xq}
Z.~Hou, C.~Reichardt, K.~Story, B.~Follin, R.~Keisler, {\em et al.},
  ``{Constraints on Cosmology from the Cosmic Microwave Background Power
  Spectrum of the 2500-square degree SPT-SZ Survey},''
  \href{http://dx.doi.org/10.1088/0004-637X/782/2/74}{{\em Astrophys.J.} {\bf
  782} (2014)  74},
\href{http://arxiv.org/abs/1212.6267}{{\tt arXiv:1212.6267 [astro-ph.CO]}}.

\bibitem{Ade:2013zuv}
{\bf Planck} Collaboration, P.~Ade {\em et al.}, ``{Planck 2013 results. XVI.
  Cosmological parameters},''
\href{http://arxiv.org/abs/1303.5076}{{\tt arXiv:1303.5076 [astro-ph.CO]}}.

\bibitem{2013ApJ...779...86S}
{\bf South Pole Telescope} Collaboration, K.~T. {Story} {\em et al.}, ``{A
  Measurement of the Cosmic Microwave Background Damping Tail from the
  2500-Square-Degree SPT-SZ Survey},''
  \href{http://dx.doi.org/10.1088/0004-637X/779/1/86}{{\em Ap. J.} {\bf 779}
  (2013)  86}, \href{http://arxiv.org/abs/1210.7231}{{\tt arXiv:1210.7231
  [astro-ph.CO]}}.

\bibitem{Sievers:2013ica}
{\bf Atacama Cosmology Telescope} Collaboration, J.~L. Sievers {\em et al.},
  ``{The Atacama Cosmology Telescope: Cosmological parameters from three
  seasons of data},''
  \href{http://dx.doi.org/10.1088/1475-7516/2013/10/060}{{\em JCAP} {\bf 1310}
  (2013)  060},
\href{http://arxiv.org/abs/1301.0824}{{\tt arXiv:1301.0824 [astro-ph.CO]}}.

\bibitem{Tegmark:2003ud}
{\bf SDSS} Collaboration, M.~Tegmark {\em et al.}, ``{Cosmological parameters
  from SDSS and WMAP},''
  \href{http://dx.doi.org/10.1103/PhysRevD.69.103501}{{\em Phys.Rev.} {\bf D69}
  (2004)  103501},
\href{http://arxiv.org/abs/astro-ph/0310723}{{\tt arXiv:astro-ph/0310723
  [astro-ph]}}.

\bibitem{Seljak:2004xh}
{\bf SDSS} Collaboration, U.~Seljak {\em et al.}, ``{Cosmological parameter
  analysis including SDSS Ly-alpha forest and galaxy bias: Constraints on the
  primordial spectrum of fluctuations, neutrino mass, and dark energy},''
  \href{http://dx.doi.org/10.1103/PhysRevD.71.103515}{{\em Phys.Rev.} {\bf D71}
  (2005)  103515},
\href{http://arxiv.org/abs/astro-ph/0407372}{{\tt arXiv:astro-ph/0407372
  [astro-ph]}}.

\bibitem{Eisenstein:2005su}
{\bf SDSS} Collaboration, D.~J. Eisenstein {\em et al.}, ``{Detection of the
  baryon acoustic peak in the large-scale correlation function of SDSS luminous
  red galaxies},'' \href{http://dx.doi.org/10.1086/466512}{{\em Astrophys.J.}
  {\bf 633} (2005)  560--574},
\href{http://arxiv.org/abs/astro-ph/0501171}{{\tt arXiv:astro-ph/0501171
  [astro-ph]}}.

\bibitem{Blake:2011en}
C.~Blake, E.~Kazin, F.~Beutler, T.~Davis, D.~Parkinson, {\em et al.}, ``{The
  WiggleZ Dark Energy Survey: mapping the distance-redshift relation with
  baryon acoustic oscillations},''
  \href{http://dx.doi.org/10.1111/j.1365-2966.2011.19592.x}{{\em
  Mon.Not.Roy.Astron.Soc.} {\bf 418} (2011)  1707--1724},
\href{http://arxiv.org/abs/1108.2635}{{\tt arXiv:1108.2635 [astro-ph.CO]}}.

\bibitem{Beutler:2011hx}
F.~Beutler, C.~Blake, M.~Colless, D.~H. Jones, L.~Staveley-Smith, {\em et al.},
  ``{The 6dF Galaxy Survey: Baryon Acoustic Oscillations and the Local Hubble
  Constant},'' \href{http://dx.doi.org/10.1111/j.1365-2966.2011.19250.x}{{\em
  Mon.Not.Roy.Astron.Soc.} {\bf 416} (2011)  3017--3032},
\href{http://arxiv.org/abs/1106.3366}{{\tt arXiv:1106.3366 [astro-ph.CO]}}.

\bibitem{Dawson:2012va}
{\bf BOSS} Collaboration, K.~S. Dawson {\em et al.}, ``{The Baryon Oscillation
  Spectroscopic Survey of SDSS-III},''
  \href{http://dx.doi.org/10.1088/0004-6256/145/1/10}{{\em Astron.J.} {\bf 145}
  (2013)  10},
\href{http://arxiv.org/abs/1208.0022}{{\tt arXiv:1208.0022 [astro-ph.CO]}}.

\bibitem{2012MNRAS.427.3435A}
L.~{Anderson} {\em et al.}, ``{The clustering of galaxies in the SDSS-III
  Baryon Oscillation Spectroscopic Survey: baryon acoustic oscillations in the
  Data Release 9 spectroscopic galaxy sample},''
  \href{http://dx.doi.org/10.1111/j.1365-2966.2012.22066.x}{{\em MNRAS} {\bf
  427} (2012)  3435--3467}, \href{http://arxiv.org/abs/1203.6594}{{\tt
  arXiv:1203.6594 [astro-ph.CO]}}.

\bibitem{Samushia:2012iq}
L.~Samushia, B.~A. Reid, M.~White, W.~J. Percival, A.~J. Cuesta, {\em et al.},
  ``{The Clustering of Galaxies in the SDSS-III DR9 Baryon Oscillation
  Spectroscopic Survey: Testing Deviations from $\Lambda$ and General
  Relativity using anisotropic clustering of galaxies},''
  \href{http://dx.doi.org/10.1093/mnras/sts443}{{\em Mon.Not.Roy.Astron.Soc.}
  {\bf 429} (2013)  1514--1528},
\href{http://arxiv.org/abs/1206.5309}{{\tt arXiv:1206.5309 [astro-ph.CO]}}.

\bibitem{Ostriker:1995rn}
J.~Ostriker and P.~J. Steinhardt, ``{Cosmic concordance},''
\href{http://arxiv.org/abs/astro-ph/9505066}{{\tt arXiv:astro-ph/9505066
  [astro-ph]}}.

\bibitem{Bahcall:1999xn}
N.~A. Bahcall, J.~P. Ostriker, S.~Perlmutter, and P.~J. Steinhardt, ``{The
  Cosmic triangle: Assessing the state of the universe},''
  \href{http://dx.doi.org/10.1126/science.284.5419.1481}{{\em Science} {\bf
  284} (1999)  1481--1488},
\href{http://arxiv.org/abs/astro-ph/9906463}{{\tt arXiv:astro-ph/9906463
  [astro-ph]}}.

\bibitem{'tHooft:1979bh}
G.~'t~Hooft, ``{Naturalness, chiral symmetry, and spontaneous chiral symmetry
  breaking},''
{\em NATO Adv.Study Inst.Ser.B Phys.} {\bf 59} (1980)  135.

\bibitem{Weinberg:1988cp}
S.~Weinberg, ``{The Cosmological Constant Problem},''
\href{http://dx.doi.org/10.1103/RevModPhys.61.1}{{\em Rev.Mod.Phys.} {\bf 61}
  (1989)  1--23}.

\bibitem{Weinberg:1987dv}
S.~Weinberg, ``{Anthropic Bound on the Cosmological Constant},''
\href{http://dx.doi.org/10.1103/PhysRevLett.59.2607}{{\em Phys.Rev.Lett.} {\bf
  59} (1987)  2607}.

\bibitem{Susskind:2003kw}
L.~Susskind, ``{The Anthropic landscape of string theory},''
\href{http://arxiv.org/abs/hep-th/0302219}{{\tt arXiv:hep-th/0302219
  [hep-th]}}.

\bibitem{Bousso:2000xa}
R.~Bousso and J.~Polchinski, ``{Quantization of four form fluxes and dynamical
  neutralization of the cosmological constant},'' {\em JHEP} {\bf 0006} (2000)
  006,
\href{http://arxiv.org/abs/hep-th/0004134}{{\tt arXiv:hep-th/0004134
  [hep-th]}}.

\bibitem{Kachru:2003aw}
S.~Kachru, R.~Kallosh, A.~D. Linde, and S.~P. Trivedi, ``{De Sitter vacua in
  string theory},'' \href{http://dx.doi.org/10.1103/PhysRevD.68.046005}{{\em
  Phys.Rev.} {\bf D68} (2003)  046005},
\href{http://arxiv.org/abs/hep-th/0301240}{{\tt arXiv:hep-th/0301240
  [hep-th]}}.

\bibitem{Douglas:2003um}
M.~R. Douglas, ``{The Statistics of string / M theory vacua},''
  \href{http://dx.doi.org/10.1088/1126-6708/2003/05/046}{{\em JHEP} {\bf 0305}
  (2003)  046},
\href{http://arxiv.org/abs/hep-th/0303194}{{\tt arXiv:hep-th/0303194
  [hep-th]}}.

\bibitem{Douglas:2006es}
M.~R. Douglas and S.~Kachru, ``{Flux compactification},''
  \href{http://dx.doi.org/10.1103/RevModPhys.79.733}{{\em Rev.Mod.Phys.} {\bf
  79} (2007)  733--796},
\href{http://arxiv.org/abs/hep-th/0610102}{{\tt arXiv:hep-th/0610102
  [hep-th]}}.

\bibitem{Papapetrou:1948jw}
A.~Papapetrou, ``{Einstein's theory of gravitation and flat space},''
{\em Proc.Roy.Irish Acad.(Sect.A)} {\bf 52A} (1948)  11--23.

\bibitem{Gupta:1952zz}
S.~Gupta, ``{Quantization of Einstein's gravitational field: general
  treatment},''
{\em Proc.Phys.Soc.} {\bf A65} (1952)  608--619.

\bibitem{Kraichnan:1955zz}
R.~H. Kraichnan, ``{Special-Relativistic Derivation of Generally Covariant
  Gravitation Theory},''
\href{http://dx.doi.org/10.1103/PhysRev.98.1118}{{\em Phys.Rev.} {\bf 98}
  (1955)  1118--1122}.

\bibitem{Weinberg:1965rz}
S.~Weinberg, ``{Photons and gravitons in perturbation theory: Derivation of
  Maxwell's and Einstein's equations},''
\href{http://dx.doi.org/10.1103/PhysRev.138.B988}{{\em Phys.Rev.} {\bf 138}
  (1965)  B988--B1002}.

\bibitem{Feynman:1996kb}
R.~Feynman, F.~Morinigo, W.~Wagner, and B.~Hatfield,
``{Feynman lectures on gravitation},''.

\bibitem{Deser:1969wk}
S.~Deser, ``{Selfinteraction and gauge invariance},''
  \href{http://dx.doi.org/10.1007/BF00759198}{{\em Gen.Rel.Grav.} {\bf 1}
  (1970)  9--18},
\href{http://arxiv.org/abs/gr-qc/0411023}{{\tt arXiv:gr-qc/0411023 [gr-qc]}}.

\bibitem{Khoury:2011ay}
J.~Khoury, G.~E. Miller, and A.~J. Tolley, ``{Spatially Covariant Theories of a
  Transverse, Traceless Graviton, Part I: Formalism},''
  \href{http://dx.doi.org/10.1103/PhysRevD.85.084002}{{\em Phys.Rev.} {\bf D85}
  (2012)  084002},
\href{http://arxiv.org/abs/1108.1397}{{\tt arXiv:1108.1397 [hep-th]}}.

\bibitem{Khoury:2013oqa}
J.~Khoury, G.~E.~J. Miller, and A.~J. Tolley, ``{On the Origin of Gravitational
  Lorentz Covariance},''
\href{http://arxiv.org/abs/1305.0822}{{\tt arXiv:1305.0822 [hep-th]}}.

\bibitem{Coleman:1977py}
S.~R. Coleman, ``{The Fate of the False Vacuum. 1. Semiclassical Theory},''
\href{http://dx.doi.org/10.1103/PhysRevD.15.2929,
  10.1103/PhysRevD.16.1248}{{\em Phys.Rev.} {\bf D15} (1977)  2929--2936}.

\bibitem{Callan:1977pt}
J.~Callan, Curtis~G. and S.~R. Coleman, ``{The Fate of the False Vacuum. 2.
  First Quantum Corrections},''
\href{http://dx.doi.org/10.1103/PhysRevD.16.1762}{{\em Phys.Rev.} {\bf D16}
  (1977)  1762--1768}.

\bibitem{Coleman:1980aw}
S.~R. Coleman and F.~De~Luccia, ``{Gravitational Effects on and of Vacuum
  Decay},''
\href{http://dx.doi.org/10.1103/PhysRevD.21.3305}{{\em Phys.Rev.} {\bf D21}
  (1980)  3305}.

\bibitem{Vilenkin:1983xq}
A.~Vilenkin, ``{The Birth of Inflationary Universes},''
\href{http://dx.doi.org/10.1103/PhysRevD.27.2848}{{\em Phys.Rev.} {\bf D27}
  (1983)  2848}.

\bibitem{Linde:1986fd}
A.~D. Linde, ``{Eternally Existing Selfreproducing Chaotic Inflationary
  Universe},''
\href{http://dx.doi.org/10.1016/0370-2693(86)90611-8}{{\em Phys.Lett.} {\bf
  B175} (1986)  395--400}.

\bibitem{Vilenkin:1994ua}
A.~Vilenkin, ``{Predictions from quantum cosmology},''
  \href{http://dx.doi.org/10.1103/PhysRevLett.74.846}{{\em Phys.Rev.Lett.} {\bf
  74} (1995)  846--849},
\href{http://arxiv.org/abs/gr-qc/9406010}{{\tt arXiv:gr-qc/9406010 [gr-qc]}}.

\bibitem{Guth:2000ka}
A.~H. Guth, ``{Inflation and eternal inflation},''
  \href{http://dx.doi.org/10.1016/S0370-1573(00)00037-5}{{\em Phys.Rept.} {\bf
  333} (2000)  555--574},
\href{http://arxiv.org/abs/astro-ph/0002156}{{\tt arXiv:astro-ph/0002156
  [astro-ph]}}.

\bibitem{Creminelli:2008es}
P.~Creminelli, S.~Dubovsky, A.~Nicolis, L.~Senatore, and M.~Zaldarriaga, ``{The
  Phase Transition to Slow-roll Eternal Inflation},''
  \href{http://dx.doi.org/10.1088/1126-6708/2008/09/036}{{\em JHEP} {\bf 0809}
  (2008)  036},
\href{http://arxiv.org/abs/0802.1067}{{\tt arXiv:0802.1067 [hep-th]}}.

\bibitem{Martinec:2014uva}
E.~J. Martinec and W.~E. Moore, ``{Modeling Quantum Gravity Effects in
  Inflation},''
\href{http://arxiv.org/abs/1401.7681}{{\tt arXiv:1401.7681 [hep-th]}}.

\bibitem{Boddy:2014eba}
K.~K. Boddy, S.~M. Carroll, and J.~Pollack, ``{De Sitter Space Without Quantum
  Fluctuations},''
\href{http://arxiv.org/abs/1405.0298}{{\tt arXiv:1405.0298 [hep-th]}}.

\bibitem{Starobinsky:1979ty}
A.~A. Starobinsky, ``{Spectrum of relict gravitational radiation and the early
  state of the universe},''
{\em JETP Lett.} {\bf 30} (1979)  682--685.

\bibitem{Guth:1980zm}
A.~H. Guth, ``{The Inflationary Universe: A Possible Solution to the Horizon
  and Flatness Problems},''
\href{http://dx.doi.org/10.1103/PhysRevD.23.347}{{\em Phys.Rev.} {\bf D23}
  (1981)  347--356}.

\bibitem{Albrecht:1982wi}
A.~Albrecht and P.~J. Steinhardt, ``{Cosmology for Grand Unified Theories with
  Radiatively Induced Symmetry Breaking},''
\href{http://dx.doi.org/10.1103/PhysRevLett.48.1220}{{\em Phys.Rev.Lett.} {\bf
  48} (1982)  1220--1223}.

\bibitem{Linde:1981mu}
A.~D. Linde, ``{A New Inflationary Universe Scenario: A Possible Solution of
  the Horizon, Flatness, Homogeneity, Isotropy and Primordial Monopole
  Problems},''
\href{http://dx.doi.org/10.1016/0370-2693(82)91219-9}{{\em Phys.Lett.} {\bf
  B108} (1982)  389--393}.

\bibitem{Fujii:1982ms}
Y.~Fujii, ``{Origin of the Gravitational Constant and Particle Masses in Scale
  Invariant Scalar - Tensor Theory},''
\href{http://dx.doi.org/10.1103/PhysRevD.26.2580}{{\em Phys.Rev.} {\bf D26}
  (1982)  2580}.

\bibitem{Ford:1987de}
L.~Ford, ``{Cosmological Constant Damping By Unstable Scalar Fields},''
\href{http://dx.doi.org/10.1103/PhysRevD.35.2339}{{\em Phys.Rev.} {\bf D35}
  (1987)  2339}.

\bibitem{Wetterich:1987fm}
C.~Wetterich, ``{Cosmology and the Fate of Dilatation Symmetry},''
\href{http://dx.doi.org/10.1016/0550-3213(88)90193-9}{{\em Nucl.Phys.} {\bf
  B302} (1988)  668}.

\bibitem{Peebles:1987ek}
P.~Peebles and B.~Ratra, ``{Cosmology with a Time Variable Cosmological
  Constant},''
{\em Astrophys.J.} {\bf 325} (1988)  L17.

\bibitem{Ratra:1987rm}
B.~Ratra and P.~Peebles, ``{Cosmological Consequences of a Rolling Homogeneous
  Scalar Field},''
\href{http://dx.doi.org/10.1103/PhysRevD.37.3406}{{\em Phys.Rev.} {\bf D37}
  (1988)  3406}.

\bibitem{Caldwell:1997ii}
R.~Caldwell, R.~Dave, and P.~J. Steinhardt, ``{Cosmological imprint of an
  energy component with general equation of state},''
  \href{http://dx.doi.org/10.1103/PhysRevLett.80.1582}{{\em Phys.Rev.Lett.}
  {\bf 80} (1998)  1582--1585},
\href{http://arxiv.org/abs/astro-ph/9708069}{{\tt arXiv:astro-ph/9708069
  [astro-ph]}}.

\bibitem{Liddle:1998xm}
A.~R. Liddle and R.~J. Scherrer, ``{A Classification of scalar field potentials
  with cosmological scaling solutions},''
  \href{http://dx.doi.org/10.1103/PhysRevD.59.023509}{{\em Phys.Rev.} {\bf D59}
  (1999)  023509},
\href{http://arxiv.org/abs/astro-ph/9809272}{{\tt arXiv:astro-ph/9809272
  [astro-ph]}}.

\bibitem{Kolda:1998wq}
C.~F. Kolda and D.~H. Lyth, ``{Quintessential difficulties},''
  \href{http://dx.doi.org/10.1016/S0370-2693(99)00657-7}{{\em Phys.Lett.} {\bf
  B458} (1999)  197--201},
\href{http://arxiv.org/abs/hep-ph/9811375}{{\tt arXiv:hep-ph/9811375
  [hep-ph]}}.

\bibitem{Amendola:1999er}
L.~Amendola, ``{Coupled quintessence},''
  \href{http://dx.doi.org/10.1103/PhysRevD.62.043511}{{\em Phys.Rev.} {\bf D62}
  (2000)  043511},
\href{http://arxiv.org/abs/astro-ph/9908023}{{\tt arXiv:astro-ph/9908023
  [astro-ph]}}.

\bibitem{Wang:1999fa}
L.-M. Wang, R.~Caldwell, J.~Ostriker, and P.~J. Steinhardt, ``{Cosmic
  concordance and quintessence},'' \href{http://dx.doi.org/10.1086/308331}{{\em
  Astrophys.J.} {\bf 530} (2000)  17--35},
\href{http://arxiv.org/abs/astro-ph/9901388}{{\tt arXiv:astro-ph/9901388
  [astro-ph]}}.

\bibitem{Cooray:1999da}
A.~R. Cooray and D.~Huterer, ``{Gravitational lensing as a probe of
  quintessence},'' \href{http://dx.doi.org/10.1086/311927}{{\em Astrophys.J.}
  {\bf 513} (1999)  L95--L98},
\href{http://arxiv.org/abs/astro-ph/9901097}{{\tt arXiv:astro-ph/9901097
  [astro-ph]}}.

\bibitem{Barreiro:1999zs}
T.~Barreiro, E.~J. Copeland, and N.~Nunes, ``{Quintessence arising from
  exponential potentials},''
  \href{http://dx.doi.org/10.1103/PhysRevD.61.127301}{{\em Phys.Rev.} {\bf D61}
  (2000)  127301},
\href{http://arxiv.org/abs/astro-ph/9910214}{{\tt arXiv:astro-ph/9910214
  [astro-ph]}}.

\bibitem{Huterer:2000mj}
D.~Huterer and M.~S. Turner, ``{Probing the dark energy: Methods and
  strategies},'' \href{http://dx.doi.org/10.1103/PhysRevD.64.123527}{{\em
  Phys.Rev.} {\bf D64} (2001)  123527},
\href{http://arxiv.org/abs/astro-ph/0012510}{{\tt arXiv:astro-ph/0012510
  [astro-ph]}}.

\bibitem{Chevallier:2000qy}
M.~Chevallier and D.~Polarski, ``{Accelerating universes with scaling dark
  matter},'' \href{http://dx.doi.org/10.1142/S0218271801000822}{{\em
  Int.J.Mod.Phys.} {\bf D10} (2001)  213--224},
\href{http://arxiv.org/abs/gr-qc/0009008}{{\tt arXiv:gr-qc/0009008 [gr-qc]}}.

\bibitem{Boyle:2001du}
L.~A. Boyle, R.~R. Caldwell, and M.~Kamionkowski, ``{Spintessence! New models
  for dark matter and dark energy},''
  \href{http://dx.doi.org/10.1016/S0370-2693(02)02590-X}{{\em Phys.Lett.} {\bf
  B545} (2002)  17--22},
\href{http://arxiv.org/abs/astro-ph/0105318}{{\tt arXiv:astro-ph/0105318
  [astro-ph]}}.

\bibitem{Melchiorri:2002ux}
A.~Melchiorri, L.~Mersini-Houghton, C.~J. Odman, and M.~Trodden, ``{The State
  of the dark energy equation of state},''
  \href{http://dx.doi.org/10.1103/PhysRevD.68.043509}{{\em Phys.Rev.} {\bf D68}
  (2003)  043509},
\href{http://arxiv.org/abs/astro-ph/0211522}{{\tt arXiv:astro-ph/0211522
  [astro-ph]}}.

\bibitem{Pilo:2003gu}
L.~Pilo, D.~Rayner, and A.~Riotto, ``{Gauge quintessence},''
  \href{http://dx.doi.org/10.1103/PhysRevD.68.043503}{{\em Phys.Rev.} {\bf D68}
  (2003)  043503},
\href{http://arxiv.org/abs/hep-ph/0302087}{{\tt arXiv:hep-ph/0302087
  [hep-ph]}}.

\bibitem{Simon:2004tf}
J.~Simon, L.~Verde, and R.~Jimenez, ``{Constraints on the redshift dependence
  of the dark energy potential},''
  \href{http://dx.doi.org/10.1103/PhysRevD.71.123001}{{\em Phys.Rev.} {\bf D71}
  (2005)  123001},
\href{http://arxiv.org/abs/astro-ph/0412269}{{\tt arXiv:astro-ph/0412269
  [astro-ph]}}.

\bibitem{Huterer:2006mv}
D.~Huterer and H.~V. Peiris, ``{Dynamical behavior of generic quintessence
  potentials: Constraints on key dark energy observables},''
  \href{http://dx.doi.org/10.1103/PhysRevD.75.083503}{{\em Phys.Rev.} {\bf D75}
  (2007)  083503},
\href{http://arxiv.org/abs/astro-ph/0610427}{{\tt arXiv:astro-ph/0610427
  [astro-ph]}}.

\bibitem{Lim:2010yk}
E.~A. Lim, I.~Sawicki, and A.~Vikman, ``{Dust of Dark Energy},''
  \href{http://dx.doi.org/10.1088/1475-7516/2010/05/012}{{\em JCAP} {\bf 1005}
  (2010)  012},
\href{http://arxiv.org/abs/1003.5751}{{\tt arXiv:1003.5751 [astro-ph.CO]}}.

\bibitem{Mortonson:2010mj}
M.~J. Mortonson, W.~Hu, and D.~Huterer, ``{Simultaneous Falsification of
  $\Lambda$CDM and Quintessence with Massive, Distant Clusters},''
  \href{http://dx.doi.org/10.1103/PhysRevD.83.023015}{{\em Phys.Rev.} {\bf D83}
  (2011)  023015},
\href{http://arxiv.org/abs/1011.0004}{{\tt arXiv:1011.0004 [astro-ph.CO]}}.

\bibitem{Marsh:2014xoa}
D.~J.~E. Marsh, P.~Bull, P.~G. Ferreira, and A.~Pontzen, ``{Quintessence in a
  quandary: On prior dependence in dark energy models},''
\href{http://arxiv.org/abs/1406.2301}{{\tt arXiv:1406.2301 [astro-ph.CO]}}.

\bibitem{Sahni:1999gb}
V.~Sahni and A.~A. Starobinsky, ``{The Case for a positive cosmological Lambda
  term},'' {\em Int.J.Mod.Phys.} {\bf D9} (2000)  373--444,
\href{http://arxiv.org/abs/astro-ph/9904398}{{\tt arXiv:astro-ph/9904398
  [astro-ph]}}.

\bibitem{Peebles:2002gy}
P.~Peebles and B.~Ratra, ``{The Cosmological constant and dark energy},''
  \href{http://dx.doi.org/10.1103/RevModPhys.75.559}{{\em Rev.Mod.Phys.} {\bf
  75} (2003)  559--606},
\href{http://arxiv.org/abs/astro-ph/0207347}{{\tt arXiv:astro-ph/0207347
  [astro-ph]}}.

\bibitem{Padmanabhan:2002ji}
T.~Padmanabhan, ``{Cosmological constant: The Weight of the vacuum},''
  \href{http://dx.doi.org/10.1016/S0370-1573(03)00120-0}{{\em Phys.Rept.} {\bf
  380} (2003)  235--320},
\href{http://arxiv.org/abs/hep-th/0212290}{{\tt arXiv:hep-th/0212290
  [hep-th]}}.

\bibitem{Copeland:2006wr}
E.~J. Copeland, M.~Sami, and S.~Tsujikawa, ``{Dynamics of dark energy},''
  \href{http://dx.doi.org/10.1142/S021827180600942X}{{\em Int.J.Mod.Phys.} {\bf
  D15} (2006)  1753--1936},
\href{http://arxiv.org/abs/hep-th/0603057}{{\tt arXiv:hep-th/0603057
  [hep-th]}}.

\bibitem{Linder:2007wa}
E.~V. Linder, ``{The Dynamics of Quintessence, The Quintessence of Dynamics},''
  \href{http://dx.doi.org/10.1007/s10714-007-0550-z}{{\em Gen.Rel.Grav.} {\bf
  40} (2008)  329--356},
\href{http://arxiv.org/abs/0704.2064}{{\tt arXiv:0704.2064 [astro-ph]}}.

\bibitem{Frieman:2008sn}
J.~Frieman, M.~Turner, and D.~Huterer, ``{Dark Energy and the Accelerating
  Universe},''
  \href{http://dx.doi.org/10.1146/annurev.astro.46.060407.145243}{{\em
  Ann.Rev.Astron.Astrophys.} {\bf 46} (2008)  385--432},
\href{http://arxiv.org/abs/0803.0982}{{\tt arXiv:0803.0982 [astro-ph]}}.

\bibitem{Li:2011sd}
M.~Li, X.-D. Li, S.~Wang, and Y.~Wang, ``{Dark Energy},''
  \href{http://dx.doi.org/10.1088/0253-6102/56/3/24}{{\em Commun.Theor.Phys.}
  {\bf 56} (2011)  525--604},
\href{http://arxiv.org/abs/1103.5870}{{\tt arXiv:1103.5870 [astro-ph.CO]}}.

\bibitem{Tsujikawa:2013fta}
S.~Tsujikawa, ``{Quintessence: A Review},''
  \href{http://dx.doi.org/10.1088/0264-9381/30/21/214003}{{\em
  Class.Quant.Grav.} {\bf 30} (2013)  214003},
\href{http://arxiv.org/abs/1304.1961}{{\tt arXiv:1304.1961 [gr-qc]}}.

\bibitem{Caldwell:1999ew}
R.~Caldwell, ``{A Phantom menace?},''
  \href{http://dx.doi.org/10.1016/S0370-2693(02)02589-3}{{\em Phys.Lett.} {\bf
  B545} (2002)  23--29},
\href{http://arxiv.org/abs/astro-ph/9908168}{{\tt arXiv:astro-ph/9908168
  [astro-ph]}}.

\bibitem{Faraoni:2001tq}
V.~Faraoni, ``{Superquintessence},''
  \href{http://dx.doi.org/10.1142/S0218271802001809}{{\em Int.J.Mod.Phys.} {\bf
  D11} (2002)  471--482},
\href{http://arxiv.org/abs/astro-ph/0110067}{{\tt arXiv:astro-ph/0110067
  [astro-ph]}}.

\bibitem{Caldwell:2003vq}
R.~R. Caldwell, M.~Kamionkowski, and N.~N. Weinberg, ``{Phantom energy and
  cosmic doomsday},''
  \href{http://dx.doi.org/10.1103/PhysRevLett.91.071301}{{\em Phys.Rev.Lett.}
  {\bf 91} (2003)  071301},
\href{http://arxiv.org/abs/astro-ph/0302506}{{\tt arXiv:astro-ph/0302506
  [astro-ph]}}.

\bibitem{Carroll:2003st}
S.~M. Carroll, M.~Hoffman, and M.~Trodden, ``{Can the dark energy equation - of
  - state parameter w be less than -1?},''
  \href{http://dx.doi.org/10.1103/PhysRevD.68.023509}{{\em Phys.Rev.} {\bf D68}
  (2003)  023509},
\href{http://arxiv.org/abs/astro-ph/0301273}{{\tt arXiv:astro-ph/0301273
  [astro-ph]}}.

\bibitem{Nojiri:2003vn}
S.~Nojiri and S.~D. Odintsov, ``{Quantum de Sitter cosmology and phantom
  matter},'' \href{http://dx.doi.org/10.1016/S0370-2693(03)00594-X}{{\em
  Phys.Lett.} {\bf B562} (2003)  147--152},
\href{http://arxiv.org/abs/hep-th/0303117}{{\tt arXiv:hep-th/0303117
  [hep-th]}}.

\bibitem{Singh:2003vx}
P.~Singh, M.~Sami, and N.~Dadhich, ``{Cosmological dynamics of phantom
  field},'' \href{http://dx.doi.org/10.1103/PhysRevD.68.023522}{{\em Phys.Rev.}
  {\bf D68} (2003)  023522},
\href{http://arxiv.org/abs/hep-th/0305110}{{\tt arXiv:hep-th/0305110
  [hep-th]}}.

\bibitem{Elizalde:2004mq}
E.~Elizalde, S.~Nojiri, and S.~D. Odintsov, ``{Late-time cosmology in (phantom)
  scalar-tensor theory: Dark energy and the cosmic speed-up},''
  \href{http://dx.doi.org/10.1103/PhysRevD.70.043539}{{\em Phys.Rev.} {\bf D70}
  (2004)  043539},
\href{http://arxiv.org/abs/hep-th/0405034}{{\tt arXiv:hep-th/0405034
  [hep-th]}}.

\bibitem{Feng:2004ad}
B.~Feng, X.-L. Wang, and X.-M. Zhang, ``{Dark energy constraints from the
  cosmic age and supernova},''
  \href{http://dx.doi.org/10.1016/j.physletb.2004.12.071}{{\em Phys.Lett.} {\bf
  B607} (2005)  35--41},
\href{http://arxiv.org/abs/astro-ph/0404224}{{\tt arXiv:astro-ph/0404224
  [astro-ph]}}.

\bibitem{Guo:2004fq}
Z.-K. Guo, Y.-S. Piao, X.-M. Zhang, and Y.-Z. Zhang, ``{Cosmological evolution
  of a quintom model of dark energy},''
  \href{http://dx.doi.org/10.1016/j.physletb.2005.01.017}{{\em Phys.Lett.} {\bf
  B608} (2005)  177--182},
\href{http://arxiv.org/abs/astro-ph/0410654}{{\tt arXiv:astro-ph/0410654
  [astro-ph]}}.

\bibitem{Vikman:2004dc}
A.~Vikman, ``{Can dark energy evolve to the phantom?},''
  \href{http://dx.doi.org/10.1103/PhysRevD.71.023515}{{\em Phys.Rev.} {\bf D71}
  (2005)  023515},
\href{http://arxiv.org/abs/astro-ph/0407107}{{\tt arXiv:astro-ph/0407107
  [astro-ph]}}.

\bibitem{Hu:2004kh}
W.~Hu, ``{Crossing the phantom divide: Dark energy internal degrees of
  freedom},'' \href{http://dx.doi.org/10.1103/PhysRevD.71.047301}{{\em
  Phys.Rev.} {\bf D71} (2005)  047301},
\href{http://arxiv.org/abs/astro-ph/0410680}{{\tt arXiv:astro-ph/0410680
  [astro-ph]}}.

\bibitem{Nojiri:2005sx}
S.~Nojiri, S.~D. Odintsov, and S.~Tsujikawa, ``{Properties of singularities in
  (phantom) dark energy universe},''
  \href{http://dx.doi.org/10.1103/PhysRevD.71.063004}{{\em Phys.Rev.} {\bf D71}
  (2005)  063004},
\href{http://arxiv.org/abs/hep-th/0501025}{{\tt arXiv:hep-th/0501025
  [hep-th]}}.

\bibitem{Nojiri:2005sr}
S.~Nojiri and S.~D. Odintsov, ``{Inhomogeneous equation of state of the
  universe: Phantom era, future singularity and crossing the phantom
  barrier},'' \href{http://dx.doi.org/10.1103/PhysRevD.72.023003}{{\em
  Phys.Rev.} {\bf D72} (2005)  023003},
\href{http://arxiv.org/abs/hep-th/0505215}{{\tt arXiv:hep-th/0505215
  [hep-th]}}.

\bibitem{Cai:2009zp}
Y.-F. Cai, E.~N. Saridakis, M.~R. Setare, and J.-Q. Xia, ``{Quintom Cosmology:
  Theoretical implications and observations},''
  \href{http://dx.doi.org/10.1016/j.physrep.2010.04.001}{{\em Phys.Rept.} {\bf
  493} (2010)  1--60},
\href{http://arxiv.org/abs/0909.2776}{{\tt arXiv:0909.2776 [hep-th]}}.

\bibitem{Freese:1990rb}
K.~Freese, J.~A. Frieman, and A.~V. Olinto, ``{Natural inflation with pseudo -
  Nambu-Goldstone bosons},''
\href{http://dx.doi.org/10.1103/PhysRevLett.65.3233}{{\em Phys.Rev.Lett.} {\bf
  65} (1990)  3233--3236}.

\bibitem{Frieman:1995pm}
J.~A. Frieman, C.~T. Hill, A.~Stebbins, and I.~Waga, ``{Cosmology with
  ultralight pseudo Nambu-Goldstone bosons},''
  \href{http://dx.doi.org/10.1103/PhysRevLett.75.2077}{{\em Phys.Rev.Lett.}
  {\bf 75} (1995)  2077--2080},
\href{http://arxiv.org/abs/astro-ph/9505060}{{\tt arXiv:astro-ph/9505060
  [astro-ph]}}.

\bibitem{Kaloper:2005aj}
N.~Kaloper and L.~Sorbo, ``{Of pngb quintessence},''
  \href{http://dx.doi.org/10.1088/1475-7516/2006/04/007}{{\em JCAP} {\bf 0604}
  (2006)  007},
\href{http://arxiv.org/abs/astro-ph/0511543}{{\tt arXiv:astro-ph/0511543
  [astro-ph]}}.

\bibitem{Peebles:1998qn}
P.~Peebles and A.~Vilenkin, ``{Quintessential inflation},''
  \href{http://dx.doi.org/10.1103/PhysRevD.59.063505}{{\em Phys.Rev.} {\bf D59}
  (1999)  063505},
\href{http://arxiv.org/abs/astro-ph/9810509}{{\tt arXiv:astro-ph/9810509
  [astro-ph]}}.

\bibitem{Peloso:1999dm}
M.~Peloso and F.~Rosati, ``{On the construction of quintessential inflation
  models},'' \href{http://dx.doi.org/10.1088/1126-6708/1999/12/026}{{\em JHEP}
  {\bf 9912} (1999)  026},
\href{http://arxiv.org/abs/hep-ph/9908271}{{\tt arXiv:hep-ph/9908271
  [hep-ph]}}.

\bibitem{Kaganovich:2000fc}
A.~Kaganovich, ``{Field theory model giving rise to 'quintessential inflation'
  without the cosmological constant and other fine tuning problems},''
  \href{http://dx.doi.org/10.1103/PhysRevD.63.025022}{{\em Phys.Rev.} {\bf D63}
  (2001)  025022},
\href{http://arxiv.org/abs/hep-th/0007144}{{\tt arXiv:hep-th/0007144
  [hep-th]}}.

\bibitem{Dimopoulos:2001ix}
K.~Dimopoulos and J.~Valle, ``{Modeling quintessential inflation},''
  \href{http://dx.doi.org/10.1016/S0927-6505(02)00115-9}{{\em Astropart.Phys.}
  {\bf 18} (2002)  287--306},
\href{http://arxiv.org/abs/astro-ph/0111417}{{\tt arXiv:astro-ph/0111417
  [astro-ph]}}.

\bibitem{Majumdar:2001mm}
A.~Majumdar, ``{From brane assisted inflation to quintessence through a single
  scalar field},'' \href{http://dx.doi.org/10.1103/PhysRevD.64.083503}{{\em
  Phys.Rev.} {\bf D64} (2001)  083503},
\href{http://arxiv.org/abs/astro-ph/0105518}{{\tt arXiv:astro-ph/0105518
  [astro-ph]}}.

\bibitem{Rosenfeld:2005mt}
R.~Rosenfeld and J.~Frieman, ``{A Simple model for quintessential inflation},''
  \href{http://dx.doi.org/10.1088/1475-7516/2005/09/003}{{\em JCAP} {\bf 0509}
  (2005)  003},
\href{http://arxiv.org/abs/astro-ph/0504191}{{\tt arXiv:astro-ph/0504191
  [astro-ph]}}.

\bibitem{Neupane:2007mu}
I.~P. Neupane, ``{Reconstructing a model of quintessential inflation},''
  \href{http://dx.doi.org/10.1088/0264-9381/25/12/125013}{{\em
  Class.Quant.Grav.} {\bf 25} (2008)  125013},
\href{http://arxiv.org/abs/0706.2654}{{\tt arXiv:0706.2654 [hep-th]}}.

\bibitem{Neupane:2007jm}
I.~P. Neupane and C.~Scherer, ``{Inflation and Quintessence: Theoretical
  Approach of Cosmological Reconstruction},''
  \href{http://dx.doi.org/10.1088/1475-7516/2008/05/009}{{\em JCAP} {\bf 0805}
  (2008)  009},
\href{http://arxiv.org/abs/0712.2468}{{\tt arXiv:0712.2468 [astro-ph]}}.

\bibitem{Bose:2008ew}
N.~Bose and A.~Majumdar, ``{A k-essence Model Of Inflation, Dark Matter and
  Dark Energy},'' \href{http://dx.doi.org/10.1103/PhysRevD.79.103517}{{\em
  Phys.Rev.} {\bf D79} (2009)  103517},
\href{http://arxiv.org/abs/0812.4131}{{\tt arXiv:0812.4131 [astro-ph]}}.

\bibitem{Wetterich:1994bg}
C.~Wetterich, ``{The Cosmon model for an asymptotically vanishing time
  dependent cosmological 'constant'},'' {\em Astron.Astrophys.} {\bf 301}
  (1995)  321--328,
\href{http://arxiv.org/abs/hep-th/9408025}{{\tt arXiv:hep-th/9408025
  [hep-th]}}.

\bibitem{Copeland:1997et}
E.~J. Copeland, A.~R. Liddle, and D.~Wands, ``{Exponential potentials and
  cosmological scaling solutions},''
  \href{http://dx.doi.org/10.1103/PhysRevD.57.4686}{{\em Phys.Rev.} {\bf D57}
  (1998)  4686--4690},
\href{http://arxiv.org/abs/gr-qc/9711068}{{\tt arXiv:gr-qc/9711068 [gr-qc]}}.

\bibitem{Ferreira:1997hj}
P.~G. Ferreira and M.~Joyce, ``{Cosmology with a primordial scaling field},''
  \href{http://dx.doi.org/10.1103/PhysRevD.58.023503}{{\em Phys.Rev.} {\bf D58}
  (1998)  023503},
\href{http://arxiv.org/abs/astro-ph/9711102}{{\tt arXiv:astro-ph/9711102
  [astro-ph]}}.

\bibitem{Ferreira:1997au}
P.~G. Ferreira and M.~Joyce, ``{Structure formation with a selftuning scalar
  field},'' \href{http://dx.doi.org/10.1103/PhysRevLett.79.4740}{{\em
  Phys.Rev.Lett.} {\bf 79} (1997)  4740--4743},
\href{http://arxiv.org/abs/astro-ph/9707286}{{\tt arXiv:astro-ph/9707286
  [astro-ph]}}.

\bibitem{Zlatev:1998tr}
I.~Zlatev, L.-M. Wang, and P.~J. Steinhardt, ``{Quintessence, cosmic
  coincidence, and the cosmological constant},''
  \href{http://dx.doi.org/10.1103/PhysRevLett.82.896}{{\em Phys.Rev.Lett.} {\bf
  82} (1999)  896--899},
\href{http://arxiv.org/abs/astro-ph/9807002}{{\tt arXiv:astro-ph/9807002
  [astro-ph]}}.

\bibitem{Steinhardt:1999nw}
P.~J. Steinhardt, L.-M. Wang, and I.~Zlatev, ``{Cosmological tracking
  solutions},'' \href{http://dx.doi.org/10.1103/PhysRevD.59.123504}{{\em
  Phys.Rev.} {\bf D59} (1999)  123504},
\href{http://arxiv.org/abs/astro-ph/9812313}{{\tt arXiv:astro-ph/9812313
  [astro-ph]}}.

\bibitem{Zimdahl:2001ar}
W.~Zimdahl and D.~Pavon, ``{Interacting quintessence},''
  \href{http://dx.doi.org/10.1016/S0370-2693(01)01174-1}{{\em Phys.Lett.} {\bf
  B521} (2001)  133--138},
\href{http://arxiv.org/abs/astro-ph/0105479}{{\tt arXiv:astro-ph/0105479
  [astro-ph]}}.

\bibitem{Chimento:2003iea}
L.~P. Chimento, A.~S. Jakubi, D.~Pavon, and W.~Zimdahl, ``{Interacting
  quintessence solution to the coincidence problem},''
  \href{http://dx.doi.org/10.1103/PhysRevD.67.083513}{{\em Phys.Rev.} {\bf D67}
  (2003)  083513},
\href{http://arxiv.org/abs/astro-ph/0303145}{{\tt arXiv:astro-ph/0303145
  [astro-ph]}}.

\bibitem{Caldwell:2005tm}
R.~Caldwell and E.~V. Linder, ``{The Limits of quintessence},''
  \href{http://dx.doi.org/10.1103/PhysRevLett.95.141301}{{\em Phys.Rev.Lett.}
  {\bf 95} (2005)  141301},
\href{http://arxiv.org/abs/astro-ph/0505494}{{\tt arXiv:astro-ph/0505494
  [astro-ph]}}.

\bibitem{Chiba:2005tj}
T.~Chiba, ``{W and w' of scalar field models of dark energy},''
  \href{http://dx.doi.org/10.1103/PhysRevD.80.129901,
  10.1103/PhysRevD.73.063501}{{\em Phys.Rev.} {\bf D73} (2006)  063501},
\href{http://arxiv.org/abs/astro-ph/0510598}{{\tt arXiv:astro-ph/0510598
  [astro-ph]}}.

\bibitem{Scherrer:2005je}
R.~J. Scherrer, ``{Dark energy models in the w-w' plane},''
  \href{http://dx.doi.org/10.1103/PhysRevD.73.043502}{{\em Phys.Rev.} {\bf D73}
  (2006)  043502},
\href{http://arxiv.org/abs/astro-ph/0509890}{{\tt arXiv:astro-ph/0509890
  [astro-ph]}}.

\bibitem{Barger:2005sb}
V.~Barger, E.~Guarnaccia, and D.~Marfatia, ``{Classification of dark energy
  models in the (w(0), w(a)) plane},''
  \href{http://dx.doi.org/10.1016/j.physletb.2006.02.018}{{\em Phys.Lett.} {\bf
  B635} (2006)  61--65},
\href{http://arxiv.org/abs/hep-ph/0512320}{{\tt arXiv:hep-ph/0512320
  [hep-ph]}}.

\bibitem{Linder:2006sv}
E.~V. Linder, ``{The paths of quintessence},''
  \href{http://dx.doi.org/10.1103/PhysRevD.73.063010}{{\em Phys.Rev.} {\bf D73}
  (2006)  063010},
\href{http://arxiv.org/abs/astro-ph/0601052}{{\tt arXiv:astro-ph/0601052
  [astro-ph]}}.

\bibitem{Scherrer:2007pu}
R.~J. Scherrer and A.~Sen, ``{Thawing quintessence with a nearly flat
  potential},'' \href{http://dx.doi.org/10.1103/PhysRevD.77.083515}{{\em
  Phys.Rev.} {\bf D77} (2008)  083515},
\href{http://arxiv.org/abs/0712.3450}{{\tt arXiv:0712.3450 [astro-ph]}}.

\bibitem{Chiba:2012cb}
T.~Chiba, A.~De~Felice, and S.~Tsujikawa, ``{Observational Constraints on
  Quintessence: Thawing, Tracker, and Scaling models},''
  \href{http://dx.doi.org/10.1103/PhysRevD.87.083505}{{\em Phys.Rev.} {\bf D87}
  (2013)  083505},
\href{http://arxiv.org/abs/1210.3859}{{\tt arXiv:1210.3859 [astro-ph.CO]}}.

\bibitem{Chiba:1999ka}
T.~Chiba, T.~Okabe, and M.~Yamaguchi, ``{Kinetically driven quintessence},''
  \href{http://dx.doi.org/10.1103/PhysRevD.62.023511}{{\em Phys.Rev.} {\bf D62}
  (2000)  023511},
\href{http://arxiv.org/abs/astro-ph/9912463}{{\tt arXiv:astro-ph/9912463
  [astro-ph]}}.

\bibitem{ArmendarizPicon:2000dh}
C.~Armendariz-Picon, V.~F. Mukhanov, and P.~J. Steinhardt, ``{A Dynamical
  solution to the problem of a small cosmological constant and late time cosmic
  acceleration},'' \href{http://dx.doi.org/10.1103/PhysRevLett.85.4438}{{\em
  Phys.Rev.Lett.} {\bf 85} (2000)  4438--4441},
\href{http://arxiv.org/abs/astro-ph/0004134}{{\tt arXiv:astro-ph/0004134
  [astro-ph]}}.

\bibitem{ArmendarizPicon:2000ah}
C.~Armendariz-Picon, V.~F. Mukhanov, and P.~J. Steinhardt, ``{Essentials of k
  essence},'' \href{http://dx.doi.org/10.1103/PhysRevD.63.103510}{{\em
  Phys.Rev.} {\bf D63} (2001)  103510},
\href{http://arxiv.org/abs/astro-ph/0006373}{{\tt arXiv:astro-ph/0006373
  [astro-ph]}}.

\bibitem{Chiba:2002mw}
T.~Chiba, ``{Tracking K-essence},''
  \href{http://dx.doi.org/10.1103/PhysRevD.66.063514}{{\em Phys.Rev.} {\bf D66}
  (2002)  063514},
\href{http://arxiv.org/abs/astro-ph/0206298}{{\tt arXiv:astro-ph/0206298
  [astro-ph]}}.

\bibitem{Padmanabhan:2002cp}
T.~Padmanabhan, ``{Accelerated expansion of the universe driven by tachyonic
  matter},'' \href{http://dx.doi.org/10.1103/PhysRevD.66.021301}{{\em
  Phys.Rev.} {\bf D66} (2002)  021301},
\href{http://arxiv.org/abs/hep-th/0204150}{{\tt arXiv:hep-th/0204150
  [hep-th]}}.

\bibitem{Malquarti:2003nn}
M.~Malquarti, E.~J. Copeland, A.~R. Liddle, and M.~Trodden, ``{A New view of
  k-essence},'' \href{http://dx.doi.org/10.1103/PhysRevD.67.123503}{{\em
  Phys.Rev.} {\bf D67} (2003)  123503},
\href{http://arxiv.org/abs/astro-ph/0302279}{{\tt arXiv:astro-ph/0302279
  [astro-ph]}}.

\bibitem{Malquarti:2003hn}
M.~Malquarti, E.~J. Copeland, and A.~R. Liddle, ``{K-essence and the
  coincidence problem},''
  \href{http://dx.doi.org/10.1103/PhysRevD.68.023512}{{\em Phys.Rev.} {\bf D68}
  (2003)  023512},
\href{http://arxiv.org/abs/astro-ph/0304277}{{\tt arXiv:astro-ph/0304277
  [astro-ph]}}.

\bibitem{Chimento:2003zf}
L.~P. Chimento and A.~Feinstein, ``{Power - law expansion in k-essence
  cosmology},'' \href{http://dx.doi.org/10.1142/S0217732304013507}{{\em
  Mod.Phys.Lett.} {\bf A19} (2004)  761--768},
\href{http://arxiv.org/abs/astro-ph/0305007}{{\tt arXiv:astro-ph/0305007
  [astro-ph]}}.

\bibitem{Silverstein:2003hf}
E.~Silverstein and D.~Tong, ``{Scalar speed limits and cosmology: Acceleration
  from D-cceleration},''
  \href{http://dx.doi.org/10.1103/PhysRevD.70.103505}{{\em Phys.Rev.} {\bf D70}
  (2004)  103505},
\href{http://arxiv.org/abs/hep-th/0310221}{{\tt arXiv:hep-th/0310221
  [hep-th]}}.

\bibitem{GonzalezDiaz:2003rf}
P.~F. Gonzalez-Diaz, ``{K-essential phantom energy: Doomsday around the
  corner?},'' \href{http://dx.doi.org/10.1016/j.physletb.2003.12.077}{{\em
  Phys.Lett.} {\bf B586} (2004)  1--4},
\href{http://arxiv.org/abs/astro-ph/0312579}{{\tt arXiv:astro-ph/0312579
  [astro-ph]}}.

\bibitem{Scherrer:2004au}
R.~J. Scherrer, ``{Purely kinetic k-essence as unified dark matter},''
  \href{http://dx.doi.org/10.1103/PhysRevLett.93.011301}{{\em Phys.Rev.Lett.}
  {\bf 93} (2004)  011301},
\href{http://arxiv.org/abs/astro-ph/0402316}{{\tt arXiv:astro-ph/0402316
  [astro-ph]}}.

\bibitem{Aguirregabiria:2004te}
J.~M. Aguirregabiria, L.~P. Chimento, and R.~Lazkoz, ``{Phantom k-essence
  cosmologies},'' \href{http://dx.doi.org/10.1103/PhysRevD.70.023509}{{\em
  Phys.Rev.} {\bf D70} (2004)  023509},
\href{http://arxiv.org/abs/astro-ph/0403157}{{\tt arXiv:astro-ph/0403157
  [astro-ph]}}.

\bibitem{Piazza:2004df}
F.~Piazza and S.~Tsujikawa, ``{Dilatonic ghost condensate as dark energy},''
  \href{http://dx.doi.org/10.1088/1475-7516/2004/07/004}{{\em JCAP} {\bf 0407}
  (2004)  004},
\href{http://arxiv.org/abs/hep-th/0405054}{{\tt arXiv:hep-th/0405054
  [hep-th]}}.

\bibitem{Rendall:2005fv}
A.~D. Rendall, ``{Dynamics of k-essence},''
  \href{http://dx.doi.org/10.1088/0264-9381/23/5/008}{{\em Class.Quant.Grav.}
  {\bf 23} (2006)  1557--1570},
\href{http://arxiv.org/abs/gr-qc/0511158}{{\tt arXiv:gr-qc/0511158 [gr-qc]}}.

\bibitem{Bonvin:2006vc}
C.~Bonvin, C.~Caprini, and R.~Durrer, ``{A no-go theorem for k-essence dark
  energy},'' \href{http://dx.doi.org/10.1103/PhysRevLett.97.081303}{{\em
  Phys.Rev.Lett.} {\bf 97} (2006)  081303},
\href{http://arxiv.org/abs/astro-ph/0606584}{{\tt arXiv:astro-ph/0606584
  [astro-ph]}}.

\bibitem{Babichev:2007dw}
E.~Babichev, V.~Mukhanov, and A.~Vikman, ``{k-Essence, superluminal
  propagation, causality and emergent geometry},''
  \href{http://dx.doi.org/10.1088/1126-6708/2008/02/101}{{\em JHEP} {\bf 0802}
  (2008)  101},
\href{http://arxiv.org/abs/0708.0561}{{\tt arXiv:0708.0561 [hep-th]}}.

\bibitem{dePutter:2007ny}
R.~de~Putter and E.~V. Linder, ``{Kinetic k-essence and Quintessence},''
  \href{http://dx.doi.org/10.1016/j.astropartphys.2007.05.011}{{\em
  Astropart.Phys.} {\bf 28} (2007)  263--272},
\href{http://arxiv.org/abs/0705.0400}{{\tt arXiv:0705.0400 [astro-ph]}}.

\bibitem{Kang:2007vs}
J.~U. Kang, V.~Vanchurin, and S.~Winitzki, ``{Attractor scenarios and
  superluminal signals in k-essence cosmology},''
  \href{http://dx.doi.org/10.1103/PhysRevD.76.083511}{{\em Phys.Rev.} {\bf D76}
  (2007)  083511},
\href{http://arxiv.org/abs/0706.3994}{{\tt arXiv:0706.3994 [gr-qc]}}.

\bibitem{Bilic:2008zk}
N.~Bilic, ``{Thermodynamics of k-essence},''
  \href{http://dx.doi.org/10.1103/PhysRevD.78.105012}{{\em Phys.Rev.} {\bf D78}
  (2008)  105012},
\href{http://arxiv.org/abs/0806.0642}{{\tt arXiv:0806.0642 [gr-qc]}}.

\bibitem{Martin:2008xw}
J.~Martin and M.~Yamaguchi, ``{DBI-essence},''
  \href{http://dx.doi.org/10.1103/PhysRevD.77.123508}{{\em Phys.Rev.} {\bf D77}
  (2008)  123508},
\href{http://arxiv.org/abs/0801.3375}{{\tt arXiv:0801.3375 [hep-th]}}.

\bibitem{Myrzakulov:2010tc}
R.~Myrzakulov, ``{F(T) gravity and k-essence},''
  \href{http://dx.doi.org/10.1007/s10714-012-1439-z}{{\em Gen.Rel.Grav.} {\bf
  44} (2012)  3059--3080},
\href{http://arxiv.org/abs/1008.4486}{{\tt arXiv:1008.4486 [physics.gen-ph]}}.

\bibitem{ArmendarizPicon:1999rj}
C.~Armendariz-Picon, T.~Damour, and V.~F. Mukhanov, ``{k - inflation},''
  \href{http://dx.doi.org/10.1016/S0370-2693(99)00603-6}{{\em Phys.Lett.} {\bf
  B458} (1999)  209--218},
\href{http://arxiv.org/abs/hep-th/9904075}{{\tt arXiv:hep-th/9904075
  [hep-th]}}.

\bibitem{Garriga:1999vw}
J.~Garriga and V.~F. Mukhanov, ``{Perturbations in k-inflation},''
  \href{http://dx.doi.org/10.1016/S0370-2693(99)00602-4}{{\em Phys.Lett.} {\bf
  B458} (1999)  219--225},
\href{http://arxiv.org/abs/hep-th/9904176}{{\tt arXiv:hep-th/9904176
  [hep-th]}}.

\bibitem{Brax:1999yv}
P.~Brax and J.~Martin, ``{The Robustness of quintessence},''
  \href{http://dx.doi.org/10.1103/PhysRevD.61.103502}{{\em Phys.Rev.} {\bf D61}
  (2000)  103502},
\href{http://arxiv.org/abs/astro-ph/9912046}{{\tt arXiv:astro-ph/9912046
  [astro-ph]}}.

\bibitem{Brax:1999gp}
P.~Brax and J.~Martin, ``{Quintessence and supergravity},''
  \href{http://dx.doi.org/10.1016/S0370-2693(99)01209-5}{{\em Phys.Lett.} {\bf
  B468} (1999)  40--45},
\href{http://arxiv.org/abs/astro-ph/9905040}{{\tt arXiv:astro-ph/9905040
  [astro-ph]}}.

\bibitem{Masiero:1999sq}
A.~Masiero, M.~Pietroni, and F.~Rosati, ``{SUSY QCD and quintessence},''
  \href{http://dx.doi.org/10.1103/PhysRevD.61.023504}{{\em Phys.Rev.} {\bf D61}
  (2000)  023504},
\href{http://arxiv.org/abs/hep-ph/9905346}{{\tt arXiv:hep-ph/9905346
  [hep-ph]}}.

\bibitem{Copeland:2000vh}
E.~J. Copeland, N.~Nunes, and F.~Rosati, ``{Quintessence models in
  supergravity},'' \href{http://dx.doi.org/10.1103/PhysRevD.62.123503}{{\em
  Phys.Rev.} {\bf D62} (2000)  123503},
\href{http://arxiv.org/abs/hep-ph/0005222}{{\tt arXiv:hep-ph/0005222
  [hep-ph]}}.

\bibitem{Kallosh:2002gf}
R.~Kallosh, A.~D. Linde, S.~Prokushkin, and M.~Shmakova, ``{Supergravity, dark
  energy and the fate of the universe},''
  \href{http://dx.doi.org/10.1103/PhysRevD.66.123503}{{\em Phys.Rev.} {\bf D66}
  (2002)  123503},
\href{http://arxiv.org/abs/hep-th/0208156}{{\tt arXiv:hep-th/0208156
  [hep-th]}}.

\bibitem{Choi:1999xn}
K.~Choi, ``{String or M theory axion as a quintessence},''
  \href{http://dx.doi.org/10.1103/PhysRevD.62.043509}{{\em Phys.Rev.} {\bf D62}
  (2000)  043509},
\href{http://arxiv.org/abs/hep-ph/9902292}{{\tt arXiv:hep-ph/9902292
  [hep-ph]}}.

\bibitem{Gasperini:2001pc}
M.~Gasperini, F.~Piazza, and G.~Veneziano, ``{Quintessence as a runaway
  dilaton},'' \href{http://dx.doi.org/10.1103/PhysRevD.65.023508}{{\em
  Phys.Rev.} {\bf D65} (2002)  023508},
\href{http://arxiv.org/abs/gr-qc/0108016}{{\tt arXiv:gr-qc/0108016 [gr-qc]}}.

\bibitem{Townsend:2001ea}
P.~Townsend, ``{Quintessence from M theory},''
  \href{http://dx.doi.org/10.1088/1126-6708/2001/11/042}{{\em JHEP} {\bf 0111}
  (2001)  042},
\href{http://arxiv.org/abs/hep-th/0110072}{{\tt arXiv:hep-th/0110072
  [hep-th]}}.

\bibitem{Damour:2002nv}
T.~Damour, F.~Piazza, and G.~Veneziano, ``{Violations of the equivalence
  principle in a dilaton runaway scenario},''
  \href{http://dx.doi.org/10.1103/PhysRevD.66.046007}{{\em Phys.Rev.} {\bf D66}
  (2002)  046007},
\href{http://arxiv.org/abs/hep-th/0205111}{{\tt arXiv:hep-th/0205111
  [hep-th]}}.

\bibitem{Damour:2002mi}
T.~Damour, F.~Piazza, and G.~Veneziano, ``{Runaway dilaton and equivalence
  principle violations},''
  \href{http://dx.doi.org/10.1103/PhysRevLett.89.081601}{{\em Phys.Rev.Lett.}
  {\bf 89} (2002)  081601},
\href{http://arxiv.org/abs/gr-qc/0204094}{{\tt arXiv:gr-qc/0204094 [gr-qc]}}.

\bibitem{Kim:2002tq}
J.~E. Kim and H.~P. Nilles, ``{A Quintessential axion},''
  \href{http://dx.doi.org/10.1016/S0370-2693(02)03148-9}{{\em Phys.Lett.} {\bf
  B553} (2003)  1--6},
\href{http://arxiv.org/abs/hep-ph/0210402}{{\tt arXiv:hep-ph/0210402
  [hep-ph]}}.

\bibitem{Kaloper:2008qs}
N.~Kaloper and L.~Sorbo, ``{Where in the String Landscape is Quintessence},''
  \href{http://dx.doi.org/10.1103/PhysRevD.79.043528}{{\em Phys.Rev.} {\bf D79}
  (2009)  043528},
\href{http://arxiv.org/abs/0810.5346}{{\tt arXiv:0810.5346 [hep-th]}}.

\bibitem{Panda:2010uq}
S.~Panda, Y.~Sumitomo, and S.~P. Trivedi, ``{Axions as Quintessence in String
  Theory},'' \href{http://dx.doi.org/10.1103/PhysRevD.83.083506}{{\em
  Phys.Rev.} {\bf D83} (2011)  083506},
\href{http://arxiv.org/abs/1011.5877}{{\tt arXiv:1011.5877 [hep-th]}}.

\bibitem{Cicoli:2012tz}
M.~Cicoli, F.~G. Pedro, and G.~Tasinato, ``{Natural Quintessence in String
  Theory},'' \href{http://dx.doi.org/10.1088/1475-7516/2012/07/044}{{\em JCAP}
  {\bf 1207} (2012)  044},
\href{http://arxiv.org/abs/1203.6655}{{\tt arXiv:1203.6655 [hep-th]}}.

\bibitem{Carroll:1998zi}
S.~M. Carroll, ``{Quintessence and the rest of the world},''
  \href{http://dx.doi.org/10.1103/PhysRevLett.81.3067}{{\em Phys.Rev.Lett.}
  {\bf 81} (1998)  3067--3070},
\href{http://arxiv.org/abs/astro-ph/9806099}{{\tt arXiv:astro-ph/9806099
  [astro-ph]}}.

\bibitem{Lue:1998mq}
A.~Lue, L.-M. Wang, and M.~Kamionkowski, ``{Cosmological signature of new
  parity violating interactions},''
  \href{http://dx.doi.org/10.1103/PhysRevLett.83.1506}{{\em Phys.Rev.Lett.}
  {\bf 83} (1999)  1506--1509},
\href{http://arxiv.org/abs/astro-ph/9812088}{{\tt arXiv:astro-ph/9812088
  [astro-ph]}}.

\bibitem{Copeland:2003cv}
E.~Copeland, N.~Nunes, and M.~Pospelov, ``{Models of quintessence coupled to
  the electromagnetic field and the cosmological evolution of alpha},''
  \href{http://dx.doi.org/10.1103/PhysRevD.69.023501}{{\em Phys.Rev.} {\bf D69}
  (2004)  023501},
\href{http://arxiv.org/abs/hep-ph/0307299}{{\tt arXiv:hep-ph/0307299
  [hep-ph]}}.

\bibitem{Anderson:1971pn}
J.~Anderson and D.~Finkelstein, ``{Cosmological constant and fundamental
  length},''
\href{http://dx.doi.org/10.1119/1.1986321}{{\em Am.J.Phys.} {\bf 39} (1971)
  901--904}.

\bibitem{Padilla:2014yea}
A.~Padilla and I.~D. Saltas, ``{A note on classical and quantum unimodular
  gravity},''
\href{http://arxiv.org/abs/1409.3573}{{\tt arXiv:1409.3573 [gr-qc]}}.

\bibitem{Fiol:2008vk}
B.~Fiol and J.~Garriga, ``{Semiclassical Unimodular Gravity},''
  \href{http://dx.doi.org/10.1088/1475-7516/2010/08/015}{{\em JCAP} {\bf 1008}
  (2010)  015},
\href{http://arxiv.org/abs/0809.1371}{{\tt arXiv:0809.1371 [hep-th]}}.

\bibitem{Kaloper:2013zca}
N.~Kaloper and A.~Padilla, ``{Sequestering the Standard Model Vacuum Energy},''
  \href{http://dx.doi.org/10.1103/PhysRevLett.112.091304}{{\em Phys.Rev.Lett.}
  {\bf 112} (2014)  091304},
\href{http://arxiv.org/abs/1309.6562}{{\tt arXiv:1309.6562 [hep-th]}}.

\bibitem{Kaloper:2014dqa}
N.~Kaloper and A.~Padilla, ``{Vacuum Energy Sequestering: The Framework and Its
  Cosmological Consequences},''
\href{http://arxiv.org/abs/1406.0711}{{\tt arXiv:1406.0711 [hep-th]}}.

\bibitem{Gabadadze:2014rwa}
G.~Gabadadze, ``{The Big Constant Out, The Small Constant In},''
\href{http://arxiv.org/abs/1406.6701}{{\tt arXiv:1406.6701 [hep-th]}}.

\bibitem{Carroll:2006jn}
S.~M. Carroll, I.~Sawicki, A.~Silvestri, and M.~Trodden, ``{Modified-Source
  Gravity and Cosmological Structure Formation},''
  \href{http://dx.doi.org/10.1088/1367-2630/8/12/323}{{\em New J.Phys.} {\bf 8}
  (2006)  323},
\href{http://arxiv.org/abs/astro-ph/0607458}{{\tt arXiv:astro-ph/0607458
  [astro-ph]}}.

\bibitem{Pani:2013qfa}
P.~Pani, T.~P. Sotiriou, and D.~Vernieri, ``{Gravity with Auxiliary Fields},''
  \href{http://dx.doi.org/10.1103/PhysRevD.88.121502}{{\em Phys.Rev.} {\bf D88}
  (2013)  121502},
\href{http://arxiv.org/abs/1306.1835}{{\tt arXiv:1306.1835 [gr-qc]}}.

\bibitem{Will:2005va}
C.~M. Will, ``{The Confrontation between general relativity and experiment},''
  {\em Living Rev.Rel.} {\bf 9} (2006)  3,
\href{http://arxiv.org/abs/gr-qc/0510072}{{\tt arXiv:gr-qc/0510072 [gr-qc]}}.

\bibitem{Will:2014kxa}
C.~M. Will, ``{The Confrontation between General Relativity and Experiment},''
\href{http://arxiv.org/abs/1403.7377}{{\tt arXiv:1403.7377 [gr-qc]}}.

\bibitem{Khoury:2003aq}
J.~Khoury and A.~Weltman, ``{Chameleon fields: Awaiting surprises for tests of
  gravity in space},''
  \href{http://dx.doi.org/10.1103/PhysRevLett.93.171104}{{\em Phys.Rev.Lett.}
  {\bf 93} (2004)  171104},
\href{http://arxiv.org/abs/astro-ph/0309300}{{\tt arXiv:astro-ph/0309300
  [astro-ph]}}.

\bibitem{Khoury:2003rn}
J.~Khoury and A.~Weltman, ``{Chameleon cosmology},''
  \href{http://dx.doi.org/10.1103/PhysRevD.69.044026}{{\em Phys.Rev.} {\bf D69}
  (2004)  044026},
\href{http://arxiv.org/abs/astro-ph/0309411}{{\tt arXiv:astro-ph/0309411
  [astro-ph]}}.

\bibitem{Koivisto:2012za}
T.~S. Koivisto, D.~F. Mota, and M.~Zumalacarregui, ``{Screening Modifications
  of Gravity through Disformally Coupled Fields},''
  \href{http://dx.doi.org/10.1103/PhysRevLett.109.241102}{{\em Phys.Rev.Lett.}
  {\bf 109} (2012)  241102},
\href{http://arxiv.org/abs/1205.3167}{{\tt arXiv:1205.3167 [astro-ph.CO]}}.

\bibitem{Noller:2012sv}
J.~Noller, ``{Derivative Chameleons},''
  \href{http://dx.doi.org/10.1088/1475-7516/2012/07/013}{{\em JCAP} {\bf 1207}
  (2012)  013},
\href{http://arxiv.org/abs/1203.6639}{{\tt arXiv:1203.6639 [gr-qc]}}.

\bibitem{Brax:2012ie}
P.~Brax, C.~Burrage, and A.-C. Davis, ``{Shining Light on Modifications of
  Gravity},'' \href{http://dx.doi.org/10.1088/1475-7516/2012/10/016}{{\em JCAP}
  {\bf 1210} (2012)  016},
\href{http://arxiv.org/abs/1206.1809}{{\tt arXiv:1206.1809 [hep-th]}}.

\bibitem{Zumalacarregui:2012us}
M.~Zumalacarregui, T.~S. Koivisto, and D.~F. Mota, ``{DBI Galileons in the
  Einstein Frame: Local Gravity and Cosmology},''
  \href{http://dx.doi.org/10.1103/PhysRevD.87.083010}{{\em Phys.Rev.} {\bf D87}
  (2013)  083010},
\href{http://arxiv.org/abs/1210.8016}{{\tt arXiv:1210.8016 [astro-ph.CO]}}.

\bibitem{vandeBruck:2013yxa}
C.~van~de Bruck, J.~Morrice, and S.~Vu, ``{Constraints on Nonconformal
  Couplings from the Properties of the Cosmic Microwave Background
  Radiation},'' \href{http://dx.doi.org/10.1103/PhysRevLett.111.161302}{{\em
  Phys.Rev.Lett.} {\bf 111} (2013)  161302},
\href{http://arxiv.org/abs/1303.1773}{{\tt arXiv:1303.1773 [astro-ph.CO]}}.

\bibitem{Brax:2013nsa}
P.~Brax, C.~Burrage, A.-C. Davis, and G.~Gubitosi, ``{Cosmological Tests of the
  Disformal Coupling to Radiation},''
  \href{http://dx.doi.org/10.1088/1475-7516/2013/11/001}{{\em JCAP} {\bf 1311}
  (2013)  001},
\href{http://arxiv.org/abs/1306.4168}{{\tt arXiv:1306.4168 [astro-ph.CO]}}.

\bibitem{Hinterbichler:2010es}
K.~Hinterbichler and J.~Khoury, ``{Symmetron Fields: Screening Long-Range
  Forces Through Local Symmetry Restoration},''
  \href{http://dx.doi.org/10.1103/PhysRevLett.104.231301}{{\em Phys.Rev.Lett.}
  {\bf 104} (2010)  231301},
\href{http://arxiv.org/abs/1001.4525}{{\tt arXiv:1001.4525 [hep-th]}}.

\bibitem{Pietroni:2005pv}
M.~Pietroni, ``{Dark energy condensation},''
  \href{http://dx.doi.org/10.1103/PhysRevD.72.043535}{{\em Phys.Rev.} {\bf D72}
  (2005)  043535},
\href{http://arxiv.org/abs/astro-ph/0505615}{{\tt arXiv:astro-ph/0505615
  [astro-ph]}}.

\bibitem{Olive:2007aj}
K.~A. Olive and M.~Pospelov, ``{Environmental dependence of masses and coupling
  constants},'' \href{http://dx.doi.org/10.1103/PhysRevD.77.043524}{{\em
  Phys.Rev.} {\bf D77} (2008)  043524},
\href{http://arxiv.org/abs/0709.3825}{{\tt arXiv:0709.3825 [hep-ph]}}.

\bibitem{Damour:1994zq}
T.~Damour and A.~M. Polyakov, ``{The String dilaton and a least coupling
  principle},'' \href{http://dx.doi.org/10.1016/0550-3213(94)90143-0}{{\em
  Nucl.Phys.} {\bf B423} (1994)  532--558},
\href{http://arxiv.org/abs/hep-th/9401069}{{\tt arXiv:hep-th/9401069
  [hep-th]}}.

\bibitem{Brax:2011ja}
P.~Brax, C.~van~de Bruck, A.-C. Davis, B.~Li, and D.~J. Shaw, ``{Nonlinear
  Structure Formation with the Environmentally Dependent Dilaton},''
  \href{http://dx.doi.org/10.1103/PhysRevD.83.104026}{{\em Phys.Rev.} {\bf D83}
  (2011)  104026},
\href{http://arxiv.org/abs/1102.3692}{{\tt arXiv:1102.3692 [astro-ph.CO]}}.

\bibitem{Babichev:2009ee}
E.~Babichev, C.~Deffayet, and R.~Ziour, ``{k-Mouflage gravity},''
  \href{http://dx.doi.org/10.1142/S0218271809016107}{{\em Int.J.Mod.Phys.} {\bf
  D18} (2009)  2147--2154},
\href{http://arxiv.org/abs/0905.2943}{{\tt arXiv:0905.2943 [hep-th]}}.

\bibitem{Babichev:2011kq}
E.~Babichev, C.~Deffayet, and G.~Esposito-Farese, ``{Improving relativistic
  MOND with Galileon k-mouflage},''
  \href{http://dx.doi.org/10.1103/PhysRevD.84.061502}{{\em Phys.Rev.} {\bf D84}
  (2011)  061502},
\href{http://arxiv.org/abs/1106.2538}{{\tt arXiv:1106.2538 [gr-qc]}}.

\bibitem{Brax:2012jr}
P.~Brax, C.~Burrage, and A.-C. Davis, ``{Screening fifth forces in k-essence
  and DBI models},''
  \href{http://dx.doi.org/10.1088/1475-7516/2013/01/020}{{\em JCAP} {\bf 1301}
  (2013)  020},
\href{http://arxiv.org/abs/1209.1293}{{\tt arXiv:1209.1293 [hep-th]}}.

\bibitem{Burrage:2014uwa}
C.~Burrage and J.~Khoury, ``{D-BIonic Screening of Scalar Fields},''
\href{http://arxiv.org/abs/1403.6120}{{\tt arXiv:1403.6120 [hep-th]}}.

\bibitem{Vainshtein:1972sx}
A.~Vainshtein, ``{To the problem of nonvanishing gravitation mass},''
\href{http://dx.doi.org/10.1016/0370-2693(72)90147-5}{{\em Phys.Lett.} {\bf
  B39} (1972)  393--394}.

\bibitem{Deffayet:2001uk}
C.~Deffayet, G.~Dvali, G.~Gabadadze, and A.~I. Vainshtein, ``{Nonperturbative
  continuity in graviton mass versus perturbative discontinuity},''
  \href{http://dx.doi.org/10.1103/PhysRevD.65.044026}{{\em Phys.Rev.} {\bf D65}
  (2002)  044026},
\href{http://arxiv.org/abs/hep-th/0106001}{{\tt arXiv:hep-th/0106001
  [hep-th]}}.

\bibitem{Nicolis:2004qq}
A.~Nicolis and R.~Rattazzi, ``{Classical and quantum consistency of the DGP
  model},'' \href{http://dx.doi.org/10.1088/1126-6708/2004/06/059}{{\em JHEP}
  {\bf 0406} (2004)  059},
\href{http://arxiv.org/abs/hep-th/0404159}{{\tt arXiv:hep-th/0404159
  [hep-th]}}.

\bibitem{Nicolis:2008in}
A.~Nicolis, R.~Rattazzi, and E.~Trincherini, ``{The Galileon as a local
  modification of gravity},''
  \href{http://dx.doi.org/10.1103/PhysRevD.79.064036}{{\em Phys.Rev.} {\bf D79}
  (2009)  064036},
\href{http://arxiv.org/abs/0811.2197}{{\tt arXiv:0811.2197 [hep-th]}}.

\bibitem{Milgrom:1983ca}
M.~Milgrom, ``{A Modification of the Newtonian dynamics as a possible
  alternative to the hidden mass hypothesis},''
\href{http://dx.doi.org/10.1086/161130}{{\em Astrophys.J.} {\bf 270} (1983)
  365--370}.

\bibitem{Sanders:2002pf}
R.~H. Sanders and S.~S. McGaugh, ``{Modified Newtonian dynamics as an
  alternative to dark matter},''
  \href{http://dx.doi.org/10.1146/annurev.astro.40.060401.093923}{{\em
  Ann.Rev.Astron.Astrophys.} {\bf 40} (2002)  263--317},
\href{http://arxiv.org/abs/astro-ph/0204521}{{\tt arXiv:astro-ph/0204521
  [astro-ph]}}.

\bibitem{vanDam:1970vg}
H.~van Dam and M.~Veltman, ``{Massive and massless Yang-Mills and gravitational
  fields},''
\href{http://dx.doi.org/10.1016/0550-3213(70)90416-5}{{\em Nucl.Phys.} {\bf
  B22} (1970)  397--411}.

\bibitem{Zakharov:1970cc}
V.~Zakharov, ``{Linearized gravitation theory and the graviton mass},''
{\em JETP Lett.} {\bf 12} (1970)  312.

\bibitem{Gruzinov:2001hp}
A.~Gruzinov, ``{On the graviton mass},''
  \href{http://dx.doi.org/10.1016/j.newast.2004.12.001}{{\em New Astron.} {\bf
  10} (2005)  311--314},
\href{http://arxiv.org/abs/astro-ph/0112246}{{\tt arXiv:astro-ph/0112246
  [astro-ph]}}.

\bibitem{Porrati:2002cp}
M.~Porrati, ``{Fully covariant van Dam-Veltman-Zakharov discontinuity, and
  absence thereof},''
  \href{http://dx.doi.org/10.1016/S0370-2693(02)01656-8}{{\em Phys.Lett.} {\bf
  B534} (2002)  209--215},
\href{http://arxiv.org/abs/hep-th/0203014}{{\tt arXiv:hep-th/0203014
  [hep-th]}}.

\bibitem{Babichev:2009jt}
E.~Babichev, C.~Deffayet, and R.~Ziour, ``{Recovering General Relativity from
  massive gravity},''
  \href{http://dx.doi.org/10.1103/PhysRevLett.103.201102}{{\em Phys.Rev.Lett.}
  {\bf 103} (2009)  201102},
\href{http://arxiv.org/abs/0907.4103}{{\tt arXiv:0907.4103 [gr-qc]}}.

\bibitem{Babichev:2009us}
E.~Babichev, C.~Deffayet, and R.~Ziour, ``{The Vainshtein mechanism in the
  Decoupling Limit of massive gravity},''
  \href{http://dx.doi.org/10.1088/1126-6708/2009/05/098}{{\em JHEP} {\bf 0905}
  (2009)  098},
\href{http://arxiv.org/abs/0901.0393}{{\tt arXiv:0901.0393 [hep-th]}}.

\bibitem{Babichev:2010jd}
E.~Babichev, C.~Deffayet, and R.~Ziour, ``{The Recovery of General Relativity
  in massive gravity via the Vainshtein mechanism},''
  \href{http://dx.doi.org/10.1103/PhysRevD.82.104008}{{\em Phys.Rev.} {\bf D82}
  (2010)  104008},
\href{http://arxiv.org/abs/1007.4506}{{\tt arXiv:1007.4506 [gr-qc]}}.

\bibitem{ArkaniHamed:2002sp}
N.~Arkani-Hamed, H.~Georgi, and M.~D. Schwartz, ``{Effective field theory for
  massive gravitons and gravity in theory space},''
  \href{http://dx.doi.org/10.1016/S0003-4916(03)00068-X}{{\em Annals Phys.}
  {\bf 305} (2003)  96--118},
\href{http://arxiv.org/abs/hep-th/0210184}{{\tt arXiv:hep-th/0210184
  [hep-th]}}.

\bibitem{Creminelli:2005qk}
P.~Creminelli, A.~Nicolis, M.~Papucci, and E.~Trincherini, ``{Ghosts in massive
  gravity},'' \href{http://dx.doi.org/10.1088/1126-6708/2005/09/003}{{\em JHEP}
  {\bf 0509} (2005)  003},
\href{http://arxiv.org/abs/hep-th/0505147}{{\tt arXiv:hep-th/0505147
  [hep-th]}}.

\bibitem{Nibbelink:2006sz}
S.~Groot~Nibbelink, M.~Peloso, and M.~Sexton, ``{Nonlinear Properties of
  Vielbein Massive Gravity},''
  \href{http://dx.doi.org/10.1140/epjc/s10052-007-0311-x}{{\em Eur.Phys.J.}
  {\bf C51} (2007)  741--752},
\href{http://arxiv.org/abs/hep-th/0610169}{{\tt arXiv:hep-th/0610169
  [hep-th]}}.

\bibitem{deRham:2010ik}
C.~de~Rham and G.~Gabadadze, ``{Generalization of the Fierz-Pauli Action},''
  \href{http://dx.doi.org/10.1103/PhysRevD.82.044020}{{\em Phys.Rev.} {\bf D82}
  (2010)  044020},
\href{http://arxiv.org/abs/1007.0443}{{\tt arXiv:1007.0443 [hep-th]}}.

\bibitem{deRham:2010kj}
C.~de~Rham, G.~Gabadadze, and A.~J. Tolley, ``{Resummation of Massive
  Gravity},'' \href{http://dx.doi.org/10.1103/PhysRevLett.106.231101}{{\em
  Phys.Rev.Lett.} {\bf 106} (2011)  231101},
\href{http://arxiv.org/abs/1011.1232}{{\tt arXiv:1011.1232 [hep-th]}}.

\bibitem{Hinterbichler:2014cwa}
K.~Hinterbichler and A.~Joyce, ``{Goldstones with Extended Shift Symmetries},''
\href{http://arxiv.org/abs/1404.4047}{{\tt arXiv:1404.4047 [hep-th]}}.

\bibitem{Ostrogradsky}
M.~Ostrogradsky, ``{},'' {\em Mem. Ac. St. Petersbourg} {\bf VI} (1850)  385.

\bibitem{Polchinski:1992ed}
J.~Polchinski, ``{Effective field theory and the Fermi surface},''
\href{http://arxiv.org/abs/hep-th/9210046}{{\tt arXiv:hep-th/9210046
  [hep-th]}}.

\bibitem{Burgess:2003jk}
C.~Burgess, ``{Quantum gravity in everyday life: General relativity as an
  effective field theory},'' \href{http://dx.doi.org/10.12942/lrr-2004-5}{{\em
  Living Rev.Rel.} {\bf 7} (2004)  5--56},
\href{http://arxiv.org/abs/gr-qc/0311082}{{\tt arXiv:gr-qc/0311082 [gr-qc]}}.

\bibitem{Kaplan:2005es}
D.~B. Kaplan, ``{Five lectures on effective field theory},''
\href{http://arxiv.org/abs/nucl-th/0510023}{{\tt arXiv:nucl-th/0510023
  [nucl-th]}}.

\bibitem{Burgess:2007pt}
C.~Burgess, ``{Introduction to Effective Field Theory},''
  \href{http://dx.doi.org/10.1146/annurev.nucl.56.080805.140508}{{\em
  Ann.Rev.Nucl.Part.Sci.} {\bf 57} (2007)  329--362},
\href{http://arxiv.org/abs/hep-th/0701053}{{\tt arXiv:hep-th/0701053
  [hep-th]}}.

\bibitem{Hawking:1991nk}
S.~Hawking, ``{The Chronology protection conjecture},''
\href{http://dx.doi.org/10.1103/PhysRevD.46.603}{{\em Phys.Rev.} {\bf D46}
  (1992)  603--611}.

\bibitem{Burrage:2011cr}
C.~Burrage, C.~de~Rham, L.~Heisenberg, and A.~J. Tolley, ``{Chronology
  Protection in Galileon Models and Massive Gravity},''
  \href{http://dx.doi.org/10.1088/1475-7516/2012/07/004}{{\em JCAP} {\bf 1207}
  (2012)  004},
\href{http://arxiv.org/abs/1111.5549}{{\tt arXiv:1111.5549 [hep-th]}}.

\bibitem{Evslin:2011rj}
J.~Evslin, ``{Stability of Closed Timelike Curves in a Galileon Model},''
  \href{http://dx.doi.org/10.1007/JHEP03(2012)009}{{\em JHEP} {\bf 1203} (2012)
   009},
\href{http://arxiv.org/abs/1112.1349}{{\tt arXiv:1112.1349 [hep-th]}}.

\bibitem{Adams:2006sv}
A.~Adams, N.~Arkani-Hamed, S.~Dubovsky, A.~Nicolis, and R.~Rattazzi,
  ``{Causality, analyticity and an IR obstruction to UV completion},''
  \href{http://dx.doi.org/10.1088/1126-6708/2006/10/014}{{\em JHEP} {\bf 0610}
  (2006)  014},
\href{http://arxiv.org/abs/hep-th/0602178}{{\tt arXiv:hep-th/0602178
  [hep-th]}}.

\bibitem{Upadhye:2012vh}
A.~Upadhye, W.~Hu, and J.~Khoury, ``{Quantum Stability of Chameleon Field
  Theories},'' \href{http://dx.doi.org/10.1103/PhysRevLett.109.041301}{{\em
  Phys.Rev.Lett.} {\bf 109} (2012)  041301},
\href{http://arxiv.org/abs/1204.3906}{{\tt arXiv:1204.3906 [hep-ph]}}.

\bibitem{Park:2010cw}
M.~Park, K.~M. Zurek, and S.~Watson, ``{A Unified Approach to Cosmic
  Acceleration},'' \href{http://dx.doi.org/10.1103/PhysRevD.81.124008}{{\em
  Phys.Rev.} {\bf D81} (2010)  124008},
\href{http://arxiv.org/abs/1003.1722}{{\tt arXiv:1003.1722 [hep-th]}}.

\bibitem{Creminelli:2008wc}
P.~Creminelli, G.~D'Amico, J.~Norena, and F.~Vernizzi, ``{The Effective Theory
  of Quintessence: the w < -1 Side Unveiled},''
  \href{http://dx.doi.org/10.1088/1475-7516/2009/02/018}{{\em JCAP} {\bf 0902}
  (2009)  018},
\href{http://arxiv.org/abs/0811.0827}{{\tt arXiv:0811.0827 [astro-ph]}}.

\bibitem{Bloomfield:2012ff}
J.~K. Bloomfield, ƒ.~ƒ. Flanagan, M.~Park, and S.~Watson, ``{Dark energy or
  modified gravity? An effective field theory approach},''
  \href{http://dx.doi.org/10.1088/1475-7516/2013/08/010}{{\em JCAP} {\bf 1308}
  (2013)  010},
\href{http://arxiv.org/abs/1211.7054}{{\tt arXiv:1211.7054 [astro-ph.CO]}}.

\bibitem{Gubitosi:2012hu}
G.~Gubitosi, F.~Piazza, and F.~Vernizzi, ``{The Effective Field Theory of Dark
  Energy},'' \href{http://dx.doi.org/10.1088/1475-7516/2013/02/032}{{\em JCAP}
  {\bf 1302} (2013)  032},
\href{http://arxiv.org/abs/1210.0201}{{\tt arXiv:1210.0201 [hep-th]}}.

\bibitem{Gleyzes:2013ooa}
J.~Gleyzes, D.~Langlois, F.~Piazza, and F.~Vernizzi, ``{Essential Building
  Blocks of Dark Energy},''
  \href{http://dx.doi.org/10.1088/1475-7516/2013/08/025}{{\em JCAP} {\bf 1308}
  (2013)  025},
\href{http://arxiv.org/abs/1304.4840}{{\tt arXiv:1304.4840 [hep-th]}}.

\bibitem{Piazza:2013coa}
F.~Piazza and F.~Vernizzi, ``{Effective Field Theory of Cosmological
  Perturbations},''
  \href{http://dx.doi.org/10.1088/0264-9381/30/21/214007}{{\em
  Class.Quant.Grav.} {\bf 30} (2013)  214007},
\href{http://arxiv.org/abs/1307.4350}{{\tt arXiv:1307.4350}}.

\bibitem{Frusciante:2013zop}
N.~Frusciante, M.~Raveri, and A.~Silvestri, ``{Effective Field Theory of Dark
  Energy: a Dynamical Analysis},''
  \href{http://dx.doi.org/10.1088/1475-7516/2014/02/026}{{\em JCAP} {\bf 1402}
  (2014)  026},
\href{http://arxiv.org/abs/1310.6026}{{\tt arXiv:1310.6026 [astro-ph.CO]}}.

\bibitem{Bloomfield:2013efa}
J.~Bloomfield, ``{A Simplified Approach to General Scalar-Tensor Theories},''
\href{http://arxiv.org/abs/1304.6712}{{\tt arXiv:1304.6712 [astro-ph.CO]}}.

\bibitem{Gergely:2014rna}
L.~ç. Gergely and S.~Tsujikawa, ``{Effective field theory of modified gravity
  with two scalar fields: dark energy and dark matter},''
  \href{http://dx.doi.org/10.1103/PhysRevD.89.064059}{{\em Phys.Rev.} {\bf D89}
  (2014)  064059},
\href{http://arxiv.org/abs/1402.0553}{{\tt arXiv:1402.0553 [hep-th]}}.

\bibitem{Creminelli:2006xe}
P.~Creminelli, M.~A. Luty, A.~Nicolis, and L.~Senatore, ``{Starting the
  Universe: Stable Violation of the Null Energy Condition and Non-standard
  Cosmologies},'' \href{http://dx.doi.org/10.1088/1126-6708/2006/12/080}{{\em
  JHEP} {\bf 0612} (2006)  080},
\href{http://arxiv.org/abs/hep-th/0606090}{{\tt arXiv:hep-th/0606090
  [hep-th]}}.

\bibitem{Cheung:2007st}
C.~Cheung, P.~Creminelli, A.~L. Fitzpatrick, J.~Kaplan, and L.~Senatore, ``{The
  Effective Field Theory of Inflation},''
  \href{http://dx.doi.org/10.1088/1126-6708/2008/03/014}{{\em JHEP} {\bf 0803}
  (2008)  014},
\href{http://arxiv.org/abs/0709.0293}{{\tt arXiv:0709.0293 [hep-th]}}.

\bibitem{Cheung:2007sv}
C.~Cheung, A.~L. Fitzpatrick, J.~Kaplan, and L.~Senatore, ``{On the consistency
  relation of the 3-point function in single field inflation},''
  \href{http://dx.doi.org/10.1088/1475-7516/2008/02/021}{{\em JCAP} {\bf 0802}
  (2008)  021},
\href{http://arxiv.org/abs/0709.0295}{{\tt arXiv:0709.0295 [hep-th]}}.

\bibitem{Carroll:1991mt}
S.~M. Carroll, W.~H. Press, and E.~L. Turner, ``{The Cosmological constant},''
\href{http://dx.doi.org/10.1146/annurev.aa.30.090192.002435}{{\em
  Ann.Rev.Astron.Astrophys.} {\bf 30} (1992)  499--542}.

\bibitem{Carroll:2000fy}
S.~M. Carroll, ``{The Cosmological constant},'' {\em Living Rev.Rel.} {\bf 4}
  (2001)  1,
\href{http://arxiv.org/abs/astro-ph/0004075}{{\tt arXiv:astro-ph/0004075
  [astro-ph]}}.

\bibitem{Weinberg:2000yb}
S.~Weinberg, ``{The Cosmological constant problems},''
\href{http://arxiv.org/abs/astro-ph/0005265}{{\tt arXiv:astro-ph/0005265
  [astro-ph]}}.

\bibitem{Silvestri:2009hh}
A.~Silvestri and M.~Trodden, ``{Approaches to Understanding Cosmic
  Acceleration},'' \href{http://dx.doi.org/10.1088/0034-4885/72/9/096901}{{\em
  Rept.Prog.Phys.} {\bf 72} (2009)  096901},
\href{http://arxiv.org/abs/0904.0024}{{\tt arXiv:0904.0024 [astro-ph.CO]}}.

\bibitem{Caldwell:2009ix}
R.~R. Caldwell and M.~Kamionkowski, ``{The Physics of Cosmic Acceleration},''
  \href{http://dx.doi.org/10.1146/annurev-nucl-010709-151330}{{\em
  Ann.Rev.Nucl.Part.Sci.} {\bf 59} (2009)  397--429},
\href{http://arxiv.org/abs/0903.0866}{{\tt arXiv:0903.0866 [astro-ph.CO]}}.

\bibitem{Jain:2010ka}
B.~Jain and J.~Khoury, ``{Cosmological Tests of Gravity},''
  \href{http://dx.doi.org/10.1016/j.aop.2010.04.002}{{\em Annals Phys.} {\bf
  325} (2010)  1479--1516},
\href{http://arxiv.org/abs/1004.3294}{{\tt arXiv:1004.3294 [astro-ph.CO]}}.

\bibitem{Tsujikawa:2010zza}
S.~Tsujikawa, ``{Modified gravity models of dark energy},''
  \href{http://dx.doi.org/10.1007/978-3-642-10598-2_3}{{\em Lect.Notes Phys.}
  {\bf 800} (2010)  99--145},
\href{http://arxiv.org/abs/1101.0191}{{\tt arXiv:1101.0191 [gr-qc]}}.

\bibitem{Clifton:2011jh}
T.~Clifton, P.~G. Ferreira, A.~Padilla, and C.~Skordis, ``{Modified Gravity and
  Cosmology},'' \href{http://dx.doi.org/10.1016/j.physrep.2012.01.001}{{\em
  Phys.Rept.} {\bf 513} (2012)  1--189},
\href{http://arxiv.org/abs/1106.2476}{{\tt arXiv:1106.2476 [astro-ph.CO]}}.

\bibitem{Brax:2012bsa}
P.~Brax, ``{Screened modified gravity},''
  \href{http://dx.doi.org/10.5506/APhysPolB.43.2307}{{\em Acta Phys.Polon.}
  {\bf B43} (2012)  2307--2329},
\href{http://arxiv.org/abs/1211.5237}{{\tt arXiv:1211.5237 [hep-th]}}.

\bibitem{Khoury:2013tda}
J.~Khoury, ``{Les Houches Lectures on Physics Beyond the Standard Model of
  Cosmology},''
\href{http://arxiv.org/abs/1312.2006}{{\tt arXiv:1312.2006 [astro-ph.CO]}}.

\bibitem{Woodard:2006nt}
R.~P. Woodard, ``{Avoiding dark energy with 1/r modifications of gravity},''
  \href{http://dx.doi.org/10.1007/978-3-540-71013-4_14}{{\em Lect.Notes Phys.}
  {\bf 720} (2007)  403--433},
\href{http://arxiv.org/abs/astro-ph/0601672}{{\tt arXiv:astro-ph/0601672
  [astro-ph]}}.

\bibitem{Sotiriou:2008rp}
T.~P. Sotiriou and V.~Faraoni, ``{f(R) Theories Of Gravity},''
  \href{http://dx.doi.org/10.1103/RevModPhys.82.451}{{\em Rev.Mod.Phys.} {\bf
  82} (2010)  451--497},
\href{http://arxiv.org/abs/0805.1726}{{\tt arXiv:0805.1726 [gr-qc]}}.

\bibitem{DeFelice:2010aj}
A.~De~Felice and S.~Tsujikawa, ``{f(R) theories},'' {\em Living Rev.Rel.} {\bf
  13} (2010)  3,
\href{http://arxiv.org/abs/1002.4928}{{\tt arXiv:1002.4928 [gr-qc]}}.

\bibitem{Nojiri:2010wj}
S.~Nojiri and S.~D. Odintsov, ``{Unified cosmic history in modified gravity:
  from F(R) theory to Lorentz non-invariant models},''
  \href{http://dx.doi.org/10.1016/j.physrep.2011.04.001}{{\em Phys.Rept.} {\bf
  505} (2011)  59--144},
\href{http://arxiv.org/abs/1011.0544}{{\tt arXiv:1011.0544 [gr-qc]}}.

\bibitem{Khoury:2013yya}
J.~Khoury, ``{Chameleon Field Theories},''
  \href{http://dx.doi.org/10.1088/0264-9381/30/21/214004}{{\em
  Class.Quant.Grav.} {\bf 30} (2013)  214004},
\href{http://arxiv.org/abs/1306.4326}{{\tt arXiv:1306.4326 [astro-ph.CO]}}.

\bibitem{Khoury:2010xi}
J.~Khoury, ``{Theories of Dark Energy with Screening Mechanisms},''
\href{http://arxiv.org/abs/1011.5909}{{\tt arXiv:1011.5909 [astro-ph.CO]}}.

\bibitem{Trodden:2011xh}
M.~Trodden and K.~Hinterbichler, ``{Generalizing Galileons},''
  \href{http://dx.doi.org/10.1088/0264-9381/28/20/204003}{{\em
  Class.Quant.Grav.} {\bf 28} (2011)  204003},
\href{http://arxiv.org/abs/1104.2088}{{\tt arXiv:1104.2088 [hep-th]}}.

\bibitem{deRham:2012az}
C.~de~Rham, ``{Galileons in the Sky},''
  \href{http://dx.doi.org/10.1016/j.crhy.2012.04.006}{{\em Comptes Rendus
  Physique} {\bf 13} (2012)  666--681},
\href{http://arxiv.org/abs/1204.5492}{{\tt arXiv:1204.5492 [astro-ph.CO]}}.

\bibitem{Deffayet:2013lga}
C.~Deffayet and D.~A. Steer, ``{A formal introduction to Horndeski and Galileon
  theories and their generalizations},''
  \href{http://dx.doi.org/10.1088/0264-9381/30/21/214006}{{\em
  Class.Quant.Grav.} {\bf 30} (2013)  214006},
\href{http://arxiv.org/abs/1307.2450}{{\tt arXiv:1307.2450 [hep-th]}}.

\bibitem{Babichev:2013usa}
E.~Babichev and C.~Deffayet, ``{An introduction to the Vainshtein mechanism},''
  \href{http://dx.doi.org/10.1088/0264-9381/30/18/184001}{{\em
  Class.Quant.Grav.} {\bf 30} (2013)  184001},
\href{http://arxiv.org/abs/1304.7240}{{\tt arXiv:1304.7240 [gr-qc]}}.

\bibitem{Hinterbichler:2011tt}
K.~Hinterbichler, ``{Theoretical Aspects of Massive Gravity},''
  \href{http://dx.doi.org/10.1103/RevModPhys.84.671}{{\em Rev.Mod.Phys.} {\bf
  84} (2012)  671--710},
\href{http://arxiv.org/abs/1105.3735}{{\tt arXiv:1105.3735 [hep-th]}}.

\bibitem{deRham:2014zqa}
C.~de~Rham, ``{Massive Gravity},''
\href{http://arxiv.org/abs/1401.4173}{{\tt arXiv:1401.4173 [hep-th]}}.

\bibitem{Sahni:2006pa}
V.~Sahni and A.~Starobinsky, ``{Reconstructing Dark Energy},''
  \href{http://dx.doi.org/10.1142/S0218271806009704}{{\em Int.J.Mod.Phys.} {\bf
  D15} (2006)  2105--2132},
\href{http://arxiv.org/abs/astro-ph/0610026}{{\tt arXiv:astro-ph/0610026
  [astro-ph]}}.

\bibitem{Weinberg:2012es}
D.~H. Weinberg, M.~J. Mortonson, D.~J. Eisenstein, C.~Hirata, A.~G. Riess, {\em
  et al.}, ``{Observational Probes of Cosmic Acceleration},''
  \href{http://dx.doi.org/10.1016/j.physrep.2013.05.001}{{\em Phys.Rept.} {\bf
  530} (2013)  87--255},
\href{http://arxiv.org/abs/1201.2434}{{\tt arXiv:1201.2434 [astro-ph.CO]}}.

\bibitem{Rubakov:2001kp}
V.~Rubakov, ``{Large and infinite extra dimensions: An Introduction},''
  \href{http://dx.doi.org/10.1070/PU2001v044n09ABEH001000}{{\em Phys.Usp.} {\bf
  44} (2001)  871--893},
\href{http://arxiv.org/abs/hep-ph/0104152}{{\tt arXiv:hep-ph/0104152
  [hep-ph]}}.

\bibitem{Langlois:2002bb}
D.~Langlois, ``{Brane cosmology: An Introduction},''
  \href{http://dx.doi.org/10.1143/PTPS.148.181}{{\em Prog.Theor.Phys.Suppl.}
  {\bf 148} (2003)  181--212},
\href{http://arxiv.org/abs/hep-th/0209261}{{\tt arXiv:hep-th/0209261
  [hep-th]}}.

\bibitem{Brax:2003fv}
P.~Brax and C.~van~de Bruck, ``{Cosmology and brane worlds: A Review},''
  \href{http://dx.doi.org/10.1088/0264-9381/20/9/202}{{\em Class.Quant.Grav.}
  {\bf 20} (2003)  R201--R232},
\href{http://arxiv.org/abs/hep-th/0303095}{{\tt arXiv:hep-th/0303095
  [hep-th]}}.

\bibitem{Maartens:2003tw}
R.~Maartens, ``{Brane world gravity},'' {\em Living Rev.Rel.} {\bf 7} (2004)
  7,
\href{http://arxiv.org/abs/gr-qc/0312059}{{\tt arXiv:gr-qc/0312059 [gr-qc]}}.

\bibitem{Brax:2004xh}
P.~Brax, C.~van~de Bruck, and A.-C. Davis, ``{Brane world cosmology},''
  \href{http://dx.doi.org/10.1088/0034-4885/67/12/R02}{{\em Rept.Prog.Phys.}
  {\bf 67} (2004)  2183--2232},
\href{http://arxiv.org/abs/hep-th/0404011}{{\tt arXiv:hep-th/0404011
  [hep-th]}}.

\bibitem{Lue:2005ya}
A.~Lue, ``{The phenomenology of dvali-gabadadze-porrati cosmologies},''
  \href{http://dx.doi.org/10.1016/j.physrep.2005.10.007}{{\em Phys.Rept.} {\bf
  423} (2006)  1--48},
\href{http://arxiv.org/abs/astro-ph/0510068}{{\tt arXiv:astro-ph/0510068
  [astro-ph]}}.

\bibitem{Maartens:2010ar}
R.~Maartens and K.~Koyama, ``{Brane-World Gravity},'' {\em Living Rev.Rel.}
  {\bf 13} (2010)  5,
\href{http://arxiv.org/abs/1004.3962}{{\tt arXiv:1004.3962 [hep-th]}}.

\bibitem{Brax:2004qh}
P.~Brax, C.~van~de Bruck, A.-C. Davis, J.~Khoury, and A.~Weltman, ``{Detecting
  dark energy in orbit - The Cosmological chameleon},''
  \href{http://dx.doi.org/10.1103/PhysRevD.70.123518}{{\em Phys.Rev.} {\bf D70}
  (2004)  123518},
\href{http://arxiv.org/abs/astro-ph/0408415}{{\tt arXiv:astro-ph/0408415
  [astro-ph]}}.

\bibitem{Brax:2008hh}
P.~Brax, C.~van~de Bruck, A.-C. Davis, and D.~J. Shaw, ``{f(R) Gravity and
  Chameleon Theories},''
  \href{http://dx.doi.org/10.1103/PhysRevD.78.104021}{{\em Phys.Rev.} {\bf D78}
  (2008)  104021},
\href{http://arxiv.org/abs/0806.3415}{{\tt arXiv:0806.3415 [astro-ph]}}.

\bibitem{Mota:2006fz}
D.~F. Mota and D.~J. Shaw, ``{Evading Equivalence Principle Violations,
  Cosmological and other Experimental Constraints in Scalar Field Theories with
  a Strong Coupling to Matter},''
  \href{http://dx.doi.org/10.1103/PhysRevD.75.063501}{{\em Phys.Rev.} {\bf D75}
  (2007)  063501},
\href{http://arxiv.org/abs/hep-ph/0608078}{{\tt arXiv:hep-ph/0608078
  [hep-ph]}}.

\bibitem{Mota:2006ed}
D.~F. Mota and D.~J. Shaw, ``{Strongly coupled chameleon fields: New horizons
  in scalar field theory},''
  \href{http://dx.doi.org/10.1103/PhysRevLett.97.151102}{{\em Phys.Rev.Lett.}
  {\bf 97} (2006)  151102},
\href{http://arxiv.org/abs/hep-ph/0606204}{{\tt arXiv:hep-ph/0606204
  [hep-ph]}}.

\bibitem{Feldman:2006wg}
B.~Feldman and A.~E. Nelson, ``{New regions for a chameleon to hide},''
  \href{http://dx.doi.org/10.1088/1126-6708/2006/08/002}{{\em JHEP} {\bf 0608}
  (2006)  002},
\href{http://arxiv.org/abs/hep-ph/0603057}{{\tt arXiv:hep-ph/0603057
  [hep-ph]}}.

\bibitem{Brax:2007ak}
P.~Brax, C.~van~de Bruck, and A.-C. Davis, ``{Compatibility of the
  chameleon-field model with fifth-force experiments, cosmology, and PVLAS and
  CAST results},'' \href{http://dx.doi.org/10.1103/PhysRevLett.99.121103}{{\em
  Phys.Rev.Lett.} {\bf 99} (2007)  121103},
\href{http://arxiv.org/abs/hep-ph/0703243}{{\tt arXiv:hep-ph/0703243
  [HEP-PH]}}.

\bibitem{Das:2008iq}
S.~Das and N.~Banerjee, ``{Brans-Dicke Scalar Field as a Chameleon},''
  \href{http://dx.doi.org/10.1103/PhysRevD.78.043512}{{\em Phys.Rev.} {\bf D78}
  (2008)  043512},
\href{http://arxiv.org/abs/0803.3936}{{\tt arXiv:0803.3936 [gr-qc]}}.

\bibitem{Davis:2009vk}
A.-C. Davis, C.~A. Schelpe, and D.~J. Shaw, ``{The Effect of a Chameleon Scalar
  Field on the Cosmic Microwave Background},''
  \href{http://dx.doi.org/10.1103/PhysRevD.80.064016}{{\em Phys.Rev.} {\bf D80}
  (2009)  064016},
\href{http://arxiv.org/abs/0907.2672}{{\tt arXiv:0907.2672 [astro-ph.CO]}}.

\bibitem{Brax:2010tj}
P.~Brax, R.~Rosenfeld, and D.~Steer, ``{Spherical Collapse in Chameleon
  Models},'' \href{http://dx.doi.org/10.1088/1475-7516/2010/08/033}{{\em JCAP}
  {\bf 1008} (2010)  033},
\href{http://arxiv.org/abs/1005.2051}{{\tt arXiv:1005.2051 [astro-ph.CO]}}.

\bibitem{Brax:2010kv}
P.~Brax, C.~van~de Bruck, D.~F. Mota, N.~J. Nunes, and H.~A. Winther,
  ``{Chameleons with Field Dependent Couplings},''
  \href{http://dx.doi.org/10.1103/PhysRevD.82.083503}{{\em Phys.Rev.} {\bf D82}
  (2010)  083503},
\href{http://arxiv.org/abs/1006.2796}{{\tt arXiv:1006.2796 [astro-ph.CO]}}.

\bibitem{Cannata:2010qd}
F.~Cannata and A.~Y. Kamenshchik, ``{Chameleon Cosmology Model Describing the
  Phantom Divide Line Crossing},''
  \href{http://dx.doi.org/10.1142/S0218271811018755}{{\em Int.J.Mod.Phys.} {\bf
  D20} (2011)  121--131},
\href{http://arxiv.org/abs/1005.1878}{{\tt arXiv:1005.1878 [gr-qc]}}.

\bibitem{Boddy:2012xs}
K.~K. Boddy, S.~M. Carroll, and M.~Trodden, ``{Dark Matter with
  Density-Dependent Interactions},''
  \href{http://dx.doi.org/10.1103/PhysRevD.86.123529}{{\em Phys.Rev.} {\bf D86}
  (2012)  123529},
\href{http://arxiv.org/abs/1208.4376}{{\tt arXiv:1208.4376 [astro-ph.CO]}}.

\bibitem{Nastase:2013los}
H.~Nastase and A.~Weltman, ``{A natural cosmological constant from
  chameleons},''
\href{http://arxiv.org/abs/1302.1748}{{\tt arXiv:1302.1748 [hep-th]}}.

\bibitem{Erickcek:2013oma}
A.~L. Erickcek, N.~Barnaby, C.~Burrage, and Z.~Huang, ``{Catastrophic
  Consequences of Kicking the Chameleon},'' {\em Phys. Rev. Lett.} {\bf
  110:171101} (2013)  ,
\href{http://arxiv.org/abs/1304.0009}{{\tt arXiv:1304.0009 [astro-ph.CO]}}.

\bibitem{Erickcek:2013dea}
A.~L. Erickcek, N.~Barnaby, C.~Burrage, and Z.~Huang, ``{Chameleons in the
  Early Universe: Kicks, Rebounds, and Particle Production},'' {\em Phys.Rev.}
  {\bf D89} (2014)  084074,
\href{http://arxiv.org/abs/1310.5149}{{\tt arXiv:1310.5149 [astro-ph.CO]}}.

\bibitem{Wei:2004rw}
H.~Wei and R.-G. Cai, ``{K-chameleon and the coincidence problem},''
  \href{http://dx.doi.org/10.1103/PhysRevD.71.043504}{{\em Phys.Rev.} {\bf D71}
  (2005)  043504},
\href{http://arxiv.org/abs/hep-th/0412045}{{\tt arXiv:hep-th/0412045
  [hep-th]}}.

\bibitem{Hinterbichler:2013we}
K.~Hinterbichler, J.~Khoury, H.~Nastase, and R.~Rosenfeld, ``{Chameleonic
  inflation},'' \href{http://dx.doi.org/10.1007/JHEP08(2013)053}{{\em JHEP}
  {\bf 1308} (2013)  053},
\href{http://arxiv.org/abs/1301.6756}{{\tt arXiv:1301.6756 [hep-th]}}.

\bibitem{Nastase:2013ik}
H.~Nastase and A.~Weltman, ``{Chameleons on the Racetrack},''
  \href{http://dx.doi.org/10.1007/JHEP08(2013)059}{{\em JHEP} {\bf 1308} (2013)
   059},
\href{http://arxiv.org/abs/1301.7120}{{\tt arXiv:1301.7120 [hep-th]}}.

\bibitem{Brax:2004ym}
P.~Brax, C.~van~de Bruck, and A.~Davis, ``{Is the radion a chameleon?},''
  \href{http://dx.doi.org/10.1088/1475-7516/2004/11/004}{{\em JCAP} {\bf 0411}
  (2004)  004},
\href{http://arxiv.org/abs/astro-ph/0408464}{{\tt arXiv:astro-ph/0408464
  [astro-ph]}}.

\bibitem{Hinterbichler:2010wu}
K.~Hinterbichler, J.~Khoury, and H.~Nastase, ``{Towards a UV Completion for
  Chameleon Scalar Theories},''
  \href{http://dx.doi.org/10.1007/JHEP06(2011)072,
  10.1007/JHEP03(2011)061}{{\em JHEP} {\bf 1103} (2011)  061},
\href{http://arxiv.org/abs/1012.4462}{{\tt arXiv:1012.4462 [hep-th]}}.

\bibitem{Brax:2011qs}
P.~Brax and A.-C. Davis, ``{Supersymmetron},''
  \href{http://dx.doi.org/10.1016/j.physletb.2011.11.060}{{\em Phys.Lett.} {\bf
  B707} (2012)  1--7},
\href{http://arxiv.org/abs/1109.0468}{{\tt arXiv:1109.0468 [hep-ph]}}.

\bibitem{Brax:2012mq}
P.~Brax, A.-C. Davis, and J.~Sakstein, ``{SUPER-Screening},''
  \href{http://dx.doi.org/10.1016/j.physletb.2013.01.044}{{\em Phys.Lett.B}
  {\bf 719} (2013)  210--217},
\href{http://arxiv.org/abs/1212.4392}{{\tt arXiv:1212.4392 [hep-th]}}.

\bibitem{Brax:2013yja}
P.~Brax, A.-C. Davis, and J.~Sakstein, ``{Dynamics of Supersymmetric
  Chameleons},'' \href{http://dx.doi.org/10.1088/1475-7516/2013/10/007}{{\em
  JCAP} {\bf 1310} (2013)  007},
\href{http://arxiv.org/abs/1302.3080}{{\tt arXiv:1302.3080 [astro-ph.CO]}}.

\bibitem{Adelberger:2006dh}
E.~Adelberger, B.~R. Heckel, S.~A. Hoedl, C.~Hoyle, D.~Kapner, {\em et al.},
  ``{Particle Physics Implications of a Recent Test of the Gravitational
  Inverse Sqaure Law},''
  \href{http://dx.doi.org/10.1103/PhysRevLett.98.131104}{{\em Phys.Rev.Lett.}
  {\bf 98} (2007)  131104},
\href{http://arxiv.org/abs/hep-ph/0611223}{{\tt arXiv:hep-ph/0611223
  [hep-ph]}}.

\bibitem{JonesSmith:2011tn}
K.~Jones-Smith and F.~Ferrer, ``{Detecting Chameleon Dark Energy via
  Electrostatic Analogy},''
  \href{http://dx.doi.org/10.1103/PhysRevLett.108.221101}{{\em Phys.Rev.Lett.}
  {\bf 108} (2012)  221101},
\href{http://arxiv.org/abs/1105.6085}{{\tt arXiv:1105.6085 [astro-ph.CO]}}.

\bibitem{Pourhasan:2011sm}
R.~Pourhasan, N.~Afshordi, R.~Mann, and A.~Davis, ``{Chameleon Gravity,
  Electrostatics, and Kinematics in the Outer Galaxy},''
  \href{http://dx.doi.org/10.1088/1475-7516/2011/12/005}{{\em JCAP} {\bf 1112}
  (2011)  005},
\href{http://arxiv.org/abs/1109.0538}{{\tt arXiv:1109.0538 [astro-ph.CO]}}.

\bibitem{Stelle:1976gc}
K.~Stelle, ``{Renormalization of Higher Derivative Quantum Gravity},''
\href{http://dx.doi.org/10.1103/PhysRevD.16.953}{{\em Phys.Rev.} {\bf D16}
  (1977)  953--969}.

\bibitem{Stelle:1977ry}
K.~S. Stelle, ``{Classical Gravity with Higher Derivatives},''
\href{http://dx.doi.org/10.1007/BF00760427}{{\em Gen.Rel.Grav.} {\bf 9} (1978)
  353--371}.

\bibitem{Starobinsky:1980te}
A.~A. Starobinsky, ``{A New Type of Isotropic Cosmological Models Without
  Singularity},''
\href{http://dx.doi.org/10.1016/0370-2693(80)90670-X}{{\em Phys.Lett.} {\bf
  B91} (1980)  99--102}.

\bibitem{Capozziello:2002rd}
S.~Capozziello, ``{Curvature quintessence},''
  \href{http://dx.doi.org/10.1142/S0218271802002025}{{\em Int.J.Mod.Phys.} {\bf
  D11} (2002)  483--492},
\href{http://arxiv.org/abs/gr-qc/0201033}{{\tt arXiv:gr-qc/0201033 [gr-qc]}}.

\bibitem{Capozziello:2003tk}
S.~Capozziello, S.~Carloni, and A.~Troisi, ``{Quintessence without scalar
  fields},'' {\em Recent Res.Dev.Astron.Astrophys.} {\bf 1} (2003)  625,
\href{http://arxiv.org/abs/astro-ph/0303041}{{\tt arXiv:astro-ph/0303041
  [astro-ph]}}.

\bibitem{Carroll:2003wy}
S.~M. Carroll, V.~Duvvuri, M.~Trodden, and M.~S. Turner, ``{Is cosmic speed -
  up due to new gravitational physics?},''
  \href{http://dx.doi.org/10.1103/PhysRevD.70.043528}{{\em Phys.Rev.} {\bf D70}
  (2004)  043528},
\href{http://arxiv.org/abs/astro-ph/0306438}{{\tt arXiv:astro-ph/0306438
  [astro-ph]}}.

\bibitem{Chiba:2003ir}
T.~Chiba, ``{1/R gravity and scalar - tensor gravity},''
  \href{http://dx.doi.org/10.1016/j.physletb.2003.09.033}{{\em Phys.Lett.} {\bf
  B575} (2003)  1--3},
\href{http://arxiv.org/abs/astro-ph/0307338}{{\tt arXiv:astro-ph/0307338
  [astro-ph]}}.

\bibitem{Nunez:2004ji}
A.~Nunez and S.~Solganik, ``{The Content of f(R) gravity},''
\href{http://arxiv.org/abs/hep-th/0403159}{{\tt arXiv:hep-th/0403159
  [hep-th]}}.

\bibitem{Soussa:2003re}
M.~Soussa and R.~Woodard, ``{The force of gravity from a Lagrangian containing
  inverse powers of the ricci scalar},''
  \href{http://dx.doi.org/10.1023/B:GERG.0000017037.92729.69}{{\em
  Gen.Rel.Grav.} {\bf 36} (2004)  855--862},
\href{http://arxiv.org/abs/astro-ph/0308114}{{\tt arXiv:astro-ph/0308114
  [astro-ph]}}.

\bibitem{Dolgov:2003px}
A.~Dolgov and M.~Kawasaki, ``{Can modified gravity explain accelerated cosmic
  expansion?},'' \href{http://dx.doi.org/10.1016/j.physletb.2003.08.039}{{\em
  Phys.Lett.} {\bf B573} (2003)  1--4},
\href{http://arxiv.org/abs/astro-ph/0307285}{{\tt arXiv:astro-ph/0307285
  [astro-ph]}}.

\bibitem{DeFelice:2006pg}
A.~De~Felice, M.~Hindmarsh, and M.~Trodden, ``{Ghosts, Instabilities, and
  Superluminal Propagation in Modified Gravity Models},''
  \href{http://dx.doi.org/10.1088/1475-7516/2006/08/005}{{\em JCAP} {\bf 0608}
  (2006)  005},
\href{http://arxiv.org/abs/astro-ph/0604154}{{\tt arXiv:astro-ph/0604154
  [astro-ph]}}.

\bibitem{Sawicki:2007tf}
I.~Sawicki and W.~Hu, ``{Stability of Cosmological Solution in f(R) Models of
  Gravity},'' \href{http://dx.doi.org/10.1103/PhysRevD.75.127502}{{\em
  Phys.Rev.} {\bf D75} (2007)  127502},
\href{http://arxiv.org/abs/astro-ph/0702278}{{\tt arXiv:astro-ph/0702278
  [astro-ph]}}.

\bibitem{Amarzguioui:2005zq}
M.~Amarzguioui, O.~Elgaroy, D.~Mota, and T.~Multamaki, ``{Cosmological
  constraints on f(r) gravity theories within the palatini approach},''
  \href{http://dx.doi.org/10.1051/0004-6361:20064994}{{\em Astron.Astrophys.}
  {\bf 454} (2006)  707--714},
\href{http://arxiv.org/abs/astro-ph/0510519}{{\tt arXiv:astro-ph/0510519
  [astro-ph]}}.

\bibitem{Amendola:2006kh}
L.~Amendola, D.~Polarski, and S.~Tsujikawa, ``{Are f(R) dark energy models
  cosmologically viable ?},''
  \href{http://dx.doi.org/10.1103/PhysRevLett.98.131302}{{\em Phys.Rev.Lett.}
  {\bf 98} (2007)  131302},
\href{http://arxiv.org/abs/astro-ph/0603703}{{\tt arXiv:astro-ph/0603703
  [astro-ph]}}.

\bibitem{Amendola:2006we}
L.~Amendola, R.~Gannouji, D.~Polarski, and S.~Tsujikawa, ``{Conditions for the
  cosmological viability of f(R) dark energy models},''
  \href{http://dx.doi.org/10.1103/PhysRevD.75.083504}{{\em Phys.Rev.} {\bf D75}
  (2007)  083504},
\href{http://arxiv.org/abs/gr-qc/0612180}{{\tt arXiv:gr-qc/0612180 [gr-qc]}}.

\bibitem{Flanagan:2003rb}
E.~E. Flanagan, ``{Palatini form of 1/R gravity},''
  \href{http://dx.doi.org/10.1103/PhysRevLett.92.071101}{{\em Phys.Rev.Lett.}
  {\bf 92} (2004)  071101},
\href{http://arxiv.org/abs/astro-ph/0308111}{{\tt arXiv:astro-ph/0308111
  [astro-ph]}}.

\bibitem{Olmo:2005zr}
G.~J. Olmo, ``{The Gravity Lagrangian according to solar system experiments},''
  \href{http://dx.doi.org/10.1103/PhysRevLett.95.261102}{{\em Phys.Rev.Lett.}
  {\bf 95} (2005)  261102},
\href{http://arxiv.org/abs/gr-qc/0505101}{{\tt arXiv:gr-qc/0505101 [gr-qc]}}.

\bibitem{Chiba:2006jp}
T.~Chiba, T.~L. Smith, and A.~L. Erickcek, ``{Solar System constraints to
  general f(R) gravity},''
  \href{http://dx.doi.org/10.1103/PhysRevD.75.124014}{{\em Phys.Rev.} {\bf D75}
  (2007)  124014},
\href{http://arxiv.org/abs/astro-ph/0611867}{{\tt arXiv:astro-ph/0611867
  [astro-ph]}}.

\bibitem{Erickcek:2006vf}
A.~L. Erickcek, T.~L. Smith, and M.~Kamionkowski, ``{Solar System tests do rule
  out 1/R gravity},'' \href{http://dx.doi.org/10.1103/PhysRevD.74.121501}{{\em
  Phys.Rev.} {\bf D74} (2006)  121501},
\href{http://arxiv.org/abs/astro-ph/0610483}{{\tt arXiv:astro-ph/0610483
  [astro-ph]}}.

\bibitem{Jin:2006if}
X.-H. Jin, D.-J. Liu, and X.-Z. Li, ``{Solar System tests disfavor f(R)
  gravities},''
\href{http://arxiv.org/abs/astro-ph/0610854}{{\tt arXiv:astro-ph/0610854
  [astro-ph]}}.

\bibitem{Multamaki:2006zb}
T.~Multamaki and I.~Vilja, ``{Spherically symmetric solutions of modified field
  equations in f(R) theories of gravity},''
  \href{http://dx.doi.org/10.1103/PhysRevD.74.064022}{{\em Phys.Rev.} {\bf D74}
  (2006)  064022},
\href{http://arxiv.org/abs/astro-ph/0606373}{{\tt arXiv:astro-ph/0606373
  [astro-ph]}}.

\bibitem{Faulkner:2006ub}
T.~Faulkner, M.~Tegmark, E.~F. Bunn, and Y.~Mao, ``{Constraining f(R) Gravity
  as a Scalar Tensor Theory},''
  \href{http://dx.doi.org/10.1103/PhysRevD.76.063505}{{\em Phys.Rev.} {\bf D76}
  (2007)  063505},
\href{http://arxiv.org/abs/astro-ph/0612569}{{\tt arXiv:astro-ph/0612569
  [astro-ph]}}.

\bibitem{Navarro:2006mw}
I.~Navarro and K.~Van~Acoleyen, ``{f(R) actions, cosmic acceleration and local
  tests of gravity},''
  \href{http://dx.doi.org/10.1088/1475-7516/2007/02/022}{{\em JCAP} {\bf 0702}
  (2007)  022},
\href{http://arxiv.org/abs/gr-qc/0611127}{{\tt arXiv:gr-qc/0611127 [gr-qc]}}.

\bibitem{Hu:2007nk}
W.~Hu and I.~Sawicki, ``{Models of f(R) Cosmic Acceleration that Evade
  Solar-System Tests},''
  \href{http://dx.doi.org/10.1103/PhysRevD.76.064004}{{\em Phys.Rev.} {\bf D76}
  (2007)  064004},
\href{http://arxiv.org/abs/0705.1158}{{\tt arXiv:0705.1158 [astro-ph]}}.

\bibitem{Starobinsky:2007hu}
A.~A. Starobinsky, ``{Disappearing cosmological constant in f(R) gravity},''
  \href{http://dx.doi.org/10.1134/S0021364007150027}{{\em JETP Lett.} {\bf 86}
  (2007)  157--163},
\href{http://arxiv.org/abs/0706.2041}{{\tt arXiv:0706.2041 [astro-ph]}}.

\bibitem{Capozziello:2007eu}
S.~Capozziello and S.~Tsujikawa, ``{Solar system and equivalence principle
  constraints on f(R) gravity by chameleon approach},''
  \href{http://dx.doi.org/10.1103/PhysRevD.77.107501}{{\em Phys.Rev.} {\bf D77}
  (2008)  107501},
\href{http://arxiv.org/abs/0712.2268}{{\tt arXiv:0712.2268 [gr-qc]}}.

\bibitem{Tsujikawa:2007xu}
S.~Tsujikawa, ``{Observational signatures of f(R) dark energy models that
  satisfy cosmological and local gravity constraints},''
  \href{http://dx.doi.org/10.1103/PhysRevD.77.023507}{{\em Phys.Rev.} {\bf D77}
  (2008)  023507},
\href{http://arxiv.org/abs/0709.1391}{{\tt arXiv:0709.1391 [astro-ph]}}.

\bibitem{Vollick:2003aw}
D.~N. Vollick, ``{1/R Curvature corrections as the source of the cosmological
  acceleration},'' \href{http://dx.doi.org/10.1103/PhysRevD.68.063510}{{\em
  Phys.Rev.} {\bf D68} (2003)  063510},
\href{http://arxiv.org/abs/astro-ph/0306630}{{\tt arXiv:astro-ph/0306630
  [astro-ph]}}.

\bibitem{Carloni:2004kp}
S.~Carloni, P.~K. Dunsby, S.~Capozziello, and A.~Troisi, ``{Cosmological
  dynamics of R**n gravity},''
  \href{http://dx.doi.org/10.1088/0264-9381/22/22/011}{{\em Class.Quant.Grav.}
  {\bf 22} (2005)  4839--4868},
\href{http://arxiv.org/abs/gr-qc/0410046}{{\tt arXiv:gr-qc/0410046 [gr-qc]}}.

\bibitem{Allemandi:2004wn}
G.~Allemandi, A.~Borowiec, and M.~Francaviglia, ``{Accelerated cosmological
  models in Ricci squared gravity},''
  \href{http://dx.doi.org/10.1103/PhysRevD.70.103503}{{\em Phys.Rev.} {\bf D70}
  (2004)  103503},
\href{http://arxiv.org/abs/hep-th/0407090}{{\tt arXiv:hep-th/0407090
  [hep-th]}}.

\bibitem{Capozziello:2006dj}
S.~Capozziello, S.~Nojiri, S.~Odintsov, and A.~Troisi, ``{Cosmological
  viability of f(R)-gravity as an ideal fluid and its compatibility with a
  matter dominated phase},''
  \href{http://dx.doi.org/10.1016/j.physletb.2006.06.034}{{\em Phys.Lett.} {\bf
  B639} (2006)  135--143},
\href{http://arxiv.org/abs/astro-ph/0604431}{{\tt arXiv:astro-ph/0604431
  [astro-ph]}}.

\bibitem{Nojiri:2006gh}
S.~Nojiri and S.~D. Odintsov, ``{Modified f(R) gravity consistent with
  realistic cosmology: From matter dominated epoch to dark energy universe},''
  \href{http://dx.doi.org/10.1103/PhysRevD.74.086005}{{\em Phys.Rev.} {\bf D74}
  (2006)  086005},
\href{http://arxiv.org/abs/hep-th/0608008}{{\tt arXiv:hep-th/0608008
  [hep-th]}}.

\bibitem{delaCruzDombriz:2006fj}
A.~de~la Cruz-Dombriz and A.~Dobado, ``{A f(R) gravity without cosmological
  constant},'' \href{http://dx.doi.org/10.1103/PhysRevD.74.087501}{{\em
  Phys.Rev.} {\bf D74} (2006)  087501},
\href{http://arxiv.org/abs/gr-qc/0607118}{{\tt arXiv:gr-qc/0607118 [gr-qc]}}.

\bibitem{Capozziello:2007ec}
S.~Capozziello and M.~Francaviglia, ``{Extended Theories of Gravity and their
  Cosmological and Astrophysical Applications},''
  \href{http://dx.doi.org/10.1007/s10714-007-0551-y}{{\em Gen.Rel.Grav.} {\bf
  40} (2008)  357--420},
\href{http://arxiv.org/abs/0706.1146}{{\tt arXiv:0706.1146 [astro-ph]}}.

\bibitem{Motohashi:2010zz}
H.~Motohashi, A.~A. Starobinsky, and J.~Yokoyama, ``{f(R) Gravity and its
  Cosmological Implications},''
  \href{http://dx.doi.org/10.1142/S0218271811019529}{{\em Int.J.Mod.Phys.} {\bf
  D20} (2011)  1347--1355},
\href{http://arxiv.org/abs/1101.0716}{{\tt arXiv:1101.0716 [astro-ph.CO]}}.

\bibitem{Motohashi:2012wc}
H.~Motohashi, A.~A. Starobinsky, and J.~Yokoyama, ``{Cosmology Based on f(R)
  Gravity Admits 1 eV Sterile Neutrinos},''
  \href{http://dx.doi.org/10.1103/PhysRevLett.110.121302}{{\em Phys.Rev.Lett.}
  {\bf 110} (2013) no.~12, 121302},
\href{http://arxiv.org/abs/1203.6828}{{\tt arXiv:1203.6828 [astro-ph.CO]}}.

\bibitem{Zhang:2005vt}
P.~Zhang, ``{Testing $f(R)$ gravity against the large scale structure of the
  universe.},'' \href{http://dx.doi.org/10.1103/PhysRevD.73.123504}{{\em
  Phys.Rev.} {\bf D73} (2006)  123504},
\href{http://arxiv.org/abs/astro-ph/0511218}{{\tt arXiv:astro-ph/0511218
  [astro-ph]}}.

\bibitem{Bean:2006up}
R.~Bean, D.~Bernat, L.~Pogosian, A.~Silvestri, and M.~Trodden, ``{Dynamics of
  Linear Perturbations in f(R) Gravity},''
  \href{http://dx.doi.org/10.1103/PhysRevD.75.064020}{{\em Phys.Rev.} {\bf D75}
  (2007)  064020},
\href{http://arxiv.org/abs/astro-ph/0611321}{{\tt arXiv:astro-ph/0611321
  [astro-ph]}}.

\bibitem{Song:2006ej}
Y.-S. Song, W.~Hu, and I.~Sawicki, ``{The Large Scale Structure of f(R)
  Gravity},'' \href{http://dx.doi.org/10.1103/PhysRevD.75.044004}{{\em
  Phys.Rev.} {\bf D75} (2007)  044004},
\href{http://arxiv.org/abs/astro-ph/0610532}{{\tt arXiv:astro-ph/0610532
  [astro-ph]}}.

\bibitem{Koivisto:2006ie}
T.~Koivisto, ``{The matter power spectrum in f(r) gravity},''
  \href{http://dx.doi.org/10.1103/PhysRevD.73.083517}{{\em Phys.Rev.} {\bf D73}
  (2006)  083517},
\href{http://arxiv.org/abs/astro-ph/0602031}{{\tt arXiv:astro-ph/0602031
  [astro-ph]}}.

\bibitem{Hu:2007pj}
W.~Hu and I.~Sawicki, ``{A Parameterized Post-Friedmann Framework for Modified
  Gravity},'' \href{http://dx.doi.org/10.1103/PhysRevD.76.104043}{{\em
  Phys.Rev.} {\bf D76} (2007)  104043},
\href{http://arxiv.org/abs/0708.1190}{{\tt arXiv:0708.1190 [astro-ph]}}.

\bibitem{Song:2007da}
Y.-S. Song, H.~Peiris, and W.~Hu, ``{Cosmological Constraints on f(R)
  Acceleration Models},''
  \href{http://dx.doi.org/10.1103/PhysRevD.76.063517}{{\em Phys.Rev.} {\bf D76}
  (2007)  063517},
\href{http://arxiv.org/abs/0706.2399}{{\tt arXiv:0706.2399 [astro-ph]}}.

\bibitem{Pogosian:2007sw}
L.~Pogosian and A.~Silvestri, ``{The pattern of growth in viable f(R)
  cosmologies},'' \href{http://dx.doi.org/10.1103/PhysRevD.77.023503,
  10.1103/PhysRevD.81.049901}{{\em Phys.Rev.} {\bf D77} (2008)  023503},
\href{http://arxiv.org/abs/0709.0296}{{\tt arXiv:0709.0296 [astro-ph]}}.

\bibitem{Carloni:2007yv}
S.~Carloni, P.~Dunsby, and A.~Troisi, ``{The Evolution of density perturbations
  in f(R) gravity},'' \href{http://dx.doi.org/10.1103/PhysRevD.77.024024}{{\em
  Phys.Rev.} {\bf D77} (2008)  024024},
\href{http://arxiv.org/abs/0707.0106}{{\tt arXiv:0707.0106 [gr-qc]}}.

\bibitem{Koyama:2009me}
K.~Koyama, A.~Taruya, and T.~Hiramatsu, ``{Non-linear Evolution of Matter Power
  Spectrum in Modified Theory of Gravity},''
  \href{http://dx.doi.org/10.1103/PhysRevD.79.123512}{{\em Phys.Rev.} {\bf D79}
  (2009)  123512},
\href{http://arxiv.org/abs/0902.0618}{{\tt arXiv:0902.0618 [astro-ph.CO]}}.

\bibitem{Motohashi:2009qn}
H.~Motohashi, A.~A. Starobinsky, and J.~Yokoyama, ``{Analytic solution for
  matter density perturbations in a class of viable cosmological f(R)
  models},'' \href{http://dx.doi.org/10.1142/S0218271809015278}{{\em
  Int.J.Mod.Phys.} {\bf D18} (2009)  1731--1740},
\href{http://arxiv.org/abs/0905.0730}{{\tt arXiv:0905.0730 [astro-ph.CO]}}.

\bibitem{Li:2011uw}
Y.~Li and W.~Hu, ``{Chameleon Halo Modeling in f(R) Gravity},''
  \href{http://dx.doi.org/10.1103/PhysRevD.84.084033}{{\em Phys.Rev.} {\bf D84}
  (2011)  084033},
\href{http://arxiv.org/abs/1107.5120}{{\tt arXiv:1107.5120 [astro-ph.CO]}}.

\bibitem{Li:2011pj}
B.~Li, G.-B. Zhao, and K.~Koyama, ``{Halos and Voids in f(R) Gravity},''
  \href{http://dx.doi.org/10.1111/j.1365-2966.2012.20573.x}{{\em
  Mon.Not.Roy.Astron.Soc.} {\bf 421} (2012)  3481},
\href{http://arxiv.org/abs/1111.2602}{{\tt arXiv:1111.2602 [astro-ph.CO]}}.

\bibitem{Nojiri:2003ni}
S.~Nojiri and S.~D. Odintsov, ``{Modified gravity with ln R terms and cosmic
  acceleration},''
  \href{http://dx.doi.org/10.1023/B:GERG.0000035950.40718.48}{{\em
  Gen.Rel.Grav.} {\bf 36} (2004)  1765--1780},
\href{http://arxiv.org/abs/hep-th/0308176}{{\tt arXiv:hep-th/0308176
  [hep-th]}}.

\bibitem{Nojiri:2003ft}
S.~Nojiri and S.~D. Odintsov, ``{Modified gravity with negative and positive
  powers of the curvature: Unification of the inflation and of the cosmic
  acceleration},'' \href{http://dx.doi.org/10.1103/PhysRevD.68.123512}{{\em
  Phys.Rev.} {\bf D68} (2003)  123512},
\href{http://arxiv.org/abs/hep-th/0307288}{{\tt arXiv:hep-th/0307288
  [hep-th]}}.

\bibitem{Nojiri:2007as}
S.~Nojiri and S.~D. Odintsov, ``{Unifying inflation with LambdaCDM epoch in
  modified f(R) gravity consistent with Solar System tests},''
  \href{http://dx.doi.org/10.1016/j.physletb.2007.10.027}{{\em Phys.Lett.} {\bf
  B657} (2007)  238--245},
\href{http://arxiv.org/abs/0707.1941}{{\tt arXiv:0707.1941 [hep-th]}}.

\bibitem{Nojiri:2007cq}
S.~Nojiri and S.~D. Odintsov, ``{Modified f(R) gravity unifying R**m inflation
  with Lambda CDM epoch},''
  \href{http://dx.doi.org/10.1103/PhysRevD.77.026007}{{\em Phys.Rev.} {\bf D77}
  (2008)  026007},
\href{http://arxiv.org/abs/0710.1738}{{\tt arXiv:0710.1738 [hep-th]}}.

\bibitem{Nojiri:2007uq}
S.~Nojiri and S.~D. Odintsov, ``{Modified non-local-F(R) gravity as the key for
  the inflation and dark energy},''
  \href{http://dx.doi.org/10.1016/j.physletb.2007.12.001}{{\em Phys.Lett.} {\bf
  B659} (2008)  821--826},
\href{http://arxiv.org/abs/0708.0924}{{\tt arXiv:0708.0924 [hep-th]}}.

\bibitem{Cognola:2007zu}
G.~Cognola, E.~Elizalde, S.~Nojiri, S.~Odintsov, L.~Sebastiani, {\em et al.},
  ``{A Class of viable modified f(R) gravities describing inflation and the
  onset of accelerated expansion},''
  \href{http://dx.doi.org/10.1103/PhysRevD.77.046009}{{\em Phys.Rev.} {\bf D77}
  (2008)  046009},
\href{http://arxiv.org/abs/0712.4017}{{\tt arXiv:0712.4017 [hep-th]}}.

\bibitem{Nojiri:2008fk}
S.~Nojiri and S.~D. Odintsov, ``{The Future evolution and finite-time
  singularities in F(R)-gravity unifying the inflation and cosmic
  acceleration},'' \href{http://dx.doi.org/10.1103/PhysRevD.78.046006}{{\em
  Phys.Rev.} {\bf D78} (2008)  046006},
\href{http://arxiv.org/abs/0804.3519}{{\tt arXiv:0804.3519 [hep-th]}}.

\bibitem{Nojiri:2008nt}
S.~Nojiri and S.~D. Odintsov, ``{Dark energy, inflation and dark matter from
  modified F(R) gravity},'' {\em TSPU Bulletin} {\bf N8(110)} (2011)  7--19,
\href{http://arxiv.org/abs/0807.0685}{{\tt arXiv:0807.0685 [hep-th]}}.

\bibitem{Bamba:2008xa}
K.~Bamba, S.~Nojiri, and S.~D. Odintsov, ``{Inflationary cosmology and the
  late-time accelerated expansion of the universe in non-minimal
  Yang-Mills-F(R) gravity and non-minimal vector-F(R) gravity},''
  \href{http://dx.doi.org/10.1103/PhysRevD.77.123532}{{\em Phys.Rev.} {\bf D77}
  (2008)  123532},
\href{http://arxiv.org/abs/0803.3384}{{\tt arXiv:0803.3384 [hep-th]}}.

\bibitem{Nojiri:2003rz}
S.~Nojiri and S.~D. Odintsov, ``{Where new gravitational physics comes from: M
  Theory?},'' \href{http://dx.doi.org/10.1016/j.physletb.2003.09.091}{{\em
  Phys.Lett.} {\bf B576} (2003)  5--11},
\href{http://arxiv.org/abs/hep-th/0307071}{{\tt arXiv:hep-th/0307071
  [hep-th]}}.

\bibitem{Biswas:2005qr}
T.~Biswas, A.~Mazumdar, and W.~Siegel, ``{Bouncing universes in string-inspired
  gravity},'' \href{http://dx.doi.org/10.1088/1475-7516/2006/03/009}{{\em JCAP}
  {\bf 0603} (2006)  009},
\href{http://arxiv.org/abs/hep-th/0508194}{{\tt arXiv:hep-th/0508194
  [hep-th]}}.

\bibitem{Flanagan:2003iw}
E.~E. Flanagan, ``{Higher order gravity theories and scalar tensor theories},''
  \href{http://dx.doi.org/10.1088/0264-9381/21/2/006}{{\em Class.Quant.Grav.}
  {\bf 21} (2003)  417--426},
\href{http://arxiv.org/abs/gr-qc/0309015}{{\tt arXiv:gr-qc/0309015 [gr-qc]}}.

\bibitem{Meng:2003uv}
X.~Meng and P.~Wang, ``{Cosmological evolution in 1/r-gravity theory},''
  \href{http://dx.doi.org/10.1088/0264-9381/21/4/015}{{\em Class.Quant.Grav.}
  {\bf 21} (2004)  951--960},
\href{http://arxiv.org/abs/astro-ph/0308031}{{\tt arXiv:astro-ph/0308031
  [astro-ph]}}.

\bibitem{Vollick:2003ic}
D.~N. Vollick, ``{On the viability of the Palatini form of 1/R gravity},''
  \href{http://dx.doi.org/10.1088/0264-9381/21/15/N01}{{\em Class.Quant.Grav.}
  {\bf 21} (2004)  3813--3816},
\href{http://arxiv.org/abs/gr-qc/0312041}{{\tt arXiv:gr-qc/0312041 [gr-qc]}}.

\bibitem{Olmo:2005hc}
G.~J. Olmo, ``{Post-Newtonian constraints on f(R) cosmologies in metric and
  Palatini formalism},''
  \href{http://dx.doi.org/10.1103/PhysRevD.72.083505}{{\em Phys.Rev.} {\bf D72}
  (2005)  083505},
\href{http://arxiv.org/abs/gr-qc/0505135}{{\tt arXiv:gr-qc/0505135 [gr-qc]}}.

\bibitem{Allemandi:2005qs}
G.~Allemandi, A.~Borowiec, M.~Francaviglia, and S.~D. Odintsov, ``{Dark energy
  dominance and cosmic acceleration in first order formalism},''
  \href{http://dx.doi.org/10.1103/PhysRevD.72.063505}{{\em Phys.Rev.} {\bf D72}
  (2005)  063505},
\href{http://arxiv.org/abs/gr-qc/0504057}{{\tt arXiv:gr-qc/0504057 [gr-qc]}}.

\bibitem{Koivisto:2005yc}
T.~Koivisto and H.~Kurki-Suonio, ``{Cosmological perturbations in the palatini
  formulation of modified gravity},''
  \href{http://dx.doi.org/10.1088/0264-9381/23/7/009}{{\em Class.Quant.Grav.}
  {\bf 23} (2006)  2355--2369},
\href{http://arxiv.org/abs/astro-ph/0509422}{{\tt arXiv:astro-ph/0509422
  [astro-ph]}}.

\bibitem{Allemandi:2005tg}
G.~Allemandi, M.~Francaviglia, M.~L. Ruggiero, and A.~Tartaglia,
  ``{Post-Newtonian parameters from alternative theories of gravity},''
  \href{http://dx.doi.org/10.1007/s10714-005-0195-8}{{\em Gen.Rel.Grav.} {\bf
  37} (2005)  1891--1904},
\href{http://arxiv.org/abs/gr-qc/0506123}{{\tt arXiv:gr-qc/0506123 [gr-qc]}}.

\bibitem{Sotiriou:2006qn}
T.~P. Sotiriou and S.~Liberati, ``{Metric-affine f(R) theories of gravity},''
  \href{http://dx.doi.org/10.1016/j.aop.2006.06.002}{{\em Annals Phys.} {\bf
  322} (2007)  935--966},
\href{http://arxiv.org/abs/gr-qc/0604006}{{\tt arXiv:gr-qc/0604006 [gr-qc]}}.

\bibitem{Sotiriou:2006hs}
T.~P. Sotiriou, ``{f(R) gravity and scalar-tensor theory},''
  \href{http://dx.doi.org/10.1088/0264-9381/23/17/003}{{\em Class.Quant.Grav.}
  {\bf 23} (2006)  5117--5128},
\href{http://arxiv.org/abs/gr-qc/0604028}{{\tt arXiv:gr-qc/0604028 [gr-qc]}}.

\bibitem{Sotiriou:2006sf}
T.~P. Sotiriou, ``{Curvature scalar instability in f(R) gravity},''
  \href{http://dx.doi.org/10.1016/j.physletb.2007.01.003}{{\em Phys.Lett.} {\bf
  B645} (2007)  389--392},
\href{http://arxiv.org/abs/gr-qc/0611107}{{\tt arXiv:gr-qc/0611107 [gr-qc]}}.

\bibitem{Iglesias:2007nv}
A.~Iglesias, N.~Kaloper, A.~Padilla, and M.~Park, ``{How (Not) to Palatini},''
  \href{http://dx.doi.org/10.1103/PhysRevD.76.104001}{{\em Phys.Rev.} {\bf D76}
  (2007)  104001},
\href{http://arxiv.org/abs/0708.1163}{{\tt arXiv:0708.1163 [astro-ph]}}.

\bibitem{Bertolami:2007gv}
O.~Bertolami, C.~G. Boehmer, T.~Harko, and F.~S. Lobo, ``{Extra force in f(R)
  modified theories of gravity},''
  \href{http://dx.doi.org/10.1103/PhysRevD.75.104016}{{\em Phys.Rev.} {\bf D75}
  (2007)  104016},
\href{http://arxiv.org/abs/0704.1733}{{\tt arXiv:0704.1733 [gr-qc]}}.

\bibitem{Fay:2007gg}
S.~Fay, R.~Tavakol, and S.~Tsujikawa, ``{f(R) gravity theories in Palatini
  formalism: Cosmological dynamics and observational constraints},''
  \href{http://dx.doi.org/10.1103/PhysRevD.75.063509}{{\em Phys.Rev.} {\bf D75}
  (2007)  063509},
\href{http://arxiv.org/abs/astro-ph/0701479}{{\tt arXiv:astro-ph/0701479
  [astro-ph]}}.

\bibitem{Tsujikawa:2007tg}
S.~Tsujikawa, K.~Uddin, and R.~Tavakol, ``{Density perturbations in f(R)
  gravity theories in metric and Palatini formalisms},''
  \href{http://dx.doi.org/10.1103/PhysRevD.77.043007}{{\em Phys.Rev.} {\bf D77}
  (2008)  043007},
\href{http://arxiv.org/abs/0712.0082}{{\tt arXiv:0712.0082 [astro-ph]}}.

\bibitem{Olmo:2011uz}
G.~J. Olmo, ``{Palatini Approach to Modified Gravity: f(R) Theories and
  Beyond},'' \href{http://dx.doi.org/10.1142/S0218271811018925}{{\em
  Int.J.Mod.Phys.} {\bf D20} (2011)  413--462},
\href{http://arxiv.org/abs/1101.3864}{{\tt arXiv:1101.3864 [gr-qc]}}.

\bibitem{Amendola:2007nt}
L.~Amendola and S.~Tsujikawa, ``{Phantom crossing, equation-of-state
  singularities, and local gravity constraints in f(R) models},''
  \href{http://dx.doi.org/10.1016/j.physletb.2007.12.041}{{\em Phys.Lett.} {\bf
  B660} (2008)  125--132},
\href{http://arxiv.org/abs/0705.0396}{{\tt arXiv:0705.0396 [astro-ph]}}.

\bibitem{Motohashi:2010qj}
H.~Motohashi, A.~A. Starobinsky, and J.~Yokoyama, ``{Phantom behaviour and
  growth index anomalous evolution in viable f(R) gravity models},''
\href{http://arxiv.org/abs/1002.0462}{{\tt arXiv:1002.0462 [astro-ph.CO]}}.

\bibitem{Motohashi:2010tb}
H.~Motohashi, A.~A. Starobinsky, and J.~Yokoyama, ``{Phantom boundary crossing
  and anomalous growth index of fluctuations in viable f(R) models of cosmic
  acceleration},'' \href{http://dx.doi.org/10.1143/PTP.123.887}{{\em
  Prog.Theor.Phys.} {\bf 123} (2010)  887--902},
\href{http://arxiv.org/abs/1002.1141}{{\tt arXiv:1002.1141 [astro-ph.CO]}}.

\bibitem{Bamba:2010iy}
K.~Bamba, C.-Q. Geng, and C.-C. Lee, ``{Generic feature of future crossing of
  phantom divide in viable $f(R)$ gravity models},''
  \href{http://dx.doi.org/10.1088/1475-7516/2010/11/001}{{\em JCAP} {\bf 1011}
  (2010)  001},
\href{http://arxiv.org/abs/1007.0482}{{\tt arXiv:1007.0482 [astro-ph.CO]}}.

\bibitem{Jacobson:1995uq}
T.~Jacobson, G.~Kang, and R.~C. Myers, ``{Increase of black hole entropy in
  higher curvature gravity},''
  \href{http://dx.doi.org/10.1103/PhysRevD.52.3518}{{\em Phys.Rev.} {\bf D52}
  (1995)  3518--3528},
\href{http://arxiv.org/abs/gr-qc/9503020}{{\tt arXiv:gr-qc/9503020 [gr-qc]}}.

\bibitem{PerezBergliaffa:2006ni}
S.~E. Perez~Bergliaffa, ``{Constraining f(R) theories with the energy
  conditions},'' \href{http://dx.doi.org/10.1016/j.physletb.2006.10.003}{{\em
  Phys.Lett.} {\bf B642} (2006)  311--314},
\href{http://arxiv.org/abs/gr-qc/0608072}{{\tt arXiv:gr-qc/0608072 [gr-qc]}}.

\bibitem{Santos:2007bs}
J.~Santos, J.~Alcaniz, M.~Reboucas, and F.~Carvalho, ``{Energy conditions in
  f(R)-gravity},'' \href{http://dx.doi.org/10.1103/PhysRevD.76.083513}{{\em
  Phys.Rev.} {\bf D76} (2007)  083513},
\href{http://arxiv.org/abs/0708.0411}{{\tt arXiv:0708.0411 [astro-ph]}}.

\bibitem{Frolov:2008uf}
A.~V. Frolov, ``{A Singularity Problem with f(R) Dark Energy},''
  \href{http://dx.doi.org/10.1103/PhysRevLett.101.061103}{{\em Phys.Rev.Lett.}
  {\bf 101} (2008)  061103},
\href{http://arxiv.org/abs/0803.2500}{{\tt arXiv:0803.2500 [astro-ph]}}.

\bibitem{Kainulainen:2007bt}
K.~Kainulainen, J.~Piilonen, V.~Reijonen, and D.~Sunhede, ``{Spherically
  symmetric spacetimes in f(R) gravity theories},''
  \href{http://dx.doi.org/10.1103/PhysRevD.76.024020}{{\em Phys.Rev.} {\bf D76}
  (2007)  024020},
\href{http://arxiv.org/abs/0704.2729}{{\tt arXiv:0704.2729 [gr-qc]}}.

\bibitem{Kobayashi:2008tq}
T.~Kobayashi and K.-i. Maeda, ``{Relativistic stars in f(R) gravity, and
  absence thereof},'' \href{http://dx.doi.org/10.1103/PhysRevD.78.064019}{{\em
  Phys.Rev.} {\bf D78} (2008)  064019},
\href{http://arxiv.org/abs/0807.2503}{{\tt arXiv:0807.2503 [astro-ph]}}.

\bibitem{Kobayashi:2008wc}
T.~Kobayashi and K.-i. Maeda, ``{Can higher curvature corrections cure the
  singularity problem in f(R) gravity?},''
  \href{http://dx.doi.org/10.1103/PhysRevD.79.024009}{{\em Phys.Rev.} {\bf D79}
  (2009)  024009},
\href{http://arxiv.org/abs/0810.5664}{{\tt arXiv:0810.5664 [astro-ph]}}.

\bibitem{Miranda:2009rs}
V.~Miranda, S.~E. Joras, I.~Waga, and M.~Quartin, ``{Viable Singularity-Free
  f(R) Gravity Without a Cosmological Constant},''
  \href{http://dx.doi.org/10.1103/PhysRevLett.102.221101}{{\em Phys.Rev.Lett.}
  {\bf 102} (2009)  221101},
\href{http://arxiv.org/abs/0905.1941}{{\tt arXiv:0905.1941 [astro-ph.CO]}}.

\bibitem{Babichev:2009td}
E.~Babichev and D.~Langlois, ``{Relativistic stars in f(R) gravity},''
  \href{http://dx.doi.org/10.1103/PhysRevD.81.069901,
  10.1103/PhysRevD.80.121501}{{\em Phys.Rev.} {\bf D80} (2009)  121501},
\href{http://arxiv.org/abs/0904.1382}{{\tt arXiv:0904.1382 [gr-qc]}}.

\bibitem{Upadhye:2009kt}
A.~Upadhye and W.~Hu, ``{The existence of relativistic stars in f(R)
  gravity},'' \href{http://dx.doi.org/10.1103/PhysRevD.80.064002}{{\em
  Phys.Rev.} {\bf D80} (2009)  064002},
\href{http://arxiv.org/abs/0905.4055}{{\tt arXiv:0905.4055 [astro-ph.CO]}}.

\bibitem{Babichev:2009fi}
E.~Babichev and D.~Langlois, ``{Relativistic stars in f(R) and scalar-tensor
  theories},'' \href{http://dx.doi.org/10.1103/PhysRevD.81.124051}{{\em
  Phys.Rev.} {\bf D81} (2010)  124051},
\href{http://arxiv.org/abs/0911.1297}{{\tt arXiv:0911.1297 [gr-qc]}}.

\bibitem{Cooney:2009rr}
A.~Cooney, S.~DeDeo, and D.~Psaltis, ``{Neutron Stars in f(R) Gravity with
  Perturbative Constraints},''
  \href{http://dx.doi.org/10.1103/PhysRevD.82.064033}{{\em Phys.Rev.} {\bf D82}
  (2010)  064033},
\href{http://arxiv.org/abs/0910.5480}{{\tt arXiv:0910.5480 [astro-ph.HE]}}.

\bibitem{Cognola:2005de}
G.~Cognola, E.~Elizalde, S.~Nojiri, S.~D. Odintsov, and S.~Zerbini, ``{One-loop
  f(R) gravity in de Sitter universe},''
  \href{http://dx.doi.org/10.1088/1475-7516/2005/02/010}{{\em JCAP} {\bf 0502}
  (2005)  010},
\href{http://arxiv.org/abs/hep-th/0501096}{{\tt arXiv:hep-th/0501096
  [hep-th]}}.

\bibitem{Cognola:2005sg}
G.~Cognola and S.~Zerbini, ``{One-loop f(R) gravitational modified models},''
  \href{http://dx.doi.org/10.1088/0305-4470/39/21/S15}{{\em J.Phys.} {\bf A39}
  (2006)  6245--6251},
\href{http://arxiv.org/abs/hep-th/0511233}{{\tt arXiv:hep-th/0511233
  [hep-th]}}.

\bibitem{Machado:2007ea}
P.~F. Machado and F.~Saueressig, ``{On the renormalization group flow of
  f(R)-gravity},'' \href{http://dx.doi.org/10.1103/PhysRevD.77.124045}{{\em
  Phys.Rev.} {\bf D77} (2008)  124045},
\href{http://arxiv.org/abs/0712.0445}{{\tt arXiv:0712.0445 [hep-th]}}.

\bibitem{Dyer:2008hb}
E.~Dyer and K.~Hinterbichler, ``{Boundary Terms, Variational Principles and
  Higher Derivative Modified Gravity},''
  \href{http://dx.doi.org/10.1103/PhysRevD.79.024028}{{\em Phys.Rev.} {\bf D79}
  (2009)  024028},
\href{http://arxiv.org/abs/0809.4033}{{\tt arXiv:0809.4033 [gr-qc]}}.

\bibitem{Schmidt:2008tn}
F.~Schmidt, M.~V. Lima, H.~Oyaizu, and W.~Hu, ``{Non-linear Evolution of f(R)
  Cosmologies III: Halo Statistics},''
  \href{http://dx.doi.org/10.1103/PhysRevD.79.083518}{{\em Phys.Rev.} {\bf D79}
  (2009)  083518},
\href{http://arxiv.org/abs/0812.0545}{{\tt arXiv:0812.0545 [astro-ph]}}.

\bibitem{Oyaizu:2008sr}
H.~Oyaizu, ``{Non-linear evolution of f(R) cosmologies I: methodology},''
  \href{http://dx.doi.org/10.1103/PhysRevD.78.123523}{{\em Phys.Rev.} {\bf D78}
  (2008)  123523},
\href{http://arxiv.org/abs/0807.2449}{{\tt arXiv:0807.2449 [astro-ph]}}.

\bibitem{Oyaizu:2008tb}
H.~Oyaizu, M.~Lima, and W.~Hu, ``{Nonlinear evolution of f(R) cosmologies. 2.
  Power spectrum},'' \href{http://dx.doi.org/10.1103/PhysRevD.78.123524}{{\em
  Phys.Rev.} {\bf D78} (2008)  123524},
\href{http://arxiv.org/abs/0807.2462}{{\tt arXiv:0807.2462 [astro-ph]}}.

\bibitem{Khoury:2009tk}
J.~Khoury and M.~Wyman, ``{N-Body Simulations of DGP and Degravitation
  Theories},'' \href{http://dx.doi.org/10.1103/PhysRevD.80.064023}{{\em
  Phys.Rev.} {\bf D80} (2009)  064023},
\href{http://arxiv.org/abs/0903.1292}{{\tt arXiv:0903.1292 [astro-ph.CO]}}.

\bibitem{Schmidt:2009sv}
F.~Schmidt, ``{Cosmological Simulations of Normal-Branch Braneworld Gravity},''
  \href{http://dx.doi.org/10.1103/PhysRevD.80.123003}{{\em Phys.Rev.} {\bf D80}
  (2009)  123003},
\href{http://arxiv.org/abs/0910.0235}{{\tt arXiv:0910.0235 [astro-ph.CO]}}.

\bibitem{Chan:2009ew}
K.~Chan and R.~Scoccimarro, ``{Large-Scale Structure in Brane-Induced Gravity
  II. Numerical Simulations},''
  \href{http://dx.doi.org/10.1103/PhysRevD.80.104005}{{\em Phys.Rev.} {\bf D80}
  (2009)  104005},
\href{http://arxiv.org/abs/0906.4548}{{\tt arXiv:0906.4548 [astro-ph.CO]}}.

\bibitem{Wyman:2013jaa}
M.~Wyman, E.~Jennings, and M.~Lima, ``{Simulations of Galileon modified
  gravity: Clustering statistics in real and redshift space},''
  \href{http://dx.doi.org/10.1103/PhysRevD.88.084029}{{\em Phys.Rev.} {\bf D88}
  (2013)  084029},
\href{http://arxiv.org/abs/1303.6630}{{\tt arXiv:1303.6630 [astro-ph.CO]}}.

\bibitem{Zhao:2010qy}
G.-B. Zhao, B.~Li, and K.~Koyama, ``{N-body Simulations for f(R) Gravity using
  a Self-adaptive Particle-Mesh Code},''
  \href{http://dx.doi.org/10.1103/PhysRevD.83.044007}{{\em Phys.Rev.} {\bf D83}
  (2011)  044007},
\href{http://arxiv.org/abs/1011.1257}{{\tt arXiv:1011.1257 [astro-ph.CO]}}.

\bibitem{Ferraro:2010gh}
S.~Ferraro, F.~Schmidt, and W.~Hu, ``{Cluster Abundance in f(R) Gravity
  Models},'' \href{http://dx.doi.org/10.1103/PhysRevD.83.063503}{{\em
  Phys.Rev.} {\bf D83} (2011)  063503},
\href{http://arxiv.org/abs/1011.0992}{{\tt arXiv:1011.0992 [astro-ph.CO]}}.

\bibitem{Li:2012by}
B.~Li, W.~A. Hellwing, K.~Koyama, G.-B. Zhao, E.~Jennings, {\em et al.}, ``{The
  nonlinear matter and velocity power spectra in f(R) gravity},''
  \href{http://dx.doi.org/10.1093/mnras/sts072}{{\em Mon.Not.Roy.Astron.Soc.}
  {\bf 428} (2013)  743--755},
\href{http://arxiv.org/abs/1206.4317}{{\tt arXiv:1206.4317 [astro-ph.CO]}}.

\bibitem{Li:2013tda}
B.~Li, A.~Barreira, C.~M. Baugh, W.~A. Hellwing, K.~Koyama, {\em et al.},
  ``{Simulating the quartic Galileon gravity model on adaptively refined
  meshes},'' \href{http://dx.doi.org/10.1088/1475-7516/2013/11/012}{{\em JCAP}
  {\bf 1311} (2013)  012},
\href{http://arxiv.org/abs/1308.3491}{{\tt arXiv:1308.3491 [astro-ph.CO]}}.

\bibitem{Baldi:2013iza}
M.~Baldi, F.~Villaescusa-Navarro, M.~Viel, E.~Puchwein, V.~Springel, {\em et
  al.}, ``{Cosmic Degeneracies I: Joint N-body Simulations of Modified Gravity
  and Massive Neutrinos},''
\href{http://arxiv.org/abs/1311.2588}{{\tt arXiv:1311.2588 [astro-ph.CO]}}.

\bibitem{He:2014eva}
J.-h. He, B.~Li, A.~J. Hawken, and B.~R. Granett, ``{Revisiting the screening
  mechanism in $f(R)$ gravity},''
\href{http://arxiv.org/abs/1406.6820}{{\tt arXiv:1406.6820 [astro-ph.CO]}}.

\bibitem{Carroll:2004de}
S.~M. Carroll, A.~De~Felice, V.~Duvvuri, D.~A. Easson, M.~Trodden, {\em et
  al.}, ``{The Cosmology of generalized modified gravity models},''
  \href{http://dx.doi.org/10.1103/PhysRevD.71.063513}{{\em Phys.Rev.} {\bf D71}
  (2005)  063513},
\href{http://arxiv.org/abs/astro-ph/0410031}{{\tt arXiv:astro-ph/0410031
  [astro-ph]}}.

\bibitem{Nunez:2004ts}
A.~Nunez and S.~Solganik, ``{Ghost constraints on modified gravity},''
  \href{http://dx.doi.org/10.1016/j.physletb.2005.01.015}{{\em Phys.Lett.} {\bf
  B608} (2005)  189--193},
\href{http://arxiv.org/abs/hep-th/0411102}{{\tt arXiv:hep-th/0411102
  [hep-th]}}.

\bibitem{Nojiri:2005jg}
S.~Nojiri and S.~D. Odintsov, ``{Modified Gauss-Bonnet theory as gravitational
  alternative for dark energy},''
  \href{http://dx.doi.org/10.1016/j.physletb.2005.10.010}{{\em Phys.Lett.} {\bf
  B631} (2005)  1--6},
\href{http://arxiv.org/abs/hep-th/0508049}{{\tt arXiv:hep-th/0508049
  [hep-th]}}.

\bibitem{Cognola:2006eg}
G.~Cognola, E.~Elizalde, S.~Nojiri, S.~D. Odintsov, and S.~Zerbini, ``{Dark
  energy in modified Gauss-Bonnet gravity: Late-time acceleration and the
  hierarchy problem},''
  \href{http://dx.doi.org/10.1103/PhysRevD.73.084007}{{\em Phys.Rev.} {\bf D73}
  (2006)  084007},
\href{http://arxiv.org/abs/hep-th/0601008}{{\tt arXiv:hep-th/0601008
  [hep-th]}}.

\bibitem{Capozziello:2006uv}
S.~Capozziello, V.~Cardone, and A.~Troisi, ``{Dark energy and dark matter as
  curvature effects},''
  \href{http://dx.doi.org/10.1088/1475-7516/2006/08/001}{{\em JCAP} {\bf 0608}
  (2006)  001},
\href{http://arxiv.org/abs/astro-ph/0602349}{{\tt arXiv:astro-ph/0602349
  [astro-ph]}}.

\bibitem{Hinterbichler:2011ca}
K.~Hinterbichler, J.~Khoury, A.~Levy, and A.~Matas, ``{Symmetron Cosmology},''
  \href{http://dx.doi.org/10.1103/PhysRevD.84.103521}{{\em Phys.Rev.} {\bf D84}
  (2011)  103521},
\href{http://arxiv.org/abs/1107.2112}{{\tt arXiv:1107.2112 [astro-ph.CO]}}.

\bibitem{Clampitt:2011mx}
J.~Clampitt, B.~Jain, and J.~Khoury, ``{Halo Scale Predictions of Symmetron
  Modified Gravity},''
  \href{http://dx.doi.org/10.1088/1475-7516/2012/01/030}{{\em JCAP} {\bf 1201}
  (2012)  030},
\href{http://arxiv.org/abs/1110.2177}{{\tt arXiv:1110.2177 [astro-ph.CO]}}.

\bibitem{Brax:2011pk}
P.~Brax, C.~van~de Bruck, A.-C. Davis, B.~Li, B.~Schmauch, {\em et al.},
  ``{Linear Growth of Structure in the Symmetron Model},''
  \href{http://dx.doi.org/10.1103/PhysRevD.84.123524}{{\em Phys.Rev.} {\bf D84}
  (2011)  123524},
\href{http://arxiv.org/abs/1108.3082}{{\tt arXiv:1108.3082 [astro-ph.CO]}}.

\bibitem{Llinares:2012ds}
C.~Llinares and D.~F. Mota, ``{Shape of Clusters of Galaxies as a Probe of
  Screening Mechanisms in Modified Gravity},''
  \href{http://dx.doi.org/10.1103/PhysRevLett.110.151104}{{\em Phys.Rev.Lett.}
  {\bf 110} (2013) no.~15, 151104},
\href{http://arxiv.org/abs/1205.5775}{{\tt arXiv:1205.5775 [astro-ph.CO]}}.

\bibitem{Taddei:2013bsk}
L.~Taddei, R.~Catena, and M.~Pietroni, ``{Spherical collapse and halo mass
  function in the symmetron model},''
\href{http://arxiv.org/abs/1310.6175}{{\tt arXiv:1310.6175 [astro-ph.CO]}}.

\bibitem{Davis:2011pj}
A.-C. Davis, B.~Li, D.~F. Mota, and H.~A. Winther, ``{Structure Formation in
  the Symmetron Model},''
  \href{http://dx.doi.org/10.1088/0004-637X/748/1/61}{{\em Astrophys.J.} {\bf
  748} (2012)  61},
\href{http://arxiv.org/abs/1108.3081}{{\tt arXiv:1108.3081 [astro-ph.CO]}}.

\bibitem{Winther:2011qb}
H.~A. Winther, D.~F. Mota, and B.~Li, ``{Environment Dependence of Dark Matter
  Halos in Symmetron Modified Gravity},''
  \href{http://dx.doi.org/10.1088/0004-637X/756/2/166}{{\em Astrophys.J.} {\bf
  756} (2012)  166},
\href{http://arxiv.org/abs/1110.6438}{{\tt arXiv:1110.6438 [astro-ph.CO]}}.

\bibitem{Olive:2010vh}
K.~A. Olive, M.~Peloso, and J.-P. Uzan, ``{The Wall of Fundamental
  Constants},'' \href{http://dx.doi.org/10.1103/PhysRevD.83.043509}{{\em
  Phys.Rev.} {\bf D83} (2011)  043509},
\href{http://arxiv.org/abs/1011.1504}{{\tt arXiv:1011.1504 [astro-ph.CO]}}.

\bibitem{Olive:2012ck}
K.~A. Olive, M.~Peloso, and A.~J. Peterson, ``{Where are the walls? Spatial
  variation in the fine-structure constant},''
  \href{http://dx.doi.org/10.1103/PhysRevD.86.043501}{{\em Phys.Rev.} {\bf D86}
  (2012)  043501},
\href{http://arxiv.org/abs/1204.4391}{{\tt arXiv:1204.4391 [astro-ph.CO]}}.

\bibitem{Dong:2013swa}
R.~Dong, W.~H. Kinney, and D.~Stojkovic, ``{Symmetron Inflation},''
  \href{http://dx.doi.org/10.1088/1475-7516/2014/01/021}{{\em JCAP} {\bf 01}
  (2014)  021},
\href{http://arxiv.org/abs/1307.4451}{{\tt arXiv:1307.4451 [astro-ph.CO]}}.

\bibitem{Wang:2012kj}
J.~Wang, L.~Hui, and J.~Khoury, ``{No-Go Theorems for Generalized Chameleon
  Field Theories},''
  \href{http://dx.doi.org/10.1103/PhysRevLett.109.241301}{{\em Phys.Rev.Lett.}
  {\bf 109} (2012)  241301},
\href{http://arxiv.org/abs/1208.4612}{{\tt arXiv:1208.4612 [astro-ph.CO]}}.

\bibitem{Brax:2011aw}
P.~Brax, A.-C. Davis, and B.~Li, ``{Modified Gravity Tomography},''
  \href{http://dx.doi.org/10.1016/j.physletb.2012.08.002}{{\em Phys.Lett.} {\bf
  B715} (2012)  38--43},
\href{http://arxiv.org/abs/1111.6613}{{\tt arXiv:1111.6613 [astro-ph.CO]}}.

\bibitem{Bamba:2012yf}
K.~Bamba, R.~Gannouji, M.~Kamijo, S.~Nojiri, and M.~Sami, ``{Spontaneous
  symmetry breaking in cosmos: The hybrid symmetron as a dark energy switching
  device},'' \href{http://dx.doi.org/10.1088/1475-7516/2013/07/017}{{\em JCAP}
  {\bf 1307} (2013)  017},
\href{http://arxiv.org/abs/1211.2289}{{\tt arXiv:1211.2289 [hep-th]}}.

\bibitem{Weinberg:1996kr}
S.~Weinberg,
``{The quantum theory of fields. Vol. 2: Modern applications},''.

\bibitem{Kapner:2006si}
D.~Kapner, T.~Cook, E.~Adelberger, J.~Gundlach, B.~R. Heckel, {\em et al.},
  ``{Tests of the gravitational inverse-square law below the dark-energy length
  scale},'' \href{http://dx.doi.org/10.1103/PhysRevLett.98.021101}{{\em
  Phys.Rev.Lett.} {\bf 98} (2007)  021101},
\href{http://arxiv.org/abs/hep-ph/0611184}{{\tt arXiv:hep-ph/0611184
  [hep-ph]}}.

\bibitem{ArkaniHamed:2003uz}
N.~Arkani-Hamed, P.~Creminelli, S.~Mukohyama, and M.~Zaldarriaga, ``{Ghost
  inflation},'' \href{http://dx.doi.org/10.1088/1475-7516/2004/04/001}{{\em
  JCAP} {\bf 0404} (2004)  001},
\href{http://arxiv.org/abs/hep-th/0312100}{{\tt arXiv:hep-th/0312100
  [hep-th]}}.

\bibitem{Senatore:2004rj}
L.~Senatore, ``{Tilted ghost inflation},''
  \href{http://dx.doi.org/10.1103/PhysRevD.71.043512}{{\em Phys.Rev.} {\bf D71}
  (2005)  043512},
\href{http://arxiv.org/abs/astro-ph/0406187}{{\tt arXiv:astro-ph/0406187
  [astro-ph]}}.

\bibitem{Mukhanov:2005bu}
V.~F. Mukhanov and A.~Vikman, ``{Enhancing the tensor-to-scalar ratio in simple
  inflation},'' \href{http://dx.doi.org/10.1088/1475-7516/2006/02/004}{{\em
  JCAP} {\bf 0602} (2006)  004},
\href{http://arxiv.org/abs/astro-ph/0512066}{{\tt arXiv:astro-ph/0512066
  [astro-ph]}}.

\bibitem{Langlois:2008wt}
D.~Langlois, S.~Renaux-Petel, D.~A. Steer, and T.~Tanaka, ``{Primordial
  fluctuations and non-Gaussianities in multi-field DBI inflation},''
  \href{http://dx.doi.org/10.1103/PhysRevLett.101.061301}{{\em Phys.Rev.Lett.}
  {\bf 101} (2008)  061301},
\href{http://arxiv.org/abs/0804.3139}{{\tt arXiv:0804.3139 [hep-th]}}.

\bibitem{Seery:2005wm}
D.~Seery and J.~E. Lidsey, ``{Primordial non-Gaussianities in single field
  inflation},'' \href{http://dx.doi.org/10.1088/1475-7516/2005/06/003}{{\em
  JCAP} {\bf 0506} (2005)  003},
\href{http://arxiv.org/abs/astro-ph/0503692}{{\tt arXiv:astro-ph/0503692
  [astro-ph]}}.

\bibitem{Chen:2006nt}
X.~Chen, M.-x. Huang, S.~Kachru, and G.~Shiu, ``{Observational signatures and
  non-Gaussianities of general single field inflation},''
  \href{http://dx.doi.org/10.1088/1475-7516/2007/01/002}{{\em JCAP} {\bf 0701}
  (2007)  002},
\href{http://arxiv.org/abs/hep-th/0605045}{{\tt arXiv:hep-th/0605045
  [hep-th]}}.

\bibitem{Chen:2006xjb}
X.~Chen, R.~Easther, and E.~A. Lim, ``{Large Non-Gaussianities in Single Field
  Inflation},'' \href{http://dx.doi.org/10.1088/1475-7516/2007/06/023}{{\em
  JCAP} {\bf 0706} (2007)  023},
\href{http://arxiv.org/abs/astro-ph/0611645}{{\tt arXiv:astro-ph/0611645
  [astro-ph]}}.

\bibitem{Huang:2006eha}
X.~Chen, M.-x. Huang, and G.~Shiu, ``{The Inflationary Trispectrum for Models
  with Large Non-Gaussianities},''
  \href{http://dx.doi.org/10.1103/PhysRevD.74.121301}{{\em Phys.Rev.} {\bf D74}
  (2006)  121301},
\href{http://arxiv.org/abs/hep-th/0610235}{{\tt arXiv:hep-th/0610235
  [hep-th]}}.

\bibitem{Langlois:2008qf}
D.~Langlois, S.~Renaux-Petel, D.~A. Steer, and T.~Tanaka, ``{Primordial
  perturbations and non-Gaussianities in DBI and general multi-field
  inflation},'' \href{http://dx.doi.org/10.1103/PhysRevD.78.063523}{{\em
  Phys.Rev.} {\bf D78} (2008)  063523},
\href{http://arxiv.org/abs/0806.0336}{{\tt arXiv:0806.0336 [hep-th]}}.

\bibitem{Arroja:2008yy}
F.~Arroja, S.~Mizuno, and K.~Koyama, ``{Non-gaussianity from the bispectrum in
  general multiple field inflation},''
  \href{http://dx.doi.org/10.1088/1475-7516/2008/08/015}{{\em JCAP} {\bf 0808}
  (2008)  015},
\href{http://arxiv.org/abs/0806.0619}{{\tt arXiv:0806.0619 [astro-ph]}}.

\bibitem{Langlois:2008mn}
D.~Langlois and S.~Renaux-Petel, ``{Perturbations in generalized multi-field
  inflation},'' \href{http://dx.doi.org/10.1088/1475-7516/2008/04/017}{{\em
  JCAP} {\bf 0804} (2008)  017},
\href{http://arxiv.org/abs/0801.1085}{{\tt arXiv:0801.1085 [hep-th]}}.

\bibitem{Khoury:2008wj}
J.~Khoury and F.~Piazza, ``{Rapidly-Varying Speed of Sound, Scale Invariance
  and Non-Gaussian Signatures},''
  \href{http://dx.doi.org/10.1088/1475-7516/2009/07/026}{{\em JCAP} {\bf 0907}
  (2009)  026},
\href{http://arxiv.org/abs/0811.3633}{{\tt arXiv:0811.3633 [hep-th]}}.

\bibitem{RenauxPetel:2008gi}
S.~Renaux-Petel and G.~Tasinato, ``{Nonlinear perturbations of cosmological
  scalar fields with non-standard kinetic terms},''
  \href{http://dx.doi.org/10.1088/1475-7516/2009/01/012}{{\em JCAP} {\bf 0901}
  (2009)  012},
\href{http://arxiv.org/abs/0810.2405}{{\tt arXiv:0810.2405 [hep-th]}}.

\bibitem{Chen:2009bc}
X.~Chen, B.~Hu, M.-x. Huang, G.~Shiu, and Y.~Wang, ``{Large Primordial
  Trispectra in General Single Field Inflation},''
  \href{http://dx.doi.org/10.1088/1475-7516/2009/08/008}{{\em JCAP} {\bf 0908}
  (2009)  008},
\href{http://arxiv.org/abs/0905.3494}{{\tt arXiv:0905.3494 [astro-ph.CO]}}.

\bibitem{Mizuno:2010ag}
S.~Mizuno and K.~Koyama, ``{Primordial non-Gaussianity from the DBI
  Galileons},'' \href{http://dx.doi.org/10.1103/PhysRevD.82.103518}{{\em
  Phys.Rev.} {\bf D82} (2010)  103518},
\href{http://arxiv.org/abs/1009.0677}{{\tt arXiv:1009.0677 [hep-th]}}.

\bibitem{RenauxPetel:2011dv}
S.~Renaux-Petel, ``{Orthogonal non-Gaussianities from Dirac-Born-Infeld
  Galileon inflation},''
  \href{http://dx.doi.org/10.1088/0264-9381/28/24/249601,
  10.1088/0264-9381/28/18/182001}{{\em Class.Quant.Grav.} {\bf 28} (2011)
  182001},
\href{http://arxiv.org/abs/1105.6366}{{\tt arXiv:1105.6366 [astro-ph.CO]}}.

\bibitem{Ribeiro:2012ar}
R.~H. Ribeiro, ``{Inflationary signatures of single-field models beyond
  slow-roll},'' \href{http://dx.doi.org/10.1088/1475-7516/2012/05/037}{{\em
  JCAP} {\bf 1205} (2012)  037},
\href{http://arxiv.org/abs/1202.4453}{{\tt arXiv:1202.4453 [astro-ph.CO]}}.

\bibitem{Alishahiha:2004eh}
M.~Alishahiha, E.~Silverstein, and D.~Tong, ``{DBI in the sky},''
  \href{http://dx.doi.org/10.1103/PhysRevD.70.123505}{{\em Phys.Rev.} {\bf D70}
  (2004)  123505},
\href{http://arxiv.org/abs/hep-th/0404084}{{\tt arXiv:hep-th/0404084
  [hep-th]}}.

\bibitem{Chen:2005ad}
X.~Chen, ``{Inflation from warped space},''
  \href{http://dx.doi.org/10.1088/1126-6708/2005/08/045}{{\em JHEP} {\bf 0508}
  (2005)  045},
\href{http://arxiv.org/abs/hep-th/0501184}{{\tt arXiv:hep-th/0501184
  [hep-th]}}.

\bibitem{ArkaniHamed:2003uy}
N.~Arkani-Hamed, H.-C. Cheng, M.~A. Luty, and S.~Mukohyama, ``{Ghost
  condensation and a consistent infrared modification of gravity},''
  \href{http://dx.doi.org/10.1088/1126-6708/2004/05/074}{{\em JHEP} {\bf 0405}
  (2004)  074},
\href{http://arxiv.org/abs/hep-th/0312099}{{\tt arXiv:hep-th/0312099
  [hep-th]}}.

\bibitem{Buchbinder:2007ad}
E.~I. Buchbinder, J.~Khoury, and B.~A. Ovrut, ``{New Ekpyrotic cosmology},''
  \href{http://dx.doi.org/10.1103/PhysRevD.76.123503}{{\em Phys.Rev.} {\bf D76}
  (2007)  123503},
\href{http://arxiv.org/abs/hep-th/0702154}{{\tt arXiv:hep-th/0702154
  [hep-th]}}.

\bibitem{Creminelli:2007aq}
P.~Creminelli and L.~Senatore, ``{A Smooth bouncing cosmology with scale
  invariant spectrum},''
  \href{http://dx.doi.org/10.1088/1475-7516/2007/11/010}{{\em JCAP} {\bf 0711}
  (2007)  010},
\href{http://arxiv.org/abs/hep-th/0702165}{{\tt arXiv:hep-th/0702165
  [hep-th]}}.

\bibitem{Dubovsky:2006vk}
S.~Dubovsky and S.~Sibiryakov, ``{Spontaneous breaking of Lorentz invariance,
  black holes and perpetuum mobile of the 2nd kind},''
  \href{http://dx.doi.org/10.1016/j.physletb.2006.05.074}{{\em Phys.Lett.} {\bf
  B638} (2006)  509--514},
\href{http://arxiv.org/abs/hep-th/0603158}{{\tt arXiv:hep-th/0603158
  [hep-th]}}.

\bibitem{Eling:2007qd}
C.~Eling, B.~Z. Foster, T.~Jacobson, and A.~C. Wall, ``{Lorentz violation and
  perpetual motion},'' \href{http://dx.doi.org/10.1103/PhysRevD.75.101502}{{\em
  Phys.Rev.} {\bf D75} (2007)  101502},
\href{http://arxiv.org/abs/hep-th/0702124}{{\tt arXiv:hep-th/0702124
  [HEP-TH]}}.

\bibitem{Mukohyama:2009rk}
S.~Mukohyama, ``{Ghost condensate and generalized second law},''
  \href{http://dx.doi.org/10.1088/1126-6708/2009/09/070}{{\em JHEP} {\bf 0909}
  (2009)  070},
\href{http://arxiv.org/abs/0901.3595}{{\tt arXiv:0901.3595 [hep-th]}}.

\bibitem{Nozawa:2013maa}
M.~Nozawa and T.~Shiromizu, ``{Modeling scalar fields consistent with positive
  mass},'' \href{http://dx.doi.org/10.1103/PhysRevD.89.023011}{{\em Phys.Rev.}
  {\bf D89} (2014)  023011},
\href{http://arxiv.org/abs/1310.1663}{{\tt arXiv:1310.1663 [gr-qc]}}.

\bibitem{Elder:2014fea}
B.~Elder, A.~Joyce, J.~Khoury, and A.~J. Tolley, ``{A Positive Energy Theorem
  for $P(X, \phi)$ Theories},''
\href{http://arxiv.org/abs/1405.7696}{{\tt arXiv:1405.7696 [hep-th]}}.

\bibitem{Khoury:2010gb}
J.~Khoury, J.-L. Lehners, and B.~Ovrut, ``{Supersymmetric P(X,$\phi$) and the
  Ghost Condensate},'' \href{http://dx.doi.org/10.1103/PhysRevD.83.125031}{{\em
  Phys.Rev.} {\bf D83} (2011)  125031},
\href{http://arxiv.org/abs/1012.3748}{{\tt arXiv:1012.3748 [hep-th]}}.

\bibitem{Baumann:2011nm}
D.~Baumann and D.~Green, ``{Supergravity for Effective Theories},''
  \href{http://dx.doi.org/10.1007/JHEP03(2012)001}{{\em JHEP} {\bf 1203} (2012)
   001},
\href{http://arxiv.org/abs/1109.0293}{{\tt arXiv:1109.0293 [hep-th]}}.

\bibitem{Babichev:2006cy}
E.~Babichev, ``{Global topological k-defects},''
  \href{http://dx.doi.org/10.1103/PhysRevD.74.085004}{{\em Phys.Rev.} {\bf D74}
  (2006)  085004},
\href{http://arxiv.org/abs/hep-th/0608071}{{\tt arXiv:hep-th/0608071
  [hep-th]}}.

\bibitem{Bazeia:2007df}
D.~Bazeia, L.~Losano, R.~Menezes, and J.~Oliveira, ``{Generalized Global Defect
  Solutions},'' \href{http://dx.doi.org/10.1140/epjc/s10052-007-0329-0}{{\em
  Eur.Phys.J.} {\bf C51} (2007)  953--962},
\href{http://arxiv.org/abs/hep-th/0702052}{{\tt arXiv:hep-th/0702052
  [hep-th]}}.

\bibitem{Babichev:2007tn}
E.~Babichev, ``{Gauge k-vortices},''
  \href{http://dx.doi.org/10.1103/PhysRevD.77.065021}{{\em Phys.Rev.} {\bf D77}
  (2008)  065021},
\href{http://arxiv.org/abs/0711.0376}{{\tt arXiv:0711.0376 [hep-th]}}.

\bibitem{Jin:2007fz}
X.-h. Jin, X.-z. Li, and D.-j. Liu, ``{Gravitating global k-monopole},''
  \href{http://dx.doi.org/10.1088/0264-9381/24/11/001}{{\em Class.Quant.Grav.}
  {\bf 24} (2007)  2773--2780},
\href{http://arxiv.org/abs/0704.1685}{{\tt arXiv:0704.1685 [gr-qc]}}.

\bibitem{Adam:2007ij}
C.~Adam, J.~Sanchez-Guillen, and A.~Wereszczynski, ``{k-defects as
  compactons},'' \href{http://dx.doi.org/10.1088/1751-8121/42/8/089801,
  10.1088/1751-8113/40/45/009}{{\em J.Phys.} {\bf A40} (2007)  13625--13643},
\href{http://arxiv.org/abs/0705.3554}{{\tt arXiv:0705.3554 [hep-th]}}.

\bibitem{Adam:2007ag}
C.~Adam, N.~Grandi, J.~Sanchez-Guillen, and A.~Wereszczynski, ``{K fields,
  compactons, and thick branes},''
  \href{http://dx.doi.org/10.1088/1751-8113/42/15/159801,
  10.1088/1751-8113/41/21/212004}{{\em J.Phys.} {\bf A41} (2008)  212004},
\href{http://arxiv.org/abs/0711.3550}{{\tt arXiv:0711.3550 [hep-th]}}.

\bibitem{Babichev:2008qv}
E.~Babichev, P.~Brax, C.~Caprini, J.~Martin, and D.~A. Steer, ``{Dirac Born
  Infeld (DBI) Cosmic Strings},''
  \href{http://dx.doi.org/10.1088/1126-6708/2009/03/091}{{\em JHEP} {\bf 0903}
  (2009)  091},
\href{http://arxiv.org/abs/0809.2013}{{\tt arXiv:0809.2013 [hep-th]}}.

\bibitem{Andrews:2010eh}
M.~Andrews, M.~Lewandowski, M.~Trodden, and D.~Wesley, ``{Distinguishing
  $k$-defects from their canonical twins},''
  \href{http://dx.doi.org/10.1103/PhysRevD.82.105006}{{\em Phys.Rev.} {\bf D82}
  (2010)  105006},
\href{http://arxiv.org/abs/1007.3438}{{\tt arXiv:1007.3438 [hep-th]}}.

\bibitem{Bazeia:2010vb}
D.~Bazeia, E.~da~Hora, R.~Menezes, H.~de~Oliveira, and C.~dos Santos,
  ``{Compact-like kinks and vortices in generalized models},''
  \href{http://dx.doi.org/10.1103/PhysRevD.81.125016}{{\em Phys.Rev.} {\bf D81}
  (2010)  125016},
\href{http://arxiv.org/abs/1004.3710}{{\tt arXiv:1004.3710 [hep-th]}}.

\bibitem{Amin:2013ika}
M.~A. Amin, ``{K-oscillons: Oscillons with noncanonical kinetic terms},''
  \href{http://dx.doi.org/10.1103/PhysRevD.87.123505}{{\em Phys.Rev.} {\bf D87}
  (2013) no.~12, 123505},
\href{http://arxiv.org/abs/1303.1102}{{\tt arXiv:1303.1102 [astro-ph.CO]}}.

\bibitem{Goon:2010xh}
G.~L. Goon, K.~Hinterbichler, and M.~Trodden, ``{Stability and superluminality
  of spherical DBI galileon solutions},''
  \href{http://dx.doi.org/10.1103/PhysRevD.83.085015}{{\em Phys.Rev.} {\bf D83}
  (2011)  085015},
\href{http://arxiv.org/abs/1008.4580}{{\tt arXiv:1008.4580 [hep-th]}}.

\bibitem{Brax:2014wla}
P.~Brax and P.~Valageas, ``{K-mouflage Cosmology: the Background Evolution},''
\href{http://arxiv.org/abs/1403.5420}{{\tt arXiv:1403.5420 [astro-ph.CO]}}.

\bibitem{ArmendarizPicon:2005nz}
C.~Armendariz-Picon and E.~A. Lim, ``{Haloes of k-essence},''
  \href{http://dx.doi.org/10.1088/1475-7516/2005/08/007}{{\em JCAP} {\bf 0508}
  (2005)  007},
\href{http://arxiv.org/abs/astro-ph/0505207}{{\tt arXiv:astro-ph/0505207
  [astro-ph]}}.

\bibitem{Brax:2014yla}
P.~Brax and P.~Valageas, ``{K-mouflage Cosmology: Formation of Large-Scale
  Structures},''
\href{http://arxiv.org/abs/1403.5424}{{\tt arXiv:1403.5424 [astro-ph.CO]}}.

\bibitem{Comer:1993zfa}
G.~L. Comer and D.~Langlois, ``{Hamiltonian formulation for multi-constituent
  relativistic perfect fluids},''
\href{http://dx.doi.org/10.1088/0264-9381/10/11/014}{{\em Class.Quant.Grav.}
  {\bf 10} (1993)  2317--2327}.

\bibitem{Dubovsky:2005xd}
S.~Dubovsky, T.~Gregoire, A.~Nicolis, and R.~Rattazzi, ``{Null energy condition
  and superluminal propagation},''
  \href{http://dx.doi.org/10.1088/1126-6708/2006/03/025}{{\em JHEP} {\bf 0603}
  (2006)  025},
\href{http://arxiv.org/abs/hep-th/0512260}{{\tt arXiv:hep-th/0512260
  [hep-th]}}.

\bibitem{DiezTejedor:2005fz}
A.~Diez-Tejedor and A.~Feinstein, ``{Relativistic hydrodynamics with sources
  for cosmological K-fluids},''
  \href{http://dx.doi.org/10.1142/S0218271805007152}{{\em Int.J.Mod.Phys.} {\bf
  D14} (2005)  1561--1576},
\href{http://arxiv.org/abs/gr-qc/0501101}{{\tt arXiv:gr-qc/0501101 [gr-qc]}}.

\bibitem{Arroja:2010wy}
F.~Arroja and M.~Sasaki, ``{A note on the equivalence of a barotropic perfect
  fluid with a K-essence scalar field},''
  \href{http://dx.doi.org/10.1103/PhysRevD.81.107301}{{\em Phys.Rev.} {\bf D81}
  (2010)  107301},
\href{http://arxiv.org/abs/1002.1376}{{\tt arXiv:1002.1376 [astro-ph.CO]}}.

\bibitem{Endlich:2010hf}
S.~Endlich, A.~Nicolis, R.~Rattazzi, and J.~Wang, ``{The Quantum mechanics of
  perfect fluids},'' \href{http://dx.doi.org/10.1007/JHEP04(2011)102}{{\em
  JHEP} {\bf 1104} (2011)  102},
\href{http://arxiv.org/abs/1011.6396}{{\tt arXiv:1011.6396 [hep-th]}}.

\bibitem{Dubovsky:2011sj}
S.~Dubovsky, L.~Hui, A.~Nicolis, and D.~T. Son, ``{Effective field theory for
  hydrodynamics: thermodynamics, and the derivative expansion},''
  \href{http://dx.doi.org/10.1103/PhysRevD.85.085029}{{\em Phys.Rev.} {\bf D85}
  (2012)  085029},
\href{http://arxiv.org/abs/1107.0731}{{\tt arXiv:1107.0731 [hep-th]}}.

\bibitem{Dubovsky:2011sk}
S.~Dubovsky, L.~Hui, and A.~Nicolis, ``{Effective field theory for
  hydrodynamics: Wess-Zumino term and anomalies in two spacetime dimensions},''
  \href{http://dx.doi.org/10.1103/PhysRevD.89.045016}{{\em Phys.Rev.} {\bf D89}
  (2014)  045016},
\href{http://arxiv.org/abs/1107.0732}{{\tt arXiv:1107.0732 [hep-th]}}.

\bibitem{Nicolis:2011cs}
A.~Nicolis, ``{Low-energy effective field theory for finite-temperature
  relativistic superfluids},''
\href{http://arxiv.org/abs/1108.2513}{{\tt arXiv:1108.2513 [hep-th]}}.

\bibitem{Nicolis:2011ey}
A.~Nicolis and D.~T. Son, ``{Hall viscosity from effective field theory},''
\href{http://arxiv.org/abs/1103.2137}{{\tt arXiv:1103.2137 [hep-th]}}.

\bibitem{Endlich:2012pz}
S.~Endlich, A.~Nicolis, and J.~Wang, ``{Solid Inflation},''
  \href{http://dx.doi.org/10.1088/1475-7516/2013/10/011}{{\em JCAP} {\bf 1310}
  (2013)  011},
\href{http://arxiv.org/abs/1210.0569}{{\tt arXiv:1210.0569 [hep-th]}}.

\bibitem{Endlich:2012vt}
S.~Endlich, A.~Nicolis, R.~A. Porto, and J.~Wang, ``{Dissipation in the
  effective field theory for hydrodynamics: First order effects},''
  \href{http://dx.doi.org/10.1103/PhysRevD.88.105001}{{\em Phys.Rev.} {\bf D88}
  (2013)  105001},
\href{http://arxiv.org/abs/1211.6461}{{\tt arXiv:1211.6461 [hep-th]}}.

\bibitem{Nicolis:2013lma}
A.~Nicolis, R.~Penco, and R.~A. Rosen, ``{Relativistic Fluids, Superfluids,
  Solids and Supersolids from a Coset Construction},''
\href{http://arxiv.org/abs/1307.0517}{{\tt arXiv:1307.0517 [hep-th]}}.

\bibitem{Gabadadze:2012sm}
G.~Gabadadze, K.~Hinterbichler, and D.~Pirtskhalava, ``{Classical Duals of
  Derivatively Self-Coupled Theories},''
  \href{http://dx.doi.org/10.1103/PhysRevD.85.125007}{{\em Phys.Rev.} {\bf D85}
  (2012)  125007},
\href{http://arxiv.org/abs/1202.6364}{{\tt arXiv:1202.6364 [hep-th]}}.

\bibitem{Dvali:2012zc}
G.~Dvali, A.~Franca, and C.~Gomez, ``{Road Signs for UV-Completion},''
\href{http://arxiv.org/abs/1204.6388}{{\tt arXiv:1204.6388 [hep-th]}}.

\bibitem{deRham:2014wfa}
C.~de~Rham and R.~H. Ribeiro, ``{Riding on irrelevant operators},''
\href{http://arxiv.org/abs/1405.5213}{{\tt arXiv:1405.5213 [hep-th]}}.

\bibitem{deRham:2010eu}
C.~de~Rham and A.~J. Tolley, ``{DBI and the Galileon reunited},''
  \href{http://dx.doi.org/10.1088/1475-7516/2010/05/015}{{\em JCAP} {\bf 1005}
  (2010)  015},
\href{http://arxiv.org/abs/1003.5917}{{\tt arXiv:1003.5917 [hep-th]}}.

\bibitem{Goon:2011uw}
G.~Goon, K.~Hinterbichler, and M.~Trodden, ``{A New Class of Effective Field
  Theories from Embedded Branes},''
  \href{http://dx.doi.org/10.1103/PhysRevLett.106.231102}{{\em Phys.Rev.Lett.}
  {\bf 106} (2011)  231102},
\href{http://arxiv.org/abs/1103.6029}{{\tt arXiv:1103.6029 [hep-th]}}.

\bibitem{Goon:2011qf}
G.~Goon, K.~Hinterbichler, and M.~Trodden, ``{Symmetries for Galileons and DBI
  scalars on curved space},''
  \href{http://dx.doi.org/10.1088/1475-7516/2011/07/017}{{\em JCAP} {\bf 1107}
  (2011)  017},
\href{http://arxiv.org/abs/1103.5745}{{\tt arXiv:1103.5745 [hep-th]}}.

\bibitem{Burrage:2011bt}
C.~Burrage, C.~de~Rham, and L.~Heisenberg, ``{de Sitter Galileon},''
  \href{http://dx.doi.org/10.1088/1475-7516/2011/05/025}{{\em JCAP} {\bf 1105}
  (2011)  025},
\href{http://arxiv.org/abs/1104.0155}{{\tt arXiv:1104.0155 [hep-th]}}.

\bibitem{Dvali:2010jz}
G.~Dvali, G.~F. Giudice, C.~Gomez, and A.~Kehagias, ``{UV-Completion by
  Classicalization},'' \href{http://dx.doi.org/10.1007/JHEP08(2011)108}{{\em
  JHEP} {\bf 1108} (2011)  108},
\href{http://arxiv.org/abs/1010.1415}{{\tt arXiv:1010.1415 [hep-ph]}}.

\bibitem{Rubakov:1983bb}
V.~Rubakov and M.~Shaposhnikov, ``{Do We Live Inside a Domain Wall?},''
\href{http://dx.doi.org/10.1016/0370-2693(83)91253-4}{{\em Phys.Lett.} {\bf
  B125} (1983)  136--138}.

\bibitem{Rubakov:1983bz}
V.~Rubakov and M.~Shaposhnikov, ``{Extra Space-Time Dimensions: Towards a
  Solution to the Cosmological Constant Problem},''
\href{http://dx.doi.org/10.1016/0370-2693(83)91254-6}{{\em Phys.Lett.} {\bf
  B125} (1983)  139}.

\bibitem{Randall:1999ee}
L.~Randall and R.~Sundrum, ``{A Large mass hierarchy from a small extra
  dimension},'' \href{http://dx.doi.org/10.1103/PhysRevLett.83.3370}{{\em
  Phys.Rev.Lett.} {\bf 83} (1999)  3370--3373},
\href{http://arxiv.org/abs/hep-ph/9905221}{{\tt arXiv:hep-ph/9905221
  [hep-ph]}}.

\bibitem{Randall:1999vf}
L.~Randall and R.~Sundrum, ``{An Alternative to compactification},''
  \href{http://dx.doi.org/10.1103/PhysRevLett.83.4690}{{\em Phys.Rev.Lett.}
  {\bf 83} (1999)  4690--4693},
\href{http://arxiv.org/abs/hep-th/9906064}{{\tt arXiv:hep-th/9906064
  [hep-th]}}.

\bibitem{Binetruy:1999ut}
P.~Binetruy, C.~Deffayet, and D.~Langlois, ``{Nonconventional cosmology from a
  brane universe},''
  \href{http://dx.doi.org/10.1016/S0550-3213(99)00696-3}{{\em Nucl.Phys.} {\bf
  B565} (2000)  269--287},
\href{http://arxiv.org/abs/hep-th/9905012}{{\tt arXiv:hep-th/9905012
  [hep-th]}}.

\bibitem{Shiromizu:1999wj}
T.~Shiromizu, K.-i. Maeda, and M.~Sasaki, ``{The Einstein equation on the
  3-brane world},'' \href{http://dx.doi.org/10.1103/PhysRevD.62.024012}{{\em
  Phys.Rev.} {\bf D62} (2000)  024012},
\href{http://arxiv.org/abs/gr-qc/9910076}{{\tt arXiv:gr-qc/9910076 [gr-qc]}}.

\bibitem{Dvali:2000hr}
G.~Dvali, G.~Gabadadze, and M.~Porrati, ``{4-D gravity on a brane in 5-D
  Minkowski space},''
  \href{http://dx.doi.org/10.1016/S0370-2693(00)00669-9}{{\em Phys.Lett.} {\bf
  B485} (2000)  208--214},
\href{http://arxiv.org/abs/hep-th/0005016}{{\tt arXiv:hep-th/0005016
  [hep-th]}}.

\bibitem{Dvali:2000xg}
G.~Dvali and G.~Gabadadze, ``{Gravity on a brane in infinite volume extra
  space},'' \href{http://dx.doi.org/10.1103/PhysRevD.63.065007}{{\em Phys.Rev.}
  {\bf D63} (2001)  065007},
\href{http://arxiv.org/abs/hep-th/0008054}{{\tt arXiv:hep-th/0008054
  [hep-th]}}.

\bibitem{Garriga:1999yh}
J.~Garriga and T.~Tanaka, ``{Gravity in the brane world},''
  \href{http://dx.doi.org/10.1103/PhysRevLett.84.2778}{{\em Phys.Rev.Lett.}
  {\bf 84} (2000)  2778--2781},
\href{http://arxiv.org/abs/hep-th/9911055}{{\tt arXiv:hep-th/9911055
  [hep-th]}}.

\bibitem{Csaki:1999jh}
C.~Csaki, M.~Graesser, C.~F. Kolda, and J.~Terning, ``{Cosmology of one extra
  dimension with localized gravity},''
  \href{http://dx.doi.org/10.1016/S0370-2693(99)00896-5}{{\em Phys.Lett.} {\bf
  B462} (1999)  34--40},
\href{http://arxiv.org/abs/hep-ph/9906513}{{\tt arXiv:hep-ph/9906513
  [hep-ph]}}.

\bibitem{Csaki:1999mp}
C.~Csaki, M.~Graesser, L.~Randall, and J.~Terning, ``{Cosmology of brane models
  with radion stabilization},''
  \href{http://dx.doi.org/10.1103/PhysRevD.62.045015}{{\em Phys.Rev.} {\bf D62}
  (2000)  045015},
\href{http://arxiv.org/abs/hep-ph/9911406}{{\tt arXiv:hep-ph/9911406
  [hep-ph]}}.

\bibitem{ArkaniHamed:2000eg}
N.~Arkani-Hamed, S.~Dimopoulos, N.~Kaloper, and R.~Sundrum, ``{A Small
  cosmological constant from a large extra dimension},''
  \href{http://dx.doi.org/10.1016/S0370-2693(00)00359-2}{{\em Phys.Lett.} {\bf
  B480} (2000)  193--199},
\href{http://arxiv.org/abs/hep-th/0001197}{{\tt arXiv:hep-th/0001197
  [hep-th]}}.

\bibitem{Gregory:2000jc}
R.~Gregory, V.~Rubakov, and S.~M. Sibiryakov, ``{Opening up extra dimensions at
  ultra large scales},''
  \href{http://dx.doi.org/10.1103/PhysRevLett.84.5928}{{\em Phys.Rev.Lett.}
  {\bf 84} (2000)  5928--5931},
\href{http://arxiv.org/abs/hep-th/0002072}{{\tt arXiv:hep-th/0002072
  [hep-th]}}.

\bibitem{Bowcock:2000cq}
P.~Bowcock, C.~Charmousis, and R.~Gregory, ``{General brane cosmologies and
  their global space-time structure},''
  \href{http://dx.doi.org/10.1088/0264-9381/17/22/313}{{\em Class.Quant.Grav.}
  {\bf 17} (2000)  4745--4764},
\href{http://arxiv.org/abs/hep-th/0007177}{{\tt arXiv:hep-th/0007177
  [hep-th]}}.

\bibitem{Deffayet:2000uy}
C.~Deffayet, ``{Cosmology on a brane in Minkowski bulk},''
  \href{http://dx.doi.org/10.1016/S0370-2693(01)00160-5}{{\em Phys.Lett.} {\bf
  B502} (2001)  199--208},
\href{http://arxiv.org/abs/hep-th/0010186}{{\tt arXiv:hep-th/0010186
  [hep-th]}}.

\bibitem{Deffayet:2001pu}
C.~Deffayet, G.~Dvali, and G.~Gabadadze, ``{Accelerated universe from gravity
  leaking to extra dimensions},''
  \href{http://dx.doi.org/10.1103/PhysRevD.65.044023}{{\em Phys.Rev.} {\bf D65}
  (2002)  044023},
\href{http://arxiv.org/abs/astro-ph/0105068}{{\tt arXiv:astro-ph/0105068
  [astro-ph]}}.

\bibitem{Deffayet:2002sp}
C.~Deffayet, S.~J. Landau, J.~Raux, M.~Zaldarriaga, and P.~Astier,
  ``{Supernovae, CMB, and gravitational leakage into extra dimensions},''
  \href{http://dx.doi.org/10.1103/PhysRevD.66.024019}{{\em Phys.Rev.} {\bf D66}
  (2002)  024019},
\href{http://arxiv.org/abs/astro-ph/0201164}{{\tt arXiv:astro-ph/0201164
  [astro-ph]}}.

\bibitem{Sahni:2002dx}
V.~Sahni and Y.~Shtanov, ``{Brane world models of dark energy},''
  \href{http://dx.doi.org/10.1088/1475-7516/2003/11/014}{{\em JCAP} {\bf 0311}
  (2003)  014},
\href{http://arxiv.org/abs/astro-ph/0202346}{{\tt arXiv:astro-ph/0202346
  [astro-ph]}}.

\bibitem{Maeda:2003ar}
K.-i. Maeda, S.~Mizuno, and T.~Torii, ``{Effective gravitational equations on
  brane world with induced gravity},''
  \href{http://dx.doi.org/10.1103/PhysRevD.68.024033}{{\em Phys.Rev.} {\bf D68}
  (2003)  024033},
\href{http://arxiv.org/abs/gr-qc/0303039}{{\tt arXiv:gr-qc/0303039 [gr-qc]}}.

\bibitem{Lue:2004rj}
A.~Lue, R.~Scoccimarro, and G.~D. Starkman, ``{Probing Newton's constant on
  vast scales: DGP gravity, cosmic acceleration and large scale structure},''
  \href{http://dx.doi.org/10.1103/PhysRevD.69.124015}{{\em Phys.Rev.} {\bf D69}
  (2004)  124015},
\href{http://arxiv.org/abs/astro-ph/0401515}{{\tt arXiv:astro-ph/0401515
  [astro-ph]}}.

\bibitem{Song:2006jk}
Y.-S. Song, I.~Sawicki, and W.~Hu, ``{Large-Scale Tests of the DGP Model},''
  \href{http://dx.doi.org/10.1103/PhysRevD.75.064003}{{\em Phys.Rev.} {\bf D75}
  (2007)  064003},
\href{http://arxiv.org/abs/astro-ph/0606286}{{\tt arXiv:astro-ph/0606286
  [astro-ph]}}.

\bibitem{Luty:2003vm}
M.~A. Luty, M.~Porrati, and R.~Rattazzi, ``{Strong interactions and stability
  in the DGP model},'' {\em JHEP} {\bf 0309} (2003)  029,
\href{http://arxiv.org/abs/hep-th/0303116}{{\tt arXiv:hep-th/0303116
  [hep-th]}}.

\bibitem{Koyama:2005tx}
K.~Koyama, ``{Are there ghosts in the self-accelerating brane universe?},''
  \href{http://dx.doi.org/10.1103/PhysRevD.72.123511}{{\em Phys.Rev.} {\bf D72}
  (2005)  123511},
\href{http://arxiv.org/abs/hep-th/0503191}{{\tt arXiv:hep-th/0503191
  [hep-th]}}.

\bibitem{Gorbunov:2005zk}
D.~Gorbunov, K.~Koyama, and S.~Sibiryakov, ``{More on ghosts in DGP model},''
  \href{http://dx.doi.org/10.1103/PhysRevD.73.044016}{{\em Phys.Rev.} {\bf D73}
  (2006)  044016},
\href{http://arxiv.org/abs/hep-th/0512097}{{\tt arXiv:hep-th/0512097
  [hep-th]}}.

\bibitem{Charmousis:2006pn}
C.~Charmousis, R.~Gregory, N.~Kaloper, and A.~Padilla, ``{DGP Specteroscopy},''
  \href{http://dx.doi.org/10.1088/1126-6708/2006/10/066}{{\em JHEP} {\bf 0610}
  (2006)  066},
\href{http://arxiv.org/abs/hep-th/0604086}{{\tt arXiv:hep-th/0604086
  [hep-th]}}.

\bibitem{Koyama:2007zz}
K.~Koyama, ``{Ghosts in the self-accelerating universe},''
  \href{http://dx.doi.org/10.1088/0264-9381/24/24/R01}{{\em Class.Quant.Grav.}
  {\bf 24} (2007)  R231--R253},
\href{http://arxiv.org/abs/0709.2399}{{\tt arXiv:0709.2399 [hep-th]}}.

\bibitem{Goon:2012dy}
G.~Goon, K.~Hinterbichler, A.~Joyce, and M.~Trodden, ``{Galileons as
  Wess-Zumino Terms},'' \href{http://dx.doi.org/10.1007/JHEP06(2012)004}{{\em
  JHEP} {\bf 1206} (2012)  004},
\href{http://arxiv.org/abs/1203.3191}{{\tt arXiv:1203.3191 [hep-th]}}.

\bibitem{Chow:2009fm}
N.~Chow and J.~Khoury, ``{Galileon Cosmology},''
  \href{http://dx.doi.org/10.1103/PhysRevD.80.024037}{{\em Phys.Rev.} {\bf D80}
  (2009)  024037},
\href{http://arxiv.org/abs/0905.1325}{{\tt arXiv:0905.1325 [hep-th]}}.

\bibitem{Silva:2009km}
F.~P. Silva and K.~Koyama, ``{Self-Accelerating Universe in Galileon
  Cosmology},'' \href{http://dx.doi.org/10.1103/PhysRevD.80.121301}{{\em
  Phys.Rev.} {\bf D80} (2009)  121301},
\href{http://arxiv.org/abs/0909.4538}{{\tt arXiv:0909.4538 [astro-ph.CO]}}.

\bibitem{Kobayashi:2009wr}
T.~Kobayashi, H.~Tashiro, and D.~Suzuki, ``{Evolution of linear cosmological
  perturbations and its observational implications in Galileon-type modified
  gravity},'' \href{http://dx.doi.org/10.1103/PhysRevD.81.063513}{{\em
  Phys.Rev.} {\bf D81} (2010)  063513},
\href{http://arxiv.org/abs/0912.4641}{{\tt arXiv:0912.4641 [astro-ph.CO]}}.

\bibitem{DeFelice:2010as}
A.~De~Felice, R.~Kase, and S.~Tsujikawa, ``{Matter perturbations in Galileon
  cosmology},'' \href{http://dx.doi.org/10.1103/PhysRevD.83.043515}{{\em
  Phys.Rev.} {\bf D83} (2011)  043515},
\href{http://arxiv.org/abs/1011.6132}{{\tt arXiv:1011.6132 [astro-ph.CO]}}.

\bibitem{DeFelice:2010gb}
A.~De~Felice, S.~Mukohyama, and S.~Tsujikawa, ``{Density perturbations in
  general modified gravitational theories},''
  \href{http://dx.doi.org/10.1103/PhysRevD.82.023524}{{\em Phys.Rev.} {\bf D82}
  (2010)  023524},
\href{http://arxiv.org/abs/1006.0281}{{\tt arXiv:1006.0281 [astro-ph.CO]}}.

\bibitem{DeFelice:2010pv}
A.~De~Felice and S.~Tsujikawa, ``{Cosmology of a covariant Galileon field},''
  \href{http://dx.doi.org/10.1103/PhysRevLett.105.111301}{{\em Phys.Rev.Lett.}
  {\bf 105} (2010)  111301},
\href{http://arxiv.org/abs/1007.2700}{{\tt arXiv:1007.2700 [astro-ph.CO]}}.

\bibitem{Gannouji:2010au}
R.~Gannouji and M.~Sami, ``{Galileon gravity and its relevance to late time
  cosmic acceleration},''
  \href{http://dx.doi.org/10.1103/PhysRevD.82.024011}{{\em Phys.Rev.} {\bf D82}
  (2010)  024011},
\href{http://arxiv.org/abs/1004.2808}{{\tt arXiv:1004.2808 [gr-qc]}}.

\bibitem{Ali:2010gr}
A.~Ali, R.~Gannouji, and M.~Sami, ``{Modified gravity a la Galileon: Late time
  cosmic acceleration and observational constraints},''
  \href{http://dx.doi.org/10.1103/PhysRevD.82.103015}{{\em Phys.Rev.} {\bf D82}
  (2010)  103015},
\href{http://arxiv.org/abs/1008.1588}{{\tt arXiv:1008.1588 [astro-ph.CO]}}.

\bibitem{Mota:2010bs}
D.~F. Mota, M.~Sandstad, and T.~Zlosnik, ``{Cosmology of the selfaccelerating
  third order Galileon},''
  \href{http://dx.doi.org/10.1007/JHEP12(2010)051}{{\em JHEP} {\bf 1012} (2010)
   051},
\href{http://arxiv.org/abs/1009.6151}{{\tt arXiv:1009.6151 [astro-ph.CO]}}.

\bibitem{Barreira:2012kk}
A.~Barreira, B.~Li, C.~M. Baugh, and S.~Pascoli, ``{Linear perturbations in
  Galileon gravity models},''
  \href{http://dx.doi.org/10.1103/PhysRevD.86.124016}{{\em Phys.Rev.} {\bf D86}
  (2012)  124016},
\href{http://arxiv.org/abs/1208.0600}{{\tt arXiv:1208.0600 [astro-ph.CO]}}.

\bibitem{Anisimov:2005ne}
A.~Anisimov, E.~Babichev, and A.~Vikman, ``{B-inflation},''
  \href{http://dx.doi.org/10.1088/1475-7516/2005/06/006}{{\em JCAP} {\bf 0506}
  (2005)  006},
\href{http://arxiv.org/abs/astro-ph/0504560}{{\tt arXiv:astro-ph/0504560
  [astro-ph]}}.

\bibitem{Kobayashi:2010cm}
T.~Kobayashi, M.~Yamaguchi, and J.~Yokoyama, ``{G-inflation: Inflation driven
  by the Galileon field},''
  \href{http://dx.doi.org/10.1103/PhysRevLett.105.231302}{{\em Phys.Rev.Lett.}
  {\bf 105} (2010)  231302},
\href{http://arxiv.org/abs/1008.0603}{{\tt arXiv:1008.0603 [hep-th]}}.

\bibitem{Burrage:2010cu}
C.~Burrage, C.~de~Rham, D.~Seery, and A.~J. Tolley, ``{Galileon inflation},''
  \href{http://dx.doi.org/10.1088/1475-7516/2011/01/014}{{\em JCAP} {\bf 1101}
  (2011)  014},
\href{http://arxiv.org/abs/1009.2497}{{\tt arXiv:1009.2497 [hep-th]}}.

\bibitem{Creminelli:2010qf}
P.~Creminelli, G.~D'Amico, M.~Musso, J.~Norena, and E.~Trincherini, ``{Galilean
  symmetry in the effective theory of inflation: new shapes of
  non-Gaussianity},''
  \href{http://dx.doi.org/10.1088/1475-7516/2011/02/006}{{\em JCAP} {\bf 1102}
  (2011)  006},
\href{http://arxiv.org/abs/1011.3004}{{\tt arXiv:1011.3004 [hep-th]}}.

\bibitem{Kamada:2010qe}
K.~Kamada, T.~Kobayashi, M.~Yamaguchi, and J.~Yokoyama, ``{Higgs
  G-inflation},'' \href{http://dx.doi.org/10.1103/PhysRevD.83.083515}{{\em
  Phys.Rev.} {\bf D83} (2011)  083515},
\href{http://arxiv.org/abs/1012.4238}{{\tt arXiv:1012.4238 [astro-ph.CO]}}.

\bibitem{DeFelice:2010nf}
A.~De~Felice and S.~Tsujikawa, ``{Generalized Galileon cosmology},''
  \href{http://dx.doi.org/10.1103/PhysRevD.84.124029}{{\em Phys.Rev.} {\bf D84}
  (2011)  124029},
\href{http://arxiv.org/abs/1008.4236}{{\tt arXiv:1008.4236 [hep-th]}}.

\bibitem{Kobayashi:2010wa}
T.~Kobayashi, ``{Cosmic expansion and growth histories in Galileon
  scalar-tensor models of dark energy},''
  \href{http://dx.doi.org/10.1103/PhysRevD.81.103533}{{\em Phys.Rev.} {\bf D81}
  (2010)  103533},
\href{http://arxiv.org/abs/1003.3281}{{\tt arXiv:1003.3281 [astro-ph.CO]}}.

\bibitem{DeFelice:2011zh}
A.~De~Felice and S.~Tsujikawa, ``{Primordial non-Gaussianities in general
  modified gravitational models of inflation},''
  \href{http://dx.doi.org/10.1088/1475-7516/2011/04/029}{{\em JCAP} {\bf 1104}
  (2011)  029},
\href{http://arxiv.org/abs/1103.1172}{{\tt arXiv:1103.1172 [astro-ph.CO]}}.

\bibitem{Kobayashi:2011pc}
T.~Kobayashi, M.~Yamaguchi, and J.~Yokoyama, ``{Primordial non-Gaussianity from
  G-inflation},'' \href{http://dx.doi.org/10.1103/PhysRevD.83.103524}{{\em
  Phys.Rev.} {\bf D83} (2011)  103524},
\href{http://arxiv.org/abs/1103.1740}{{\tt arXiv:1103.1740 [hep-th]}}.

\bibitem{RenauxPetel:2011uk}
S.~Renaux-Petel, S.~Mizuno, and K.~Koyama, ``{Primordial fluctuations and
  non-Gaussianities from multifield DBI Galileon inflation},''
  \href{http://dx.doi.org/10.1088/1475-7516/2011/11/042}{{\em JCAP} {\bf 1111}
  (2011)  042},
\href{http://arxiv.org/abs/1108.0305}{{\tt arXiv:1108.0305 [astro-ph.CO]}}.

\bibitem{Fasiello:2013dla}
M.~Fasiello, ``{Trispectrum from Co-dimension 2(n) Galileons},''
  \href{http://dx.doi.org/10.1088/1475-7516/2013/12/033}{{\em JCAP} {\bf 1312}
  (2013)  033},
\href{http://arxiv.org/abs/1303.5015}{{\tt arXiv:1303.5015 [hep-th]}}.

\bibitem{Creminelli:2010ba}
P.~Creminelli, A.~Nicolis, and E.~Trincherini, ``{Galilean Genesis: An
  Alternative to inflation},''
  \href{http://dx.doi.org/10.1088/1475-7516/2010/11/021}{{\em JCAP} {\bf 1011}
  (2010)  021},
\href{http://arxiv.org/abs/1007.0027}{{\tt arXiv:1007.0027 [hep-th]}}.

\bibitem{Hinterbichler:2011qk}
K.~Hinterbichler and J.~Khoury, ``{The Pseudo-Conformal Universe: Scale
  Invariance from Spontaneous Breaking of Conformal Symmetry},''
  \href{http://dx.doi.org/10.1088/1475-7516/2012/04/023}{{\em JCAP} {\bf 1204}
  (2012)  023},
\href{http://arxiv.org/abs/1106.1428}{{\tt arXiv:1106.1428 [hep-th]}}.

\bibitem{LevasseurPerreault:2011mw}
L.~Perreault~Levasseur, R.~Brandenberger, and A.-C. Davis, ``{Defrosting in an
  Emergent Galileon Cosmology},''
  \href{http://dx.doi.org/10.1103/PhysRevD.84.103512}{{\em Phys.Rev.} {\bf D84}
  (2011)  103512},
\href{http://arxiv.org/abs/1105.5649}{{\tt arXiv:1105.5649 [astro-ph.CO]}}.

\bibitem{Wang:2012bq}
Y.~Wang and R.~Brandenberger, ``{Scale-Invariant Fluctuations from Galilean
  Genesis},'' \href{http://dx.doi.org/10.1088/1475-7516/2012/10/021}{{\em JCAP}
  {\bf 1210} (2012)  021},
\href{http://arxiv.org/abs/1206.4309}{{\tt arXiv:1206.4309 [hep-th]}}.

\bibitem{Liu:2011ns}
Z.-G. Liu, J.~Zhang, and Y.-S. Piao, ``{A Galileon Design of Slow Expansion},''
  \href{http://dx.doi.org/10.1103/PhysRevD.84.063508}{{\em Phys.Rev.} {\bf D84}
  (2011)  063508},
\href{http://arxiv.org/abs/1105.5713}{{\tt arXiv:1105.5713 [astro-ph.CO]}}.

\bibitem{Hinterbichler:2012mv}
K.~Hinterbichler, A.~Joyce, and J.~Khoury, ``{Non-linear Realizations of
  Conformal Symmetry and Effective Field Theory for the Pseudo-Conformal
  Universe},'' \href{http://dx.doi.org/10.1088/1475-7516/2012/06/043}{{\em
  JCAP} {\bf 1206} (2012)  043},
\href{http://arxiv.org/abs/1202.6056}{{\tt arXiv:1202.6056 [hep-th]}}.

\bibitem{Hinterbichler:2012fr}
K.~Hinterbichler, A.~Joyce, J.~Khoury, and G.~E. Miller, ``{DBI Realizations of
  the Pseudo-Conformal Universe and Galilean Genesis Scenarios},''
  \href{http://dx.doi.org/10.1088/1475-7516/2012/12/030}{{\em JCAP} {\bf 1212}
  (2012)  030},
\href{http://arxiv.org/abs/1209.5742}{{\tt arXiv:1209.5742 [hep-th]}}.

\bibitem{Hinterbichler:2012yn}
K.~Hinterbichler, A.~Joyce, J.~Khoury, and G.~E. Miller, ``{DBI Genesis: An
  Improved Violation of the Null Energy Condition},''
  \href{http://dx.doi.org/10.1103/PhysRevLett.110.241303}{{\em Phys.Rev.Lett.}
  {\bf 110} (2013)  241303},
\href{http://arxiv.org/abs/1212.3607}{{\tt arXiv:1212.3607 [hep-th]}}.

\bibitem{Creminelli:2012my}
P.~Creminelli, K.~Hinterbichler, J.~Khoury, A.~Nicolis, and E.~Trincherini,
  ``{Subluminal Galilean Genesis},''
  \href{http://dx.doi.org/10.1007/JHEP02(2013)006}{{\em JHEP} {\bf 1302} (2013)
   006},
\href{http://arxiv.org/abs/1209.3768}{{\tt arXiv:1209.3768 [hep-th]}}.

\bibitem{Nicolis:2009qm}
A.~Nicolis, R.~Rattazzi, and E.~Trincherini, ``{Energy's and amplitudes'
  positivity},'' \href{http://dx.doi.org/10.1007/JHEP05(2010)095,
  10.1007/JHEP11(2011)128}{{\em JHEP} {\bf 1005} (2010)  095},
\href{http://arxiv.org/abs/0912.4258}{{\tt arXiv:0912.4258 [hep-th]}}.

\bibitem{Qiu:2011cy}
T.~Qiu, J.~Evslin, Y.-F. Cai, M.~Li, and X.~Zhang, ``{Bouncing Galileon
  Cosmologies},'' \href{http://dx.doi.org/10.1088/1475-7516/2011/10/036}{{\em
  JCAP} {\bf 1110} (2011)  036},
\href{http://arxiv.org/abs/1108.0593}{{\tt arXiv:1108.0593 [hep-th]}}.

\bibitem{Easson:2011zy}
D.~A. Easson, I.~Sawicki, and A.~Vikman, ``{G-Bounce},''
  \href{http://dx.doi.org/10.1088/1475-7516/2011/11/021}{{\em JCAP} {\bf 1111}
  (2011)  021},
\href{http://arxiv.org/abs/1109.1047}{{\tt arXiv:1109.1047 [hep-th]}}.

\bibitem{Easson:2013bda}
D.~A. Easson, I.~Sawicki, and A.~Vikman, ``{When Matter Matters},''
  \href{http://dx.doi.org/10.1088/1475-7516/2013/07/014}{{\em JCAP} {\bf 1307}
  (2013)  014},
\href{http://arxiv.org/abs/1304.3903}{{\tt arXiv:1304.3903 [hep-th]}}.

\bibitem{Rubakov:2013kaa}
V.~Rubakov, ``{Consistent null-energy-condition violation: Towards creating a
  universe in the laboratory},''
  \href{http://dx.doi.org/10.1103/PhysRevD.88.044015}{{\em Phys.Rev.} {\bf D88}
  (2013)  044015},
\href{http://arxiv.org/abs/1305.2614}{{\tt arXiv:1305.2614 [hep-th]}}.

\bibitem{Elder:2013gya}
B.~Elder, A.~Joyce, and J.~Khoury, ``{From Satisfying to Violating the Null
  Energy Condition},'' \href{http://dx.doi.org/10.1103/PhysRevD.89.044027}{{\em
  Phys.Rev.} {\bf D89} (2014)  044027},
\href{http://arxiv.org/abs/1311.5889}{{\tt arXiv:1311.5889 [hep-th]}}.

\bibitem{Rubakov:2014jja}
V.~Rubakov, ``{The Null Energy Condition and its violation},''
\href{http://arxiv.org/abs/1401.4024}{{\tt arXiv:1401.4024 [hep-th]}}.

\bibitem{Battefeld:2014uga}
D.~Battefeld and P.~Peter, ``{A Critical Review of Classical Bouncing
  Cosmologies},''
\href{http://arxiv.org/abs/1406.2790}{{\tt arXiv:1406.2790 [astro-ph.CO]}}.

\bibitem{Deffayet:2009wt}
C.~Deffayet, G.~Esposito-Farese, and A.~Vikman, ``{Covariant Galileon},''
  \href{http://dx.doi.org/10.1103/PhysRevD.79.084003}{{\em Phys.Rev.} {\bf D79}
  (2009)  084003},
\href{http://arxiv.org/abs/0901.1314}{{\tt arXiv:0901.1314 [hep-th]}}.

\bibitem{Deffayet:2009mn}
C.~Deffayet, S.~Deser, and G.~Esposito-Farese, ``{Generalized Galileons: All
  scalar models whose curved background extensions maintain second-order field
  equations and stress-tensors},''
  \href{http://dx.doi.org/10.1103/PhysRevD.80.064015}{{\em Phys.Rev.} {\bf D80}
  (2009)  064015},
\href{http://arxiv.org/abs/0906.1967}{{\tt arXiv:0906.1967 [gr-qc]}}.

\bibitem{Deffayet:2010qz}
C.~Deffayet, O.~Pujolas, I.~Sawicki, and A.~Vikman, ``{Imperfect Dark Energy
  from Kinetic Gravity Braiding},''
  \href{http://dx.doi.org/10.1088/1475-7516/2010/10/026}{{\em JCAP} {\bf 1010}
  (2010)  026},
\href{http://arxiv.org/abs/1008.0048}{{\tt arXiv:1008.0048 [hep-th]}}.

\bibitem{Deffayet:2011gz}
C.~Deffayet, X.~Gao, D.~Steer, and G.~Zahariade, ``{From k-essence to
  generalised Galileons},''
  \href{http://dx.doi.org/10.1103/PhysRevD.84.064039}{{\em Phys.Rev.} {\bf D84}
  (2011)  064039},
\href{http://arxiv.org/abs/1103.3260}{{\tt arXiv:1103.3260 [hep-th]}}.

\bibitem{Pujolas:2011he}
O.~Pujolas, I.~Sawicki, and A.~Vikman, ``{The Imperfect Fluid behind Kinetic
  Gravity Braiding},'' \href{http://dx.doi.org/10.1007/JHEP11(2011)156}{{\em
  JHEP} {\bf 1111} (2011)  156},
\href{http://arxiv.org/abs/1103.5360}{{\tt arXiv:1103.5360 [hep-th]}}.

\bibitem{Gabadadze:2012tr}
G.~Gabadadze, K.~Hinterbichler, J.~Khoury, D.~Pirtskhalava, and M.~Trodden,
  ``{A Covariant Master Theory for Novel Galilean Invariant Models and Massive
  Gravity},'' \href{http://dx.doi.org/10.1103/PhysRevD.86.124004}{{\em
  Phys.Rev.} {\bf D86} (2012)  124004},
\href{http://arxiv.org/abs/1208.5773}{{\tt arXiv:1208.5773 [hep-th]}}.

\bibitem{Andrews:2013ora}
M.~Andrews, G.~Goon, K.~Hinterbichler, J.~Stokes, and M.~Trodden, ``{Massive
  Gravity Coupled to Galileons is Ghost-Free},''
  \href{http://dx.doi.org/10.1103/PhysRevLett.111.061107}{{\em Phys.Rev.Lett.}
  {\bf 111} (2013) no.~6, 061107},
\href{http://arxiv.org/abs/1303.1177}{{\tt arXiv:1303.1177 [hep-th]}}.

\bibitem{Goon:2014ywa}
G.~Goon, A.~E. Gumrukcuoglu, K.~Hinterbichler, S.~Mukohyama, and M.~Trodden,
  ``{Galileons Coupled to Massive Gravity: General Analysis and Cosmological
  Solutions},''
\href{http://arxiv.org/abs/1402.5424}{{\tt arXiv:1402.5424 [hep-th]}}.

\bibitem{Khoury:2011da}
J.~Khoury, J.-L. Lehners, and B.~A. Ovrut, ``{Supersymmetric Galileons},''
  \href{http://dx.doi.org/10.1103/PhysRevD.84.043521}{{\em Phys.Rev.} {\bf D84}
  (2011)  043521},
\href{http://arxiv.org/abs/1103.0003}{{\tt arXiv:1103.0003 [hep-th]}}.

\bibitem{Koehn:2013hk}
M.~Koehn, J.-L. Lehners, and B.~Ovrut, ``{Supersymmetric Galileons Have
  Ghosts},'' \href{http://dx.doi.org/10.1103/PhysRevD.88.023528}{{\em
  Phys.Rev.} {\bf D88} (2013)  023528},
\href{http://arxiv.org/abs/1302.0840}{{\tt arXiv:1302.0840 [hep-th]}}.

\bibitem{Koehn:2013upa}
M.~Koehn, J.-L. Lehners, and B.~A. Ovrut, ``{A Cosmological Super-Bounce},''
\href{http://arxiv.org/abs/1310.7577}{{\tt arXiv:1310.7577 [hep-th]}}.

\bibitem{Farakos:2013fne}
F.~Farakos, C.~Germani, and A.~Kehagias, ``{On ghost-free supersymmetric
  galileons},'' \href{http://dx.doi.org/10.1007/JHEP11(2013)045}{{\em JHEP}
  {\bf 1311} (2013)  045},
\href{http://arxiv.org/abs/1306.2961}{{\tt arXiv:1306.2961 [hep-th]}}.

\bibitem{Deffayet:2010zh}
C.~Deffayet, S.~Deser, and G.~Esposito-Farese, ``{Arbitrary $p$-form
  Galileons},'' \href{http://dx.doi.org/10.1103/PhysRevD.82.061501}{{\em
  Phys.Rev.} {\bf D82} (2010)  061501},
\href{http://arxiv.org/abs/1007.5278}{{\tt arXiv:1007.5278 [gr-qc]}}.

\bibitem{Padilla:2010de}
A.~Padilla, P.~M. Saffin, and S.-Y. Zhou, ``{Bi-galileon theory I: Motivation
  and formulation},'' \href{http://dx.doi.org/10.1007/JHEP12(2010)031}{{\em
  JHEP} {\bf 1012} (2010)  031},
\href{http://arxiv.org/abs/1007.5424}{{\tt arXiv:1007.5424 [hep-th]}}.

\bibitem{Padilla:2010tj}
A.~Padilla, P.~M. Saffin, and S.-Y. Zhou, ``{Bi-galileon theory II:
  Phenomenology},'' \href{http://dx.doi.org/10.1007/JHEP01(2011)099}{{\em JHEP}
  {\bf 1101} (2011)  099},
\href{http://arxiv.org/abs/1008.3312}{{\tt arXiv:1008.3312 [hep-th]}}.

\bibitem{Padilla:2010ir}
A.~Padilla, P.~M. Saffin, and S.-Y. Zhou, ``{Multi-galileons, solitons and
  Derrick's theorem},''
  \href{http://dx.doi.org/10.1103/PhysRevD.83.045009}{{\em Phys.Rev.} {\bf D83}
  (2011)  045009},
\href{http://arxiv.org/abs/1008.0745}{{\tt arXiv:1008.0745 [hep-th]}}.

\bibitem{Hinterbichler:2010xn}
K.~Hinterbichler, M.~Trodden, and D.~Wesley, ``{Multi-field galileons and
  higher co-dimension branes},''
  \href{http://dx.doi.org/10.1103/PhysRevD.82.124018}{{\em Phys.Rev.} {\bf D82}
  (2010)  124018},
\href{http://arxiv.org/abs/1008.1305}{{\tt arXiv:1008.1305 [hep-th]}}.

\bibitem{Andrews:2010km}
M.~Andrews, K.~Hinterbichler, J.~Khoury, and M.~Trodden, ``{Instabilities of
  Spherical Solutions with Multiple Galileons and SO(N) Symmetry},''
  \href{http://dx.doi.org/10.1103/PhysRevD.83.044042}{{\em Phys.Rev.} {\bf D83}
  (2011)  044042},
\href{http://arxiv.org/abs/1008.4128}{{\tt arXiv:1008.4128 [hep-th]}}.

\bibitem{Garcia-Saenz:2013gya}
S.~Garcia-Saenz, ``{Behavior of perturbations on spherically symmetric
  backgrounds in multi-Galileon theory},''
  \href{http://dx.doi.org/10.1103/PhysRevD.87.104012}{{\em Phys.Rev.} {\bf D87}
  (2013) no.~10, 104012},
\href{http://arxiv.org/abs/1303.2905}{{\tt arXiv:1303.2905 [hep-th]}}.

\bibitem{Zhou:2011ix}
S.-Y. Zhou and E.~J. Copeland, ``{Galileons with Gauge Symmetries},''
  \href{http://dx.doi.org/10.1103/PhysRevD.85.065002}{{\em Phys.Rev.} {\bf D85}
  (2012)  065002},
\href{http://arxiv.org/abs/1112.0968}{{\tt arXiv:1112.0968 [hep-th]}}.

\bibitem{Goon:2012mu}
G.~Goon, K.~Hinterbichler, A.~Joyce, and M.~Trodden, ``{Gauged Galileons From
  Branes},'' \href{http://dx.doi.org/10.1016/j.physletb.2012.06.065}{{\em
  Phys.Lett.} {\bf B714} (2012)  115--119},
\href{http://arxiv.org/abs/1201.0015}{{\tt arXiv:1201.0015 [hep-th]}}.

\bibitem{Endlich:2010zj}
S.~Endlich, K.~Hinterbichler, L.~Hui, A.~Nicolis, and J.~Wang, ``{Derrick's
  theorem beyond a potential},''
  \href{http://dx.doi.org/10.1007/JHEP05(2011)073}{{\em JHEP} {\bf 1105} (2011)
   073},
\href{http://arxiv.org/abs/1002.4873}{{\tt arXiv:1002.4873 [hep-th]}}.

\bibitem{Masoumi:2012np}
A.~Masoumi and X.~Xiao, ``{Moving Stable Solitons in Galileon Theory},''
  \href{http://dx.doi.org/10.1016/j.physletb.2012.07.019}{{\em Phys.Lett.} {\bf
  B715} (2012)  214--218},
\href{http://arxiv.org/abs/1201.3132}{{\tt arXiv:1201.3132 [hep-th]}}.

\bibitem{Babichev:2012qs}
E.~Babichev, ``{Plane waves in the generalized Galileon theory},''
  \href{http://dx.doi.org/10.1103/PhysRevD.86.084037}{{\em Phys.Rev.} {\bf D86}
  (2012)  084037},
\href{http://arxiv.org/abs/1207.4764}{{\tt arXiv:1207.4764 [gr-qc]}}.

\bibitem{Zhou:2012fk}
S.-Y. Zhou, ``{Note on the Stabilities of the Light-like Galileon Solutions},''
  \href{http://dx.doi.org/10.1103/PhysRevD.85.104005}{{\em Phys.Rev.} {\bf D85}
  (2012)  104005},
\href{http://arxiv.org/abs/1202.5769}{{\tt arXiv:1202.5769 [hep-th]}}.

\bibitem{Tasinato:2014eka}
G.~Tasinato, ``{Cosmic Acceleration from Abelian Symmetry Breaking},''
  \href{http://dx.doi.org/10.1007/JHEP04(2014)067}{{\em JHEP} {\bf 1404} (2014)
   067},
\href{http://arxiv.org/abs/1402.6450}{{\tt arXiv:1402.6450 [hep-th]}}.

\bibitem{Heisenberg:2014rta}
L.~Heisenberg, ``{Generalization of the Proca Action},''
  \href{http://dx.doi.org/10.1088/1475-7516/2014/05/015}{{\em JCAP} {\bf 1405}
  (2014)  015},
\href{http://arxiv.org/abs/1402.7026}{{\tt arXiv:1402.7026 [hep-th]}}.

\bibitem{Tasinato:2014mia}
G.~Tasinato, ``{A small cosmological constant from Abelian symmetry
  breaking},''
\href{http://arxiv.org/abs/1404.4883}{{\tt arXiv:1404.4883 [hep-th]}}.

\bibitem{Hiramatsu:2012xj}
T.~Hiramatsu, W.~Hu, K.~Koyama, and F.~Schmidt, ``{Equivalence Principle
  Violation in Vainshtein Screened Two-Body Systems},''
  \href{http://dx.doi.org/10.1103/PhysRevD.87.063525}{{\em Phys.Rev.} {\bf D87}
  (2013) no.~6, 063525},
\href{http://arxiv.org/abs/1209.3364}{{\tt arXiv:1209.3364 [hep-th]}}.

\bibitem{Li:2013nua}
B.~Li, G.-B. Zhao, and K.~Koyama, ``{Exploring Vainshtein mechanism on
  adaptively refined meshes},''
  \href{http://dx.doi.org/10.1088/1475-7516/2013/05/023}{{\em JCAP} {\bf 1305}
  (2013)  023},
\href{http://arxiv.org/abs/1303.0008}{{\tt arXiv:1303.0008 [astro-ph.CO]}}.

\bibitem{Lovelock:1971yv}
D.~Lovelock, ``{The Einstein tensor and its generalizations},''
\href{http://dx.doi.org/10.1063/1.1665613}{{\em J.Math.Phys.} {\bf 12} (1971)
  498--501}.

\bibitem{Goon:2011xf}
G.~Goon, K.~Hinterbichler, and M.~Trodden, ``{Galileons on Cosmological
  Backgrounds},'' \href{http://dx.doi.org/10.1088/1475-7516/2011/12/004}{{\em
  JCAP} {\bf 1112} (2011)  004},
\href{http://arxiv.org/abs/1109.3450}{{\tt arXiv:1109.3450 [hep-th]}}.

\bibitem{Bloomfield:2014zfa}
J.~K. Bloomfield, C.~Burrage, and A.-C. Davis, ``{The Shape Dependence of
  Vainshtein Screening},''
\href{http://arxiv.org/abs/1408.4759}{{\tt arXiv:1408.4759 [gr-qc]}}.

\bibitem{Brito:2014ifa}
R.~Brito, A.~Terrana, M.~Johnson, and V.~Cardoso, ``{The nonlinear dynamical
  stability of infrared modifications of gravity},''
\href{http://arxiv.org/abs/1409.0886}{{\tt arXiv:1409.0886 [hep-th]}}.

\bibitem{Babichev:2012re}
E.~Babichev and G.~Esposito-Farse, ``{Time-Dependent Spherically Symmetric
  Covariant Galileons},''
  \href{http://dx.doi.org/10.1103/PhysRevD.87.044032}{{\em Phys.Rev.} {\bf D87}
  (2013) no.~4, 044032},
\href{http://arxiv.org/abs/1212.1394}{{\tt arXiv:1212.1394 [gr-qc]}}.

\bibitem{Berezhiani:2013dw}
L.~Berezhiani, G.~Chkareuli, and G.~Gabadadze, ``{Restricted Galileons},''
  \href{http://dx.doi.org/10.1103/PhysRevD.88.124020}{{\em Phys.Rev.} {\bf D88}
  (2013)  124020},
\href{http://arxiv.org/abs/1302.0549}{{\tt arXiv:1302.0549 [hep-th]}}.

\bibitem{Berezhiani:2013dca}
L.~Berezhiani, G.~Chkareuli, C.~de~Rham, G.~Gabadadze, and A.~Tolley, ``{Mixed
  Galileons and Spherically Symmetric Solutions},''
  \href{http://dx.doi.org/10.1088/0264-9381/30/18/184003}{{\em
  Class.Quant.Grav.} {\bf 30} (2013)  184003},
\href{http://arxiv.org/abs/1305.0271}{{\tt arXiv:1305.0271 [hep-th]}}.

\bibitem{Hinterbichler:2009kq}
K.~Hinterbichler, A.~Nicolis, and M.~Porrati, ``{Superluminality in DGP},''
  \href{http://dx.doi.org/10.1088/1126-6708/2009/09/089}{{\em JHEP} {\bf 0909}
  (2009)  089},
\href{http://arxiv.org/abs/0905.2359}{{\tt arXiv:0905.2359 [hep-th]}}.

\bibitem{deFromont:2013iwa}
P.~de~Fromont, C.~de~Rham, L.~Heisenberg, and A.~Matas, ``{Superluminality in
  the Bi- and Multi- Galileon},''
  \href{http://dx.doi.org/10.1007/JHEP07(2013)067}{{\em JHEP} {\bf 1307} (2013)
   067},
\href{http://arxiv.org/abs/1303.0274}{{\tt arXiv:1303.0274 [hep-th]}}.

\bibitem{deRham:2013hsa}
C.~de~Rham, M.~Fasiello, and A.~J. Tolley, ``{Galileon Duality},''
\href{http://arxiv.org/abs/1308.2702}{{\tt arXiv:1308.2702 [hep-th]}}.

\bibitem{deRham:2014lqa}
C.~de~Rham, L.~Keltner, and A.~J. Tolley, ``{Generalized Galileon Duality},''
\href{http://arxiv.org/abs/1403.3690}{{\tt arXiv:1403.3690 [hep-th]}}.

\bibitem{deRham:2012fw}
C.~de~Rham, A.~J. Tolley, and D.~H. Wesley, ``{Vainshtein Mechanism in Binary
  Pulsars},'' \href{http://dx.doi.org/10.1103/PhysRevD.87.044025}{{\em
  Phys.Rev.} {\bf D87} (2013) no.~4, 044025},
\href{http://arxiv.org/abs/1208.0580}{{\tt arXiv:1208.0580 [gr-qc]}}.

\bibitem{Chu:2012kz}
Y.-Z. Chu and M.~Trodden, ``{Retarded Green's Function Of A Vainshtein System
  And Galileon Waves},''
  \href{http://dx.doi.org/10.1103/PhysRevD.87.024011}{{\em Phys.Rev.} {\bf D87}
  (2013)  024011},
\href{http://arxiv.org/abs/1210.6651}{{\tt arXiv:1210.6651 [astro-ph.CO]}}.

\bibitem{deRham:2012fg}
C.~de~Rham, A.~Matas, and A.~J. Tolley, ``{Galileon Radiation from Binary
  Systems},'' \href{http://dx.doi.org/10.1103/PhysRevD.87.064024}{{\em
  Phys.Rev.} {\bf D87} (2013)  064024},
\href{http://arxiv.org/abs/1212.5212}{{\tt arXiv:1212.5212 [hep-th]}}.

\bibitem{Fairlie:1991qe}
D.~Fairlie, J.~Govaerts, and A.~Morozov, ``{Universal field equations with
  covariant solutions},''
  \href{http://dx.doi.org/10.1016/0550-3213(92)90455-K}{{\em Nucl.Phys.} {\bf
  B373} (1992)  214--232},
\href{http://arxiv.org/abs/hep-th/9110022}{{\tt arXiv:hep-th/9110022
  [hep-th]}}.

\bibitem{Fairlie:1992nb}
D.~Fairlie and J.~Govaerts, ``{Euler hierarchies and universal equations},''
  \href{http://dx.doi.org/10.1063/1.529904}{{\em J.Math.Phys.} {\bf 33} (1992)
  3543--3566},
\href{http://arxiv.org/abs/hep-th/9204074}{{\tt arXiv:hep-th/9204074
  [hep-th]}}.

\bibitem{Fairlie:1992he}
D.~Fairlie and J.~Govaerts, ``{Universal field equations with reparametrization
  invariance},'' \href{http://dx.doi.org/10.1016/0370-2693(92)90273-7}{{\em
  Phys.Lett.} {\bf B281} (1992)  49--53},
\href{http://arxiv.org/abs/hep-th/9202056}{{\tt arXiv:hep-th/9202056
  [hep-th]}}.

\bibitem{Curtright:2012gx}
T.~L. Curtright and D.~B. Fairlie, ``{A Galileon Primer},''
\href{http://arxiv.org/abs/1212.6972}{{\tt arXiv:1212.6972 [hep-th]}}.

\bibitem{Ivanov:1975zq}
E.~Ivanov and V.~Ogievetsky, ``{The Inverse Higgs Phenomenon in Nonlinear
  Realizations},''
{\em Teor.Mat.Fiz.} {\bf 25} (1975)  164--177.

\bibitem{Volkov:1973vd}
D.~V. Volkov, ``{Phenomenological Lagrangians},''
{\em Fiz.Elem.Chast.Atom.Yadra} {\bf 4} (1973)  3--41.

\bibitem{Nielsen:1975hm}
H.~B. Nielsen and S.~Chadha, ``{On How to Count Goldstone Bosons},''
\href{http://dx.doi.org/10.1016/0550-3213(76)90025-0}{{\em Nucl.Phys.} {\bf
  B105} (1976)  445}.

\bibitem{Low:2001bw}
I.~Low and A.~V. Manohar, ``{Spontaneously broken space-time symmetries and
  Goldstone's theorem},''
  \href{http://dx.doi.org/10.1103/PhysRevLett.88.101602}{{\em Phys.Rev.Lett.}
  {\bf 88} (2002)  101602},
\href{http://arxiv.org/abs/hep-th/0110285}{{\tt arXiv:hep-th/0110285
  [hep-th]}}.

\bibitem{McArthur:2010zm}
I.~McArthur, ``{Nonlinear realizations of symmetries and unphysical Goldstone
  bosons},'' \href{http://dx.doi.org/10.1007/JHEP11(2010)140}{{\em JHEP} {\bf
  1011} (2010)  140},
\href{http://arxiv.org/abs/1009.3696}{{\tt arXiv:1009.3696 [hep-th]}}.

\bibitem{Hidaka:2012ym}
Y.~Hidaka, ``{Counting rule for Nambu-Goldstone modes in nonrelativistic
  systems},'' \href{http://dx.doi.org/10.1103/PhysRevLett.110.091601}{{\em
  Phys.Rev.Lett.} {\bf 110} (2013)  091601},
\href{http://arxiv.org/abs/1203.1494}{{\tt arXiv:1203.1494 [hep-th]}}.

\bibitem{Watanabe:2012hr}
H.~Watanabe and H.~Murayama, ``{Unified Description of Nambu-Goldstone Bosons
  without Lorentz Invariance},''
  \href{http://dx.doi.org/10.1103/PhysRevLett.108.251602}{{\em Phys.Rev.Lett.}
  {\bf 108} (2012)  251602},
\href{http://arxiv.org/abs/1203.0609}{{\tt arXiv:1203.0609 [hep-th]}}.

\bibitem{Nicolis:2012vf}
A.~Nicolis and F.~Piazza, ``{A relativistic non-relativistic Goldstone theorem:
  gapped Goldstones at finite charge density},''
  \href{http://dx.doi.org/10.1103/PhysRevLett.110.011602,
  10.1103/PhysRevLett.110.039901}{{\em Phys.Rev.Lett.} {\bf 110} (2013)
  011602},
\href{http://arxiv.org/abs/1204.1570}{{\tt arXiv:1204.1570 [hep-th]}}.

\bibitem{Nicolis:2013sga}
A.~Nicolis, R.~Penco, F.~Piazza, and R.~A. Rosen, ``{More on gapped Goldstones
  at finite density: More gapped Goldstones},''
\href{http://arxiv.org/abs/1306.1240}{{\tt arXiv:1306.1240 [hep-th]}}.

\bibitem{Coleman:1969sm}
S.~R. Coleman, J.~Wess, and B.~Zumino, ``{Structure of phenomenological
  Lagrangians. 1.},''
\href{http://dx.doi.org/10.1103/PhysRev.177.2239}{{\em Phys.Rev.} {\bf 177}
  (1969)  2239--2247}.

\bibitem{Callan:1969sn}
J.~Callan, Curtis~G., S.~R. Coleman, J.~Wess, and B.~Zumino, ``{Structure of
  phenomenological Lagrangians. 2.},''
\href{http://dx.doi.org/10.1103/PhysRev.177.2247}{{\em Phys.Rev.} {\bf 177}
  (1969)  2247--2250}.

\bibitem{Wess:1971yu}
J.~Wess and B.~Zumino, ``{Consequences of anomalous Ward identities},''
\href{http://dx.doi.org/10.1016/0370-2693(71)90582-X}{{\em Phys.Lett.} {\bf
  B37} (1971)  95}.

\bibitem{Witten:1983tw}
E.~Witten, ``{Global Aspects of Current Algebra},''
\href{http://dx.doi.org/10.1016/0550-3213(83)90063-9}{{\em Nucl.Phys.} {\bf
  B223} (1983)  422--432}.

\bibitem{D'Hoker:1994ti}
E.~D'Hoker and S.~Weinberg, ``{General effective actions},''
  \href{http://dx.doi.org/10.1103/PhysRevD.50.R6050}{{\em Phys.Rev.} {\bf D50}
  (1994)  6050--6053},
\href{http://arxiv.org/abs/hep-ph/9409402}{{\tt arXiv:hep-ph/9409402
  [hep-ph]}}.

\bibitem{Chevalley:1948zz}
C.~Chevalley and S.~Eilenberg, ``{Cohomology Theory of Lie Groups and Lie
  Algebras},''
{\em Trans.Am.Math.Soc.} {\bf 63} (1948)  85--124.

\bibitem{deAzcarraga:1995jw}
J.~A. de~Azcarraga and J.~M. Izquierdo,
``{Lie groups, Lie algebras, cohomology and some applications in physics},''.

\bibitem{deAzcarraga:1998uy}
J.~de~Azcarraga, J.~Izquierdo, and J.~Perez~Bueno, ``{An Introduction to some
  novel applications of Lie algebra cohomology in mathematics and physics},''
  {\em Rev.R.Acad.Cien.Exactas Fis.Nat.Ser.A Mat.} {\bf 95} (2001)  225--248,
\href{http://arxiv.org/abs/physics/9803046}{{\tt arXiv:physics/9803046
  [physics]}}.

\bibitem{deAzcarraga:1997gn}
J.~de~Azcarraga, A.~Macfarlane, and J.~Perez~Bueno, ``{Effective actions,
  relative cohomology and Chern Simons forms},''
  \href{http://dx.doi.org/10.1016/S0370-2693(97)01434-2}{{\em Phys.Lett.} {\bf
  B419} (1998)  186--194},
\href{http://arxiv.org/abs/hep-th/9711064}{{\tt arXiv:hep-th/9711064
  [hep-th]}}.

\bibitem{Fasiello:2013woa}
M.~Fasiello and A.~J. Tolley, ``{Cosmological Stability Bound in Massive
  Gravity and Bigravity},''
\href{http://arxiv.org/abs/1308.1647}{{\tt arXiv:1308.1647 [hep-th]}}.

\bibitem{Kampf:2014rka}
K.~Kampf and J.~Novotny, ``{Unification of Galileon Dualities},''
\href{http://arxiv.org/abs/1403.6813}{{\tt arXiv:1403.6813 [hep-th]}}.

\bibitem{Bellucci:2002ji}
S.~Bellucci, E.~Ivanov, and S.~Krivonos, ``{AdS / CFT equivalence
  transformation},'' \href{http://dx.doi.org/10.1103/PhysRevD.66.086001,
  10.1103/PhysRevD.67.049901}{{\em Phys.Rev.} {\bf D66} (2002)  086001},
\href{http://arxiv.org/abs/hep-th/0206126}{{\tt arXiv:hep-th/0206126
  [hep-th]}}.

\bibitem{Creminelli:2013fxa}
P.~Creminelli, M.~Serone, and E.~Trincherini, ``{Non-linear Representations of
  the Conformal Group and Mapping of Galileons},''
  \href{http://dx.doi.org/10.1007/JHEP10(2013)040}{{\em JHEP} {\bf 1310} (2013)
   040},
\href{http://arxiv.org/abs/1306.2946}{{\tt arXiv:1306.2946 [hep-th]}}.

\bibitem{Creminelli:2014zxa}
P.~Creminelli, M.~Serone, G.~Trevisan, and E.~Trincherini, ``{Inequivalence of
  Coset Constructions for Spacetime Symmetries},''
\href{http://arxiv.org/abs/1403.3095}{{\tt arXiv:1403.3095 [hep-th]}}.

\bibitem{Boulware:1973my}
D.~Boulware and S.~Deser, ``{Can gravitation have a finite range?},''
\href{http://dx.doi.org/10.1103/PhysRevD.6.3368}{{\em Phys.Rev.} {\bf D6}
  (1972)  3368--3382}.

\bibitem{Gabadadze:2009ja}
G.~Gabadadze, ``{General Relativity With An Auxiliary Dimension},''
  \href{http://dx.doi.org/10.1016/j.physletb.2009.10.002}{{\em Phys.Lett.} {\bf
  B681} (2009)  89--95},
\href{http://arxiv.org/abs/0908.1112}{{\tt arXiv:0908.1112 [hep-th]}}.

\bibitem{deRham:2009rm}
C.~de~Rham, ``{Massive gravity from Dirichlet boundary conditions},''
  \href{http://dx.doi.org/10.1016/j.physletb.2010.04.005}{{\em Phys.Lett.} {\bf
  B688} (2010)  137--141},
\href{http://arxiv.org/abs/0910.5474}{{\tt arXiv:0910.5474 [hep-th]}}.

\bibitem{deRham:2010gu}
C.~de~Rham and G.~Gabadadze, ``{Selftuned Massive Spin-2},''
  \href{http://dx.doi.org/10.1016/j.physletb.2010.08.043}{{\em Phys.Lett.} {\bf
  B693} (2010)  334--338},
\href{http://arxiv.org/abs/1006.4367}{{\tt arXiv:1006.4367 [hep-th]}}.

\bibitem{Hassan:2011zr}
S.~Hassan and R.~A. Rosen, ``{Exact Solution to the 'Auxiliary Extra Dimension'
  Model of Massive Gravity},''
  \href{http://dx.doi.org/10.1016/j.physletb.2011.06.056}{{\em Phys.Lett.} {\bf
  B702} (2011)  90--93},
\href{http://arxiv.org/abs/1104.1373}{{\tt arXiv:1104.1373 [hep-th]}}.

\bibitem{Berezhiani:2011nc}
L.~Berezhiani and M.~Mirbabayi, ``{Generalized Framework for Auxiliary Extra
  Dimensions},'' \href{http://dx.doi.org/10.1016/j.physletb.2011.06.036}{{\em
  Phys.Lett.} {\bf B701} (2011)  654--659},
\href{http://arxiv.org/abs/1104.5279}{{\tt arXiv:1104.5279 [hep-th]}}.

\bibitem{deRham:2012ew}
C.~de~Rham, G.~Gabadadze, L.~Heisenberg, and D.~Pirtskhalava,
  ``{Non-Renormalization and Naturalness in a Class of Scalar-Tensor
  Theories},''
\href{http://arxiv.org/abs/1212.4128}{{\tt arXiv:1212.4128 [hep-th]}}.

\bibitem{deRham:2013qqa}
C.~de~Rham, L.~Heisenberg, and R.~H. Ribeiro, ``{Quantum Corrections in Massive
  Gravity},'' \href{http://dx.doi.org/10.1103/PhysRevD.88.084058}{{\em
  Phys.Rev.} {\bf D88} (2013)  084058},
\href{http://arxiv.org/abs/1307.7169}{{\tt arXiv:1307.7169 [hep-th]}}.

\bibitem{Fierz:1939ix}
M.~Fierz and W.~Pauli, ``{On relativistic wave equations for particles of
  arbitrary spin in an electromagnetic field},''
\href{http://dx.doi.org/10.1098/rspa.1939.0140}{{\em Proc.Roy.Soc.Lond.} {\bf
  A173} (1939)  211--232}.

\bibitem{Hassan:2011hr}
S.~Hassan and R.~A. Rosen, ``{Resolving the Ghost Problem in non-Linear Massive
  Gravity},'' \href{http://dx.doi.org/10.1103/PhysRevLett.108.041101}{{\em
  Phys.Rev.Lett.} {\bf 108} (2012)  041101},
\href{http://arxiv.org/abs/1106.3344}{{\tt arXiv:1106.3344 [hep-th]}}.

\bibitem{deRham:2010tw}
C.~de~Rham, G.~Gabadadze, L.~Heisenberg, and D.~Pirtskhalava, ``{Cosmic
  Acceleration and the Helicity-0 Graviton},''
  \href{http://dx.doi.org/10.1103/PhysRevD.83.103516}{{\em Phys.Rev.} {\bf D83}
  (2011)  103516},
\href{http://arxiv.org/abs/1010.1780}{{\tt arXiv:1010.1780 [hep-th]}}.

\bibitem{D'Amico:2011jj}
G.~D'Amico, C.~de~Rham, S.~Dubovsky, G.~Gabadadze, D.~Pirtskhalava, {\em et
  al.}, ``{Massive Cosmologies},''
  \href{http://dx.doi.org/10.1103/PhysRevD.84.124046}{{\em Phys.Rev.} {\bf D84}
  (2011)  124046},
\href{http://arxiv.org/abs/1108.5231}{{\tt arXiv:1108.5231 [hep-th]}}.

\bibitem{Koyama:2011yg}
K.~Koyama, G.~Niz, and G.~Tasinato, ``{Strong interactions and exact solutions
  in non-linear massive gravity},''
  \href{http://dx.doi.org/10.1103/PhysRevD.84.064033}{{\em Phys.Rev.} {\bf D84}
  (2011)  064033},
\href{http://arxiv.org/abs/1104.2143}{{\tt arXiv:1104.2143 [hep-th]}}.

\bibitem{Nieuwenhuizen:2011sq}
T.~Nieuwenhuizen, ``{Exact Schwarzschild-de Sitter black holes in a family of
  massive gravity models},''
  \href{http://dx.doi.org/10.1103/PhysRevD.84.024038}{{\em Phys.Rev.} {\bf D84}
  (2011)  024038},
\href{http://arxiv.org/abs/1103.5912}{{\tt arXiv:1103.5912 [gr-qc]}}.

\bibitem{Chamseddine:2011bu}
A.~H. Chamseddine and M.~S. Volkov, ``{Cosmological solutions with massive
  gravitons},'' \href{http://dx.doi.org/10.1016/j.physletb.2011.09.085}{{\em
  Phys.Lett.} {\bf B704} (2011)  652--654},
\href{http://arxiv.org/abs/1107.5504}{{\tt arXiv:1107.5504 [hep-th]}}.

\bibitem{Gumrukcuoglu:2011ew}
A.~E. Gumrukcuoglu, C.~Lin, and S.~Mukohyama, ``{Open FRW universes and
  self-acceleration from nonlinear massive gravity},''
  \href{http://dx.doi.org/10.1088/1475-7516/2011/11/030}{{\em JCAP} {\bf 1111}
  (2011)  030},
\href{http://arxiv.org/abs/1109.3845}{{\tt arXiv:1109.3845 [hep-th]}}.

\bibitem{Gratia:2012wt}
P.~Gratia, W.~Hu, and M.~Wyman, ``{Self-accelerating Massive Gravity: Exact
  solutions for any isotropic matter distribution},''
  \href{http://dx.doi.org/10.1103/PhysRevD.86.061504}{{\em Phys.Rev.} {\bf D86}
  (2012)  061504},
\href{http://arxiv.org/abs/1205.4241}{{\tt arXiv:1205.4241 [hep-th]}}.

\bibitem{Kobayashi:2012fz}
T.~Kobayashi, M.~Siino, M.~Yamaguchi, and D.~Yoshida, ``{New Cosmological
  Solutions in Massive Gravity},''
  \href{http://dx.doi.org/10.1103/PhysRevD.86.061505}{{\em Phys.Rev.} {\bf D86}
  (2012)  061505},
\href{http://arxiv.org/abs/1205.4938}{{\tt arXiv:1205.4938 [hep-th]}}.

\bibitem{Gumrukcuoglu:2012aa}
A.~E. Gumrukcuoglu, C.~Lin, and S.~Mukohyama, ``{Anisotropic
  Friedmann-Robertson-Walker universe from nonlinear massive gravity},''
  \href{http://dx.doi.org/10.1016/j.physletb.2012.09.049}{{\em Phys.Lett.} {\bf
  B717} (2012)  295--298},
\href{http://arxiv.org/abs/1206.2723}{{\tt arXiv:1206.2723 [hep-th]}}.

\bibitem{Langlois:2012hk}
D.~Langlois and A.~Naruko, ``{Cosmological solutions of massive gravity on de
  Sitter},'' \href{http://dx.doi.org/10.1088/0264-9381/29/20/202001}{{\em
  Class.Quant.Grav.} {\bf 29} (2012)  202001},
\href{http://arxiv.org/abs/1206.6810}{{\tt arXiv:1206.6810 [hep-th]}}.

\bibitem{Motohashi:2012jd}
H.~Motohashi and T.~Suyama, ``{Self-accelerating Solutions in Massive Gravity
  on Isotropic Reference Metric},''
  \href{http://dx.doi.org/10.1103/PhysRevD.86.081502}{{\em Phys.Rev.} {\bf D86}
  (2012)  081502},
\href{http://arxiv.org/abs/1208.3019}{{\tt arXiv:1208.3019 [hep-th]}}.

\bibitem{Maeda:2013bha}
K.-i. Maeda and M.~S. Volkov, ``{Anisotropic universes in the ghost-free
  bigravity},'' \href{http://dx.doi.org/10.1103/PhysRevD.87.104009}{{\em
  Phys.Rev.} {\bf D87} (2013)  104009},
\href{http://arxiv.org/abs/1302.6198}{{\tt arXiv:1302.6198 [hep-th]}}.

\bibitem{Berezhiani:2011mt}
L.~Berezhiani, G.~Chkareuli, C.~de~Rham, G.~Gabadadze, and A.~Tolley, ``{On
  Black Holes in Massive Gravity},''
  \href{http://dx.doi.org/10.1103/PhysRevD.85.044024}{{\em Phys.Rev.} {\bf D85}
  (2012)  044024},
\href{http://arxiv.org/abs/1111.3613}{{\tt arXiv:1111.3613 [hep-th]}}.

\bibitem{Deffayet:2011rh}
C.~Deffayet and T.~Jacobson, ``{On horizon structure of bimetric spacetimes},''
  \href{http://dx.doi.org/10.1088/0264-9381/29/6/065009}{{\em
  Class.Quant.Grav.} {\bf 29} (2012)  065009},
\href{http://arxiv.org/abs/1107.4978}{{\tt arXiv:1107.4978 [gr-qc]}}.

\bibitem{Volkov:2012wp}
M.~S. Volkov, ``{Hairy black holes in the ghost-free bigravity theory},''
  \href{http://dx.doi.org/10.1103/PhysRevD.85.124043}{{\em Phys.Rev.} {\bf D85}
  (2012)  124043},
\href{http://arxiv.org/abs/1202.6682}{{\tt arXiv:1202.6682 [hep-th]}}.

\bibitem{Mirbabayi:2013sva}
M.~Mirbabayi and A.~Gruzinov, ``{Black hole discharge in massive
  electrodynamics and black hole disappearance in massive gravity},''
  \href{http://dx.doi.org/10.1103/PhysRevD.88.064008}{{\em Phys.Rev.} {\bf D88}
  (2013)  064008},
\href{http://arxiv.org/abs/1303.2665}{{\tt arXiv:1303.2665 [hep-th]}}.

\bibitem{Brito:2013wya}
R.~Brito, V.~Cardoso, and P.~Pani, ``{Massive spin-2 fields on black hole
  spacetimes: Instability of the Schwarzschild and Kerr solutions and bounds on
  the graviton mass},''
  \href{http://dx.doi.org/10.1103/PhysRevD.88.023514}{{\em Phys.Rev.} {\bf D88}
  (2013) no.~2, 023514},
\href{http://arxiv.org/abs/1304.6725}{{\tt arXiv:1304.6725 [gr-qc]}}.

\bibitem{Babichev:2013una}
E.~Babichev and A.~Fabbri, ``{Instability of black holes in massive gravity},''
  \href{http://dx.doi.org/10.1088/0264-9381/30/15/152001}{{\em
  Class.Quant.Grav.} {\bf 30} (2013)  152001},
\href{http://arxiv.org/abs/1304.5992}{{\tt arXiv:1304.5992 [gr-qc]}}.

\bibitem{Zumino:1970tu}
B.~Zumino, ``{Effective Lagrangians and broken symmetries},''
{\em Brandeis Univ. 1970, Lectures On Elementary Particles And Quantum Field
  Theory} {\bf 2} (1970)  437--500.

\bibitem{Chamseddine:2011mu}
A.~H. Chamseddine and V.~Mukhanov, ``{Massive Gravity Simplified: A Quadratic
  Action},'' \href{http://dx.doi.org/10.1007/JHEP08(2011)091}{{\em JHEP} {\bf
  1108} (2011)  091},
\href{http://arxiv.org/abs/1106.5868}{{\tt arXiv:1106.5868 [hep-th]}}.

\bibitem{Hinterbichler:2012cn}
K.~Hinterbichler and R.~A. Rosen, ``{Interacting Spin-2 Fields},''
  \href{http://dx.doi.org/10.1007/JHEP07(2012)047}{{\em JHEP} {\bf 1207} (2012)
   047},
\href{http://arxiv.org/abs/1203.5783}{{\tt arXiv:1203.5783 [hep-th]}}.

\bibitem{Deffayet:2012nr}
C.~Deffayet, J.~Mourad, and G.~Zahariade, ``{Covariant constraints in ghost
  free massive gravity},''
  \href{http://dx.doi.org/10.1088/1475-7516/2013/01/032}{{\em JCAP} {\bf 1301}
  (2013)  032},
\href{http://arxiv.org/abs/1207.6338}{{\tt arXiv:1207.6338 [hep-th]}}.

\bibitem{Deffayet:2012zc}
C.~Deffayet, J.~Mourad, and G.~Zahariade, ``{A note on 'symmetric' vielbeins in
  bimetric, massive, perturbative and non perturbative gravities},''
  \href{http://dx.doi.org/10.1007/JHEP03(2013)086}{{\em JHEP} {\bf 1303} (2013)
   086},
\href{http://arxiv.org/abs/1208.4493}{{\tt arXiv:1208.4493 [gr-qc]}}.

\bibitem{Hassan:2011zd}
S.~Hassan and R.~A. Rosen, ``{Bimetric Gravity from Ghost-free Massive
  Gravity},'' \href{http://dx.doi.org/10.1007/JHEP02(2012)126}{{\em JHEP} {\bf
  1202} (2012)  126},
\href{http://arxiv.org/abs/1109.3515}{{\tt arXiv:1109.3515 [hep-th]}}.

\bibitem{Rubakov:2004eb}
V.~Rubakov, ``{Lorentz-violating graviton masses: Getting around ghosts, low
  strong coupling scale and VDVZ discontinuity},''
\href{http://arxiv.org/abs/hep-th/0407104}{{\tt arXiv:hep-th/0407104
  [hep-th]}}.

\bibitem{Dubovsky:2004sg}
S.~Dubovsky, ``{Phases of massive gravity},''
  \href{http://dx.doi.org/10.1088/1126-6708/2004/10/076}{{\em JHEP} {\bf 0410}
  (2004)  076},
\href{http://arxiv.org/abs/hep-th/0409124}{{\tt arXiv:hep-th/0409124
  [hep-th]}}.

\bibitem{Dubovsky:2004ud}
S.~Dubovsky, P.~Tinyakov, and I.~Tkachev, ``{Massive graviton as a testable
  cold dark matter candidate},''
  \href{http://dx.doi.org/10.1103/PhysRevLett.94.181102}{{\em Phys.Rev.Lett.}
  {\bf 94} (2005)  181102},
\href{http://arxiv.org/abs/hep-th/0411158}{{\tt arXiv:hep-th/0411158
  [hep-th]}}.

\bibitem{Gabadadze:2004iv}
G.~Gabadadze and L.~Grisa, ``{Lorentz-violating massive gauge and gravitational
  fields},'' \href{http://dx.doi.org/10.1016/j.physletb.2005.04.064}{{\em
  Phys.Lett.} {\bf B617} (2005)  124--132},
\href{http://arxiv.org/abs/hep-th/0412332}{{\tt arXiv:hep-th/0412332
  [hep-th]}}.

\bibitem{Kirsch:2005st}
I.~Kirsch, ``{A Higgs mechanism for gravity},''
  \href{http://dx.doi.org/10.1103/PhysRevD.72.024001}{{\em Phys.Rev.} {\bf D72}
  (2005)  024001},
\href{http://arxiv.org/abs/hep-th/0503024}{{\tt arXiv:hep-th/0503024
  [hep-th]}}.

\bibitem{Libanov:2005vu}
M.~Libanov and V.~Rubakov, ``{More about spontaneous Lorentz-violation and
  infrared modification of gravity},''
  \href{http://dx.doi.org/10.1088/1126-6708/2005/08/001}{{\em JHEP} {\bf 0508}
  (2005)  001},
\href{http://arxiv.org/abs/hep-th/0505231}{{\tt arXiv:hep-th/0505231
  [hep-th]}}.

\bibitem{ArkaniHamed:2005gu}
N.~Arkani-Hamed, H.-C. Cheng, M.~A. Luty, S.~Mukohyama, and T.~Wiseman,
  ``{Dynamics of gravity in a Higgs phase},''
  \href{http://dx.doi.org/10.1088/1126-6708/2007/01/036}{{\em JHEP} {\bf 0701}
  (2007)  036},
\href{http://arxiv.org/abs/hep-ph/0507120}{{\tt arXiv:hep-ph/0507120
  [hep-ph]}}.

\bibitem{Cheng:2006us}
H.-C. Cheng, M.~A. Luty, S.~Mukohyama, and J.~Thaler, ``{Spontaneous Lorentz
  breaking at high energies},''
  \href{http://dx.doi.org/10.1088/1126-6708/2006/05/076}{{\em JHEP} {\bf 0605}
  (2006)  076},
\href{http://arxiv.org/abs/hep-th/0603010}{{\tt arXiv:hep-th/0603010
  [hep-th]}}.

\bibitem{Berezhiani:2007zf}
Z.~Berezhiani, D.~Comelli, F.~Nesti, and L.~Pilo, ``{Spontaneous Lorentz
  Breaking and Massive Gravity},''
  \href{http://dx.doi.org/10.1103/PhysRevLett.99.131101}{{\em Phys.Rev.Lett.}
  {\bf 99} (2007)  131101},
\href{http://arxiv.org/abs/hep-th/0703264}{{\tt arXiv:hep-th/0703264
  [HEP-TH]}}.

\bibitem{Blas:2007ep}
D.~Blas, C.~Deffayet, and J.~Garriga, ``{Bigravity and Lorentz-violating
  Massive Gravity},'' \href{http://dx.doi.org/10.1103/PhysRevD.76.104036}{{\em
  Phys.Rev.} {\bf D76} (2007)  104036},
\href{http://arxiv.org/abs/0705.1982}{{\tt arXiv:0705.1982 [hep-th]}}.

\bibitem{Dubovsky:2007zi}
S.~Dubovsky, P.~Tinyakov, and M.~Zaldarriaga, ``{Bumpy black holes from
  spontaneous Lorentz violation},''
  \href{http://dx.doi.org/10.1088/1126-6708/2007/11/083}{{\em JHEP} {\bf 0711}
  (2007)  083},
\href{http://arxiv.org/abs/0706.0288}{{\tt arXiv:0706.0288 [hep-th]}}.

\bibitem{Rubakov:2008nh}
V.~Rubakov and P.~Tinyakov, ``{Infrared-modified gravities and massive
  gravitons},'' \href{http://dx.doi.org/10.1070/PU2008v051n08ABEH006600}{{\em
  Phys.Usp.} {\bf 51} (2008)  759--792},
\href{http://arxiv.org/abs/0802.4379}{{\tt arXiv:0802.4379 [hep-th]}}.

\bibitem{Grisa:2008um}
L.~Grisa, ``{Lorentz-Violating Massive Gravity in Curved Space},''
  \href{http://dx.doi.org/10.1088/1126-6708/2008/11/023}{{\em JHEP} {\bf 0811}
  (2008)  023},
\href{http://arxiv.org/abs/0803.1137}{{\tt arXiv:0803.1137 [hep-th]}}.

\bibitem{Blas:2009my}
D.~Blas, D.~Comelli, F.~Nesti, and L.~Pilo, ``{Lorentz Breaking Massive Gravity
  in Curved Space},'' \href{http://dx.doi.org/10.1103/PhysRevD.80.044025}{{\em
  Phys.Rev.} {\bf D80} (2009)  044025},
\href{http://arxiv.org/abs/0905.1699}{{\tt arXiv:0905.1699 [hep-th]}}.

\bibitem{Lin:2013aha}
C.~Lin, ``{SO(3) massive gravity},''
  \href{http://dx.doi.org/10.1016/j.physletb.2013.10.031}{{\em Phys.Lett.} {\bf
  B727} (2013)  31--36},
\href{http://arxiv.org/abs/1305.2069}{{\tt arXiv:1305.2069 [hep-th]}}.

\bibitem{Langlois:2014jba}
D.~Langlois, S.~Mukohyama, R.~Namba, and A.~Naruko, ``{Cosmology in
  rotation-invariant massive gravity with non-trivial fiducial metric},''
\href{http://arxiv.org/abs/1405.0358}{{\tt arXiv:1405.0358 [hep-th]}}.

\bibitem{Higuchi:1986py}
A.~Higuchi, ``{Forbidden Mass Range for Spin-2 Field Theory in De Sitter
  Space-time},''
\href{http://dx.doi.org/10.1016/0550-3213(87)90691-2}{{\em Nucl.Phys.} {\bf
  B282} (1987)  397}.

\bibitem{Kogan:2000uy}
I.~I. Kogan, S.~Mouslopoulos, and A.~Papazoglou, ``{The m > 0 limit for massive
  graviton in dS(4) and AdS(4): How to circumvent the van Dam-Veltman-Zakharov
  discontinuity},'' \href{http://dx.doi.org/10.1016/S0370-2693(01)00209-X}{{\em
  Phys.Lett.} {\bf B503} (2001)  173--180},
\href{http://arxiv.org/abs/hep-th/0011138}{{\tt arXiv:hep-th/0011138
  [hep-th]}}.

\bibitem{Porrati:2000cp}
M.~Porrati, ``{No van Dam-Veltman-Zakharov discontinuity in AdS space},''
  \href{http://dx.doi.org/10.1016/S0370-2693(00)01380-0}{{\em Phys.Lett.} {\bf
  B498} (2001)  92--96},
\href{http://arxiv.org/abs/hep-th/0011152}{{\tt arXiv:hep-th/0011152
  [hep-th]}}.

\bibitem{Folkerts:2011ev}
S.~Folkerts, A.~Pritzel, and N.~Wintergerst, ``{On ghosts in theories of
  self-interacting massive spin-2 particles},''
\href{http://arxiv.org/abs/1107.3157}{{\tt arXiv:1107.3157 [hep-th]}}.

\bibitem{Hinterbichler:2013eza}
K.~Hinterbichler, ``{Ghost-Free Derivative Interactions for a Massive
  Graviton},''
\href{http://arxiv.org/abs/1305.7227}{{\tt arXiv:1305.7227 [hep-th]}}.

\bibitem{Zinoviev:2013hac}
Y.~Zinoviev, ``{All spin-2 cubic vertices with two derivatives},''
  \href{http://dx.doi.org/10.1016/j.nuclphysb.2013.03.013}{{\em Nucl.Phys.}
  {\bf B872} (2013)  21--37},
\href{http://arxiv.org/abs/1302.1983}{{\tt arXiv:1302.1983 [hep-th]}}.

\bibitem{Gao:2014jja}
X.~Gao, ``{On derivative interactions for a spin-2 field at cubic order},''
\href{http://arxiv.org/abs/1403.6781}{{\tt arXiv:1403.6781 [hep-th]}}.

\bibitem{Kimura:2013ika}
R.~Kimura and D.~Yamauchi, ``{Derivative interactions in de
  Rham-Gabadadze-Tolley massive gravity},''
  \href{http://dx.doi.org/10.1103/PhysRevD.88.084025}{{\em Phys.Rev.} {\bf D88}
  (2013)  084025},
\href{http://arxiv.org/abs/1308.0523}{{\tt arXiv:1308.0523 [gr-qc]}}.

\bibitem{deRham:2013tfa}
C.~de~Rham, A.~Matas, and A.~J. Tolley, ``{New Kinetic Interactions for Massive
  Gravity?},''
\href{http://arxiv.org/abs/1311.6485}{{\tt arXiv:1311.6485 [hep-th]}}.

\bibitem{Koyama:2011wx}
K.~Koyama, G.~Niz, and G.~Tasinato, ``{The Self-Accelerating Universe with
  Vectors in Massive Gravity},''
  \href{http://dx.doi.org/10.1007/JHEP12(2011)065}{{\em JHEP} {\bf 1112} (2011)
   065},
\href{http://arxiv.org/abs/1110.2618}{{\tt arXiv:1110.2618 [hep-th]}}.

\bibitem{Gabadadze:2013ria}
G.~Gabadadze, K.~Hinterbichler, D.~Pirtskhalava, and Y.~Shang, ``{On the
  Potential for General Relativity and its Geometry},''
  \href{http://dx.doi.org/10.1103/PhysRevD.88.084003}{{\em Phys.Rev.} {\bf D88}
  (2013)  084003},
\href{http://arxiv.org/abs/1307.2245}{{\tt arXiv:1307.2245}}.

\bibitem{Ondo:2013wka}
N.~A. Ondo and A.~J. Tolley, ``{Complete Decoupling Limit of Ghost-free Massive
  Gravity},'' \href{http://dx.doi.org/10.1007/JHEP11(2013)059}{{\em JHEP} {\bf
  1311} (2013)  059},
\href{http://arxiv.org/abs/1307.4769}{{\tt arXiv:1307.4769 [hep-th]}}.

\bibitem{Hill:2000mu}
C.~T. Hill, S.~Pokorski, and J.~Wang, ``{Gauge invariant effective Lagrangian
  for Kaluza-Klein modes},''
  \href{http://dx.doi.org/10.1103/PhysRevD.64.105005}{{\em Phys.Rev.} {\bf D64}
  (2001)  105005},
\href{http://arxiv.org/abs/hep-th/0104035}{{\tt arXiv:hep-th/0104035
  [hep-th]}}.

\bibitem{ArkaniHamed:2001ca}
N.~Arkani-Hamed, A.~G. Cohen, and H.~Georgi, ``{(De)constructing dimensions},''
  \href{http://dx.doi.org/10.1103/PhysRevLett.86.4757}{{\em Phys.Rev.Lett.}
  {\bf 86} (2001)  4757--4761},
\href{http://arxiv.org/abs/hep-th/0104005}{{\tt arXiv:hep-th/0104005
  [hep-th]}}.

\bibitem{deRham:2013awa}
C.~de~Rham, A.~Matas, and A.~J. Tolley, ``{Deconstructing Dimensions and
  Massive Gravity},''
  \href{http://dx.doi.org/10.1088/0264-9381/31/2/025004}{{\em
  Class.Quant.Grav.} {\bf 31} (2014)  025004},
\href{http://arxiv.org/abs/1308.4136}{{\tt arXiv:1308.4136 [hep-th]}}.

\bibitem{Gruzinov:2011sq}
A.~Gruzinov, ``{All Fierz-Paulian massive gravity theories have ghosts or
  superluminal modes},''
\href{http://arxiv.org/abs/1106.3972}{{\tt arXiv:1106.3972 [hep-th]}}.

\bibitem{Deser:2012qx}
S.~Deser and A.~Waldron, ``{Acausality of Massive Gravity},''
  \href{http://dx.doi.org/10.1103/PhysRevLett.110.111101}{{\em Phys.Rev.Lett.}
  {\bf 110} (2013) no.~11, 111101},
\href{http://arxiv.org/abs/1212.5835}{{\tt arXiv:1212.5835 [hep-th]}}.

\bibitem{Deser:2013eua}
S.~Deser, K.~Izumi, Y.~Ong, and A.~Waldron, ``{Massive Gravity Acausality
  Redux},'' \href{http://dx.doi.org/10.1016/j.physletb.2013.09.001}{{\em
  Phys.Lett.} {\bf B726} (2013)  544--548},
\href{http://arxiv.org/abs/1306.5457}{{\tt arXiv:1306.5457 [hep-th]}}.

\bibitem{Deser:2013qza}
S.~Deser, K.~Izumi, Y.~Ong, and A.~Waldron, ``{Superluminal Propagation and
  Acausality of Nonlinear Massive Gravity},''
\href{http://arxiv.org/abs/1312.1115}{{\tt arXiv:1312.1115 [hep-th]}}.

\bibitem{Yu:2013owa}
S.~Yu, ``{Superluminal Vector in Ghost-free Massive Gravity},''
\href{http://arxiv.org/abs/1310.6469}{{\tt arXiv:1310.6469 [hep-th]}}.

\bibitem{deRham:2011pt}
C.~de~Rham, G.~Gabadadze, and A.~J. Tolley, ``{Comments on
  (super)luminality},''
\href{http://arxiv.org/abs/1107.0710}{{\tt arXiv:1107.0710 [hep-th]}}.

\bibitem{Hassan:2011vm}
S.~Hassan and R.~A. Rosen, ``{On Non-Linear Actions for Massive Gravity},''
  \href{http://dx.doi.org/10.1007/JHEP07(2011)009}{{\em JHEP} {\bf 1107} (2011)
   009},
\href{http://arxiv.org/abs/1103.6055}{{\tt arXiv:1103.6055 [hep-th]}}.

\bibitem{Hassan:2011ea}
S.~Hassan and R.~A. Rosen, ``{Confirmation of the Secondary Constraint and
  Absence of Ghost in Massive Gravity and Bimetric Gravity},''
  \href{http://dx.doi.org/10.1007/JHEP04(2012)123}{{\em JHEP} {\bf 1204} (2012)
   123},
\href{http://arxiv.org/abs/1111.2070}{{\tt arXiv:1111.2070 [hep-th]}}.

\bibitem{Golovnev:2011aa}
A.~Golovnev, ``{On the Hamiltonian analysis of non-linear massive gravity},''
  \href{http://dx.doi.org/10.1016/j.physletb.2011.12.064}{{\em Phys.Lett.} {\bf
  B707} (2012)  404--408},
\href{http://arxiv.org/abs/1112.2134}{{\tt arXiv:1112.2134 [gr-qc]}}.

\bibitem{Comelli:2012vz}
D.~Comelli, M.~Crisostomi, F.~Nesti, and L.~Pilo, ``{Degrees of Freedom in
  Massive Gravity},'' \href{http://dx.doi.org/10.1103/PhysRevD.86.101502}{{\em
  Phys.Rev.} {\bf D86} (2012)  101502},
\href{http://arxiv.org/abs/1204.1027}{{\tt arXiv:1204.1027 [hep-th]}}.

\bibitem{Kluson:2012wf}
J.~Kluson, ``{Non-Linear Massive Gravity with Additional Primary Constraint and
  Absence of Ghosts},''
  \href{http://dx.doi.org/10.1103/PhysRevD.86.044024}{{\em Phys.Rev.} {\bf D86}
  (2012)  044024},
\href{http://arxiv.org/abs/1204.2957}{{\tt arXiv:1204.2957 [hep-th]}}.

\bibitem{deRham:2011rn}
C.~de~Rham, G.~Gabadadze, and A.~J. Tolley, ``{Ghost free Massive Gravity in
  the St\'uckelberg language},''
  \href{http://dx.doi.org/10.1016/j.physletb.2012.03.081}{{\em Phys.Lett.} {\bf
  B711} (2012)  190--195},
\href{http://arxiv.org/abs/1107.3820}{{\tt arXiv:1107.3820 [hep-th]}}.

\bibitem{Mirbabayi:2011aa}
M.~Mirbabayi, ``{A Proof Of Ghost Freedom In de Rham-Gabadadze-Tolley Massive
  Gravity},'' \href{http://dx.doi.org/10.1103/PhysRevD.86.084006}{{\em
  Phys.Rev.} {\bf D86} (2012)  084006},
\href{http://arxiv.org/abs/1112.1435}{{\tt arXiv:1112.1435 [hep-th]}}.

\bibitem{deRham:2011qq}
C.~de~Rham, G.~Gabadadze, and A.~J. Tolley, ``{Helicity Decomposition of
  Ghost-free Massive Gravity},''
  \href{http://dx.doi.org/10.1007/JHEP11(2011)093}{{\em JHEP} {\bf 1111} (2011)
   093},
\href{http://arxiv.org/abs/1108.4521}{{\tt arXiv:1108.4521 [hep-th]}}.

\bibitem{Hassan:2012qv}
S.~Hassan, A.~Schmidt-May, and M.~von Strauss, ``{Proof of Consistency of
  Nonlinear Massive Gravity in the St\'uckelberg Formulation},''
  \href{http://dx.doi.org/10.1016/j.physletb.2012.07.018}{{\em Phys.Lett.} {\bf
  B715} (2012)  335--339},
\href{http://arxiv.org/abs/1203.5283}{{\tt arXiv:1203.5283 [hep-th]}}.

\bibitem{deRham:2011by}
C.~de~Rham and L.~Heisenberg, ``{Cosmology of the Galileon from Massive
  Gravity},'' \href{http://dx.doi.org/10.1103/PhysRevD.84.043503}{{\em
  Phys.Rev.} {\bf D84} (2011)  043503},
\href{http://arxiv.org/abs/1106.3312}{{\tt arXiv:1106.3312 [hep-th]}}.

\bibitem{Heisenberg:2014kea}
L.~Heisenberg, R.~Kimura, and K.~Yamamoto, ``{Cosmology of the proxy theory to
  massive gravity},'' \href{http://dx.doi.org/10.1103/PhysRevD.89.103008}{{\em
  Phys.Rev.} {\bf D89} (2014)  103008},
\href{http://arxiv.org/abs/1403.2049}{{\tt arXiv:1403.2049 [hep-th]}}.

\bibitem{Tasinato:2012ze}
G.~Tasinato, K.~Koyama, and G.~Niz, ``{Vector instabilities and
  self-acceleration in the decoupling limit of massive gravity},''
  \href{http://dx.doi.org/10.1103/PhysRevD.87.064029}{{\em Phys.Rev.} {\bf D87}
  (2013) no.~6, 064029},
\href{http://arxiv.org/abs/1210.3627}{{\tt arXiv:1210.3627 [hep-th]}}.

\bibitem{Gumrukcuoglu:2011zh}
A.~E. Gumrukcuoglu, C.~Lin, and S.~Mukohyama, ``{Cosmological perturbations of
  self-accelerating universe in nonlinear massive gravity},''
  \href{http://dx.doi.org/10.1088/1475-7516/2012/03/006}{{\em JCAP} {\bf 1203}
  (2012)  006},
\href{http://arxiv.org/abs/1111.4107}{{\tt arXiv:1111.4107 [hep-th]}}.

\bibitem{D'Amico:2012pi}
G.~D'Amico, ``{Cosmology and perturbations in massive gravity},''
  \href{http://dx.doi.org/10.1103/PhysRevD.86.124019}{{\em Phys.Rev.} {\bf D86}
  (2012)  124019},
\href{http://arxiv.org/abs/1206.3617}{{\tt arXiv:1206.3617 [hep-th]}}.

\bibitem{Fasiello:2012rw}
M.~Fasiello and A.~J. Tolley, ``{Cosmological perturbations in Massive Gravity
  and the Higuchi bound},''
  \href{http://dx.doi.org/10.1088/1475-7516/2012/11/035}{{\em JCAP} {\bf 1211}
  (2012)  035},
\href{http://arxiv.org/abs/1206.3852}{{\tt arXiv:1206.3852 [hep-th]}}.

\bibitem{Chiang:2012vh}
C.-I. Chiang, K.~Izumi, and P.~Chen, ``{Spherically symmetric analysis on open
  FLRW solution in non-linear massive gravity},''
  \href{http://dx.doi.org/10.1088/1475-7516/2012/12/025}{{\em JCAP} {\bf 1212}
  (2012)  025},
\href{http://arxiv.org/abs/1208.1222}{{\tt arXiv:1208.1222 [hep-th]}}.

\bibitem{Wyman:2012iw}
M.~Wyman, W.~Hu, and P.~Gratia, ``{Self-accelerating Massive Gravity: Time for
  Field Fluctuations},''
  \href{http://dx.doi.org/10.1103/PhysRevD.87.084046}{{\em Phys.Rev.} {\bf D87}
  (2013) no.~8, 084046},
\href{http://arxiv.org/abs/1211.4576}{{\tt arXiv:1211.4576 [hep-th]}}.

\bibitem{Khosravi:2013axa}
N.~Khosravi, G.~Niz, K.~Koyama, and G.~Tasinato, ``{Stability of the
  Self-accelerating Universe in Massive Gravity},''
  \href{http://dx.doi.org/10.1088/1475-7516/2013/08/044}{{\em JCAP} {\bf 1308}
  (2013)  044},
\href{http://arxiv.org/abs/1305.4950}{{\tt arXiv:1305.4950 [hep-th]}}.

\bibitem{Kuhnel:2012gh}
F.~Kuhnel, ``{Instability of certain bimetric and massive-gravity theories},''
  \href{http://dx.doi.org/10.1103/PhysRevD.88.064024}{{\em Phys.Rev.} {\bf D88}
  (2013) no.~6, 064024},
\href{http://arxiv.org/abs/1208.1764}{{\tt arXiv:1208.1764 [gr-qc]}}.

\bibitem{DeFelice:2012mx}
A.~De~Felice, A.~E. Gumrukcuoglu, and S.~Mukohyama, ``{Massive gravity:
  nonlinear instability of the homogeneous and isotropic universe},''
  \href{http://dx.doi.org/10.1103/PhysRevLett.109.171101}{{\em Phys.Rev.Lett.}
  {\bf 109} (2012)  171101},
\href{http://arxiv.org/abs/1206.2080}{{\tt arXiv:1206.2080 [hep-th]}}.

\bibitem{DeFelice:2013awa}
A.~De~Felice, A.~E. Gumrukcuoglu, C.~Lin, and S.~Mukohyama, ``{Nonlinear
  stability of cosmological solutions in massive gravity},''
  \href{http://dx.doi.org/10.1088/1475-7516/2013/05/035}{{\em JCAP} {\bf 1305}
  (2013)  035},
\href{http://arxiv.org/abs/1303.4154}{{\tt arXiv:1303.4154 [hep-th]}}.

\bibitem{Baccetti:2012bk}
V.~Baccetti, P.~Martin-Moruno, and M.~Visser, ``{Massive gravity from bimetric
  gravity},'' \href{http://dx.doi.org/10.1088/0264-9381/30/1/015004}{{\em
  Class.Quant.Grav.} {\bf 30} (2013)  015004},
\href{http://arxiv.org/abs/1205.2158}{{\tt arXiv:1205.2158 [gr-qc]}}.

\bibitem{Hassan:2012wr}
S.~Hassan, A.~Schmidt-May, and M.~von Strauss, ``{On Consistent Theories of
  Massive Spin-2 Fields Coupled to Gravity},''
  \href{http://dx.doi.org/10.1007/JHEP05(2013)086}{{\em JHEP} {\bf 1305} (2013)
   086},
\href{http://arxiv.org/abs/1208.1515}{{\tt arXiv:1208.1515 [hep-th]}}.

\bibitem{Hassan:2012wt}
S.~Hassan, A.~Schmidt-May, and M.~von Strauss, ``{Metric Formulation of
  Ghost-Free Multivielbein Theory},''
\href{http://arxiv.org/abs/1204.5202}{{\tt arXiv:1204.5202 [hep-th]}}.

\bibitem{Khosravi:2011zi}
N.~Khosravi, N.~Rahmanpour, H.~R. Sepangi, and S.~Shahidi, ``{Multi-Metric
  Gravity via Massive Gravity},''
  \href{http://dx.doi.org/10.1103/PhysRevD.85.024049}{{\em Phys.Rev.} {\bf D85}
  (2012)  024049},
\href{http://arxiv.org/abs/1111.5346}{{\tt arXiv:1111.5346 [hep-th]}}.

\bibitem{Noller:2013yja}
J.~Noller, J.~H.~C. Scargill, and P.~G. Ferreira, ``{Interacting spin-2 fields
  in the StŸckelberg picture},''
  \href{http://dx.doi.org/10.1088/1475-7516/2014/02/007}{{\em JCAP} {\bf 1402}
  (2014)  007},
\href{http://arxiv.org/abs/1311.7009}{{\tt arXiv:1311.7009 [hep-th]}}.

\bibitem{Comelli:2011zm}
D.~Comelli, M.~Crisostomi, F.~Nesti, and L.~Pilo, ``{FRW Cosmology in Ghost
  Free Massive Gravity},'' \href{http://dx.doi.org/10.1007/JHEP06(2012)020,
  10.1007/JHEP03(2012)067}{{\em JHEP} {\bf 1203} (2012)  067},
\href{http://arxiv.org/abs/1111.1983}{{\tt arXiv:1111.1983 [hep-th]}}.

\bibitem{Comelli:2011wq}
D.~Comelli, M.~Crisostomi, F.~Nesti, and L.~Pilo, ``{Spherically Symmetric
  Solutions in Ghost-Free Massive Gravity},''
  \href{http://dx.doi.org/10.1103/PhysRevD.85.024044}{{\em Phys.Rev.} {\bf D85}
  (2012)  024044},
\href{http://arxiv.org/abs/1110.4967}{{\tt arXiv:1110.4967 [hep-th]}}.

\bibitem{vonStrauss:2011mq}
M.~von Strauss, A.~Schmidt-May, J.~Enander, E.~Mortsell, and S.~Hassan,
  ``{Cosmological Solutions in Bimetric Gravity and their Observational
  Tests},'' \href{http://dx.doi.org/10.1088/1475-7516/2012/03/042}{{\em JCAP}
  {\bf 1203} (2012)  042},
\href{http://arxiv.org/abs/1111.1655}{{\tt arXiv:1111.1655 [gr-qc]}}.

\bibitem{Volkov:2011an}
M.~S. Volkov, ``{Cosmological solutions with massive gravitons in the bigravity
  theory},'' \href{http://dx.doi.org/10.1007/JHEP01(2012)035}{{\em JHEP} {\bf
  1201} (2012)  035},
\href{http://arxiv.org/abs/1110.6153}{{\tt arXiv:1110.6153 [hep-th]}}.

\bibitem{Volkov:2012cf}
M.~S. Volkov, ``{Exact self-accelerating cosmologies in the ghost-free
  bigravity and massive gravity},''
  \href{http://dx.doi.org/10.1103/PhysRevD.86.061502}{{\em Phys.Rev.} {\bf D86}
  (2012)  061502},
\href{http://arxiv.org/abs/1205.5713}{{\tt arXiv:1205.5713 [hep-th]}}.

\bibitem{Volkov:2012zb}
M.~S. Volkov, ``{Exact self-accelerating cosmologies in the ghost-free massive
  gravity -- the detailed derivation},''
  \href{http://dx.doi.org/10.1103/PhysRevD.86.104022}{{\em Phys.Rev.} {\bf D86}
  (2012)  104022},
\href{http://arxiv.org/abs/1207.3723}{{\tt arXiv:1207.3723 [hep-th]}}.

\bibitem{Comelli:2012db}
D.~Comelli, M.~Crisostomi, and L.~Pilo, ``{Perturbations in Massive Gravity
  Cosmology},'' \href{http://dx.doi.org/10.1007/JHEP06(2012)085}{{\em JHEP}
  {\bf 1206} (2012)  085},
\href{http://arxiv.org/abs/1202.1986}{{\tt arXiv:1202.1986 [hep-th]}}.

\bibitem{Berg:2012kn}
M.~Berg, I.~Buchberger, J.~Enander, E.~Mortsell, and S.~Sjors, ``{Growth
  Histories in Bimetric Massive Gravity},''
  \href{http://dx.doi.org/10.1088/1475-7516/2012/12/021}{{\em JCAP} {\bf 1212}
  (2012)  021},
\href{http://arxiv.org/abs/1206.3496}{{\tt arXiv:1206.3496 [gr-qc]}}.

\bibitem{Akrami:2012vf}
Y.~Akrami, T.~S. Koivisto, and M.~Sandstad, ``{Accelerated expansion from
  ghost-free bigravity: a statistical analysis with improved generality},''
  \href{http://dx.doi.org/10.1007/JHEP03(2013)099}{{\em JHEP} {\bf 1303} (2013)
   099},
\href{http://arxiv.org/abs/1209.0457}{{\tt arXiv:1209.0457 [astro-ph.CO]}}.

\bibitem{Volkov:2013roa}
M.~S. Volkov, ``{Self-accelerating cosmologies and hairy black holes in
  ghost-free bigravity and massive gravity},''
  \href{http://dx.doi.org/10.1088/0264-9381/30/18/184009}{{\em
  Class.Quant.Grav.} {\bf 30} (2013)  184009},
\href{http://arxiv.org/abs/1304.0238}{{\tt arXiv:1304.0238 [hep-th]}}.

\bibitem{Tamanini:2013xia}
N.~Tamanini, E.~N. Saridakis, and T.~S. Koivisto, ``{The Cosmology of
  Interacting Spin-2 Fields},''
  \href{http://dx.doi.org/10.1088/1475-7516/2014/02/015}{{\em JCAP} {\bf 1402}
  (2014)  015},
\href{http://arxiv.org/abs/1307.5984}{{\tt arXiv:1307.5984 [hep-th]}}.

\bibitem{Akrami:2013ffa}
Y.~Akrami, T.~S. Koivisto, D.~F. Mota, and M.~Sandstad, ``{Bimetric gravity
  doubly coupled to matter: theory and cosmological implications},''
  \href{http://dx.doi.org/10.1088/1475-7516/2013/10/046}{{\em JCAP} {\bf 1310}
  (2013)  046},
\href{http://arxiv.org/abs/1306.0004}{{\tt arXiv:1306.0004 [hep-th]}}.

\bibitem{DeFelice:2014nja}
A.~De~Felice, A.~E. Gumrukcuoglu, S.~Mukohyama, N.~Tanahashi, and T.~Tanaka,
  ``{Viable cosmology in bimetric theory},''
\href{http://arxiv.org/abs/1404.0008}{{\tt arXiv:1404.0008 [hep-th]}}.

\bibitem{Solomon:2014dua}
A.~R. Solomon, Y.~Akrami, and T.~S. Koivisto, ``{Cosmological perturbations in
  massive bigravity: I. Linear growth of structures},''
\href{http://arxiv.org/abs/1404.4061}{{\tt arXiv:1404.4061 [astro-ph.CO]}}.

\bibitem{Huang:2012pe}
Q.-G. Huang, Y.-S. Piao, and S.-Y. Zhou, ``{Mass-Varying Massive Gravity},''
  \href{http://dx.doi.org/10.1103/PhysRevD.86.124014}{{\em Phys.Rev.} {\bf D86}
  (2012)  124014},
\href{http://arxiv.org/abs/1206.5678}{{\tt arXiv:1206.5678 [hep-th]}}.

\bibitem{Hinterbichler:2013dv}
K.~Hinterbichler, J.~Stokes, and M.~Trodden, ``{Cosmologies of extended massive
  gravity},'' \href{http://dx.doi.org/10.1016/j.physletb.2013.07.009}{{\em
  Phys. Lett.} {\bf B725} (2013)  1},
\href{http://arxiv.org/abs/1301.4993}{{\tt arXiv:1301.4993 [astro-ph.CO]}}.

\bibitem{Gumrukcuoglu:2013nza}
A.~E. Gumrukcuoglu, K.~Hinterbichler, C.~Lin, S.~Mukohyama, and M.~Trodden,
  ``{Cosmological Perturbations in Extended Massive Gravity},''
  \href{http://dx.doi.org/10.1103/PhysRevD.88.024023}{{\em Phys.Rev.} {\bf D88}
  (2013) no.~2, 024023},
\href{http://arxiv.org/abs/1304.0449}{{\tt arXiv:1304.0449 [hep-th]}}.

\bibitem{Huang:2013mha}
Q.-G. Huang, K.-C. Zhang, and S.-Y. Zhou, ``{Generalized massive gravity in
  arbitrary dimensions and its Hamiltonian formulation},''
  \href{http://dx.doi.org/10.1088/1475-7516/2013/08/050}{{\em JCAP} {\bf 1308}
  (2013)  050},
\href{http://arxiv.org/abs/1306.4740}{{\tt arXiv:1306.4740 [hep-th]}}.

\bibitem{Kluson:2013yaa}
J.~Kluso?, S.~Nojiri, and S.~D. Odintsov, ``{New proposal for non-linear
  ghost-free massive $F(R)$ gravity: Cosmic acceleration and Hamiltonian
  analysis},'' \href{http://dx.doi.org/10.1016/j.physletb.2013.10.003}{{\em
  Phys.Lett.} {\bf B726} (2013)  918--925},
\href{http://arxiv.org/abs/1309.2185}{{\tt arXiv:1309.2185 [hep-th]}}.

\bibitem{Bamba:2013aca}
K.~Bamba, M.~W. Hossain, R.~Myrzakulov, S.~Nojiri, and M.~Sami, ``{Cosmological
  investigations of (extended) nonlinear massive gravity schemes with
  non-minimal coupling},''
  \href{http://dx.doi.org/10.1103/PhysRevD.89.083518}{{\em Phys.Rev.} {\bf D89}
  (2014)  083518},
\href{http://arxiv.org/abs/1309.6413}{{\tt arXiv:1309.6413 [hep-th]}}.

\bibitem{Cai:2014upa}
Y.-F. Cai and E.~N. Saridakis, ``{Cosmology of F(R) nonlinear massive
  gravity},''
\href{http://arxiv.org/abs/1401.4418}{{\tt arXiv:1401.4418 [astro-ph.CO]}}.

\bibitem{Wu:2014hva}
D.-J. Wu, ``{Cosmological evolutions of $F(R)$ nonlinear massive gravity},''
\href{http://arxiv.org/abs/1403.4442}{{\tt arXiv:1403.4442 [hep-th]}}.

\bibitem{Andrews:2013uca}
M.~Andrews, K.~Hinterbichler, J.~Stokes, and M.~Trodden, ``{Cosmological
  perturbations of massive gravity coupled to DBI Galileons},''
  \href{http://dx.doi.org/10.1088/0264-9381/30/18/184006}{{\em
  Class.Quant.Grav.} {\bf 30} (2013)  184006},
\href{http://arxiv.org/abs/1306.5743}{{\tt arXiv:1306.5743 [hep-th]}}.

\bibitem{D'Amico:2012zv}
G.~D'Amico, G.~Gabadadze, L.~Hui, and D.~Pirtskhalava, ``{Quasidilaton: Theory
  and cosmology},'' \href{http://dx.doi.org/10.1103/PhysRevD.87.064037}{{\em
  Phys.Rev.} {\bf D87} (2013) no.~6, 064037},
\href{http://arxiv.org/abs/1206.4253}{{\tt arXiv:1206.4253 [hep-th]}}.

\bibitem{D'Amico:2013kya}
G.~DÕAmico, G.~Gabadadze, L.~Hui, and D.~Pirtskhalava, ``{On Cosmological
  Perturbations of Quasidilaton},''
  \href{http://dx.doi.org/10.1088/0264-9381/30/18/184005}{{\em
  Class.Quant.Grav.} {\bf 30} (2013)  184005},
\href{http://arxiv.org/abs/1304.0723}{{\tt arXiv:1304.0723 [hep-th]}}.

\bibitem{Gannouji:2013rwa}
R.~Gannouji, M.~W. Hossain, M.~Sami, and E.~N. Saridakis, ``{Quasidilaton
  nonlinear massive gravity: Investigations of background cosmological
  dynamics},'' \href{http://dx.doi.org/10.1103/PhysRevD.87.123536}{{\em
  Phys.Rev.} {\bf D87} (2013) no.~12, 123536},
\href{http://arxiv.org/abs/1304.5095}{{\tt arXiv:1304.5095 [gr-qc]}}.

\bibitem{DeFelice:2013tsa}
A.~De~Felice and S.~Mukohyama, ``{Towards consistent extension of quasidilaton
  massive gravity},''
  \href{http://dx.doi.org/10.1016/j.physletb.2013.12.041}{{\em Phys.Lett.} {\bf
  B728} (2014)  622--625},
\href{http://arxiv.org/abs/1306.5502}{{\tt arXiv:1306.5502 [hep-th]}}.

\bibitem{DeFelice:2013dua}
A.~De~Felice, A.~E. Gumrukcuoglu, and S.~Mukohyama, ``{Generalized
  quasi-dilaton theory},''
  \href{http://dx.doi.org/10.1103/PhysRevD.88.124006}{{\em Phys.Rev.} {\bf D88}
  (2013)  124006},
\href{http://arxiv.org/abs/1309.3162}{{\tt arXiv:1309.3162 [hep-th]}}.

\bibitem{Gabadadze:2014kaa}
G.~Gabadadze, R.~Kimura, and D.~Pirtskhalava, ``{Selfacceleration with
  Quasidilaton},''
\href{http://arxiv.org/abs/1401.5403}{{\tt arXiv:1401.5403 [hep-th]}}.

\bibitem{Deser:2001pe}
S.~Deser and A.~Waldron, ``{Gauge invariances and phases of massive higher
  spins in (A)dS},''
  \href{http://dx.doi.org/10.1103/PhysRevLett.87.031601}{{\em Phys.Rev.Lett.}
  {\bf 87} (2001)  031601},
\href{http://arxiv.org/abs/hep-th/0102166}{{\tt arXiv:hep-th/0102166
  [hep-th]}}.

\bibitem{Deser:2001us}
S.~Deser and A.~Waldron, ``{Partial masslessness of higher spins in (A)dS},''
  \href{http://dx.doi.org/10.1016/S0550-3213(01)00212-7}{{\em Nucl.Phys.} {\bf
  B607} (2001)  577--604},
\href{http://arxiv.org/abs/hep-th/0103198}{{\tt arXiv:hep-th/0103198
  [hep-th]}}.

\bibitem{deRham:2012kf}
C.~de~Rham and S.~Renaux-Petel, ``{Massive Gravity on de Sitter and Unique
  Candidate for Partially Massless Gravity},''
  \href{http://dx.doi.org/10.1088/1475-7516/2013/01/035}{{\em JCAP} {\bf 1301}
  (2013)  035},
\href{http://arxiv.org/abs/1206.3482}{{\tt arXiv:1206.3482 [hep-th]}}.

\bibitem{Hassan:2012gz}
S.~Hassan, A.~Schmidt-May, and M.~von Strauss, ``{On Partially Massless
  Bimetric Gravity},''
  \href{http://dx.doi.org/10.1016/j.physletb.2013.09.021}{{\em Phys.Lett.} {\bf
  B726,} (2013)  834--838},
\href{http://arxiv.org/abs/1208.1797}{{\tt arXiv:1208.1797 [hep-th]}}.

\bibitem{Deser:2012qg}
S.~Deser, E.~Joung, and A.~Waldron, ``{Partial Masslessness and Conformal
  Gravity},'' \href{http://dx.doi.org/10.1088/1751-8113/46/21/214019}{{\em
  J.Phys.} {\bf A46} (2013)  214019},
\href{http://arxiv.org/abs/1208.1307}{{\tt arXiv:1208.1307 [hep-th]}}.

\bibitem{Hassan:2012rq}
S.~Hassan, A.~Schmidt-May, and M.~von Strauss, ``{Bimetric theory and partial
  masslessness with LanczosÐLovelock terms in arbitrary dimensions},''
  \href{http://dx.doi.org/10.1088/0264-9381/30/18/184010}{{\em
  Class.Quant.Grav.} {\bf 30} (2013)  184010},
\href{http://arxiv.org/abs/1212.4525}{{\tt arXiv:1212.4525 [hep-th]}}.

\bibitem{deRham:2013wv}
C.~de~Rham, K.~Hinterbichler, R.~A. Rosen, and A.~J. Tolley, ``{Evidence for
  and obstructions to nonlinear partially massless gravity},''
  \href{http://dx.doi.org/10.1103/PhysRevD.88.024003}{{\em Phys.Rev.} {\bf D88}
  (2013) no.~2, 024003},
\href{http://arxiv.org/abs/1302.0025}{{\tt arXiv:1302.0025 [hep-th]}}.

\bibitem{Deser:2013uy}
S.~Deser, M.~Sandora, and A.~Waldron, ``{Nonlinear Partially Massless from
  Massive Gravity?},'' \href{http://dx.doi.org/10.1103/PhysRevD.87.101501}{{\em
  Phys.Rev.} {\bf D87} (2013)  101501},
\href{http://arxiv.org/abs/1301.5621}{{\tt arXiv:1301.5621 [hep-th]}}.

\bibitem{Joung:2014aba}
E.~Joung, W.~Li, and M.~Taronna, ``{No unitary theory of PM spin two and
  gravity},''
\href{http://arxiv.org/abs/1406.2335}{{\tt arXiv:1406.2335 [hep-th]}}.

\bibitem{Chkareuli:2011te}
G.~Chkareuli and D.~Pirtskhalava, ``{Vainshtein Mechanism In $\Lambda_3$ -
  Theories},'' \href{http://dx.doi.org/10.1016/j.physletb.2012.05.030}{{\em
  Phys.Lett.} {\bf B713} (2012)  99--103},
\href{http://arxiv.org/abs/1105.1783}{{\tt arXiv:1105.1783 [hep-th]}}.

\bibitem{Koyama:2011xz}
K.~Koyama, G.~Niz, and G.~Tasinato, ``{Analytic solutions in non-linear massive
  gravity},'' \href{http://dx.doi.org/10.1103/PhysRevLett.107.131101}{{\em
  Phys.Rev.Lett.} {\bf 107} (2011)  131101},
\href{http://arxiv.org/abs/1103.4708}{{\tt arXiv:1103.4708 [hep-th]}}.

\bibitem{Sjors:2011iv}
S.~Sjors and E.~Mortsell, ``{Spherically Symmetric Solutions in Massive Gravity
  and Constraints from Galaxies},''
  \href{http://dx.doi.org/10.1007/JHEP02(2013)080}{{\em JHEP} {\bf 1302} (2013)
   080},
\href{http://arxiv.org/abs/1111.5961}{{\tt arXiv:1111.5961 [gr-qc]}}.

\bibitem{Sbisa:2012zk}
F.~Sbisa, G.~Niz, K.~Koyama, and G.~Tasinato, ``{Characterising Vainshtein
  Solutions in Massive Gravity},''
  \href{http://dx.doi.org/10.1103/PhysRevD.86.024033}{{\em Phys.Rev.} {\bf D86}
  (2012)  024033},
\href{http://arxiv.org/abs/1204.1193}{{\tt arXiv:1204.1193 [hep-th]}}.

\bibitem{Burrage:2012ja}
C.~Burrage, N.~Kaloper, and A.~Padilla, ``{Strong Coupling and Bounds on the
  Spin-2 Mass in Massive Gravity},''
  \href{http://dx.doi.org/10.1103/PhysRevLett.111.021802}{{\em Phys.Rev.Lett.}
  {\bf 111} (2013) no.~2, 021802},
\href{http://arxiv.org/abs/1211.6001}{{\tt arXiv:1211.6001 [hep-th]}}.

\bibitem{Tasinato:2013rza}
G.~Tasinato, K.~Koyama, and G.~Niz, ``{Exact Solutions in Massive Gravity},''
  \href{http://dx.doi.org/10.1088/0264-9381/30/18/184002}{{\em
  Class.Quant.Grav.} {\bf 30} (2013)  184002},
\href{http://arxiv.org/abs/1304.0601}{{\tt arXiv:1304.0601 [hep-th]}}.

\bibitem{Babichev:2013pfa}
E.~Babichev and M.~Crisostomi, ``{Restoring general relativity in massive
  bigravity theory},'' \href{http://dx.doi.org/10.1103/PhysRevD.88.084002}{{\em
  Phys.Rev.} {\bf D88} (2013) no.~8, 084002},
\href{http://arxiv.org/abs/1307.3640}{{\tt arXiv:1307.3640}}.

\bibitem{Dvali:2002pe}
G.~Dvali, G.~Gabadadze, and M.~Shifman, ``{Diluting cosmological constant in
  infinite volume extra dimensions},''
  \href{http://dx.doi.org/10.1103/PhysRevD.67.044020}{{\em Phys.Rev.} {\bf D67}
  (2003)  044020},
\href{http://arxiv.org/abs/hep-th/0202174}{{\tt arXiv:hep-th/0202174
  [hep-th]}}.

\bibitem{ArkaniHamed:2002fu}
N.~Arkani-Hamed, S.~Dimopoulos, G.~Dvali, and G.~Gabadadze, ``{Nonlocal
  modification of gravity and the cosmological constant problem},''
\href{http://arxiv.org/abs/hep-th/0209227}{{\tt arXiv:hep-th/0209227
  [hep-th]}}.

\bibitem{Dvali:2007kt}
G.~Dvali, S.~Hofmann, and J.~Khoury, ``{Degravitation of the cosmological
  constant and graviton width},''
  \href{http://dx.doi.org/10.1103/PhysRevD.76.084006}{{\em Phys.Rev.} {\bf D76}
  (2007)  084006},
\href{http://arxiv.org/abs/hep-th/0703027}{{\tt arXiv:hep-th/0703027
  [HEP-TH]}}.

\bibitem{Dvali:2006su}
G.~Dvali, ``{Predictive Power of Strong Coupling in Theories with Large
  Distance Modified Gravity},''
  \href{http://dx.doi.org/10.1088/1367-2630/8/12/326}{{\em New J.Phys.} {\bf 8}
  (2006)  326},
\href{http://arxiv.org/abs/hep-th/0610013}{{\tt arXiv:hep-th/0610013
  [hep-th]}}.

\bibitem{Charmousis:2011bf}
C.~Charmousis, E.~J. Copeland, A.~Padilla, and P.~M. Saffin, ``{General second
  order scalar-tensor theory, self tuning, and the Fab Four},''
  \href{http://dx.doi.org/10.1103/PhysRevLett.108.051101}{{\em Phys.Rev.Lett.}
  {\bf 108} (2012)  051101},
\href{http://arxiv.org/abs/1106.2000}{{\tt arXiv:1106.2000 [hep-th]}}.

\bibitem{Charmousis:2011ea}
C.~Charmousis, E.~J. Copeland, A.~Padilla, and P.~M. Saffin, ``{Self-tuning and
  the derivation of a class of scalar-tensor theories},''
  \href{http://dx.doi.org/10.1103/PhysRevD.85.104040}{{\em Phys.Rev.} {\bf D85}
  (2012)  104040},
\href{http://arxiv.org/abs/1112.4866}{{\tt arXiv:1112.4866 [hep-th]}}.

\bibitem{Copeland:2012qf}
E.~J. Copeland, A.~Padilla, and P.~M. Saffin, ``{The cosmology of the
  Fab-Four},'' \href{http://dx.doi.org/10.1088/1475-7516/2012/12/026}{{\em
  JCAP} {\bf 1212} (2012)  026},
\href{http://arxiv.org/abs/1208.3373}{{\tt arXiv:1208.3373 [hep-th]}}.

\bibitem{deRham:2007rw}
C.~de~Rham, S.~Hofmann, J.~Khoury, and A.~J. Tolley, ``{Cascading Gravity and
  Degravitation},'' \href{http://dx.doi.org/10.1088/1475-7516/2008/02/011}{{\em
  JCAP} {\bf 0802} (2008)  011},
\href{http://arxiv.org/abs/0712.2821}{{\tt arXiv:0712.2821 [hep-th]}}.

\bibitem{Dubovsky:2002jm}
S.~Dubovsky and V.~Rubakov, ``{Brane induced gravity in more than one extra
  dimensions: Violation of equivalence principle and ghost},''
  \href{http://dx.doi.org/10.1103/PhysRevD.67.104014}{{\em Phys.Rev.} {\bf D67}
  (2003)  104014},
\href{http://arxiv.org/abs/hep-th/0212222}{{\tt arXiv:hep-th/0212222
  [hep-th]}}.

\bibitem{Gabadadze:2003ck}
G.~Gabadadze and M.~Shifman, ``{Softly massive gravity},''
  \href{http://dx.doi.org/10.1103/PhysRevD.69.124032}{{\em Phys.Rev.} {\bf D69}
  (2004)  124032},
\href{http://arxiv.org/abs/hep-th/0312289}{{\tt arXiv:hep-th/0312289
  [hep-th]}}.

\bibitem{Berkhahn:2012wg}
F.~Berkhahn, S.~Hofmann, and F.~Niedermann, ``{Brane Induced Gravity: From a
  No-Go to a No-Ghost Theorem},''
  \href{http://dx.doi.org/10.1103/PhysRevD.86.124022}{{\em Phys.Rev.} {\bf D86}
  (2012)  124022},
\href{http://arxiv.org/abs/1205.6801}{{\tt arXiv:1205.6801 [hep-th]}}.

\bibitem{FlorianRobertinprogress}
S.~Hofmann, J.~Khoury, F.~Niedermann, and R.~Schneider, ``{Cosmology of
  Co-Dimension 2 Brane-Induced Gravity},'' {\em to appear}  .

\bibitem{deRham:2008zz}
C.~de~Rham, G.~Dvali, S.~Hofmann, J.~Khoury, O.~Pujolas, {\em et al.},
  ``{Cascading gravity: Extending the Dvali-Gabadadze-Porrati model to higher
  dimension},'' \href{http://dx.doi.org/10.1103/PhysRevLett.100.251603}{{\em
  Phys.Rev.Lett.} {\bf 100} (2008)  251603},
\href{http://arxiv.org/abs/0711.2072}{{\tt arXiv:0711.2072 [hep-th]}}.

\bibitem{deRham:2009wb}
C.~de~Rham, J.~Khoury, and A.~J. Tolley, ``{Flat 3-Brane with Tension in
  Cascading Gravity},''
  \href{http://dx.doi.org/10.1103/PhysRevLett.103.161601}{{\em Phys.Rev.Lett.}
  {\bf 103} (2009)  161601},
\href{http://arxiv.org/abs/0907.0473}{{\tt arXiv:0907.0473 [hep-th]}}.

\bibitem{deRham:2010rw}
C.~de~Rham, J.~Khoury, and A.~J. Tolley, ``{Cascading Gravity is Ghost Free},''
  \href{http://dx.doi.org/10.1103/PhysRevD.81.124027}{{\em Phys.Rev.} {\bf D81}
  (2010)  124027},
\href{http://arxiv.org/abs/1002.1075}{{\tt arXiv:1002.1075 [hep-th]}}.

\bibitem{Kaloper:2007ap}
N.~Kaloper and D.~Kiley, ``{Charting the landscape of modified gravity},''
  \href{http://dx.doi.org/10.1088/1126-6708/2007/05/045}{{\em JHEP} {\bf 0705}
  (2007)  045},
\href{http://arxiv.org/abs/hep-th/0703190}{{\tt arXiv:hep-th/0703190
  [hep-th]}}.

\bibitem{Corradini:2007cz}
O.~Corradini, K.~Koyama, and G.~Tasinato, ``{Induced gravity on intersecting
  brane-worlds. Part I. Maximally symmetric solutions},''
  \href{http://dx.doi.org/10.1103/PhysRevD.77.084006}{{\em Phys.Rev.} {\bf D77}
  (2008)  084006},
\href{http://arxiv.org/abs/0712.0385}{{\tt arXiv:0712.0385 [hep-th]}}.

\bibitem{Corradini:2008tu}
O.~Corradini, K.~Koyama, and G.~Tasinato, ``{Induced gravity on intersecting
  brane-worlds. Part II. Cosmology},''
  \href{http://dx.doi.org/10.1103/PhysRevD.78.124002}{{\em Phys.Rev.} {\bf D78}
  (2008)  124002},
\href{http://arxiv.org/abs/0803.1850}{{\tt arXiv:0803.1850 [hep-th]}}.

\bibitem{Sbisa':2014uza}
F.~Sbisa and K.~Koyama, ``{Perturbations of Nested Branes With Induced
  Gravity},''
\href{http://arxiv.org/abs/1404.0712}{{\tt arXiv:1404.0712 [hep-th]}}.

\bibitem{Agarwal:2009gy}
N.~Agarwal, R.~Bean, J.~Khoury, and M.~Trodden, ``{Cascading Cosmology},''
  \href{http://dx.doi.org/10.1103/PhysRevD.81.084020}{{\em Phys.Rev.} {\bf D81}
  (2010)  084020},
\href{http://arxiv.org/abs/0912.3798}{{\tt arXiv:0912.3798 [hep-th]}}.

\bibitem{Agarwal:2011mg}
N.~Agarwal, R.~Bean, J.~Khoury, and M.~Trodden, ``{Screening bulk curvature in
  the presence of large brane tension},''
  \href{http://dx.doi.org/10.1103/PhysRevD.83.124004}{{\em Phys.Rev.} {\bf D83}
  (2011)  124004},
\href{http://arxiv.org/abs/1102.5091}{{\tt arXiv:1102.5091 [hep-th]}}.

\bibitem{Minamitsuji:2008fz}
M.~Minamitsuji, ``{Self-accelerating solutions in cascading DGP braneworld},''
  \href{http://dx.doi.org/10.1016/j.physletb.2010.01.010}{{\em Phys.Lett.} {\bf
  B684} (2010)  92--95},
\href{http://arxiv.org/abs/0806.2390}{{\tt arXiv:0806.2390 [gr-qc]}}.

\bibitem{Horndeski:1974wa}
G.~W. Horndeski, ``{Second-order scalar-tensor field equations in a
  four-dimensional space},''
\href{http://dx.doi.org/10.1007/BF01807638}{{\em Int.J.Theor.Phys.} {\bf 10}
  (1974)  363--384}.

\bibitem{Kobayashi:2011nu}
T.~Kobayashi, M.~Yamaguchi, and J.~Yokoyama, ``{Generalized G-inflation:
  Inflation with the most general second-order field equations},''
  \href{http://dx.doi.org/10.1143/PTP.126.511}{{\em Prog.Theor.Phys.} {\bf 126}
  (2011)  511--529},
\href{http://arxiv.org/abs/1105.5723}{{\tt arXiv:1105.5723 [hep-th]}}.

\bibitem{DeFelice:2011th}
A.~De~Felice, R.~Kase, and S.~Tsujikawa, ``{Vainshtein mechanism in
  second-order scalar-tensor theories},''
  \href{http://dx.doi.org/10.1103/PhysRevD.85.044059}{{\em Phys.Rev.} {\bf D85}
  (2012)  044059},
\href{http://arxiv.org/abs/1111.5090}{{\tt arXiv:1111.5090 [gr-qc]}}.

\bibitem{Koyama:2013paa}
K.~Koyama, G.~Niz, and G.~Tasinato, ``{Effective theory for the Vainshtein
  mechanism from the Horndeski action},''
  \href{http://dx.doi.org/10.1103/PhysRevD.88.021502}{{\em Phys.Rev.} {\bf D88}
  (2013) no.~2, 021502},
\href{http://arxiv.org/abs/1305.0279}{{\tt arXiv:1305.0279 [hep-th]}}.

\bibitem{Appleby:2012rx}
S.~A. Appleby, A.~De~Felice, and E.~V. Linder, ``{Fab 5: Noncanonical Kinetic
  Gravity, Self Tuning, and Cosmic Acceleration},''
  \href{http://dx.doi.org/10.1088/1475-7516/2012/10/060}{{\em JCAP} {\bf 1210}
  (2012)  060},
\href{http://arxiv.org/abs/1208.4163}{{\tt arXiv:1208.4163 [astro-ph.CO]}}.

\bibitem{Bruneton:2012zk}
J.-P. Bruneton, M.~Rinaldi, A.~Kanfon, A.~Hees, S.~Schlogel, {\em et al.},
  ``{Fab Four: When John and George play gravitation and cosmology},''
  \href{http://dx.doi.org/10.1155/2012/430694}{{\em Adv.Astron.} {\bf 2012}
  (2012)  430694},
\href{http://arxiv.org/abs/1203.4446}{{\tt arXiv:1203.4446 [gr-qc]}}.

\bibitem{Linder:2013zoa}
E.~V. Linder, ``{How Fabulous Is Fab 5 Cosmology?},''
  \href{http://dx.doi.org/10.1088/1475-7516/2013/12/032}{{\em JCAP} {\bf 1312}
  (2013)  032},
\href{http://arxiv.org/abs/1310.7597}{{\tt arXiv:1310.7597 [astro-ph.CO]}}.

\bibitem{Kaloper:2014xya}
N.~Kaloper and M.~Sandora, ``{Spherical cows in the sky with fab four},''
  \href{http://dx.doi.org/10.1088/1475-7516/2014/05/028}{{\em JCAP} {\bf 1405}
  (2014)  028},
\href{http://arxiv.org/abs/1310.5058}{{\tt arXiv:1310.5058}}.

\bibitem{Charmousis:2014mia}
C.~Charmousis, ``{From Lovelock to Horndeski's generalised scalar-tensor
  theory},''
\href{http://arxiv.org/abs/1405.1612}{{\tt arXiv:1405.1612 [gr-qc]}}.

\bibitem{Gao:2011vs}
X.~Gao, T.~Kobayashi, M.~Yamaguchi, and J.~Yokoyama, ``{Primordial
  non-Gaussianities of gravitational waves in the most general single-field
  inflation model},''
  \href{http://dx.doi.org/10.1103/PhysRevLett.107.211301}{{\em Phys.Rev.Lett.}
  {\bf 107} (2011)  211301},
\href{http://arxiv.org/abs/1108.3513}{{\tt arXiv:1108.3513 [astro-ph.CO]}}.

\bibitem{Gao:2011qe}
X.~Gao and D.~A. Steer, ``{Inflation and primordial non-Gaussianities of
  'generalized Galileons'},''
  \href{http://dx.doi.org/10.1088/1475-7516/2011/12/019}{{\em JCAP} {\bf 1112}
  (2011)  019},
\href{http://arxiv.org/abs/1107.2642}{{\tt arXiv:1107.2642 [astro-ph.CO]}}.

\bibitem{Burrage:2011hd}
C.~Burrage, R.~H. Ribeiro, and D.~Seery, ``{Large slow-roll corrections to the
  bispectrum of noncanonical inflation},''
  \href{http://dx.doi.org/10.1088/1475-7516/2011/07/032}{{\em JCAP} {\bf 1107}
  (2011)  032},
\href{http://arxiv.org/abs/1103.4126}{{\tt arXiv:1103.4126 [astro-ph.CO]}}.

\bibitem{Ribeiro:2011ax}
R.~H. Ribeiro and D.~Seery, ``{Decoding the bispectrum of single-field
  inflation},'' \href{http://dx.doi.org/10.1088/1475-7516/2011/10/027}{{\em
  JCAP} {\bf 1110} (2011)  027},
\href{http://arxiv.org/abs/1108.3839}{{\tt arXiv:1108.3839 [astro-ph.CO]}}.

\bibitem{DeFelice:2011uc}
A.~De~Felice and S.~Tsujikawa, ``{Inflationary non-Gaussianities in the most
  general second-order scalar-tensor theories},''
  \href{http://dx.doi.org/10.1103/PhysRevD.84.083504}{{\em Phys.Rev.} {\bf D84}
  (2011)  083504},
\href{http://arxiv.org/abs/1107.3917}{{\tt arXiv:1107.3917 [gr-qc]}}.

\bibitem{RenauxPetel:2011sb}
S.~Renaux-Petel, ``{On the redundancy of operators and the bispectrum in the
  most general second-order scalar-tensor theory},''
  \href{http://dx.doi.org/10.1088/1475-7516/2012/02/020}{{\em JCAP} {\bf 1202}
  (2012)  020},
\href{http://arxiv.org/abs/1107.5020}{{\tt arXiv:1107.5020 [astro-ph.CO]}}.

\bibitem{DeFelice:2013ar}
A.~De~Felice and S.~Tsujikawa, ``{Shapes of primordial non-Gaussianities in the
  Horndeski's most general scalar-tensor theories},''
  \href{http://dx.doi.org/10.1088/1475-7516/2013/03/030}{{\em JCAP} {\bf 1303}
  (2013)  030},
\href{http://arxiv.org/abs/1301.5721}{{\tt arXiv:1301.5721 [hep-th]}}.

\bibitem{DeFelice:2011hq}
A.~De~Felice, T.~Kobayashi, and S.~Tsujikawa, ``{Effective gravitational
  couplings for cosmological perturbations in the most general scalar-tensor
  theories with second-order field equations},''
  \href{http://dx.doi.org/10.1016/j.physletb.2011.11.028}{{\em Phys.Lett.} {\bf
  B706} (2011)  123--133},
\href{http://arxiv.org/abs/1108.4242}{{\tt arXiv:1108.4242 [gr-qc]}}.

\bibitem{DeFelice:2011bh}
A.~De~Felice and S.~Tsujikawa, ``{Conditions for the cosmological viability of
  the most general scalar-tensor theories and their applications to extended
  Galileon dark energy models},''
  \href{http://dx.doi.org/10.1088/1475-7516/2012/02/007}{{\em JCAP} {\bf 1202}
  (2012)  007},
\href{http://arxiv.org/abs/1110.3878}{{\tt arXiv:1110.3878 [gr-qc]}}.

\bibitem{Leon:2012mt}
G.~Leon and E.~N. Saridakis, ``{Dynamical analysis of generalized Galileon
  cosmology},'' \href{http://dx.doi.org/10.1088/1475-7516/2013/03/025}{{\em
  JCAP} {\bf 1303} (2013)  025},
\href{http://arxiv.org/abs/1211.3088}{{\tt arXiv:1211.3088 [astro-ph.CO]}}.

\bibitem{Bloomfield:2011np}
J.~K. Bloomfield and E.~E. Flanagan, ``{A Class of Effective Field Theory
  Models of Cosmic Acceleration},''
  \href{http://dx.doi.org/10.1088/1475-7516/2012/10/039}{{\em JCAP} {\bf 1210}
  (2012)  039},
\href{http://arxiv.org/abs/1112.0303}{{\tt arXiv:1112.0303 [gr-qc]}}.

\bibitem{Mueller:2012kb}
E.-M. Mueller, R.~Bean, and S.~Watson, ``{Cosmological Implications of the
  Effective Field Theory of Cosmic Acceleration},''
  \href{http://dx.doi.org/10.1103/PhysRevD.87.083504}{{\em Phys.Rev.} {\bf D87}
  (2013)  083504},
\href{http://arxiv.org/abs/1209.2706}{{\tt arXiv:1209.2706 [astro-ph.CO]}}.

\bibitem{Tsujikawa:2014mba}
S.~Tsujikawa, ``{The effective field theory of inflation/dark energy and the
  Horndeski theory},''
\href{http://arxiv.org/abs/1404.2684}{{\tt arXiv:1404.2684 [gr-qc]}}.

\bibitem{Charmousis:2012dw}
C.~Charmousis, B.~Gouteraux, and E.~Kiritsis, ``{Higher-derivative
  scalar-vector-tensor theories: black holes, Galileons, singularity cloaking
  and holography},'' \href{http://dx.doi.org/10.1007/JHEP09(2012)011}{{\em
  JHEP} {\bf 1209} (2012)  011},
\href{http://arxiv.org/abs/1206.1499}{{\tt arXiv:1206.1499 [hep-th]}}.

\bibitem{Babichev:2013cya}
E.~Babichev and C.~Charmousis, ``{Dressing a black hole with a time-dependent
  Galileon},''
\href{http://arxiv.org/abs/1312.3204}{{\tt arXiv:1312.3204 [gr-qc]}}.

\bibitem{Anabalon:2013oea}
A.~Anabalon, A.~Cisterna, and J.~Oliva, ``{Asymptotically locally AdS and flat
  black holes in Horndeski theory},''
  \href{http://dx.doi.org/10.1103/PhysRevD.89.084050}{{\em Phys.Rev.} {\bf D89}
  (2014)  084050},
\href{http://arxiv.org/abs/1312.3597}{{\tt arXiv:1312.3597 [gr-qc]}}.

\bibitem{Sotiriou:2013qea}
T.~P. Sotiriou and S.-Y. Zhou, ``{Black hole hair in generalized scalar-tensor
  gravity},''
\href{http://arxiv.org/abs/1312.3622}{{\tt arXiv:1312.3622 [gr-qc]}}.

\bibitem{Kobayashi:2012kh}
T.~Kobayashi, H.~Motohashi, and T.~Suyama, ``{Black hole perturbation in the
  most general scalar-tensor theory with second-order field equations I: the
  odd-parity sector},''
  \href{http://dx.doi.org/10.1103/PhysRevD.85.084025}{{\em Phys.Rev.} {\bf D85}
  (2012)  084025},
\href{http://arxiv.org/abs/1202.4893}{{\tt arXiv:1202.4893 [gr-qc]}}.

\bibitem{Kobayashi:2014wsa}
T.~Kobayashi, H.~Motohashi, and T.~Suyama, ``{Black hole perturbation in the
  most general scalar-tensor theory with second-order field equations II: the
  even-parity sector},''
  \href{http://dx.doi.org/10.1103/PhysRevD.89.084042}{{\em Phys.Rev.} {\bf D89}
  (2014)  084042},
\href{http://arxiv.org/abs/1402.6740}{{\tt arXiv:1402.6740 [gr-qc]}}.

\bibitem{Charmousis:2014zaa}
C.~Charmousis, T.~Kolyvaris, E.~Papantonopoulos, and M.~Tsoukalas, ``{Black
  Holes in Bi-scalar Extensions of Horndeski Theories},''
\href{http://arxiv.org/abs/1404.1024}{{\tt arXiv:1404.1024 [gr-qc]}}.

\bibitem{Padilla:2012dx}
A.~Padilla and V.~Sivanesan, ``{Covariant multi-galileons and their
  generalisation},'' \href{http://dx.doi.org/10.1007/JHEP04(2013)032}{{\em
  JHEP} {\bf 1304} (2013)  032},
\href{http://arxiv.org/abs/1210.4026}{{\tt arXiv:1210.4026 [gr-qc]}}.

\bibitem{Sivanesan:2013tba}
V.~Sivanesan, ``{Proof of the most general multiple-scalar field theory in
  Minkowski space-time free of Ostrogradski Ghost},''
\href{http://arxiv.org/abs/1307.8081}{{\tt arXiv:1307.8081}}.

\bibitem{Padilla:2013jza}
A.~Padilla, D.~Stefanyszyn, and M.~Tsoukalas, ``{Generalised Scale Invariant
  Theories},'' \href{http://dx.doi.org/10.1103/PhysRevD.89.065009}{{\em
  Phys.Rev.} {\bf D89} (2014)  065009},
\href{http://arxiv.org/abs/1312.0975}{{\tt arXiv:1312.0975 [hep-th]}}.

\bibitem{Chandrasekhar:1931ih}
S.~Chandrasekhar, ``{The maximum mass of ideal white dwarfs},''
\href{http://dx.doi.org/10.1086/143324}{{\em Astrophys.J.} {\bf 74} (1931)
  81--82}.

\bibitem{Hillebrandt:2000ga}
W.~Hillebrandt and J.~C. Niemeyer, ``{Type Ia supernova explosion models},''
  \href{http://dx.doi.org/10.1146/annurev.astro.38.1.191}{{\em
  Ann.Rev.Astron.Astrophys.} {\bf 38} (2000)  191--230},
\href{http://arxiv.org/abs/astro-ph/0006305}{{\tt arXiv:astro-ph/0006305
  [astro-ph]}}.

\bibitem{Phillips:1993ng}
M.~Phillips, ``{The absolute magnitudes of Type IA supernovae},''
\href{http://dx.doi.org/10.1086/186970}{{\em Astrophys.J.} {\bf 413} (1993)
  L105--L108}.

\bibitem{Riess:1996pa}
A.~G. Riess, W.~H. Press, and R.~P. Kirshner, ``{A Precise distance indicator:
  Type Ia supernova multicolor light curve shapes},''
  \href{http://dx.doi.org/10.1086/178129}{{\em Astrophys.J.} {\bf 473} (1996)
  88},
\href{http://arxiv.org/abs/astro-ph/9604143}{{\tt arXiv:astro-ph/9604143
  [astro-ph]}}.

\bibitem{Hamuy:1996sq}
M.~Hamuy, M.~Phillips, R.~A. Schommer, N.~B. Suntzeff, J.~Maza, {\em et al.},
  ``{The Absolute luminosities of the Calan/Tololo type IA supernovae},''
  \href{http://dx.doi.org/10.1086/118190}{{\em Astron.J.} {\bf 112} (1996)
  2391},
\href{http://arxiv.org/abs/astro-ph/9609059}{{\tt arXiv:astro-ph/9609059
  [astro-ph]}}.

\bibitem{Perlmutter:1996ds}
{\bf Supernova Cosmology Project} Collaboration, S.~Perlmutter {\em et al.},
  ``{Measurements of the cosmological parameters Omega and Lambda from the
  first 7 supernovae at z $\geq$ 0.35},''
  \href{http://dx.doi.org/10.1086/304265}{{\em Astrophys.J.} {\bf 483} (1997)
  565},
\href{http://arxiv.org/abs/astro-ph/9608192}{{\tt arXiv:astro-ph/9608192
  [astro-ph]}}.

\bibitem{Weinberg:2008zzc}
S.~Weinberg,
``{Cosmology},''.

\bibitem{Hu:1996qs}
W.~Hu, N.~Sugiyama, and J.~Silk, ``{The Physics of microwave background
  anisotropies},'' \href{http://dx.doi.org/10.1038/386037a0}{{\em Nature} {\bf
  386} (1997)  37--43},
\href{http://arxiv.org/abs/astro-ph/9604166}{{\tt arXiv:astro-ph/9604166
  [astro-ph]}}.

\bibitem{Trodden:2004st}
M.~Trodden and S.~M. Carroll, ``{TASI lectures: Introduction to cosmology},''
\href{http://arxiv.org/abs/astro-ph/0401547}{{\tt arXiv:astro-ph/0401547
  [astro-ph]}}.

\bibitem{Kamionkowski:1993aw}
M.~Kamionkowski, D.~N. Spergel, and N.~Sugiyama, ``{Small scale cosmic
  microwave background anisotropies as a probe of the geometry of the
  universe},'' \href{http://dx.doi.org/10.1086/187339}{{\em Astrophys.J.} {\bf
  426} (1994)  L57},
\href{http://arxiv.org/abs/astro-ph/9401003}{{\tt arXiv:astro-ph/9401003
  [astro-ph]}}.

\bibitem{Brans:1961sx}
C.~Brans and R.~Dicke, ``{Mach's principle and a relativistic theory of
  gravitation},''
\href{http://dx.doi.org/10.1103/PhysRev.124.925}{{\em Phys.Rev.} {\bf 124}
  (1961)  925--935}.

\bibitem{Haugan:2001ix}
M.~P. Haugan and C.~Lammerzahl, ``{Principles of equivalence: their role in
  gravitation physics and experiments that test them},'' {\em Lect.Notes Phys.}
  {\bf 562} (2001)  195--212,
\href{http://arxiv.org/abs/gr-qc/0103067}{{\tt arXiv:gr-qc/0103067 [gr-qc]}}.

\bibitem{Shapiro:1964uw}
I.~I. Shapiro, ``{Fourth Test of General Relativity},''
\href{http://dx.doi.org/10.1103/PhysRevLett.13.789}{{\em Phys.Rev.Lett.} {\bf
  13} (1964)  789--791}.

\bibitem{Hulse:1974eb}
R.~Hulse and J.~Taylor, ``{Discovery of a pulsar in a binary system},''
\href{http://dx.doi.org/10.1086/181708}{{\em Astrophys.J.} {\bf 195} (1975)
  L51--L53}.

\bibitem{Wagner:2012ui}
T.~Wagner, S.~Schlamminger, J.~Gundlach, and E.~Adelberger, ``{Torsion-balance
  tests of the weak equivalence principle},''
  \href{http://dx.doi.org/10.1088/0264-9381/29/18/184002}{{\em
  Class.Quant.Grav.} {\bf 29} (2012)  184002},
\href{http://arxiv.org/abs/1207.2442}{{\tt arXiv:1207.2442 [gr-qc]}}.

\bibitem{Will:2014xja}
C.~M. Will, ``{The Confrontation between General Relativity and Experiment},''
{\em Living Rev.Rel.} {\bf 17} (2014)  4.

\bibitem{Nordtvedt:1968qr}
K.~Nordtvedt, ``{Equivalence Principle for Massive Bodies. 1. Phenomenology},''
\href{http://dx.doi.org/10.1103/PhysRev.169.1014}{{\em Phys.Rev.} {\bf 169}
  (1968)  1014--1016}.

\bibitem{Baessler:1999iv}
S.~Baessler, B.~R. Heckel, E.~Adelberger, J.~Gundlach, U.~Schmidt, {\em et
  al.}, ``{Improved Test of the Equivalence Principle for Gravitational
  Self-Energy},''
\href{http://dx.doi.org/10.1103/PhysRevLett.83.003585}{{\em Phys.Rev.Lett.}
  {\bf 83} (1999)  3585}.

\bibitem{Williams:2004qba}
J.~G. Williams, S.~G. Turyshev, and D.~H. Boggs, ``{Progress in lunar laser
  ranging tests of relativistic gravity},''
  \href{http://dx.doi.org/10.1103/PhysRevLett.93.261101}{{\em Phys.Rev.Lett.}
  {\bf 93} (2004)  261101},
\href{http://arxiv.org/abs/gr-qc/0411113}{{\tt arXiv:gr-qc/0411113 [gr-qc]}}.

\bibitem{Williams:2003wu}
J.~G. Williams, S.~G. Turyshev, and J.~Murphy, Thomas~W., ``{Improving LLR
  tests of gravitational theory},''
  \href{http://dx.doi.org/10.1142/S0218271804004682}{{\em Int.J.Mod.Phys.} {\bf
  D13} (2004)  567--582},
\href{http://arxiv.org/abs/gr-qc/0311021}{{\tt arXiv:gr-qc/0311021 [gr-qc]}}.

\bibitem{Fischbach:1999bc}
E.~Fischbach and C.~Talmadge,
``{The search for nonNewtonian gravity},''.

\bibitem{Schlamminger:2007ht}
S.~Schlamminger, K.-Y. Choi, T.~Wagner, J.~Gundlach, and E.~Adelberger, ``{Test
  of the equivalence principle using a rotating torsion balance},''
  \href{http://dx.doi.org/10.1103/PhysRevLett.100.041101}{{\em Phys.Rev.Lett.}
  {\bf 100} (2008)  041101},
\href{http://arxiv.org/abs/0712.0607}{{\tt arXiv:0712.0607 [gr-qc]}}.

\bibitem{Long:1998dk}
J.~C. Long, H.~W. Chan, and J.~C. Price, ``{Experimental status of
  gravitational strength forces in the subcentimeter regime},''
  \href{http://dx.doi.org/10.1016/S0550-3213(98)00711-1}{{\em Nucl.Phys.} {\bf
  B539} (1999)  23--34},
\href{http://arxiv.org/abs/hep-ph/9805217}{{\tt arXiv:hep-ph/9805217
  [hep-ph]}}.

\bibitem{Hoyle:2000cv}
C.~Hoyle, U.~Schmidt, B.~R. Heckel, E.~Adelberger, J.~Gundlach, {\em et al.},
  ``{Submillimeter tests of the gravitational inverse square law: a search for
  'large' extra dimensions},''
  \href{http://dx.doi.org/10.1103/PhysRevLett.86.1418}{{\em Phys.Rev.Lett.}
  {\bf 86} (2001)  1418--1421},
\href{http://arxiv.org/abs/hep-ph/0011014}{{\tt arXiv:hep-ph/0011014
  [hep-ph]}}.

\bibitem{Chiaverini:2002cb}
J.~Chiaverini, S.~Smullin, A.~Geraci, D.~Weld, and A.~Kapitulnik, ``{New
  experimental constraints on nonNewtonian forces below 100 microns},''
  \href{http://dx.doi.org/10.1103/PhysRevLett.90.151101}{{\em Phys.Rev.Lett.}
  {\bf 90} (2003)  151101},
\href{http://arxiv.org/abs/hep-ph/0209325}{{\tt arXiv:hep-ph/0209325
  [hep-ph]}}.

\bibitem{Long:2002wn}
J.~C. Long, H.~W. Chan, A.~B. Churnside, E.~A. Gulbis, M.~C. Varney, {\em et
  al.}, ``{New experimental limits on macroscopic forces below 100 microns},''
\href{http://arxiv.org/abs/hep-ph/0210004}{{\tt arXiv:hep-ph/0210004
  [hep-ph]}}.

\bibitem{Hoyle:2004cw}
C.~Hoyle, D.~Kapner, B.~R. Heckel, E.~Adelberger, J.~Gundlach, {\em et al.},
  ``{Sub-millimeter tests of the gravitational inverse-square law},''
  \href{http://dx.doi.org/10.1103/PhysRevD.70.042004}{{\em Phys.Rev.} {\bf D70}
  (2004)  042004},
\href{http://arxiv.org/abs/hep-ph/0405262}{{\tt arXiv:hep-ph/0405262
  [hep-ph]}}.

\bibitem{Geraci:2008hb}
A.~A. Geraci, S.~J. Smullin, D.~M. Weld, J.~Chiaverini, and A.~Kapitulnik,
  ``{Improved constraints on non-Newtonian forces at 10 microns},''
  \href{http://dx.doi.org/10.1103/PhysRevD.78.022002}{{\em Phys.Rev.} {\bf D78}
  (2008)  022002},
\href{http://arxiv.org/abs/0802.2350}{{\tt arXiv:0802.2350 [hep-ex]}}.

\bibitem{Bezerra:2011xc}
V.~Bezerra, G.~Klimchitskaya, V.~Mostepanenko, and C.~Romero, ``{Constraints on
  non-Newtonian gravity from measuring the Casimir force in a configuration
  with nanoscale rectangular corrugations},''
  \href{http://dx.doi.org/10.1103/PhysRevD.83.075004}{{\em Phys.Rev.} {\bf D83}
  (2011)  075004},
\href{http://arxiv.org/abs/1103.0993}{{\tt arXiv:1103.0993 [hep-ph]}}.

\bibitem{Klimchitskaya:2013rwd}
G.~Klimchitskaya, U.~Mohideen, and V.~Mostepanenko, ``{Constraints on
  corrections to Newtonian gravity from two recent measurements of the Casimir
  interaction between metallic surfaces},''
  \href{http://dx.doi.org/10.1103/PhysRevD.87.125031}{{\em Phys.Rev.} {\bf D87}
  (2013) no.~12, 125031},
\href{http://arxiv.org/abs/1306.4979}{{\tt arXiv:1306.4979 [gr-qc]}}.

\bibitem{Adelberger:2003zx}
E.~Adelberger, B.~R. Heckel, and A.~Nelson, ``{Tests of the gravitational
  inverse square law},''
  \href{http://dx.doi.org/10.1146/annurev.nucl.53.041002.110503}{{\em
  Ann.Rev.Nucl.Part.Sci.} {\bf 53} (2003)  77--121},
\href{http://arxiv.org/abs/hep-ph/0307284}{{\tt arXiv:hep-ph/0307284
  [hep-ph]}}.

\bibitem{Nordtvedt:1968qs}
K.~Nordtvedt, ``{Equivalence Principle for Massive Bodies. 2. Theory},''
\href{http://dx.doi.org/10.1103/PhysRev.169.1017}{{\em Phys.Rev.} {\bf 169}
  (1968)  1017--1025}.

\bibitem{Will:1971zzb}
C.~M. Will, ``{Theoretical Frameworks for Testing Relativistic Gravity. 2.
  Parametrized Post-Newtonian Hydrodynamics, and the Nordtvedt Effect},''
\href{http://dx.doi.org/10.1086/150804}{{\em Astrophys.J.} {\bf 163} (1971)
  611--627}.

\bibitem{Will:1972zz}
C.~M. Will and J.~Nordtvedt, Kenneth, ``{Conservation Laws and Preferred Frames
  in Relativistic Gravity. I. Preferred-Frame Theories and an Extended PPN
  Formalism},''
\href{http://dx.doi.org/10.1086/151754}{{\em Astrophys.J.} {\bf 177} (1972)
  757}.

\bibitem{Bertotti:2003rm}
B.~Bertotti, L.~Iess, and P.~Tortora, ``{A test of general relativity using
  radio links with the Cassini spacecraft},''
\href{http://dx.doi.org/10.1038/nature01997}{{\em Nature} {\bf 425} (2003)
  374}.

\bibitem{Shapiro:2004zz}
S.~Shapiro, J.~Davis, D.~Lebach, and J.~Gregory, ``{Measurement of the Solar
  Gravitational Deflection of Radio Waves using Geodetic Very-Long-Baseline
  Interferometry Data, 1979-1999},''
\href{http://dx.doi.org/10.1103/PhysRevLett.92.121101}{{\em Phys.Rev.Lett.}
  {\bf 92} (2004)  121101}.

\bibitem{mercury}
I.~I. Shapiro, ``{Solar system tests of general relativity: Recent results and
  present plans},'' {\em General Relativity and Gravitation} {\bf Proceedings
  of the 12th International Conference on General Relativity and Gravitation,
  University of Colorado at Boulder, July 2 - 8} (1989)  313Ð330.

\bibitem{Gubser:2004uf}
S.~S. Gubser and J.~Khoury, ``{Scalar self-interactions loosen constraints from
  fifth force searches},''
  \href{http://dx.doi.org/10.1103/PhysRevD.70.104001}{{\em Phys.Rev.} {\bf D70}
  (2004)  104001},
\href{http://arxiv.org/abs/hep-ph/0405231}{{\tt arXiv:hep-ph/0405231
  [hep-ph]}}.

\bibitem{Jain:2012tn}
B.~Jain, V.~Vikram, and J.~Sakstein, ``{Astrophysical Tests of Modified
  Gravity: Constraints from Distance Indicators in the Nearby Universe},''
\href{http://arxiv.org/abs/1204.6044}{{\tt arXiv:1204.6044 [astro-ph.CO]}}.

\bibitem{Vikram:2013uba}
V.~Vikram, A.~CabrŽ, B.~Jain, and J.~VanderPlas, ``{Astrophysical Tests of
  Modified Gravity: the Morphology and Kinematics of Dwarf Galaxies},''
  \href{http://dx.doi.org/10.1088/1475-7516/2013/08/020}{{\em JCAP} {\bf 1308}
  (2013)  020},
\href{http://arxiv.org/abs/1303.0295}{{\tt arXiv:1303.0295 [astro-ph.CO]}}.

\bibitem{Brax:2011hb}
P.~Brax and G.~Pignol, ``{Strongly Coupled Chameleons and the Neutronic Quantum
  Bouncer},'' \href{http://dx.doi.org/10.1103/PhysRevLett.107.111301}{{\em
  Phys.Rev.Lett.} {\bf 107} (2011)  111301},
\href{http://arxiv.org/abs/1105.3420}{{\tt arXiv:1105.3420 [hep-ph]}}.

\bibitem{Pokotilovski:2013gta}
Y.~Pokotilovski, ``{Constraints on strongly coupled chameleon fields from the
  experimental test of the weak equivalence principle for the neutron},''
\href{http://dx.doi.org/10.1134/S0021364012240095}{{\em JETP Lett.} {\bf 96}
  (2013)  751--753}.

\bibitem{Jenke:2014yel}
T.~Jenke, G.~Cronenberg, J.~Burgdorfer, L.~Chizhova, P.~Geltenbort, {\em et
  al.}, ``{Gravity Resonance Spectroscopy Constrains Dark Energy and Dark
  Matter Scenarios},''
  \href{http://dx.doi.org/10.1103/PhysRevLett.112.151105}{{\em Phys.Rev.Lett.}
  {\bf 112} (2014)  151105},
\href{http://arxiv.org/abs/1404.4099}{{\tt arXiv:1404.4099 [gr-qc]}}.

\bibitem{Brax:2013cfa}
P.~Brax, G.~Pignol, and D.~Roulier, ``{Probing Strongly Coupled Chameleons with
  Slow Neutrons},'' \href{http://dx.doi.org/10.1103/PhysRevD.88.083004}{{\em
  Phys.Rev.} {\bf D88} (2013)  083004},
\href{http://arxiv.org/abs/1306.6536}{{\tt arXiv:1306.6536 [quant-ph]}}.

\bibitem{Brax:2007vm}
P.~Brax, C.~van~de Bruck, A.-C. Davis, D.~F. Mota, and D.~J. Shaw, ``{Detecting
  chameleons through Casimir force measurements},''
  \href{http://dx.doi.org/10.1103/PhysRevD.76.124034}{{\em Phys.Rev.} {\bf D76}
  (2007)  124034},
\href{http://arxiv.org/abs/0709.2075}{{\tt arXiv:0709.2075 [hep-ph]}}.

\bibitem{Upadhye:2006vi}
A.~Upadhye, S.~S. Gubser, and J.~Khoury, ``{Unveiling chameleons in tests of
  gravitational inverse-square law},''
  \href{http://dx.doi.org/10.1103/PhysRevD.74.104024}{{\em Phys.Rev.} {\bf D74}
  (2006)  104024},
\href{http://arxiv.org/abs/hep-ph/0608186}{{\tt arXiv:hep-ph/0608186
  [hep-ph]}}.

\bibitem{Chou:2008gr}
{\bf GammeV} Collaboration, A.~S. Chou {\em et al.}, ``{A Search for chameleon
  particles using a photon regeneration technique},''
  \href{http://dx.doi.org/10.1103/PhysRevLett.102.030402}{{\em Phys.Rev.Lett.}
  {\bf 102} (2009)  030402},
\href{http://arxiv.org/abs/0806.2438}{{\tt arXiv:0806.2438 [hep-ex]}}.

\bibitem{Steffen:2009sc}
J.~H. Steffen and A.~Upadhye, ``{The GammeV suite of experimental searches for
  axion-like particles},''
  \href{http://dx.doi.org/10.1142/S0217732309031727}{{\em Mod.Phys.Lett.} {\bf
  A24} (2009)  2053--2068},
\href{http://arxiv.org/abs/0908.1529}{{\tt arXiv:0908.1529 [hep-ex]}}.

\bibitem{Upadhye:2009iv}
A.~Upadhye, J.~Steffen, and A.~Weltman, ``{Constraining chameleon field
  theories using the GammeV afterglow experiments},''
  \href{http://dx.doi.org/10.1103/PhysRevD.81.015013}{{\em Phys.Rev.} {\bf D81}
  (2010)  015013},
\href{http://arxiv.org/abs/0911.3906}{{\tt arXiv:0911.3906 [hep-ph]}}.

\bibitem{Steffen:2010ze}
{\bf GammeV} Collaboration, J.~H. Steffen {\em et al.}, ``{Laboratory
  constraints on chameleon dark energy and power-law fields},''
  \href{http://dx.doi.org/10.1103/PhysRevLett.105.261803}{{\em Phys.Rev.Lett.}
  {\bf 105} (2010)  261803},
\href{http://arxiv.org/abs/1010.0988}{{\tt arXiv:1010.0988 [astro-ph.CO]}}.

\bibitem{Upadhye:2012ar}
A.~Upadhye, J.~H. Steffen, and A.~S. Chou, ``{Designing dark energy afterglow
  experiments},'' \href{http://dx.doi.org/10.1103/PhysRevD.86.035006}{{\em
  Phys.Rev.} {\bf D86} (2012)  035006},
\href{http://arxiv.org/abs/1204.5476}{{\tt arXiv:1204.5476 [hep-ph]}}.

\bibitem{Steffen:2012rw}
{\bf GammeV} Collaboration, J.~H. Steffen, A.~Baumbaugh, A.~S. Chou, R.~Tomlin,
  and A.~Upadhye, ``{On the anomalous afterglow seen in a chameleon afterglow
  search},'' \href{http://dx.doi.org/10.1103/PhysRevD.86.012003}{{\em
  Phys.Rev.} {\bf D86} (2012)  012003},
\href{http://arxiv.org/abs/1205.6495}{{\tt arXiv:1205.6495 [physics.ins-det]}}.

\bibitem{Brax:2013tsa}
P.~Brax and A.~Upadhye, ``{Chameleon Fragmentation},''
  \href{http://dx.doi.org/10.1088/1475-7516/2014/02/018}{{\em JCAP} {\bf 1402}
  (2014)  018},
\href{http://arxiv.org/abs/1312.2747}{{\tt arXiv:1312.2747 [hep-ph]}}.

\bibitem{Rybka:2010ah}
G.~Rybka, M.~Hotz, L.~Rosenberg, S.~Asztalos, G.~Carosi, {\em et al.}, ``{A
  Search for Scalar Chameleons with ADMX},''
  \href{http://dx.doi.org/10.1103/PhysRevLett.105.051801}{{\em Phys.Rev.Lett.}
  {\bf 105} (2010)  051801},
\href{http://arxiv.org/abs/1004.5160}{{\tt arXiv:1004.5160 [astro-ph.CO]}}.

\bibitem{Brax:2010xq}
P.~Brax and K.~Zioutas, ``{Solar Chameleons},''
  \href{http://dx.doi.org/10.1103/PhysRevD.82.043007}{{\em Phys.Rev.} {\bf D82}
  (2010)  043007},
\href{http://arxiv.org/abs/1004.1846}{{\tt arXiv:1004.1846 [astro-ph.SR]}}.

\bibitem{Burrage:2008ii}
C.~Burrage, A.-C. Davis, and D.~J. Shaw, ``{Detecting Chameleons: The
  Astronomical Polarization Produced by Chameleon-like Scalar Fields},''
  \href{http://dx.doi.org/10.1103/PhysRevD.79.044028}{{\em Phys.Rev.} {\bf D79}
  (2009)  044028},
\href{http://arxiv.org/abs/0809.1763}{{\tt arXiv:0809.1763 [astro-ph]}}.

\bibitem{Brax:2009ey}
P.~Brax, C.~Burrage, A.-C. Davis, D.~Seery, and A.~Weltman, ``{Higgs production
  as a probe of Chameleon Dark Energy},''
  \href{http://dx.doi.org/10.1103/PhysRevD.81.103524}{{\em Phys.Rev.} {\bf D81}
  (2010)  103524},
\href{http://arxiv.org/abs/0911.1267}{{\tt arXiv:0911.1267 [hep-ph]}}.

\bibitem{Brax:2009aw}
P.~Brax, C.~Burrage, A.-C. Davis, D.~Seery, and A.~Weltman, ``{Collider
  constraints on interactions of dark energy with the Standard Model},''
  \href{http://dx.doi.org/10.1088/1126-6708/2009/09/128}{{\em JHEP} {\bf 0909}
  (2009)  128},
\href{http://arxiv.org/abs/0904.3002}{{\tt arXiv:0904.3002 [hep-ph]}}.

\bibitem{Upadhye:2012rc}
A.~Upadhye, ``{Symmetron dark energy in laboratory experiments},''
  \href{http://dx.doi.org/10.1103/PhysRevLett.110.031301}{{\em Phys.Rev.Lett.}
  {\bf 110} (2013)  031301},
\href{http://arxiv.org/abs/1210.7804}{{\tt arXiv:1210.7804 [hep-ph]}}.

\bibitem{Nordtvedt:2003pj}
K.~Nordtvedt, ``{Lunar laser ranging: A Comprehensive probe of postNewtonian
  gravity},''
\href{http://arxiv.org/abs/gr-qc/0301024}{{\tt arXiv:gr-qc/0301024 [gr-qc]}}.

\bibitem{Murphy:2012rea}
J.~Murphy, T.W., E.~Adelberger, J.~Battat, C.~Hoyle, N.~Johnson, {\em et al.},
  ``{APOLLO: millimeter lunar laser ranging},''
\href{http://dx.doi.org/10.1088/0264-9381/29/18/184005}{{\em Class.Quant.Grav.}
  {\bf 29} (2012)  184005}.

\bibitem{Dvali:2002vf}
G.~Dvali, A.~Gruzinov, and M.~Zaldarriaga, ``{The Accelerated universe and the
  moon},'' \href{http://dx.doi.org/10.1103/PhysRevD.68.024012}{{\em Phys.Rev.}
  {\bf D68} (2003)  024012},
\href{http://arxiv.org/abs/hep-ph/0212069}{{\tt arXiv:hep-ph/0212069
  [hep-ph]}}.

\bibitem{Lue:2002sw}
A.~Lue and G.~Starkman, ``{Gravitational leakage into extra dimensions: Probing
  dark energy using local gravity},''
  \href{http://dx.doi.org/10.1103/PhysRevD.67.064002}{{\em Phys.Rev.} {\bf D67}
  (2003)  064002},
\href{http://arxiv.org/abs/astro-ph/0212083}{{\tt arXiv:astro-ph/0212083
  [astro-ph]}}.

\bibitem{Afshordi:2008rd}
N.~Afshordi, G.~Geshnizjani, and J.~Khoury, ``{Do observations offer evidence
  for cosmological-scale extra dimensions?},''
  \href{http://dx.doi.org/10.1088/1475-7516/2009/08/030}{{\em JCAP} {\bf 0908}
  (2009)  030},
\href{http://arxiv.org/abs/0812.2244}{{\tt arXiv:0812.2244 [astro-ph]}}.

\bibitem{Battat:2008bu}
J.~B. Battat, C.~W. Stubbs, and J.~F. Chandler, ``{Solar system constraints on
  the Dvali-Gabadadze-Porrati braneworld theory of gravity},''
  \href{http://dx.doi.org/10.1103/PhysRevD.78.022003}{{\em Phys.Rev.} {\bf D78}
  (2008)  022003},
\href{http://arxiv.org/abs/0805.4466}{{\tt arXiv:0805.4466 [gr-qc]}}.

\bibitem{Andrews:2013qva}
M.~Andrews, Y.-Z. Chu, and M.~Trodden, ``{Galileon forces in the Solar
  System},'' \href{http://dx.doi.org/10.1103/PhysRevD.88.084028}{{\em
  Phys.Rev.} {\bf D88} (2013)  084028},
\href{http://arxiv.org/abs/1305.2194}{{\tt arXiv:1305.2194 [astro-ph.CO]}}.

\bibitem{Brax:2011sv}
P.~Brax, C.~Burrage, and A.-C. Davis, ``{Laboratory Tests of the Galileon},''
  \href{http://dx.doi.org/10.1088/1475-7516/2011/09/020}{{\em JCAP} {\bf 1109}
  (2011)  020},
\href{http://arxiv.org/abs/1106.1573}{{\tt arXiv:1106.1573 [hep-ph]}}.

\bibitem{Babichev:2011iz}
E.~Babichev, C.~Deffayet, and G.~Esposito-Farese, ``{Constraints on
  Shift-Symmetric Scalar-Tensor Theories with a Vainshtein Mechanism from
  Bounds on the Time Variation of G},''
  \href{http://dx.doi.org/10.1103/PhysRevLett.107.251102}{{\em Phys.Rev.Lett.}
  {\bf 107} (2011)  251102},
\href{http://arxiv.org/abs/1107.1569}{{\tt arXiv:1107.1569 [gr-qc]}}.

\bibitem{Jain:2013wgs}
B.~Jain, A.~Joyce, R.~Thompson, A.~Upadhye, J.~Battat, {\em et al.}, ``{Novel
  Probes of Gravity and Dark Energy},''
\href{http://arxiv.org/abs/1309.5389}{{\tt arXiv:1309.5389 [astro-ph.CO]}}.

\bibitem{Chang:2010xh}
P.~Chang and L.~Hui, ``{Stellar Structure and Tests of Modified Gravity},''
  \href{http://dx.doi.org/10.1088/0004-637X/732/1/25}{{\em Astrophys.J.} {\bf
  732} (2011)  25},
\href{http://arxiv.org/abs/1011.4107}{{\tt arXiv:1011.4107 [astro-ph.CO]}}.

\bibitem{Davis:2011qf}
A.-C. Davis, E.~A. Lim, J.~Sakstein, and D.~Shaw, ``{Modified Gravity Makes
  Galaxies Brighter},''
  \href{http://dx.doi.org/10.1103/PhysRevD.85.123006}{{\em Phys.Rev.} {\bf D85}
  (2012)  123006},
\href{http://arxiv.org/abs/1102.5278}{{\tt arXiv:1102.5278 [astro-ph.CO]}}.

\bibitem{Paxton:2010ji}
B.~Paxton, L.~Bildsten, A.~Dotter, F.~Herwig, P.~Lesaffre, {\em et al.},
  ``{Modules for Experiments in Stellar Astrophysics (MESA)},''
  \href{http://dx.doi.org/10.1088/0067-0049/192/1/3}{{\em Astrophys.J.Suppl.}
  {\bf 192} (2011)  3},
\href{http://arxiv.org/abs/1009.1622}{{\tt arXiv:1009.1622 [astro-ph.SR]}}.

\bibitem{Sakstein:2013pda}
J.~Sakstein, ``{Stellar Oscillations in Modified Gravity},''
  \href{http://dx.doi.org/10.1103/PhysRevD.88.124013}{{\em Phys.Rev.} {\bf D88}
  (2013)  124013},
\href{http://arxiv.org/abs/1309.0495}{{\tt arXiv:1309.0495 [astro-ph.CO]}}.

\bibitem{Upadhye:2013nfa}
A.~Upadhye and J.~H. Steffen, ``{Monopole radiation in modified gravity},''
\href{http://arxiv.org/abs/1306.6113}{{\tt arXiv:1306.6113 [astro-ph.CO]}}.

\bibitem{Giannantonio:2009gi}
T.~Giannantonio, M.~Martinelli, A.~Silvestri, and A.~Melchiorri, ``{New
  constraints on parametrised modified gravity from correlations of the CMB
  with large scale structure},''
  \href{http://dx.doi.org/10.1088/1475-7516/2010/04/030}{{\em JCAP} {\bf 1004}
  (2010)  030},
\href{http://arxiv.org/abs/0909.2045}{{\tt arXiv:0909.2045 [astro-ph.CO]}}.

\bibitem{Schmidt:2009am}
F.~Schmidt, A.~Vikhlinin, and W.~Hu, ``{Cluster Constraints on f(R) Gravity},''
  \href{http://dx.doi.org/10.1103/PhysRevD.80.083505}{{\em Phys.Rev.} {\bf D80}
  (2009)  083505},
\href{http://arxiv.org/abs/0908.2457}{{\tt arXiv:0908.2457 [astro-ph.CO]}}.

\bibitem{Lombriser:2011zw}
L.~Lombriser, F.~Schmidt, T.~Baldauf, R.~Mandelbaum, U.~Seljak, {\em et al.},
  ``{Cluster Density Profiles as a Test of Modified Gravity},''
  \href{http://dx.doi.org/10.1103/PhysRevD.85.102001}{{\em Phys.Rev.} {\bf D85}
  (2012)  102001},
\href{http://arxiv.org/abs/1111.2020}{{\tt arXiv:1111.2020 [astro-ph.CO]}}.

\bibitem{Levshakov:2010cw}
S.~Levshakov, P.~Molaro, M.~Kozlov, A.~Lapinov, C.~Henkel, {\em et al.},
  ``{Searching for Chameleon-like Scalar Fields},''
\href{http://arxiv.org/abs/1012.0642}{{\tt arXiv:1012.0642 [astro-ph.CO]}}.

\bibitem{Kaloper:2011qc}
N.~Kaloper, A.~Padilla, and N.~Tanahashi, ``{Galileon Hairs of Dyson Spheres,
  Vainshtein's Coiffure and Hirsute Bubbles},''
  \href{http://dx.doi.org/10.1007/JHEP10(2011)148}{{\em JHEP} {\bf 1110} (2011)
   148},
\href{http://arxiv.org/abs/1106.4827}{{\tt arXiv:1106.4827 [hep-th]}}.

\bibitem{Hui:2012qt}
L.~Hui and A.~Nicolis, ``{A no-hair theorem for the galileon},''
  \href{http://dx.doi.org/10.1103/PhysRevLett.110.241104}{{\em Phys.Rev.Lett.}
  {\bf 110} (2013)  241104},
\href{http://arxiv.org/abs/1202.1296}{{\tt arXiv:1202.1296 [hep-th]}}.

\bibitem{Hui:2012jb}
L.~Hui and A.~Nicolis, ``{Proposal for an Observational Test of the Vainshtein
  Mechanism},'' \href{http://dx.doi.org/10.1103/PhysRevLett.109.051304}{{\em
  Phys.Rev.Lett.} {\bf 109} (2012)  051304},
\href{http://arxiv.org/abs/1201.1508}{{\tt arXiv:1201.1508 [astro-ph.CO]}}.

\bibitem{Song:2007wd}
Y.-S. Song, ``{Large Scale Structure Formation of normal branch in DGP brane
  world model},'' \href{http://dx.doi.org/10.1103/PhysRevD.77.124031}{{\em
  Phys.Rev.} {\bf D77} (2008)  124031},
\href{http://arxiv.org/abs/0711.2513}{{\tt arXiv:0711.2513 [astro-ph]}}.

\bibitem{Lombriser:2009xg}
L.~Lombriser, W.~Hu, W.~Fang, and U.~Seljak, ``{Cosmological Constraints on DGP
  Braneworld Gravity with Brane Tension},''
  \href{http://dx.doi.org/10.1103/PhysRevD.80.063536}{{\em Phys.Rev.} {\bf D80}
  (2009)  063536},
\href{http://arxiv.org/abs/0905.1112}{{\tt arXiv:0905.1112 [astro-ph.CO]}}.

\bibitem{Barreira:2014ija}
A.~Barreira, B.~Li, C.~Baugh, and S.~Pascoli, ``{$\nu$Galileon: modified
  gravity with massive neutrinos as a testable alternative to $\Lambda$CDM},''
\href{http://arxiv.org/abs/1404.1365}{{\tt arXiv:1404.1365 [astro-ph.CO]}}.

\bibitem{Wyman:2010jp}
M.~Wyman and J.~Khoury, ``{Enhanced Peculiar Velocities in Brane-Induced
  Gravity},'' \href{http://dx.doi.org/10.1103/PhysRevD.82.044032}{{\em
  Phys.Rev.} {\bf D82} (2010)  044032},
\href{http://arxiv.org/abs/1004.2046}{{\tt arXiv:1004.2046 [astro-ph.CO]}}.

\bibitem{Zu:2013joa}
Y.~Zu, D.~Weinberg, E.~Jennings, B.~Li, and M.~Wyman, ``{Galaxy Infall
  Kinematics as a Test of Modified Gravity},''
\href{http://arxiv.org/abs/1310.6768}{{\tt arXiv:1310.6768 [astro-ph.CO]}}.

\bibitem{Wyman:2011mp}
M.~Wyman, ``{Galilean-invariant scalar fields can strengthen gravitational
  lensing},'' \href{http://dx.doi.org/10.1103/PhysRevLett.106.201102}{{\em
  Phys.Rev.Lett.} {\bf 106} (2011)  201102},
\href{http://arxiv.org/abs/1101.1295}{{\tt arXiv:1101.1295 [astro-ph.CO]}}.

\bibitem{Saini:1999ba}
T.~D. Saini, S.~Raychaudhury, V.~Sahni, and A.~A. Starobinsky,
  ``{Reconstructing the cosmic equation of state from supernova distances},''
  \href{http://dx.doi.org/10.1103/PhysRevLett.85.1162}{{\em Phys.Rev.Lett.}
  {\bf 85} (2000)  1162--1165},
\href{http://arxiv.org/abs/astro-ph/9910231}{{\tt arXiv:astro-ph/9910231
  [astro-ph]}}.

\bibitem{Rapetti:2009ri}
D.~Rapetti, S.~W. Allen, A.~Mantz, and H.~Ebeling, ``{The Observed Growth of
  Massive Galaxy Clusters III: Testing General Relativity on Cosmological
  Scales},'' \href{http://dx.doi.org/10.1111/j.1365-2966.2010.16799.x}{{\em
  Mon.Not.Roy.Astron.Soc.} {\bf 406} (2010)  1796--1804},
\href{http://arxiv.org/abs/0911.1787}{{\tt arXiv:0911.1787 [astro-ph.CO]}}.

\bibitem{Bean:2010zq}
R.~Bean and M.~Tangmatitham, ``{Current constraints on the cosmic growth
  history},'' \href{http://dx.doi.org/10.1103/PhysRevD.81.083534}{{\em
  Phys.Rev.} {\bf D81} (2010)  083534},
\href{http://arxiv.org/abs/1002.4197}{{\tt arXiv:1002.4197 [astro-ph.CO]}}.

\bibitem{Daniel:2010ky}
S.~F. Daniel, E.~V. Linder, T.~L. Smith, R.~R. Caldwell, A.~Cooray, {\em et
  al.}, ``{Testing General Relativity with Current Cosmological Data},''
  \href{http://dx.doi.org/10.1103/PhysRevD.81.123508}{{\em Phys.Rev.} {\bf D81}
  (2010)  123508},
\href{http://arxiv.org/abs/1002.1962}{{\tt arXiv:1002.1962 [astro-ph.CO]}}.

\bibitem{Zhao:2010dz}
G.-B. Zhao, T.~Giannantonio, L.~Pogosian, A.~Silvestri, D.~J. Bacon, {\em et
  al.}, ``{Probing modifications of General Relativity using current
  cosmological observations},''
  \href{http://dx.doi.org/10.1103/PhysRevD.81.103510}{{\em Phys.Rev.} {\bf D81}
  (2010)  103510},
\href{http://arxiv.org/abs/1003.0001}{{\tt arXiv:1003.0001 [astro-ph.CO]}}.

\bibitem{Dossett:2011tn}
J.~N. Dossett, M.~Ishak, and J.~Moldenhauer, ``{Testing General Relativity at
  Cosmological Scales: Implementation and Parameter Correlations},''
  \href{http://dx.doi.org/10.1103/PhysRevD.84.123001}{{\em Phys.Rev.} {\bf D84}
  (2011)  123001},
\href{http://arxiv.org/abs/1109.4583}{{\tt arXiv:1109.4583 [astro-ph.CO]}}.

\bibitem{Nesseris:2010pc}
S.~Nesseris, A.~De~Felice, and S.~Tsujikawa, ``{Observational constraints on
  Galileon cosmology},''
  \href{http://dx.doi.org/10.1103/PhysRevD.82.124054}{{\em Phys.Rev.} {\bf D82}
  (2010)  124054},
\href{http://arxiv.org/abs/1010.0407}{{\tt arXiv:1010.0407 [astro-ph.CO]}}.

\bibitem{Appleby:2011aa}
S.~Appleby and E.~V. Linder, ``{The Paths of Gravity in Galileon Cosmology},''
  \href{http://dx.doi.org/10.1088/1475-7516/2012/03/043}{{\em JCAP} {\bf 1203}
  (2012)  043},
\href{http://arxiv.org/abs/1112.1981}{{\tt arXiv:1112.1981 [astro-ph.CO]}}.

\bibitem{Appleby:2012ba}
S.~A. Appleby and E.~V. Linder, ``{Trial of Galileon gravity by cosmological
  expansion and growth observations},''
  \href{http://dx.doi.org/10.1088/1475-7516/2012/08/026}{{\em JCAP} {\bf 1208}
  (2012)  026},
\href{http://arxiv.org/abs/1204.4314}{{\tt arXiv:1204.4314 [astro-ph.CO]}}.

\bibitem{Neveu:2013mfa}
J.~Neveu, V.~Ruhlmann-Kleider, A.~Conley, N.~Palanque-Delabrouille, P.~Astier,
  {\em et al.}, ``{Experimental constraints on the uncoupled Galileon model
  from SNLS3 data and other cosmological probes},''
  \href{http://dx.doi.org/10.1051/0004-6361/201321256}{{\em Astron.Astrophys.}
  {\bf 555} (2013)  A53},
\href{http://arxiv.org/abs/1302.2786}{{\tt arXiv:1302.2786 [gr-qc]}}.

\bibitem{Barreira:2013jma}
A.~Barreira, B.~Li, A.~Sanchez, C.~M. Baugh, and S.~Pascoli, ``{Parameter space
  in Galileon gravity models},''
  \href{http://dx.doi.org/10.1103/PhysRevD.87.103511}{{\em Phys.Rev.} {\bf D87}
  (2013) no.~10, 103511},
\href{http://arxiv.org/abs/1302.6241}{{\tt arXiv:1302.6241 [astro-ph.CO]}}.

\bibitem{Barreira:2014jha}
A.~Barreira, B.~Li, C.~Baugh, and S.~Pascoli, ``{The observational status of
  Galileon gravity after Planck},''
\href{http://arxiv.org/abs/1406.0485}{{\tt arXiv:1406.0485 [astro-ph.CO]}}.

\bibitem{Neveu:2014vua}
J.~Neveu, V.~Ruhlmann-Kleider, P.~Astier, M.~Besanon, A.~Conley, {\em et al.},
  ``{First experimental constraints on the disformally-coupled Galileon
  model},''
\href{http://arxiv.org/abs/1403.0854}{{\tt arXiv:1403.0854 [gr-qc]}}.

\bibitem{White:2001kt}
M.~J. White and C.~Kochanek, ``{Constraints on the long range properties of
  gravity from weak gravitational lensing},''
  \href{http://dx.doi.org/10.1086/323074}{{\em Astrophys.J.} {\bf 560} (2001)
  539--543},
\href{http://arxiv.org/abs/astro-ph/0105227}{{\tt arXiv:astro-ph/0105227
  [astro-ph]}}.

\bibitem{Sealfon:2004gz}
C.~Sealfon, L.~Verde, and R.~Jimenez, ``{Limits on deviations from the inverse
  - square law on megaparsec scales},''
  \href{http://dx.doi.org/10.1103/PhysRevD.71.083004}{{\em Phys.Rev.} {\bf D71}
  (2005)  083004},
\href{http://arxiv.org/abs/astro-ph/0404111}{{\tt arXiv:astro-ph/0404111
  [astro-ph]}}.

\bibitem{Shirata:2005yr}
A.~Shirata, T.~Shiromizu, N.~Yoshida, and Y.~Suto, ``{Galaxy clustering
  constraints on deviations from Newtonian gravity at cosmological scales},''
  \href{http://dx.doi.org/10.1103/PhysRevD.71.064030}{{\em Phys.Rev.} {\bf D71}
  (2005)  064030},
\href{http://arxiv.org/abs/astro-ph/0501366}{{\tt arXiv:astro-ph/0501366
  [astro-ph]}}.

\bibitem{Stabenau:2006td}
H.~F. Stabenau and B.~Jain, ``{N-Body Simulations of Alternate Gravity
  Models},'' \href{http://dx.doi.org/10.1103/PhysRevD.74.084007}{{\em
  Phys.Rev.} {\bf D74} (2006)  084007},
\href{http://arxiv.org/abs/astro-ph/0604038}{{\tt arXiv:astro-ph/0604038
  [astro-ph]}}.

\bibitem{Sereno:2006qu}
M.~Sereno and J.~Peacock, ``{Imprints of deviations from the gravitational
  inverse-square law on the power spectrum of mass fluctuations},''
  \href{http://dx.doi.org/10.1111/j.1365-2966.2006.10703.x}{{\em
  Mon.Not.Roy.Astron.Soc.} {\bf 371} (2006)  719--726},
\href{http://arxiv.org/abs/astro-ph/0605498}{{\tt arXiv:astro-ph/0605498
  [astro-ph]}}.

\bibitem{Shirata:2007qk}
A.~Shirata, Y.~Suto, C.~Hikage, T.~Shiromizu, and N.~Yoshida, ``{Galaxy
  clustering constraints on deviations from Newtonian gravity at cosmological
  scales II: Perturbative and numerical analyses of power spectrum and
  bispectrum},'' \href{http://dx.doi.org/10.1103/PhysRevD.76.044026}{{\em
  Phys.Rev.} {\bf D76} (2007)  044026},
\href{http://arxiv.org/abs/0705.1311}{{\tt arXiv:0705.1311 [astro-ph]}}.

\bibitem{Wang:2007fsa}
S.~Wang, L.~Hui, M.~May, and Z.~Haiman, ``{Is Modified Gravity Required by
  Observations? An Empirical Consistency Test of Dark Energy Models},''
  \href{http://dx.doi.org/10.1103/PhysRevD.76.063503}{{\em Phys.Rev.} {\bf D76}
  (2007)  063503},
\href{http://arxiv.org/abs/0705.0165}{{\tt arXiv:0705.0165 [astro-ph]}}.

\bibitem{Skordis:2005xk}
C.~Skordis, D.~Mota, P.~Ferreira, and C.~Boehm, ``{Large Scale Structure in
  Bekenstein's theory of relativistic Modified Newtonian Dynamics},''
  \href{http://dx.doi.org/10.1103/PhysRevLett.96.011301}{{\em Phys.Rev.Lett.}
  {\bf 96} (2006)  011301},
\href{http://arxiv.org/abs/astro-ph/0505519}{{\tt arXiv:astro-ph/0505519
  [astro-ph]}}.

\bibitem{Skordis:2005eu}
C.~Skordis, ``{Teves cosmology : covariant formalism for the background
  evolution and linear perturbation theory},''
  \href{http://dx.doi.org/10.1103/PhysRevD.74.103513}{{\em Phys.Rev.} {\bf D74}
  (2006)  103513},
\href{http://arxiv.org/abs/astro-ph/0511591}{{\tt arXiv:astro-ph/0511591
  [astro-ph]}}.

\bibitem{Dodelson:2006zt}
S.~Dodelson and M.~Liguori, ``{Can Cosmic Structure form without Dark
  Matter?},'' \href{http://dx.doi.org/10.1103/PhysRevLett.97.231301}{{\em
  Phys.Rev.Lett.} {\bf 97} (2006)  231301},
\href{http://arxiv.org/abs/astro-ph/0608602}{{\tt arXiv:astro-ph/0608602
  [astro-ph]}}.

\bibitem{Knox:2005rg}
L.~Knox, Y.-S. Song, and J.~A. Tyson, ``{Distance-redshift and growth-redshift
  relations as two windows on acceleration and gravitation: Dark energy or new
  gravity?},'' \href{http://dx.doi.org/10.1103/PhysRevD.74.023512}{{\em
  Phys.Rev.} {\bf D74} (2006)  023512},
\href{http://arxiv.org/abs/astro-ph/0503644}{{\tt arXiv:astro-ph/0503644
  [astro-ph]}}.

\bibitem{Ishak:2005zs}
M.~Ishak, A.~Upadhye, and D.~N. Spergel, ``{Probing cosmic acceleration beyond
  the equation of state: Distinguishing between dark energy and modified
  gravity models},'' \href{http://dx.doi.org/10.1103/PhysRevD.74.043513}{{\em
  Phys.Rev.} {\bf D74} (2006)  043513},
\href{http://arxiv.org/abs/astro-ph/0507184}{{\tt arXiv:astro-ph/0507184
  [astro-ph]}}.

\bibitem{Koyama:2005kd}
K.~Koyama and R.~Maartens, ``{Structure formation in the dgp cosmological
  model},'' \href{http://dx.doi.org/10.1088/1475-7516/2006/01/016}{{\em JCAP}
  {\bf 0601} (2006)  016},
\href{http://arxiv.org/abs/astro-ph/0511634}{{\tt arXiv:astro-ph/0511634
  [astro-ph]}}.

\bibitem{Li:2006vi}
B.~Li and M.-C. Chu, ``{CMB and Matter Power Spectra of Early f(R) Cosmology in
  Palatini Formalism},''
  \href{http://dx.doi.org/10.1103/PhysRevD.74.104010}{{\em Phys.Rev.} {\bf D74}
  (2006)  104010},
\href{http://arxiv.org/abs/astro-ph/0610486}{{\tt arXiv:astro-ph/0610486
  [astro-ph]}}.

\bibitem{Li:2007xn}
B.~Li and J.~D. Barrow, ``{The Cosmology of f(R) gravity in metric variational
  approach},'' \href{http://dx.doi.org/10.1103/PhysRevD.75.084010}{{\em
  Phys.Rev.} {\bf D75} (2007)  084010},
\href{http://arxiv.org/abs/gr-qc/0701111}{{\tt arXiv:gr-qc/0701111 [gr-qc]}}.

\bibitem{Linder:2005in}
E.~V. Linder, ``{Cosmic growth history and expansion history},''
  \href{http://dx.doi.org/10.1103/PhysRevD.72.043529}{{\em Phys.Rev.} {\bf D72}
  (2005)  043529},
\href{http://arxiv.org/abs/astro-ph/0507263}{{\tt arXiv:astro-ph/0507263
  [astro-ph]}}.

\bibitem{Huterer:2006mva}
D.~Huterer and E.~V. Linder, ``{Separating Dark Physics from Physical Darkness:
  Minimalist Modified Gravity vs. Dark Energy},''
  \href{http://dx.doi.org/10.1103/PhysRevD.75.023519}{{\em Phys.Rev.} {\bf D75}
  (2007)  023519},
\href{http://arxiv.org/abs/astro-ph/0608681}{{\tt arXiv:astro-ph/0608681
  [astro-ph]}}.

\bibitem{Uzan:2006mf}
J.-P. Uzan, ``{The acceleration of the universe and the physics behind it},''
  \href{http://dx.doi.org/10.1007/s10714-006-0385-z}{{\em Gen.Rel.Grav.} {\bf
  39} (2007)  307--342},
\href{http://arxiv.org/abs/astro-ph/0605313}{{\tt arXiv:astro-ph/0605313
  [astro-ph]}}.

\bibitem{Caldwell:2007cw}
R.~Caldwell, A.~Cooray, and A.~Melchiorri, ``{Constraints on a New Post-General
  Relativity Cosmological Parameter},''
  \href{http://dx.doi.org/10.1103/PhysRevD.76.023507}{{\em Phys.Rev.} {\bf D76}
  (2007)  023507},
\href{http://arxiv.org/abs/astro-ph/0703375}{{\tt arXiv:astro-ph/0703375
  [ASTRO-PH]}}.

\bibitem{Amendola:2007rr}
L.~Amendola, M.~Kunz, and D.~Sapone, ``{Measuring the dark side (with weak
  lensing)},'' \href{http://dx.doi.org/10.1088/1475-7516/2008/04/013}{{\em
  JCAP} {\bf 0804} (2008)  013},
\href{http://arxiv.org/abs/0704.2421}{{\tt arXiv:0704.2421 [astro-ph]}}.

\bibitem{Baker:2012zs}
T.~Baker, P.~G. Ferreira, and C.~Skordis, ``{The Parameterized Post-Friedmann
  Framework for Theories of Modified Gravity: Concepts, Formalism and
  Examples},'' \href{http://dx.doi.org/10.1103/PhysRevD.87.024015}{{\em
  Phys.Rev.} {\bf D87} (2013)  024015},
\href{http://arxiv.org/abs/1209.2117}{{\tt arXiv:1209.2117 [astro-ph.CO]}}.

\bibitem{Bardeen:1980kt}
J.~M. Bardeen, ``{Gauge Invariant Cosmological Perturbations},''
\href{http://dx.doi.org/10.1103/PhysRevD.22.1882}{{\em Phys.Rev.} {\bf D22}
  (1980)  1882--1905}.

\bibitem{Mukhanov:1990me}
V.~F. Mukhanov, H.~Feldman, and R.~H. Brandenberger, ``{Theory of cosmological
  perturbations. Part 1. Classical perturbations. Part 2. Quantum theory of
  perturbations. Part 3. Extensions},''
\href{http://dx.doi.org/10.1016/0370-1573(92)90044-Z}{{\em Phys.Rept.} {\bf
  215} (1992)  203--333}.

\bibitem{Ma:1995ey}
C.-P. Ma and E.~Bertschinger, ``{Cosmological perturbation theory in the
  synchronous and conformal Newtonian gauges},''
  \href{http://dx.doi.org/10.1086/176550}{{\em Astrophys.J.} {\bf 455} (1995)
  7--25},
\href{http://arxiv.org/abs/astro-ph/9506072}{{\tt arXiv:astro-ph/9506072
  [astro-ph]}}.

\bibitem{Dodelson:2003ft}
S.~Dodelson,
``{Modern cosmology},''.

\bibitem{Mukhanov:2005sc}
V.~Mukhanov,
``{Physical foundations of cosmology},''.

\bibitem{Baumann:2009ds}
D.~Baumann, ``{TASI Lectures on Inflation},''
\href{http://arxiv.org/abs/0907.5424}{{\tt arXiv:0907.5424 [hep-th]}}.

\bibitem{Bertschinger:2006aw}
E.~Bertschinger, ``{On the Growth of Perturbations as a Test of Dark Energy},''
  \href{http://dx.doi.org/10.1086/506021}{{\em Astrophys.J.} {\bf 648} (2006)
  797--806},
\href{http://arxiv.org/abs/astro-ph/0604485}{{\tt arXiv:astro-ph/0604485
  [astro-ph]}}.

\bibitem{Zhao:2008bn}
G.-B. Zhao, L.~Pogosian, A.~Silvestri, and J.~Zylberberg, ``{Searching for
  modified growth patterns with tomographic surveys},''
  \href{http://dx.doi.org/10.1103/PhysRevD.79.083513}{{\em Phys.Rev.} {\bf D79}
  (2009)  083513},
\href{http://arxiv.org/abs/0809.3791}{{\tt arXiv:0809.3791 [astro-ph]}}.

\bibitem{Silvestri:2013ne}
A.~Silvestri, L.~Pogosian, and R.~V. Buniy, ``{A practical approach to
  cosmological perturbations in modified gravity},''
  \href{http://dx.doi.org/10.1103/PhysRevD.87.104015}{{\em Phys.Rev.} {\bf D87}
  (2013)  104015},
\href{http://arxiv.org/abs/1302.1193}{{\tt arXiv:1302.1193 [astro-ph.CO]}}.

\bibitem{Barreira:2014zza}
A.~Barreira, B.~Li, W.~A. Hellwing, L.~Lombriser, C.~M. Baugh, {\em et al.},
  ``{Halo model and halo properties in Galileon gravity cosmologies},''
\href{http://arxiv.org/abs/1401.1497}{{\tt arXiv:1401.1497 [astro-ph.CO]}}.

\bibitem{Carroll:2004st}
S.~M. Carroll,
``{Spacetime and geometry: An introduction to general relativity},''.

\bibitem{Jain:2007yk}
B.~Jain and P.~Zhang, ``{Observational Tests of Modified Gravity},''
  \href{http://dx.doi.org/10.1103/PhysRevD.78.063503}{{\em Phys.Rev.} {\bf D78}
  (2008)  063503},
\href{http://arxiv.org/abs/0709.2375}{{\tt arXiv:0709.2375 [astro-ph]}}.

\bibitem{Guzik:2009cm}
J.~Guzik, B.~Jain, and M.~Takada, ``{Tests of Gravity from Imaging and
  Spectroscopic Surveys},''
  \href{http://dx.doi.org/10.1103/PhysRevD.81.023503}{{\em Phys.Rev.} {\bf D81}
  (2010)  023503},
\href{http://arxiv.org/abs/0906.2221}{{\tt arXiv:0906.2221 [astro-ph.CO]}}.

\bibitem{Limber1953}
D.~N. Limber, ``{The Analysis of Counts of the Extragalactic Nebulae in Terms
  of a Fluctuating Density Field},''
  \href{http://dx.doi.org/10.1086/145672}{{\em Astrophys.J.} {\bf 117} (1953)
  134}.

\bibitem{Hu:1999ek}
W.~Hu, ``{Power spectrum tomography with weak lensing},''
  \href{http://dx.doi.org/10.1086/312210}{{\em Astrophys.J.} {\bf 522} (1999)
  L21--L24},
\href{http://arxiv.org/abs/astro-ph/9904153}{{\tt arXiv:astro-ph/9904153
  [astro-ph]}}.

\bibitem{Hoekstra:2008db}
H.~Hoekstra and B.~Jain, ``{Weak Gravitational Lensing and its Cosmological
  Applications},''
  \href{http://dx.doi.org/10.1146/annurev.nucl.58.110707.171151}{{\em
  Ann.Rev.Nucl.Part.Sci.} {\bf 58} (2008)  99--123},
\href{http://arxiv.org/abs/0805.0139}{{\tt arXiv:0805.0139 [astro-ph]}}.

\bibitem{Simpson:2012ra}
F.~Simpson, C.~Heymans, D.~Parkinson, C.~Blake, M.~Kilbinger, {\em et al.},
  ``{CFHTLenS: Testing the Laws of Gravity with Tomographic Weak Lensing and
  Redshift Space Distortions},''
\href{http://arxiv.org/abs/1212.3339}{{\tt arXiv:1212.3339 [astro-ph.CO]}}.

\bibitem{Bruneton:2007si}
J.-P. Bruneton and G.~Esposito-Farese, ``{Field-theoretical formulations of
  MOND-like gravity},'' \href{http://dx.doi.org/10.1103/PhysRevD.76.129902,
  10.1103/PhysRevD.76.124012}{{\em Phys.Rev.} {\bf D76} (2007)  124012},
\href{http://arxiv.org/abs/0705.4043}{{\tt arXiv:0705.4043 [gr-qc]}}.

\bibitem{Heavens:2007ka}
A.~F. Heavens, T.~Kitching, and L.~Verde, ``{On model selection forecasting,
  Dark Energy and modified gravity},''
  \href{http://dx.doi.org/10.1111/j.1365-2966.2007.12134.x}{{\em
  Mon.Not.Roy.Astron.Soc.} {\bf 380} (2007)  1029--1035},
\href{http://arxiv.org/abs/astro-ph/0703191}{{\tt arXiv:astro-ph/0703191
  [astro-ph]}}.

\bibitem{Sachs:1967er}
R.~Sachs and A.~Wolfe, ``{Perturbations of a cosmological model and angular
  variations of the microwave background},''
\href{http://dx.doi.org/10.1007/s10714-007-0448-9}{{\em Astrophys.J.} {\bf 147}
  (1967)  73--90}.

\bibitem{Lewis:2006fu}
A.~Lewis and A.~Challinor, ``{Weak gravitational lensing of the cmb},''
  \href{http://dx.doi.org/10.1016/j.physrep.2006.03.002}{{\em Phys.Rept.} {\bf
  429} (2006)  1--65},
\href{http://arxiv.org/abs/astro-ph/0601594}{{\tt arXiv:astro-ph/0601594
  [astro-ph]}}.

\bibitem{Cabre:2006qm}
A.~Cabre, E.~Gaztanaga, M.~Manera, P.~Fosalba, and F.~Castander,
  ``{Cross-correlation of wmap 3rd year and the sdss dr4 galaxy survey: new
  evidence for dark energy},''
  \href{http://dx.doi.org/10.1111/j.1745-3933.2006.00218.x}{{\em
  Mon.Not.Roy.Astron.Soc.} {\bf 372} (2006)  L23--L27},
\href{http://arxiv.org/abs/astro-ph/0603690}{{\tt arXiv:astro-ph/0603690
  [astro-ph]}}.

\bibitem{Pietrobon:2006gh}
D.~Pietrobon, A.~Balbi, and D.~Marinucci, ``{Integrated Sachs-Wolfe effect from
  the cross-correlation of WMAP 3 year and NVSS: new results and constraints on
  dark energy},'' \href{http://dx.doi.org/10.1103/PhysRevD.74.043524}{{\em
  Phys.Rev.} {\bf D74} (2006)  043524},
\href{http://arxiv.org/abs/astro-ph/0606475}{{\tt arXiv:astro-ph/0606475
  [astro-ph]}}.

\bibitem{Giannantonio:2006du}
T.~Giannantonio, R.~G. Crittenden, R.~C. Nichol, R.~Scranton, G.~T. Richards,
  {\em et al.}, ``{A high redshift detection of the integrated Sachs-Wolfe
  effect},'' \href{http://dx.doi.org/10.1103/PhysRevD.74.063520}{{\em
  Phys.Rev.} {\bf D74} (2006)  063520},
\href{http://arxiv.org/abs/astro-ph/0607572}{{\tt arXiv:astro-ph/0607572
  [astro-ph]}}.

\bibitem{Ho:2008bz}
S.~Ho, C.~Hirata, N.~Padmanabhan, U.~Seljak, and N.~Bahcall, ``{Correlation of
  CMB with large-scale structure: I. ISW Tomography and Cosmological
  Implications},'' \href{http://dx.doi.org/10.1103/PhysRevD.78.043519}{{\em
  Phys.Rev.} {\bf D78} (2008)  043519},
\href{http://arxiv.org/abs/0801.0642}{{\tt arXiv:0801.0642 [astro-ph]}}.

\bibitem{vanEngelen:2012va}
A.~van Engelen, R.~Keisler, O.~Zahn, K.~Aird, B.~Benson, {\em et al.}, ``{A
  measurement of gravitational lensing of the microwave background using South
  Pole Telescope data},''
  \href{http://dx.doi.org/10.1088/0004-637X/756/2/142}{{\em Astrophys.J.} {\bf
  756} (2012)  142},
\href{http://arxiv.org/abs/1202.0546}{{\tt arXiv:1202.0546 [astro-ph.CO]}}.

\bibitem{Das:2011ak}
S.~Das, B.~D. Sherwin, P.~Aguirre, J.~W. Appel, J.~R. Bond, {\em et al.},
  ``{Detection of the Power Spectrum of Cosmic Microwave Background Lensing by
  the Atacama Cosmology Telescope},''
  \href{http://dx.doi.org/10.1103/PhysRevLett.107.021301}{{\em Phys.Rev.Lett.}
  {\bf 107} (2011)  021301},
\href{http://arxiv.org/abs/1103.2124}{{\tt arXiv:1103.2124 [astro-ph.CO]}}.

\bibitem{Ade:2013tyw}
{\bf Planck} Collaboration, P.~Ade {\em et al.}, ``{Planck 2013 results. XVII.
  Gravitational lensing by large-scale structure},''
\href{http://arxiv.org/abs/1303.5077}{{\tt arXiv:1303.5077 [astro-ph.CO]}}.

\bibitem{Munshi:2014tua}
D.~Munshi, B.~Hu, A.~Renzi, A.~Heavens, and P.~Coles, ``{Probing Modified
  Gravity Theories with ISW and CMB Lensing},''
\href{http://arxiv.org/abs/1403.0852}{{\tt arXiv:1403.0852 [astro-ph.CO]}}.

\bibitem{Namikawa:2014xga}
T.~Namikawa, ``{Cosmology from weak lensing of CMB},''
\href{http://arxiv.org/abs/1403.3569}{{\tt arXiv:1403.3569 [astro-ph.CO]}}.

\bibitem{Macaulay:2013swa}
E.~Macaulay, I.~K. Wehus, and H.~K. Eriksen, ``{A Lower Growth Rate from Recent
  Redshift Space Distortions than Expected from Planck},''
\href{http://arxiv.org/abs/1303.6583}{{\tt arXiv:1303.6583 [astro-ph.CO]}}.

\bibitem{Kaiser:1987qv}
N.~Kaiser, ``{Clustering in real space and in redshift space},''
{\em Mon.Not.Roy.Astron.Soc.} {\bf 227} (1987)  1--27.

\bibitem{Hamilton1998}
A.~J.~S. Hamilton, ``{The Evolving Universe},'' {\em Astrophysics and Space
  Science Library Series} {\bf 231} (1998)  185.

\bibitem{Scoccimarro:2004tg}
R.~Scoccimarro, ``{Redshift-space distortions, pairwise velocities and
  nonlinearities},'' \href{http://dx.doi.org/10.1103/PhysRevD.70.083007}{{\em
  Phys.Rev.} {\bf D70} (2004)  083007},
\href{http://arxiv.org/abs/astro-ph/0407214}{{\tt arXiv:astro-ph/0407214
  [astro-ph]}}.

\bibitem{Beutler:2013yhm}
{\bf BOSS} Collaboration, F.~Beutler {\em et al.}, ``{The clustering of
  galaxies in the SDSS-III Baryon Oscillation Spectroscopic Survey: Testing
  gravity with redshift-space distortions using the power spectrum
  multipoles},''
\href{http://arxiv.org/abs/1312.4611}{{\tt arXiv:1312.4611 [astro-ph.CO]}}.

\bibitem{Sanchez:2013tga}
A.~G. Sanchez, F.~Montesano, E.~A. Kazin, E.~Aubourg, F.~Beutler, {\em et al.},
  ``{The clustering of galaxies in the SDSS-III Baryon Oscillation
  Spectroscopic Survey: cosmological implications of the full shape of the
  clustering wedges},'' \href{http://dx.doi.org/10.1093/mnras/stt799}{{\em
  Mon.Not.Roy.Astron.Soc.} {\bf 433} (2013)  1202--1222},
\href{http://arxiv.org/abs/1312.4854}{{\tt arXiv:1312.4854 [astro-ph.CO]}}.

\bibitem{Samushia:2013yga}
L.~Samushia, B.~A. Reid, M.~White, W.~J. Percival, A.~J. Cuesta, {\em et al.},
  ``{The Clustering of Galaxies in the SDSS-III Baryon Oscillation
  Spectroscopic Survey (BOSS): measuring growth rate and geometry with
  anisotropic clustering},''
\href{http://arxiv.org/abs/1312.4899}{{\tt arXiv:1312.4899 [astro-ph.CO]}}.

\bibitem{Chuang:2013wga}
C.-H. Chuang, F.~Prada, F.~Beutler, D.~J. Eisenstein, S.~Escoffier, {\em et
  al.}, ``{The clustering of galaxies in the SDSS-III Baryon Oscillation
  Spectroscopic Survey: single-probe measurements from CMASS and LOWZ
  anisotropic galaxy clustering},''
\href{http://arxiv.org/abs/1312.4889}{{\tt arXiv:1312.4889 [astro-ph.CO]}}.

\bibitem{Hojjati:2013xqa}
A.~Hojjati, L.~Pogosian, A.~Silvestri, and G.-B. Zhao, ``{Observable physical
  modes of modified gravity},''
\href{http://arxiv.org/abs/1312.5309}{{\tt arXiv:1312.5309 [astro-ph.CO]}}.

\bibitem{Bellini:2014fua}
E.~Bellini and I.~Sawicki, ``{Maximal freedom at minimum cost: linear
  large-scale structure in general modifications of gravity},''
\href{http://arxiv.org/abs/1404.3713}{{\tt arXiv:1404.3713 [astro-ph.CO]}}.

\bibitem{Huterer:2002hy}
D.~Huterer and G.~Starkman, ``{Parameterization of dark-energy properties: A
  Principal-component approach},''
  \href{http://dx.doi.org/10.1103/PhysRevLett.90.031301}{{\em Phys.Rev.Lett.}
  {\bf 90} (2003)  031301},
\href{http://arxiv.org/abs/astro-ph/0207517}{{\tt arXiv:astro-ph/0207517
  [astro-ph]}}.

\bibitem{Zhao:2009fn}
G.-B. Zhao, L.~Pogosian, A.~Silvestri, and J.~Zylberberg, ``{Cosmological Tests
  of General Relativity with Future Tomographic Surveys},''
  \href{http://dx.doi.org/10.1103/PhysRevLett.103.241301}{{\em Phys.Rev.Lett.}
  {\bf 103} (2009)  241301},
\href{http://arxiv.org/abs/0905.1326}{{\tt arXiv:0905.1326 [astro-ph.CO]}}.

\bibitem{Hojjati:2011xd}
A.~Hojjati, G.-B. Zhao, L.~Pogosian, A.~Silvestri, R.~Crittenden, {\em et al.},
  ``{Cosmological tests of General Relativity: a principal component
  analysis},'' \href{http://dx.doi.org/10.1103/PhysRevD.85.043508}{{\em
  Phys.Rev.} {\bf D85} (2012)  043508},
\href{http://arxiv.org/abs/1111.3960}{{\tt arXiv:1111.3960 [astro-ph.CO]}}.

\bibitem{Asaba:2013xql}
S.~Asaba, C.~Hikage, K.~Koyama, G.-B. Zhao, A.~Hojjati, {\em et al.},
  ``{Principal Component Analysis of Modified Gravity using Weak Lensing and
  Peculiar Velocity Measurements},''
  \href{http://dx.doi.org/10.1088/1475-7516/2013/08/029}{{\em JCAP} {\bf 1308}
  (2013)  029},
\href{http://arxiv.org/abs/1306.2546}{{\tt arXiv:1306.2546 [astro-ph.CO]}}.

\bibitem{Baker:2013hia}
T.~Baker, P.~G. Ferreira, and C.~Skordis, ``{A Fast Route to Modified
  Gravitational Growth},''
  \href{http://dx.doi.org/10.1103/PhysRevD.89.024026}{{\em Phys.Rev.} {\bf D89}
  (2014)  024026},
\href{http://arxiv.org/abs/1310.1086}{{\tt arXiv:1310.1086 [astro-ph.CO]}}.

\bibitem{Battye:2013aaa}
R.~A. Battye and J.~A. Pearson, ``{Parametrizing dark sector perturbations via
  equations of state},''
  \href{http://dx.doi.org/10.1103/PhysRevD.88.061301}{{\em Phys.Rev.} {\bf D88}
  (2013) no.~6, 061301},
\href{http://arxiv.org/abs/1306.1175}{{\tt arXiv:1306.1175 [astro-ph.CO]}}.

\bibitem{Battye:2013ida}
R.~A. Battye and J.~A. Pearson, ``{Computing model independent perturbations in
  dark energy and modified gravity},''
  \href{http://dx.doi.org/10.1088/1475-7516/2014/03/051}{{\em JCAP} {\bf 1403}
  (2014)  051},
\href{http://arxiv.org/abs/1311.6737}{{\tt arXiv:1311.6737 [astro-ph.CO]}}.

\bibitem{Hu:2008zd}
W.~Hu, ``{Parametrized Post-Friedmann Signatures of Acceleration in the CMB},''
  \href{http://dx.doi.org/10.1103/PhysRevD.77.103524}{{\em Phys.Rev.} {\bf D77}
  (2008)  103524},
\href{http://arxiv.org/abs/0801.2433}{{\tt arXiv:0801.2433 [astro-ph]}}.

\bibitem{Fang:2008sn}
W.~Fang, W.~Hu, and A.~Lewis, ``{Crossing the Phantom Divide with Parameterized
  Post-Friedmann Dark Energy},''
  \href{http://dx.doi.org/10.1103/PhysRevD.78.087303}{{\em Phys.Rev.} {\bf D78}
  (2008)  087303},
\href{http://arxiv.org/abs/0808.3125}{{\tt arXiv:0808.3125 [astro-ph]}}.

\bibitem{Hu:2013twa}
B.~Hu, M.~Raveri, N.~Frusciante, and A.~Silvestri, ``{Effective Field Theory of
  Cosmic Acceleration: an implementation in CAMB},''
  \href{http://dx.doi.org/10.1103/PhysRevD.89.103530}{{\em Phys.Rev.} {\bf D89}
  (2014)  103530},
\href{http://arxiv.org/abs/1312.5742}{{\tt arXiv:1312.5742 [astro-ph.CO]}}.

\bibitem{Zhang:2007nk}
P.~Zhang, M.~Liguori, R.~Bean, and S.~Dodelson, ``{Probing Gravity at
  Cosmological Scales by Measurements which Test the Relationship between
  Gravitational Lensing and Matter Overdensity},''
  \href{http://dx.doi.org/10.1103/PhysRevLett.99.141302}{{\em Phys.Rev.Lett.}
  {\bf 99} (2007)  141302},
\href{http://arxiv.org/abs/0704.1932}{{\tt arXiv:0704.1932 [astro-ph]}}.

\bibitem{Acquaviva:2008qp}
V.~Acquaviva, A.~Hajian, D.~N. Spergel, and S.~Das, ``{Next Generation Redshift
  Surveys and the Origin of Cosmic Acceleration},''
  \href{http://dx.doi.org/10.1103/PhysRevD.78.043514}{{\em Phys.Rev.} {\bf D78}
  (2008)  043514},
\href{http://arxiv.org/abs/0803.2236}{{\tt arXiv:0803.2236 [astro-ph]}}.

\bibitem{Song:2008xd}
Y.-S. Song and O.~Dore, ``{A step towards testing general relativity using weak
  gravitational lensing and redshift surveys},''
  \href{http://dx.doi.org/10.1088/1475-7516/2009/03/025}{{\em JCAP} {\bf 0903}
  (2009)  025},
\href{http://arxiv.org/abs/0812.0002}{{\tt arXiv:0812.0002 [astro-ph]}}.

\bibitem{Reyes:2010tr}
R.~Reyes, R.~Mandelbaum, U.~Seljak, T.~Baldauf, J.~E. Gunn, {\em et al.},
  ``{Confirmation of general relativity on large scales from weak lensing and
  galaxy velocities},'' \href{http://dx.doi.org/10.1038/nature08857}{{\em
  Nature} {\bf 464} (2010)  256--258},
\href{http://arxiv.org/abs/1003.2185}{{\tt arXiv:1003.2185 [astro-ph.CO]}}.

\bibitem{Bolton:2005nf}
A.~S. Bolton, S.~Burles, L.~V. Koopmans, T.~Treu, and L.~A. Moustakas, ``{The
  sloan lens acs survey. 1. a large spectroscopically selected sample of
  massive early-type lens galaxies},''
  \href{http://dx.doi.org/10.1086/498884}{{\em Astrophys.J.} {\bf 638} (2006)
  703--724},
\href{http://arxiv.org/abs/astro-ph/0511453}{{\tt arXiv:astro-ph/0511453
  [astro-ph]}}.

\bibitem{Schmidt:2010jr}
F.~Schmidt, ``{Dynamical Masses in Modified Gravity},''
  \href{http://dx.doi.org/10.1103/PhysRevD.81.103002}{{\em Phys.Rev.} {\bf D81}
  (2010)  103002},
\href{http://arxiv.org/abs/1003.0409}{{\tt arXiv:1003.0409 [astro-ph.CO]}}.

\bibitem{Bolton:2006yz}
A.~S. Bolton, S.~Rappaport, and S.~Burles, ``{Constraint on the Post-Newtonian
  Parameter gamma on Galactic Size Scales},''
  \href{http://dx.doi.org/10.1103/PhysRevD.74.061501}{{\em Phys.Rev.} {\bf D74}
  (2006)  061501},
\href{http://arxiv.org/abs/astro-ph/0607657}{{\tt arXiv:astro-ph/0607657
  [astro-ph]}}.

\bibitem{Schwab:2009nz}
J.~Schwab, A.~S. Bolton, and S.~A. Rappaport, ``{Galaxy-Scale Strong Lensing
  Tests of Gravity and Geometric Cosmology: Constraints and Systematic
  Limitations},'' \href{http://dx.doi.org/10.1088/0004-637X/708/1/750}{{\em
  Astrophys.J.} {\bf 708} (2010)  750--757},
\href{http://arxiv.org/abs/0907.4992}{{\tt arXiv:0907.4992 [astro-ph.CO]}}.

\bibitem{Schmidt:2009sg}
F.~Schmidt, ``{Self-Consistent Cosmological Simulations of DGP Braneworld
  Gravity},'' \href{http://dx.doi.org/10.1103/PhysRevD.80.043001}{{\em
  Phys.Rev.} {\bf D80} (2009)  043001},
\href{http://arxiv.org/abs/0905.0858}{{\tt arXiv:0905.0858 [astro-ph.CO]}}.

\bibitem{Schmidt:2009yj}
F.~Schmidt, W.~Hu, and M.~Lima, ``{Spherical Collapse and the Halo Model in
  Braneworld Gravity},''
  \href{http://dx.doi.org/10.1103/PhysRevD.81.063005}{{\em Phys.Rev.} {\bf D81}
  (2010)  063005},
\href{http://arxiv.org/abs/0911.5178}{{\tt arXiv:0911.5178 [astro-ph.CO]}}.

\bibitem{Scoccimarro:2009eu}
R.~Scoccimarro, ``{Large-Scale Structure in Brane-Induced Gravity I.
  Perturbation Theory},''
  \href{http://dx.doi.org/10.1103/PhysRevD.80.104006}{{\em Phys.Rev.} {\bf D80}
  (2009)  104006},
\href{http://arxiv.org/abs/0906.4545}{{\tt arXiv:0906.4545 [astro-ph.CO]}}.

\bibitem{Martino:2008ae}
M.~C. Martino, H.~F. Stabenau, and R.~K. Sheth, ``{Spherical Collapse and
  Modified Gravity},'' \href{http://dx.doi.org/10.1103/PhysRevD.79.084013}{{\em
  Phys.Rev.} {\bf D79} (2009)  084013},
\href{http://arxiv.org/abs/0812.0200}{{\tt arXiv:0812.0200 [astro-ph]}}.

\bibitem{Huterer:2013xky}
D.~Huterer, D.~Kirkby, R.~Bean, A.~Connolly, K.~Dawson, {\em et al.}, ``{Growth
  of Cosmic Structure: Probing Dark Energy Beyond Expansion},''
\href{http://arxiv.org/abs/1309.5385}{{\tt arXiv:1309.5385 [astro-ph.CO]}}.

\bibitem{Lombriser:2010mp}
L.~Lombriser, A.~Slosar, U.~Seljak, and W.~Hu, ``{Constraints on f(R) gravity
  from probing the large-scale structure},''
  \href{http://dx.doi.org/10.1103/PhysRevD.85.124038}{{\em Phys.Rev.} {\bf D85}
  (2012)  124038},
\href{http://arxiv.org/abs/1003.3009}{{\tt arXiv:1003.3009 [astro-ph.CO]}}.

\bibitem{deUrries:1998bi}
F.~de~Urries and J.~Julve, ``{Ostrogradski formalism for higher derivative
  scalar field theories},''
  \href{http://dx.doi.org/10.1088/0305-4470/31/33/006}{{\em J.Phys.} {\bf A31}
  (1998)  6949--6964},
\href{http://arxiv.org/abs/hep-th/9802115}{{\tt arXiv:hep-th/9802115
  [hep-th]}}.

\bibitem{deUrries:1995ty}
F.~de~Urries and J.~Julve, ``{Degrees of freedom of arbitrarily higher
  derivative field theories},''
\href{http://arxiv.org/abs/gr-qc/9506009}{{\tt arXiv:gr-qc/9506009 [gr-qc]}}.

\bibitem{Pais:1950za}
A.~Pais and G.~Uhlenbeck, ``{On Field theories with nonlocalized action},''
\href{http://dx.doi.org/10.1103/PhysRev.79.145}{{\em Phys.Rev.} {\bf 79} (1950)
   145--165}.

\bibitem{Mannheim:2004qz}
P.~D. Mannheim and A.~Davidson, ``{Dirac quantization of the Pais-Uhlenbeck
  fourth order oscillator},''
  \href{http://dx.doi.org/10.1103/PhysRevA.71.042110}{{\em Phys.Rev.} {\bf A71}
  (2005)  042110},
\href{http://arxiv.org/abs/hep-th/0408104}{{\tt arXiv:hep-th/0408104
  [hep-th]}}.

\bibitem{Bender:2007wu}
C.~M. Bender and P.~D. Mannheim, ``{No-ghost theorem for the fourth-order
  derivative Pais-Uhlenbeck oscillator model},''
  \href{http://dx.doi.org/10.1103/PhysRevLett.100.110402}{{\em Phys.Rev.Lett.}
  {\bf 100} (2008)  110402},
\href{http://arxiv.org/abs/0706.0207}{{\tt arXiv:0706.0207 [hep-th]}}.

\bibitem{Chen:2012au}
T.-j. Chen, M.~Fasiello, E.~A. Lim, and A.~J. Tolley, ``{Higher derivative
  theories with constraints: Exorcising Ostrogradski's Ghost},''
  \href{http://dx.doi.org/10.1088/1475-7516/2013/02/042}{{\em JCAP} {\bf 1302}
  (2013)  042},
\href{http://arxiv.org/abs/1209.0583}{{\tt arXiv:1209.0583 [hep-th]}}.

\bibitem{Cline:2003gs}
J.~M. Cline, S.~Jeon, and G.~D. Moore, ``{The Phantom menaced: Constraints on
  low-energy effective ghosts},''
  \href{http://dx.doi.org/10.1103/PhysRevD.70.043543}{{\em Phys.Rev.} {\bf D70}
  (2004)  043543},
\href{http://arxiv.org/abs/hep-ph/0311312}{{\tt arXiv:hep-ph/0311312
  [hep-ph]}}.

\bibitem{Peskin:1995ev}
M.~E. Peskin and D.~V. Schroeder,
``{An Introduction to quantum field theory},''.

\bibitem{Dubovsky:2007ac}
S.~Dubovsky, A.~Nicolis, E.~Trincherini, and G.~Villadoro, ``{Microcausality in
  curved space-time},''
  \href{http://dx.doi.org/10.1103/PhysRevD.77.084016}{{\em Phys.Rev.} {\bf D77}
  (2008)  084016},
\href{http://arxiv.org/abs/0709.1483}{{\tt arXiv:0709.1483 [hep-th]}}.

\bibitem{Moore:2001bv}
G.~D. Moore and A.~E. Nelson, ``{Lower bound on the propagation speed of
  gravity from gravitational Cherenkov radiation},''
  \href{http://dx.doi.org/10.1088/1126-6708/2001/09/023}{{\em JHEP} {\bf 0109}
  (2001)  023},
\href{http://arxiv.org/abs/hep-ph/0106220}{{\tt arXiv:hep-ph/0106220
  [hep-ph]}}.

\bibitem{Wald:1984rg}
R.~M. Wald,
``{General Relativity},''.

\bibitem{Bruneton:2006gf}
J.-P. Bruneton, ``{On causality and superluminal behavior in classical field
  theories: Applications to k-essence theories and MOND-like theories of
  gravity},'' \href{http://dx.doi.org/10.1103/PhysRevD.75.085013}{{\em
  Phys.Rev.} {\bf D75} (2007)  085013},
\href{http://arxiv.org/abs/gr-qc/0607055}{{\tt arXiv:gr-qc/0607055 [gr-qc]}}.

\bibitem{Froissart:1961ux}
M.~Froissart, ``{Asymptotic behavior and subtractions in the Mandelstam
  representation},''
\href{http://dx.doi.org/10.1103/PhysRev.123.1053}{{\em Phys.Rev.} {\bf 123}
  (1961)  1053--1057}.

\bibitem{Martin:1962rt}
A.~Martin, ``{Unitarity and high-energy behavior of scattering amplitudes},''
\href{http://dx.doi.org/10.1103/PhysRev.129.1432}{{\em Phys.Rev.} {\bf 129}
  (1963)  1432--1436}.

\bibitem{Komargodski:2011xv}
Z.~Komargodski, ``{The Constraints of Conformal Symmetry on RG Flows},''
  \href{http://dx.doi.org/10.1007/JHEP07(2012)069}{{\em JHEP} {\bf 1207} (2012)
   069},
\href{http://arxiv.org/abs/1112.4538}{{\tt arXiv:1112.4538 [hep-th]}}.

\bibitem{Luty:2012ww}
M.~A. Luty, J.~Polchinski, and R.~Rattazzi, ``{The $a$-theorem and the
  Asymptotics of 4D Quantum Field Theory},''
  \href{http://dx.doi.org/10.1007/JHEP01(2013)152}{{\em JHEP} {\bf 1301} (2013)
   152},
\href{http://arxiv.org/abs/1204.5221}{{\tt arXiv:1204.5221 [hep-th]}}.

\bibitem{Elvang:2012st}
H.~Elvang, D.~Z. Freedman, L.-Y. Hung, M.~Kiermaier, R.~C. Myers, {\em et al.},
  ``{On renormalization group flows and the a-theorem in 6d},''
  \href{http://dx.doi.org/10.1007/JHEP10(2012)011}{{\em JHEP} {\bf 1210} (2012)
   011},
\href{http://arxiv.org/abs/1205.3994}{{\tt arXiv:1205.3994 [hep-th]}}.

\bibitem{Bellazzini:2014waa}
B.~Bellazzini, L.~Martucci, and R.~Torre, ``{Symmetries, Sum Rules and
  Constraints on Effective Field Theories},''
  \href{http://dx.doi.org/10.1007/JHEP09(2014)100}{{\em JHEP} {\bf 1409} (2014)
   100},
\href{http://arxiv.org/abs/1405.2960}{{\tt arXiv:1405.2960 [hep-th]}}.

\bibitem{Shore:2007um}
G.~Shore, ``{Superluminality and UV completion},''
  \href{http://dx.doi.org/10.1016/j.nuclphysb.2007.03.034}{{\em Nucl.Phys.}
  {\bf B778} (2007)  219--258},
\href{http://arxiv.org/abs/hep-th/0701185}{{\tt arXiv:hep-th/0701185
  [hep-th]}}.

\bibitem{fastlightslowlight}
P.~W. Milonni, ``{Fast Light, Slow Light and Left-Handed Light},''.

\end{thebibliography}\endgroup
\end{document}